\documentclass[12pt]{JHEP3}
\newcommand{\intd}{\mathrm{d}}
\newcommand{\ex}{\mathrm{e}}
\newcommand {\non}{\nonumber}

\usepackage{amsmath,epsfig}
\usepackage{placeins}
\usepackage{amssymb,amsfonts}
\usepackage{latexsym}
\usepackage{slashed}
\usepackage{empheq}
\numberwithin{equation}{section}
\usepackage{dsfont}
\usepackage{cancel}
\usepackage{overpic}
\usepackage{subcaption}
\usepackage{subeqnarray}
\usepackage{diagbox}
\usepackage{xcolor}
\let\normalcolor\relax
\usepackage{graphicx}
\usepackage{amsmath}
\usepackage{longtable}
\usepackage{color}
\usepackage{bbm}

\usepackage{longtable}
\usepackage{caption}

\usepackage[force]{feynmp-auto}

\allowdisplaybreaks

\usepackage[multiple]{footmisc}

\usepackage{booktabs}

\newcommand{\exclude}[1]{}
\def\nn{\nonumber}

\def\d{\mathrm{d}}

\def\a#1{\alpha_{#1}}

\def\beq{\begin{equation}}
\def\eeq{\end{equation}}
\def\be{\begin{equation}}
\def\ee{\end{equation}}
\def\bea{\begin{eqnarray}}
\def\eea{\end{eqnarray}}
\def\bal{\begin{align}}
\def\eal{\end{align}}

\def\Im{\textrm{Im}}
\def\Re{\textrm{Re}}

\def\2b2[#1,#2][#3,#4]{\left( \begin{array}{cc} #1 & #2 \\ #3 & #4 \end{array}
\right)}
\def\3b3[#1,#2,#3][#4,#5,#6][#7,#8,#9]{\left( \begin{array}{ccc} #1 & #2 #3 \\
#4 & #5 & #6\\#7&#8&#9\end{array} \right)}

\newcommand\fverb{\setbox\pippobox=\hbox\bgroup\verb}
\newcommand\fverbdo{\egroup\medskip\noindent%
                        \fbox{\unhbox\pippobox}\ }
\newcommand\fverbit{\egroup\item[\fbox{\unhbox\pippobox}]}

\usepackage{float}

\newcommand{\bear}{\begin{eqnarray}}
\usepackage{comment}

\newcommand{\eear}{\end{eqnarray}}

\newcommand{\bsea}{\begin{subeqnarray}}
\newcommand{\esea}{\end{subeqnarray}}
\newbox\pippobox

\def\d{\delta}

\def\6{\partial}
\def\a{\alpha}
\def\nn{\nonumber}

\def\pa{\partial}

\def\e{\epsilon}
\def\m{\mu}
\def\n{\nu}

\def\s{\sigma}

\def\sp{\;\;\;,\;\;\;}

\def\sq
\def\a{\alpha}

\def\l{\lambda}

\def\hri#1#2{\href{http://arxiv.org/abs/#1}{[ArXiv:#1]#2}}
\def\hre#1#2{\href{http://arxiv.org/abs/#1/#2}{[ArXiv:#1/#2]}}
\def\hrj#1#2{\href{https://doi.org/#1}{#2}}

\def\e{\epsilon}

\def\d{\delta}

\def\D{\Delta}

\def\AA{{\cal A}}

\def\CC{{\cal C}}

\def\GG{{\cal G}}

\def\NN{{\cal N}}
\def\OO{{\cal O}}
\def\PP{{\cal P}}

\def\TT{{\cal T}}

\title{Flavour current correlators and the non-Abelian hydrodynamic approximation: the charged sector}
\author{
 T. Apostolidis$^\natural$, M. J\"arvinen$^*$, E. Kiritsis$^{\natural,\flat}$,  F. Nitti$^\natural$, A. Olzi$^{\natural}$, E. Pr\'eau$^\ddagger$
~\\
$^\natural$ \href{http://www.apc.univ-paris7.fr}{Universit\'e  Paris Cit\'e, CNRS, Astroparticule et Cosmologie,  F-75006 Paris, France}\\
$^*$ \href{http://www.itp.cas.cn}{Institute of Theoretical Physics, Chinese Academy of Sciences}, Beijing 100190, China \\
$^\flat$ \href{http://hep.physics.uoc.gr}{Crete Center for Theoretical Physics}, Institute for Theoretical and Computational Physics,
Department of Physics,\\
University of Crete, 70013, Heraklion, Greece\\
$\ddagger$ Institute for Theoretical Physics, Utrecht University, 3584 CC Utrecht, The Netherlands
}

\preprint{
CCTP-2026-13\\
ITCP-2026/13}

\abstract{ Flavor-current correlators are studied in strongly-coupled dense (holographic) matter, at finite quark chemical potential $\mu_q$ and finite isospin asymmetry. The non-Abelian hydrodynamic description of the charged currents is derived in the presence of an isospin chemical potential $\mu_3$.
The two-point correlators of charged currents are then computed  holographically at finite quark and isospin chemical potentials. In the near-extremal hydrodynamic regime, $\omega,k,T,\mu_3\ll\mu \equiv\sqrt{\mu_q^2+\mu_3^2}$, relevant for cold strongly coupled matter, the IR properties of the correlators are studied. It is shown that in this regime, the correlators agree with the non-Abelian hydrodynamic predictions. Therefore, the traditional regime of validity of standard hydrodynamics extends beyond $\omega,k\ll T\ll\mu$ to the so-called extended hydrodynamic regime $T\ll\omega,k\ll \mu$. The holographic product formula is applied  to the present non-Abelian system, and is used to propose  an extended hydrodynamic approximation capturing both hydrodynamic-like poles and the leading effect of AdS$_2$ poles, by resumming the low-$\omega$ logarithms.  The results are verified through a detailed numerical analysis of the exact correlators and quasi-normal mode spectrum.
 }

\begin{document}
\maketitle

\section{Introduction}

Holographic studies of cold strongly-coupled systems indicate the emergence of a hydrodynamic-like behavior over long times and distances, which is valid up to a hard scale $\mu$, (in the present paper it is  a chemical potential), much larger than the temperature $T$ \cite{Edalati:2010hk,Edalati:2010pn,Edalati:2010pn2,Davison:2011uk,Brattan:2010pq,DP13,DP14,MST,neutrinopaper,Gouteraux:2025kta,Preau:2025rex}.
This is in contrast with standard thermodynamics that is typically valid up to scales of order the temperature, \cite{Kovtun}. 

This behavior is generic for all systems described by a near-extremal black-hole, which suggests that it could be shared by many strongly-coupled dense systems, also beyond holography. Such systems are all characterised by a near-horizon geometry isometric to the AdS$_2$  space. This type of hydrodynamics may be relevant for transport in neutron stars, which are filled with strongly-coupled QCD matter, with a temperature much smaller than the quark number chemical potential $\mu_q$. For example, the toy model analysis of \cite{neutrinopaper}, indicates that hydrodynamics may give a good approximation to neutrino radiative coefficients in neutron star conditions, well beyond its  traditional validity regime. A very similar analysis is performed in \cite{Hoyos:2024pkl}, where the same model was used to compute the weak reaction rates of cold quark matter which are relevant to determine the bulk viscosity in neutron stars.

In addition to hydrodynamic-like features, another interesting property of near-extremal holographic systems is the emergence of a lower-dimensional conformal description in the infra-red, and the associated non-trivial scaling behaviour of retarded Green's functions and transport coefficients \cite{LI08,Lee:2008xf,Liu:2009dm,Lee:2009epi,Faulkner:2009wj,Edalati:2010pn,Goldstein:2009cv,CGKKM,Iizuka:2011hg,Gouteraux:2011ce}. This emergent conformal behaviour is in apparent contradiction with the hydrodynamic features, since it implies, in particular, the presence of a branch cut at zero-temperature for retarded correlators \cite{Faulkner:2009wj,Gouteraux:2025kta}, whereas hydrodynamic correlators are meromorphic.

This paradox was resolved in \cite{MST,Gouteraux:2025kta}, where it was shown that the branch cut, although not visible at leading order in the near-extremal hydrodynamic expansion of the correlators, $\omega,k,T\ll\mu$,\footnote{In our notation, $\omega$ stands for the cyclic frequency while $k\equiv |\vec k|$ is the magnitude of the momentum vector.}  appears at higher order. Based on the holographic product formula \cite{Dodelson:2023vrw}, it was further shown in \cite{Preau:2025rex} that, in the limit $T\ll \omega,k\ll \mu$, the correlators can be written in a simple factorized form, where both the infra-red scaling behaviour and the hydrodynamic-like (or simply \textit{hydro-like}) poles appear explicitly.
On the quantitative side, numerical evaluation of two-point functions of currents in \cite{neutrinopaper} has indicated that indeed hydrodynamics provides reasonably good approximations to them in the extended hydrodynamic region
$T\ll \omega,k\ll \mu$, which is not part of the standard hydrodynamic regime $\omega,k\ll T\ll \mu$.

It should be stressed that, although at the linearized level we have a reasonably good understanding of the issues involved, the non-linear description of the dynamics across the near-extremal hydrodynamic region, in particular in the extended hydrodynamic regime $T\ll \omega,k\ll \mu$, is not settled yet. The standard procedure that yields standard hydrodynamics, was applied in this case, and the logarithmic  contributions were identified,
 \cite{MST}.
The presence of logarithms in $\omega$,  indicate non-local contributions and it is not yet clear how to package them into a non-linear system of integro-differential equations.
This is an important problem for both theoretical and practical reasons, as we explain below.

In the near horizon region, there are more phenomena at play, as was realized in the last ten years, due to studies on the SYK Model, \cite{Kitaev}-\cite{Malda-SYK-2}, and its gravity dual, JT Gravity, \cite{Jensen,Malda-SYK-2,Mertens}. This duality is relevant for AdS$_2$ physics and therefore, for any near extremal black hole in higher dimensions. It was found that at sufficiently small temperatures, $T\ll E_{gap}$, there are quantum gravitational corrections that are unsuppressed and affect non-trivially the semi-classical physics, \cite{Turiaci}. The relevant modes are known as Schwarzian modes, due to the fact that their action is given by the Schwarzian expression.  $E_{gap}$ is an energy scale that controls the linear in temperature behaviour of the entropy as $T\to 0$, and is inversely proportional to the (large) dual number of degrees of freedom.
Such quantum corrections affect non-trivially the AdS$_2$ correlators, \cite{Mertens}, and therefore can affect the low-energy near extremal (hydro)dynamics. The quantum corrected $\eta/s$ ratio  was computed in \cite{Kanargias}\footnote{See also \cite{Ge}-\cite{Gouteraux}.}  and shown to increase enormously as $T\ll E_{gap}$, giving credence to the expectation of a similarity between the low-temperature dynamics of the Reissner-Nordstr\"om (RN) black holes and that of glasses, \cite{Kurchan}\footnote{See also \cite{anninos} for a different realization of glassy dynamics in the context of near-extremal black holes.}. Moreover, the scaling behaviour of AdS$_{2}$ correlators in the extended hydrodynamic regime, $T\ll \omega,k\ll \mu$ was connected, \cite{Gralla}, to the Aretakis instability, \cite{Aretakis}.
Given the fact that the quantum effects erase the exponential fall-off, due to temperature of AdS$_2$ correlators, \cite{Moore} it is expected that the Aretakis behaviour will be enhanced due to the quantum effects even at non-zero (bur small enough)  temperature.

The issues discussed above, concerning near-extremal black holes, are relevant to the holographic description of cold holographic systems at finite density, and in particular to applications to the concrete problem of equilibrium and dynamical properties of neutron stars.
Since the appearance of the holographic correspondence, several holographic models have been proposed for QCD. Some are top-down models, with the more prominent one being the Witten-Sakai-Sugimoto model, \cite{Witten,SS}. There are also bottom up models for Yang-Mills (YM), \cite{I1}-\cite{I4}, and QCD, like V-QCD, \cite{V1}-\cite{V7} or simplified models involving a tachyon, \cite{tachyon1,tachyon2}. Such models have provided insights into several QCD phases. They have an important advantage
compared to otherwise reliable descriptions, like the chiral effective theory, or lattice QCD, in deconfined or partially-deconfined phases at finite density.

Returning to the physics of cold nuclear matter and neutron stars,
in addition to the quark (or equivalently baryon) number density, neutron star matter is also characterized by the relative abundance of neutrons compared to protons, which is quantified by the isospin density $n_3$ -- or equivalently, the isospin chemical potential $\mu_3$. Note that in general, neutron star matter composition at the highest densities may be characterized by several chemical potentials, both baryonic (like isospin, and possibly strangeness), and leptonic (like the muon number) \cite{HaenselBook}. Working with two chemical potentials $\mu_q$ and $\mu_3$ corresponds to the simplest model of neutron star composition.

Our aim in this work is to explore how the structure of the correlators is modified by the isospin asymmetry and if some kind of hydrodynamic theory well approximates the correlators in the near-extremal hydrodynamic regime. We shall consider for this analysis, the same holographic toy model as in \cite{neutrinopaper}, which corresponds to a dense deconfined matter\footnote{The strong (glue) dynamics of such matter in the absence of flavor is assumed to be scale invariant. This is at odds with YM, but at the level of a toy model, it has given valuable insights into strong dynamics, \cite{Gubser}.}, where apart from a quark chemical potential, $\mu_q$,  we also add an isospin chemical potential $\mu_3$\footnote{This chemical potential is known to affect non-trivially the phase structure of nuclear matter, \cite{IsoSon}.}. The flavor symmetry of this model is U(1)$_V\times$SU(2)$_V$ and corresponds to the unbroken flavour symmetry of QCD. Moreover there are no pions\footnote{All QCD ingredients appear in the V-QCD model, \cite{V1}-\cite{V7} which is however more involved. This is the reason that here we chose a simpler theory to do our analysis.}. The main object of interest will be the retarded correlator for the flavor currents $J^\mu_a$, which are dual to the bulk flavor gauge fields $A_\mu^a$ (with $\mu$ the space-time index and $a$ the flavor index). As explained in \cite{neutrinopaper}, these are the correlators which determine, in particular, how neutrinos are transported in the strongly-coupled medium.

The ground states of the holographic theory with U(1)$\times$SU(2) global symmetry at finite $T$ and $\mu_q,\mu_3$ has been investigated recently in \cite{JKNP}\footnote{The vacuum structure of a CFT with SU(2) global symmetry at finite density has been analysed earlier in \cite{G1,G2}. Superfluid states were found, but the true ground state was analysed only in the probe approximation.}.
It was found that for sufficiently small $\mu_3$, the dominant ground state is the RN black hole with total chemical potential $\mu=\sqrt{\m_q^2+\mu_3^2}$. However, for larger values of $\mu_3$, the analogue of the $\rho$-meson condenses, (as studied in the WSS model in \cite{Aha})
giving rise to a superfluid with a vector order parameter. There are two inequivalent channels of condensation, namely p and p+ip. The most stable one is the p+ip channel. The transition is second order and it becomes first order at sufficiently large $\mu_3$, \cite{JKNP}. In this paper, we analyse the dynamics of the asymptotically-AdS$_5$ Reissner-Nordstr\"om ground state of the theory.

We shall focus in this work on the so-called charged currents, whose name comes from the electromagnetic charge that they carry in the Standard Model. These currents are also charged under isospin, so that the corresponding correlators are most affected by the background isospin asymmetry. The other class of currents---the so-called neutral currents---will be discussed in future work.

For $\mu_3 = 0$, the holographic system of interest is of the type mentioned above, which admits a hydrodynamic-like description at low temperature $T\ll \mu_q$ \cite{neutrinopaper}. This means that, over times and distances much longer than the density scale $\mu_q$, the charged current dynamics reduces to the conservation equations\footnote{Since the charged currents vanish in the background, their dynamics decouple from stress-energy. In the neutral sector, the coupling of baryon number density to energy density and the existence of mixed  correlators has to be taken into account.} of their expectation values, $\pa_\mu J^\mu_a = 0$. One of our main goals will be to understand how this property generalizes for an isospin asymmetric medium.

The first issue is that, in the presence of a non-Abelian source like an isospin chemical potential, ordinary conservation laws are replaced by covariant conservation laws $(D_\mu J^\mu)_a = 0$. The effective theory whose equations of motion are given by these covariant conservation equations is referred to as \emph{non-Abelian hydrodynamics}. The literature on the topic contains  early formulations describing the canonical and Lagrangian structure of non-Abelian fluids, the appearance of group-valued fluid variables, and the fluid analogue of Wong equations for locally precessing non-Abelian charges \cite{Jackiw:2000qj,Bistrovic:2002nv,Jackiw:2004nm}. Later works developed relativistic and holographic formulations, including dissipative transport, multiple charge sectors, and the effects of non-Abelian external gauge fields \cite{Son:1999ChiralHydrodynamics}-\cite{NS}.
In parallel, triangle anomalies were shown to fix additional first-order transport coefficients, such as chiral magnetic and vortical terms, for both Abelian and non-Abelian currents \cite{Son:2009tf,Neiman:2010zi}.

In principle, the QCD charged currents' transport is also affected by the QCD chiral anomalies, whose effects are captured in holography by topological Chern-Simons terms. We shall not include however this type of terms here, leaving their analysis for future work (as discussed in the outlook section).

In the case of interest here, where the non-Abelian charge is an isospin chemical potential $\mu_3$, the commutator terms in the covariant conservation equation imply that the hydrodynamics of the charged currents are not governed by an ordinary diffusion equation: even at zero spatial momentum, they undergo flavor precession with a frequency fixed by $\mu_3$. This is the hydrodynamic counterpart of the familiar splitting of charged modes in a medium with finite isospin density, as also observed in holographic studies of flavor-current correlators and mesonic spectral functions at finite isospin chemical potential \cite{Erdmenger:2007ja,Erdmenger:2007cm}.

\subsection{Outline and results\label{results}}

We now present the structure of this paper, together with a summary of our main results. We caution that, for the sake of maximal simplicity, some notations below are somewhat simplified compared to the main text that starts at section 2.  This summary and the main text should therefore be read independently when making detailed comparisons.

Before presenting the main results of this work, we should make some precise definition of terms introduced in the discussion above, and that we  use recurrently in the rest of the paper. We remind the reader that we always assume in this paper that $T\ll \mu$.

\vspace{0.4cm}

First, we define three dynamical regimes (or regions).

\begin{itemize}
	
	\item  The  {\em (standard) hydrodynamic regime} is the dynamical regime where \\
	$|\vec k|,|\omega|,|\mu_3| \ll T$.
	
	\item The {\em near-extremal hydrodynamic regime} is the dynamical regime where \\
	$|\vec k|,|\omega|, |\mu_3|\ll \mu$.
	
	\item  The {\em extended hydrodynamic regime} is the dynamical regime where\\
	$T\ll|\vec k,|\omega|,|\mu_3|\ll \mu$.
	
\end{itemize}

Then, we name the {\em near-extremal effective theory}, the effective theory that captures the non-linear dynamics of the massless and the AdS$_2$-related IR (massive) poles, in the near-extremal hydrodynamic region, $|\vec k|,|\omega|,|\mu_3|\ll \mu$. We also define two different approximations to such effective theory.

\begin{itemize}
	
	\item The {\em near-extremal hydrodynamics} describes standard dynamical equations of hydrodynamics applied to the near-extremal hydrodynamic region.
	
	\item The {\em extended hydrodynamics} describes the improved dynamical equations of hydrodynamics, (whose linearized version is introduced in section \ref{sec:IRcorr-approx}), applied to the near-extremal hydrodynamic region.

\end{itemize}

We now discuss the main results of this work. Section \ref{sec:NAcorr} presents a review of non-Abelian hydrodynamics, and its application to charged current transport in the presence of an isospin chemical potential. The non-Abelian hydrodynamic constitutive relations are expansions in derivatives, where ordinary and covariant derivatives are treated as the same order \cite{Neiman:2010zi}. These expansions are therefore valid for derivatives, but also $\mu_3$, much smaller than the microscopic scale given by the temperature (or eventually $\mu$). We then apply this framework to derive an expression for the charged current correlators at leading order\footnote{That is with constitutive relations for the currents up to first order in derivatives.} in the standard hydrodynamic regime, $\omega,|\vec{k}|,\mu_3\ll T$
\begin{equation}
	\label{su1} \left<J^{\pm}_\l J^{\mp}_\s\right>^R(\omega,\vec{k}) = P_{\l\s}^{\perp,\pm}(\omega,\vec{k}) i\Pi^{\perp,\pm}(\omega,\vec{k}) + P_{\l\s}^{\parallel,\pm}(\omega,\vec{k}) i\Pi^{\parallel,\pm}(\omega,\vec{k}) \mp i\delta_\l^0\delta_\s^0\dfrac{n_3}{\omega\pm\mu_3} \, ,
\end{equation}
\begin{equation}
	\label{su2}
	\Pi^{\parallel,\pm}(\omega,\vec k) = -\frac{i\Sigma \,\omega((\omega \pm\m_3)^2-\vec k^2)}{(\omega \pm\m_3+ iD\vec{k}^2)(\omega\pm\m_3)}\,,
\end{equation}
\begin{equation}
	\label{su3} \Pi^{\perp,\pm}(\omega,\vec k) = -i  \Sigma\,\omega  \,,
\end{equation}
with $n_3$ the isospin density, $\Sigma$ the flavor conductivity, and $P_{\mu\nu}^{\perp,\pm},P_{\mu\nu}^{\parallel,\pm}$ the transverse and longitudinal projectors to the spatial momentum $\vec{k}$. Their explicit form can be found in \eqref{Po}-\eqref{Pp}. The expression for the longitudinal correlator, \eqref{su2}, implies the existence of a hydrodynamic-like mode, with dispersion relation
\begin{equation}
	\label{su2b} \omega^\pm_D(\mu_3,k) = \mp \mu_3 - i D\vec{k}^2 + \OO(\mu_3^2,\mu_3\vec{k}^2,\vec{k}^4) \, .
\end{equation}
In this work, we are interested in studying charged current correlators in the holographic model, where we expect to recover this type of expressions \eqref{su2}-\eqref{su3} in the appropriate regime.

The holographic model considered in this work, is introduced in section \ref{sec:holomodel}. It takes the form of a five-dimensional Einstein-Yang-Mills theory, with action
\begin{equation}
	\label{su4} S = \NN\int\mathrm{d}^5x\sqrt{-g}\, \left(R + \frac{12}{\ell^2} - \frac{1}{4g_f^2}\text{Tr}F_{MN}F^{MN} \right) \, ,
\end{equation}
where $R$ is the bulk Ricci scalar, $\NN$ a normalization constant, $\ell$ the AdS length, $g_f$ the bulk flavor Yang-Mills coupling, and $F_{MN}$ the field strength for the gauge fields $A_M$ dual to the U(1)$\times$SU(2) boundary flavor currents. Our goal is to compute charged current correlators for this model, in a state described in the bulk by an AdS$_5$-RN black hole with two charges $\mu_q$ and $\mu_3$. The metric of this solution takes the form
\begin{equation}
	\label{su5} \intd s^2 = \frac{\ell^2}{r^2}\left(\frac{\intd r^2}{f(r)} - f(r)\intd t^2 + \intd\vec{x}^2\right) \, ,
\end{equation}
where $r$ is the holographic coordinate, equal to zero at the AdS boundary, and $\vec{x}$ is the vector of spatial boundary coordinates. $f(r)$ is the blackening function, which vanishes at the location of the horizon $r_H$; its explicit expression is given by \eqref{bmet}. In addition to the metric, the RN solution has two types of gauge fields turned on, that are respectively sourced by the quark number and isospin chemical potentials, such that the total U(1)$\times$SU(2)  gauge field 1-form is given by
\begin{equation}
	\label{su6a} \mathbf{A} = \frac{1}{2}(\Phi(r) \mathbb{I}_2  + \Phi_3(r)\s_3)\intd t
\end{equation}
\begin{equation}
	\label{su6b} \Phi(r) = \mu_q\left(1- \frac{r^2}{r_H^2}\right) \sp \Phi_3(r) = \mu_3\left(1- \frac{r^2}{r_H^2}\right) \, ,
\end{equation}
with $\mathbb{I}_2 = \text{diag}(1,1)$ and $\s_3 = \text{diag}(1,-1)$ the third Pauli matrix. As mentioned in the introduction, the AdS$_5$ RN solution is not always the dominant saddle for different values of the parameters $\mu_q/T,\mu_3/\mu_q$ and $g_f$ \cite{JKNP}. However, it was found in \cite{JKNP} that realistic neutron-star like conditions are more likely to correspond to the RN phase.

The holographic calculation of the charged current correlators \eqref{su1} is then detailed in section \ref{sec:holoccc}. These correlators can be computed by solving the fluctuation equations for gauge field perturbations $\delta A_M^\pm$ in the bulk, with the label $\pm$ referring to the charge of the dual current\footnote{The precise definition is $\d A_M^\pm \equiv \text{Tr}\left(\d A_M(\s_1\mp i\s_2)\right)$.}. These fluctuations obey the linearized Yang-Mills equations\footnote{The charged gauge fields (dual to the charged currents) do not couple to metric fluctuations. The neutral gauge fields do couple to metric perturbations, but these will be studied elsewhere. }, that can be written in Fourier space and in the radial gauge $A_r = 0$, as two independent decoupled equations
\begin{equation}
	\label{su7} rf(r)\pa_r\left(\frac{f(r)}{r}\pa_r\d A_\perp^\pm\right) + (\Omega_\pm(r)^2 - f(r)\vec{k}^2) \d A_\perp^\pm = 0 \, ,
\end{equation}
\begin{equation}
	\label{su8} rf(r)\pa_r\left(\frac{f(r)}{r}\frac{\Omega_\pm(r)^2}{\Omega_\pm(r)^2 - f(r)\vec{k}^2}\pa_r\d\varphi^\pm\right) + \Omega_\pm(r)^2\d\varphi^\pm = 0 \, ,
\end{equation}
with
\begin{equation}
	\label{su9} \Omega_\pm(r) \equiv  \omega \pm \Phi_3(r) \, ,
\end{equation}
where $k^\mu=(\omega,\vec{k})$ is the boundary four-momentum of the Fourier mode. The fields $\d A_\perp$ and $\d\varphi$ are respectively the transverse fluctuation of the gauge field, and a gauge-invariant combination of longitudinal fluctuations, defined as
\begin{equation}
	\label{su10} \d A^{\perp,\pm}_\mu \equiv P_{\mu\nu}^{\perp,\pm} \d A^{\pm,\nu} \sp \d\varphi^\pm \equiv \Omega_\pm(r)^{-1}k^\mu\d F_{t\mu} \, .
\end{equation}
The transverse projector $P_{\mu\nu}^{\perp,\pm}$ is the same that appears in the decomposition of the current two-point function \eqref{su1}. After computing the on-shell action, we finally find that the polarization functions $\Pi$ in \eqref{su1}, can be expressed in terms of the near-boundary data for the solutions to \eqref{su7}-\eqref{su8}, as
\begin{equation}
	\label{su11} \Pi^\pm_\perp = -\frac{\mathcal{N}}{8g_f^2} \underset{r\to 0}{\lim}\left(\frac{1}{r}\frac{\pa_r\d A_\perp^\pm}{\d A_\perp^\pm}\right) \sp \Pi^\pm_\parallel = -\frac{\mathcal{N}}{8g_f^2} \underset{r\to 0}{\lim}\left(\frac{1}{r}\frac{\pa_r\d \varphi^\pm}{\d \varphi^\pm}\right) \, ,
\end{equation}
with $\mathcal{N}$ the overall normalization of the action \eqref{su4}, and $g_f$ the flavor Yang-Mills coupling. As is clear from the Fefferman-Graham near-boundary expansions, \eqref{Ltnb}-\eqref{Elnb}, the expressions \eqref{su11} are actually logarithmically divergent. This divergence can be taken care of, via a standard holographic renormalization procedure, as detailed in appendix \ref{app:holoren}. However, this is not even needed for the subsequent analyses in this work, since we focus on the imaginary parts of the polarization functions \eqref{su11} (i.e. the spectral functions),
which are free of UV divergences.

Section \ref{sec:holoNAcomputation} is dedicated to the analysis of the holographic polarization functions \eqref{su11} in the regime of near-extremal hydrodynamics, mentioned at the beginning of this section, corresponding to $\omega,|\vec{k}|,\mu_3,T\ll \mu$. We generalize the calculation of \cite{DP13,neutrinopaper} to the non-Abelian case, showing that in the regime of near-extremal hydrodynamics, the fluctuation equations \eqref{su7}-\eqref{su8} can be split into two regions where they simplify:
\begin{itemize}
	\item The inner region close to the horizon, where at leading order the fluctuation equations \eqref{su7}-\eqref{su8} reduce to the equations of motion obeyed by a scalar field in AdS$_2$-Schwarzschild, \eqref{HT7}.
	\item The outer region, which connects the inner region to the boundary. This region is such that \eqref{su7}-\eqref{su8} become conservation equations at leading order in the regime of near-extremal hydrodynamics
	\begin{equation}
		\nn \pa_r\left(\frac{f(r)}{r}\pa_r\d A_\perp^\pm\right) = 0 \sp \pa_r\left(\frac{f(r)}{r}\frac{\Omega_\pm(r)^2}{\Omega_\pm(r)^2 - f(r)\vec{k}^2}\pa_r\d \varphi^\pm\right) = 0 \, .
	\end{equation}
\end{itemize}
The solution to \eqref{su7}-\eqref{su8} can then be computed at leading order in the near-extremal hydrodynamic expansion by solving the equations in each region, and matching them where they overlap. Substituting the solution thus obtained into \eqref{su11}, we find that the holographic spectral functions, $\Im\Pi^\pm$, agree with the non-Abelian hydrodynamic expression \eqref{su2}, at leading order in the near-extremal hydrodynamic regime, with the conductivity and diffusivity given by
\begin{equation}
	\label{su11b} \Sigma =\frac{\mathcal{N}}{8g_f^2}\; r_H^{-1} \sp D = \frac{1}{2}r_H \, .
\end{equation}
$\NN$ is the same normalization constant as in \eqref{su11}. We therefore find that, at least at the level of the charged current two-point functions, near-extremal hydrodynamics generalizes to the non-Abelian case.

In section \ref{sec:QNMt}, we go further in our analysis of the polarization functions $\Pi^\pm$ in the near-extremal hydrodynamic regime: we  provide a detailed numerical analysis of their  quasi-normal modes. In addition to the hydrodynamic-like mode \eqref{su2b} exhibited in section \ref{sec:holoNAcomputation}, we emphasize the emergence of the poles associated with the infra-red conformal description at low temperature. These modes are closely related to the poles of the IR-AdS$_2$ correlator for charged scalar fields (with the charge corresponding to the isospin charge)
\begin{align}\label{su12}
	\GG^\pm_\text{IR}(\omega, k) =&(4\pi T)^{2\D( k,\mu_3)-1}\frac{\Gamma(1\!-\!2\D(k,\mu_3))\Gamma(\D(k,\mu_3)\!-\!\frac{i\omega}{2\pi T}\pm \frac{i}{6}r_H\mu_3)}{\Gamma(2\D(k,\mu_3)\!-\!1)\Gamma(1\!-\!\D(k,\mu_3)\!-\!\frac{i\omega}{2\pi T}\pm \frac{i}{6}r_H\mu_3)}\times\\ \non
	&\times\dfrac{\Gamma(\D(k,\mu_3)\mp \frac{i}{6}r_H\mu_3)}{\Gamma(1\!-\!\D(k,\mu_3)\mp \frac{i}{6}r_H\mu_3)},
\end{align}
where $\D(k,\mu_3)$ is the IR conformal dimension
\begin{equation}
	\label{su13} \D(k,\mu_3) = \frac{1}{2} + \frac{1}{2}\sqrt{1+\frac{1}{3}r_H^2 k^2 - \frac{1}{9}r_H^2\mu_3^2} \, .
\end{equation}
The corresponding \emph{AdS$_2$ poles} are given by
\begin{equation}
	\label{su14} \omega^{\pm,(n)}_{\text{AdS}_2} = -i2\pi T(\D(k,\mu_3) + n) \pm \frac{1}{3} r_H \pi T \mu_3 \sp n \in \mathbb{N} \, .
\end{equation}
We show that the transverse and longitudinal poles have different behaviours in the near-extremal regime: while the transverse poles approach the AdS$_2$ poles \eqref{su14} uniformly in $k$ and $\mu_3$ as the temperature goes to zero (see figures \ref{fig:wt4} and \ref{fig:wt5}), the longitudinal sector is characterized by non-trivial interactions between the IR modes and the hydrodynamic-like pole \eqref{su2b} (see figures \ref{fig:wl2} and \ref{fig:wl3}).

At $\mu_3=0$ (figures \ref{fig:wl2} and \ref{fig:wl3} top-left), these interactions are similar to what was observed in \cite{Gouteraux:2025kta,Preau:2025rex}: after a sequence of collisions\footnote{For the example shown in figures \ref{fig:wl2} and \ref{fig:wl3}, there is actually a single collision.} between the hydro-like pole and the first IR poles, two hydro-like poles eventually emerge at sufficiently large momentum, with non-zero opposite real parts. The IR poles are close to the AdS$_2$ poles $\omega_{\text{AdS}_2}^{\pm,(n)}$ in \eqref{su14}, except for values of momentum where the imaginary part of the hydrodynamic-like poles approaches the corresponding AdS$_2$ level $n$.

Extending the analysis to non-zero $\mu_3$, we find novel features in the longitudinal mode dynamics.  Perhaps unexpectedly, even though the hydro-like pole moves away from the IR poles in agreement with \eqref{su2b} as $\mu_3$ is increased, the hydro-like---IR interactions are still large in some range of momenta. More precisely, the IR modes get more and more affected by crossings with the hydro-like pole imaginary part as momentum is increased, until a second hydro-like mode emerges, similarly to $\mu_3 = 0$. The presence of $\mu_3$ however affects the hydro-like poles in two different ways: their real parts are not opposite anymore (although they still have opposite signs), and their imaginary parts also differ by a quantity proportional to $\mu_3$. We expect these features to be visible from a generalization of the matching calculation in section \ref{sec:holoNAcomputation}, to next-to-leading order in the near-extremal-hydrodynamic expansion (following \cite{Gouteraux:2025kta}).

Based on the results of section \ref{sec:QNMt} for the quasi-normal spectrum, we discuss in section \ref{sec:IRcorr-approx} the possibility to improve on the near-extremal hydrodynamic approximation to the charged current correlators \eqref{su2}-\eqref{su3}, in the near-extremal hydrodynamic regime. The idea is to follow \cite{Preau:2025rex}, and use the product formula of \cite{Dodelson:2023vrw} to rephrase our knowledge of the QNM spectrum into an approximation to the spectral functions $\text{Im}\Pi^\pm$. Applying the results of \cite{Preau:2025rex} to the non-Abelian case\footnote{As mentioned in section \ref{sec:IRcorr-approx}, a rigorous proof that the results of \cite{Preau:2025rex} apply to the longitudinal sector in the non-Abelian case would require further investigation.}, we obtain in particular in the limit $T\ll \omega,|\vec{k}|,\mu_3\ll \mu$, what we call the \emph{extended hydrodynamic approximation}
\begin{align}
	\label{su15}&\text{Im}\Pi^{\perp,\pm}_\text{ext-hydro} = -\Sigma\,\mu(\omega/\mu)^{2\Delta(k,\mu_3)-1}\,,\\
	\label{su16}&\text{Im}\Pi^{\parallel,\pm}_\text{ext-hydro} = -\Sigma\,\mu(\omega/\mu)^{2\Delta(k,\mu_3)-1}\dfrac{(\omega\pm\mu_3)^2-\vec k^2}{(\omega\pm\mu_3)^2+D\,\vec k^4}\,.
\end{align}
These expressions feature both the hydrodynamic-like behaviour -- in particular the presence of the diffusive-like pole \eqref{su2b} -- and the IR conformal behaviour, characterized by the low-energy power-law $\omega^{2\D -1}$. The isospin chemical potential $\mu_3$ is seen to have the following effects on the approximations: on the one hand it modifies the IR dimension $\D(k,\mu_3)$, and on the other hand it shifts the zeroes and poles of the longitudinal spectral function \eqref{su16}.

The approximations \eqref{su15}-\eqref{su16} are compared with the exact numerical result for the spectral functions in our last section \ref{sec:exactresults}, where we also consider the near-extremal hydrodynamic approximations \eqref{su2}-\eqref{su3}. These comparisons verify that both types of approximations are good in the near-extremal hydrodynamic regime $\omega,k,T,\mu_3\ll\mu$, and the extended hydrodynamic approximation does better in the low energy regime $\omega\to 0$.

The appendices contain several technical details and supplementary analyses. Appendices \ref{app:notationpm}, \ref{Sec:TwoPointDec},  \ref{app:WI} and \ref{app:Ons} collect additional details and properties of the current two-point function. In appendix \ref{app:thermo}, we provide some details about the holographic model, and appendix \ref{app:holoren} shows how to renormalize the divergent part of the current correlators studied in this work. Appendices \ref{app:IRcorr-approx} and \ref{app:residues} contain a detailed study of the IR-AdS$_2$ correlator \eqref{su12}. In particular, appendix \ref{app:IRcorr-approx} discusses several other approximations to the exact correlator, based on more refined implementations of the IR-AdS$_2$ conformal correlator.  Appendix \ref{app:residues} gives the detailed computation of the residues of the AdS$_2$ poles of such correlator.

Then, appendices \ref{app:otherapprox65}, \ref{app:otherexactresults}, \ref{app:coarsegrained}, and \ref{app:finegrained} contains a large amount of additional numerical data for the correlators. The other analytic approximations based on the IR-AdS$_2$ correlator presented in appendix \ref{app:IRcorr-approx} are tested against the exact correlator. This analysis shows that, as expected, these other approximations have a similar accuracy as the extended hydrodynamic approximation \eqref{su15}-\eqref{su16}. Moreover, such analysis is extended to different values of density and isospin asymmetry than the ones studied in section \ref{sec:exactresults}.

\subsection{Outlook}

We discuss here open problems that would be interesting to address as a continuation of this work.

A natural continuation of this work is the study of the neutral-current sector. In contrast with the charged currents considered here, neutral currents couple to energy and baryon-density fluctuations, leading to a richer hydrodynamic structure involving coupled diffusive and sound-like modes. Understanding how near-extremal hydrodynamics generalizes in this sector is an important open problem.

An important motivation for studying neutral and charged currents is the description of neutrino transport in dense QCD matter under neutron-star conditions, \cite{neutrinopaper,Hoyos:2024pkl}. Since neutrino interaction rates are determined by charged and neutral current correlators, extending the present analysis to finite isospin asymmetry provides a first step toward understanding neutrino transport in realistic isospin-asymmetric matter.

An equally important direction is the inclusion of Chern--Simons terms in the holographic model in order to correctly reproduce chiral anomalies. On the hydrodynamic side, this would lead to anomalous transport effects such as chiral magnetic and vortical contributions to the constitutive relations. It would then be interesting to investigate the interplay between anomaly-induced transport, non-Abelian hydrodynamics, and the infrared conformal dynamics characteristic of near-extremal holographic systems.

It would also be interesting to consider the extension of the flavor action from YM to the full DBI, which is the natural effective action arising from string theoretic models \cite{SS,V1,Karch02}. In this direction, there is already an interesting question at zero $\mu_3$. In the regime where the back-reaction of the flavor fields to the geometry is small (the so-called D3/D7 setup \cite{Karch02}), it is known that there is a zero-sound mode emerging instead of the diffusive mode at low temperature \cite{Karch:2008fa,Davison:2011ek}. It will be interesting to observe what happens when back-reaction is included, that is for transport on the DBI black-brane background \cite{V4}. The same question can be asked at finite $\mu_3$ for a non-Abelian generalization of the DBI action (like the symmetrized trace prescription \cite{ckp,V1}).\footnote{See \cite{Hoyos:2025qwd} for a recent discussion on how to deal with the non-Abelian generalization of the DBI action at the quadratic level.} In the latter case, even the background solutions and phase structure remain to be investigated.
Eventually, the impact of all of the above to the merger dynamics of neutron stars, \cite{Jarvinen} must be investigated.

\section{Non-Abelian hydrodynamics for the charged currents}\label{sec:NAcorr}

In this section, the framework of non-Abelian hydrodynamics is first introduced (section \ref{sec:NAhydro}), and then used to derive the long range approximation to the charged current correlators in the presence of isospin asymmetry (section \ref{sec:expr2pt}).

\subsection{Non-Abelian hydrodynamics}
\label{sec:NAhydro}

Non-Abelian hydrodynamics is the effective long-wavelength, late-time description of a system with non-Abelian global symmetries. It generically applies to the regime where all gradients are small compared with the microscopic scale set by the temperature, namely $\omega,k\ll T$ in Fourier space. The symmetry that will of interest to us, is the unbroken flavor symmetry U(1)$\times$ SU(2) where the U(1) is associated to the baryon number while the SU(2) is associated to isospin symmetry. In the presence of background non-Abelian chemical potentials, the condition of validity is  that covariant gradients are also small, $\omega, k,\mu_3\ll T$ (with $\mu_3$ being the isospin chemical potential). In this regime, the dynamics is governed entirely by the conserved quantities associated with the symmetries of the system, namely energy, momentum and non-Abelian charges. In this work, we focus on the hydrodynamics of conserved currents associated with the flavor symmetry.

We start by setting up the notation that we  use in this work. The Greek indices $\mu,\nu,\rho,\ldots \in\{0,1,2,3\}$, denote four-dimensional Minkowski space-time indices and they are raised and lowered with the flat Minkowskian space-time metric $\eta_{\mu\nu}$, which in four dimensions is defined as
\begin{equation}
	\eta_{\mu\nu}=\operatorname{diag}(-,+,+,+)\,.
\end{equation}
The Latin indices $i,j,k,\ldots\in\{1,2,3\}$ are reserved for spatial components.

The flavor currents satisfy a conservation equation which for the U(1) baryonic current, that we denote by a hat, reads
\begin{equation}
	\partial_\mu \hat J^\mu =0\,.
\end{equation}
The SU(2) current is Lie-algebra-valued and is denoted by\footnote{The sum over repeated indices in understood.}
\begin{equation}
	J_\mu=t_aJ^a_{\mu}\,,
\end{equation}
with $a\in\{1,2,3\}$, the group indices, and the generators of the SU(2) flavor group being $t_a = \sigma_a/2$, with $\sigma_a$ the Pauli matrices. In this basis, the SU(2) generators are normalized according to
\begin{equation}
	\text{Tr}(t_at_b) = \dfrac{\delta_{ab}}{2}\,.
\end{equation}
The Pauli matrices have a non-trivial commutator defined by the structure constants $f_{ab}\,^c$\,:
\be
[\sigma_a,\sigma_b]=2i\,f_{ab}\,^c\sigma_c.
\ee
The gauge group metric used to raise and lower the group indices is defined as
\be
\mathbf{g}_{ab} = 2\,\text{Tr}(t_at_b) = \dfrac{1}{2}\text{Tr}(\sigma_a\sigma_b).
\ee
In the Cartesian basis associated with $a,b\in\{1,2,3\}$, it simply reads $\mathbf{g}_{ab}=\delta_{ab}$, while $f_{abc}=\epsilon_{abc}$ the Levi-Civita symbol.\footnote{See appendix \ref{app:notationpm} for more details.}

The flavor SU(2) currents, $J^{\mu}$, obey a covariant conservation equation
\begin{equation}
	\label{DJ}	D_{\mu}J^{\mu}=\partial_\mu J^\mu-i[A_\mu,J^\mu]=0\,,
\end{equation}
where $A_\mu = t_aA^a_\mu$ is the external SU(2) gauge field that couples to $J^\mu$, and the group indices are implicit.

We can group together the Abelian and the non-Abelian components of the current and gauge field, in fields associated with the group U(1)$\times$ SU(2):
\begin{align}	\label{eq:Jdefinition}
	&\mathcal{J}_\mu = \hat t\,\hat J_\mu +  t_aJ^a_\mu = \dfrac{\mathbb{I}_2}{2}\hat J_\mu + \dfrac{1}{2}\sigma_aJ^a_\mu,
	\\ 	\label{eq:Adefinition}
	&\mathcal{A}_\mu = \hat t\,\hat A_\mu + t_aA^a_\mu = \dfrac{\mathbb{I}_2}{2}\hat A_\mu + \dfrac{1}{2} \sigma_aA^a_\mu,
\end{align}
where we have introduced $\hat t=\mathbb{I}_2/2$ as the U(1) generator, with $\mathbb{I}_2$ is the two-dimensional unit matrix.

For completeness, we also give the conservation equation for the energy-momentum tensor
\begin{equation}
	\partial_\mu T^{\mu\nu} = \hat F^{\nu\rho}\hat J_\rho + F^{a,\nu\rho}J_{a,\rho}\,
	\label{Tcon}\end{equation}
where $a$ is again the group index, $\hat F_{\mu\nu}$ is the field strength tensor for the Abelian external field $\hat A_\mu$, and $F^{a}_{\mu\nu}$ is the field strength tensor for the SU(2) field $A_\mu^a$. They are defined as
\begin{align}
	& \hat F_{\m\nu} = \partial_\mu \hat A_\nu -\partial_\nu \hat A_\mu\,,\\
	& F^a_{\m\nu} = \partial_\mu A_\nu^a -\partial_\nu A_\mu^a + f^a{}_{bc}A_\mu^bA_\nu^c\,.
\end{align}
The right-hand side of equation \eqref{Tcon} is the force-density term due to the external Abelian and non-Abelian field strengths.

The conservation equations (\ref{DJ}) and (\ref{Tcon}), are supplemented by constitutive relations, that express the conserved currents in terms of the local thermodynamic variables and fluid velocity. The constitutive relations are organized as a derivative expansion, whose coefficients determine the transport properties of the medium. In this work, we restrict to first order in derivatives and linear order in fluctuations around an equilibrium state, with non-zero baryon and isospin density.

At zeroth (ideal) order, the constitutive equations for the SU(2) chiral current $J_{a}^{\mu}$ and the baryon current $\hat{J}^{\mu}$ are given by
\begin{equation}
	\label{consteqJ} \left(J_a^{\mu}\right)^{(0)}  =n_a u^\mu \sp (\hat{J}^{\mu})^{(0)} = \hat n \,u^\mu
\end{equation}
where the four-velocity $u^\mu$ satisfies $u_\mu u^\mu=-1$, and $n_a$, $\hat n$ are the non-Abelian and the Abelian charge densities respectively. At first order, and working in the Landau frame,\footnote{In relativistic hydrodynamics, the definition of the fluid thermodynamic variables is not unique beyond ideal order, since one can redefine $u^\mu$, $T$, and $\mu_a,\,\hat \mu$ by terms of higher order in gradients without affecting physical observables. This freedom is known as \emph{frame ambiguity}. The Landau frame is defined by the condition that the energy flux vanishes in the local rest frame of the fluid, \textit{i.e.} $u_\mu T^{\mu\nu} = -\varepsilon\, u^\nu$, which implies that there are no dissipative corrections to the energy density or energy flow. This choice fixes the velocity field uniquely order by order in the derivative expansion and is commonly used in relativistic hydrodynamics.} the constitutive relations of non-Abelian hydrodynamics are given by \cite{Neiman:2010zi,Kovtun12}
\begin{equation}
\label{consteq1J}
\left(J_a^{\mu}\right)^{(1)}  = \Sigma_a\,^b\left(E_b^\mu-T P^{\mu \nu} D_\nu \frac{\mu_b}{T}\right)+\xi_a \omega^\mu+\xi_{a b}^{(B)} B^{b \mu} \,,
\end{equation}
\begin{equation}
\label{consteq1Jh}
(\hat{J}^{ \mu})^{(1)}  =\hat{\Sigma}\left(\hat{E}^\mu-T P^{\mu \nu} \partial_\nu \frac{\hat \mu}{T}\right) + \xi \omega^\mu + \xi^{(B)}B^\mu\,,
\end{equation}
where $\hat\mu$ and $\mu_a$ are the Abelian and non-Abelian chemical potentials, while $D_{\mu}$ is the gauge-covariant derivative that acts on the non-Abelian components of the chemical potential as
\be
D_\nu \mu_a=\partial_\nu \mu_a+f_{a b c} A_\nu^b \mu^c\;,
\ee
with $f_{abc}$ the structure constants of the group. Moreover,
\be
{P_\mu}^\nu={\delta_\mu}^\nu+u_\mu u^\nu
\ee
 is the projector orthogonal to $u^\mu$. The non-Abelian and Abelian conductivity matrices are indicated as $\Sigma_{ab}$ and $\hat \Sigma$ respectively. The electric and magnetic fields are expressed in terms of the field strength and velocity vector as
\begin{equation}
E_a^\mu \equiv F_a^{\mu \nu} u_\nu \sp  \hat{E}^\mu \equiv \hat{F}^{\mu \nu} u_\nu\, ,
\end{equation}
\begin{equation}
B^{a \mu} \equiv \frac{1}{2} \epsilon^{\mu \nu \rho \sigma} u_\nu F_{\rho \sigma}^a \sp \hat B^{\mu} \equiv \frac{1}{2} \epsilon^{\mu \nu \rho \sigma} u_\nu \hat F_{\rho \sigma}\,,
\end{equation}
and the vorticity density $\omega_{\mu}$ is defined as
\begin{equation}
	\omega^\mu \equiv \frac{1}{2} \epsilon^{\mu \nu \rho \sigma} u_\nu \partial_\rho u_\sigma\, .
\end{equation}
The vortical and magnetic coefficients can be written as
\begin{equation}
	\begin{aligned} \xi_a & =C_{a b c} \mu^b \mu^c-\frac{2 n_a}{\varepsilon+p}\left(\frac{1}{3} C_{b c d} \mu^b \mu^c \mu^d+\gamma T^3\right)\,, \\ \xi_{a b}^{(B)} & =C_{a b c} \mu^c-\frac{n_a}{\varepsilon+p}\frac{1}{2} C_{b c d} \mu^c \mu^d\, ,
	\end{aligned}
\end{equation}
\begin{equation}
	\begin{aligned} \xi & =C \hat\mu^2 +2\beta\,T^2-\frac{2 \hat n}{\varepsilon+p}\left(\frac{1}{3} C\hat\mu^3+2\beta\,\hat\mu T^2+\gamma T^3\right)\,, \\ \xi^{(B)} & =C \hat\mu-\frac{\hat n}{\varepsilon+p}\left(\frac{1}{2} C \hat \mu^2+\beta\,T^2\right)\, ,
	\end{aligned}
\end{equation}
where $\varepsilon$ and $p$ are the energy density and the pressure.
\be
C_{abc} = \dfrac{1}{4\pi^2}\;\text{tr}(T_{(a}T_bT_{c)})
\label{sym}\ee
is a symmetric tensor\footnote{The parentheses denote normalized symmetrization over the three adjoint indices. Explicitly
 	\begin{equation}
 		T_{(a}T_bT_{c)}
 		\equiv
 		\frac{1}{3!}
 		\sum_{\pi\in S_3}
 		T_{\pi(a)}T_{\pi(b)}T_{\pi(c)}\,,
 \end{equation}
 where $S_3$ is the group of the six permutations of the group indices $a,b,c$. With this convention, $C_{abc}$ is totally symmetric in its indices.} of anomaly coefficients. Here $T_a$ is the generator associated with the current $J_a^\mu$, represented on the microscopic fields that carry the corresponding global charge. The trace in \eqref{sym} is taken over this microscopic representation. Therefore, in a theory containing chiral fermions, $T_a$ acts on the fermion multiplets, charged under the global symmetry, and the anomaly coefficient is obtained by summing their chiral contributions. In addition
\be\label{eq:defCanomaly}
C = \dfrac{1}{4\pi^2}\sum\limits_i\left(q_{i,(R)}^3-q_{i,(L)}^3\right)\,,
\ee
 is the U(1) anomaly coefficient. The subscripts $(R)$, and $(L)$ label the right- and left-handed Weyl fermions while $q_{i,(R,L)}$ are the charges of the fermion $i$ under the Abelian axial symmetry. Note that, for a genuinely axial symmetry of Dirac fermions, the right- and left-handed components have opposite charges and therefore
\be
C = \dfrac{1}{2\pi^2}\sum\limits_i\,q_{i}^3\,.
\ee

 The constant $\gamma$ parametrizes a possible temperature-dependent contribution to the vortical transport coefficients. In the entropy-current analysis of anomalous hydrodynamics, it appears as an integration constant in the parity-odd sector and is not fixed by the triangle anomaly coefficients \cite{EHKY,SoS,Land,Neiman:2010zi}. Although parity breaking is necessary for $\gamma$ to be non-zero, it is not sufficient: in \cite{Bhattacharya:2011tra} it was shown that a non-zero $\gamma$ violates CPT invariance. Therefore, in the standard class of local CPT-invariant quantum field theories, the corresponding $T^3$ contribution is absent \cite{Landsteiner:2011iq,Banerjee:2012iz}. We therefore set $\gamma=0$ in the present CPT-invariant setup.

 The constant $\beta$ parametrizes additional $T^2$ contributions to the Abelian vortical and magnetic conductivities. In the notation of \cite{Neiman:2010zi}, such terms can occur only for an Abelian axial charge, namely for a U(1) current with axial rather than vector transformation properties. In microscopic chiral theories, the $T^2$ vortical contribution is associated with the mixed gauge-gravitational anomaly \cite{Landsteiner:2011cp,Landsteiner:2011iq}. Since the Abelian U(1) charge considered here is vector-like baryon number, rather than an axial U(1) charge, we set $\beta=0$. For completeness, we provide a formula for the coefficient $\beta$ according to the conventions of \cite{Neiman:2010zi}
 \begin{equation}
 	\beta = \dfrac{1}{24}\sum\limits_i\left(q_{i,(R)}-q_{i,(L)}\right)\,.
 \end{equation}
As in formula \eqref{eq:defCanomaly}, $(R)$, and $(L)$ label the right- and left-handed Weyl fermions while $q_{i,(R,L)}$ are the charges of the fermion $i$ under the Abelian axial symmetry. For an axial symmetry built from Dirac fermions, the above formula reduces to
 \begin{equation}
   \beta = \dfrac{1}{12}\sum\limits_i\,q_{i}\,.
 \end{equation}

 In the holographic model introduced in the next section, we do not include parity-odd Chern-Simons interactions. In particular, we do not include gauge Chern-Simons terms, which holographically encode flavor triangle anomalies, or mixed gauge-gravitational Chern-Simons terms, which encode the mixed gauge-gravitational anomaly and generate the corresponding $T^2$ vortical contribution \cite{Landsteiner:2011iq}. Consequently, the other anomaly-induced coefficients $C$ and $C_{abc}$ also vanish in the model studied in this work. In summary,
 \begin{equation}
 	C_{abc}=0\,,
 	\qquad
 	C=0\,,
 	\qquad
 	\beta=0\,,
 	\qquad
 	\gamma=0\,.
 \end{equation}
 We leave the inclusion of anomaly-induced transport effects for future work.

For completeness, we also write the constitutive relations for the energy-momentum tensor, which at the zeroth and first order are
\begin{align}\label{eq:constT0}
	&(T_{\mu\nu})^{(0)} = \epsilon\,u_\mu\, u_\nu +p\,P_{\mu\nu}\,,\\ \label{eq:constT1}
	&(T_{\mu\nu})^{(1)} = 2\eta\,\pi_{\mu\nu}+\zeta\,P_{\mu\nu}\partial_\rho\,u^\rho\,,
\end{align}
where $\eta\geq 0$ and $\zeta\geq 0$ are two additional transport coefficients namely the shear and bulk viscosities. The shear tensor $\pi_{\mu\nu}$ is defined as
\begin{align}
	\pi_{\mu\nu}=P^\rho\,_\mu P^\sigma\,_\nu\partial_{(\rho}u_{\sigma)}-\dfrac{1}{3}P_{\mu\nu}\partial_\rho u^\rho\,.
\end{align}
The energy-momentum tensor constitutive relations \eqref{eq:constT0}, \eqref{eq:constT1} together with the conservation equation \eqref{Tcon} provide the set of hydrodynamic equations associated with $T_{\mu\nu}$. As we explain in the next section, in this work we  focus on the hydrodynamics of current components which decouple from the energy-momentum fluctuations. The analysis of the hydrodynamic sector involving the energy-momentum sector will be the topic of an upcoming work.

\subsection{The linearized hydrodynamic relations for the charged currents}
In this section, we provide the linearized hydrodynamic equations for the so-called \textit{charged} currents. As it will be clear in a few lines, the terminology charged refers to charge under the residual Cartan subgroup generated by $t_3$. At the linearized level, these current components decouple from the energy-momentum fluctuations and therefore from now on we focus only on the current fluctuations.

We derive the linearized version of the conservation equations. These are obtained by writing both the current and the gauge field as the equilibrium expectation value, $\bar{\mathcal{J}}^\mu$, $\bar{\mathcal{A}}_\mu$, plus a perturbation
\begin{equation}
	\mathcal{J}^\mu = \bar{\mathcal{J}}^\mu+\delta \mathcal{J}^\mu \quad,\quad \mathcal{A}_\mu = \bar{\mathcal{A}}_\mu+\delta \mathcal{A}_\mu \, ,
\end{equation}
where
\begin{align} \label{eq:bkgJ}
	&\bar{\mathcal{J}}^\mu = \dfrac{\delta^\mu\,_0}{2}\left(\bar{\hat n}\,\mathbb{I}_2+\bar n_3\sigma_3\right)\,,\\ \label{eq:bkgA}
	&\bar{\mathcal{A}}_\mu = \dfrac{\delta^0\,_{\mu}}{2}\left(\bar{\hat \mu}\,\mathbb{I}_2+\bar\mu_3\sigma_3\right)\,,
\end{align}
with $\bar{\hat n},\,\bar n_3$ the non-zero background densities, and $\bar{\hat\mu},\,\bar \mu_3$ the associated chemical potentials. Note that the background density and chemical potential in the other non-Abelian directions are zero at the background level $\bar n_{1,2} =0$, $\bar \mu_{1,2} =0$. Also, we write the various thermodynamic fields as a constant background value plus a small perturbation
\begin{align}
	&u_\mu = \bar u_\mu+\delta u_\mu \sp T = \bar T+\delta T,\\
	&\mu_a = \bar\mu_a+\delta\mu_a \sp \hat\mu= \bar{\hat \mu}+\delta\hat\mu,\\
	&n_a = \bar n_a+\delta n_a \sp \hat n = \bar{\hat n}+\delta\hat n.
\end{align}

Now, we perform the following change of basis involving the generators of the SU(2) flavor group. In particular, we define the \textit{charged} generators $t_\pm$ as a linear combination of $t_{1,2}$ as (see Appendix \ref{app:notationpm} for more details)
\begin{equation}
	t_\pm = \dfrac{t_1\pm i\,t_2}{2}\,.
\end{equation}
In this new basis, the flavor indices run over $a\in\{+,-,3\}$ and the charged components of the flavor current and the gauge field are written as
\begin{equation}\label{eq:chargedbasis}
	J^{\pm,\mu} = J^{1,\mu}\mp i\, J^{2,\mu} \sp A_\mu^\pm = A_\mu^1\mp i\, A_\mu^2\,.
\end{equation}
The terminology ``charged'' refers to charge under the residual Cartan subgroup U(1)$_{I_3}\subset$ SU(2) generated by $t_3$. Indeed, the charged generators are eigenvectors of the adjoint action of $t_3$,
\begin{equation}
	[t_3,t_\pm]=\pm t_\pm\,.
\end{equation}
Therefore the fluctuations along $t_+$ and $t_-$ carry charge $+1$ and $-1$, respectively, under U(1)$_{I_3}$. Note that in this basis, the explicit form of the gauge group metric $\mathbf{g}_{ab}$ is
\begin{align}\label{eq:groupmetric}
	\mathbf{g}_{ab}=2\,\text{Tr}(t_at_b)= \begin{pmatrix}
		0 \quad& 1/2 \quad& 0\\
		1/2 \quad& 0 \quad& 0\\
		0 \quad& 0 \quad& 1
	\end{pmatrix}\, .
\end{align}
where we recall that now the flavor indices run over $a,b\in\{+,-,3\}$ (see Appendix \ref{app:notationpm} for a derivation).

Then, the linearized covariant conservation equation \eqref{DJ} for the non-Abelian current, written in components, reads
\begin{equation}
	\label{eq:consdeltaJ}
	\partial_\mu\delta J_a^{\mu}
	+
	f_a{}^{bc}
	\left(
	\bar A_{b,\mu}\delta J_c^{\mu}
	+
	\delta A_{b,\mu}\bar J_c^{\mu}
	\right)
	=
	0\,.
\end{equation}
For $a=3$ this reduces to
\begin{equation}
\partial_\mu\delta J^{\mu}_3 = 0,
\end{equation}
which is a standard conservation equation. For the charged components, the terms proportional to the background fields are not zero:
\begin{equation}\label{eq:DJpm}
	\partial_\mu \delta J^{\mu}_\pm \pm \,i \,(\bar{\mu}_3 \delta J^{0}_\pm- \bar n_3  \delta A_{\pm,0})=0\,,
\end{equation}
where we have used the explicit expressions for the background fields \eqref{eq:bkgJ}, \eqref{eq:bkgA}, and the expression of the structure constants in the charged basis \eqref{eq:structpm}. From the above equation, one can observe that the charged current fluctuations are not conserved in the ordinary sense, unlike the neutral fluctuation $\delta J_3^{\mu}$. The charged fluctuations $\delta J_\pm^{\mu}$ acquire a precession term due to the commutator with the non-zero background gauge field $\bar A_{3,\mu}$.

The conservation equation \eqref{eq:consdeltaJ}, together with the constitutive relations \eqref{consteqJ}-\eqref{consteq1Jh}, can be used to extract the retarded Green function for the charged currents. We work in the rest frame $\bar u_i=0$, and as a consequence of the constraint $u_\mu u^\mu=-1$, at zeroth order, we have $\bar u^0=1$,\footnote{This choice corresponds to a future-directed timelike vector.} while at the linear order we have $\delta u_0=0$.

From the linearized constitutive relations \eqref{consteqJ}-\eqref{consteq1J}, up to first order in derivatives the current perturbations can be written as
\begin{align}
	&\delta J^0_a = \delta n_a,\\
	&\delta J^i_a = \bar n_a\delta u^i - \Sigma^b{}_a\delta F^{i0}_b-\Sigma^b{}_a\delta^{ij}\,\left(\partial_j\delta \mu_b-\dfrac{\bar\mu_b}{\bar T}\partial_j\delta T\right)+\\ \nonumber
	&\quad\quad\quad -\Sigma_a{}^bf_{bcd}\delta^{ij}\delta A^c_j\bar\mu^d,\\
	&\delta \hat J^0= \delta \hat n,\\
	&\delta \hat J^i = \bar{\hat n}\,\delta u^i - \hat \Sigma \, \delta\hat F^{i0}-\hat \Sigma\,\delta^{ij}\left(\partial_j\delta \hat\mu-\dfrac{\bar{\hat\mu}}{\bar T}\partial_j\delta T\right),
\end{align}
where we have introduced $\eta^{ij}=\delta^{ij}$ as the flat metric in the spatial directions. Moreover, we recall that the SU(2) indices are denoted by $a,b,c,d$ and run over $(+,-,3)$ after the change to the charged basis. The linearized Abelian and non-Abelian field strengths are
\begin{align}
	&\delta F_b^{i0} = \partial^i\delta A^0_b-\partial^0\delta A^i_b+ f_{b}{}^{cd}(\delta A_c^i \bar A_d^0+\bar A_c^i\delta A_d^0),\\
	&\delta\hat F^{i0} = \partial^i\delta \hat A^0-\partial^0\delta\hat A^i\,.
\end{align}

In the following, we relate the densities $\hat n,n_a,\varepsilon$ to the thermodynamic conjugate variables $\hat \mu,\mu_a,T$. Since we are interested in fluctuations around a background with non-zero Abelian density and non-zero density in the third SU(2) direction, it is useful to organize the thermodynamic variables directly in the charged basis. We define the vector of extensive thermodynamic variables,
\begin{equation}
	X^I=(\hat n,n_3,\varepsilon,n_+,n_-)\,,
\end{equation}
and the corresponding conjugate (intensive) thermodynamic variables
\begin{equation}
	Y^I=(\hat \mu,\mu_3,T,\mu_+,\mu_-)\,.
\end{equation}
We recall that here $\hat n$ and $\hat \mu$ refer to the Abelian U(1) sector, $n_3$ and $\mu_3$ to the neutral SU(2) direction generated by $t_3$, and $n_\pm$, $\mu_\pm$ to the charged SU(2) directions. The index $I$ labels thermodynamic variables and should not be confused with a space-time index or with an adjoint SU(2) index.

The equilibrium state considered here satisfies
\begin{equation}
	\bar n_\pm=0\,,
	\qquad
	\bar\mu_\pm=0\,,
\end{equation}
while $\bar{\hat n}$, $\bar n_3$, $\bar{\hat \mu}$ and $\bar\mu_3$ are non-zero. All thermodynamic derivatives below are evaluated on this equilibrium background.

Locally, the equation of state says that the densities are functions of the chemical potentials and temperature: $X^I =X^I(Y^J)$. At linear order, the fluctuations are related by the susceptibility matrix
\begin{equation}
	\delta X^I
	=
	\mathcal X^I{}_{J}\,\delta Y^J\,,
	\qquad
	\mathcal X^I{}_{J}
	=
	\left.
	\frac{\partial X^I}{\partial Y^J}
	\right|_{\mathrm{eq}}\,.
\end{equation}
The above notation $...|_\text{eq}$ means $\hat\mu=\bar{\hat \mu}$, $\mu_3=\bar \mu_3$, $T = \bar T$, and $\mu_\pm =0$. Equivalently,
\begin{equation}
	\delta Y^I
	=
	\mathcal M^I{}_{J}\,\delta X^J\,,
	\qquad
	\mathcal M=\mathcal X^{-1}\,.
\end{equation}

The form of $\mathcal X$ is constrained by the symmetry left unbroken by the background. Since the background points only in the $t_3$ direction, it is invariant under the Cartan subgroup U(1)$_{I_3}\subset$ SU(2) generated by $t_3$. The variables $n_q$, $n_3$, $\varepsilon$, $\mu_q$, $\mu_3$ and $T$ are neutral under this U(1)$_{I_3}$, while $n_\pm$ and $\mu_\pm$ carry charge $\pm1$. A linear susceptibility coefficient evaluated on the equilibrium background is itself neutral under U(1)$_{I_3}$. Therefore it can be non-zero only if it relates two fluctuations with the same U(1)$_{I_3}$ charge. In particular, neutral fluctuations cannot mix linearly with charged fluctuations, and the $+$ and $-$ sectors cannot mix with each other.

With the ordering chosen above, the susceptibility matrix therefore takes the block-diagonal form
\begin{equation}
	\mathcal X
	=
	\begin{pmatrix}
		\mathcal X_{\mathrm{N}} & 0 & 0 \\
		0 & \chi_+ & 0 \\
		0 & 0 & \chi_-
	\end{pmatrix}_{\mathrm{eq}}\,.
\end{equation}
Here $\mathcal X_{\mathrm{N}}$ is the neutral susceptibility matrix,
\begin{equation}
	\mathcal X_{\mathrm{N}}
	=
	\left[
	\frac{\partial(n_q,n_3,\varepsilon)}
	{\partial(\mu_q,\mu_3,T)}
	\right]_{\mathrm{eq}}\,,
\end{equation}
and $\chi_\pm$ are the charged susceptibilities
\begin{equation}
	\chi_\pm = \dfrac{\partial n_\pm}{\partial\mu_\pm}\bigg|_\text{eq}\,.
\end{equation}
In components, the charged part of the linearized equation of state is
\begin{equation}
	\delta n_\pm
	=
	\chi_\pm\,\delta\mu_\pm\,.
\end{equation}
Equivalently, using the inverse susceptibility matrix,
\begin{equation}
	\delta\mu_\pm
	=
	\dfrac{1}{\chi_\pm}\,\delta n_\pm\,.
\end{equation}
In addition, the residual rotational symmetry in the plane transverse to the third isospin direction forces the charged susceptibilities to be equal
\begin{equation}
	\chi_+ = \chi_- \equiv \chi_\text{C}\,.
\end{equation}

Since a non-zero $\mu_3$ breaks the SU(2) symmetry, non-Abelian hydrodynamics is an expansion around a background with $\mu_3=0$, for which the full SU(2) symmetry is preserved. In other words, the hydrodynamic approximation is an expansion in gradients and, since the charged currents are covariantly conserved \eqref{eq:DJpm}, we have to assume that the background value for $\mu_3$ is also small. In such a case,  the conductivity matrix is constrained by symmetry to be proportional to the invariant group metric,
\begin{equation}
	\label{sigma_ab}
	\Sigma_{ab}=\mathbf{g}_{ab}\,\Sigma \, ,
\end{equation}
where $\mathbf{g}_{ab}$ is defined in \eqref{eq:groupmetric}.

Therefore, we can write the charged current fluctuations derived from the constitutive relations as
\begin{align}
	&\delta J^0_\pm = \delta n_\pm,\\
	&\delta J^i_\pm = -i\,\Sigma\left(k^i\,\delta A^0_\pm+\dfrac{k^i}{\chi_{\mathrm{C}}}\delta n_\pm-\omega\,\delta A^i_\pm\right)\,,
\end{align}
where we used that $\Sigma_\pm{}^\pm =\Sigma$ using equation \eqref{sigma_ab}.

\subsection{Charged current correlators in the hydrodynamic limit}
\label{sec:expr2pt}

In this subsection, we shall use the linearized non-Abelian hydrodynamic equations, derived in the previous subsection, to compute the hydrodynamic approximation to the charged-current retarded correlator. We first introduce the general form of the two-point function, as determined by the symmetries of our system, before proceeding to the calculation in the hydrodynamic regime.

The general tensor structure of the flavor current two-point function, can be inferred from the symmetries of the background. The finite temperature plasma is invariant under SO(3) spatial rotations, as well as flavor transformations, which imply that the current correlators obey the corresponding Ward identities.

In the presence of general boundary sources $\bar{A}^a_\mu$, the flavor Ward identities correspond to covariant conservation \eqref{DJ}
\begin{equation}
	\label{WI} \pa_{\mu}J^{\mu} - i [\bar{A}_\m,J^\m] = 0 \, ,
\end{equation}
which is valid in correlators up to contact terms. The full Ward identity for the current two-point function including contact terms is derived in appendix \ref{app:WI}.

For the background of interest, written explicitly in \eqref{eq:bkgJ} and \eqref{eq:bkgA}, the equation \eqref{WI} yields in Fourier space,
\begin{equation}
	\label{WIf} k_{\mu}J^{\mu}=\frac{\m_3}{2}\left[\sigma_3,J^{0}\right] \, ,
\end{equation}
since the only non-Abelian source is $\bar{A}_{0}^3 = \m_3$.\footnote{We can generalize the above expressions to a general isospin source $\bar A_\mu = A_\mu^3\sigma_3$, for which the Ward identity will look as \eqref{ccc} but with
	\begin{equation}
		k_\mu^\pm = k_\mu\mp \bar A_\mu^3\,.
\end{equation}} For the charged currents,\footnote{Details on the notations for the charged fields can be found in appendix \ref{app:notationpm}.} \eqref{WI} can be written as
\begin{equation}\label{ccc}
	k_{\mu}^{\pm}J^{\mu,\pm}=0\, ,
\end{equation}
where we recall that the charged currents $J^{\m,\pm}$  are defined in \eqref{eq:chargedbasis}, and we have defined
\be\label{eq:shiftedk}
k_{\mu}^{\pm}=k_{\mu}\mp\m_3\delta^{0}\,_{\mu}\sp k_\mu = (-\omega,\vec k)\, .
\ee
Note that the index $\pm$ of the momentum vector is not a group index, but it refers to the $\mu_3$-dependent shift.

Taking into account the Ward identities \eqref{ccc}, the charged current two-point function can then be decomposed into a longitudinal and a part transverse to the spatial three-momentum $\vec{k}$ (see \ref{Sec:TwoPointDec}  for a derivation) as
\begin{align}
	\nn\label{ltcR} \left<J^{\pm}_\l J^{\mp}_\s\right>^R(\omega,\vec{k}) =& \left(P_{\l\s}^{\perp,\pm}(\omega,\vec{k}) i\Pi^{\perp,\pm}(\omega,\vec{k}) + P_{\l\s}^{\parallel,\pm}(\omega,\vec{k}) i\Pi^{\parallel,\pm}(\omega,\vec{k}) \right) \mp\\ &\mp i\delta_\l^0\delta_\s^0\dfrac{n_3}{\omega\pm\mu_3} \, .
\end{align}
where $n_3$ is the background isospin density. The last term is the contact term due to the contact contribution in the Ward identity for the current-current correlator. A derivation of this contact term can be found in appendix \ref{app:WI}.

The transverse and longitudinal projectors are defined with respect to the shifted momentum $k_\mu^\pm$, defined in \eqref{eq:shiftedk}.  We give a derivation of this decomposition in appendix \ref{Sec:TwoPointDec}. The non-zero components of $P^{\perp,\pm}_{\lambda\sigma}$ and $P^ {\parallel,\pm}_{\lambda\sigma}$ are
\be
\label{Po}  P^{\perp,\pm}_{ij}(\omega,\vec{k})=P^\perp_{ij}(\omega,\vec{k}) = \d_{ij} - \frac{k_ik_j}{\vec{k}^2} \, ,
\ee
\be
\label{Pp} P^{\parallel,\pm}_{00} = \frac{\vec{k}^2}{\left(\omega\pm\m_3\right)^2-\vec{k}^2} \,\, , \,\, \,\,\,\,P^{\parallel,\pm}_{0i}= P^{\parallel,\pm}_{i0} =- \frac{\left(\omega\pm\m_3\right) k_i}{\left(\omega\pm\m_3\right)^2-\vec{k}^2} \,\, , \,\,
\ee
\be
P^{\parallel,\pm}_{ij} = \frac{k_ik_j}{\vec{k}^2}\frac{\left(\omega\pm\m_3\right)^2}{\left(\omega\pm\m_3\right)^2-\vec{k}^2} \, .
\ee
The other components are zero
\begin{align}
	P^{\perp,\pm}_{00} = P^{\perp,\pm}_{0i}=P^{\perp,\pm}_{i0}=0.
\end{align}
The sum of the two projectors, gives the flat four-dimensional projector transverse to $k^{\pm}_\mu$
\be
\label{P4} P_{\mu\nu}^{\perp,\pm} + P_{\mu\nu}^{\parallel,\pm} = \eta_{\mu\nu} - \frac{k^{\pm}_\mu k^{\pm}_\nu}{\left(k^{\pm}\right)^2} \equiv P_{\mu\nu}^{\pm}\,,
\ee
where we recall that
\be
k^{\pm}_\mu = \left(-\omega\mp\m_3,\vec{k}\right) \, .
\ee
The presence of non-zero $\mu_3$ does not introduce a parity-breaking term, so the action is still symmetric under parity $(\vec{x}\leftrightarrow -\vec{x})$. Therefore,  we did not include any term involving the Levi-Civita tensor in \eqref{ltcR}. Also, the polarization functions have the following properties :
\begin{itemize}
	\item At $\vec{k}=0$, the transverse and longitudinal directions cannot be distinguished anymore, so that the two-point function has to be written as
	\be
	\label{eqpt1} \left<J^{\pm}_\l J^{\mp}_\s\right>^R(\omega,0) =  P_{\l\s}^{\pm}\, i\Pi^{\pm}(\omega) + \text{contact term} \, ,
	\ee
	which implies that
	\be
	\label{eqpt2} \Pi^{\parallel,\pm}(\omega,0) = \Pi^{\perp,\pm}(\omega,0) \, .
	\ee
	\item Near the special points where $\vec k^2=(\omega\pm\m_3)^2$, the longitudinal polarization function $\Pi^{\parallel,\pm}$ has to vanish as
	\be\label{eqpt3}
	\Pi^{\parallel,\pm}\big|_{(k_0^{\pm})^2\to\vec{k}^2}\sim(\omega\pm\m_3)^2-\vec k^2\, .
	\ee
	This is due to the divergence of $P^{\parallel,\pm}_{\mu\nu}\big|_{(k_0^{\pm})^2\to\vec{k}^2}$, while the retarded two-point function should stay regular.
	\item The retarded correlator obeys the Onsager reciprocity relations (we give a derivation in appendix \ref{app:Ons}) from which we can infer the following relations for the polarization functions
	\begin{align}
		&\text{Im}\Pi^{\perp,\pm}(\mu_3;\omega) = - \text{Im}\Pi^{\perp,\mp}(\mu_3;-\omega)\,,\\
		&\text{Im}\Pi^{\parallel,\pm}(\mu_3;\omega) = - \text{Im}\Pi^{\parallel,\mp}(\mu_3;-\omega)\,.
	\end{align}
\end{itemize}

\subsubsection{The hydrodynamic limit}
We shall now use the linearized hydrodynamic equations, derived in section \ref{sec:NAhydro} to derive the analytic expression for the polarization functions in the hydrodynamic limit.

First, we focus on the longitudinal sector of the retarded correlator. We remind the reader  that the charged current fluctuations, derived from the constitutive relations read
\begin{align}
	&\delta J^0_\pm = \delta n_\pm,\\
	&\delta J^i_\pm = -i\,\Sigma\left(k^i\,\delta A^0_\pm+\dfrac{k^i}{\chi_{\mathrm{C}}}\delta n_\pm-\omega\,\delta A^i_\pm\right)\,,
\end{align}
We can now substitute them in the covariant conservation equation \eqref{eq:consdeltaJ} for the charged current fluctuation
\begin{equation}\label{eq:timefluct}
	-(\omega\mp\bar \mu_3)\delta J^0_\pm-i\,\Sigma\,\vec k^2\delta A^0_\pm+i\Sigma\,\omega k_i\delta A^i_\pm-i\vec k^2D\,\delta J^0_\pm\pm\bar n_3\delta A^0_\pm=0,
\end{equation}
where we have defined the diffusivity
\begin{equation}
	D = \dfrac{\Sigma}{\chi_{\mathrm{C}}}\,.
\end{equation}

Note that we can rewrite the background isospin density in terms of $\Sigma$ and $D$ as follows
\begin{equation}\label{eq:n3DS}
	\bar n_3=\chi_{\mathrm C}\bar\mu_3=\dfrac{\Sigma\,\bar\mu_3}{D}\,.
\end{equation}
This relation follows from the SU(2)-invariance of the equation of state. Before selecting the equilibrium direction, the pressure is an SU(2)-invariant function of the isospin chemical-potential vector and can therefore be written as $p=p(T,\hat \mu,\mu_a\mu^a)$. The density $\bar n_3$ is obtained by varying $\mu_3$ while holding $T$, $\hat\mu$, and $\mu_\pm$ fixed, which gives $\bar n_3=2\bar\mu_3\,\partial p/\partial(\mu_a\mu^a)$. Since the charged susceptibility can be computed as $\chi_{\mathrm C}=2\partial p/\partial(\mu_a\mu^a)$, we can write $\bar n_3=\chi_{\mathrm C}\bar\mu_3$. Using $D=\Sigma/\chi_{\mathrm C}$, this can also be written as $\bar n_3=\Sigma\,\bar\mu_3/D$.

The time-time component of the retarded two-point function can be extracted from equation \eqref{eq:timefluct}:
\be\label{eq:hydro2pt}
\langle J^{0,\pm}J^{0,\mp}\rangle^R = \dfrac{1}{2i}\dfrac{\delta J^{0,\pm}}{\delta A_{0,\mp}} = i\dfrac{\delta J^{0}_\mp}{\delta A^0_{\mp}} = -i\,\dfrac{\pm\bar n_3-i\,\Sigma\,\vec k^2}{\omega\pm\bar\mu_3+i\,D \vec k^2}\,,
\ee
where we used the group metric \eqref{eq:groupmetric} to lower flavor indices. Equation \eqref{eq:hydro2pt} has a pole at
\begin{equation}
	\omega_\pm = \mp\bar\mu_3-iD\,\vec k^2\, ,
\end{equation}
which corresponds to a diffusive hydrodynamic mode. Note that the isospin chemical potential $\bar{\m}_3$, implies a real shift of the diffusive pole, therefore implying slow oscillations together with the standard diffusive behaviour.

Now, the leading hydrodynamic approximation to the longitudinal polarization function, defined in \eqref{ltcR}, can finally be computed, by subtracting the contact term from \eqref{eq:hydro2pt}. This  gives
\be\label{H2}
\Pi^{\parallel,\pm}(\omega,\vec k) = -\frac{i\,\Sigma\, \omega((\omega \pm\bar{\m}_3)^2-\vec k^2)}{(\omega \pm\bar{\m}_3+ iD\,\vec{k}^2)(\omega\pm\bar\m_3)}\,,
\ee
where we have used \eqref{eq:n3DS}.

The calculation of the transverse polarization function is much simpler than the longitudinal case, since it can be directly identified from the constitutive relation
\begin{equation}  \delta J^{\perp}_\pm=i\omega\Sigma\, \delta A^{\perp}_{\pm}\,,
\end{equation}
which gives
\be
\label{H3} \Pi^{\perp,\pm}(\omega,\vec k) = -i \omega \Sigma \,.
\ee

The equilibrium current correlators are fully determined by the spectral function, which is proportional to the imaginary parts of the charged current polarization functions (see e.g. section 2 of \cite{neutrinopaper}). Therefore, our focus will be on the imaginary parts, which are given by
\begin{equation}
	\label{ImPiL}
	\text{Im}\Pi^{\parallel,\pm}(\omega,\vec k) =\frac{\Sigma\,\omega(\vec{k}^2-(\omega\pm\bar{\mu}_3)^2)}{(\omega\pm\bar{\m}_3)^2+ (D\,\vec{k}^2)^2}\,,
\end{equation}
\begin{equation}
	\label{ImPiT}
	\text{Im}\Pi^{\perp,\pm}(\omega,\vec k) =	-\Sigma\, \omega\,.
\end{equation}

Note, that when $\vec{k}=0$, the transverse and longitudinal polarization functions are equal, in agreement with \eqref{eqpt2}. Also, in the limit of $\bar{\m}_3\to0$, the polarization functions reduce to the U(1) case \cite{neutrinopaper}.

The shape of the correlators \eqref{H2} and \eqref{H3} is determined by hydrodynamics, but the transport coefficients $D$ and $\Sigma$ are computed from the microscopic theory. They will be computed for the holographic system of interest in the next section.

\section{The holographic theory}
\label{sec:holomodel}
In this section, we introduce the holographic model, that is used in this work to compute charged current-current correlators. It is the simplest holographic model describing the dynamics of chiral current operators, \cite{JKNP}.\footnote{The diagonal U(2) model that we are analysing here is a consistent truncation of the holographic theory dual to N=4 sYM, where the global symmetry is a subgroup of the R-symmetry.}

We consider a strongly interacting medium in a theory with $N_f$ quarks and U$(N_f)_L\times \text{U}(N_f)_R$ chiral flavor symmetry. According to the holographic duality, this theory is dual to a five-dimensional gravitational theory that lives on an asymptotically Anti-de Sitter space AdS$_5$, which is a constant negative curvature space with a four-dimensional time-like boundary.

The field content of the bulk theory is dictated by the types of operators that we want the dual (boundary) quantum field theory to include. In the present case, in addition to the metric dual to the stress-energy tensor, the operators of interest are the chiral flavor currents $\mathcal{J}^{(L/R)}_\m$, which are dual to chiral gauge fields in the five-dimensional bulk $\mathbf{L}_M$ and $\mathbf{R}_M$, with  $M$ the five-dimensional space-time index. These gauge fields, are elements of the Lie algebra of the chiral group U($N_f$)$_L\times$ U($N_f$)$_R$.

The bulk gravitational action controlling the dynamics of these fields is then constructed as the sum of a color and a flavor part
\be
\label{eq:Sb} S = S_c + S_f \, .
\ee
The action for the color sector is the five-dimensional Einstein-Hilbert action
\be
\label{eq:SEH} S_{\text{c}} = M^3N_c^2 \int\mathrm{d}^5x\sqrt{-g}\, \left(R + \frac{12}{\ell^2}\right) \, ,
\ee
where $R$ is the five-dimensional Ricci scalar, $M$ the five-dimensional Planck mass, $\ell$ the AdS$_5$ radius and $N_c$ the number of colors. We choose to describe the flavor sector  with a quadratic Yang-Mills action for the chiral gauge fields
\be
\label{eq:SYM} S_f = - \frac{1}{8\ell}(M\ell)^3w_0^2 N_c \int\mathrm{d}^5x\sqrt{-g}\,\text{Tr}\left(\mathbf{F}_{MN}^{(L)}\mathbf{F}^{MN,(L)} + \mathbf{F}_{MN}^{(R)}\mathbf{F}^{MN,(R)} \right) \, ,
\ee
where $w_0$ controls the (inverse) flavor Yang-Mills coupling, and $\mathbf{F}^{(L/R)}$ are the field strengths of the gauge fields $\mathbf{L}$ and $\mathbf{R}$ defined as
\be
\label{eq:defF} \mathbf{F}^{(L)} \equiv \mathrm{d}\mathbf{L} - i\mathbf{L}\wedge \mathbf{L} \sp \mathbf{F}^{(R)} \equiv \mathrm{d}\mathbf{R} - i\mathbf{R}\wedge \mathbf{R} \, .
\ee

As is standard in holographic setups, we assume a large number of colors $N_c$ so that the bulk theory admits a semi-classical description. Since our goal is to describe strongly interacting dense matter, the back-reaction of the flavor sector on the gluonic sector cannot be neglected. To ensure that this back-reaction remains finite, we work in the Veneziano large-$N_c$ limit~\cite{veneziano}
\be
\label{eq:V1}
N_c \to \infty \,, \qquad N_f \to \infty \,, \qquad \frac{N_f}{N_c} \ \text{fixed} \, .
\ee
Although this model is a toy model, it is a simple first step to doing the calculations we do, before we go to the most appropriate  bottom-up model for holographic QCD, the V-QCD model, \cite{V1}.
The model here has conformal glue dynamics, and no chiral symmetry breaking. However, it provides a good toy model for the finite density effects we are after. In appendix \ref{app:thermo}, we described how the parameters $M$, $\ell$ and $w_0$ are determined by comparing to QCD.

Although both $N_c$ and $N_f$ are assumed to be large, after calculations, the results are extrapolated to $N_c=3$. Moreover we can always look at the subsection with $N_f=2$ although there may be more flavors.

For a chiral symmetry group $\mathrm{U}(2)_L \times \mathrm{U}(2)_R$, the chiral currents and their dual bulk gauge fields can be expanded in the Pauli basis $\{\sigma_a\}$, with $a=1,2,3$,
\be\label{eq:JsP}
\mathcal{J}^{(L)}_\mu = \frac{1}{2}\hat{J}^{(L)}_\mu\mathbb{I}_2 + \frac{1}{2}\sum_{a=1}^3 J^{a,(L)}_\mu \s_a \sp \mathbf{L}_M = \frac{1}{2}\hat{L}_M\mathbb{I}_2 + \frac{1}{2}\sum_{a=1}^3 L^{a}_M \s_a \, ,
\ee
\be
\mathcal{J}^{(R)}_\mu = \frac{1}{2}\hat{J}^{(R)}_\mu\mathbb{I}_2 + \frac{1}{2}\sum_{a=1}^3 J^{a,(R)}_\mu \s_a \sp \mathbf{R}_M = \frac{1}{2}\hat{R}_M\mathbb{I}_2 + \frac{1}{2}\sum_{a=1}^3 R^{a}_M \s_a \, .
\ee

From now, on we shall only consider the vector currents, $\mathcal{J}_{\mu}^{(V)}$, defined in (\ref{eq:Jdefinition}) and their dual bulk vector gauge fields, $\mathbf{V}_M$.
\be
\mathcal{J}^{(V)}_\mu = \frac{1}{2}(\mathcal{J}_{\mu}^{(L)}+\mathcal{J}_{\mu}^{(R)}) \sp \mathbf{V}_M =\dfrac{1}{2}(\mathbf{L}_M+\mathbf{R}_M)\, .
\ee
They are associated with the vector symmetry group U(1)$_V$ $\times$ SU(2)$_V$, which is the unbroken symmetry group in QCD. This symmetry group is exactly the one discussed in the previous section. We stress that our calculations in this theory, are also valid individually for left and right handed currents.

\subsection{Background solution}
\label{sec:bkgsol}
We now present the background solution for the bulk action \eqref{eq:Sb}, at finite temperature and density. The state of matter that it describes in the dual boundary theory, corresponds to a plasma of deconfined (generalized) quarks and gluons. Introducing a finite density of deconfined baryonic matter, is equivalent to sourcing the bulk baryon number gauge field with a baryon number chemical potential
\be
\label{eq:smq} \left.\hat V_0\right|_{\text {boundary }}=\mu_q \, ,
\ee
where $\mu_q$ is the quark number chemical potential, related to the baryon number chemical potential by $\mu_B=N_c \mu_q$. We shall also be interested in states featuring isospin asymmetry, which is introduced by sourcing the isospin gauge field, with an isospin chemical potential $\mu_3$
\be
\label{eq:sm3} \left.V^3_0\right|_{\text {boundary }}=\mu_3 .
\ee

The background solution then corresponds to the solution of the Einstein-Yang-Mills equations of motion from \eqref{eq:Sb}-\eqref{eq:SYM}, with the UV boundary conditions \eqref{eq:smq} and \eqref{eq:sm3}, together with appropriate regularity conditions in the IR. The solution takes the form of an asymptotically $\operatorname{AdS}_{5}$ Reissner-Nordström (RN) black-hole, which can be computed by starting from the following ansatz for the metric and gauge fields
\be
\label{eq:ds2RN} \mathrm{ds}^2 = \ex^{2A(r)}\left( -f(r)\mathrm{dt}^2 + f(r)^{-1}\mathrm{dr}^2 + \vec{\mathrm{d x}}^2 \right) \, ,
\ee
\be
\label{eq:Ba5} \mathbf{V}_M = \frac{1}{2}\,\d^{0}\,_M\, \left(\Phi(r)\mathbb{I}_2+\Phi_3(r)\sigma_3\right) \, .
\ee
The holographic coordinate $r$ is defined such that $r=0$ is the UV boundary, and there is a horizon at $r=r_H$, where the blackening function vanishes ($f(r_H)=0$).
The functions $A(r)$, $f(r)$, $\Phi(r)$ and $\Phi_3(r)$ obey the following system of equations
\be
\partial^2_r A - (\partial_r A)^2 = 0 \, ,
\ee
\be
\partial_r A\big(\partial_r f + 4\partial_r A f(r)\big) - \frac{4}{\ell^2}\ex^{2A(r)} + \frac{w_0^2\ell^2}{48N_c}\ex^{-2A(r)} \left((\partial_r \Phi(r))^2+(\partial_r \Phi_3(r))^2\right) = 0 \, ,
\ee
\be
 \partial_r\Big(\ex^{A(r)} \partial_r \Phi(r)\Big) = 0  \,,\,\quad \partial_r\Big(\ex^{A(r)} \partial_r \Phi_3(r)\Big) = 0\, ,
\ee
which is solved by
\be\label{bmet}
\ex^{A(r)} = \frac{\ell}{r} \sp f(r) = 1 - \left(\frac{r}{r_H}\right)^4\left(1 + 2\left(1 - \pi T r_H\right)\right) + 2\left(1 - \pi T r_H\right)\left(\frac{r}{r_H}\right)^6  \, ,
\ee
\be
\label{eq3:Phiasrh}
\Phi(r)=\mu_q\left(1-\left(\frac{r}{r_H}\right)^2\right) \quad, \quad \Phi_3(r)=\mu_3\left(1-\left(\frac{r}{r_H}\right)^2\right)\, .
\ee
Importantly, note that the dependence of the metric on $\mu_3$ reduces to the dependence of the horizon radius $r_H = r_H(T,\mu_q,\mu_3)$, with $T$ the temperature, which is identified in the usual way from the horizon surface gravity. Defining the total chemical potential $\m$ as
\be
\mu \equiv \sqrt{\mu_q^2+\mu_3^2}\, ,
\ee
the horizon radius takes the form
\be
\label{eq:rHRN} r_H = \frac{2}{\pi T} \left( 1 + \sqrt{1 + \frac{w_0^2}{3N_c}\frac{\m^2}{\pi^2T^2}} \right)^{-1}\simeq \dfrac{2\sqrt{3}\sqrt{N_c}}{w_0}\;\dfrac{1}{ \mu} \, .
\ee
where the approximate relation is valid in the near horizon limit $T\ll \mu$. For the values that are obtained from comparison to QCD, as discussed in appendix  \ref{app:thermo}, $w_0/\sqrt{N_c}\simeq 2.15$.

The background solution for the metric at finite $\mu_3$ is therefore AdS$_5$ RN, but the effective chemical potential entering the expression for the horizon radius is $\mu = \sqrt{\mu_q^2+\mu_3^2} >\mu_q$. In addition, there is a non-zero background isospin gauge field \eqref{eq3:Phiasrh} which will enter the fluctuation equations of motion for the charged gauge field, as we shall observe in the following sections. More details on the thermodynamics of this solution are provided in appendix \ref{app:thermo}.

\section{Holographic charged current correlators}

\label{sec:holoccc}

In this section, we present the calculation of the retarded two-point function for the charged flavor current, $J^{\pm}_{(V)}$, in the holographic model presented in section \ref{sec:holomodel}. We follow the standard prescription of \cite{Son:2002sd}, and study the linearized field equations for small perturbations $\d V^{1,2}$ of the bulk gauge fields dual to the flavor vector currents $J_{(V)}^{1,2}$
\be
\label{eq:EoML} \partial_M\left(\sqrt{-g} \,\delta F_{(V)}^{M N}\right)-i\sqrt{-g}\left[\delta V_{M},F_{(V)}^{MN}\right]-i\sqrt{-g}\left[ V_{M},\delta F_{(V)}^{MN}\right] = 0 \, ,
\ee
where
\be
\label{eq:deltaF}
\delta F^{(V)}_{MN}=\partial_M\delta V_N -\partial_N \delta V_M-i\left[\delta V_{M},V_N\right]-i\left[\delta V_{N},V_M\right]\, .
\ee
We choose the axial gauge
\be
\label{eq:Ag} V^{a}_r=0 \, ,
\ee
and define the four-dimensional Fourier transform of the perturbation as
\be
\label{eq:ansL} \d V^a_\mu(r;t,\vec{x}) = \int \frac{\intd k^4}{(2\pi)^4} \ex^{-i(\omega t - \vec{k}.\vec{x})} \mathcal{L}^a_{\mu,k}(r) \, .
\ee
From now on, we drop the $k$ subscript on the gauge field fluctuation to simplify the notation.

It will be useful to decompose the spatial part of the gauge field perturbation into a transverse $\mathcal{L}_i^\perp$, and longitudinal part $\mathcal{L}_i^\parallel$ to the 3-momentum $\vec{k}$ as follows
\begin{equation}
 \mathcal{L}_i^{\parallel} = \dfrac{k_i}{\sqrt{\vec k^2}}\mathcal{L}_i \quad,\quad \mathcal{L}_i^{\perp} = \mathcal{L}_i -\dfrac{k_i}{\sqrt{\vec k^2}}\mathcal{L}_i\,.
\end{equation}
Now, we write the equations of motion \eqref{eq:EoML} in the basis \eqref{eq:chargedbasis} for the charged fields
\be
\label{pm_coords}
\mathcal{L}_M^{\pm}=\mathcal{L}^1_{M} \mp i \mathcal{L}^2_{M}\, ,
\ee
which receive contributions from commutators with the non-zero background isospin gauge field. Note that we do not write the equations of motion for the Abelian and the isospin components of the fluctuations, which couple to the metric fluctuations. This will be the focus of a follow-up work.

In the axial gauge \eqref{eq:Ag}, the $N=r$ component of the equation of motion \eqref{eq:EoML} implies the constraint
\be
\label{EoMLrpm} \partial_r\mathcal{L}_0^{\pm} + \frac{f(r)}{\omega}\partial_r\left(k_i\mathcal{L}_i^{\pm}\right)\pm\frac{1}{\omega}\left(\Phi_3\partial_r\mathcal{L}_0^{\pm}-\Phi_3^{\prime}\mathcal{L}_0^{\pm}\right) = 0 \, ,
\ee
while the other components of the equation of motion \eqref{eq:EoML} are, for the transverse part
\be
\label{EoMVhtpm} \partial_r\left(\frac{f(r)}{r}\partial_r\mathcal{L}_i^{\perp,\pm}\right) + \frac{1}{rf(r)}\left(\Omega_\pm(r)^2-\vec{k}^2f(r)\right)\mathcal{L}_i^{\perp,\pm}=0\, ,
\ee
and for the longitudinal part
\be
\label{EoMVh0pm} \partial_r\left(\frac{1}{r} \partial_r\mathcal{L}_0^{\pm}\right) - \frac{\sqrt{\vec{k}^2}}{rf(r)} E^{\parallel,\pm} = 0 \, ,
\ee
\be
\label{EoMVhpipm} \partial_r\left(\frac{f(r)}{r}\partial_r\mathcal{L}_i^{\parallel,\pm}\right) + \frac{\omega}{rf(r)}\frac{k_i}{\sqrt{\vec{k}^2}}E^{\parallel,\pm}+\frac{\Phi_3}{rf}\left(\pm k_i\mathcal{L}_0^{\pm}+\left(\Phi_3\pm\omega\right)\mathcal{L}_i^{\parallel,\pm}\right)= 0 \, ,.
\ee
In the equations above, we have denoted the fluctuation of the longitudinal electric field as
\be
\label{defEpm} E^{\parallel,\pm} \equiv i \frac{2k_i}{\sqrt{\vec{k}^2}}\delta F_{0i}^{\pm}= \sqrt{\vec{k}^2}\mathcal{L}_0^{\pm} + \frac{\omega}{\sqrt{\vec{k}^2}}\left(k_j\mathcal{L}_j^{\pm}\right)\pm\frac{\Phi_3}{\sqrt{\vec{k}^2}}k_j\mathcal{L}_j^{\pm} \, ,
\ee
and defined
\begin{equation}
\label{defOpm} \Omega_\pm(r) \equiv \omega \pm \Phi_3(r) \, .
\end{equation}
Equations \eqref{EoMLrpm} and \eqref{EoMVh0pm} can be combined into a second order equation for $E^{\parallel,\pm}$
\begin{equation}
\label{EpEoM} \partial_r\left(\frac{ f(r) \partial_rE^{\parallel,\pm}}{r \big(\Omega_\pm(r)^2-f(r)k^2\big)}\right)+ \frac{E^{\parallel,\pm}}{r f(r)}\left(1 + f(r) \Phi_3'(r)\frac{2 f(r) \Phi_3'(r)\mp f'(r)\Omega_\pm(r)}{\big(\Omega_\pm(r)^2-f(r)k^2\big)^2}\right)=0\, ,
\end{equation}
which may be further simplified by considering the field redefinition
\begin{equation}
\label{fr} E^{\parallel\pm}(r) \equiv \Omega_\pm(r) \varphi^\pm(r).
\end{equation}
We obtain
\begin{equation}
\label{Eqvf} \partial_r\left(\frac{f(r) \Omega_\pm(r)^2 \pa_r\varphi^\pm}{r \big(\Omega_\pm(r)^2- f(r)k^2\big)}\right) + \frac{\Omega_\pm(r)^2}{rf(r)}\varphi^\pm = 0 \, .
\end{equation}
Note that the equations of motion for respectively $\mathcal{L}_i^{\perp,\pm}$ and $E^{\parallel,\pm}$ are also satisfied by $\mathcal{L}_i^{\perp,\mp}$ and $E^{\parallel,\mp}$ after the replacement $\mu_3\to - \mu_3$. As a consequence, we can solve for one charged fluctuation and obtain the other by flipping the sign of the parameter $\mu_3$.

The charged current retarded two-point function, is extracted from the solution to the equations of motion \eqref{EoMVhtpm} and \eqref{EpEoM}, with infalling boundary conditions at the horizon \cite{Son:2002sd}. Focusing on the vector gauge fields, the Yang-Mills action to quadratic order in the perturbations takes the form
\begin{equation}
\label{pert_quad_action}
 \delta^{(2)}S_f = - \frac{1}{16\ell}(M\ell)^3w_0^2 N_c \int\mathrm{d}^5x\sqrt{-g} \left(\delta F^{(V)}_{MN,a}\delta F^{MN,a}_{(V)} \right) \, ,
\end{equation}
where $\delta F^{MN}_{(V)}= g^{MP}g^{NQ}\delta F^{(V)}_{PQ}$ and the linearized Yang-Mills field strength with explicit group indices is given by
\begin{equation}
\delta F^{(V)a}_{MN}=\partial_M\delta V^{a}_{N}-\partial_N\delta V_{M}^{a}+\epsilon^{abc}\left(\delta V^{b}_{M}V^{c}_{N}+V^{b}_{M}\delta V^{c}_{N}\right)\, .
\end{equation}
Upon integrating the action \eqref{pert_quad_action} by parts, and focusing on the charged fluctuations, we obtain
\begin{align}
\nn \delta^{(2)}S^{\text{on-shell}}_f&\bigg|_\text{charged} = -\frac{1}{16\ell}(M\ell)^3w_0^2 N_c \int \frac{\intd^4k}{(2\pi)^4}\bigg[\frac{\ell f(r)}{r} \bigg\{\mathcal{L}_i^{\perp,\pm}(-k_\mu)\partial_r\mathcal{L}_i^{\perp,\mp}(k_\mu)+ \\
&+\mathcal{L}_i^{\parallel,\pm}(-k_\mu)\partial_r\mathcal{L}_i^{\parallel,\mp}(k_\mu)-\frac{1}{f(r)}\mathcal{L}_0^{\pm}(-k_\mu)\partial_r\mathcal{L}_0^{\mp}(k_\mu)\bigg\}\bigg]\bigg|^{r=r_H}_{r=0}\, ,
\end{align}
where we have isolated the transverse and longitudinal components of $\mathcal{L}_i^\pm$. We have adopted the following compact notation
\be
\mathcal{L}^\pm(-k_\mu)\partial_r\mathcal{L}^\mp(k_\mu)= \mathcal{L}^+(-k_\mu)\partial_r\mathcal{L}^-(k_\mu)+ \mathcal{L}^-(-k_\mu)\partial_r\mathcal{L}^+(k_\mu)\,.
\ee

It is then useful to write the action for the charged, longitudinal fluctuations in terms of the electric field $E^{\parallel,\pm}$ and $\mathcal{L}^\pm_0$. Using
\begin{align}
&\mathcal{L}_i^{\parallel,\pm}(k)=\dfrac{k_i}{\sqrt{\vec k^2 }}\dfrac{1}{\Omega_\pm}\left(E^{\parallel,\pm}(k)-\sqrt{\vec k^2 }\,\mathcal{L}_0^\pm(k)\right),\\
&\mathcal{L}_i^{\parallel,\pm}(-k)=\dfrac{k_i}{\sqrt{\vec k^2 }}\dfrac{1}{\Omega_\mp}\left(E^{\parallel,\mp}(-k)-\sqrt{\vec k^2 }\,\mathcal{L}_0^\mp(-k)\right)\,,
\end{align}
the action can be rewritten as
\begin{align}\nonumber
\delta^{(2)}S^{\text{on-shell}}_f&\bigg|_\text{charged}=\dfrac{1}{16\ell}(M\ell)^3w_0^2N_c\int\frac{\intd^4k}{(2\pi)^4}\bigg[\dfrac{\ell}{r}\bigg\{\mathcal{L}_i^{\perp,\pm}(-k_\mu)\partial_r\mathcal{L}^{\perp,\mp}_i(k_\mu)+\\ \nonumber
&+\dfrac{E^{\parallel,\pm}(-k_\mu)\partial_rE^{\parallel,\mp}(k_\mu)}{\omega_\mp^2-\vec k^2}+\\ \nonumber
&+\partial_r\Phi_3\left(\dfrac{E^{\parallel,+}(-k_\mu)E^{\parallel,-}(k_\mu)}{\omega_-(\omega_-^2-\vec k^2)}-\dfrac{E^{\parallel,-}(-k_\mu)E^{\parallel,+}(k_\mu)}{\omega_+(\omega_+^2-\vec k^2)}\right)+\\
&+\partial_r\Phi_3\left(\dfrac{\mathcal{L}_0^+(-k_\mu)\mathcal{L}_0^-(k_\mu)}{\omega_-}-\dfrac{\mathcal{L}_0^-(-k_\mu)\mathcal{L}_0^+(k_\mu)}{\omega_+}\right)\bigg\}\bigg]\bigg|_{r=\epsilon}\,,
\end{align}
where $\e\ll 1$ is a UV cutoff.

The above boundary action can also be rewritten as
\be\label{Sos2}
\delta^{(2)}S^{\text{on-shell}}_f\bigg|_\text{charged} = S_\text{nl}+S_\text{c} \, ,
\ee
where the non-local part $ S_\text{nl} $ is given by
\begin{align}
\nn S_\text{nl} = \frac{1}{16\ell}(M\ell)^3w_0^2 N_c &\int\frac{\intd^4k}{(2\pi)^4} \bigg[\frac{\ell}{r}\mathcal{L}^{\l,\mp}(-k_\mu)\left( P_{\l\s}^{\perp,\pm}(k_\mu) \frac{\partial_r\mathcal{L}_i^{\perp,\pm}(k_\mu)}{\mathcal{L}^{\perp,\pm}_i(k_\mu)}\right.+\\
&\left.+P_{\l\s}^{\parallel,\pm}(k_\mu) \frac{\partial_rE^{\parallel,\pm}(k_\mu)}{E^{\parallel,\pm}(k_\mu)}
 \right)\mathcal{L}^{\sigma,\pm}(k_\mu)\bigg]\bigg|_{r=\e}\, ,
\end{align}
and $S_\text{c}$ is the contact term action given by
\begin{align}\label{Snp}
 S_\text{c} = \frac{1}{16\ell}(M\ell)^3w_0^2 N_c\! \int\frac{\intd^4k}{(2\pi)^4} \bigg[\!\mp\!\frac{\ell}{r}\,\partial_r\Phi_3\,\mathcal{L}^{\l,\mp}(-k_\mu)\!\left(  \frac{P_{\l\s}^{\parallel,\pm}(k_\mu)+\delta^{0}_{\l}\delta^{0}_{\sigma}}{\omega\pm\m_3}\right)\!\mathcal{L}^{\sigma,\pm}(k_\mu)\bigg]\bigg|_{r=\e}.
\end{align}
Note that, in the contact term action $S_\text{c}$, the apparent singularity at $\omega \pm\mu_3=0$ cancels against zeroes in the numerator for each component of the longitudinal projector \eqref{Pp} so that the full expression is not singular.

According to the prescription of \cite{Son:2002sd}, this implies that the polarization functions for the flavor currents in \eqref{ltcR} are given by

\be
\label{PLt} \Pi^{\perp,\pm}(\omega,\vec{k}) = -\frac{1}{8}(M\ell)^3w_0^2 N_c \frac{1}{r} \left.\frac{\partial_r\mathcal{L}_i^{\perp,\pm}}{\mathcal{L}_i^{\perp,\pm}} \right|_{r=\e} \, ,
\ee
\begin{align}
\label{PLl} \Pi^{\parallel,\pm}(\omega,\vec{k}) &= -\frac{1}{8}(M\ell)^3w_0^2 N_c \frac{1}{r} \left.\left(\frac{\partial_rE^{\parallel,\pm}}{E^{\parallel,\pm}}\mp\frac{\partial_r\Phi_3}{\omega\pm\m_3}\right) \right|_{r=\e} \\ \non
&=-\frac{1}{8}(M\ell)^3w_0^2 N_c \frac{1}{r} \left.\frac{\partial_r\varphi^{\pm}}{\varphi^{\pm}}\right|_{r=\e}
\end{align}
Note that the contribution to the longitudinal polarization function, coming from the contact term action $S_\text{c}$ is purely real. In the last line,  we used the field $\varphi^\pm$ defined in \eqref{fr} to rewrite the polarization in a more compact way.

Whether the expressions \eqref{PLt} and \eqref{PLl} give finite results or need to be regularized and renormalized,  depends on the near-boundary behaviour of the solutions.

 The near-boundary behavior is obtained by solving the equations of motion for $\mathcal{L}_i^{\perp,\pm}$ \eqref{EoMVhtpm} and for $E^{\parallel,\pm}$ \eqref{EpEoM} around $r\to 0$ (the background is discussed in section \ref{sec:bkgsol}).

\begin{align}\label{Ltnb}
&\mathcal{L}^{\perp,\pm}_i = \mathcal{L}^{\perp,\pm,(0)}_i + r^2\left(\mathcal{L}^{\perp,\pm,(2)}_i - \frac{1}{2}\left((\omega\pm\mu_3)^2-\vec{k}^2\right)\mathcal{L}^{\perp,\pm,(0)}_i \log(r/\ell)\right)\Big(1 + \OO\big(r^2\big)\Big) \, ,\\ \label{Elnb}
&E^{\parallel,\pm} = E^{\parallel,\pm,(0)} + r^2\left(E^{\parallel,\pm,(2)} - \frac{1}{2}\left((\omega\pm\mu_3)^2-\vec{k}^2\right)E^{\parallel,\pm,(0)} \log(r/\ell)\right)\Big(1 + \OO\big(r^2\big)\Big) \, ,
\end{align}
with the two independent integration constants given by the source $\mathcal{L}^{\perp,\pm,(0)}_i,\, E^{\parallel,\pm,(0)}$ and vev terms $\mathcal{L}^{\perp,\pm,(2)}_i, E^{\parallel,\pm,(2)}$. To connect this language to section \ref{sec:NAcorr}, we recall that the source is intended as the boundary source for the dual charged current $J^\pm$ and with vev the one-point function  $\left<J^\pm\right>$.

The near-boundary behavior of the fluctuation fields implies that the polarization functions are subject to a logarithmic UV divergence. However, the log-term contributes only to the real part of the polarization functions, whereas the imaginary part (the spectral function) does not need to be renormalized, as it is finite. Since equilibrium two-point functions are fully determined by the spectral function (see e.g. section 2 of \cite{neutrinopaper}), we focus in this work on the imaginary part of the polarization functions, for which no renormalization  is required. Nonetheless, for completeness, in appendix \ref{app:holoren} we give the explicit expression of the required counterterms to regularize the polarization functions and we present their renormalized expression.

In terms of the near-boundary data, the imaginary part of the polarization functions is given by
\be
\label{PVtim} \text{Im}\Pi^{\perp,\pm}(\omega,\vec{k}) = -\frac{1}{4}(M\ell)^3w_0^2 N_c \text{Im}\frac{\mathcal{L}^{\perp,\pm,(2)}}{\mathcal{L}^{\perp,\pm,(0)}} \, ,
\ee
\be
\label{PLcim} \text{Im}\Pi^{\parallel,\pm}(\omega,\vec{k}) = -\frac{1}{4}(M\ell)^3w_0^2 N_c \text{Im}\frac{E^{\parallel,\pm,(2)}}{E^{\parallel,\pm,(0)}}  \, .
\ee

\section{Holographic charged current correlators in the hydrodynamic limit near extremality}\label{sec:holoNAcomputation}

In section \ref{sec:NAcorr}, we derived the low-energy approximation to the charged current correlators from non-Abelian hydrodynamics. The corresponding expressions for the imaginary part of the polarization functions are given by \eqref{ImPiL} and \eqref{ImPiT}, that we reproduce here for convenience,
\begin{equation}
\label{h2}
\text{Im}\Pi^{\parallel,\pm}(\omega,\vec k) =\frac{\Sigma\,\omega(\vec{k}^2-(\omega\pm\bar{\mu}_3)^2)}{(\omega\pm\bar{\m}_3)^2+ (D\vec{k}^2)^2}\,,
\end{equation}
\begin{equation}
\label{h3}
\text{Im}\Pi^{\perp,\pm}(\omega,\vec k) =	-\Sigma\, \omega\,.
\end{equation}
As mentioned before, the transport coefficients $\Sigma$ (the conductivity) and $D$ (the diffusivity) can only be computed from the full microscopic description of the theory. The expressions \eqref{h2}-\eqref{h3} were derived in \ref{sec:NAcorr} at leading order in the standard hydrodynamic regime $\omega,|\vec{k}|,\m_3\ll T$. However, we shall also be interested in a different kind of parametric  regime, for which the temperature is much smaller than the chemical potential $\mu$, and $\omega$, $|\vec{k}|$ and $\mu_3$ are not necessarily smaller than temperature, but remain negligible compared to the hard scale $\mu$. In other words, we consider the regime of so-called near-extremal hydrodynamics: $\omega,|\vec{k}|,T\ll \mu$ \cite{Edalati:2010pn,Davison:2011ek,DP13,Gouteraux:2025kta,Preau:2025rex}, with the additional condition that $\mu_3\ll \mu$ in our non-Abelian setup.

As already stressed in the introduction, holographic states, described in terms of a near-extremal black-hole, show the emergence of a hydrodynamic-like behavior (as in \eqref{h2}-\eqref{h3}) over long times and distances, which is valid up to the hard scale $\mu\gg T$ even when $T \ll \omega,|\vec{k}| \ll \mu$ \cite{Edalati:2010hk,Edalati:2010pn,Edalati:2010pn2,Davison:2011uk,Brattan:2010pq,DP13,MST,neutrinopaper,Gouteraux:2025kta,Preau:2025rex}. This suggests that the effective hydrodynamic description can be reorganized as an expansion in $\omega/\mu$ and $|\vec{k}|/\mu$, rather than strictly in $\omega/T$ and $|\vec{k}|/T$, with the chemical potential playing the role of the dominant infrared scale.

Our calculation follows the same steps as \cite{DP13,neutrinopaper}, and extends these results to the case of non-Abelian hydrodynamics in an isospin asymmetric medium. Specifically, we find that the near-extremal hydrodynamics generalizes to the non-Abelian case, in the sense that the current-current correlators are described by \eqref{h2}-\eqref{h3} in the regime $\omega,|\vec{k}|,\mu_3,T \ll \mu$, regardless whether $\omega\ll T$ (as in the standard hydrodynamics regime) or $\omega\geq T$.

To compute the polarization functions in the near-extremal hydrodynamic regime, we solve the equations of motion \eqref{EoMVhtpm} and \eqref{Eqvf} at $\omega,k, T, \mu_3 \ll \mu$. To do this, a small parameter $\e\ll 1$ is introduced, and we consider the following scaling of the energy, momentum, and temperature
\begin{align}
\label{HT1}
&r_H\,\omega \to \e\, r_H\,\omega,& &r_H\,\vec k \to \e^a\, r_H\,\vec k,& &r_H\,T \to \e^b\, r_H\,T,& &r_H\,\mu_3 \to \e^c\, r_H\,\mu_3,&
\end{align}
where $a,b,c>0$. The value of $c$ is set to 1 if we want to recover the $\omega_\pm = \omega\pm\mu_3$ dependence of the correlator, meaning that the frequency $\omega$ and the isospin chemical potential $\mu_3$ enter at the same order. The ratio between the frequency and the temperature scales as
\begin{equation}
	\frac{\omega}{T}\sim \e^{1-b},.
\end{equation}
and therefore two different regimes are possible depending on the value of $b$:
\begin{itemize}
\item $b<1$ corresponds to the standard hydrodynamic regime $\omega \ll T$ ;
\item $b\geq 1$ corresponds to the (extended) regime where $\omega \geq T$, but the leading order hydrodynamic approximation remains valid as long as $\omega \ll \m$.
\end{itemize}
We perform the matching calculation at $b=1$ for which the ratio $\omega/T$ remains finite in the $\epsilon\to 0$ limit. This choice allows us to retain the full dependence on $\omega/T$ in the calculation. The regimes $\omega\ll T$ and $T\ll\omega$ can then be recovered by taking the corresponding limits of the final result. The leading-order correlators are independent of the finite ratio $\omega/T$. The exponent $a>0$ in \eqref{HT1} controls the scaling of the spatial momentum and will be constrained below by the existence of an overlap region.

The calculation is based on a separation of the bulk into outer and inner regions, in each of which the equations of motion take a simpler form. These regions are defined as follows
\begin{itemize}
\item \emph{The outer region} is where the holographic coordinate $r$ is sufficiently far from the horizon for
\begin{equation}
\label{H5}
\partial_r^2\mathcal{L}\gg\dfrac{\Omega_\pm(r)^2}{f(r)^2}\mathcal{L} \quad,\quad \partial_r^2\mathcal{L}\gg\dfrac{\vec k^2}{f(r)}\mathcal{L},
\end{equation}
to be obeyed by the generic fluctuation $\mathcal{L}$ in \eqref{eq:ansL}. Then, at leading order in $\epsilon$, the transverse and longitudinal fluctuation equations reduce to radial conservation equations. For the transverse fluctuation, one finds
\begin{equation}\label{HT3}
	\partial_r\left(\frac{f(r)}{r}\partial_r\mathcal{L}_{i,\mathrm{out}}^{\perp,\pm}\right)=0,.
\end{equation}
For the longitudinal sector, it is convenient to use the field $\varphi^\pm$ defined in \eqref{fr}, for which the equation becomes
\begin{equation}\label{HL1phi}
	\partial_r\left(
	\frac{f(r)}{r}
	\frac{\Omega_\pm(r)^2}
	{\Omega_\pm(r)^2-\vec{k}^2f(r)}
	\partial_r\varphi^\pm_{\mathrm{out}}
	\right)=0,.
\end{equation}
These equations hold throughout the outer region, which includes the asymptotic boundary and ends when
\begin{equation}
\label{H6}  u \equiv 1-\frac{r}{r_H} \simeq  \mathcal{O}(\e)\, .
\end{equation}

\item \emph{The inner region} is where the holographic coordinate $r$ is sufficiently close to the horizon for
\be
\label{H7} u\ll 1 \sp f(r)\vec k^2 \ll \omega^2 \, ,
\ee
to be obeyed. To determine the range of validity of the inner approximation in the longitudinal channel, let $u\sim\epsilon^l$. For $b=1$ and $l<1$, the near-horizon blackening function is dominated by its quadratic term, so that $f(r)\sim u^2\sim\epsilon^{2l}$. Using \eqref{HT1}, the condition $f(r)\vec{k}^2\ll\omega^2$ reads
\begin{equation}
	\frac{f(r)\vec{k}^2}{\omega^2}
	\sim
	\epsilon^{2(l+a-1)}
	\ll 1\,.
\end{equation}
It is therefore satisfied for $l>1-a$. The inner region may consequently be extended up to $u\sim\epsilon^p$, with
\begin{equation}
	\max(1-a,0)<p<1\,.
\end{equation}
Since the outer region is valid for $u\gg\epsilon$, the longitudinal inner and outer regions overlap for
\begin{equation}
	p<l<1\,.
\end{equation}
Such a choice of $p$ exists whenever $a>0$, namely whenever $|\vec{k}|\ll\mu_q$.

For the transverse channel, the momentum term does not contribute at leading order. The inner approximation is therefore valid throughout the near-horizon regime $u\ll1$, and the transverse overlap region is simply characterized by $0<l<1$.

\end{itemize}

We now describe how the perturbation equations are solved at leading order in the hydrodynamic expansion, which is done separately for the transverse and longitudinal cases.

\subsubsection*{The transverse correlator}
We start by holographically computing the hydrodynamic limit of the transverse polarization function, whose leading hydrodynamic approximation is given by \eqref{h3}.

\begin{itemize}
\item In the outer region, the equation of motion \eqref{EoMVhtpm} for the transverse fluctuation reduces to the conservation equation \eqref{HT3}. The solution takes the form
\be
\label{HT4} \mathcal{L}^{\perp,\pm}_{i,\text{out}} = A_i^\pm + B_i^\pm \int_0^r\intd r' \frac{r'}{f(r')} \, ,
\ee
where $A_i^\pm$ and $B_i^\pm$ are two integration constants.

\item
The solution in the inner region, is better analysed by zooming on the near-horizon geometry, which is done by defining
\be
\label{HT6} u = 1-\dfrac{r}{r_H}\equiv\e\, \zeta \, ,
\ee
where $\zeta$ is the radial coordinate that describes the AdS$_2$-Schwarzschild factor of the near-horizon geometry (which has an additional $\mathbb{R}^3$ factor). Here, $\epsilon$ is the same infinitesimal parameter as in \eqref{HT1}.

The equation of motion \eqref{EoMVhtpm} reduces in the inner region to the equation for a massless scalar field in AdS$_2$-Schwarzschild
\be
\label{HT7} \partial_\zeta\left((4\pi r_H T\, \zeta + 12\zeta^2)\partial_\zeta\mathcal{L}_{i,\text{inn}}^{\perp,\pm}\right) + \frac{(r_H \omega)^2}{4\pi r_H T \zeta + 12\zeta^2}\mathcal{L}_{i,\text{inn}}^{\perp,\pm} = 0 \, .
\ee
The infalling solution in the inner region is then given by
\be
\label{HT8} \mathcal{L}^{\perp,\pm}_{i,\text{inn}} = C_i^\pm \left( \frac{3\zeta}{3\zeta + \pi r_H T} \right)^{-\frac{i\omega}{4\pi T}} \, ,
\ee
with $C_i^\pm$ an integration constant.
\end{itemize}

We match the above expression~\eqref{HT8} with the outer solution \eqref{HT4} expanded close to the horizon
\be
\mathcal{L}_{i,\text{out}}^{\perp,\pm} = A_i^\pm-\dfrac{B_i^\pm}{4\pi Tr_H}\log\left(1-\dfrac{r}{r_H}\right)+\mathcal{O}(r_H-r).
\ee
The matching implies
\be
\label{HT5b} B_i^\pm = ir_H\omega A_i^\pm \, .
\ee

The full solution at leading order in $\e$ is finally obtained by imposing the matching condition \eqref{HT5b}. Proceeding as such, the solution to \eqref{EoMVhtpm} in the outer region is given by
\be
\label{HT12} \mathcal{L}^{\perp,\pm}_{i,\text{out}}(r) = A_i^\pm\left(1 + ir_H\omega \int_0^{\frac{r}{r_H}} \intd x\frac{x}{f(x)} + \mathcal{O}\big(\e^2,\e^{1+2a}\big) \right) \, ,
\ee
Remarkably, the leading matching condition \eqref{HT5b} is independent of the finite ratio $\omega/T$. As explained after \eqref{HT1}, the choice $b=1$ retains this ratio explicitly; the result therefore has smooth limits for both $\omega\ll T$ and $T\ll\omega$.

Then, from \eqref{PVtim}, the transverse polarization function is found to follow the hydrodynamic behavior \eqref{h3}
\begin{equation} \label{HT13}\text{Im}\Pi^{\perp,\pm}(\omega,\vec{k}) = -\frac{1}{8} (M\ell)^3 w_0^2 N_c\,\dfrac{\omega}{r_H}\left(1 + \OO\big((r_H\omega)^2,(r_H\vec{k})^2,(r_H\mu_3)^2\big)\right) \, .
\end{equation}
Note that this expression is valid as long as $\omega,|\vec k|,\mu_3,T\ll \m_q $, including both $\omega,|\vec k|,\mu_3\ll T \ll \m_q$ and $T \ll \omega,|\vec k|,\mu_3 \ll \m_q$.

Comparing \eqref{HT13} with \eqref{h3}, the DC conductivity is identified to be
\be
\label{sf} \Sigma = \frac{1}{8r_H}N_c (M\ell)^3w_0^2 \, .
\ee
This result is formally the same as for zero $\mu_3$ \cite{neutrinopaper}, which is expected since the non-Abelian hydrodynamic expansion is also an expansion in $\mu_3$ and the presence of $\mu_3$ in the equation of motion completely drops-out.

\subsubsection*{The longitudinal correlator}

We now turn to the computation of the longitudinal polarization function. For this, we need to solve in the inner and outer regions the equation of motion \eqref{Eqvf} for the variable $\varphi^{\pm}$ in the hydrodynamic limit.

\begin{itemize}
\item In the outer region, the equation of motion \eqref{Eqvf} reduces to equation \eqref{HL1phi}, whose solution takes the form
\be
\varphi_\text{out}^\pm = A^\pm + B^\pm\int^r_0dr'\,\dfrac{r'}{f(r')}\dfrac{\Omega_\pm(r')^2-\vec k^2f(r)}{\Omega_\pm(r')^2}\,.
\ee

The correlator \eqref{PLcim} is expressed in terms of the electric field $E^{\parallel,\pm}(r) = \Omega_\pm(r)\varphi(r)^\pm$, for which the outer solution reads
\be\label{Eout}
E_\text{out}^{\parallel,\pm} = \Omega_\pm(r)\left(A^\pm + B^\pm\int^r_0dr'\,\dfrac{r'}{f(r')}\dfrac{\Omega_\pm(r')^2-\vec k^2f(r)}{\Omega_\pm(r')^2}\right)\,.
\ee
In order to have the correct $\mu_3\to 0$ limit for the solution, we rescale
\be
A^\pm\to(\omega\pm\mu_3)^{-1} A^\pm, \qquad\qquad B^\pm\to(\omega\pm\mu_3) B^\pm.
\ee

From \eqref{PLcim}, the imaginary part of the longitudinal polarization function is then identified as
\begin{equation}\label{IM1}
\text{Im}\Pi^{\parallel,\pm}(\omega,\vec k) = -\Sigma\,r_H\left((\omega \pm\mu_3)^2-\vec k^2\right)\text{Im}\dfrac{B^\pm}{A^\pm}\,,
\end{equation}
with $\Sigma$ the conductivity \eqref{sf}.

\item As for the transverse equation of motion, the solution in the inner region is better analysed by zooming on the near-horizon geometry, thus using $\zeta$ \eqref{HT6}, the radial coordinate that describes the AdS$_2$-Schwarzschild factor of the near-horizon geometry.

The equation of motion \eqref{EpEoM} reduces in the inner region to the equation for a massless scalar field in AdS$_2$-Schwarzschild
\be
\label{HL7} \partial_\zeta\left((4\pi r_H T\, \zeta + 12\zeta^2)\partial_\zeta E^{\parallel,\pm}_\text{inn}\right) + \frac{(r_H \omega)^2}{4\pi r_H T \zeta + 12\zeta^2}E^{\parallel,\pm}_\text{inn} = 0 \, .
\ee
The infalling solution in the inner region is then given by
\be
\label{HL8} E^{\parallel,\pm}_{\text{inn}} = C^\pm \left( \frac{3\zeta}{3\zeta + \pi r_H T} \right)^{-\frac{i\omega}{4\pi T}} \, ,
\ee
with $C^\pm$ an integration constant.
\end{itemize}

We need to match the above expression, ~\eqref{HL8}, with the outer solution, \eqref{Eout}, expanded near the horizon. After some algebra, we end up with
\be
\dfrac{B^\pm}{A^\pm} = \dfrac{1}{r_H}\dfrac{1}{\omega\pm\mu_3}\dfrac{i\,\omega}{\omega\pm\mu_3+\frac{i}{2}r_H\vec k^2}.
\ee

After matching, we can therefore write the solution in the outer region as
\begin{align} \nonumber E^{\parallel,\pm}_{\text{out}}(r) = &A^\pm\,\dfrac{\Omega_\pm(r)}{\omega\pm\mu_3}\bigg(1 +\dfrac{i\,\omega}{\omega\pm\mu_3+\frac{i}{2}r_h\vec k^2}\times\\
&\times\int_0^{\frac{r}{r_H}} \intd x\frac{x}{f(x)}\dfrac{\Omega_\pm(x)^2-\vec k^2 f(x)}{\Omega_\pm(x)^2} + \mathcal{O}\big(\e^2,\e^{1+b},\e^{1+2a}\big) \bigg) \, ,
\end{align}

Substituting this expression into \eqref{IM1}, we obtain
\be
\text{Im}\Pi^{\parallel,\pm}(\omega,\vec k)=-\Sigma\,\dfrac{\omega\left((\omega \pm\mu_3)^2-\vec k^2\right)}{(\omega\pm\mu_3)^2+\vec k^4(r_H/2)^2}\left(1+\mathcal{O}\left((r_H\omega)^2,(r_H\vec k)^2,(r_H\mu_3)^2\right)\right),
\ee
which agrees with \eqref{h2}. The diffusivity can be identified to be
\be
D = \dfrac{r_H}{2},
\ee
as in the case of $\mu_3=0$.

\section{Quasinormal modes in the charged sector}\label{sec:QNMt}

In this section, we compute the quasinormal modes (QNMs) associated with the charged fluctuations $\mathcal{L}^{\perp,\pm}_i$ and $E^{\parallel,\pm}$, whose equations of motion are given by \eqref{EoMVhtpm} and \eqref{EpEoM} respectively. These modes are computed numerically using the Mathematica package developed in \cite{Jansen:2017oag}, which is an efficient tool ready to use for all kinds of QNMs calculations on analytic backgrounds. In order to use the Mathematica code, the equations of motion for the charged fluctuations have to be rewritten in Eddington-Finkelstein coordinates. For completeness, we provide their expressions in appendix \ref{app:EF}.

Our analysis discusses the QNMs' dependence on all three dimensionless parameters: the chemical potentials $\mu_q/T$ and $\mu_3/T$, and the momentum $k/T$. According to \eqref{Or11}, the $+$ and $-$ modes can be mapped to each other by the $\mathbb{Z}_2$ transformation $\mu_3\to -\mu_3$. It will therefore be sufficient to focus on the equation associated with  one sign of the charge,  which is fixed in the following to correspond to the $+$ fluctuations of the SU(2) current.

As mentioned in the introduction, the low frequency behavior of the theory at high density $\mu/T\gg 1$, is controlled by the near-horizon AdS$_2$ geometry, corresponding to a CFT$_1$ behavior \cite{Faulkner:2009wj}. Since we are particularly interested in this near-extremal regime (as stated in the previous section), we find it useful to analyse here the consequences of the emergent IR conformal description for the quasi-normal spectrum. Specifically, we present in the first subsection, the expression for the IR-AdS$_2$ correlator of charged currents in the presence of a finite isospin chemical potential, and extract the corresponding poles. As shown by our numerical results below, the low-lying modes of the full correlator, are close to these IR-AdS$_2$ poles in the low temperature regime.

\subsection{The infrared-AdS$_2$ correlator}\label{sec:IRAdS2CORR}
The IR limit of correlator at finite isospin chemical potential, can be computed by taking the near-extremal, low-energy, near-horizon limit
\begin{equation}
	\label{eq2:1} T \to \epsilon \,T \sp \omega \to \e \,\omega \sp r = r_H(1-\epsilon\,\zeta) \, ,
\end{equation}
with $\epsilon\ll 1$ a small parameter, and $\zeta$ the radial coordinate that describes the AdS$_2$ Schwarzschild factor of the near-horizon geometry. In this regime, the equations of motion for the charged transverse and longitudinal gauge field perturbation \eqref{EoMVhtpm} and \eqref{Eqvf} become identical and take the form
\begin{equation}
	\label{eq:q2} \partial_\zeta\left((4\pi r_H T\, \zeta + 12\zeta^2)\partial_\zeta y\right) + \frac{r_H^2(\omega\pm 2\mu_3\zeta)^2}{4\pi r_H T \zeta + 12\zeta^2}y = 0 \quad,\quad y \in \{\mathcal{L}_{i}^{\perp,\pm},\varphi^\pm\} \, .
\end{equation}
This is the equation obeyed by a charged scalar field in AdS$_2$-Schwarzschild space-time with a constant electric field, which can be solved exactly to reproduce the expected thermal CFT$_1$ retarded correlator\footnote{Note that the $\pm$ in the right-hand side of \eqref{eq:q3} corresponds to the charge label of the correlator, and not to the notation sometimes found in the literature $\Gamma(\pm)=\Gamma(+)\Gamma(-)$.} \cite{Faulkner:2009wj}
\begin{align}\label{eq:q3}
	\GG^\pm_\text{IR}(\omega, k) =&(4\pi T)^{2\D( k,\mu_3)-1}\frac{\Gamma(1\!-\!2\D(k,\mu_3))\Gamma(\D(k,\mu_3)\!-\!\frac{i\omega}{2\pi T}\pm \frac{i}{6}r_H\mu_3)}{\Gamma(2\D(k,\mu_3)\!-\!1)\Gamma(1\!-\!\D(k,\mu_3)\!-\!\frac{i\omega}{2\pi T}\pm \frac{i}{6}r_H\mu_3)}\times\\ \non
	&\times\dfrac{\Gamma(\D(k,\mu_3)\mp \frac{i}{6}r_H\mu_3)}{\Gamma(1\!-\!\D(k,\mu_3)\mp \frac{i}{6}r_H\mu_3)},
\end{align}
with $\D(k,\mu_3)$ the IR conformal dimension computed from the near-boundary ($\zeta\to\infty$) asymptotics of \eqref{eq:q2}
\begin{equation}
	\label{eq:q4} \D(k,\mu_3) = \frac{1}{2} + \frac{1}{2}\sqrt{1+\frac{1}{3}r_H^2 k^2 - \frac{1}{9}r_H^2\mu_3^2} \, .
\end{equation}
We shall call these correlators from now on, \textit{IR-AdS$_2$ correlators}. More details on the IR-AdS$_2$ correlators can be found in appendix \ref{app:IRcorr-approx}.

The AdS$_2$ poles then correspond to the poles of the second $\Gamma$-function in the numerator of \eqref{eq:q3}. They are given by
\begin{equation}
	\label{eq:q5} \omega^{\pm,(n)}_{\text{AdS}_2} = -i2\pi T(\D(k,\mu_3) + n) \pm \frac{1}{3} r_H \pi T \mu_3 \sp n \in \mathbb{N} \, ,
\end{equation}
where the superscript $(n)$ corresponds to the $n^\text{th}$ pole. The residues associated with these poles for a generic $\Delta(k,\mu_3)$ are given by the following formula
\begin{align}\non
	\underset{\omega=\omega^{\pm,(n)}_\text{IR}}{\mathrm{Res}}\,
	\GG^\pm_\text{IR}(\omega,k)
	=
	&i\,2\pi T\,(4\pi T)^{2\D(k,\mu_3)-1}
	\frac{
		(-1)^n\Gamma\left(1-2\D(k,\mu_3)\right)
	}{
		n!\,\Gamma\left(2\D(k,\mu_3)-1\right)
		\Gamma\left(1-2\D(k,\mu_3)-n\right)
	}\times
	\\ \label{eq:6}
	&\times\frac{
		\Gamma\left(\D(k,\mu_3)\mp\frac{i}{6}r_H\mu_3\right)
	}{
		\Gamma\left(1-\D(k,\mu_3)\mp\frac{i}{6}r_H\mu_3\right)
	}\,.
\end{align}
The detailed derivation of this formula is given in appendix \ref{app:residues}.

\subsection{Transverse QNMs}

\begin{figure}[h]
	\begin{center}
		\includegraphics[scale=0.49]{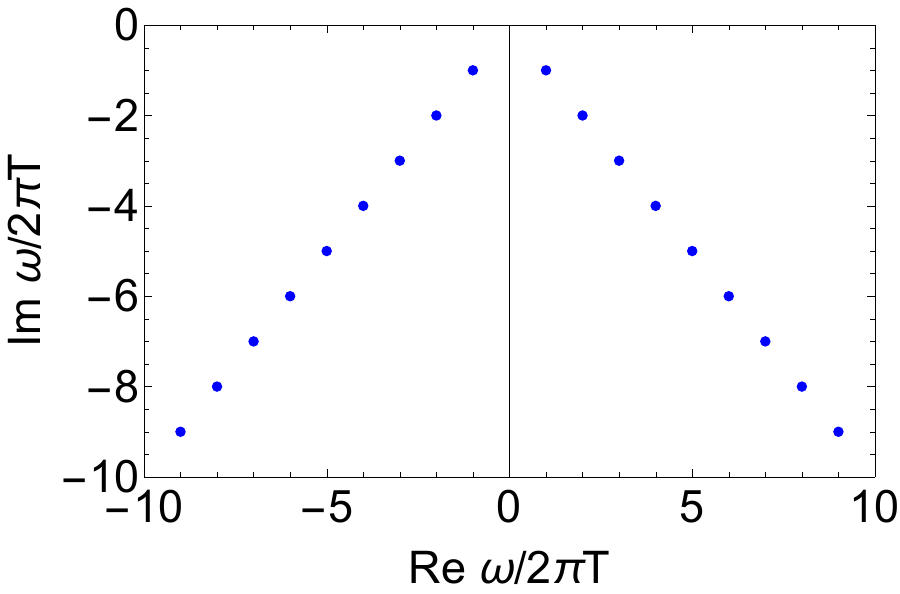}
		\hfill
		\includegraphics[scale=0.49]{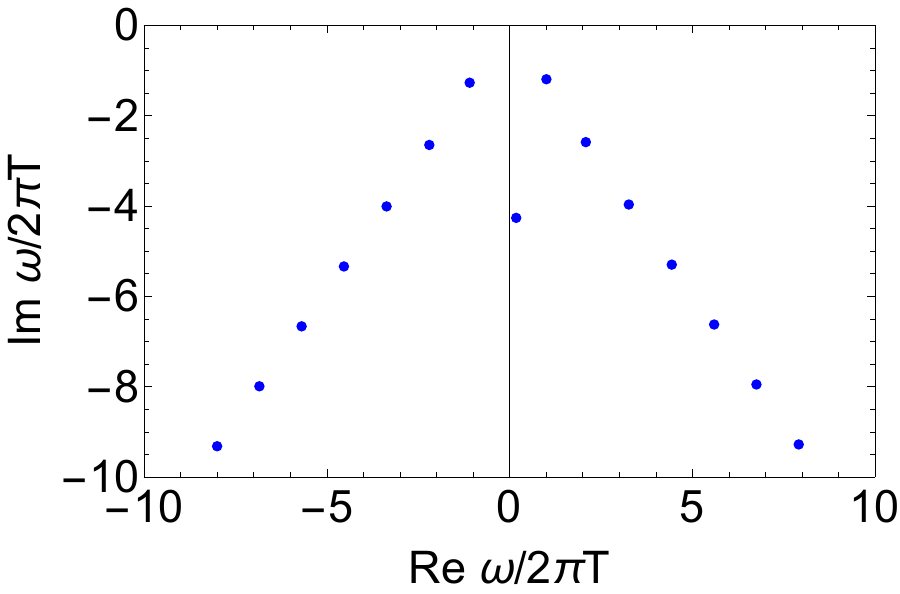}
		\includegraphics[scale=0.49]{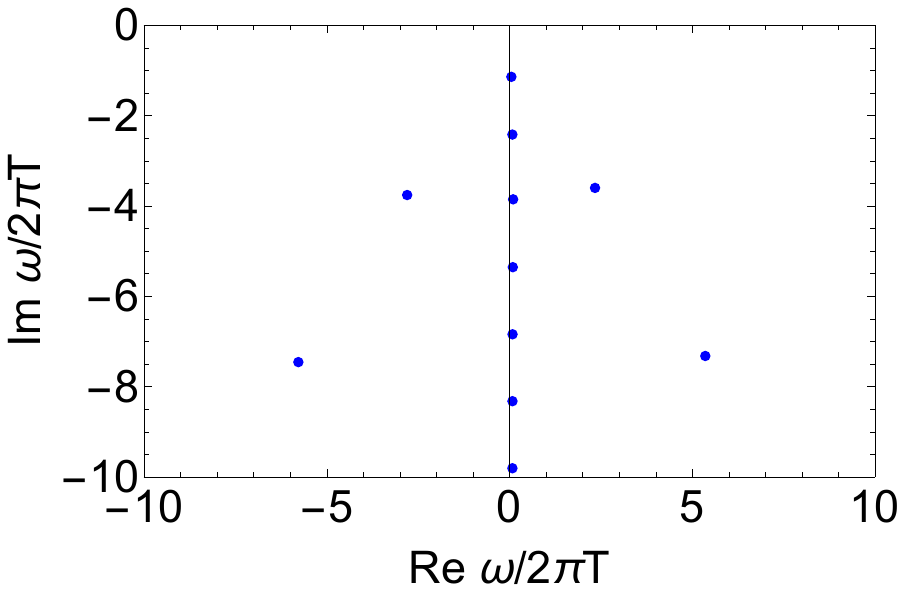}
		\hfill
		\includegraphics[scale=0.49]{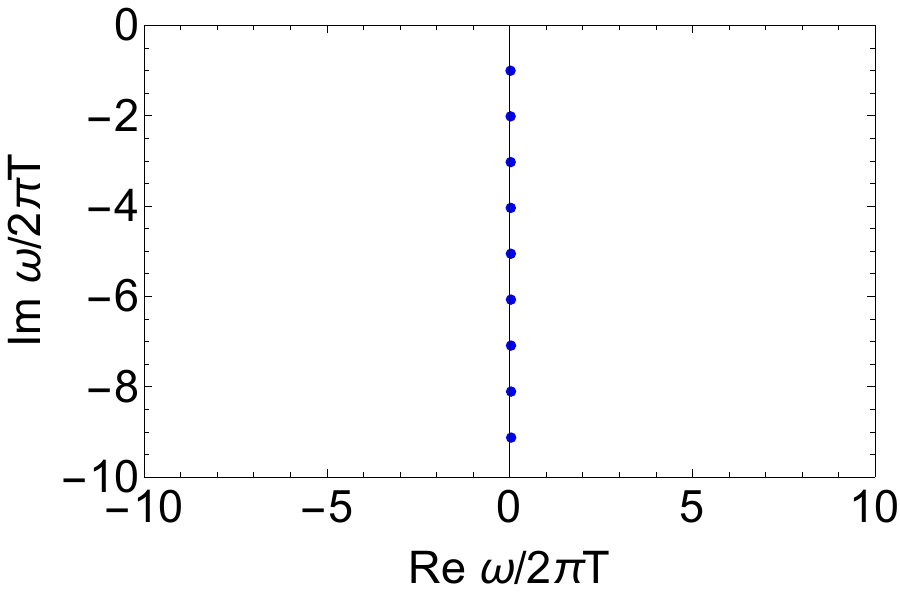}
		\caption{Lowest lying transverse QNMs at fixed $k=0$ and $\mu_3/\mu_q=0.1$, for four different values of $\mu_q/T$. From top-left to bottom-right: $\mu_q/T = 0$, 2, 10, 100. The frequency is measured in units of $2\pi T$.}
		\label{fig:wt1}
	\end{center}
\end{figure}

We now present our numerical results for the first quasinormal modes of the charged current correlators. We start with the transverse sector of fluctuations $\mathcal{L}_i^{+,\perp}$, associated with equation \eqref{EoMVhtpm}, which is not expected to feature any hydro-like mode which is truly gapless, such that its frequency vanishes at zero spatial momentum.

The typical structure of QNMs in the transverse sector, is visible in figure \ref{fig:wt1}, which shows the lowest-lying transverse QNMs for four different values of $\mu_q/T$, at fixed $k=0$ and $\mu_3/\mu_q=0.1$. The evolution of the poles, as the temperature is lowered,  is seen to follow the same typical behavior that was observed in other contexts (see e.g. \cite{Edalati:2010hk,Edalati:2010pn}): They start from the Schwarzschild ``Christmas tree" structure at zero chemical potentials. Then, some poles start appearing on the imaginary axis as $\mu_q/T$ increases, whereas the off-axis poles -- which are of order $\mu_q/T$ at high density -- move deeper into the complex plane. In the limit of vanishing temperature, only the purely imaginary poles remain at finite $\omega/T$, which approach the QNMs \eqref{eq:q5} associated with the AdS$_2$ geometry emerging near the extremal horizon \cite{Faulkner:2009wj}.

\begin{figure}[h]
	\begin{center}
		\includegraphics[scale=0.8]{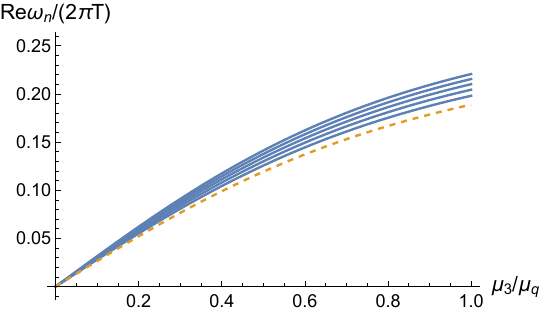}
		\hfill
		\includegraphics[scale=0.8]{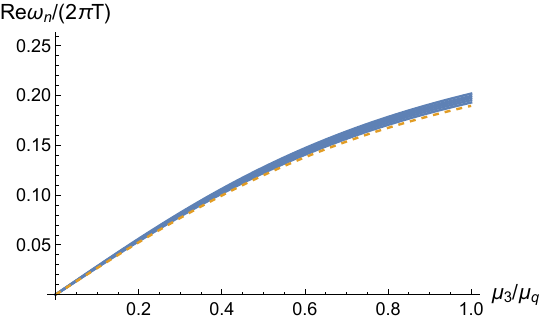}
		\includegraphics[scale=0.8]{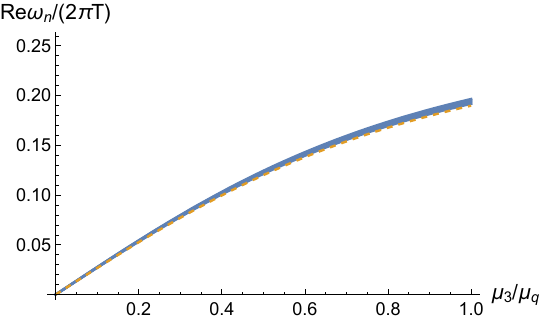}
		\caption{Real part of the first five transverse poles as a function of the ratio of chemical potentials $\mu_3/\mu_q$, for three different values of $\mu_q/T$. From top-left to bottom, $\mu_q/T=300$, 1000 and 2000. The frequency is measured in units of $2\pi T$. The blue curves are the numerical results, whereas the orange dashed curves show the analytic near-extremal result from \eqref{eq:q5}.}
		\label{fig:wt2}
	\end{center}
\end{figure}

This general behavior remains qualitatively the same, for generic values of momentum $k$ and isospin chemical potential $\mu_3$. Increasing $k$ tends to push further down the poles that are purely imaginary at $\mu_3 = 0$, and to make the complex (``Christmas tree") poles move towards the real axis with increasing real parts, as can be observed on figure \ref{fig:wt7}. Increasing $\mu_3$, introduces more asymmetry between positive and negative real parts of $\omega$.

\begin{figure}[h]
	\begin{center}
		\hspace{-2cm}\includegraphics[scale=1.2]{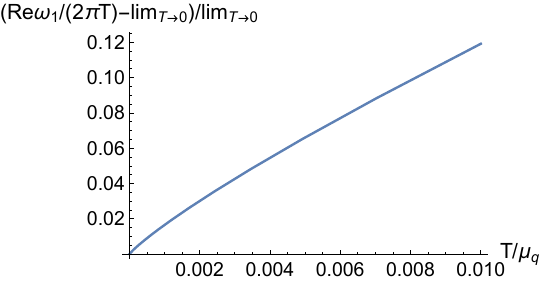}
		\caption{Relative difference of the real part of the first transverse pole in units of $2\pi T$ with its low temperature limit, as a function of $T/\mu_q$, for $\mu_3=\mu_q/10$.}
		\label{fig:wt3}
	\end{center}
\end{figure}

In the present work, there is one difference with previous examples in the literature. The low-temperature modes actually have non-zero real parts. This is a consequence of having non-zero $\mu_3$. These real parts are not clearly visible at the scale of figure \ref{fig:wt1}, but can be seen  on figure \ref{fig:wt2},
together with the analytical expression \eqref{eq:q5} for the AdS$_2$ poles (the orange dashed curves).
This clearly shows that the numerical results converge to the AdS$_2$ poles in the limit of small temperature.
This is even more explicitly confirmed by figure \ref{fig:wt3}, which shows how the real part of the first
pole approaches the $T\to 0$ limit at low $T/\mu_q$.

\begin{figure}[h]
	\begin{center}
		\centering
		\hspace{-0.5cm}
		\includegraphics[scale=0.55]{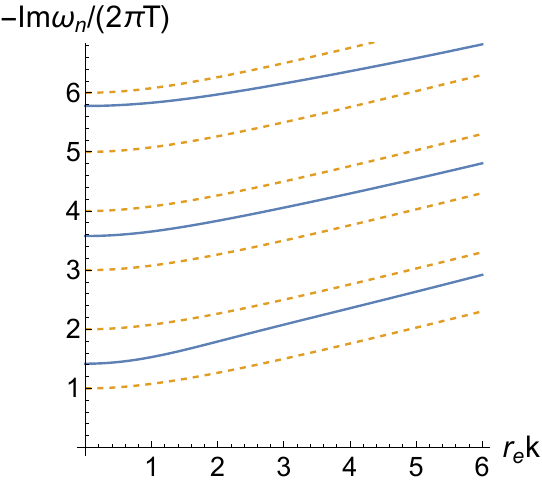}
		\hfill
		\includegraphics[scale=0.55]{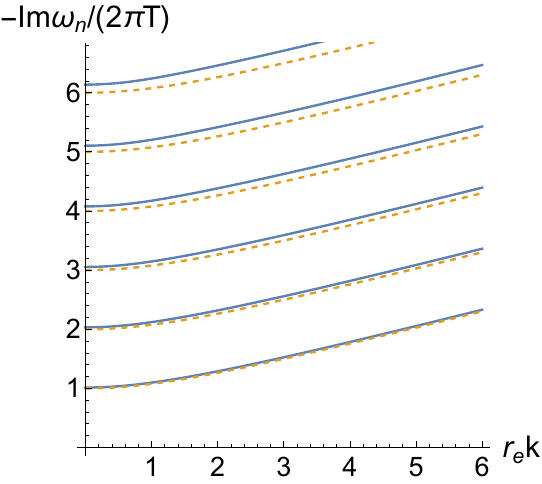}
		\includegraphics[scale=0.55]{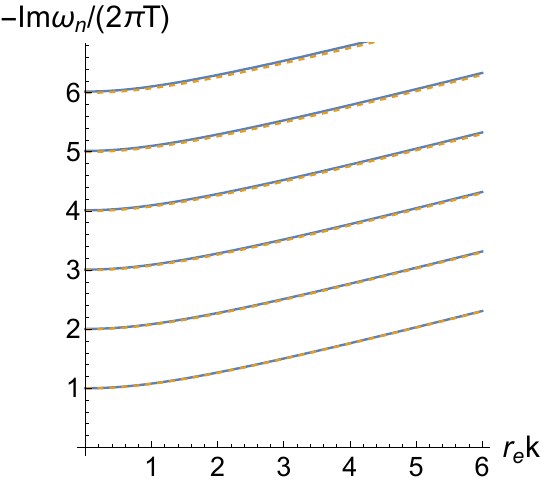}
		\centering
		\caption{Imaginary parts of the nearly imaginary transverse poles (which become purely imaginary at $\mu_3 = 0$), as a function of momentum in units of the extremal radius $r_e$, for $\mu_q/T=5$ (left), 65 (middle) and 300 (right), with $\mu_3/\mu_q=0.1$. The blue curves are the numerical results, whereas the orange dashed lines correspond to the AdS$_2$ poles \eqref{eq:q5}. As indicated by figure \ref{fig:wt1}, for $\mu_q/T=5$ there are also poles with larger real parts whose imaginary parts are in the range shown in the figures. These poles are shown below in figure \ref{fig:wt7}.}\label{fig:wt4}
	\end{center}
\end{figure}
\begin{figure}[h]
	\begin{center}
		\includegraphics[scale=0.53]{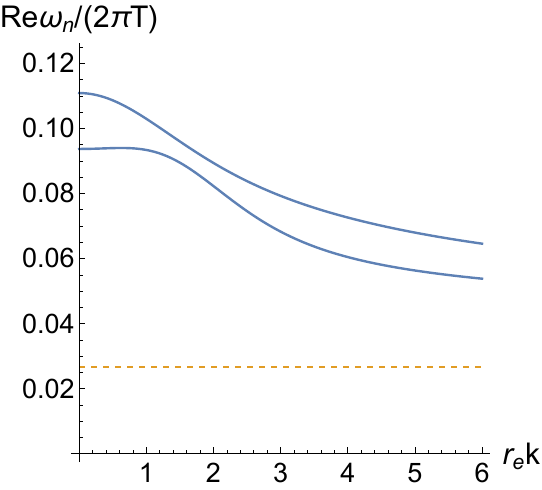}
		\includegraphics[scale=0.53]{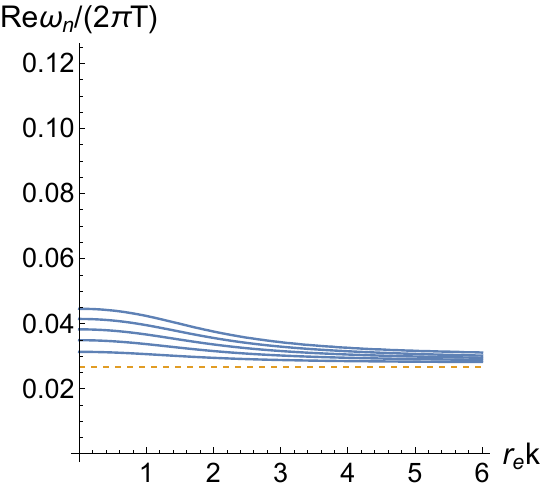}
		\includegraphics[scale=0.53]{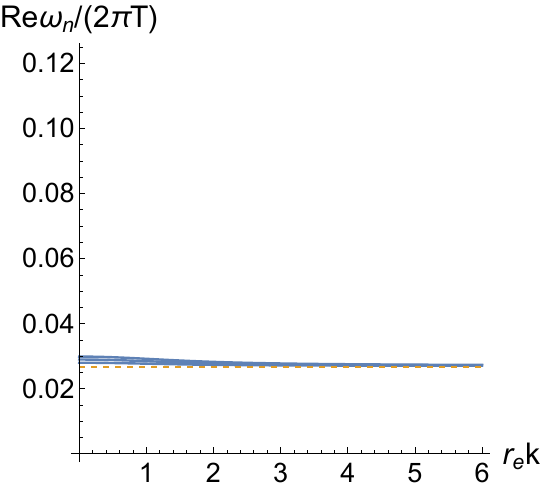}
		\caption{Real parts of the first few poles of figure \ref{fig:wt4}. For each value of $\mu_q/T$, we show only the poles for which we managed to compute the real parts with good numerical accuracy (the first two poles for $\mu_q/T =5$, the first four for $\mu_q/T =65$ and the first three for $\mu_q/T = 300$).}
		\label{fig:wt5}
	\end{center}
\end{figure}
\begin{figure}[h]
	\begin{center}
		\includegraphics[scale=0.81]{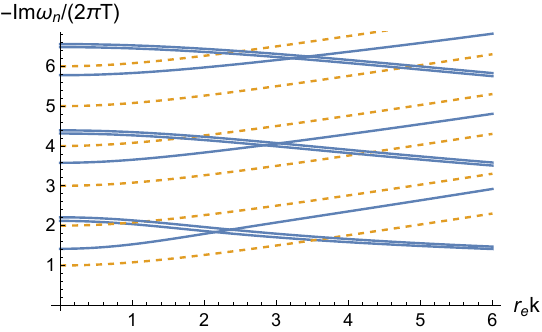}
		\hfill
		\includegraphics[scale=0.81]{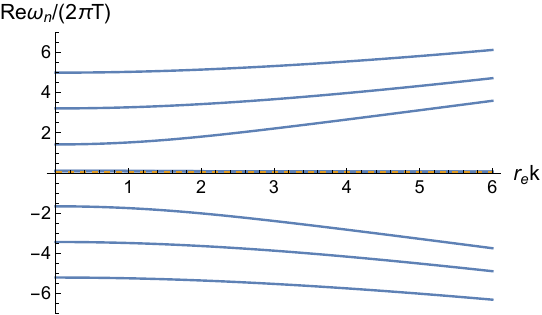}
		\caption{For $\mu_q/T = 5$ and $\mu_3/\mu_q = 0.1$, imaginary (left) and real (right) parts of the first few transverse poles as a function of momentum.}
		\label{fig:wt7}
	\end{center}
\end{figure}

The last element of our analysis in the transverse sector is about the momentum dependence of the low-lying poles, which is presented in figures \ref{fig:wt4} to \ref{fig:wt7}, at fixed $\mu_3/\mu_q=0.1$ and for three values of chemical potential, $\mu_q/T = 5,65$ and 300. Figures \ref{fig:wt4} and \ref{fig:wt5} focus on the nearly imaginary poles, that have small real parts proportional to $\mu_3$. From the left figures ($\mu_q/T=5$), we observe  that, even if not quantitatively accurate, the AdS$_2$ poles \eqref{eq:q5} (shown in dashed orange) already give the right order of magnitude for the nearly imaginary poles at $\mu_q/T$ slightly larger than 1. Moreover, we observe on figure \ref{fig:wt4} that the momentum dependence of the imaginary parts also closely follows that of the AdS$_2$ poles. For higher densities ($\mu_q/T = 65$ and 300), the middle and right plots in figures \ref{fig:wt4} and \ref{fig:wt5} show that the nearly imaginary poles become quantitatively close to the AdS$_2$ poles \eqref{eq:q5}. Also note from figure \ref{fig:wt5} that a higher momentum tends to make the real parts closer to the AdS$_2$ result.

Finally, we show for completeness, in figure \ref{fig:wt7}, the full set of poles at $\mu_q/T = 5$ within the range of imaginary parts analysed here. These include some ``Christmas tree" poles, whose real parts are of the same order as their imaginary parts (see the bottom left plot in figure \ref{fig:wt1}). Unlike the nearly imaginary poles, the imaginary parts of these poles are observed to decrease with momentum, whereas their real parts increase. Also note the small splitting between the pairs of poles induced by $\mu_3$, which is most visible at the level of the imaginary parts in left figure \ref{fig:wt7}.

\subsection{Longitudinal QNMs}
\label{sec:QNMl}

\begin{figure}[h]
	\begin{center}
		\includegraphics[scale=0.49]{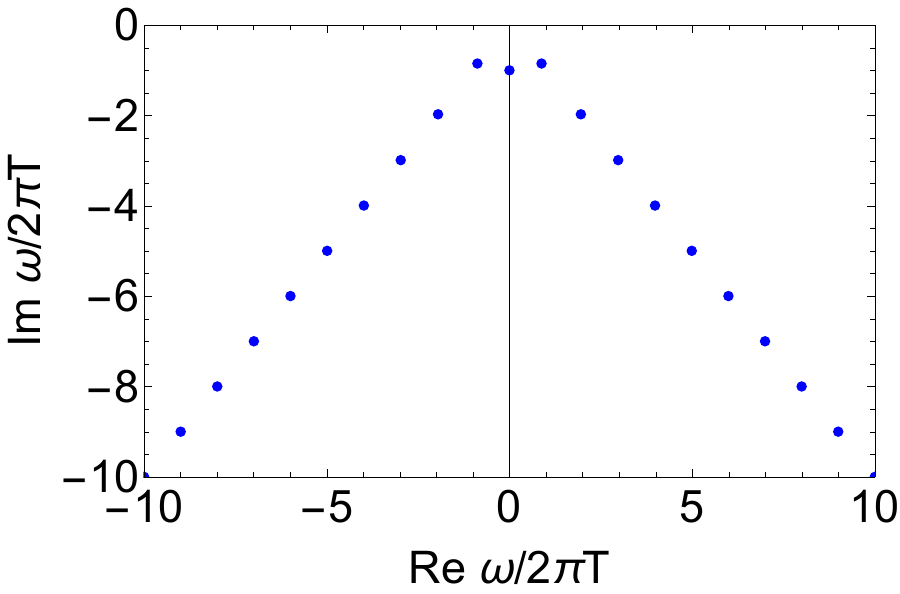}
		\hfill
		\includegraphics[scale=0.49]{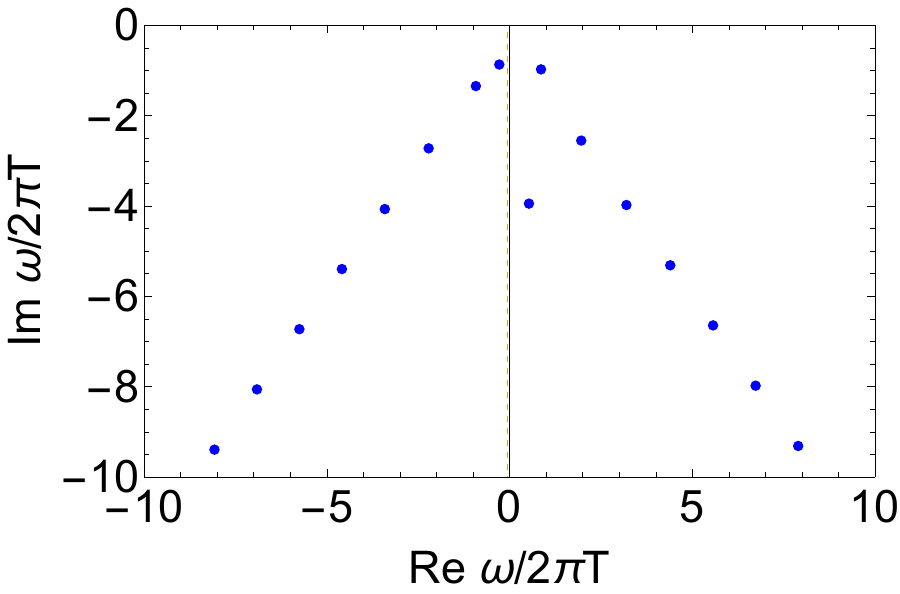}
		\includegraphics[scale=0.49]{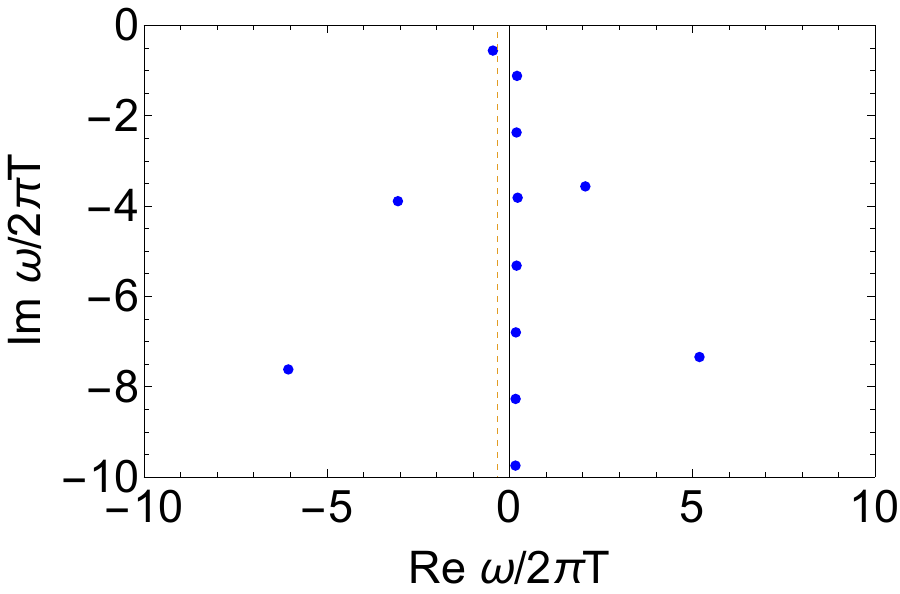}
		\hfill
		\includegraphics[scale=0.49]{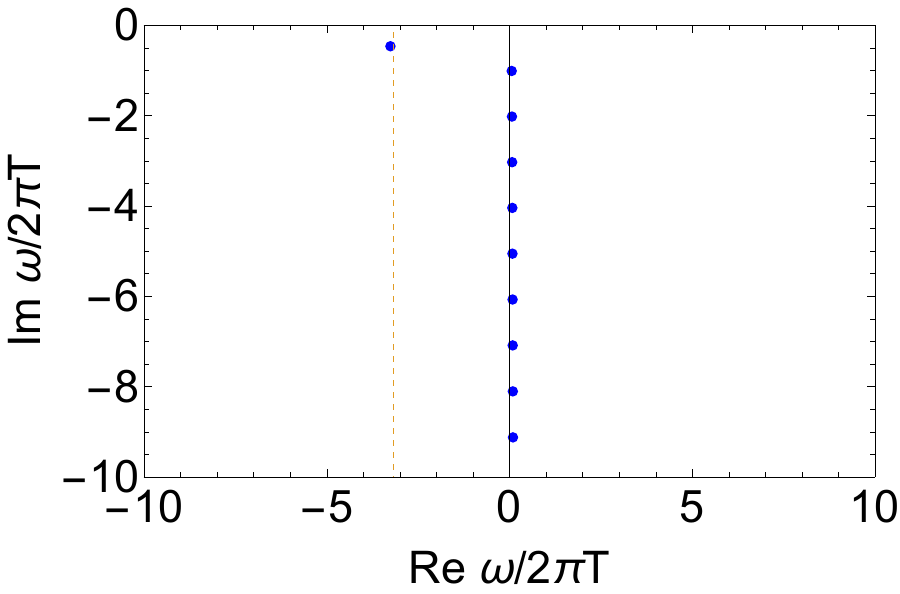}
		\caption{Lowest lying longitudinal QNMs at fixed $r_Hk^2=2\pi T$ and $\mu_3/\mu_q=0.1$, for four different values of $\mu_q/T$. From top-left to bottom-right: $\mu_q/T = 0$, 2, 10, 100. The frequency is measured in units of $2\pi T$. The orange dashed lines show the location of $-\mu_3/(2\pi T)$ on the real axis.}
		\label{fig:wl1}
	\end{center}
\end{figure}
We now discuss the quasinormal modes for the longitudinal fluctuation $\varphi^+$, which obeys \eqref{Eqvf}. For this correlator, we expect, and find, a {\em hydro-like pole}, that behaves similarly to hydrodynamic poles as momentum is varied down to $k=0$ at $\mu_3=0$.

The general pole structure in this case is shown in figure \ref{fig:wl1}, for different values of $\mu_q/T$ and at fixed $r_H k^2 = 2\pi T^{\,}$\footnote{This choice is such that the imaginary part of the diffusive pole remains visible for all values of $\mu_q/T$.} and $\mu_3/\mu_q=0.1$. The figures are similar to the transverse case (figure \ref{fig:wt1}), with the notable difference that there is an additional pole $\omega_D(k)$, which is purely imaginary at $\mu_q=\mu_3=0$ and develops a negative real part as $\mu_q/T$ is increased (with fixed $\mu_3/\mu_q$). This pole corresponds to the (hydrodynamic) diffusive-like pole $\omega_D(k) = -iDk^2 - \mu_3 + \OO(r_H^3k^4,r_H\mu_3^2)$, which was shown to exist for $r_Hk \ll 1$ in section \ref{sec:holoNAcomputation}.

\begin{figure}[h]
	\begin{center}
		\includegraphics[scale=0.98]{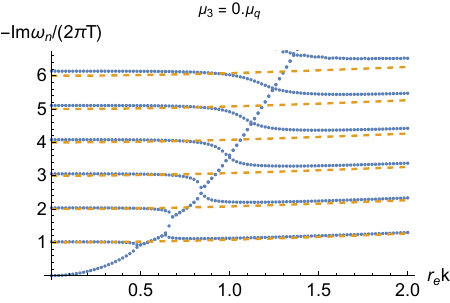}
		\hfill
		\includegraphics[scale=0.98]{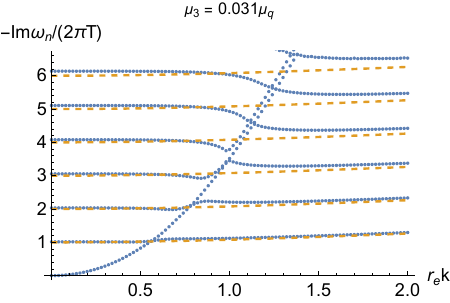}
		\includegraphics[scale=0.98]{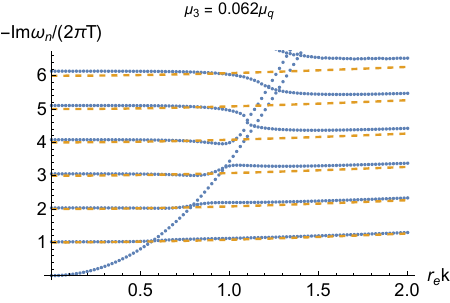}
		\hfill
		\includegraphics[scale=0.98]{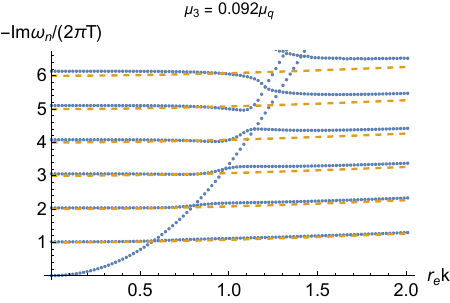}
		\caption{Imaginary part of the first seven longitudinal QNMs as a function of momentum in units of the extremal horizon radius $r_e$, for $\mu_q/T=65$ and four different values of $\mu_3/\mu_q$. From top-left to bottom-right: $\mu_3/\mu_q = 0$, $0.4/13$, $0.8/13$, $1.2/13$. The orange dashed lines represent the imaginary parts of the AdS$_2$ poles \eqref{eq:q5}.
		}
		\label{fig:wl2}
	\end{center}
\end{figure}

\begin{figure}[h]
	\begin{center}
		\includegraphics[scale=0.98]{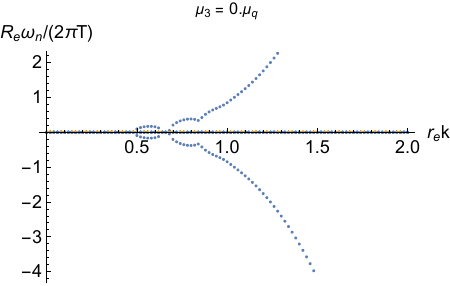}
		\hfill
		\includegraphics[scale=0.98]{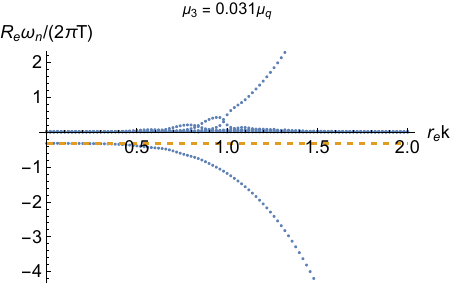}
		\includegraphics[scale=0.98]{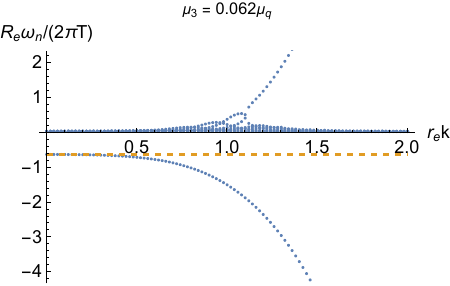}
		\hfill
		\includegraphics[scale=0.98]{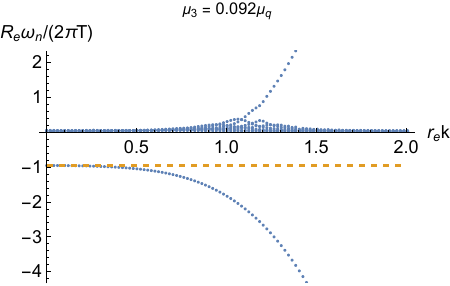}
		\caption{Real part of the first seven longitudinal QNMs as a function of momentum in units of the extremal horizon radius $r_e$, for $\mu_q/T=65$ and four different values of $\mu_3/\mu_q$. From top-left to bottom-right: $\mu_3/\mu_q = 0$, $0.4/13$, $0.8/13$, $1.2/13$. The orange dashed line shows the leading-order real part of the first hydrodynamic pole \eqref{q6}.}
		\label{fig:wl3}
	\end{center}
\end{figure}

At low temperature, $\mu_q/T\gg 1$, the interaction of the hydro-like pole with the IR poles generates an interesting structure for the evolution of the low-lying modes with momentum $k$. This is visible in figures \ref{fig:wl2} (imaginary parts) and \ref{fig:wl3} (real parts), which show the momentum dependence of the first few poles at $\mu_q/T = 65$, and for different values of $\mu_3/\mu_q$.
\begin{figure}[h]
	\begin{center}
		\includegraphics[scale=0.81]{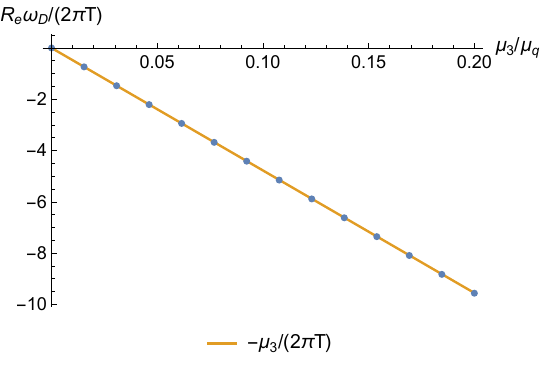}
		\caption{Real part of the diffusive pole at $\mu_q/T=300$ and $k=0$, as a function of $\mu_3/\mu_q$. Blue points are numerical results, whereas the orange line corresponds to the linear hydrodynamic prediction.}
		\label{fig:wl4}
	\end{center}
\end{figure}

At $\mu_3 = 0$, the result is similar to what was computed in \cite{Preau:2025rex} (which is also very close to \cite{Gouteraux:2025kta}): as the momentum increases, the purely imaginary diffusive pole eventually collides with the first IR pole, and both acquire (opposite) real parts before recombining on the imaginary axis. The hydro-like pole then collides with the second IR pole, beyond which two conjugate poles survive with finite real parts that increase with momentum. Even after the hydrodynamic-like poles leave the imaginary axis, they still displace the IR poles each time their (common) imaginary part approaches an AdS$_2$ level. Specifically, the corresponding level $n$ (see \eqref{eq:q5}) is lowered to the previous level $n-1$. Away from such crossings, the IR poles are well approximated by the AdS$_2$ poles \eqref{eq:q5} (shown by orange dashed curves in figure \ref{fig:wl2}).

At finite isospin chemical potential, the two hydrodynamic-like modes discussed above survive, but they are modified in distinct ways:
\begin{itemize}
	\item The pole with negative\footnote{We should keep in mind that all signs are flipped when considering the other charged field $\varphi^-$. Here for example, negative would be replaced by positive.} real part acquires a finite real part at zero momentum, in such a way that it agrees with the leading order hydrodynamic result (see \eqref{H2})
	\begin{equation}
		\label{q6} \omega_D(k) = -\mu_3 - iDk^2 + \OO(k^4,\mu_3 k^2) \, ,
	\end{equation}
	as shown in figures \ref{fig:wl2}, \ref{fig:wl3} and \ref{fig:wl4}.
	\item As can be observed on figure \ref{fig:wl2}, the pole with positive real part still starts from one of the nearly-imaginary IR poles (for example the fourth pole in the top right plot of figure \ref{fig:wl2}), but the order of this pole increases with $\mu_3$. Once the first pole \eqref{q6} crosses the corresponding AdS$_2$ level, the charged correlators feature two hydrodynamic-like poles, whose imaginary parts differ by a quantity of order $\OO(\mu_3)$. The precise dispersion relations for these two poles can be analysed analytically by generalizing the calculation of section \ref{sec:holoNAcomputation} to higher order in the hydrodynamic expansion (along the lines of \cite{Gouteraux:2025kta}). We plan to present such results in future work.
\end{itemize}

\begin{figure}[h]
	\begin{center}
		\hspace{-0.5cm}
		\includegraphics[scale=0.53]{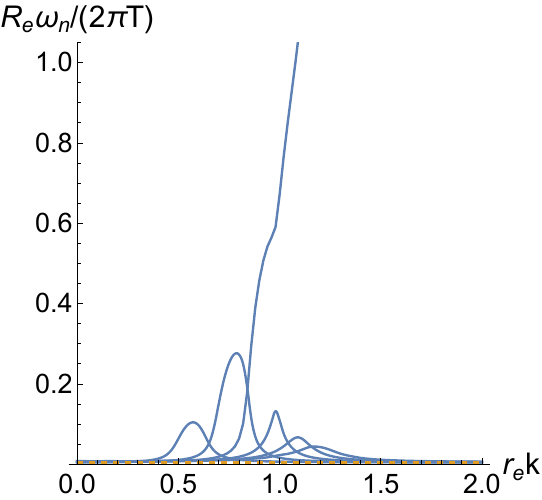}
		\includegraphics[scale=0.53]{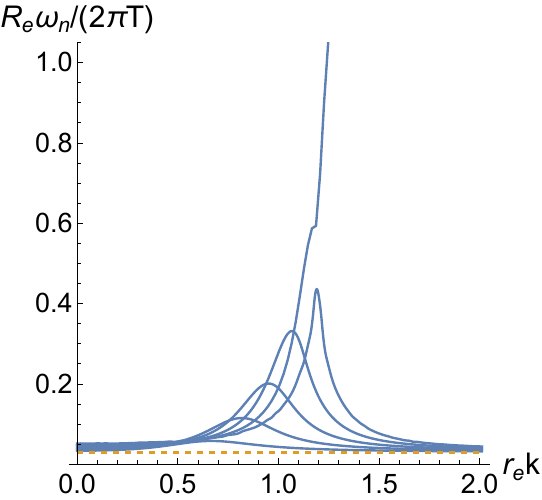}
		\includegraphics[scale=0.53]{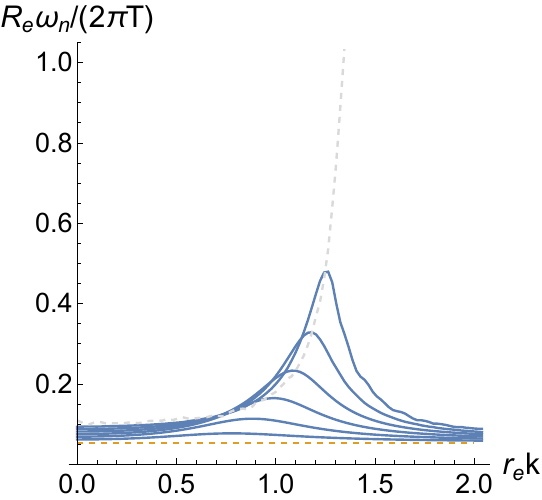}
		\caption{Real parts of the first six longitudinal QNMs with positive real parts, as a function of momentum, for $\mu_q/T=65$ and three different values of $\mu_3/\mu_q$: $\mu_3/\mu_q = 0.2/13$ (left), $1.2/13$ (middle) and $0.2$ (right). The orange dashed lines represent the real part of the AdS$_2$ poles \eqref{eq:q5}. For $\mu_3/\mu_q=0.2$ (right), the numerical result for the sixth pole has significant noise, which was somewhat smoothened with a Gaussian filter (but it remains visible). Also, the secondary hydro-like pole (see figure \ref{fig:wl2} and text) starts in this case from the seventh pole, which has a large amount of noise in our numerical calculation. We show the Gaussian-filtered seventh pole in dashed gray for demonstration.}
		\label{fig:wl5}
	\end{center}
\end{figure}

The effects of $\mu_3$ on the hydro-like poles, changes the dynamics of the low-lying poles completely, as momentum is increased. Since the first pole \eqref{q6} is now off the imaginary axis, it does not collide with the IR poles anymore, as is clear from figure \ref{fig:wl3}. There is, however, a remnant of interaction when the imaginary part of the hydro-like pole approaches an IR mode, visible in figures \ref{fig:wl2} and \ref{fig:wl3}, with a close-up on the real parts in figure \ref{fig:wl5}. The effect on the IR poles can be described as follows:
\begin{itemize}
	\item Away from hydro-like-pole crossings, the IR poles are close to the AdS$_2$ result \eqref{eq:q5} (orange dashed lines).
	\item When the imaginary part of the first pole \eqref{q6} crosses an AdS$_2$ level, the real part of the corresponding pole receives a positive shift, before going back down to its original position, whereas the imaginary parts experience a small oscillation. These shifts become larger and larger for increasing AdS$_2$ level, until we reach the level where the second hydrodynamic-like pole starts.
	\item The higher IR modes keep receiving positive real shifts after the departure of the second hydrodynamic-like pole, but the effect becomes smaller and smaller with increasing AdS$_2$ level. On the other hand, their imaginary parts do not merely oscillate, but instead shift to the previous AdS$_2$ level. Overall, the behaviour after the departure of the second hydrodynamic-like mode is very similar to the $\mu_3 = 0$ case (top left figure \ref{fig:wl2}).
\end{itemize}

Figures \ref{fig:wl2}, \ref{fig:wl3} and \ref{fig:wl5} also show how the poles dynamics depends on $\mu_3$. In particular, figure \ref{fig:wl2} indicates that the AdS$_2$ level $n_2$ that the second hydrodynamic-like pole starts from becomes higher and higher as $\mu_3$ increases, with the splitting in imaginary parts with the first pole \eqref{q6} also becoming larger. As for the effect on the IR modes, it is observed  to be strongest near the level $n_2$, whereas fixed levels $n<n_2$ become less and less affected.

\begin{figure}[h]
	\begin{center}
		\includegraphics[scale=0.65]{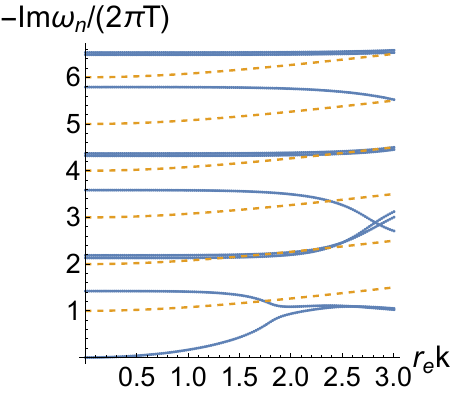}
		\includegraphics[scale=0.65]{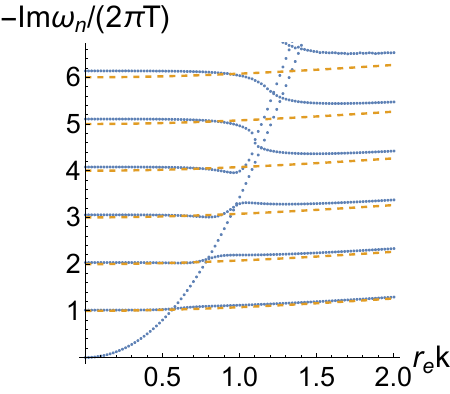}
		\includegraphics[scale=0.65]{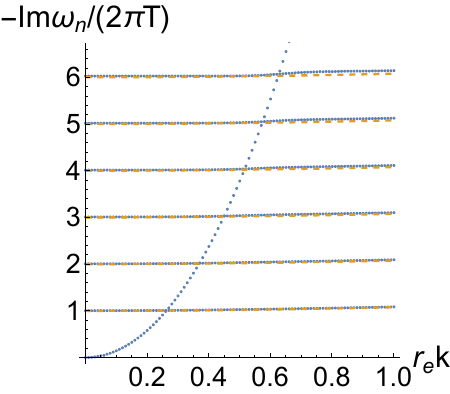}
		\caption{Imaginary parts of the first few longitudinal QNMs as a function of momentum, for $\mu_3/\mu_q=0.8/13$ and three different values of $\mu_q/T$: $\mu_q/T = 5$ (left), $65$ (middle) and $300$ (right). The orange dashed lines represent the imaginary parts of the AdS$_2$ poles \eqref{eq:q5}. Note that the momentum range differs from one plot to another, whereas the frequency range is fixed.}
		\label{fig:wl6}
	\end{center}
\end{figure}
\begin{figure}[h]
	\begin{center}
		\includegraphics[scale=0.65]{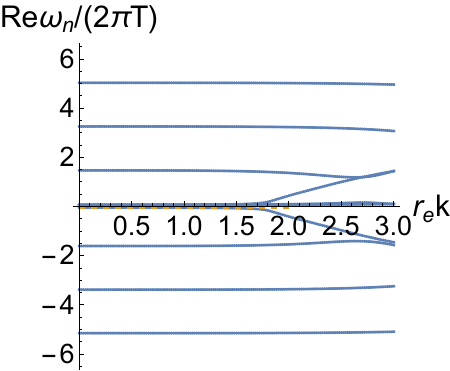}
		\includegraphics[scale=0.65]{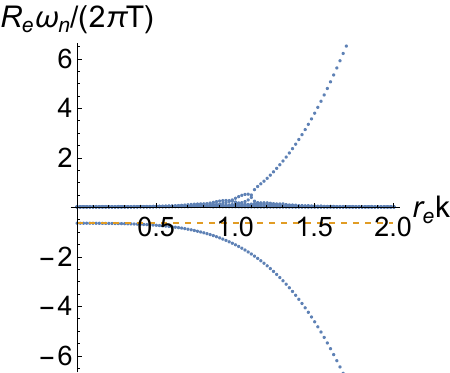}
		\includegraphics[scale=0.65]{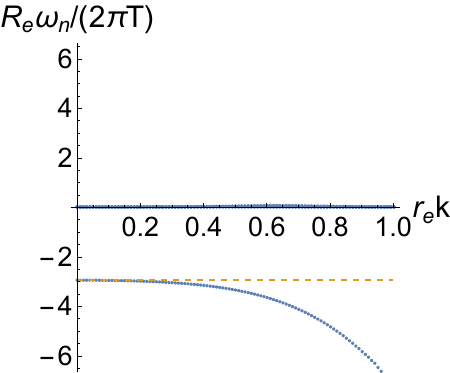}
		\caption{Real parts of the poles shown in figure \ref{fig:wl6}. The orange dashed line shows the leading-order real part of the first hydrodynamic pole \eqref{q6}. For $\mu_q/T = 300$ (right), the AdS$_2$ poles have a finite positive real part of the order of \eqref{eq:q5}, but it is too small to observe  on the scale of the figure.}
		\label{fig:wl7}
	\end{center}
\end{figure}

We conclude the analysis of this section by discussing the dependence of the previous results on the ratio of chemical potential to temperature $\mu_q/T$. For this, we compare in figure \ref{fig:wl6} and \ref{fig:wl7} the plots above for the real and imaginary parts of the longitudinal modes at $\mu_q/T = 65$, to two other values of $\mu_q/T$: 5 and 300, with $\mu_3/\mu_q = 0.8/13$ fixed. First of all, observe that, as in the transverse case, the results for $\mu_q/T = 5$ include a few ``Christmas tree" poles, that have finite real parts of the same order as their imaginary parts even at $\mu_3 = k = 0$ (see left figure \ref{fig:wl7}). These poles appear as doubled lines increasing with momentum\footnote{It can be checked that this increasing behaviour is only transitory, as the imaginary parts of these poles eventually go down at large momenta.} in left figure \ref{fig:wl6}.

We now focus on the hydrodynamic-like modes, and denote by $n_2$ the AdS$_2$ level that the second pole starts from. We then observe  in figure \ref{fig:wl6} that $n_2$ increases with $\mu_q/T$: $n_2 = 1$ for $\mu_q/T=5$, $n_2 = 4$ for $\mu_q/T = 65$, and the second mode is not even visible on the plot for $\mu_q/T = 300$. We still expect a second hydrodynamic-like pole in that case, but starting from a higher level (our numerics indicate $n_2\geq 10$).

Both for $\mu_q/T = 5$ and 300, we do not observe any significant interaction between the hydrodynamic-like poles and the nearly imaginary poles (that we call IR poles at large $\mu_q/T$). For $\mu_q/T=5$, this happens because the imaginary parts of the hydrodynamic-like modes start going down with momentum\footnote{In fact, the second pole appears at $r_e k \simeq 1.8$, for which it does not really behave like a hydrodynamic pole at all.} before they can reach the next pole. For $\mu_q/T = 300$, the AdS$_2$ levels that are shown here, with $n\leq 6$, are far from $n_2\geq 10$, which is why they are not much affected by the hydro-like pole, according to the above discussion.

\section{The extended hydrodynamic approximation}

\label{sec:IRcorr-approx}

Near-extremal holographic backgrounds of the type considered in this work, develop an emergent AdS$_2$ region in the near-horizon geometry. Low-energy correlators are then controlled not only by the hydro-like poles, but also by the non-analytic infrared conformal response inherited from the near-horizon region (see section \ref{sec:QNMt} for a detailed analysis of the quasinormal spectrum). The low-temperature IR theory is characterized by an IR Green's function $\mathcal{G}_\text{IR}(\omega,k)$ (see equation \eqref{eq:q3}), with a momentum-dependent scaling dimension $\Delta(k,\mu_3)$, whose explicit expression is given by \eqref{eq:q4}.

Based on the holographic product formula of \cite{Dodelson:2023vrw}, thermal spectral functions\footnote{See \cite{Dodelson:2023vrw} and \cite{Preau:2025rex} for the precise conditions under which the formula applies.} can be expressed in terms of the quasinormal spectrum of the dual black hole geometry. In our case, the spectral functions of interest correspond to the imaginary part of the polarization functions $\Pi^\pm$, which from \eqref{Or12} can be written as
\begin{equation}
	\label{ImP} \text{Im}\Pi^\pm(\omega,k) \equiv \frac{\Pi^\pm(\omega,k)-\Pi^\mp(-\omega,k)}{2i} \, .
\end{equation}
The product formula of \cite{Dodelson:2023vrw} is expressed in terms of the two-sided Wightman correlators,\footnote{The name ``two-sided" comes from the fact that this correlator is computed holographically by putting sources at the boundaries of the two asymptotically AdS regions, in the maximally extended AdS-RN geometry. The subscript $12$ then refers to the two sides; in this nomenclature, a one-sided correlator on the first side would for example be denoted as $11$. See \cite{Dodelson:2023vrw} and references therein for more details.} defined in position space as
\begin{equation}
	\label{pf1} i\Pi_{12}^{\pm,l}(t) =  \frac{P^{\mu\nu}_l}{{(P_l)^\l}_\l} \left<J_\mu^\pm\left(t-\frac{i}{2T}\right) J_\nu^\mp(0)\right> \sp l\in\{\perp,\parallel\} \, ,
\end{equation}
with $P^{\mu\nu}_l$ the transverse and longitudinal projectors defined in \eqref{Po}-\eqref{Pp}. The product formula obeyed by $\Pi_{12}$ is then expressed as \cite{Dodelson:2023vrw}
\begin{equation}
	\label{pf2}
	\Pi_{12}^{\pm,l}(\omega,k)
	= \dfrac{\gamma_l(\pm\mu_3, k)}{\prod_{n=1}^\infty\left(1-\dfrac{\omega}{\omega_n^l(\pm\mu_3, k)}\right)\left(1+\dfrac{\omega}{\omega_n^l(\mp\mu_3,k)}\right)} \, ,
\end{equation}
where $\omega_n^l(\mu_3,k)$ are the poles of the two-sided correlator, and $\gamma_l(\mu_3,\vec k)$ is a function free of zeroes and poles. We used the Onsager relation \eqref{Or11}, and took into account that if $\omega_n^l(\mu_3,k)$ is a pole, then so is $-\omega_n^l(-\mu_3,k)$, as can be inferred from \eqref{ImP} and the relation between the two-sided correlator and the spectral function (see below in \eqref{pf5}).

In fact, \cite{Dodelson:2023vrw} demonstrated \eqref{pf2} in the case where the two-sided correlator is free of zeroes. This is easily generalizable to the case where $\Pi_{12}$ admits a finite number of zeroes, but an infinite number of zeroes may be an obstruction to the product formula \cite{Preau:2025rex}. It was shown in \cite{Dodelson:2023vrw} that the two-sided correlator is free of zeroes in the case of regular Schr\"odinger potentials \cite{Dodelson:2023vrw,Preau:2025rex} (once the second order fluctuation equations \eqref{EoMVhtpm} and \eqref{Eqvf} are put in Schr\"odinger form). Here, it can be checked that, although the transverse potential arising from \eqref{EoMVhtpm} is regular, the longitudinal potential from \eqref{Eqvf} admits singularities in the bulk. The expressions for the corresponding Schr\"odinger potentials can be found in appendix \ref{sec:SP}. It would be interesting to analyse further the consequences of these singularities for the structure of the longitudinal correlator. However, for now we  only assume that the longitudinal two-sided correlator has at most a finite number of zeroes, for which we know that it obeys a product formula (with potentially zeroes in the numerator). The transverse and longitudinal two-sided correlators are therefore written as
\begin{equation}
	\label{pf3}
	\Pi_{12}^{\perp,\pm}(\omega,k)
	= \dfrac{\gamma_\perp(\pm\mu_3, k)}{\prod_{n=1}^\infty\left(1-\dfrac{\omega}{\omega^\perp_n(\pm\mu_3, k)}\right)\left(1+\dfrac{\omega}{\omega^\perp_n(\mp\mu_3, k)}\right)} \, ,
\end{equation}
\begin{equation}
	\label{pf4}
	\Pi_{12}^{\parallel,\pm}(\omega,k)
	= \dfrac{\gamma_\parallel(\omega,\pm\mu_3, k)}{\prod_{n=1}^\infty\left(1-\dfrac{\omega}{\omega^\parallel_n(\pm\mu_3, k)}\right)\left(1+\dfrac{\omega}{\omega^\parallel_n(\mp\mu_3, k)}\right)} \, ,
\end{equation}
where $\gamma_\perp$ is free of zeroes and $\gamma_\parallel$ may admit a finite number of zeroes.

We now come back to the original object of interest, that is the spectral function \eqref{ImP}. The latter is related to the two-sided correlator $\Pi_{12}$ via the KMS relation
\begin{equation}
\label{pf5} \text{Im}\Pi^\pm_l(\omega,k) = \sinh\left(\frac{\omega}{2T}\right) \Pi_{12}^{\pm,l}(\omega,k) \, .
\end{equation}
Note that this form of the KMS relation assumes that the time-translation operator associated with the boundary time $t$ is $K \equiv H - \mu_3 Q_3$, with $H = \int\intd x^3 T^{00}$ the Hamiltonian\footnote{If the time translation was instead generated by the Hamiltonian $H$, associated with frequency $\omega_0$, the argument of the $\sinh$ in \eqref{pf5} would instead be $\omega_0 \pm \mu_3$.} and $Q_3 = \int \intd x^3 J^0_3$ the isospin charge. This corresponds to the canonical boundary time-translation operator in presence of $\mu_3$ \cite{Papadimitriou:2005ii}. With this definition of the frequency, \eqref{pf5} implies that  the spectral functions vanish at $\omega = 0$, which agrees with the hydrodynamic results \eqref{ImPiL}-\eqref{ImPiT}. The definition of the boundary time in the holographic setup is therefore the same as in the hydrodynamic calculation.

Now, \eqref{pf5} implies that the spectral functions are fully determined by the product formulae obeyed by the two-sided correlators \eqref{pf3}-\eqref{pf4}. In the near-extremal hydrodynamic regime $\omega,k,T,\mu_3\ll\mu$, the quasinormal spectrum admits a natural separation into {\em hard poles} with gaps of order $\mu$, {\em soft poles} whose gaps are controlled by the temperature $T$, and {\em hydro-like poles}, that behave similarly to hydrodynamic poles as functions of momentum. To define more precisely the hydro-like poles, we recall the pole structure that was observed in section \ref{sec:QNMt}:
\begin{itemize}
\item In the limit of vanishing $k$ and $\mu_3$, there is a single pole going to zero, with a diffusive-like behaviour $\omega \sim -iDk^2 \pm \mu_3$.
\item As momentum increases, the gapless pole crosses IR modes (which have gaps of order $\OO(T)$) as described in section \ref{sec:QNMt}, which happens when $k$ is of order $\sqrt{\mu T}$. One of these crossings pushes the corresponding IR mode to move away from the associated AdS$_2$ level, and to start following closely the hydrodynamic-like behaviour of the gapless pole.
\item For momenta $k\gg \sqrt{\mu T}$, the two poles (gapless and its partner) approach their zero-temperature dispersion relations, that describe two gapless poles.
\end{itemize}
The hydro-like poles are then defined as the two special poles mentioned above. Note that, while one of them is gapless, the other one actually has a gap of order $\OO(T)$. In the $T\to 0$ limit, only the hydro-like poles survive below the hard scale $\mu$ (unlike the soft poles that merge into a branch cut \cite{Faulkner:2009wj,Gursoy:2021vpu,Gouteraux:2025kta,Preau:2025rex}). In this limit, the two poles are gapless, in the sense that they go to zero at $k=\mu_3=0$.

In the standard hydrodynamic regime, $\omega,k,\mu_3\ll T\ll \mu$, the correlators are dominated by the gapless hydro-like poles, while the contribution of the hard and the soft poles is invisible, as they are far in this regime. Therefore, the spectral functions can be written as
\begin{equation}
	\text{Im}\Pi^\pm_l(\omega,k)
	=
	\dfrac{\tilde \gamma_l(\omega,\pm\mu_3,k)\omega/T}{\prod_{n\in\text{gapless}}\left(\omega-\omega_n^l(\pm\mu_3, k)\right)\left(\omega+\omega_n^l(\mp\mu_3, k)\right)} \sp l\in \{\perp,\parallel\} \, ,
\end{equation}
where $\tilde \gamma(\omega,k)$ is analytic in $\omega$ and $k$. Note that we do have access to the explicit expression of $\tilde{\gamma}$ at leading order in the hydrodynamic limit. It is encoded in the results of section \ref{sec:holoNAcomputation}, where the imaginary part of the transverse/longitudinal charged current polarization functions have been computed at leading order in the hydrodynamic limit. We recall here their expressions
\be
\label{pf6}\text{Im}\Pi^\pm_\perp = -\Sigma\,\omega\sp\text{Im}\Pi^\pm_\parallel = -\Sigma\dfrac{\omega\left((\omega\pm\mu_3)^2- k^2\right)}{(\omega\pm\mu_3)^2+D^2 k^4}\,.
\ee
Therefore, the associated $\tilde{\gamma}_{\perp,\parallel}$ can be derived in the hydrodynamic limit as
\be
\tilde{\gamma}_{\perp} = -T\,\Sigma\sp\tilde{\gamma}_{\parallel} = -T\,\Sigma\left((\omega\pm\mu_3)^2- k^2\right)\,,
\ee
where the explicit formula for the conductivity $\Sigma$ is given in \eqref{sf}. Note that \eqref{pf6} implies that $\tilde\gamma_\parallel$ in \eqref{pf4} does have zeroes in the hydrodynamic limit, corresponding to $(\omega\pm\mu_3)^2 = k^2$. These zeroes are known to be present also outside the hydrodynamic regime, according to \eqref{eqpt3}.

As already pointed out at the beginning of this section, in the general near-extremal limit, $\omega,k,T,\mu_3\ll\mu$, both hydro-like poles and soft poles contribute to the correlator
\begin{equation}\label{eq:productformula}
	\text{Im}\Pi^\pm_l(\omega,k)
	=
	\sinh\left(\dfrac{\omega}{2T}\right)\dfrac{\gamma_l^{\text{ne}}(\omega,\pm\mu_3, k)\,\mathcal{G}^s_l(\omega,\pm\mu_3, k)}{\prod_{n\in\text{hydro-like}}\left(\omega-\omega_n^l(\pm\mu_3, k)\right)\left(\omega+\omega_n^l(\mp\mu_3, k)\right)}\,,
\end{equation}
where $l\in\{\perp,\parallel\}$ and  $\gamma_l^{\text{ne}}(\omega,\pm\mu_3,k)$ is analytic for $\omega,k,\mu_3\ll \mu$. The function $\mathcal{G}_l^s$ is the inverse product over the soft poles, which most of the time can be approximated (up to a factor) by the IR-AdS$_2$ correlator \eqref{eq:q3}, in the limit of small temperature and small frequencies. There are however cases where this reduction does not happen, as the IR poles are not very near the AdS$_2$ poles. An example can be seen in figure \ref{fig:wl2}, where the longitudinal poles differ near the areas of collisions. We refer to the formula \eqref{eq:productformula} as the {\em near-extremal EFT product formula}\footnote{Note that this name differs from \cite{Preau:2025rex}, where it was called the ``near-extremal hydrodynamic product formula". In this work, as explained at the beginning of section \ref{results}, we reserve the name ``near-extremal hydrodynamics" to the approximation consisting in applying standard hydrodynamics to describe the near-extremal low energy EFT.}.

The near-extremal EFT  formula, valid in the near-extremal hydrodynamic region,  takes a simpler form in the limit of vanishing temperature. In this regime, the soft poles condense into a branch cut, and the soft factor $\GG^s_l$ resums into a power-law behaviour $\omega^{2\Delta-1}$ \cite{Preau:2025rex}, with $\Delta$ the momentum-dependent scaling dimension of the emergent infrared theory (see \eqref{eq:q4}). In this limit of $T\to 0$, one can write the {\em extremal EFT product formula}
\begin{equation}\label{eq:exthydroproduct}
	\text{Im}\Pi^\pm_l(\omega,k)
	=
	\dfrac{ \gamma^{\text{e}}_l(\omega,\pm\mu_3, k)\,(\omega/\mu)^{2\Delta-1}}{\prod_{n\in\text{hydro-like}}\left(\omega-\omega_n^l(\pm\mu_3, k)\right)\left(\omega+\omega_n^l(\mp\mu_3, k)\right)}\, .
\end{equation}

In \cite{Preau:2025rex}, the formula \eqref{eq:exthydroproduct} was applied to the charged current polarization functions $\Pi^{\pm}_\perp$ and $\Pi^{\pm}_\parallel$ at zero chemical potential. We conclude this section by writing the generalization of these results to the case of finite isospin chemical potential. For reference, we also write the expression of the polarization functions in the hydrodynamic limit as derived in section \ref{sec:NAcorr} in standard hydrodynamics, and in section \ref{sec:holoNAcomputation} in the holographic model. We refer to these expressions as the {\em (near-extremal) hydrodynamic approximation}, while the one including the IR behaviour of the correlator \eqref{eq:exthydroproduct} is named {\em extended hydrodynamic approximation}:
\begin{itemize}
	\item (Near-extremal) hydrodynamic approximation
	\begin{align}\label{eq:trhydroapprox}
		&\text{Im}\Pi^{\perp,\pm}_\text{hydro} = -\Sigma\,\omega\,,\\ \label{eq:longhydroapprox}
		&\text{Im}\Pi^{\parallel,\pm}_\text{hydro} = -\Sigma\,\omega\dfrac{(\omega\pm\mu_3)^2- k^2}{(\omega\pm\mu_3)^2+D k^4}\,.
	\end{align}
	\item Extended hydrodynamic approximation
	\begin{align}\label{eq:trexthydroapprox}
		&\text{Im}\Pi^{\perp,\pm}_\text{ext-hydro} = -\Sigma\,\mu(\omega/\mu)^{2\Delta(k,\mu_3)-1}\,,\\ \label{eq:longexthydroapprox}
		&\text{Im}\Pi^{\parallel,\pm}_\text{ext-hydro} = -\Sigma\,\mu(\omega/\mu)^{2\Delta(k,\mu_3)-1}\dfrac{(\omega\pm\mu_3)^2- k^2}{(\omega\pm\mu_3)^2+D k^4}\,.
	\end{align}
\end{itemize}
In both cases, the approximations are valid at leading order in the non-Abelian hydrodynamic expansion, i.e.  in the   $\omega,\mu_3,k\ll \mu$ expansion.
This means that the generalized conductivities $\tilde{\gamma},\gamma^{\text{e}}$ are approximated by their leading order limit at  $\omega,\mu_3,k\ll \mu$ (given by the conductivity\footnote{For the longitudinal sector, the zeroes are actually factored out, so that $\gamma_\parallel = -\Sigma T((\omega\pm\mu_3)^2-\vec{k}^2)(1+\dots)$.} $\Sigma$), and the hydro-like poles reduce to their leading expansion at $\mu_3,k\ll\mu$.
It should be remarked, that as $\omega/\mu \to 0$ the two approximations become different, as the second one is resuming all the $k^2\log(\omega/ \m)$ terms that become important in this limit\footnote{Note that the exponent of $\omega$ in the extended hydrodynamic approximation \eqref{eq:trexthydroapprox} follows the expansion $2\Delta(k,\mu_3)-1 = 1 + \mathcal{O}((k^2/\mu^2),(\mu_3^2/\mu^2))$.}.

The modifications due to the presence of the non-zero isospin chemical potential are the real shift of the diffusive hydrodynamic pole, together with the $\mu_3$-dependence of the IR conformal dimension \eqref{eq:q4}. Note that the additional hydrodynamic-like pole visible in figures \ref{fig:wl2} and \ref{fig:wl3} indicates that there should be four factors in the denominator of \eqref{eq:exthydroproduct} instead of two. Although not visible at leading order in the gradient expansion, we expect these additional factors to appear at next-to-leading order (where they already appear for $\mu_3 = 0$ \cite{Gouteraux:2025kta}).

In section \ref{sec:exactresults}, we confront the exact numerical result for the charged current polarization functions with the hydrodynamic and extended hydrodynamic approximations, \eqref{eq:trhydroapprox} and \eqref{eq:trexthydroapprox}. As we shall show, the extended approximation gives a better description of the full holographic correlator over a significantly wider kinematic range than the (near-extremal) hydrodynamic approximation. This was explicitly demonstrated for probe current correlators in \cite{Preau:2025rex} at zero isospin chemical potential. In this work, we extend this analysis to non-zero isospin chemical potential, showing that the extended approximation of the correlator continues to provide a significantly better description.

\section{Exact numerical results for the charged correlators and comparison with hydrodynamic approximations}\label{sec:exactresults}

In this section, we present the results of the full numerical computation of the imaginary part of the transverse and longitudinal charged-current polarization functions; details of the numerical procedure are provided in appendix~\ref{app:numerical}. The calculation is performed in a background characterized by $\mu_q/T=65$. Although we do not attempt a direct application of our results to neutron-star phenomenology in the present work, this value has been chosen because it lies in the strongly degenerate regime relevant to matter in compact-star interiors. For typical quark chemical potentials expected at supranuclear densities, it corresponds to temperatures of a few MeV and may therefore be regarded as representative of the late-time evolution of a proto-neutron star or of a young neutron star~\cite{Roark:2018uls,Most:2018hfd,Malfatti:2019tpg}.\footnote{The value $\mu_q/T=65$ is the value used in \cite{neutrinopaper}, where these particular polarization functions were computed at zero isospin chemical potential.} We shall be particularly interested to compare these results with the (near-extremal) hydrodynamic and extended hydrodynamic approximations introduced in  formulae \eqref{eq:trhydroapprox}-\eqref{eq:longexthydroapprox} of the previous section. First we review the known results at zero isospin chemical potential and then we move to the case of non-zero isospin chemical potential, for which we focus on $\mu_3/\mu_q=-0.1$. Again, even if we do not apply our results to neutron star physics here we have picked a negative value for the isospin chemical potential in accordance with neutron star literature for which $\mu_3/\mu_q=-0.1$ represents a typical value \cite{Steiner:2004fi,Alford:2018lhf}. Also, this value of isospin chemical potential displays the qualitative features common for generic $\mu_3$.

Before discussing the analysis in detail, we give an overview of the structure and notation of the various figures and tables presented in this section and in the appendix.
First, we present in figures \ref{fig:transversepol} and \ref{fig:longitudinalpol}, respectively, the plots of the transverse and longitudinal correlators themselves as a function of frequency $\omega$ and momentum $k$. Then, we compare our exact, numerical results with the hydrodynamic (formulae \eqref{eq:trhydroapprox}-\eqref{eq:longhydroapprox}) and extended hydrodynamic (formulae \eqref{eq:trexthydroapprox}-\eqref{eq:longexthydroapprox}) approximations discussed in section \ref{sec:IRcorr-approx}.
In particular, we plot the relative difference, which  is defined as the absolute value of the difference between the exact, numerical result and a given approximation, divided by the numerical result:
\be\label{eq:percrelfdiff}
\text{relative difference} = 100\bigg|1- \dfrac{\text{Im}\Pi^{\perp,\parallel,\pm}_i}{\text{Im}\Pi^{\perp,\parallel,\pm}_\text{numerical}}\bigg|\,,
\ee
which is expressed as a percentage since it is multiplied by 100. The label $i$ is one of the two approximations of interest, hydrodynamic or extended hydrodynamic.\footnote{Other approximations are discussed in appendix \ref{app:IRcorr-approx} which are then analysed in the same way as in this section, in appendices \ref{app:otherapprox65}, \ref{app:coarsegrained}, and \ref{app:finegrained}.}

As already stated, in \cite{Preau:2025rex} (where  $\mu_3=0$), the so-called extended hydrodynamic approximation \eqref{eq:trexthydroapprox}-\eqref{eq:longexthydroapprox}, improves the hydrodynamic approximation \eqref{eq:trhydroapprox}-\eqref{eq:longhydroapprox}. In this work, we shall verify that this extends to non-zero (and small) isospin chemical potential.  Overall, the near-extremal hydrodynamic region, that reaches  up to $\omega/\mu\sim 1$ and $k/\mu\sim 1$ (see section \ref{results}), is reasonably well-described  by the extended hydrodynamic approximation, also in the case of (small) non-zero isospin chemical potential $\mu_3$.

The plots showing these relative differences are presented as follows. Figures \ref{fig:RelDiffNum_tr_mu65mu30} and \ref{fig:RelDiffNum_tr_mu65mu301} show the relative differences for the transverse polarization function respectively for $\mu_3= 0$ and $\mu_3/\mu_q=-0.1$, while figures \ref{fig:RelDiffNum_long_mu65mu30} and \ref{fig:RelDiffNum_long_mu65mu301} show the relative differences for the longitudinal polarization function respectively for $\mu_3= 0$ and $\mu_3/\mu_q=-0.1$.
We recall that in this section we consider $\mu_q/T =65$.

 In all the figures showing relative differences, the green line represents the locus $\omega = k + \mu_3$. This line corresponds to the shifted relativistic dispersion relation for charged excitations, in the presence of the isospin chemical potential. In both the standard hydrodynamic and ultraviolet regimes, the spectral functions depend predominantly on combinations of the form $(\omega-\mu_3)^2-k^2$, so that many of the dominant features of the relative-difference plots tend to align along the locus $\omega=k+\mu_3$. The red and pink contours represent the $10\%$ and $20\%$ lines, and the brown and dark brown contours the $1\%$ and $2\%$ lines. In every figure, the first row of plots displays the relative difference of the polarization function with respect to the hydrodynamic approximation, while the second row of plots is associated with the extended hydrodynamic approximation. Each row has two plots, the left one extends over the full near-extremal hydrodynamic region where $\omega,k$ reach $\mu$ , while the right one covers a smaller region closer to the standard hydrodynamic window identified by the dashed square $\omega/\mu,k/\mu  \in [0,T/\mu]$, where as usual $\mu\equiv \sqrt{\mu_q^2+\mu_3^2} = \mu_q + \OO(\mu_3/\mu_q)^2$.

In order to make our discussion more quantitative, we compute the matrix of coarse-grained and fine-grained relative differences respectively defined as
\be\label{eq:chiRL}
\chi_{ij}^{(l)} =\dfrac{1}{n_i\cdot n_j}\,\int_0^{n_i}\intd(\omega/\mu) \int_0^{n_j}\intd (k/\mu)\,\bigg|1- \dfrac{\text{Im}\Pi^{\perp,\parallel,\pm}_l}{\text{Im}\Pi^{\perp,\parallel,\pm}_\text{numerical}}\bigg|\,,
\ee
\be\label{eq:chiCELL}
s_{ij}^{(l)} =64\,\int_{n_i}^{n_i+1/8}\intd(\omega/\mu) \int_{n_j}^{n_j+1/8}\intd (k/\mu)\,\bigg|1- \dfrac{\text{Im}\Pi^{\perp,\parallel,\pm}_l}{\text{Im}\Pi^{\perp,\parallel,\pm}_\text{numerical}}\bigg|\,,
\ee
where $n_i$ and $n_j$ are the lattice points defining the region over which the integrals are performed, and $l\in\{\text{hydro},\text{ext-hydro}\}$ refers to which approximation we are considering. We recall that the  two approximations for the transverse and longitudinal correlators were defined in formulae \eqref{eq:trhydroapprox}-\eqref{eq:longexthydroapprox}. We consider $n_{i} =i/8$ with $i=1,2,3,4$ and the same for $n_j$.

In particular, the observable $\chi^{(l)}_{ij}$ is a $4\times 4$ matrix where each entry returns the value of the integrated relative difference computed inside rectangles on the $(\omega/\mu,k/\mu)$-plane defined by
\be
0\leq {\omega\over \mu}\leq n_i\sp 0\leq {k\over \mu}\leq n_j\,.
\ee
The observable $s^{(l)}_{ij}$ is a $4\times 4$ matrix where each entry returns the value of the integrated relative difference computed inside a single cell defined by
\be
n_i\leq {\omega\over \mu}\leq n_i+\dfrac{1}{8}\sp n_j\leq {k\over \mu}\leq n_j+\dfrac{1}{8}\,.
\ee
Note that in all the tables we present in this work, the horizontal axis refers to $\omega/\mu$ and the vertical axis to $k/\mu$. The four values on each axis correspond to the values of $n_{i}$ and $n_j$ respectively.

These observables will provide clear numerical results supporting what can be argued for, by looking at the plots showing the various relative differences: for the values of $\mu_q/T$ and $\mu_3/\mu_q$ presented in this section, the extended hydrodynamic approximation approximates the correlators better than the (near-extremal) hydrodynamic results over all the near-extremal hydrodynamic region. The improvement is especially significant near the $\omega=0$ axis.

In addition, appendix \ref{app:otherapprox65} is devoted to the discussion of other approximations based on the full IR-AdS$_2$ correlator \eqref{eq:q3} for $\mu_q/T=65$ and $\mu_3=0$. These approximations (discussed in appendix \ref{app:IRcorr-approx}) are shown to have a similar accuracy to the extended hydrodynamic one, leaving the latter as the simplest approximation to the charged current spectral functions. Moreover, in appendix~\ref{app:otherexactresults}, we present the full numerical results for the charged-current correlators for additional values of the dimensionless ratios $\mu_q/T$ and $\mu_3/\mu_q$. These results are compared with both the hydrodynamic and extended-hydrodynamic approximations, allowing us to assess how their domains of validity depend on the temperature and on the isospin imbalance.
In addition to the reference value $\mu_q/T=65$, we consider the other two representative regimes $\mu_q/T\in\{10^4,5\}$. These choices approximately span the range of thermodynamic conditions relevant to compact-star physics~\cite{Glendenning:1992vb,Jaikumar:2002vg,Buballa:2003qv,Steiner:2004fi,Jaikumar:2005hy,Alford:2007xm,Dexheimer:2008ax,Belvedere:2012qn,Alford:2018lhf,Roark:2018uls}. The value $\mu_q/T=10^4$ describes an extremely degenerate, effectively zero-temperature regime and may therefore be regarded as representative of cold neutron-star matter. By contrast, $\mu_q/T=5$ corresponds to a much hotter regime of the type that can be reached locally during a binary neutron-star merger. From the holographic perspective, the latter choice is also useful because it allows us to investigate the departure from the near-extremal limit and to determine how finite-temperature corrections modify the spectral functions and the range of applicability of the hydrodynamic description.
For each of the three temperature regimes, $\mu_q/T\in\{10^4,65,5\}$, we also extend the analysis to $\mu_3/\mu_q=-0.5$.  This deliberately large isospin chemical potential makes the limitations of the hydrodynamic and extended-hydrodynamic approximations particularly transparent, providing a useful stress test of the corresponding expansions. Although such a large value is not expected to characterize homogeneous, charge-neutral matter in weak equilibrium inside a neutron star, it may be viewed as a controlled proxy for large local isospin imbalances that could arise transiently during out-of-equilibrium processes, before weak interactions restore chemical equilibrium.

Finally, in appendices \ref{app:coarsegrained} and \ref{app:finegrained}, we present the results for the observables $\chi_{ij}$ (formula \eqref{eq:chiRL}) and $s_{ij}$ (formula \eqref{eq:chiCELL}) for all the values of quark and isospin chemical potentials $\mu_q/T\in\{10^4,65,5\}$, $\mu_3/\mu_q\in\{0,-0.1,-0.5\}$ and all the various analytic approximations to the exact correlator presented in appendix \ref{app:IRcorr-approx}.

\subsection{Transverse correlator}
In this section, we present the exact, numerical results for the transverse charged current correlator and compare them with the hydrodynamic and extended hydrodynamic approximations.

\begin{figure}[htb]
	\centering
	\begin{subfigure}{0.48\textwidth}
		\centering
		\includegraphics[width=\textwidth]{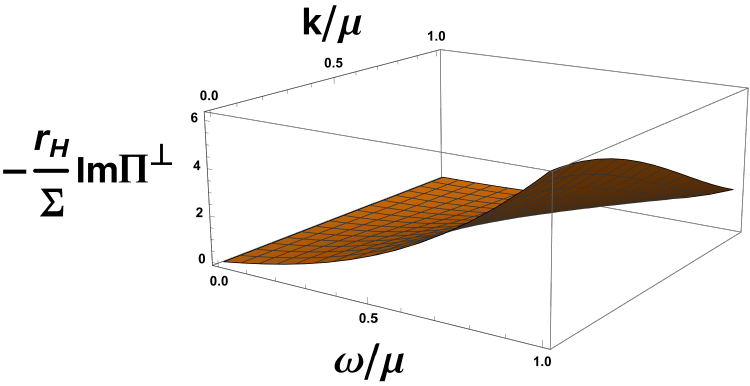}
		\caption{}\label{fig:transversepol650}
	\end{subfigure}
	\hfill
	\begin{subfigure}{0.48\textwidth}
		\centering
		\includegraphics[width=\textwidth]{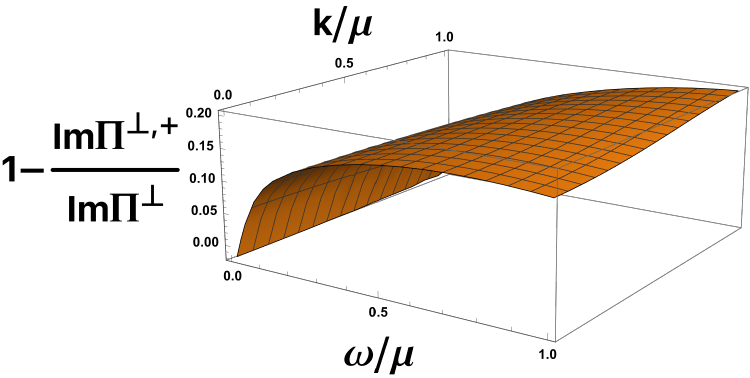}
		\caption{}\label{fig:transversepol65RelDiff}
		\end{subfigure}
	\caption{In the left panel \ref{fig:transversepol650}, we plot the
		imaginary part of the transverse charged current retarded polarization functions at zero isospin chemical potential for $\mu_q/T = 65$ (the  $\pm$ notation has been suppressed since we are working at $\mu_3=0$). The energy and momentum are expressed
		in units of $\mu$, and the polarization function is normalized by $-\Sigma/r_H$. The right panel \ref{fig:transversepol65RelDiff} shows the plot of the relative difference $1-\text{Im}\Pi^{\perp,+}/\text{Im}\Pi^{\perp}$ between the imaginary part of the polarization function at finite isospin chemical potential $\mu_3/\mu_q=-0.1$ and the one at $\mu_3=0$, as a function of $\omega/\mu$ and $k/\mu$ for $\mu_q/T = 65$.} \label{fig:transversepol}
\end{figure}

We start by plotting the imaginary part of the transverse polarization function for $\mu_q/T =65$ and $\mu_3 = 0$ in the left panel \ref{fig:transversepol650} of figure \ref{fig:transversepol}. Since the plot of the same quantity for $\mu_q/T =65$ and $\mu_3/\mu_q = -0.1$ is very similar to figure \ref{fig:transversepol650}, it is more instructive to look at their relative difference instead. In the right panel \ref{fig:transversepol65RelDiff} of figure \ref{fig:transversepol}, we plot the relative difference $1-\text{Im}\Pi^{\perp,+}/\text{Im}\Pi^{\perp}$ between the imaginary part of the polarization function at finite isospin chemical potential $\mu_3/\mu_q=-0.1$ (numerator) and the one at $\mu_3=0$, for $\mu_q/T = 65$ (denominator).

The plots show that, for both values of $\mu_3/\mu_q$, the behaviour of the spectral function is qualitatively the same: the spectral function  asymptotes  to zero for small frequencies while it grows as the frequency increases, and decreases with momentum. The magnitude of the spectral function at non-zero isospin chemical potential, is smaller with respect to the $\mu_3=0$ case.\footnote{We have checked that, by increasing the magnitude of a negative isospin chemical potential, this effect is more prominent. See appendix \ref{app:otherexactresults} where we consider $\mu_3/\mu_q=-0.5$. We have also checked that the effect is reversed for positive $\mu_3/\mu_q$.} Moreover, note that the relative difference is of the same order as $\mu_3/\mu_q$, as expected from the fluctuation equation \eqref{EoMVhtpm}.

The subsequent plots compare the numerical results shown in figure \ref{fig:transversepol} with the hydrodynamic and extended hydrodynamic approximations. We first review in subsection \ref{sec:tr_mu30} the case of zero isospin chemical potential already discussed in \cite{Preau:2025rex}, before discussing the case of non-zero $\mu_3$ in subsection \ref{sec:tr_mu301}.

\subsubsection{Review of results at zero isospin chemical potential}\label{sec:tr_mu30}

\begin{figure}[htb]
	\begin{subfigure}{\textwidth}
		\centering
		\begin{subfigure}{0.4\textwidth}
			\centering
			\includegraphics[width=\textwidth]{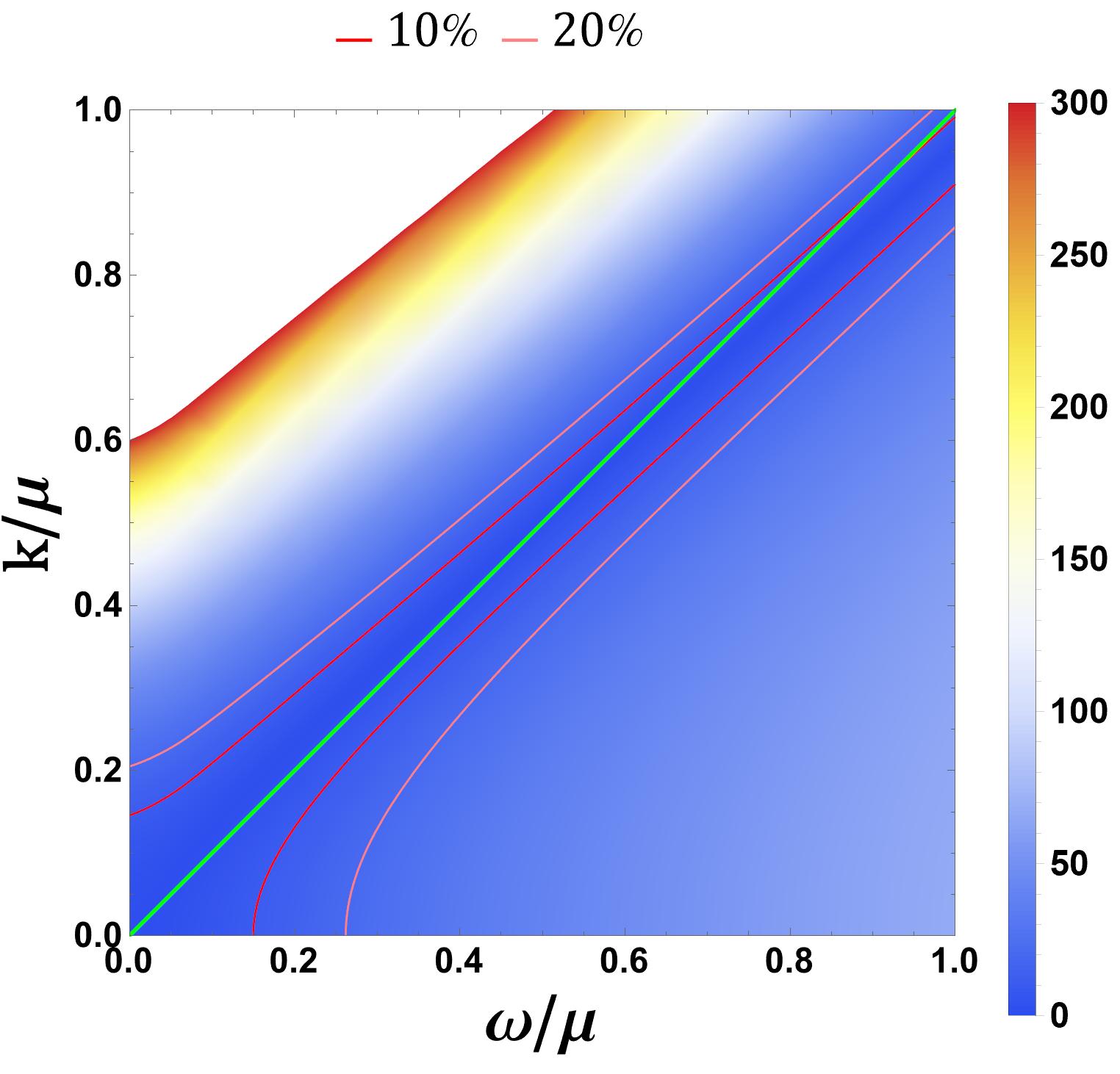}
		\end{subfigure}
		\hspace{1cm}
		\begin{subfigure}{0.4\textwidth}
			\centering
			\includegraphics[width=\textwidth]{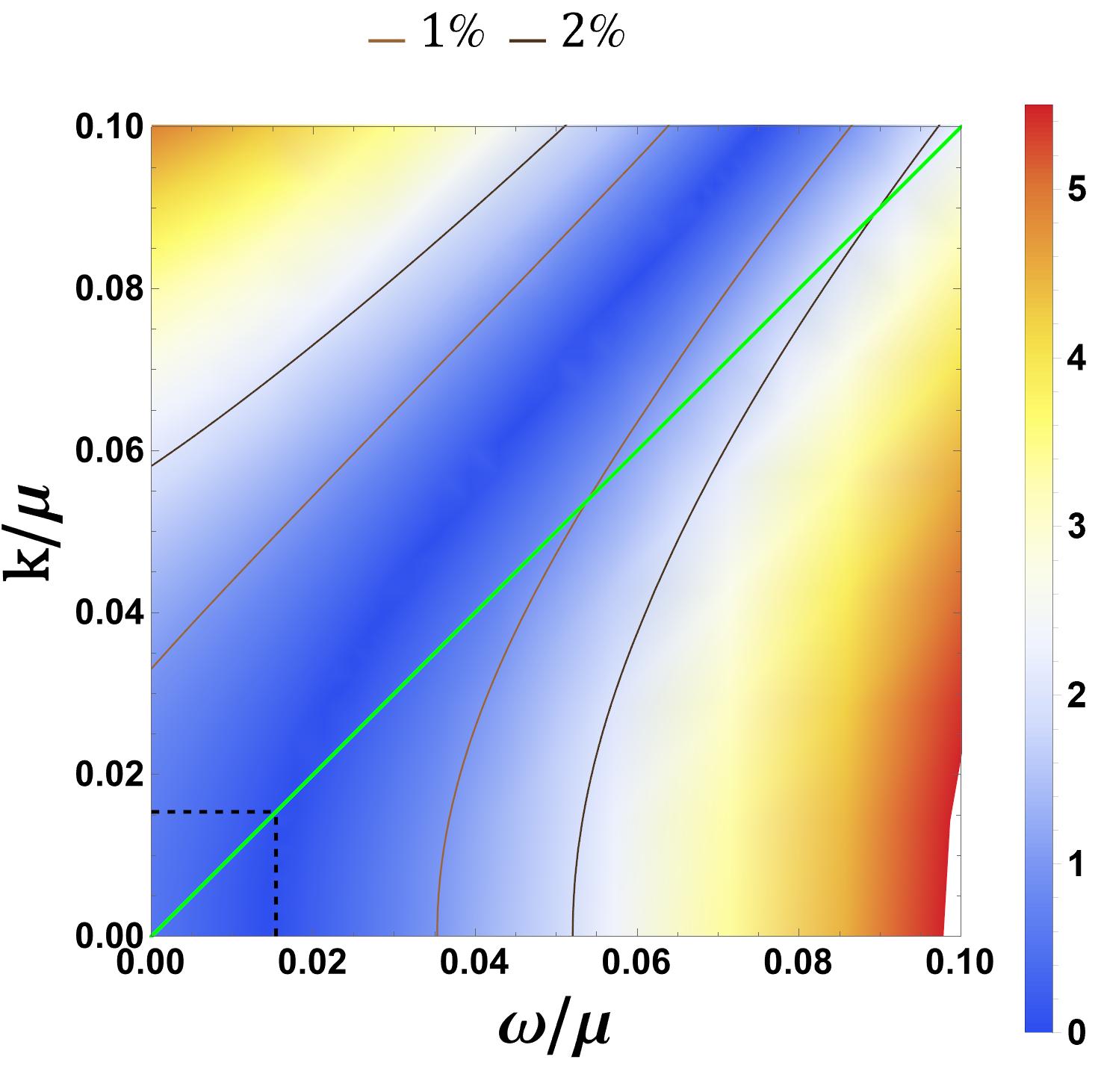}
		\end{subfigure}
	\end{subfigure}

	\begin{subfigure}{\textwidth}
		\centering
		\begin{subfigure}{0.4\textwidth}
			\centering
			\includegraphics[width=\textwidth]{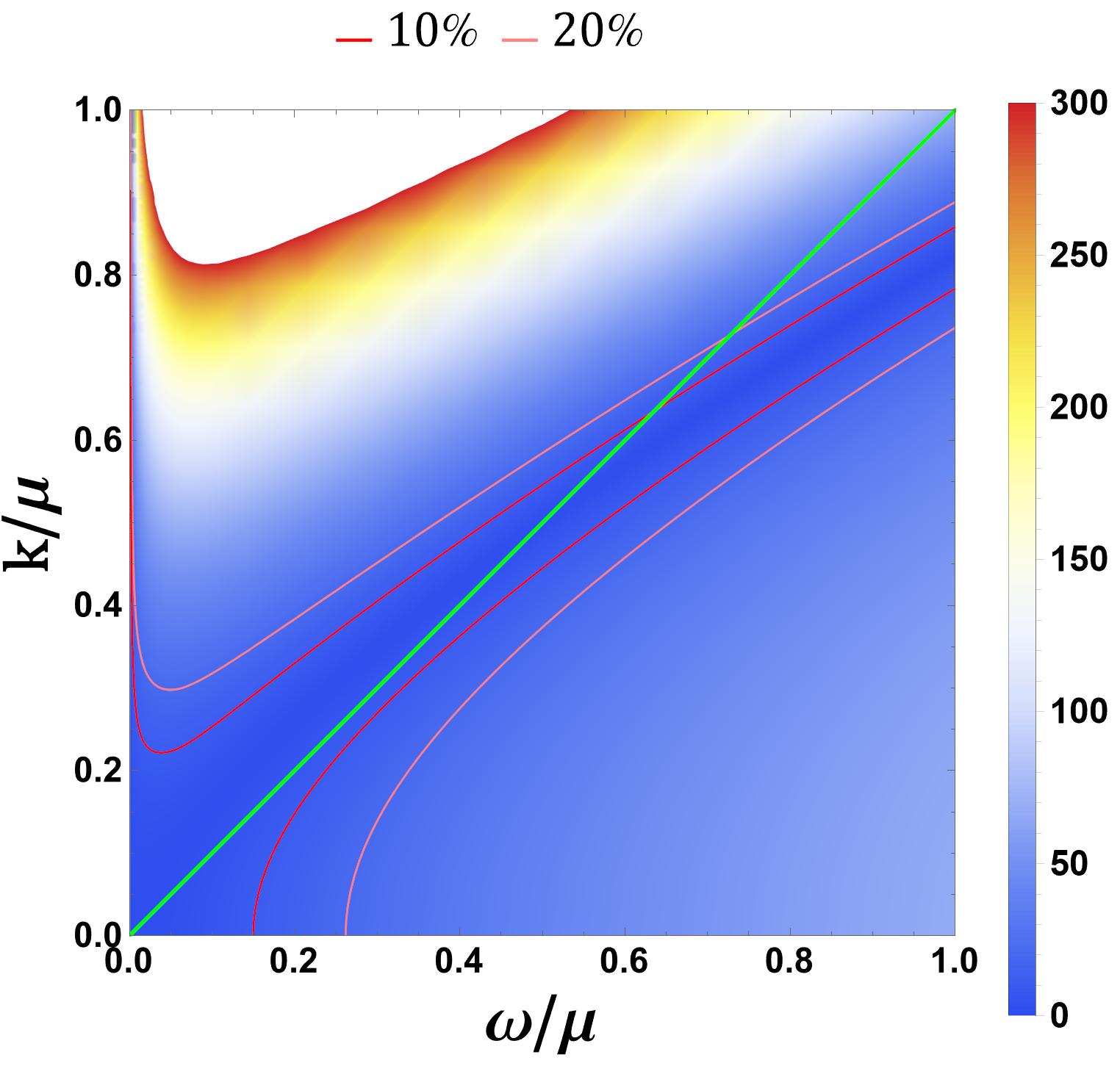}
		\end{subfigure}
		\hspace{1cm}
		\begin{subfigure}{0.4\textwidth}
			\centering
			\includegraphics[width=\textwidth]{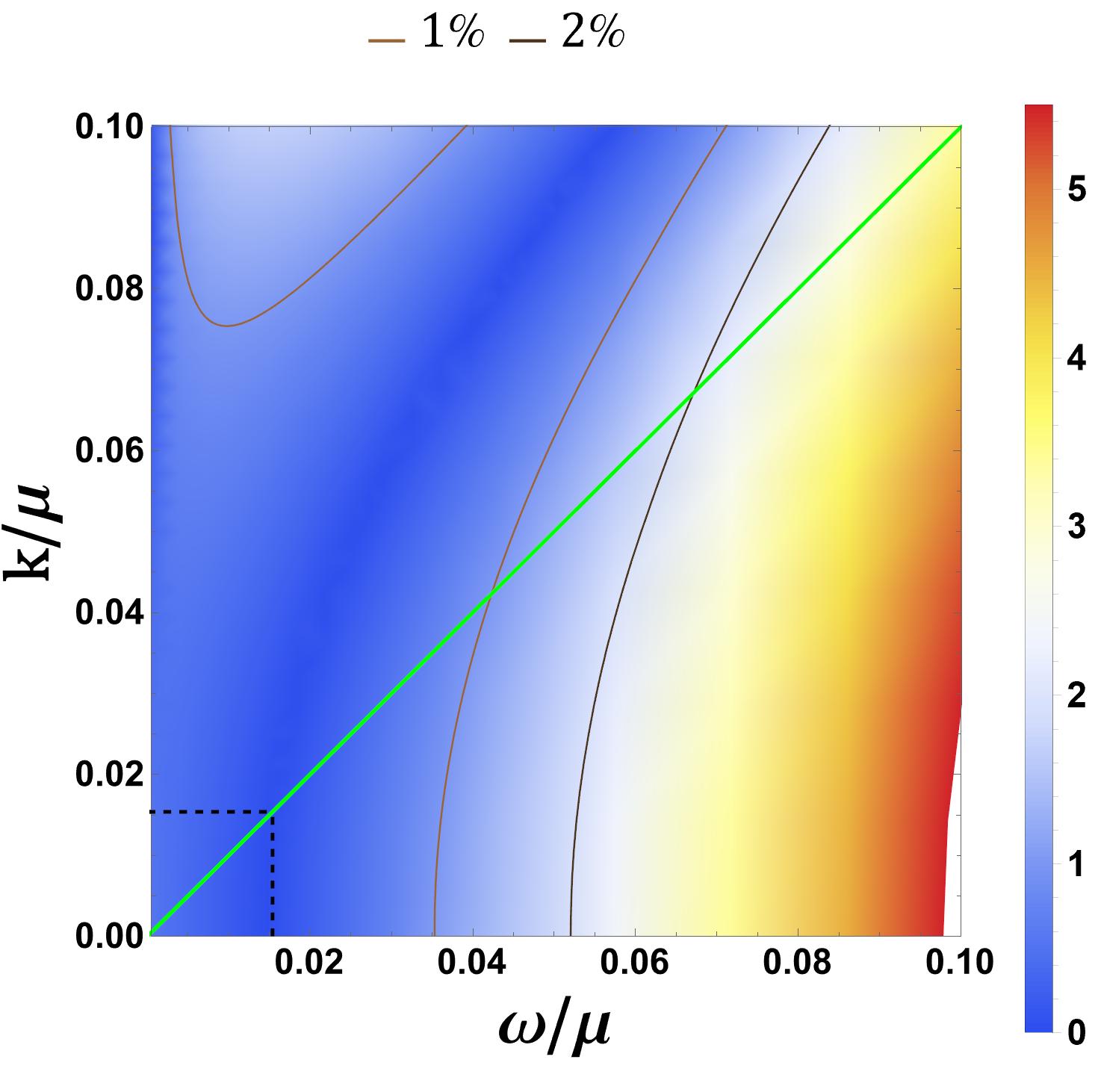}
		\end{subfigure}
	\end{subfigure}
	
	\caption{The percentage relative difference \eqref{eq:percrelfdiff} of the imaginary part of the transverse charged current polarization function with respect to the hydrodynamic approximation \eqref{eq:trhydroapprox} (top row) and the extended hydrodynamic approximation \eqref{eq:trexthydroapprox} (bottom row), for $\mu_q/T= 65$ and $\mu_3=0$. The green line shows the locus $\omega = k$. The right plots shows a subregion of the left
		plots. The dashed square in these right plots represents the standard hydrodynamic region $\omega/\mu,k/\mu \in [0,T/\mu]$.} \label{fig:RelDiffNum_tr_mu65mu30}
\end{figure}

We present in figure \ref{fig:RelDiffNum_tr_mu65mu30} the plots of the relative differences \eqref{eq:percrelfdiff} for the transverse spectral function at zero isospin chemical potential.

The first general feature, common to all panels, is that the relative difference is small in the near-extremal hydrodynamic regime, namely for small $\omega$ and $k$ much smaller than $\mu$, and then increases as one moves away from the origin towards larger frequencies and momenta. This is the behaviour expected from the analysis of section \ref{sec:holoNAcomputation}.

Overall, the message of figure \ref{fig:RelDiffNum_tr_mu65mu30} can be summarized as follows. (Near-extremal) hydrodynamics represents a good approximation (as predicted by the results of section \ref{sec:holoNAcomputation}) in the range of small frequencies and momenta. However, the extended hydrodynamic approximation, which takes into account the leading effect of the IR branch cut, approximate better the exact correlator especially, as expected, in the region of small frequencies.

It is worth mentioning that for the hydrodynamic approximation (top row of figure \ref{fig:RelDiffNum_tr_mu65mu30}) the region of small relative difference is not confined to the immediate hydrodynamic corner, but extends along a slanted band in the $(\omega/\mu,k/\mu)$ plane, approximately following the line $\omega\sim k$. From the plots, this band lies mostly within the $10\%$ and $20\%$ contours, reflecting the fact that at sufficiently large frequencies and momenta the correlators depend predominantly on combinations of the form $\omega^2-k^2$.

In the comparison with the extended hydrodynamic approximation (bottom row of figure \ref{fig:RelDiffNum_tr_mu65mu30}), the $10\%$ and $20\%$ lines are bent with respect to the hydrodynamic plots where they were following the line $\omega\sim k$. This happens because, since the approximation behaves as $\omega^{2\Delta-1}$, the frequency dependence of the approximation relative to the exact correlator is modified. As a consequence, the contours of constant relative difference are no longer approximately aligned with the shifted relativistic scaling direction, and bend at larger frequencies.

We support the analysis of the plots, by presenting the results for the observables $\chi_{ij}$ and $s_{ij}$ respectively defined in formulae \eqref{eq:chiRL} and \eqref{eq:chiCELL}, which are shown in tables \ref{tab:chi650_tr} and \ref{tab:s650_tr}.
\begin{table}[htb]
	\centering
	{
		\small
		\begin{tabular}{c|cccc}
			\multicolumn{5}{c}{(a) hydrodynamic}\\
			$1/2$ & 0.38 & 0.30 & 0.26 & 0.25 \\
			$3/8$ & 0.18 & 0.15 & 0.15 & 0.18 \\
			$1/4$ & 0.07 & 0.08 & 0.12 & 0.17 \\
			$1/8$ & 0.03 & 0.07 & 0.13 & 0.18 \\\hline
			& $1/8$ & $1/4$ & $3/8$ & $1/2$ \\
		\end{tabular}\hspace{1cm}
		\begin{tabular}{c|cccc}
			\multicolumn{5}{c}{(b) extended-hydrodynamic}\\
			$1/2$ & 0.18 & 0.17 & 0.17 & 0.18 \\
			$3/8$ & 0.09 & 0.09 & 0.12 & 0.16 \\
			$1/4$ & 0.04 & 0.06 & 0.11 & 0.16 \\
			$1/8$ & 0.02 & 0.07 & 0.13 & 0.18 \\\hline
			& $1/8$ & $1/4$ & $3/8$ & $1/2$ \\
		\end{tabular}
	}
	\caption{Values $\chi_{ij}$ (formula \eqref{eq:chiRL}) in the transverse sector for $\mu_q/T=65$ and $\mu_3=0$.}
	\label{tab:chi650_tr}
\end{table}

\begin{table}[htb]
	\centering
	{
		\small
		\begin{tabular}{c|cccc}
			\multicolumn{5}{c}{(a) hydrodynamic}\\
			$1/2$ & 0.97 & 0.54 & 0.22 & 0.08 \\
			$3/8$ & 0.41 & 0.17 & 0.07 & 0.19 \\
			$1/4$ & 0.12 & 0.06 & 0.17 & 0.29 \\
			$1/8$ & 0.03 & 0.12 & 0.23 & 0.34 \\\hline
			& $1/8$ & $1/4$ & $3/8$ & $1/2$ \\
		\end{tabular}\hspace{1cm}
		\begin{tabular}{c|cccc}
			\multicolumn{5}{c}{(b) extended-hydrodynamic}\\
			$1/2$ & 0.45 & 0.35 & 0.16 & 0.07 \\
			$3/8$ & 0.20 & 0.10 & 0.07 & 0.20 \\
			$1/4$ & 0.05 & 0.06 & 0.18 & 0.30 \\
			$1/8$ & 0.02 & 0.12 & 0.24 & 0.34 \\\hline
			& $1/8$ & $1/4$ & $3/8$ & $1/2$ \\
		\end{tabular}
	}
	\caption{Values $s_{ij}$ (formula \eqref{eq:chiCELL}) in the transverse sector for $\mu_q/T=65$ and $\mu_3=0$.}
	\label{tab:s650_tr}
\end{table}

We can observe that for both $\chi_{ij}$ (table \ref{tab:chi650_tr}) and $s_{ij}$ (table \ref{tab:s650_tr}) the values associated with the extended hydrodynamic approximation (right tables) are overall smaller than the ones displayed for the hydrodynamic approximation (left tables). The improvement is more substantial at small $\omega/\mu$ (\textit{i.e.} the first column of all the tables) and it increases at larger values of $k/\mu$.

 We note, in particular, that in this case, in units of $\mu$, $T\simeq 0.015$ and therefore the hydrodynamic approximation has an average 3\% error when $k,\omega$ go up to 8 times the temperature, and an average error of about 8\% when $k,\omega$ go up to almost 16 times the temperature. For the extended hydrodynamic approximation, the respective associated errors are 2\% and 6\%. In this case, the error rises to 18\%,  when $k,\omega$ go up to 32 times the temperature.

\subsubsection{Results at non-zero isospin chemical potential $\mu_3/\mu_q=-0.1$}\label{sec:tr_mu301}

\begin{figure}[htb]
	\begin{subfigure}{\textwidth}
		\centering
		\begin{subfigure}{0.4\textwidth}
			\centering
			\includegraphics[width=\textwidth]{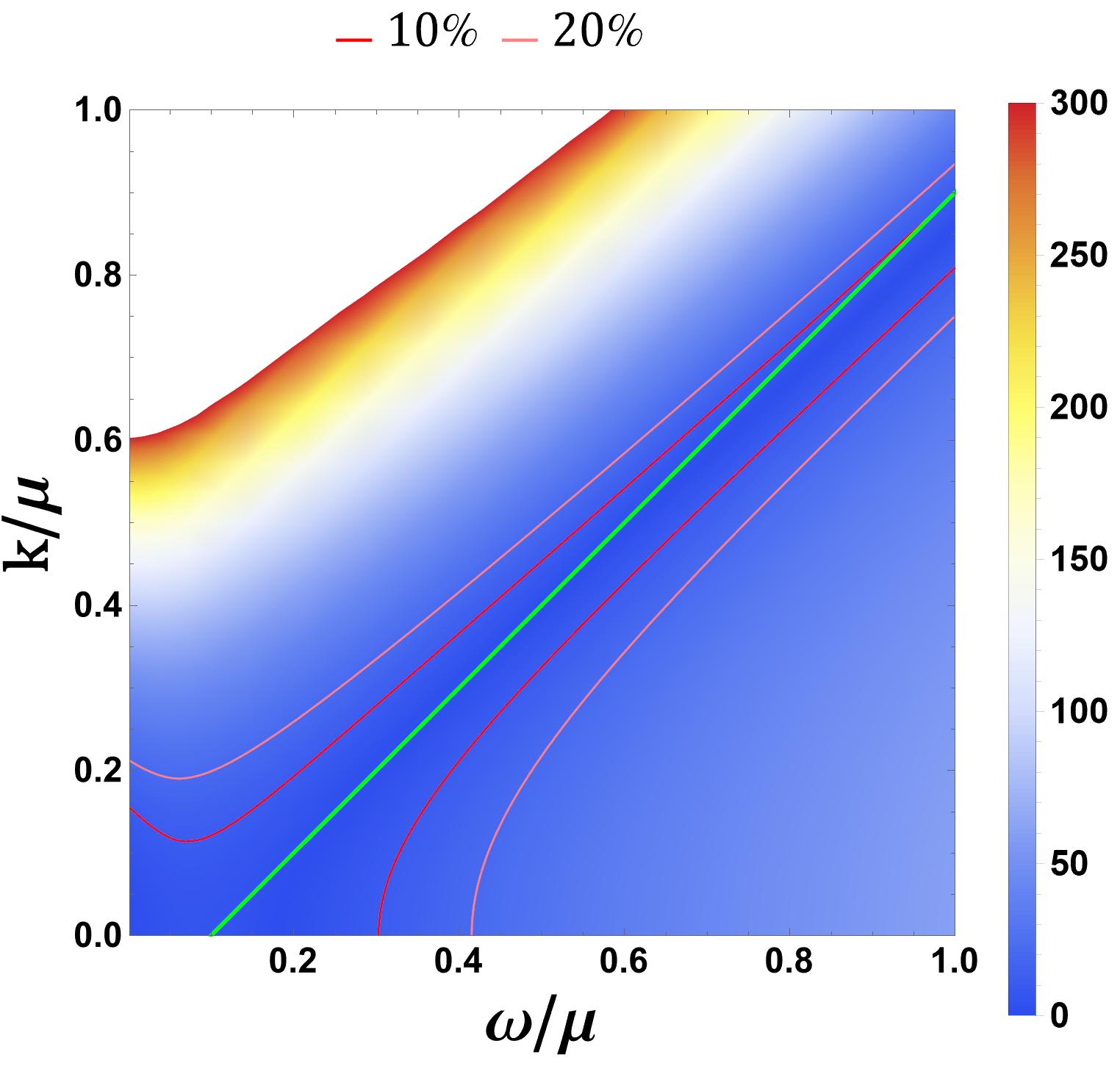}
		\end{subfigure}
		\hspace{1cm}
		\begin{subfigure}{0.4\textwidth}
			\centering
			\includegraphics[width=\textwidth]{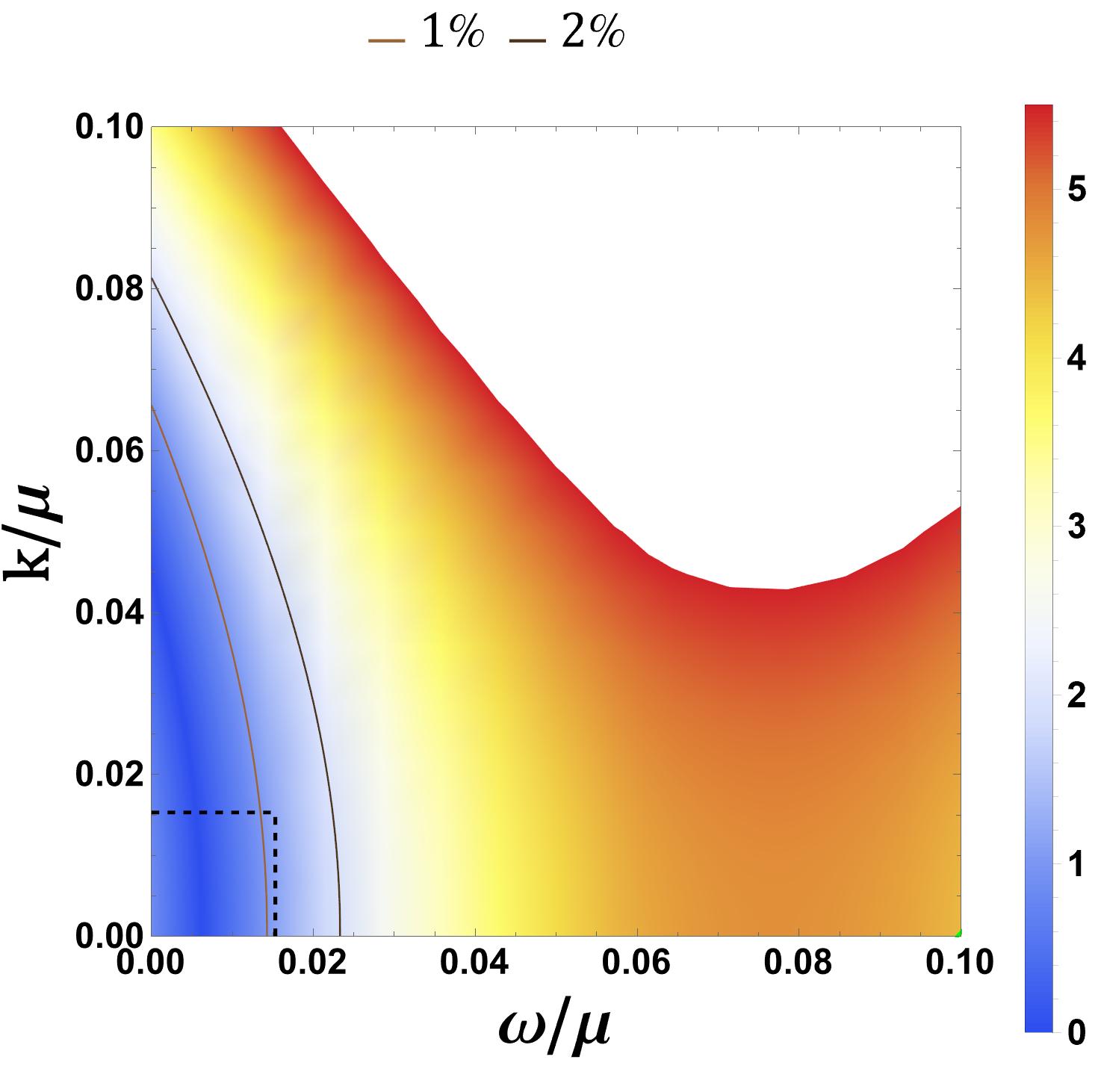}
		\end{subfigure}
	\end{subfigure}

	\begin{subfigure}{\textwidth}
		\centering
		\begin{subfigure}{0.4\textwidth}
			\centering
			\includegraphics[width=\textwidth]{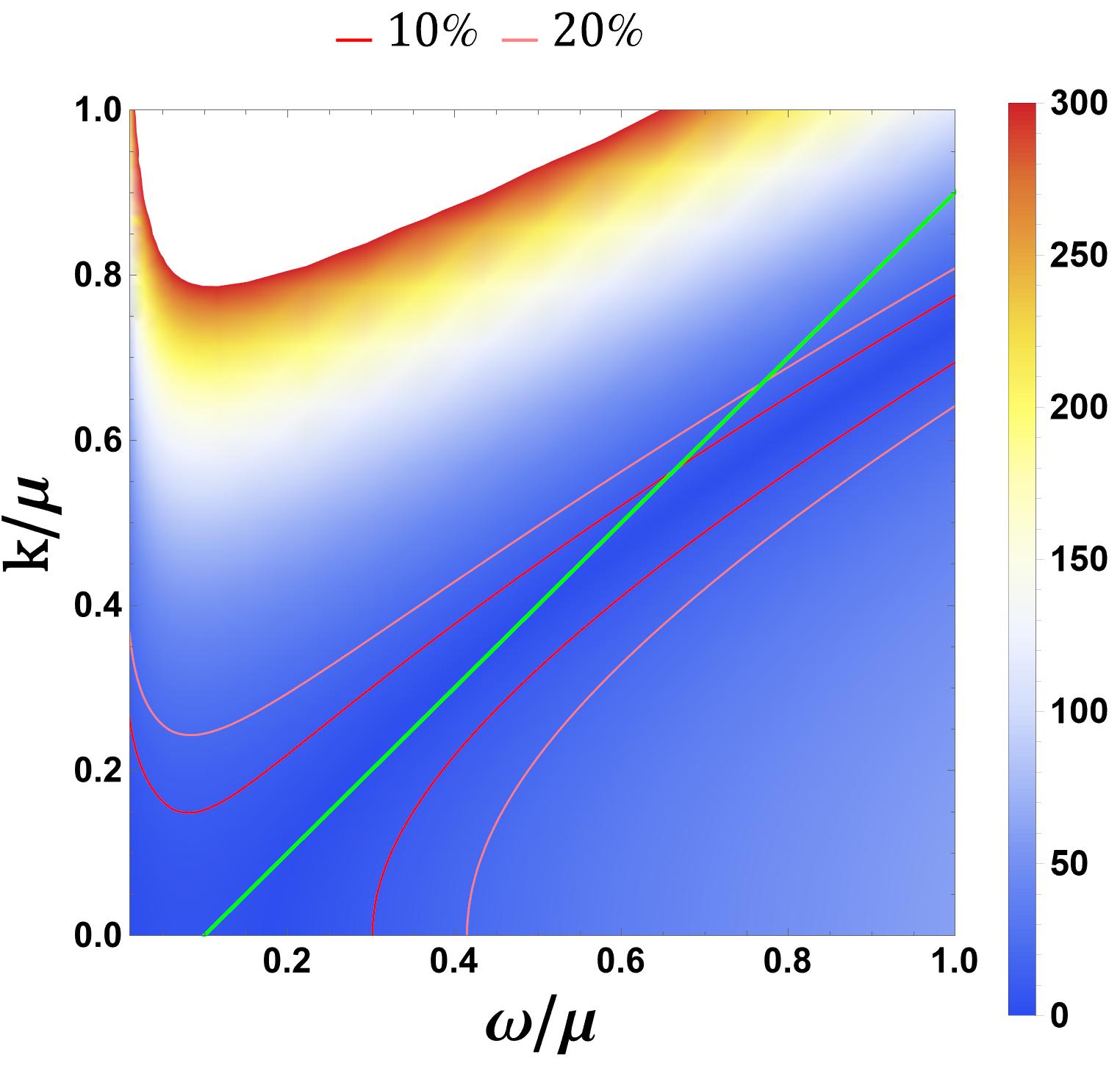}
		\end{subfigure}
		\hspace{1cm}
		\begin{subfigure}{0.4\textwidth}
			\centering
			\includegraphics[width=\textwidth]{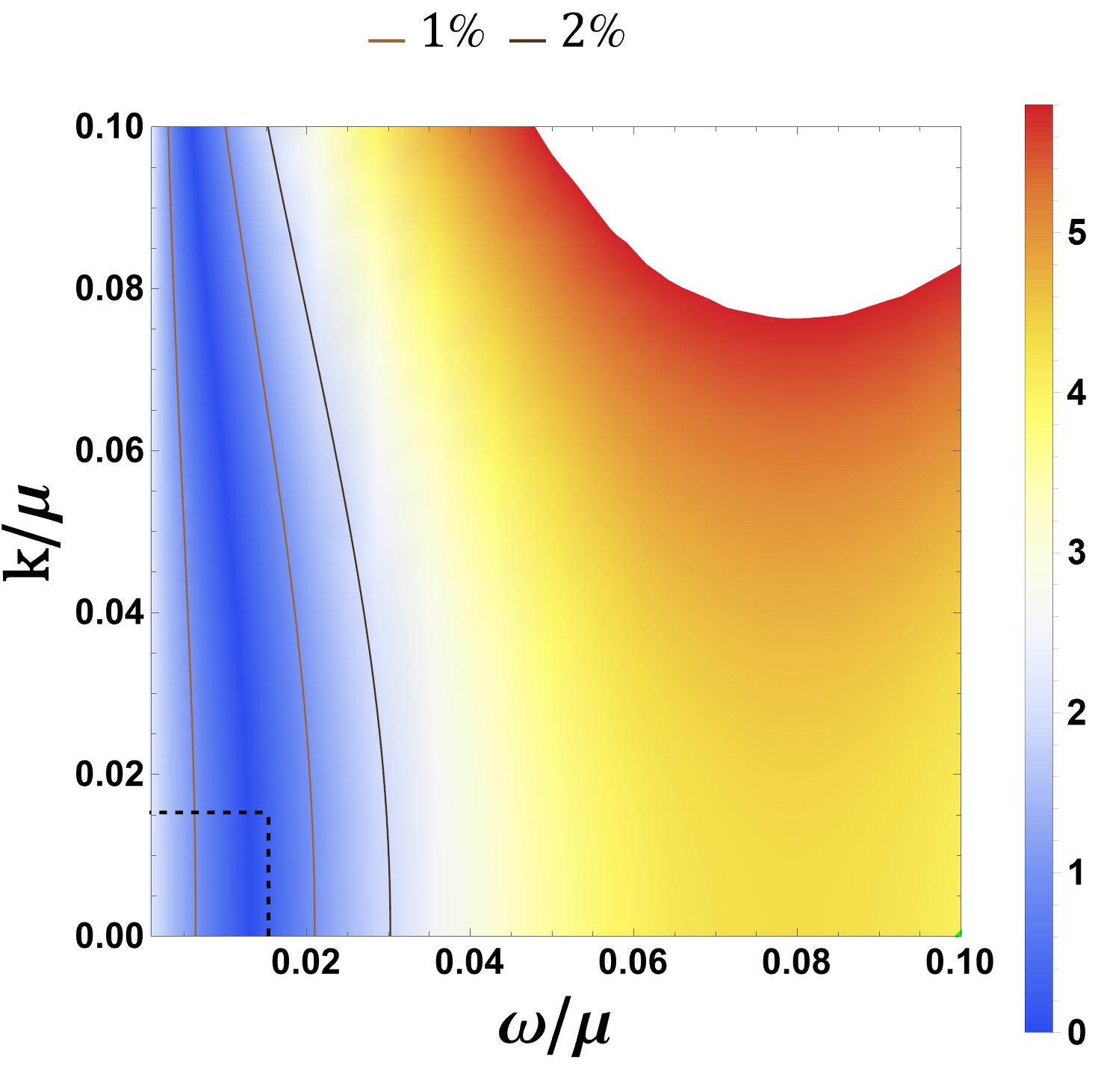}
		\end{subfigure}
	\end{subfigure}

	\caption{The percentage relative difference \eqref{eq:percrelfdiff} of the imaginary part of the transverse charged current polarization function with respect to the hydrodynamic approximation \eqref{eq:trhydroapprox} (top row) and the extended hydrodynamic approximation \eqref{eq:trexthydroapprox} (bottom row) for $\mu_q/T= 65$ and $\mu_3/\mu_q=-0.1$. The green line describes the locus $\omega=k+\mu_3$. The right plots shows a subregion of the left
		plots. The dashed square in these right plots represents the standard hydrodynamic region $\omega/\mu,k/\mu \in [0,T/\mu]$.}\label{fig:RelDiffNum_tr_mu65mu301}
\end{figure}

Figure \ref{fig:RelDiffNum_tr_mu65mu301} shows the plots of the relative difference \eqref{eq:percrelfdiff} for the transverse polarization function at $\mu_q/T=65$ and non-zero isospin chemical potential $\mu_3/\mu_q=-0.1$. Many features are similar to the previous figure \ref{fig:RelDiffNum_tr_mu65mu30}. For example, as in the $\mu_3 = 0$ case, for the extended hydrodynamic approximation (bottom row of figure \ref{fig:RelDiffNum_tr_mu65mu301}), the $10\%$ and $20\%$ contours are slightly bent with respect to the line $\omega = k + \mu_3$, while for the hydrodynamic approximation (top row of figure \ref{fig:RelDiffNum_tr_mu65mu301}) they follow more closely the line $\omega = k + \mu_3$.

A key qualitative difference of figure \ref{fig:RelDiffNum_tr_mu65mu301} with respect to the $\mu_3=0$ case (figure \ref{fig:RelDiffNum_tr_mu65mu30}) is that the region of small relative difference is no longer centered around $\omega\sim k$, but instead aligns along the shifted line, $\omega = k + \mu_3$, shown by the green line in the plots. As we already argued in the introduction of this section, since the chemical potential acts as a background temporal gauge field and shifts the frequency of charged modes according to $\omega\to\omega+\mu_3$, the correlators depend approximately on the combination $(\omega+\mu_3)^2-k^2$ instead of $\omega^2-k^2$.

In the case in which we approximate the exact result with the hydrodynamic approximation, the $10\%$ and $20\%$ contours closely follow this direction, indicating that the agreement
between the approximations and the exact result is maximal along this kinematic locus. In particular, the presence of a finite
isospin chemical potential, effectively displaces the relevant low-energy kinematics.

For the hydrodynamics at finite isospin chemical potential to be a good approximation, $\mu_3$ has to be small compared to the hard scale $\mu$. In the present case, $\mu_3$ is not too large and we observe a slight shift of the relevant low-energy excitations around $\omega \sim -\mu_3$ rather than $\omega \sim 0$. This can be inferred from the 10$\%$ and 20$\%$ lines which move to right in the left column of figure \ref{fig:RelDiffNum_tr_mu65mu301} if compared to the left column of \ref{fig:RelDiffNum_tr_mu65mu30}. However, in the appendix \ref{app:otherexactresults}, we consider the case of $\mu_3/\mu_q=-0.5$ for which the various approximations deteriorate even for small $\omega/\mu$ and $k/\mu$ due to the not so small isospin chemical potential (see section \ref{app:mu305}).

Like for the $\mu_3=0$ case, the extended hydrodynamic approximation (bottom row of figure \ref{fig:RelDiffNum_tr_mu65mu301})
significantly enlarges the region of validity close to $\omega = 0$ with respect to the hydrodynamic approximation (top row of figure \ref{fig:RelDiffNum_tr_mu65mu301}). This is again supported by the results of the relative differences \eqref{eq:chiRL} and \eqref{eq:chiCELL}, which are shown in tables \ref{tab:chi6501_tr} and \ref{tab:s6501_tr}. Also in this case we observe that the extended hydrodynamic approximation does particularly better than the hydrodynamic one at small $\omega/\mu$,  as $k/\mu$ increases (first column of all the tables). We can also compare the  tables above with the corresponding ones for $\mu_3=0$ (tables \ref{tab:chi650_tr} and \ref{tab:s650_tr}). We observe  that the integrated relative difference is bigger in the case of non-zero isospin chemical potential, especially for small frequencies. This supports the previous observation that increasing $\mu_3$ makes the hydrodynamic approximation worse.

\begin{table}[htb]
	\centering
	{
		\small
		\begin{tabular}{c|cccc}
			\multicolumn{5}{c}{(a) hydrodynamic}\\
			$1/2$ & 0.46 & 0.38 & 0.32 & 0.28 \\
			$3/8$ & 0.25 & 0.20 & 0.17 & 0.16 \\
			$1/4$ & 0.12 & 0.09 & 0.09 & 0.11 \\
			$1/8$ & 0.05 & 0.04 & 0.06 & 0.10 \\\hline
			& $1/8$ & $1/4$ & $3/8$ & $1/2$ \\
		\end{tabular}\hspace{1cm}
		\begin{tabular}{c|cccc}
			\multicolumn{5}{c}{(b) extended-hydrodynamic}\\
			$1/2$ & 0.24 & 0.23 & 0.21 & 0.19 \\
			$3/8$ & 0.14 & 0.13 & 0.11 & 0.12 \\
			$1/4$ & 0.08 & 0.06 & 0.07 & 0.10 \\
			$1/8$ & 0.05 & 0.03 & 0.05 & 0.09 \\\hline
			& $1/8$ & $1/4$ & $3/8$ & $1/2$ \\
		\end{tabular}
	}
	\caption{Values $\chi_{ij}$ (formula \eqref{eq:chiRL}) in the transverse sector for $\mu_q/T=65$ and $\mu_3=-0.1$.}
	\label{tab:chi6501_tr}
\end{table}

\begin{table}[htb]
	\centering
	{
		\small
		\begin{tabular}{c|cccc}
			\multicolumn{5}{c}{(a) hydrodynamic}\\
			$1/2$ & 1.10 & 0.76 & 0.44 & 0.19 \\
			$3/8$ & 0.50 & 0.34 & 0.14 & 0.06 \\
			$1/4$ & 0.19 & 0.11 & 0.05 & 0.15 \\
			$1/8$ & 0.05 & 0.03 & 0.10 & 0.21 \\\hline
			& $1/8$ & $1/4$ & $3/8$ & $1/2$ \\
		\end{tabular}\hspace{1cm}
		\begin{tabular}{c|cccc}
			\multicolumn{5}{c}{(b) extended-hydrodynamic}\\
			$1/2$ & 0.55 & 0.55 & 0.36 & 0.17 \\
			$3/8$ & 0.28 & 0.25 & 0.11 & 0.06 \\
			$1/4$ & 0.12 & 0.08 & 0.05 & 0.16 \\
			$1/8$ & 0.05 & 0.03 & 0.10 & 0.21 \\\hline
			& $1/8$ & $1/4$ & $3/8$ & $1/2$ \\
		\end{tabular}
	}
	\caption{Values $s_{ij}$ (formula \eqref{eq:chiCELL}) in the transverse sector for $\mu_q/T=65$ and $\mu_3=-0.1$.}
	\label{tab:s6501_tr}
\end{table}

\subsection{Longitudinal correlator}

\begin{figure}[htb]
	\centering
	\begin{subfigure}{0.48\textwidth}
		\centering
		\includegraphics[width=\textwidth]{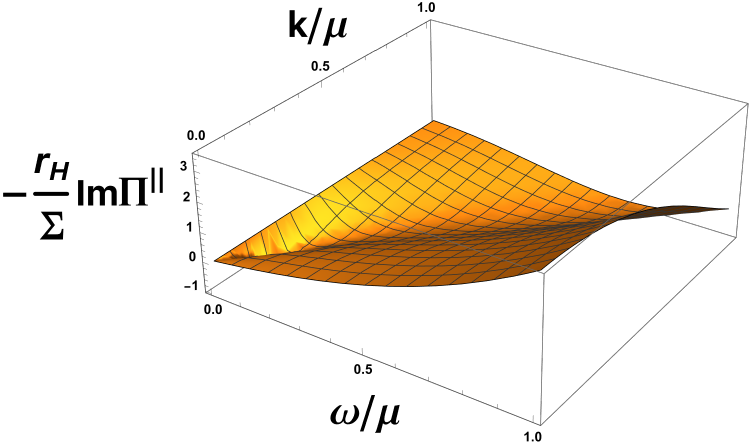}
		\caption{}\label{fig:longitudinalpol650}
	\end{subfigure}
	\hfill
	\begin{subfigure}{0.51\textwidth}
		\centering
		\includegraphics[width=\textwidth]{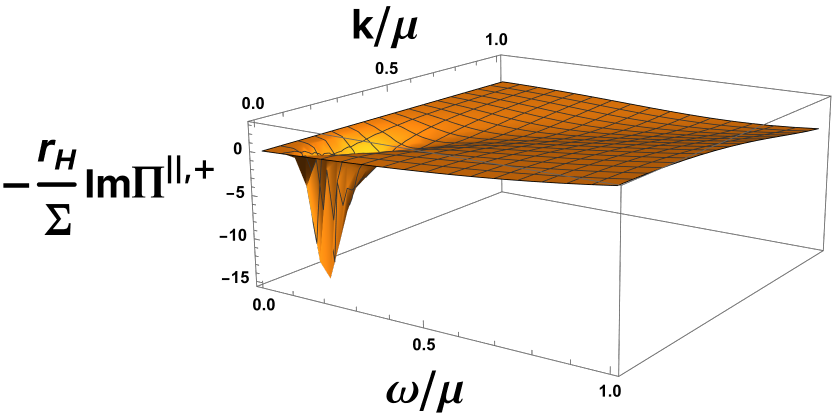}
		\caption{}\label{fig:longitudinalpol6501}
	\end{subfigure}
	\caption{Imaginary part of the longitudinal charged current retarded polarization function normalized by $-\Sigma/r_H$, as a function of $\omega/\mu$ and $k/\mu$. The background space-time is characterized by $\mu_q/T = 65$, $\mu_3=0$ (left panel \ref{fig:longitudinalpol650}, where the  $\pm$ notation has been suppressed since we are working at zero $\mu_3$) and $\mu_3/\mu_q=-0.1$ (left panel \ref{fig:longitudinalpol6501}).} \label{fig:longitudinalpol}
\end{figure}

We now present the numerical results for the longitudinal correlator, and their comparison with the hydrodynamic approximations of interest. We start by plotting the imaginary part of the longitudinal polarization function, for $\mu_q/T =65$ and $\mu_3/\mu_q\in\{0,-0.1\}$. This is the content of figure \ref{fig:longitudinalpol}: the left panel \ref{fig:longitudinalpol650} has $\mu_3/\mu_q=0$, while the right panel \ref{fig:longitudinalpol6501} is for $\mu_3/\mu_q=-0.1$. Figure \ref{fig:longitudinalpol650} shows the numerical result at $\mu_3=0$ already present in the literature \cite{neutrinopaper}, and it is characterized by an imaginary diffusive pole $\omega_D(k) = -iDk^2 + \OO(k^4)$. At finite isospin (figure \ref{fig:longitudinalpol6501}), the polarization function develops a pole at the real frequency $\omega=-\mu_3$, consistently with the non-Abelian hydrodynamic result \eqref{eq:longhydroapprox}.

The next plots show the comparison of these numerical results for the exact spectral function with the approximations of interest, \eqref{eq:longhydroapprox} and \eqref{eq:longexthydroapprox}. As for the transverse case, we first review the case of $\mu_3 = 0$, and then we move to discussing the effects of introducing a non-zero isospin chemical potential.

\subsubsection{Review of results at zero isospin chemical potential}\label{sec:long_mu30}

\begin{figure}[htb]
	\begin{subfigure}{\textwidth}
		\centering
		\begin{subfigure}{0.4\textwidth}
			\centering
			\includegraphics[width=\textwidth]{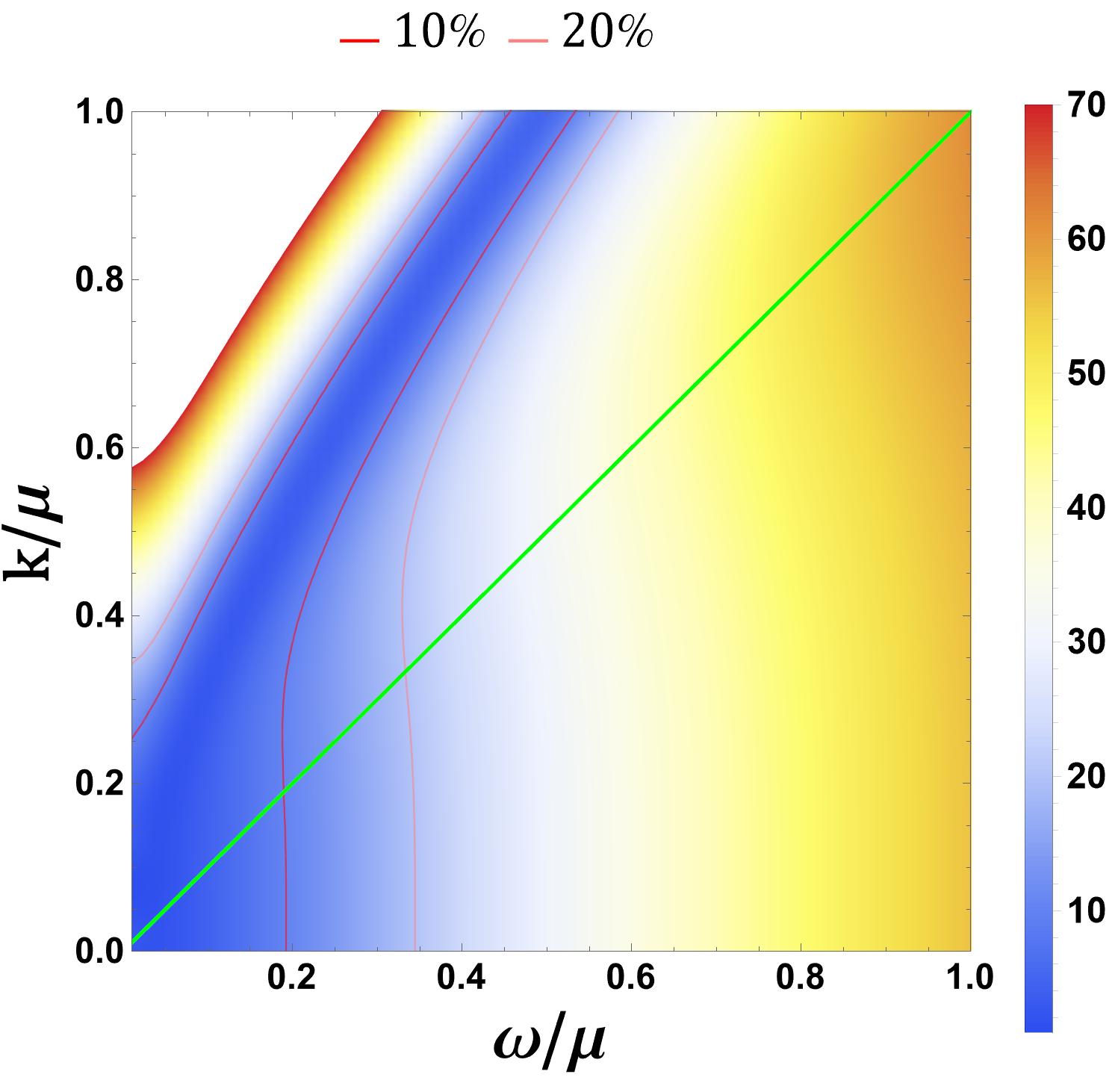}
		\end{subfigure}
		\hspace{1cm}
		\begin{subfigure}{0.4\textwidth}
			\centering
			\includegraphics[width=\textwidth]{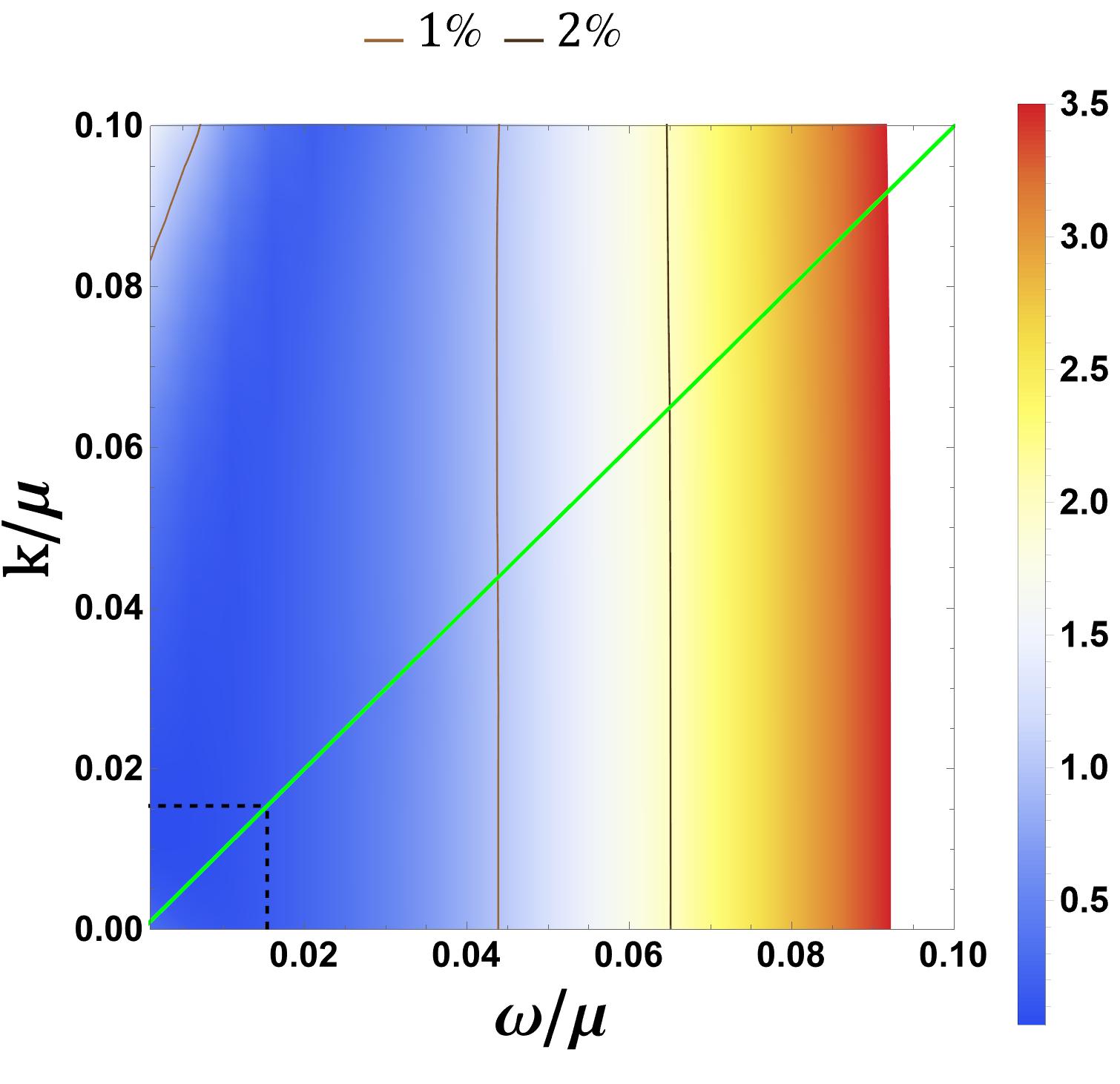}
		\end{subfigure}
	\end{subfigure}

	\begin{subfigure}{\textwidth}
		\centering
		\begin{subfigure}{0.4\textwidth}
			\centering
			\includegraphics[width=\textwidth]{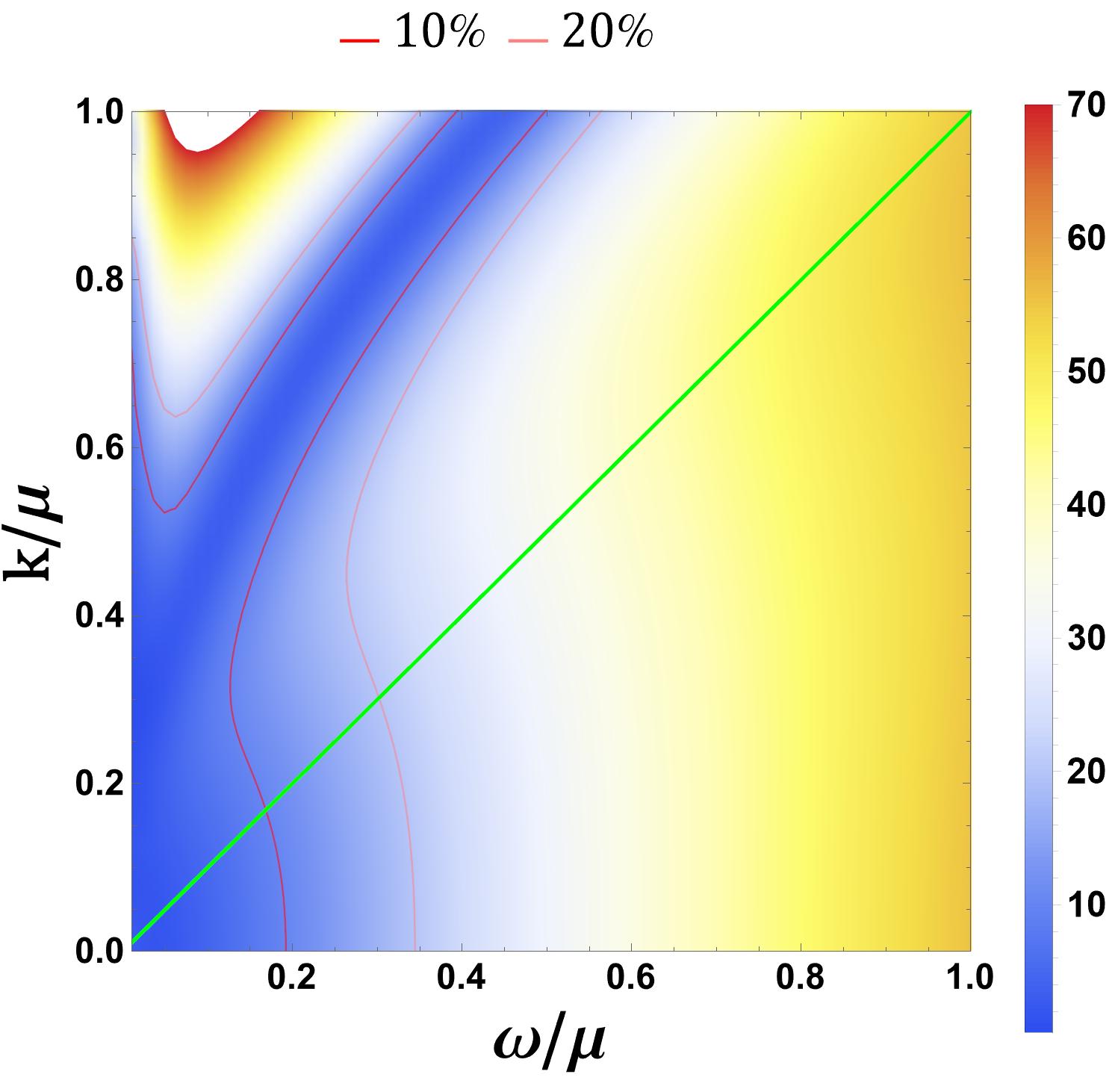}
		\end{subfigure}
		\hspace{1cm}
		\begin{subfigure}{0.4\textwidth}
			\centering
			\includegraphics[width=\textwidth]{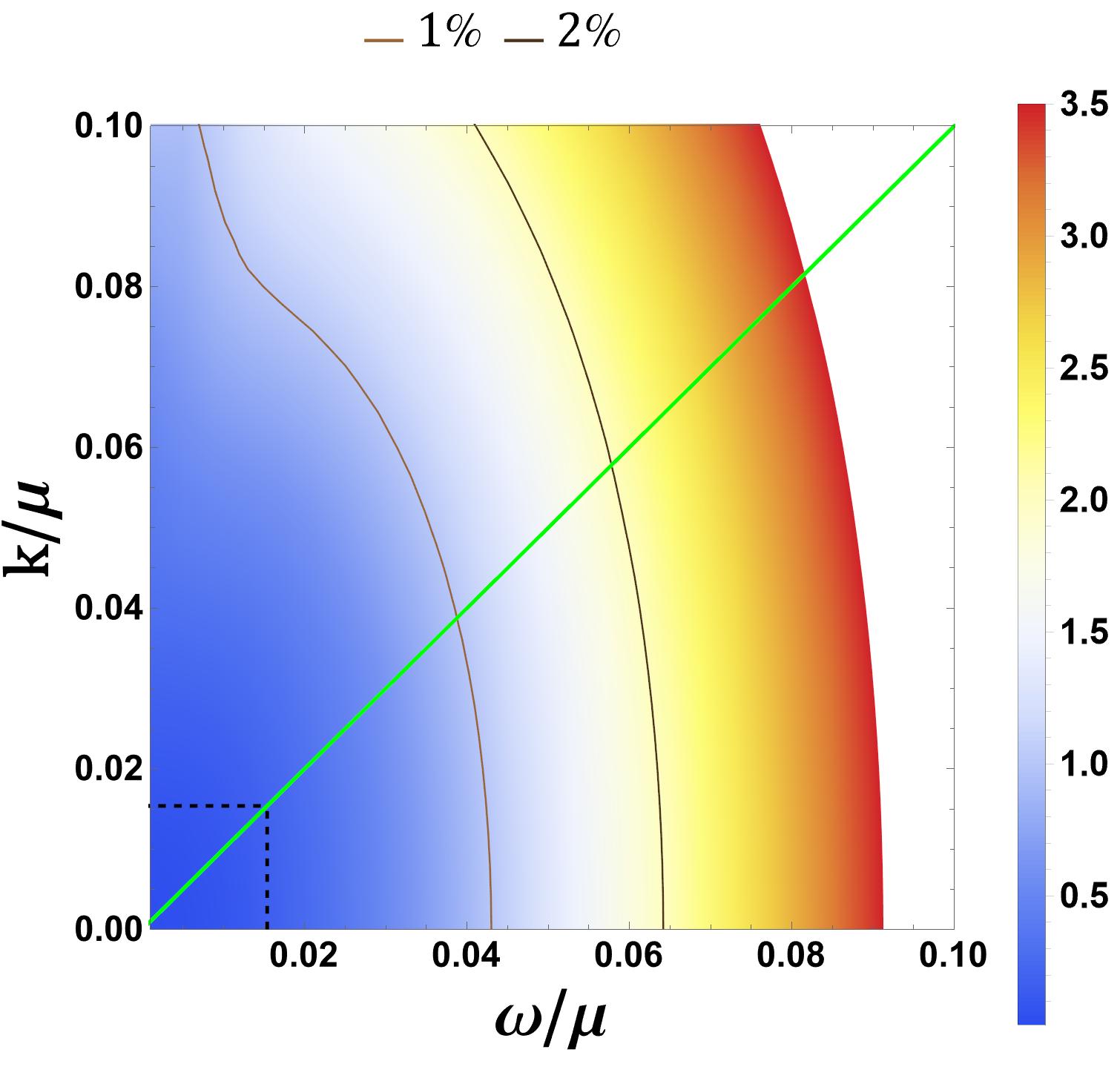}
		\end{subfigure}
	\end{subfigure}

	\caption{The percentage relative difference \eqref{eq:percrelfdiff} of the imaginary part of the longitudinal charged current polarization function with respect to the hydrodynamic approximation \eqref{eq:longhydroapprox} (top row) and the extended hydrodynamic approximation \eqref{eq:longexthydroapprox} (bottom row) for $\mu_q/T= 65$ and $\mu_3=0$. The green line describes the locus $\omega = k$. The right plots shows a subregion of the left
		plots. The dashed square in these right plots represents the standard hydrodynamic region $\omega/\mu,k/\mu \in [0,T/\mu]$.}\label{fig:RelDiffNum_long_mu65mu30}
\end{figure}

\begin{table}[htb]
	\centering
	{
		\small
		\begin{tabular}{c|cccc}
			\multicolumn{5}{c}{(a) hydrodynamic}\\
			$1/2$ & 0.09 & 0.09 & 0.12 & 0.16 \\
			$3/8$ & 0.04 & 0.07 & 0.11 & 0.15 \\
			$1/4$ & 0.03 & 0.06 & 0.10 & 0.14 \\
			$1/8$ & 0.02 & 0.06 & 0.10 & 0.14 \\\hline
			& $1/8$ & $1/4$ & $3/8$ & $1/2$ \\
		\end{tabular}\hspace{1cm}
		\begin{tabular}{c|cccc}
			\multicolumn{5}{c}{(b) extended-hydrodynamic}\\
			$1/2$ & 0.04 & 0.08 & 0.12 & 0.16 \\
			$3/8$ & 0.04 & 0.08 & 0.12 & 0.15 \\
			$1/4$ & 0.03 & 0.07 & 0.11 & 0.15 \\
			$1/8$ & 0.02 & 0.06 & 0.10 & 0.14 \\\hline
			& $1/8$ & $1/4$ & $3/8$ & $1/2$ \\
		\end{tabular}
	}
	\caption{Values $\chi_{ij}$ (formula \eqref{eq:chiRL}) in the longitudinal sector for $\mu_q/T=65$ and $\mu_3=0$.}
	\label{tab:chi650_long}
\end{table}

\begin{table}[htb]
	\centering
	{
		\small
		\begin{tabular}{c|cccc}
			\multicolumn{5}{c}{(a) hydrodynamic}\\
			$1/2$ & 0.23 & 0.07 & 0.18 & 0.27 \\
			$3/8$ & 0.08 & 0.09 & 0.18 & 0.27 \\
			$1/4$ & 0.03 & 0.10 & 0.18 & 0.26 \\
			$1/8$ & 0.02 & 0.10 & 0.18 & 0.26 \\\hline
			& $1/8$ & $1/4$ & $3/8$ & $1/2$ \\
		\end{tabular}\hspace{1cm}
		\begin{tabular}{c|cccc}
			\multicolumn{5}{c}{(b) extended-hydrodynamic}\\
			$1/2$ & 0.04 & 0.12 & 0.18 & 0.27 \\
			$3/8$ & 0.04 & 0.10 & 0.2 & 0.27 \\
			$1/4$ & 0.03 & 0.10 & 0.18 & 0.26 \\
			$1/8$ & 0.02 & 0.10 & 0.18 & 0.26 \\\hline
			& $1/8$ & $1/4$ & $3/8$ & $1/2$ \\
		\end{tabular}
	}
	\caption{Values $s_{ij}$ (formula \eqref{eq:chiCELL}) in the longitudinal sector for $\mu_q/T=65$ and $\mu_3=0$.}
	\label{tab:s650_long}
\end{table}

We now discuss the relative differences between the exact numerical results and the analytic approximations for the longitudinal polarization function at vanishing isospin chemical potential \eqref{eq:longhydroapprox} and \eqref{eq:longexthydroapprox}. The corresponding density plots are shown in figure \ref{fig:RelDiffNum_long_mu65mu30}, where the first row refers to the hydrodynamic approximation, while the second row corresponds to the extended hydrodynamic approximation.

The longitudinal sector at $\mu_3=0$ exhibits a behaviour qualitatively similar to the transverse one. The (non-extremal) hydrodynamic approximation already provides a good description of the exact correlator in the low-frequency and low-momentum region. However, comparing the two rows of figure \ref{fig:RelDiffNum_long_mu65mu30}, one observes that the extended hydrodynamic approximation substantially enlarges the region of small relative difference, especially at small $\omega/\mu$. This shows that incorporating the leading infrared contribution significantly improves the agreement with the exact correlator throughout a much larger portion of the near-extremal regime, although deviations eventually grow at sufficiently large frequencies and momenta.

Also in this case, we support these statements presenting the results from the integrated relative difference \eqref{eq:chiRL}, \eqref{eq:chiCELL} in tables \ref{tab:chi650_long}, and \ref{tab:s650_long}. We observe that the extended hydrodynamic approximations, apart from very few entries (like the second entry of the third column), does overall better than the hydrodynamic approximation, especially at small frequencies and the improvement grows with $k/\mu$. We also observe that the entries where the hydrodynamic approximation does better are aligned along the diagonal entries of the tables.

\subsubsection{Results at isospin chemical potential $\mu_3/\mu_q=-0.1$}\label{sec:long_mu301}

\begin{figure}[htb]
	\centering
	
	\begin{subfigure}{\textwidth}
		\centering
		\begin{subfigure}{0.4\textwidth}
			\includegraphics[width=\textwidth]{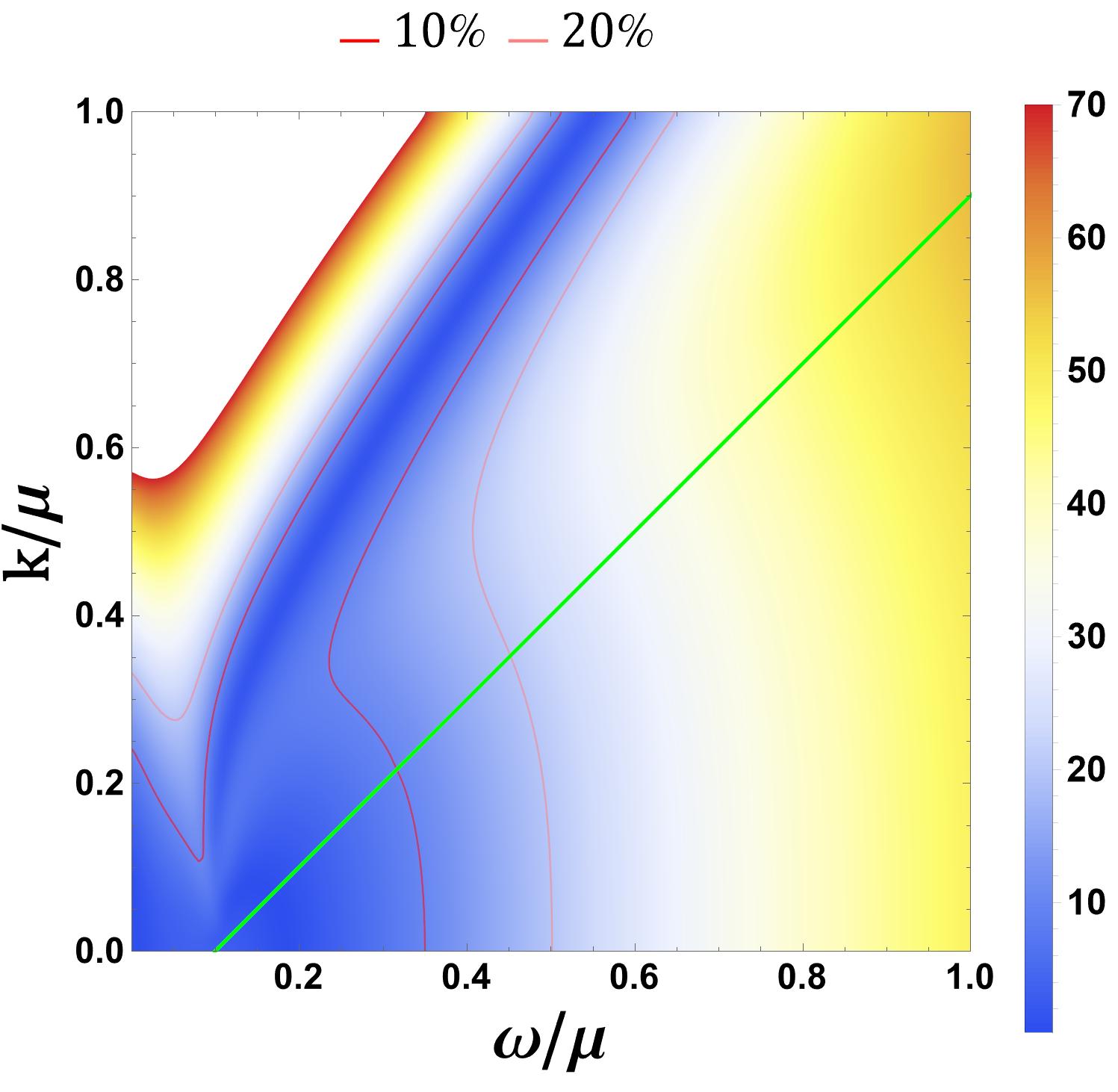}
		\end{subfigure}
		\hspace{1cm}
		\begin{subfigure}{0.4\textwidth}
			\includegraphics[width=\textwidth]{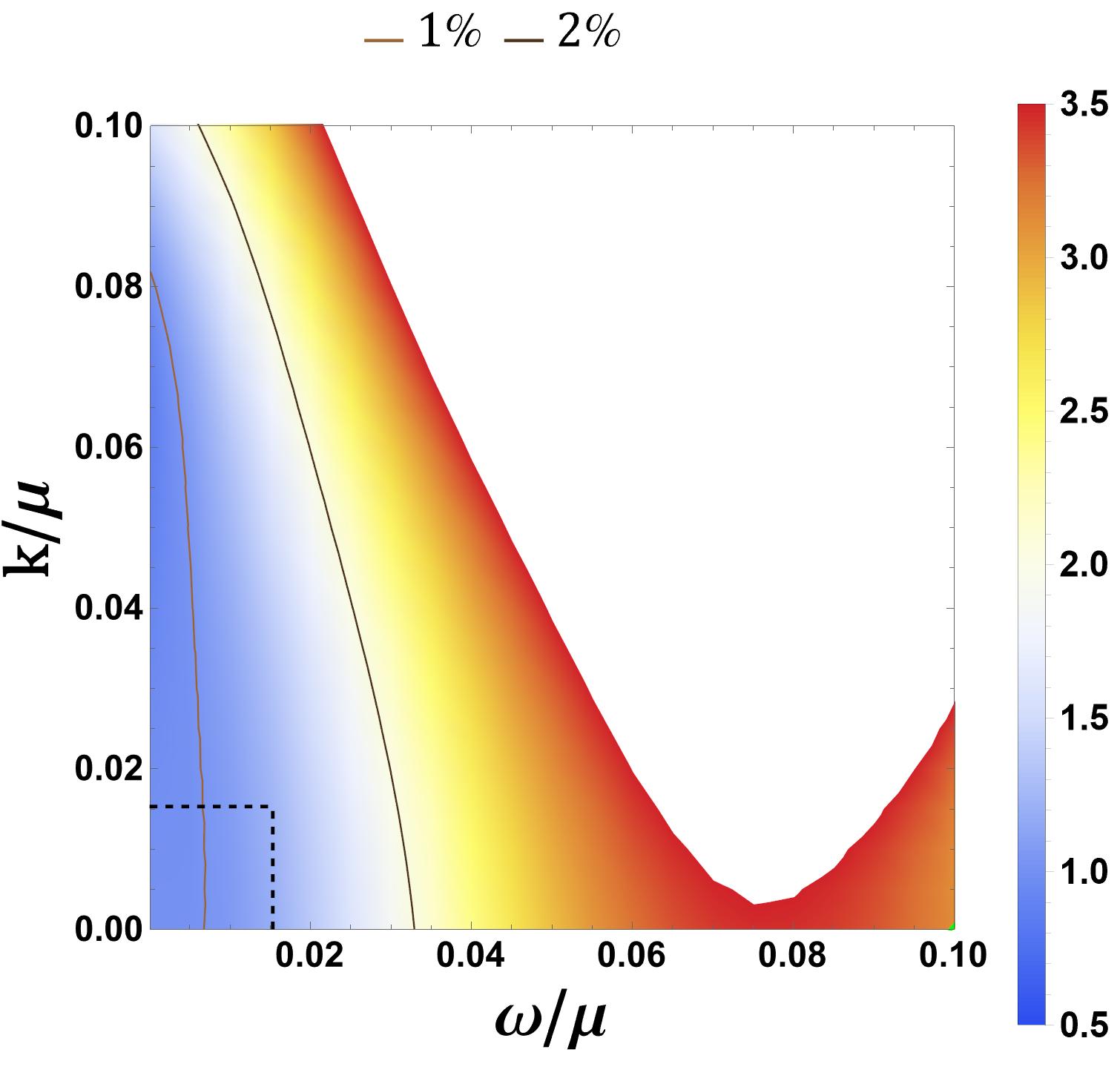}
		\end{subfigure}
	\end{subfigure}
	
	\begin{subfigure}{\textwidth}
		\centering
		\begin{subfigure}{0.4\textwidth}
			\includegraphics[width=\textwidth]{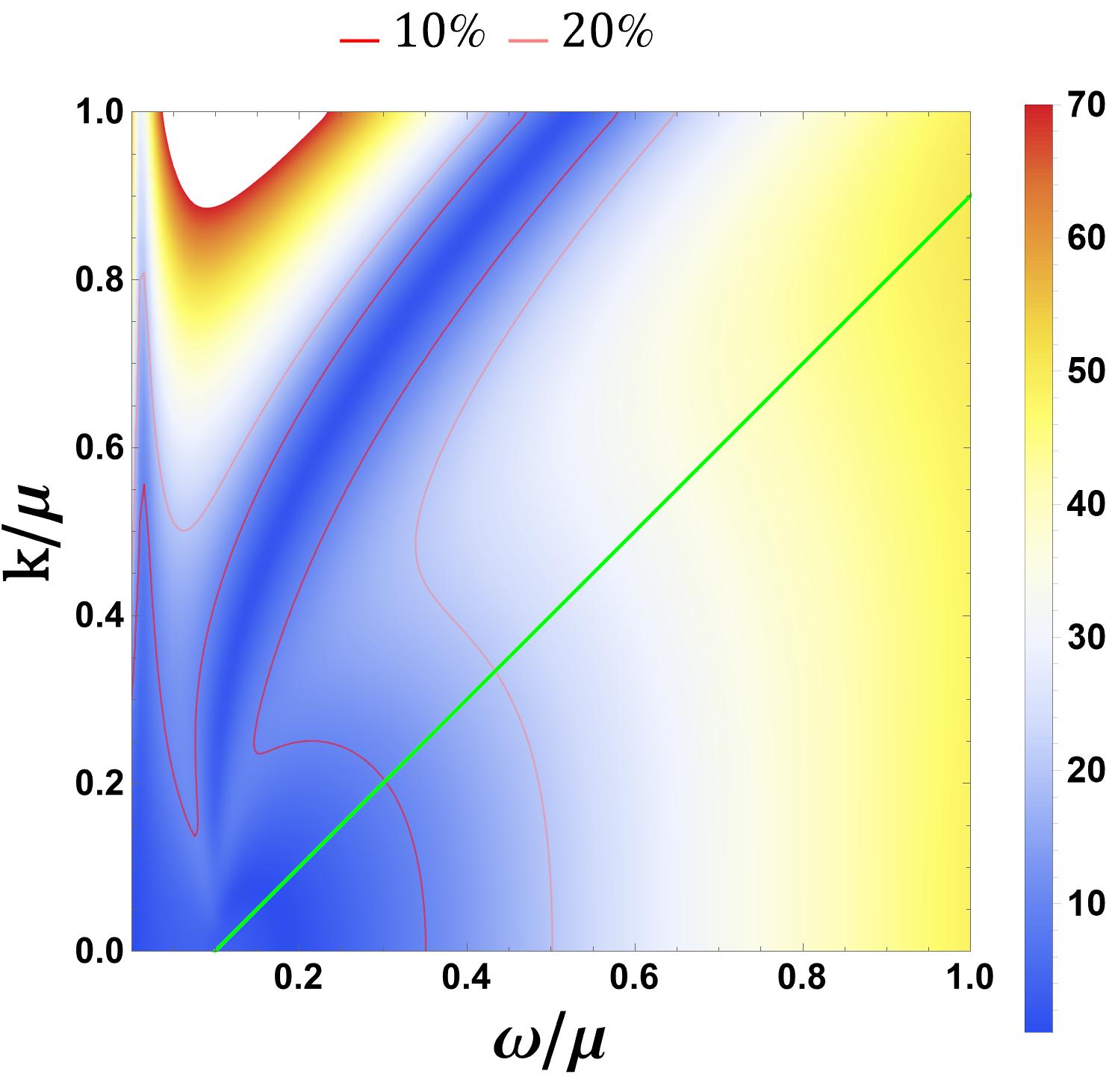}
		\end{subfigure}
		\hspace{1cm}
		\begin{subfigure}{0.4\textwidth}
			\includegraphics[width=\textwidth]{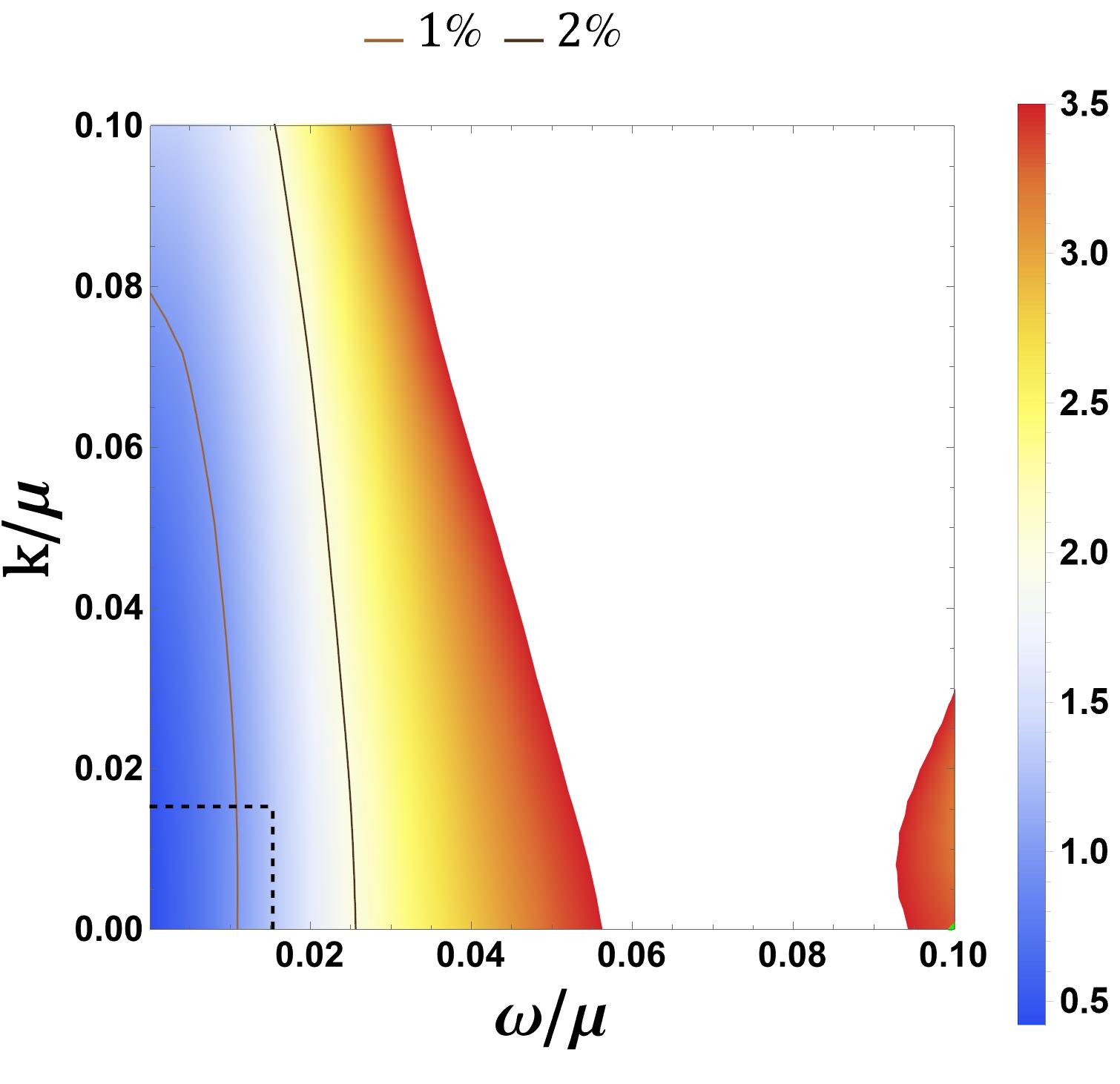}
		\end{subfigure}
	\end{subfigure}
	
	\caption{The percentage difference \eqref{eq:percrelfdiff} of the imaginary part of the longitudinal charged current polarization function with respect to the hydrodynamic approximation \eqref{eq:longhydroapprox} (top row) and the extended hydrodynamic approximation \eqref{eq:longexthydroapprox} (bottom row) for $\mu_q/T= 65$ and $\mu_3/\mu_q=-0.1$. The green line describes the locus $\omega = k+\mu_3$. The right plots shows a subregion of the left
		plots. The dashed square in these right plots represents the standard hydrodynamic region $\omega/\mu,k/\mu \in [0,T/\mu]$.} \label{fig:RelDiffNum_long_mu65mu301}
	
\end{figure}

\begin{table}[htb]
	\centering
	{
		\small
		\begin{tabular}{c|cccc}
			\multicolumn{5}{c}{(a) hydrodynamic}\\
			$1/2$ & 0.17 & 0.11 & 0.11 & 0.13 \\
			$3/8$ & 0.11 & 0.08 & 0.08 & 0.11 \\
			$1/4$ & 0.07 & 0.05 & 0.06 & 0.09 \\
			$1/8$ & 0.04 & 0.03 & 0.05 & 0.07 \\\hline
			& $1/8$ & $1/4$ & $3/8$ & $1/2$ \\
		\end{tabular}\hspace{1cm}
		\begin{tabular}{c|cccc}
			\multicolumn{5}{c}{(b) extended-hydrodynamic}\\
			$1/2$ & 0.08 & 0.07 & 0.09 & 0.12 \\
			$3/8$ & 0.06 & 0.06 & 0.08 & 0.10 \\
			$1/4$ & 0.05 & 0.05 & 0.06 & 0.09 \\
			$1/8$ & 0.04 & 0.03 & 0.05 & 0.07 \\\hline
			& $1/8$ & $1/4$ & $3/8$ & $1/2$ \\
		\end{tabular}
	}
	\caption{Values $\chi_{ij}$ (formula \eqref{eq:chiRL}) in the longitudinal sector for $\mu_q/T=65$ and $\mu_3=-0.1$.}
	\label{tab:chi6501_long}
\end{table}

\begin{table}[htb]
	\centering
	{
		\small
		\begin{tabular}{c|cccc}
			\multicolumn{5}{c}{(a) hydrodynamic}\\
			$1/2$ & 0.35 & 0.07 & 0.14 & 0.21 \\
			$3/8$ & 0.18 & 0.07 & 0.12 & 0.19 \\
			$1/4$ & 0.09 & 0.05 & 0.09 & 0.17 \\
			$1/8$ & 0.04 & 0.02 & 0.08 & 0.16 \\\hline
			& $1/8$ & $1/4$ & $3/8$ & $1/2$ \\
		\end{tabular}\hspace{1cm}
		\begin{tabular}{c|cccc}
			\multicolumn{5}{c}{(b) extended-hydrodynamic}\\
			$1/2$ & 0.12 & 0.07 & 0.14 & 0.21 \\
			$3/8$ & 0.08 & 0.07 & 0.11 & 0.19 \\
			$1/4$ & 0.07 & 0.05 & 0.09 & 0.17 \\
			$1/8$ & 0.04 & 0.02 & 0.08 & 0.16 \\\hline
			& $1/8$ & $1/4$ & $3/8$ & $1/2$ \\
		\end{tabular}
	}
	\caption{Values $s_{ij}$ (formula \eqref{eq:chiCELL}) in the longitudinal sector for $\mu_q/T=65$ and $\mu_3=-0.1$.}
	\label{tab:s6501_long}
\end{table}

We now discuss the case of a non-vanishing isospin chemical potential $\mu_3/\mu_q=-0.1$, shown in figure \ref{fig:RelDiffNum_long_mu65mu301}. As in the previous sections, the first row corresponds to the hydrodynamic approximation, while the second row refers to the extended hydrodynamic approximation.

For both approximations, the region of best agreement is again consistently found near the origin. The extended hydrodynamic approximation (second row of \ref{fig:RelDiffNum_long_mu65mu301}) leads to a clear improvement also in this case, as the region of small relative difference expands near $\omega=0$, and the contour lines are displaced further away from the origin. These observations are supported with the results for the integrated relative differences \eqref{eq:chiRL} and \eqref{eq:chiCELL} presented in tables \ref{tab:chi6501_long} and \ref{tab:s6501_long}. These tables show the same behaviour seen at $\mu_3=0$ of tables \ref{tab:chi650_long} and \ref{tab:s650_long} where, apart from few entries, the extended hydrodynamic approximation was overall the best one. As in the other cases studied in this section, the extended hydrodynamic approximation mostly improves the results at small $\omega/\mu$ (see the first column of all the tables) for both $\chi_{ij}$ and $s_{ij}$.

The presence of a non-zero $\mu_3$ does not modify the large error region at $\omega\geq 0.6$ in the left column plots. However, the $10\%$ and $20\%$ contours are now centered around $\omega\simeq -\mu_3$. As for the transverse case, the presence of $\mu_3$ tends to enhance the sensitivity of the longitudinal correlator to non-hydrodynamic effects. This is especially visible in the right column plots of figure \ref{fig:RelDiffNum_long_mu65mu301} which display a bigger relative difference than the corresponding plots at $\mu_3=0$ (see the right column plots of figure \ref{fig:RelDiffNum_long_mu65mu30}). This can also be seen by confronting the values of the first two column of tables \ref{tab:chi6501_long} and \ref{tab:s6501_long} which are larger than the ones in tables \ref{tab:chi650_long} and \ref{tab:s650_long} associated with $\mu_3=0$. As a consequence, the domain of validity of both approximations, while enlarged in the extended case, remains more constrained than in the zero chemical potential case.

\newpage

\section*{Acknowledgements}\label{ACKNOWL}
\addcontentsline{toc}{section}{Acknowledgements}

We thank  D. Arean, R. Davison, G. Fournodaylos, B. Gouteraux, C. Hoyos, N. Jokela, L. Li, S. Mathur, O. Papadoulaki, J. M. Penin-Ascariz, A. Porfyriadis, C. Rosen, H. Sun and M. Usatyuk for useful discussions.

This work was partially supported by  the H.F.R.I. call ``Basic research Financing" (Horizontal support of all Sciences)
under the National Recovery and Resilience Plan ``Greece 2.0" funded by the European Union -NextGenerationEU
(H.F.R.I. Project Number: 15384), by the In2p3 grant ``Extreme Dynamics", the ANR grant ``XtremeHolo"
(ANR project n.284452), by the H.F.R.I. Project Number: 23770  of the H.F.R.I call
``3rd Call for H.F.R.I.'s Research Projects to Support Faculty Members \& Researchers", the ERC starting grant 101078061 SINGinGR, under the European Union's Horizon Europe program for research and innovation"
and the UoC grant number 12030.

A.O. wants to thank the University of Turku and the University of Helsinki for the kind hospitality during the preparation of this work.

EP is supported by the European Union’s Horizon 2024 research and innovation program under the Marie Sklodowska-Curie grant agreement No 101210184.

\newpage

\appendix
\renewcommand{\theequation}{\thesection.\arabic{equation}}
\addcontentsline{toc}{section}{Appendix\label{app}}
\section*{Appendix}

\section{Notation for the charged fields}\label{app:notationpm}
In this appendix, we define the notation used for the charged currents or bulk fields. The general expression for the current reads
\begin{align}
	\mathcal{J}_\mu = \hat t\,\hat J_\mu + \sum\limits_a t_aJ^a_\mu = \dfrac{\mathbb{I}_2}{2}\hat J_\mu + \dfrac{1}{2}\sum\limits_a \sigma_aJ^a_\mu,
\label{eqA:Jdef}\end{align}
where $\hat t = \mathbb{I}_2/2$ and $t_a = \sigma_a/2$ with $a\in\{1,2,3\}$. The Pauli matrices have a non-trivial commutator defined by the structure constants $f_{ab}\,^c$:
\be
[\sigma_a,\sigma_b]=2i\,f_{ab}\,^c\sigma_c.
\ee
The gauge group metric $\mathbf{g}_{ab}$ used to raise and lower the group indices is defined as
\be
\mathbf{g}_{ab} = 2\,\text{Tr}(t_at_b) = \dfrac{1}{2}\text{Tr}(\sigma_a\sigma_b).
\ee
In Cartesian coordinates, $\{\sigma_1,\sigma_2,\sigma_3\}$, we end up with the usual result $\mathbf{g}_{ab}=\delta_{ab}$ and $f_{ab}\,^c = \epsilon_{abc}$.

Now, we introduce the charged currents
\be
J_\mu^\pm = J_\mu^1\mp i\,J_\mu^2.
\ee
In order to write the current as in \eqref{eqA:Jdef}, we define
\be
\sigma_\pm = \dfrac{\sigma_1\pm i\,\sigma_2}{2},
\ee
with the associated generators being $t_\pm =\sigma_\pm/2$.

We compute the structure constants in this new basis where the indices run over $a\in\{+,-,3\}$:
\begin{align}
	&[\sigma_+,\sigma_-]=\sigma_3,& &[\sigma_3,\sigma_\pm]=\pm 2\sigma_\pm,&
\end{align}
leading to
\begin{align}\label{eq:structpm}
	&f_{\pm\mp}\,^3=\mp\dfrac{i}{2},& &f_{\pm 3}\,^\pm=\pm i.&
\end{align}
Then, the gauge group metric $\mathbf{g}_{ab}$ that we use to raise and lower group indices reads now
\begin{align}
	\mathbf{g}_{ab}=2\,\text{Tr}(t_at_b)= \begin{pmatrix}
		0 \quad& 1/2 \quad& 0\\
		1/2 \quad& 0 \quad& 0\\
		0 \quad& 0 \quad& 1
	\end{pmatrix}\, .
\end{align}
As a consequence, the structure constants with all lowered indices are
\begin{align}
	&f_{\pm\mp 3}=\mp\dfrac{i}{2},& &f_{\pm 3\mp}=\pm\dfrac{i}{2}.&
\end{align}
This notation also applies to the bulk gauge fields.

\section{Two-point function decomposition}
\label{Sec:TwoPointDec}

In this appendix, we provide details on the two-point function decomposition for currents.

If $J_{\mu}$ is a U(1) conserved current, the conservation equation implies $k_{\m}J^{\m}$=0. The corresponding quantum statement is the Ward Identity $\left< J_\l J_\s\right> k^\s = 0$. In the non-Abelian case the Ward identity needs to be modified since the currents are now covariantly conserved $D_{\mu}J^{\mu}=0$. Expanding the covariant derivative yields
\be
\partial_{\mu}J^{\mu}-i\left[A_{\mu},J^{\mu}\right]=0\, ,
\ee
After Fourier transforming this reduces to
\begin{equation}
	k_{\mu}J^{\mu}=\frac{\m_3}{2}\left[\sigma_3,J^{0}\right] \, ,
\end{equation}
since only the zero component of $A_{\mu}$ is sourced. We use the charged components for the currents introduced in appendix \ref{app:notationpm} so that we can rewrite the Ward identity as
\begin{equation}
	\label{kshift}
	k_{\mu}^{\pm}J^{\mu\pm}=0\, ,
\end{equation}
where
\begin{equation}
	k_{\mu}^{\pm}=k_{\mu}\mp\m_3\delta^{0}_{\mu}\sp k_\mu=(-\omega,\vec k)\, .
\end{equation}

The two-point function can be written as
\begin{equation}
	\left<J^\pm_\l J^\mp_\s\right>(\omega,\vec{k})=\Pi^\pm_{\l\s}(\omega,\vec{k}) \, ,
\end{equation}
where $\Pi^\pm_{\l\s}(\omega,\vec{k})$ is a symmetric tensor satisfying $k^{\pm,\l}\Pi^\pm_{\l\s}=0$. In general, it can contains contact terms, as it is shown in Appendix \ref{app:WI}. We can expand the equation $k^{\pm,\l}\Pi^\pm_{\l\s}(\omega,\vec{k})=0$ as
\be
(\omega\pm\mu_3)\Pi^\pm_{00}+k^{i}\Pi_{i0}=0 \sp (\omega\pm\mu_3)\Pi_{0i}+k^{j}\Pi_{ij}=0\, .
\ee
Combining the two equations we find
\be
\label{Pirel}
\Pi_{00}^\pm=\frac{k^ik^j}{(\omega\pm\mu_3)^2}\,\Pi_{ij}^\pm \sp \Pi_{0i}^\pm=-\frac{k^j}{\omega\pm\mu_3}\,\Pi_{ij}^\pm\, .
\ee
$\Pi_{ij}^\pm$ is a symmetric tensor which is a function of $(\omega,\vec{k})$. Therefore the most general form it can take is
\be
\label{Piij}
\Pi_{ij}^\pm=a(\omega,\vec{k})\delta_{ij}+b(\omega,\vec{k}^2)\frac{k_{i}k_{j}}{\vec{k}^2}.
\ee
Then the other components of $\Pi_{\m\n}^\pm$ are from \eqref{Pirel}
\begin{equation}
	\label{Pio}
	\Pi_{00}^\pm=\left(a(\omega,\vec{k})+b(\omega,\vec{k})\right)\frac{\vec{k}^2}{(\omega\pm\mu_3)^2}\, , \quad \Pi_{0i}=-\left(a(\omega,\vec{k})+b(\omega,\vec{k})\right)\frac{k_i}{\omega\pm\mu_3}\, .
\end{equation}
We can group \eqref{Piij} and \eqref{Pio} into one equation for $\Pi_{\m\n}^\pm$
\be
\label{PiFull}
\Pi_{\m\n}=a(\omega,\vec{k})P_{\pm,\m\n}^{\perp}+\frac{\omega^2-\vec{k}^2}{\omega^2}(a(\omega,\vec{k})+b(\omega,\vec{k}))P_{\m\n}^{\pm,\parallel}\, ,
\ee
where
\be
\label{APo}  P^{\perp,\pm}_{ij}(\omega,\vec{k})=P^\perp_{ij}(\omega,\vec{k}) = \d_{ij} - \frac{k_ik_j}{\vec{k}^2} \, , \,\,\, P_{00}^{\perp,\pm} = P_{0i}^{\perp,\pm}= P_{i0}^{\perp,\pm} =0
\ee
\be
\label{APp} P^{\parallel,\pm}_{00} = \frac{\vec{k}^2}{\left(\omega\pm\m_3\right)^2-\vec{k}^2} \,\, , \,\, P^{\parallel,\pm}_{0i}= P^{\parallel,\pm}_{i0} =- \frac{\left(\omega\pm\m_3\right) k_i}{\left(\omega\pm\m_3\right)^2-\vec{k}^2} \,\, , \,\,
\ee
$$
P_{ij}^{\parallel,\pm} = \frac{k_ik_j}{\vec{k}^2}\frac{\left(\omega\pm\m_3\right)^2}{\left(\omega\pm\\
	\m_3\right)^2-\vec{k}^2} \, .
$$
Here $P_{ij}^{\perp,\pm}$ and $P_{ij}^{\parallel,\pm}$ are the projectors in the direction perpendicular and transverse to the three-momentum $\vec{k}$ respectively. Note that the sum of the two projectors is
\be
\label{AP4} P_{\mu\nu}^{\perp,\pm} + P_{\mu\nu}^{\parallel,\pm} = \eta_{\mu\nu} - \frac{k^{\pm}_\mu k^{\pm}_\nu}{k^{\pm,\rho}k_\rho^\pm} \equiv P_{\mu\nu}^{\pm}  \, .
\ee
Moreover, they satisfy
\begin{equation}
	P_{\mu\nu}^{\perp,\pm}P^{\perp,\pm,\nu}{}_\rho = P_{\mu\rho}^{\perp,\pm} \sp 	P_{\mu\nu}^{\parallel,\pm}P^{\parallel,\pm,\nu}{}_\rho = P_{\mu\rho}^{\parallel,\pm} \sp 	P_{\mu\nu}^{\perp,\pm}P^{\parallel,\pm,\nu}{}_\rho =0\,.
\end{equation}

The field $J^{\mu,\pm}$ can be decomposed into transverse and longitudinal components
\begin{equation}
	J^{\mu,\pm}=J^{\mu,\pm,\perp}+J^{\mu,\pm,\parallel} \, ,
\end{equation}
where longitudinal and transverse projectors act on $J^{\mu}$ as
\be
\label{projAct}
P_{\m\n}^{\pm,\perp}J^{\mu,\pm}=J^{\nu,\pm,\perp}\, , \quad P_{\m\n}^{\pm,\parallel}J^{\mu,\pm}=J^{\nu,\pm,\parallel}\, .
\ee
This formalism can then be applied to calculating the correlation functions for the charged currents. The transverse correlator can be written as:
\be
\label{parJ}
\left<J_\l^{\pm,\perp} J_\s^{\mp,\perp}\right>=a(\omega,\vec{k})P_{\l\s}^{\pm,\perp}\equiv P_{\l\s}^{\pm,\perp}i\Pi^{\pm,\perp}(\omega,\vec{k})\, ,
\ee
where we have used \eqref{projAct} and \eqref{APo} along with \eqref{PiFull} and we defined the transverse polarization function as $a(\omega,\vec{k})\equiv i\Pi^{\pm\perp}(\omega,\vec{k})$. Similarly we find
\be
\label{crosJ}
\left<J_\l^{\pm\perp} J_\s^{\mp,\parallel}\right>=0.
\ee
as well as
\begin{align}
	\label{perpJ}
	& \big<J_\l^{\pm,\parallel} J_\s^{\mp,\parallel}\big>=P_{\l\s}^{\pm,\parallel}\frac{(\omega\pm\mu_3)^2-\vec{k}^2}{(\omega\pm\mu_3)^2}(a(\omega,\vec{k})+b(\omega,\vec{k}))\equiv P_{\l\s}^{\pm,\parallel}i \Pi^{\pm,\parallel}(\omega,\vec{k})\, ,
\end{align}
where we defined the longitudinal polarization function as $\frac{(\omega\pm\mu_3)^2-\vec{k}^2}{(\omega\pm\mu_3)^2}(a(\omega,\vec{k})+b(\omega,\vec{k}))\equiv i\Pi^{\pm,\parallel}(\omega,\vec{k})$. The full correlation function is then given by
\be
\big<J_\l^\pm J_\s^\mp\big>=\big<J_\l^{\pm,\parallel} J_\s^{\mp\parallel}\big>+\left<J_\l^{\pm,\perp} J_\s^{\mp,\parallel}\right>\,,
\ee
so from \eqref{parJ}, \eqref{crosJ} and \eqref{perpJ} we have shown that the 2-point function can be decomposed into a longitudinal and transverse part to the 3-momentum $\vec{k}$, according to
\be
\label{AltcR} \left<J^{\pm}_\l J^{\mp}_\s\right>(\omega,\vec{k}) = P_{\l\s}^{\perp,\pm}(\omega,\vec{k}) i\Pi^{\pm,\perp}(\omega,\vec{k}) + P_{\l\s}^{\pm,\parallel}(\omega,\vec{k}) i\Pi^{\pm,\parallel}(\omega,\vec{k})\, .
\ee

\section{Ward identities for the charged currents}\label{app:WI}

In this appendix, we check that the charged current correlator, which is computed from the on-shell action, is compatible with the Ward identities for the charged currents $J_\m^{\pm}$. For the one-point function, the Ward identities read
\begin{equation}
	\label{W1} D_\m \left<J^{\m,\pm}(x)\right> = 0 \, .
\end{equation}
The covariant conservation also holds within higher-point correlation functions
\begin{equation}
	\label{W2} \left<J^{\m,\pm}(x)\mathcal{O}_1(x_1)\dots\mathcal{O}_n(x_n)\right> \, ,
\end{equation}
as long as the operator insertion points $\{x_1,\dots,x_n\}$ do not coincide with $x$.

The current $J^{\mu,\pm}$ is sourced by the $\mp$ chiral gauge field, which means that the associated symmetry is a $\mp$ chiral rotation, generated by $t_\mp = \sigma_\mp/2 = (\sigma_1\mp i\sigma_2)/2$ defined in the appendix \ref{app:notationpm}. If the operators $\mathcal{O}_i$ are assumed to transform under an infinitesimal space-time dependent $\mp$ chiral rotation with parameter $\e_\mp(x)$ as
\begin{equation}
	\label{W3} \d\mathcal{O}_i(x_i)= \e_\mp(x) X_i(x_i) \, ,
\end{equation}
then the general Ward identity for the correlator \eqref{W2} takes the form
\begin{align}
	\nn &D_\m\left<J^{\m,\pm}(x)\mathcal{O}_1(x_1)\dots\mathcal{O}_n(x_n)\right> = \\
	\label{W4} &\qquad\sum_i \d(x-x_i) \left<\mathcal{O}_1(x_1)\dots\mathcal{O}_{i-1}(x_{i-1})X_i(x_i)\mathcal{O}_{i+1}(x_{i+1})\dots\mathcal{O}_n(x_n)\right> \, .
\end{align}

We now consider the correlator that we are interested in
\begin{equation}
	\label{W5} \left<J^{\m,\pm}(x)J^{\n,\mp}(x')\right> \, ,
\end{equation}
for which we want to write down the Ward identity. We recall that, according to our definitions,
\begin{align}
	J^\mp_\mu = 2\text{Tr}\left(J_\mu\,\sigma_\pm\right).
\end{align}
An infinitesimal $\mp$ chiral transformation reads
\begin{align}
	\delta J^{\nu}(x') = \epsilon_\mp(x)[t_\mp,J^\nu(x')].
\end{align}

We want to analyse how the $\mp$ components of the current change under a chiral transformation. Therefore
\begin{align}\label{eq:transfchiral}
	\delta J^{\nu,\mp}(x') = 2\text{Tr}\left(\delta J^\nu(x')\,\sigma_\pm\right) = \epsilon_\mp(x)\text{Tr}\left([\sigma_\mp,J^\nu(x')]\,\sigma_\pm\right) = \mp \epsilon_\mp(x)J^{\nu,3}(x').
\end{align}
We use this result to write the Ward identity for the two-point function of charged currents
\begin{align}
	D_\m\left<J^{\m,\pm}(x)J^{\n,\mp}(x')\right> = \mp \,\d(x-x') \left<J^{\n,3}(x')\right> = \mp\,\d(x-x') \d^\n_0 \,n_3 \, ,
\end{align}
where we have used $\left<J^{0,3}(x')\right>=n_3$.\footnote{Note that we can raise the group index $a=3$ without any additional factor coming from the group metric.} In momentum space, the Ward identity becomes
\begin{equation}\label{eq:WIk}
	ik_\m^\pm \left<J^{\m,\pm}J^{\n,\mp}\right>(k) = \mp \d^\n_0 n_3  \, .
\end{equation}
To check whether \eqref{eq:WIk} is obeyed by the correlator computed from \eqref{Sos2}, we note that the projectors $P^{\parallel,\pm}$ and $P^{\perp,\pm}$ are both transverse to $k_\m^\pm$ (by construction), so that the only term contributing to the left-hand side of \eqref{eq:WIk} is the one not proportional to the projectors. This gives
\begin{equation}
	\label{eq:WIcheck} ik_\m^\pm \left<J^{\m,\pm}J^{\n,\mp}\right>(k) = -i\omega_\pm \left<J^{0,\pm}J^{\n,\mp}\right>(k) = \mp \d^\n_0 \left(-\frac{1}{8r}(M\ell)^3w_0^2N_c\partial_r\Phi_3\right)\bigg|_{r=0} \, ,
\end{equation}
where we have used that $k_\mu^\pm = (-\omega_\pm,k_i)$, with $\omega_\pm = \omega\pm\mu_3$, and the explicit expression of the correlator in equation \eqref{PLl}. The term in the parentheses is precisely the isospin density $n_3$, so that the Ward identity \eqref{eq:WIk} is indeed satisfied by our correlator.

\section{Onsager reciprocity relations}\label{app:Ons}

In this appendix, we present some general results about the retarded correlators of charged currents, focusing on the Onsager reciprocity relations.

For a system at thermodynamic equilibrium, obeying an antiunitary symmetry $\Theta$ that implements microscopic time reversal---which may be time reversal $\TT$ itself or a combination of $\TT$ with unitary symmetries such as charge conjugation $\CC$ or parity $\PP$---the retarded correlators obey reciprocity relations of the form
\begin{equation}
	\label{Or1} \left<AB\right>^R = \left<B_\Theta^\dagger A_\Theta^\dagger \right>^R \, .
\end{equation}
Here the expectation value is obtained in general via a density matrix. The only assumptions on $\Theta$ needed below are that it is antiunitary, so that $\Theta^\dagger=\Theta^{-1}$, and that it leaves the equilibrium density matrix invariant. We do not need to assume that $\Theta^2=1$; in many applications $\Theta^2=\pm 1$, or more generally it may square to a unitary symmetry that leaves the relevant equilibrium state and observables invariant. The operators $A$ and $B$ transform to $A_\Theta$ and $B_\Theta$ under $\Theta$, meaning
\begin{equation}
	\label{Or2} A_\Theta(t) \equiv \Theta A(-t) \Theta^\dagger \sp  B_\Theta(t) \equiv \Theta B(-t) \Theta^\dagger\,.
\end{equation}

The invariance of the system under $\Theta$ means that the density matrix $\rho$ of the system is preserved by $\Theta$
\begin{equation}
	\label{Or3} \Theta \rho \Theta^\dagger = \rho \, .
\end{equation}

For completeness, we reproduce here the steps that lead to \eqref{Or1}. An important aspect of the derivation is that $\Theta$, since it implements time reversal, is an \emph{antiunitary operator}: it is anti-linear and preserves inner products, namely
\begin{equation}
	\left<\Theta \phi \middle| \Theta \psi \right>
	=
	\left<\psi \middle| \phi \right> \, .
\end{equation}
Equivalently, $\Theta^\dagger \Theta=\Theta\Theta^\dagger=1$. In a chosen basis, any antiunitary operator can be written as the product of a unitary operator and complex conjugation.\footnote{Given a state $\left|\phi\right> = a\left|1\right> + b\left|2\right>$ with $a,b \in \mathbb{C}$, an anti-linear operator $\Theta$ acts as $\Theta\left|\phi\right> = a^*\,\Theta\left|1\right> + b^*\,\Theta\left|2\right>. $} Unlike linear operators, anti-linear operators complex-conjugate scalar coefficients, and therefore require some care when used in bra-ket notation. In particular, the action of an anti-linear operator on a bra is not equivalent to its action on a ket. For an anti-linear operator $\mathcal{A}$, one has
\begin{equation}
	\label{Or4}
	\left(\left<\phi\right|\mathcal{A}\right)\left|\psi\right>
	= \left[\left<\phi\right|\left(\mathcal{A}\left|\psi\right>\right)\right]^* .
\end{equation}
The adjoint $\mathcal{A}^\dagger$ is defined through its action on kets by
\begin{equation}
	\label{Or5}
	\left(\left<\phi\right|\mathcal{A}\right)^\dagger = \mathcal{A}^\dagger \left|\phi\right> \, .
\end{equation}
With this definition, one finds the relation between matrix elements of $\mathcal{A}$ and $\mathcal{A}^\dagger$:
\begin{equation}
	\label{Or6}
	\left<\psi\right| \left(\mathcal{A}^\dagger \left|\phi\right>\right)
	= \left[\left(\left<\phi\right|\mathcal{A}\right)\left|\psi\right>\right]^*
	= \left<\phi\right|\left(\mathcal{A}\left|\psi\right>\right) \, ,
\end{equation}
where in the last step we used equation \eqref{Or4}. This shows that the adjoint of an anti-linear operator satisfies
\begin{equation}
	\left<\psi\right| \mathcal{A}^\dagger \left|\phi\right>
	= \left<\phi\right| \mathcal{A} \left|\psi\right> \, ,
\end{equation}
which replaces the usual relation for linear operators.

With these results, we can now consider rewriting the retarded correlator \eqref{Or1} by having $\Theta$ appear. The correlator can be expressed formally as a sum over the Hilbert space
\begin{equation}
\label{Or6b} i\left<AB\right>^R(t) = \theta(t) \mathrm{Tr}\left(\rho \left[A(t),B(0)\right]\right) \, ,
\end{equation}
with
\begin{align}
	\nn \mathrm{Tr}\left(\rho \left[A(t),B(0)\right]\right) &= \sum_n \left<\phi_n\right| \rho \left[A(t),B(0)\right] \left|\phi_n\right>\\
	\nn &= \sum_n \left<\phi_n\right| \big(\Theta^\dagger \Theta \rho \left[A(t),B(0)\right] \Theta^\dagger \Theta \left|\phi_n\right>\big)\\
	\nn &= \sum_n \left<\phi_n\right| \big(\Theta^\dagger \rho \left[A_\Theta(-t),B_\Theta(0)\right] \Theta \left|\phi_n\right>\big)\\
	\label{Or7} &= \sum_n \left<\phi_n\right| \big(\Theta^\dagger \rho \left[A_\Theta(-t),B_\Theta(0)\right] \left|\phi_n\right>_\Theta\big) \, ,
\end{align}
where we denoted
\begin{equation}
	\label{Or8} \left|\phi_n\right>_\Theta \equiv \Theta \left|\phi_n\right>   \, ,
\end{equation}
which defines another basis of the Hilbert space. Now using \eqref{Or6} with $\AA = \Theta$ and $\left|\phi\right>$ = $\rho [A_\Theta(-t),B_\Theta(0)] \left|\phi_n\right>_\Theta$, the retarded correlator can be written as
\begin{equation}
	\label{Or9} \sum_n \left<\phi_n\right|_\Theta \rho [B_\Theta^\dagger(0),A_\Theta^\dagger(-t)] \left|\phi_n\right>_\Theta = \sum_n \left<\phi_n\right|_\Theta \rho [B_\Theta^\dagger(t),A_\Theta^\dagger(0)] \left|\phi_n\right>_\Theta \, ,
\end{equation}
where we also used time-translation invariance. Combining equations \eqref{Or9} and~\eqref{Or7} we obtain the reciprocity relation \eqref{Or1}.

To find the appropriate $\Theta$ relevant to our problem---\textit{i.e.} $\Theta$ such that \eqref{Or3} is satisfied---we list in table \ref{tabOr1} how the difference choices act on space-time, momenta and gauge fields (or currents).
\begin{table}[h!]
	\centering
	\begin{tabular}{ |c||p{2.4cm}|p{2.4cm}|p{2.4cm}|p{2.4cm}|  }
		\hline
		& $\TT$ & $\CC \TT$ & $\PP \TT$ & $\CC \PP \TT$ \\
		\hline
		$x^\mu$  & $(-x^0,x^i)$  & $(-x^0,x^i)$ & $(-x^0,-x^i)$ & $(-x^0,-x^i)$ \\
		$k^\mu$ & $(\omega,-k^i)$ & $(\omega,-k^i)$ & $(\omega,k^i)$ & $(\omega,k^i)$ \\
		$\hat V^\mu$ & $(\hat V^0,-\hat V^i)$  & $(-\hat V^0,\hat V^i)$ & $(\hat V^0,\hat V^i)$ & $(-\hat V^0,-\hat V^i_3)$ \\
		$V^\mu_3$ & $(V^0_3,-V^i_3)$  & $(-V^0_3,V^i_3)$ & $(V^0_3,V^i_3)$ & $(-V^0_3,-V^i_3)$ \\
		$V^\mu_\pm$ & $(V^0_\mp,-V^i_\mp)$  & $(-V^0_\mp,V^i_\mp)$ & $(V^0_\mp,V^i_\mp)$ & $(-V^0_\mp,-V^i_\mp)$ \\
		\hline
	\end{tabular}
	\caption{Transformation properties under discrete space-time symmetries involving time reversal. The vectors $\hat V^\mu$ and $V^\mu_a$ stands for both Abelian and non-Abelian gauge fields and currents respectively.}
	\label{tabOr1}
\end{table}
This indicates that we should choose $\Theta = \CC\TT$ in our case, which is most easily seen by considering the action on the Ward identities \eqref{W1}. The corresponding Onsager relations for the charged current correlators now take the form
\begin{equation}
	\label{Or10} \left<J^\mu_\pm J^\nu_\mp\right>^R(\mu_3;\omega,k^i) = \e_{(\m\n)} \left<J^\nu_\mp J^\mu_\pm\right>^R(-\mu_3;\omega,-k^i) \, ,
\end{equation}
with the symbol $\e_{(\m\n)} = -1$ for $(\m\n) = (0i)$ or $(i0)$, and 1 otherwise. We may ignore the momentum dependence and tensor structure, which are already constrained by the background isotropy, as in \eqref{ltcR}, to write \eqref{Or10} in a condensed form
\begin{equation}
	\label{Or11} G^R_\pm(\m_3;\omega) = G^R_\mp(-\m_3;\omega) \, .
\end{equation}

The retarded correlator follows the well-known relations (following from invariance under space-time translations) with advanced correlators
\begin{align}
	&G^R_\pm(\mu_3;\omega)^* = G^A_\mp(\mu_3,\omega)\,,& &G^R_\pm(\mu_3;\omega) = G^A_\pm(\mu_3;-\omega)\,.&
\end{align}
Combining the two relations we obtain
\begin{equation}\label{Or12}
	G^R_\pm(\mu_3;\omega)^* = G^R_\mp(\mu_3;-\omega)\,.
\end{equation}
This relation can be translated into a relation on the imaginary part of the polarization function. Having in mind the structure of the retarded correlator \eqref{ltcR} and the structure of the projectors in the charged sector \eqref{Po}-\eqref{Pp}, we can write
\begin{align}
	&\text{Im}\Pi^{\perp,\pm}(\mu_3;\omega) = - \text{Im}\Pi^{\perp,\mp}(\mu_3;-\omega)\,,\\
	&\text{Im}\Pi^{\parallel,\pm}(\mu_3;\omega) = - \text{Im}\Pi^{\parallel,\mp}(\mu_3;-\omega)\,.
\end{align}

Eventually, we can combine \eqref{Or11} and \eqref{Or12} to obtain
\begin{equation}
	\label{Or13} G^R_\pm(\m_3;\omega)^* = G^R_\pm(-\m_3;-\omega) \, .
\end{equation}

\section{Thermodynamics of the holographic model and parameters of the bulk action}\label{app:thermo}

In this appendix, we review the thermodynamics of the five-dimensional AdS Reissner-Nordstr\"om space-time, and we relate the parameters of the bulk action to QCD data.

The bottom-up model describing the dynamics of chiral current operators at the background level is realized by an AdS Reissner-Nordstrom solution in five dimensions. From the Euclidean on-shell action of the model \eqref{eq:Sb}, \eqref{eq:SEH}, \eqref{eq:SYM}, we can compute the grand-canonical potential $\Omega$ \cite{RN1,RN2,neutrinopaper}:
\be\label{eq:OmegaGC}
\Omega = - T\,S_\text{on-shell}^E = -V_3(M\ell)^3\left(N_cr_H^{-4}+\dfrac{1}{6}N_cw_0^2\mu^2r_H^{-2}\right),
\ee
where $V_3$ is the volume of three space and where the horizon radius $r_H$ is given in (\ref{eq:rHRN}).

From the grand-canonical potential, we can compute the thermodynamic quantities such as the pressure density $p$, the energy density $\varepsilon$, the entropy density $s$, and the charge density $n$. The pressure is simply
\be\label{eq:pressure}
p = -\Omega/V_3 = (M\ell)^3\left(N_cr_H^{-4}+\dfrac{1}{6}N_cw_0^2\mu^2r_H^{-2}\right),
\ee
Then, using the thermodynamic equation
\be
dp = n\,d\mu + s\,dT,
\ee
the energy density $s$, and the charge density $n$ read
\be
s = \left(\dfrac{\partial p}{\partial T}\right)\bigg|_\mu, \qquad\qquad n = \left(\dfrac{\partial p}{\partial \mu}\right)\bigg|_T.
\ee

The relation between $T$, $\mu$, and the horizon radius $r_H$ is
\be
T = \dfrac{1}{\pi r_H}-\dfrac{r_H^2w_0^2\mu^2}{12\pi N_c},
\ee
and in order to perform the partial derivatives, we use the chain rule
\be
s = \left(\dfrac{\partial p}{\partial T}\right)\bigg|_{r_H}\left(\dfrac{\partial r_H}{\partial T}\right)\bigg|_{\mu}, \qquad\qquad n = \left(\dfrac{\partial p}{\partial \mu}\right)\bigg|_{r_H}+ \left(\dfrac{\partial p}{\partial r_H}\right)\bigg|_{\mu}\left(\dfrac{\partial r_H}{\partial T}\right)\bigg|_{T}.
\ee

It is a straightforward exercise to show that the computation leads to
\begin{align}
	&s = 4\pi(M\ell)^3N_c^2r_H^{-3},\\
	&n =(M\ell)^3w_0^2N_cr_H^{-2}\mu.
\end{align}
From  the entropy density, we can  compute $E_\text{gap}$, that is, the energy which sets the scale for near-extremal quantum effects, \cite{Mertens}. We expand the total entropy close to extremality, that is, for small $T$
\be
S = \left(\dfrac{\pi}{\sqrt{3N_c}}(M\ell)^3w_0^3V_3\mu^3 + 2\pi^2\dfrac{T}{E_\text{gap}}\right)(1 + \mathcal{O}(T/\mu)),
\ee
where we have defined
\be
E_\text{gap} = \dfrac{4}{(M\ell)^3w_0^2V_3\mu^2}\,.
\label{gap}\ee

We can compute the energy density using the following thermodynamic identity
\begin{align}
	\varepsilon = -p +T\,s+\mu\,n = 3(M\ell)^3\left(N_cr_H^{-4}+\dfrac{1}{6}N_cw_0^2\mu^2r_H^{-2}\right).
\end{align}
As it happens in a conformally invariant theory, we observe that
\be
\varepsilon = 3p.
\ee

The bulk action \eqref{eq:Sb} possesses two parameters: the five-dimensional Planck mass in units of the AdS scale, $M\ell$, which controls the overall normalization of the action, and the flavor coupling $w_0$. We detail in this appendix, how the values of these parameters are fixed by approximately matching to QCD parameters.

$M\ell$ is fixed by imposing that the zero-chemical potential limit of the pressure be that of a free quark-gluon plasma
\be
\label{Pa1} p = \frac{\pi^2 N_c^2}{45}T^4 \left(1 + \frac{7N_f}{4N_c}\right) \, .
\ee
Lattice results \cite{Borsanyi13} indicate that, for temperatures equal to a few times the deconfining temperature, the pressure in the quark-gluon plasma is already close (within about $20\%$) to the ideal result \eqref{Pa1}. Setting $(M\ell)$ to match \eqref{Pa1} will therefore ensure that the thermodynamics of the holographic model is close to that of QCD in the deconfined phase. The pressure of the holographic model is computed from the grand-canonical potential \eqref{eq:OmegaGC} and it is given by \eqref{eq:pressure}. At $\mu \ll T$, it is given by
\be
\label{Pa2} p = (M\ell)^3\left[N_c^2(\pi T)^4  + \frac{1}{4}N_fN_c w_0^2(\pi T)^2\m^2  + \OO(\mu^4)\right]  \, .
\ee
Therefore \eqref{Pa2} matches \eqref{Pa1} at $\mu = 0$ if $(M\ell)^3$ is equal to
\be
\label{Pa3} (M\ell)_{\text{free}}^3 = \frac{13}{6}\frac{1}{45\pi^2} \, ,
\ee
where the number of flavors was set to $N_f=2$, and that of colors to $N_c=3$.

As far as the parameter $w_0$ is concerned, it can be fixed such that the baryon number susceptibility at zero density agrees with the ideal Fermi gas result. As for the pressure, it was observed to give a good approximation to the exact result for the quark-gluon plasma on the lattice \cite{Borsanyi11,Borsanyi13}. The baryon number susceptibility is defined as the first non-trivial cumulant of the pressure at $\mu = 0$
\be
\label{Pa4} \chi_B = \frac{\partial^2 p}{\partial^2 \mu_B}\bigg|_{\mu_B = 0} \, .
\ee
From \eqref{Pa2} it is equal to
\be
\label{Pa4b} \chi_B = \frac{N_f}{2N_c}w_0^2(M\ell)^3(\pi T)^2 \, ,
\ee
whereas the ideal Fermi gas result is
\be
\label{Pa5} \chi_{B,\text{free}} = \frac{N_f}{3N_c} T^2  \, .
\ee
Matching the two results fixes the value of $w_0$ to be
\be
\label{w0lat} \big(w_{0}^2(M\ell)^3\big)_{\text{free}} = \frac{2}{3\pi^2} \, .
\ee

In the numerical calculations done in this paper, we used the values of the parameters given by \eqref{Pa3} and \eqref{w0lat}.

\section{Holographic renormalization}\label{app:holoren}
In this Appendix, we present the counterterms required to regularize the divergence of the real part of the polarization functions \eqref{PLt}, \eqref{PLl}. If we substitute the near-boundary expansion of the gauge field fluctuations
\begin{align}
	&\mathcal{L}^{\perp,\pm}_i = \mathcal{L}^{\perp,\pm,(0)}_i + r^2\left(\mathcal{L}^{\perp,\pm,(2)}_i - \frac{1}{2}\left((\omega\pm\mu_3)^2-\vec{k}^2\right)\mathcal{L}_i^{\perp,\pm,(0)} \log(r/\ell)\right)\Big(1 + \OO\big(r^2\big)\Big) \, ,\\
	&E^{\parallel,\pm} = E^{\parallel,\pm,(0)} + r^2\left(E^{\parallel,\pm,(2)} - \frac{1}{2}\left((\omega\pm\mu_3)^2-\vec{k}^2\right)E^{\parallel,\pm,(0)} \log(r/\ell)\right)\Big(1 + \OO\big(r^2\big)\Big) \, ,
\end{align}
into the expression for the transverse and longitudinal polarization functions \eqref{PLt} and \eqref{PLl} we obtain
\begin{align}
	\Pi^{\perp,\pm}(\omega,\vec{k}) = &-\frac{1}{8}(M\ell)^3w_0^2 N_c\bigg(2\dfrac{\mathcal{L}_i^{\perp,\pm,(2)}}{\mathcal{L}_i^{\perp,\pm,(0)}} -((\omega\pm\mu_3)^2-\vec k^2)\left(\log(\epsilon/\ell)-\dfrac{1}{2}\right)\bigg)\,,\\ \non
	\Pi^{\parallel,\pm}(\omega,\vec{k}) = &-\frac{1}{8}(M\ell)^3w_0^2 N_c\bigg(2\dfrac{E^{\parallel,\pm,(2)}}{E^{\parallel,\pm,(0)}} -((\omega\pm\mu_3)^2-\vec k^2)\left(\log(\epsilon/\ell)-\dfrac{1}{2}\right)+\\
	&\pm\dfrac{2}{r_H^2}\dfrac{\mu_3}{\omega\pm\mu_3}\bigg)\,,
\end{align}
where the last term in the longitudinal polarization function is a real contact term.

Both the transverse and longitudinal polarization functions have the same logarithmic UV divergence which we want to regularize. We can do it with a single counterterm which takes the following form
\be
S_\text{c.t.} = -\dfrac{1}{16}(M\ell)^3w_0^2N_c\log(\epsilon\Lambda_\text{UV})\int\intd^4x\sqrt{-\gamma}\,\gamma^{\mu\rho}\gamma^{\nu\sigma}\d F^{(V),a}_{\mu\nu}\d F^{(V),a}_{\rho\sigma}\,,
\ee
where $\gamma$ is the induced metric at the boundary which in the present case reduces to the Minkowski metric $\gamma_{\mu\nu}=\eta_{\mu\nu}$.
The prefactor has a logarithmic divergence in the limit $\epsilon\to 0$ with $\epsilon$ being the small-distance cutoff. Moreover, we introduce $\Lambda_\text{UV}$, the energy scale of the theory, that determines the coefficient of the finite counterterms. $\Lambda_\text{UV}$  can be chosen arbitrarily. Its value is scheme dependent and, as it will be shown, it enters only in the real part of the polarization functions.

The counterterm action, by using the explicit formulae of the field strength of the fluctuations and focusing on the charged degrees of freedom, can be rewritten as
\begin{align}\non
	S_\text{c.t.}\Big|_\text{charged} = &-\dfrac{1}{16}(M\ell)^3w_0^2N_c\log(\epsilon\Lambda_\text{UV})((\omega\pm\mu_3)^2-\vec k^2)\times\\
	&\times\int\dfrac{\intd^4k}{(2\pi)^4}\,\mathcal{L}^{\l,\mp}(-k_\mu)\!\left(  P_{\l\s}^{\perp,\pm}(k_\mu)+P_{\l\s}^{\parallel,\pm}(k_\mu)\right)\!\mathcal{L}^{\sigma,\pm}(k_\mu)\,
\end{align}
This counterterm removes the UV logarithmic divergence in the polarization functions. We conclude this appendix by presenting their renormalized version
\begin{align}
	\Pi^{\perp,\pm}(\omega,\vec{k})\Big|_\text{ren.} = &-\frac{1}{8}(M\ell)^3w_0^2 N_c\bigg(2\dfrac{\mathcal{L}_i^{\perp,\pm,(2)}}{\mathcal{L}_i^{\perp,\pm,(0)}} +((\omega\pm\mu_3)^2-\vec k^2)\left(\log(\Lambda_\text{UV}\ell)+\dfrac{1}{2}\right)\bigg)\,,\\ \non
	\Pi^{\parallel,\pm}(\omega,\vec{k})\Big|_\text{ren.} = &-\frac{1}{8}(M\ell)^3w_0^2 N_c\bigg(2\dfrac{E^{\parallel,\pm,(2)}}{E^{\parallel,\pm,(0)}} +((\omega\pm\mu_3)^2-\vec k^2)\left(\log(\Lambda_\text{UV}\ell)+\dfrac{1}{2}\right)+\\
	&\pm\dfrac{2}{r_H^2}\dfrac{\mu_3}{\omega\pm\mu_3}\bigg)\,.
\end{align}

\section{Eddington-Finkelstein coordinates for the charged fluctuations}\label{app:EF}
In this appendix, we adopt the Eddington-Finkelstein coordinates to rewrite the equations of motion for the charged fluctuations $\mathcal{L}^{\perp,\pm}$ and $E^{\parallel,\pm}$, which are then used to compute the associated quasinormal modes and correlator. As usual, we employ infalling Eddington-Finkelstein coordinates defined by the following change of coordinates
\begin{align}
	&x^i\to x^i,&  &x^0\to u = x^0-z(r),&  &r\to r,&
\end{align}
where the tortoise coordinate $z(r)$ is defined in terms of the blackening factor $f(r)$ as
\be
\label{EF4} \frac{\intd z}{\intd r} = \frac{1}{f(r)} \, .
\ee
Then, the Fourier transform of the gauge field perturbation transforms as
\be
\label{EF5} \mathcal{L}_{\mu}(r) \to \ex^{i\omega z(r)} \mathcal{L}_{\mu'}(r) \, ,
\ee
where now $\mu'$ runs over $\{u,x^i\}$ instead of $\{x^0,x^i\}$. We decompose the general fluctuation into the transverse and longitudinal parts with respect to the spatial momentum
\be
\label{EF6} \mathcal{L}^{\perp,\pm}_i(r) \to \ex^{i\omega z} \mathcal{L}^{\perp,\pm}_i(r) \sp E^{\parallel,\pm}(r) \to \ex^{i\omega z} E^{\parallel,\pm}(r) \, .
\ee

This way, we select infalling boundary conditions for the solutions, and we solve for fields that do not oscillate very fast near the horizon for non-zero $\omega$. Indeed, since the tortoise coordinate near the horizon behaves as
\be
z(r) \sim -\dfrac{1}{4\pi T}\log\left(1-\dfrac{r}{r_H}\right),
\ee
the rescaled fluctuations do not oscillate near the horizon, and are instead analytic at $r = r_H$.\footnote{The infalling Eddington-Finkelstein coordinates are well-defined beyond the horizon, and there exists a solution to the equations of motion which is perfectly regular at $r=r_H$ in these coordinates.} Moreover, in order to select the normalizable solution in the numerical routine to find the quasinormal modes, we rescale the fluctuations as
\be
\label{EFr} \mathcal{L}^{\perp,\pm}_i(r) \to r\, \mathcal{L}^{\perp,\pm}_i(r) \sp E^{\parallel,\pm}(r) \to r\, E^{\parallel,\pm}(r) \, .
\ee
This way, the non-normalizable solution automatically blows up at the boundary.

Applying the transformations \eqref{EF5} and \eqref{EFr} to the equations of motion \eqref{EoMVhtpm} and \eqref{EpEoM} gives the differential equations
\begin{align}
	\non&\partial_r^2\mathcal{L}^{\perp,\pm}_i + \left(\dfrac{1}{r}+\dfrac{\partial_rf(r)+2i\omega}{f(r)}\right)\partial_r\mathcal{L}^{\perp,\pm}_i +\\ \label{EF8}
	&+\left(\dfrac{\Omega_\pm(r)^2-\vec k^2 f(r)-\omega^2}{f(r)^2}+\dfrac{\partial_rf(r)+i\,\omega}{r\,f(r)}-\dfrac{1}{r^2}\right)\mathcal{L}^{\perp,\pm}_i = 0 \, ,
\end{align}
\begin{align}
	\nn\label{EF9}& \partial_r^2E^{\parallel,\pm} + \Bigg( \dfrac{\Omega_\pm(r)^2 \left(\partial_rf(r)+2 i \omega \right)}{f(r) \left(\Omega_\pm(r)^2-\vec k^2 f(r)\right)}- \dfrac{2 i \vec k^2 r \omega +2 r \Omega_\pm(r) \Omega_\pm'(r)-\Omega_\pm(r)^2}{r \left(\Omega_\pm(r)^2-\vec k^2 f(r)\right)}\\ \nonumber
	&-\dfrac{\vec k^2 f(r)}{r \left(\Omega_\pm(r)^2-\vec k^2 f(r)\right)} \Bigg)\partial_rE^{\parallel,\pm} + \dfrac{E^{\parallel,\pm}}{r^2 f(r)^2 \left(\vec k^2 f(r)-\Omega_\pm(r)^2\right)}\times\\ \nonumber
	&\times \bigg(-r f(r) \big[\Omega_\pm(r)^2 \left(\partial_rf(r)-2 \vec k^2 r+i \omega \right)+\vec k^2 r \omega  \left(\omega +i \partial_rf(r)\right)+\\ \nonumber
	&+r \Omega_\pm(r) \left(\mp \partial_rf(r) \partial_r\Phi_3(r)-2 i \omega  \partial_r\Omega_\pm(r)\right)\big]-\vec k^2 f(r)^3+\\ \nonumber
	&+f(r)^2 \left(-r \left(\vec k^4 r-i \vec k^2 \omega +2 r \partial_r\Phi_3(r)^2\right)+2 r \Omega_\pm(r) \partial_r\Omega_\pm(r)+\Omega_\pm(r)^2\right)+\\
	&+r^2 \Omega_\pm(r)^2 \left(\omega ^2-\Omega_\pm(r)^2\right)\bigg)=0\,.
\end{align}

\section{Analysis of the IR-AdS$_2$ correlator and the various hydrodynamic approximations}\label{app:IRcorr-approx}

In this appendix, we study in detail the behaviour of the IR-AdS$_2$ correlator. We give, in addition to the extended hydrodynamic approximation, presented in section \ref{sec:IRcorr-approx}, other IR-based approximations for the exact full current correlator. We recall the explicit formula for the IR-AdS$_2$ correlator \eqref{eq:q3}\footnote{Recall that the $\pm$ in the right-hand side of \eqref{eq:q3} corresponds to the charge label of the correlator, and not to the notation sometimes found in the literature $\Gamma(\pm)=\Gamma(+)\Gamma(-)$.}
\begin{align}\label{eq:qq3}
	\GG^\pm_\text{IR}(\omega, k) =&(4\pi T)^{2\D( k,\mu_3)-1}\frac{\Gamma(1\!-\!2\D(k,\mu_3))\Gamma(\D(k,\mu_3)\!-\!\frac{i\omega}{2\pi T}\pm \frac{i}{6}r_H\mu_3)}{\Gamma(2\D(k,\mu_3)\!-\!1)\Gamma(1\!-\!\D(k,\mu_3)\!-\!\frac{i\omega}{2\pi T}\pm \frac{i}{6}r_H\mu_3)}\times\\ \non
	&\times\dfrac{\Gamma(\D(k,\mu_3)\mp \frac{i}{6}r_H\mu_3)}{\Gamma(1\!-\!\D(k,\mu_3)\mp \frac{i}{6}r_H\mu_3)},
\end{align}
with $\D(k,\mu_3)$ the IR conformal dimension \eqref{eq:q4}
\begin{equation}
	\D(k,\mu_3) = \frac{1}{2} + \frac{1}{2}\sqrt{1+\frac{1}{3}r_H^2 k^2 - \frac{1}{9}r_H^2\mu_3^2} \, .
\end{equation}

We note that the IR-AdS$_2$ correlator \eqref{eq:qq3} has a finite limit for vanishing spatial momentum and isospin chemical potential
\be
\GG^\pm_\text{IR}(\omega,0)\underset{\mu_3\to 0}{\longrightarrow} -2\pi T\frac{\Gamma(1-\frac{i\omega}{2\pi T})}{\Gamma(-\frac{i\omega}{2\pi T})}\,.
\ee

It is instructive to study the behavior of the IR-AdS$_2$ correlator when $\omega/T\ll 1$ and $\omega/T\gg 1$. The former limit describes the hydrodynamic region in $\omega$, while the latter, provided that $\omega \ll \mu$, reaches the extended hydrodynamic region.

When $\omega/T\ll 1$, the correlator behaves as
\begin{align}\nonumber
	&\mathcal{G}^\pm_\text{IR}(\omega,k) =\frac{(4 \pi
		T)^{2 \D(k,\mu_3)-1} \Gamma (1-2 \D(k,\mu_3))  |\Gamma \left(\D(k,\mu_3)\pm\frac{i}{6}r_H\mu_3\right)|^2}{\Gamma (2\D(k,\mu_3)-1) |\Gamma \left(1-\D(k,\mu_3)\pm\frac{i}{6}r_H\mu_3\right)|^2}+\\ \nonumber
	&+ i\,\omega\,\frac{2^{4 \D(k,\mu_3)-3} (\pi T)^{2 \D(k,\mu_3)-2} \left(\psi(\pm\frac{i}{6}r_H\mu_3+1-\D(k,\mu_3))-\psi(\pm\frac{i}{6}r_H\mu_3+\D(k,\mu_3))\right)}{\Gamma (2 \D(k,\mu_3)-1) |\Gamma \left(1-\D(k,\mu_3)\pm\frac{i}{6}r_H\mu_3\right)|^2}\times\\
	&\times \Gamma (1-2 \D(k,\mu_3))\bigg|\Gamma \left(\D(k,\mu_3)\pm\frac{i}{6}r_H\mu_3\right)\bigg|^2 +\mathcal{O}\left(\left(\dfrac{\omega}{T}\right)^2\right),
\end{align}
where $\psi(x)$ is the digamma function.  We can take the imaginary part of the IR-AdS$_2$ correlator, expanded for small $\omega/T$, and we recover the well-known hydrodynamic behaviour of the imaginary part of the correlator, (i.e. linearity in $\omega$)
\begin{align} \non
	\text{Im}\,\mathcal{G}^\pm_{\mathrm{IR}}(\omega,k) = &\frac{ (4\pi T)^{2 \Delta(k,\mu_3)-1} \Gamma \!\bigl(1-2 \Delta(k,\mu_3)\bigr)}{4\Gamma \!\bigl(2 \Delta(k,\mu_3)-1\bigr)}\frac{
		|\Gamma \!\left(\Delta(k,\mu_3)\pm \frac{i}{6}r_H\mu_3\right)|^2
	}
	{
		|\Gamma \!\left(1-\Delta(k,\mu_3)\pm \frac{i}{6}r_H\mu_3\right)|^2
	}\times \\ \label{eq2:GIRlinearomega}
	&\times 2\dfrac{\omega}{T}\Re\Bigg[\cot\!\left(\pi\left(\Delta(k,\mu_3)\pm \frac{i}{6}r_H\mu_3\right)\right)
	\Bigg]
	+\mathcal{O}\!\left(\left(\frac{\omega}{T}\right)^2\right).
\end{align}

We expand the IR-AdS$_2$ correlator, for large values of the ratio $\omega/T$, using the asymptotic Gamma expansion
\begin{equation}
	\Gamma(a-i\,x) = \sqrt{2\pi}\,x^{a-\frac{1}{2}}e^{-\frac{\pi\,x}{2}} e^{-i\,x(\log x-1)}e^{-i\frac{\pi (2a-1)}{4}}\left(1+\dfrac{i}{12x}(6a^2-6a+1)+\mathcal{O}\left(\dfrac{1}{x^2}\right)\right)\,.
\end{equation}

From the above formula, we can read off the leading term of the Gamma function for a large imaginary part. The relative corrections organize themselves in a power series in $1/x$. If we apply this formula to the IR-AdS$_2$ correlator \eqref{eq:qq3} in the limit $\omega/T\to \infty$, we obtain
\begin{align}\nonumber
	\mathcal{G}^\pm_\text{IR}(\omega,k) = &2^{2\D(k,\mu_3)-1}\frac{\Gamma(1\!-\!2\D(k,\mu_3))\Gamma(\D(k,\mu_3)\mp \frac{i}{6}r_H\mu_3)}{\Gamma(2\D(k,\mu_3)\!-\!1)\Gamma(1\!-\!\D(k,\mu_3)\mp \frac{i}{6}r_H\mu_3)} e^{-i\frac{\pi}{2}(2\D(k,\mu_3)-1)}\times\\
	&\times \omega^{2\D(k,\mu_3)-1}\left(1+ \mathcal{O}\left(\dfrac{T}{\omega}\right)\right),
\end{align}
i.e., the correlator $\sim\omega^{2\D(k,\mu_3)-1}$ in the IR. For completeness, we also present the imaginary part of the IR-AdS$_2$ correlator in this limit, which is given by taking the imaginary part of the formula above
\begin{align}\nonumber
	\text{Im}\mathcal{G}^\pm_\text{IR}(\omega,k)= &\frac{4^{\D(k,\mu_3)-1}\Gamma (1-2 \D(k,\mu_3))}{\Gamma (2 \D(k,\mu_3)-1)}\left(\frac{e^{i \pi  \D(k,\mu_3)} \Gamma \left(\D(k,\mu_3)\pm\frac{i}{6}r_H\mu_3\right)}{\Gamma \left(1-\D(k,\mu_3)\pm\frac{i}{6}r_H\mu_3\right)}+\text{c.c.}\right)\times\\ \label{eq2:GIRpowerlaw}
	&\times \omega^{2\D(k,\mu_3)-1}\left(1+ \mathcal{O}\left(\dfrac{T}{\omega}\right)\right).
\end{align}

For later convenience, we give a name to the leading term of the $\omega/T\gg 1$ expansion of the IR-AdS$_2$ correlator
\be\label{eq2:improvedIRcorr}
\mathcal{G}_\text{IR}^{\pm,\text{impr.}}(\omega,k) =(2\omega)^{2\D(k,\mu_3)-1}\frac{\Gamma(1\!-\!2\D(k,\mu_3))\Gamma(\D(k,\mu_3)\mp \frac{i}{6}r_H\mu_3)}{\Gamma(2\D(k,\mu_3)\!-\!1)\Gamma(1\!-\!\D(k,\mu_3)\mp \frac{i}{6}r_H\mu_3)} e^{-i\frac{\pi}{2}(2\D(k,\mu_3)-1)}\,.
\ee

It is instructive to compute the ratio between the coefficient of the small $\omega$ (the coefficient of $\omega/T$ in formula \eqref{eq2:GIRlinearomega}) and large $\omega$ (the coefficient of $\omega^{2\Delta -1}$ in formula \eqref{eq2:GIRpowerlaw}) expansion of the imaginary part of the correlator. We do this because we want to investigate the transition between the hydrodynamic, linear in $\omega$ behavior of the correlator and the extended-hydrodynamic, power-law behavior $\omega^{2\Delta-1}$ of the correlator. This goes in the direction of analysing  the  possible ways of approximating the exact numerical result for the charged current-current correlator in the extended hydrodynamic region. This ratio, after some algebra, can be written as
\begin{align}\non
	\varrho^\pm(k,\mu_3)&\equiv \dfrac{\underset{\omega/T\to 0}{\lim}\text{Im}\mathcal{G}^\pm_\text{IR}/\omega}{\underset{\omega/T\to\infty}{\lim}\text{Im}\mathcal{G}^\pm_\text{IR}\,\omega^{1-2\Delta(k,\mu_3)}}=\\ \label{eq2:ratioIR}
	&=(2 \pi T)^{2(\D(k,\mu_3)-1)} e^{\mp\frac{1}{6}r_H\mu_3} \bigg|\Gamma \left(\D(k,\mu_3)\pm\frac{i}{6}r_H\mu_3\right)\bigg|^2\,.
\end{align}

This is equal to 1 in the limit of $\{\mu_3\to 0, k\to 0\}$, \textit{i.e.} $\varrho^\pm(0,0)=1$. The indices $\pm$ in \eqref{eq2:ratioIR} corresponds to the plus and minus IR-AdS$_2$ correlators $\mathcal{G}^\pm_\text{IR}$. The ratios $\rho^+$ and $\rho^-$, as the correlator itself can be mapped one into the other by the transformation $\mu_3\to -\mu_3$.

We shall now investigate the limits of this ratio, for $k$ and $\mu_3$ asymptoting  to zero separately.  First, we consider the ratio at zero isospin chemical potential:
\begin{align}\label{eq:ratiozeromu3}
	& \varrho^\pm(k,0)=\varrho_0(k)=(2 \pi T)^{(2\D_0(k)-1)}\Gamma(\Delta_0(k))^2,
\end{align}
where now $\Delta_0(k)$ is the IR conformal dimension at $\mu_3=0$:
\be
\Delta(k,0) =\Delta_0(k) = \dfrac{1}{2}+\dfrac{1}{2}\sqrt{1+\dfrac{1}{3}r_H^2 k^2}.
\ee
Moreover, at zero momentum, the ratio reads
\begin{align}\label{eq:ratiozerok}
	\varrho^\pm(0,\mu_3)=&(2 \pi T)^{-1+\sqrt{1-\frac{1}{9}r_H^2\mu_3^2}} e^{\mp\frac{1}{6}r_H\mu_3}\bigg|\Gamma \left(\dfrac{1}{2}+\dfrac{1}{2}\sqrt{1-\dfrac{1}{9}r_H^2\mu_3^2}\,\pm\frac{i}{6}r_H\mu_3\right)\bigg|^2\,.
\end{align}

\begin{figure}[h]
	\begin{center}
		\begin{subfigure}{0.48\textwidth}
			\centering
			\includegraphics[width=\textwidth]{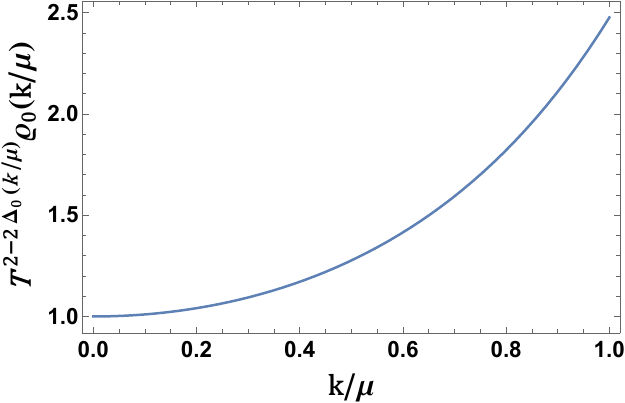}
			\caption{} \label{fig:ratiomuq104mu30}
		\end{subfigure}

		\begin{subfigure}{0.48\textwidth}
			\centering
			\includegraphics[width=\textwidth]{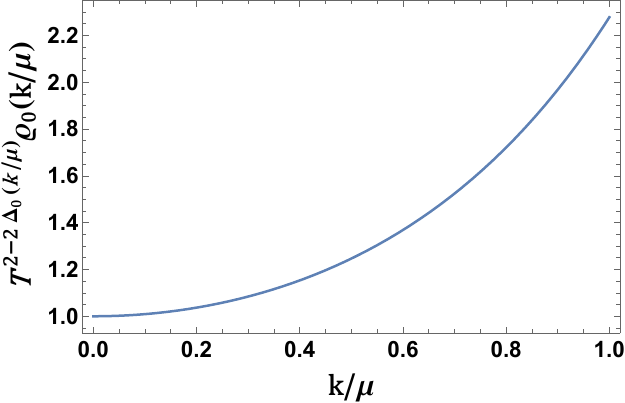}
			\caption{} \label{fig:ratiomuq65mu30}
		\end{subfigure}
		\hfill
		\begin{subfigure}{0.48\textwidth}
			\centering
			\includegraphics[width=\textwidth]{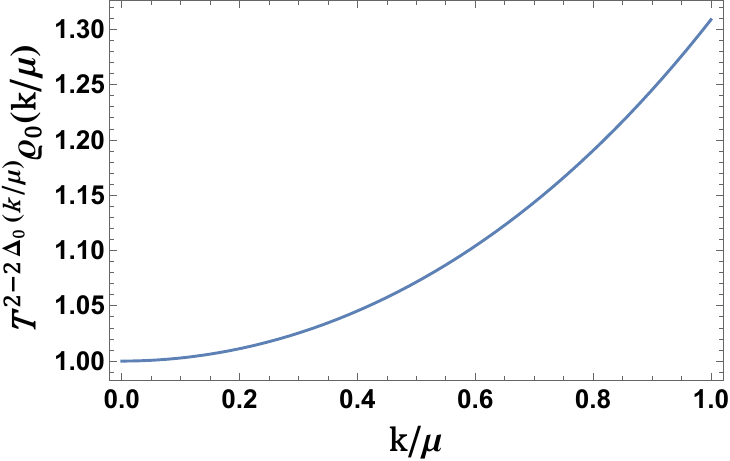}
			\caption{} \label{fig:ratiomuq5mu30}
		\end{subfigure}
		\caption{The $\mu_3=0$ ratio $\varrho_0$ \eqref{eq:ratiozeromu3} as a function of $k/\mu$ for $\mu_q/T =10^4$ (panel \ref{fig:ratiomuq104mu30}), for $\mu_q/T =65$ (panel \ref{fig:ratiomuq65mu30}), and for $\mu_q/T =5$ (panel \ref{fig:ratiomuq5mu30}).}
		\label{fig:ratioIR}
	\end{center}
\end{figure}

We plot the ratio $\rho^\pm$ as a function of $k/\mu$ at $\mu_3/T = 0$, see formula \eqref{eq:ratiozeromu3}, in figure \ref{fig:ratioIR}. We do this for $\mu_q/T =10^4$ (panel \ref{fig:ratiomuq104mu30}), for $\mu_q/T =65$ (panel \ref{fig:ratiomuq65mu30}), and for $\mu_q/T =5$ (panel \ref{fig:ratiomuq5mu30}). The ratio is somewhat sensitive to the changes of $\mu_q/T$ as it decreases as $\mu_q/T$ decreases. In all cases, as $k/\mu$ grows, the ratio grows.

\begin{figure}[h]
	\begin{center}
		\begin{subfigure}{0.48\textwidth}
			\centering
			\includegraphics[width=\textwidth]{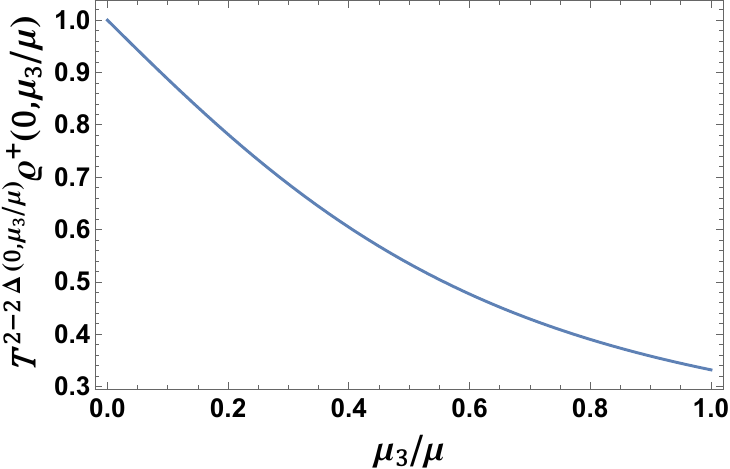}
			\caption{} \label{fig:ratiomuq104funcmu3}
		\end{subfigure}

		\begin{subfigure}{0.48\textwidth}
			\centering
			\includegraphics[width=\textwidth]{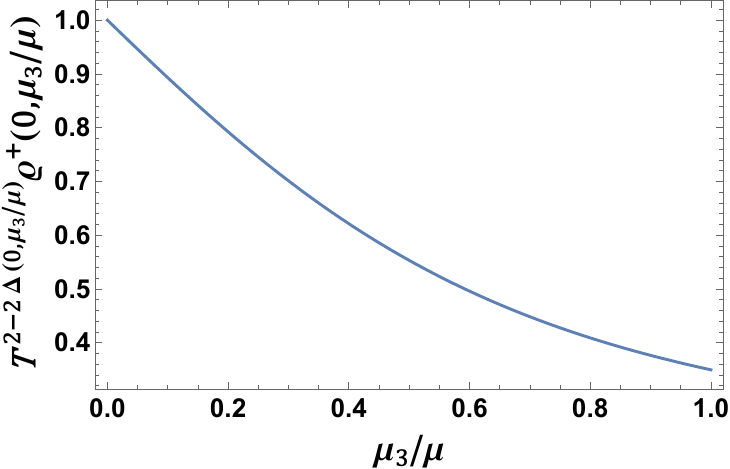}
			\caption{} \label{fig:ratiomuq65funcmu3}
		\end{subfigure}
		\hfill
		\begin{subfigure}{0.48\textwidth}
			\centering
			\includegraphics[width=\textwidth]{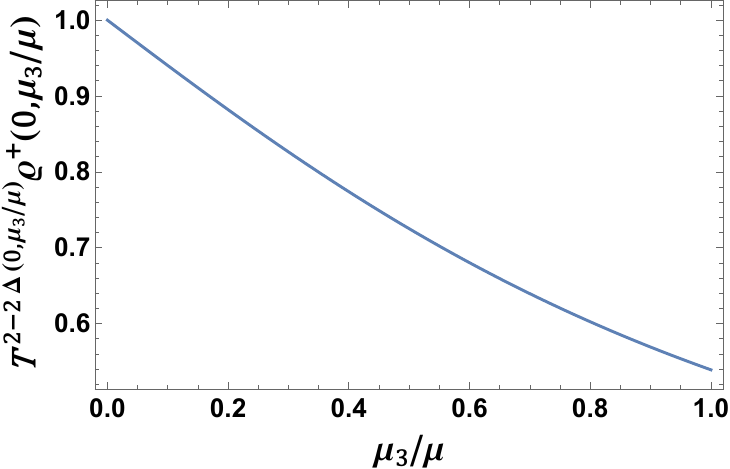}
			\caption{} \label{fig:ratiomuq5funcmu3}
		\end{subfigure}
		\caption{The ratio $\varrho^+$ at $k = 0$ \eqref{eq:ratiozerok} as a function of $\mu_3/\mu$, for $\mu_q/T =10^4$ (panel \ref{fig:ratiomuq104funcmu3}), for $\mu_q/T =65$ (panel \ref{fig:ratiomuq65funcmu3}), and for $\mu_q/T =5$ (panel \ref{fig:ratiomuq5funcmu3}).}
		\label{fig:ratioIRmu3}
	\end{center}
\end{figure}

In figure \ref{fig:ratioIRmu3}, we plot the ratio $\varrho^+$ at zero spatial momentum $k=0$ as a function of $\mu_3$, given in \eqref{eq:ratiozerok}. We do this for $\mu_q/T =10^4$ (panel \ref{fig:ratiomuq104funcmu3}), for $\mu_q/T =65$ (panel \ref{fig:ratiomuq65funcmu3}), and for $\mu_q/T =5$ (panel \ref{fig:ratiomuq5funcmu3}). Here, we observe  that at zero $k/\mu$, the ratio is a monotonic function of the isospin chemical potential for all the values $\mu_q/T$ presented.

\begin{figure}[h]
	\begin{center}
		\begin{subfigure}{0.48\textwidth}
			\centering
			\includegraphics[width=\textwidth]{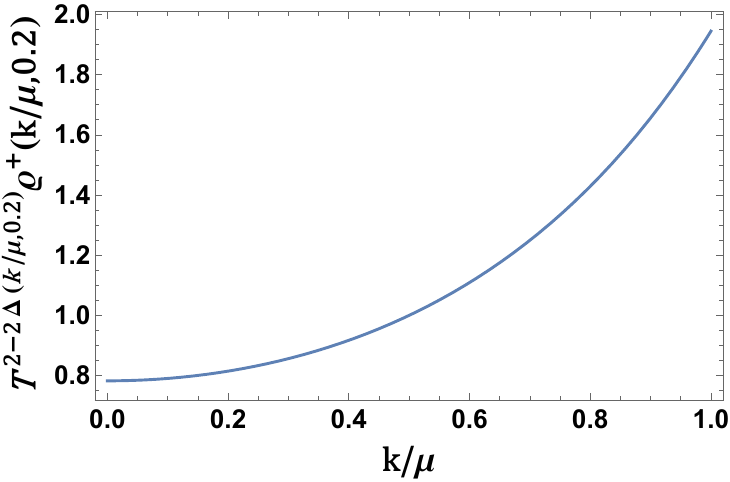}
			\caption{} \label{fig:ratiomuq104mu302}
		\end{subfigure}

		\begin{subfigure}{0.48\textwidth}
			\centering
			\includegraphics[width=\textwidth]{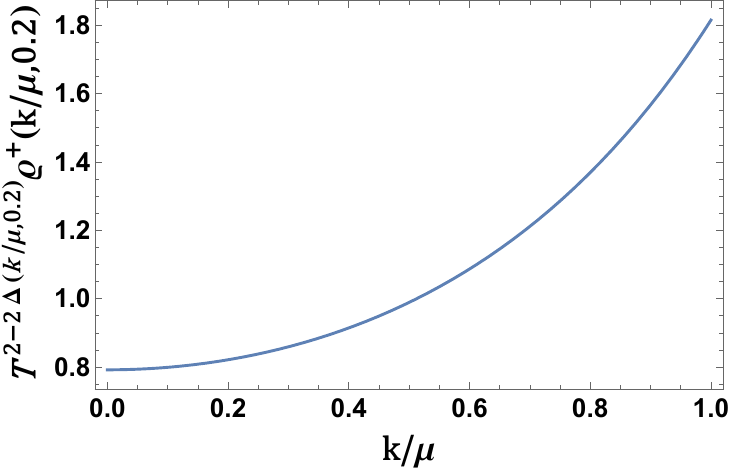}
			\caption{} \label{fig:ratiomuq65mu302}
		\end{subfigure}
		\hfill
		\begin{subfigure}{0.48\textwidth}
			\centering
			\includegraphics[width=\textwidth]{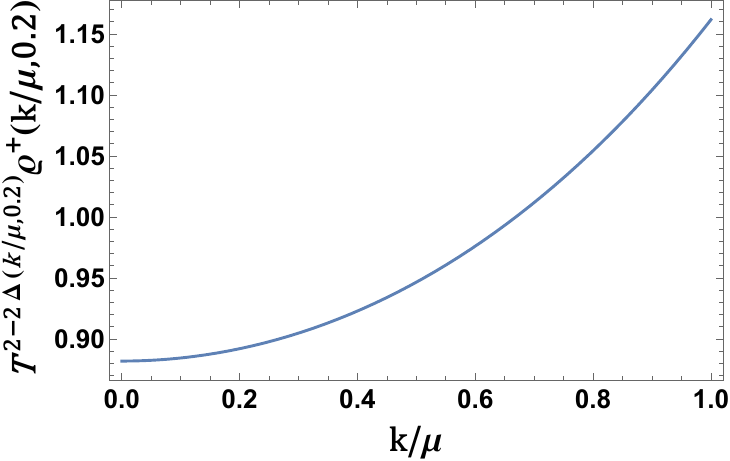}
			\caption{} \label{fig:ratiomuq5mu302}
		\end{subfigure}
		\caption{The ratio $\varrho^+$ \eqref{eq2:ratioIR} at $\mu_3/\mu_q = 0.2$ as a function of $k/\mu$ for $\mu_q/T =10^4$ (panel \ref{fig:ratiomuq104mu302}), for $\mu_q/T =65$ (panel \ref{fig:ratiomuq65mu302}), and for $\mu_q/T =5$ (panel \ref{fig:ratiomuq5mu302}).}
		\label{fig:ratioIRnonzeromu3}
	\end{center}
\end{figure}

We conclude the analysis of the ratio $\varrho^\pm$ by plotting it as a function of $k/\mu$ for non-zero $\mu_3$. In particular, in figure \ref{fig:ratioIRnonzeromu3} we plot the ratio $\varrho^+$ for $\mu_3/\mu_q=0.2$, and for $\mu_q/T =10^4$ (panel \ref{fig:ratiomuq104mu302}), for $\mu_q/T =65$ (panel \ref{fig:ratiomuq65mu302}), and for $\mu_q/T =5$ (panel \ref{fig:ratiomuq5mu302}). This confirms that the presence of non-zero isospin chemical potential does not change qualitatively the conclusions drawn at $\mu_3=0$ from figure \ref{fig:ratioIR}.

\begin{figure}[htbp]
	\begin{center}
		\includegraphics[scale=0.7]{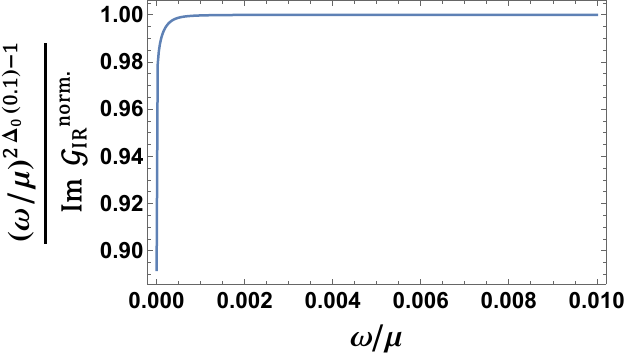}
		\hfill
		\includegraphics[scale=0.7]{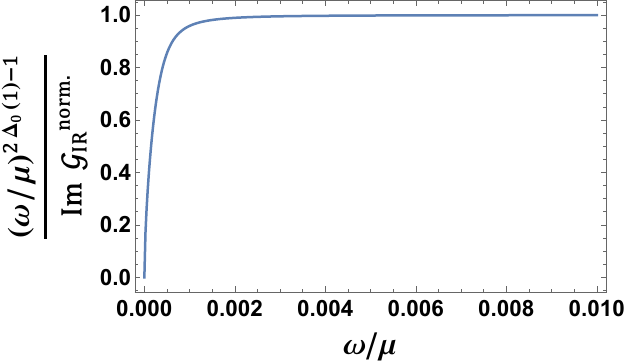}\\
		\includegraphics[scale=0.7]{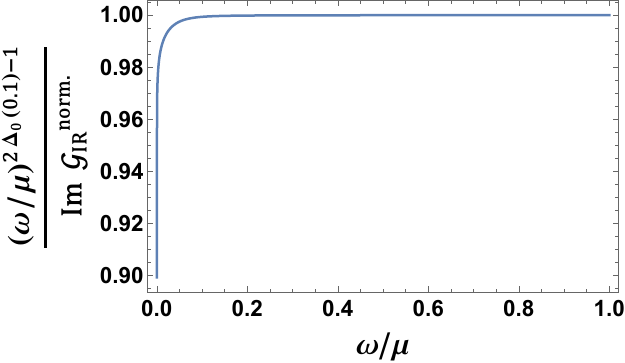}
		\hfill
		\includegraphics[scale=0.7]{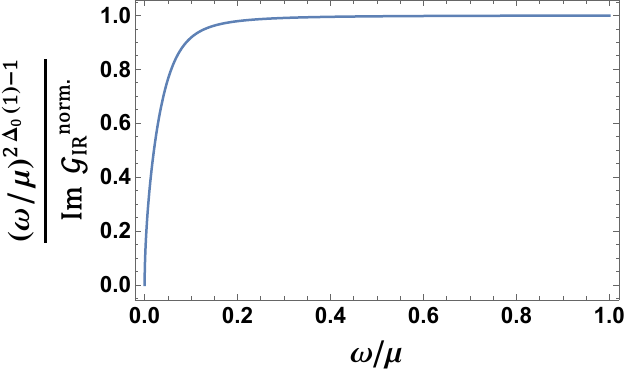}
		\includegraphics[scale=0.7]{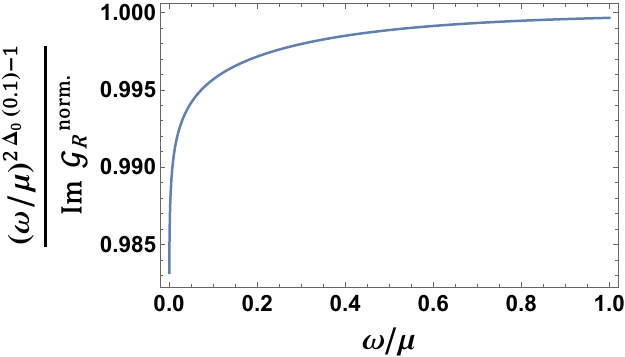}
		\hfill
		\includegraphics[scale=0.7]{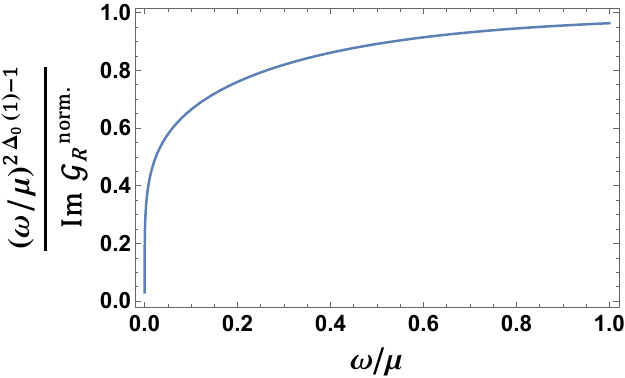}
		\caption{The ratio $\omega^{2\Delta_0(k)-1}/\text{Im}\mathcal{G}_{IR}^\text{norm.}$ where $\omega$ and $k$ are in units of $\mu$. We plot this ratio as a function of $\omega$. The top row shows the results for $\mu_q/T = 10^4$ and $\mu_3 = 0$. The middle row shows the results for $\mu_q/T = 65$ and $\mu_3 = 0$. The bottom row shows the results for $\mu_q/T = 5$ and $\mu_3 = 0$. For each row, the left and right panels are associated with $k/\mu=0.1$ and $k/\mu=1$, respectively. The $\pm$ notation is suppressed since we are working at $\mu_3=0$.}
		\label{fig2:checkcoefficient}
	\end{center}
\end{figure}

We can choose to normalize the full AdS$_2$ correlator \eqref{eq:qq3} with the coefficient of the large $\omega/T$ expansion in \eqref{eq2:GIRpowerlaw}. This way, the normalized correlator for $\omega/T\gg 1$ will approach the $\omega^{2\Delta(k,\mu_3)-1}$ behavior with coefficient 1. Such \textit{normalized} IR-AdS$_2$ correlator reads
\be\label{eq2:normIRcorr}
\GG_\text{IR}^{\pm,\text{norm.}}(\omega,k) = (2\pi\,T)^{2\Delta(k)-1} e^{i\pi(\Delta(k,\mu_3)-1)}\frac{\Gamma(\D(k)\!-\!\frac{i\omega}{2\pi T}\pm \frac{i}{6}r_H\mu_3)}{\Gamma(1\!-\!\D(k)\!-\!\frac{i\omega}{2\pi T}\pm \frac{i}{6}r_H\mu_3)}\,.
\ee
This new expression for the IR-AdS$_2$ correlator is normalized in such a way that for large $\omega/T$, its imaginary part reads
\be
\text{Im}\mathcal{G}_\text{IR}^{\pm,\text{norm.}}(\omega,k) = \omega^{2\Delta(k,\mu_3)-1}\left(1+ \mathcal{O}\left(\dfrac{T}{\omega}\right)\right).
\ee
We check this by plotting the ratio of $\omega^{2\Delta(k,\mu_3)-1}$ with $\text{Im}\mathcal{G}_\text{IR}^{\pm,\text{norm.}}(\omega,k)$, in figure \ref{fig2:checkcoefficient}, as a function of $\omega/\mu$ for some fixed values and $k/\mu$. We do this for $\mu_q/T = 10^4$ (top row), for $\mu_q/T = 65$ (middle row) and $\mu_q/T =5$ (bottom row). The plots in figure \ref{fig2:checkcoefficient} show that raising $\mu_q/T$ the ratio $\omega^{2\Delta-1}/(\text{Im}\mathcal{G}_\text{IR}^{\pm,\text{norm.}})$ approaches 1 for smaller and smaller values of $\omega/\mu$, and $k/\mu$.

Now, we present several ways of approximating the imaginary part of the transverse and longitudinal polarization functions $\Pi^{\perp,\pm}$, $\Pi^{\parallel,\pm}$. Following the discussion of section \ref{sec:IRcorr-approx}, we can approximate the polarization functions using the near-extremal hydrodynamic product formula \eqref{eq:productformula}. In the main text, we considered the limit of small temperature in which the formula simplifies to \eqref{eq:exthydroproduct}, the so-called extremal hydrodynamic product formula.

In this appendix, we have analyzed in detail the nature of the IR-AdS$_2$ correlator, which is responsible for the AdS$_2$ soft poles of the full correlator. Therefore, we wish to investigate the possibility of approximating the full correlator knowing the analytic expression of the IR-AdS$_2$ correlator. In the following, we list the various approximations for the imaginary part of the polarization functions
\begin{itemize}
	\item Standard hydrodynamic approximation.
	\begin{align}\label{app:eq:trhydroapprox}
		&\text{Im}\Pi^{\perp,\pm}_\text{hydro} = -\Sigma\,\omega\,,\\ \label{app:eq:longhydroapprox}
		&\text{Im}\Pi^{\parallel,\pm}_\text{hydro} = -\Sigma\,\omega\dfrac{\omega^2-\vec k^2}{(\omega\pm\mu_3)^2+D\,\vec k^4}\,.
	\end{align}
	\item Extended hydrodynamic approximation.
	\begin{align}\label{app:eq:trexthydroapprox}
		&\text{Im}\Pi^{\perp,\pm}_\text{ext-hydro} = -\Sigma\,\mu(\omega/\mu)^{2\Delta-1}\,,\\ \label{app:eq:longexthydroapprox}
		&\text{Im}\Pi^{\parallel,\pm}_\text{ext-hydro} = -\Sigma\,\mu(\omega/\mu)^{2\Delta-1}\dfrac{\omega^2-\vec k^2}{(\omega\pm\mu_3)^2+D\,\vec k^4}\,.
	\end{align}
	\item Improved extended hydrodynamic approximation.
	\begin{align}\label{eq:trimprexthydroapprox}
		&\text{Im}\Pi^{\perp,\pm}_\text{impr.} = -\Sigma\,\mu^{2\Delta-2}\text{Im}\GG_\text{IR}^{\pm,\text{impr.}}\,,\\ \label{eq:longimprexthydroapprox}
		&\text{Im}\Pi^{\parallel,\pm}_{\pm,\text{impr.}} = -\Sigma\,\mu^{2\Delta-2}\text{Im}\GG_\text{IR}^{\pm,\text{impr.}}\dfrac{\omega^2-\vec k^2}{(\omega\pm\mu_3)^2+D\,\vec k^4}\,.
	\end{align}
	where the expression for $\GG_\text{IR}^{\pm,\text{impr.}}$ is given in \eqref{eq2:improvedIRcorr}.
	\item Approximation using the normalized IR-AdS$_2$ correlator \eqref{eq2:normIRcorr}.
	\begin{align}\label{eq:trnormapprox}
		&\text{Im}\Pi^{\perp,\pm}_\text{norm.} = -\Sigma\,\mu^{2\Delta-2}\text{Im}\GG_\text{IR}^{\pm,\text{norm.}}\,,\\ \label{eq:longnormapprox}
		&\text{Im}\Pi^{\parallel,\pm}_\text{norm.} = -\Sigma\,\mu^{2\Delta-2}\text{Im}\GG_\text{IR}^{\pm,\text{norm.}}\dfrac{\omega^2-\vec k^2}{(\omega\pm\mu_3)^2+D\,\vec k^4}\,,
	\end{align}
	where the expression for $\GG_\text{IR}^\text{norm.}$ is given in \eqref{eq2:normIRcorr}.
	\item Approximation using the full IR-AdS$_2$ correlator \eqref{eq:q3}.
	\begin{align}\label{eq:trfullapprox}
		&\text{Im}\Pi^{\perp,\pm}_\text{full-IR} = -\Sigma\,\mu^{2\Delta-2}\text{Im}\GG^\pm_\text{IR}\,,\\ \label{eq:longfullapprox}
		&\text{Im}\Pi^{\parallel,\pm}_\text{full-IR} = -\Sigma\,\mu^{2\Delta-2}\text{Im}\GG^\pm_\text{IR}\dfrac{\omega^2-\vec k^2}{(\omega\pm\mu_3)^2+D\,\vec k^4}\,,
	\end{align}
	where the expression for $\GG^\pm_\text{IR}$ is given in \eqref{eq:q3}.
\end{itemize}

Besides the (near-extremal) hydrodynamic and the extended hydrodynamic approximations presented in the main text, we have introduced above three new approximations based on the form of the IR-AdS$_2$ correlator.

\begin{figure}[h]
	\centering
	\begin{subfigure}{0.32\textwidth}
		\centering
		\includegraphics[width=\textwidth]{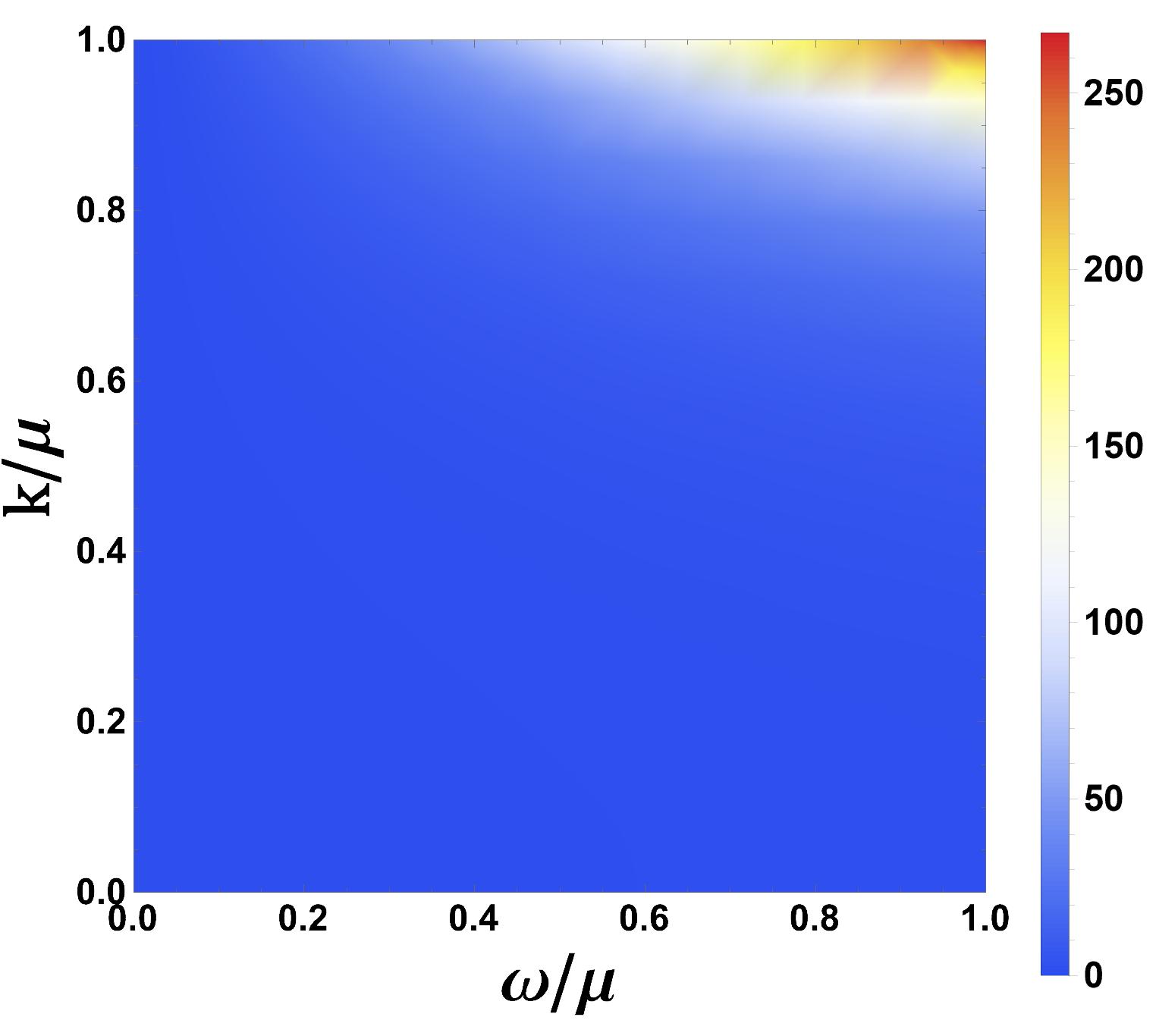}
		\caption{} \label{fig2:Plot_fullGIR_104}
	\end{subfigure}
	\hfill
	\begin{subfigure}{0.32\textwidth}
		\centering
		\includegraphics[width=\textwidth]{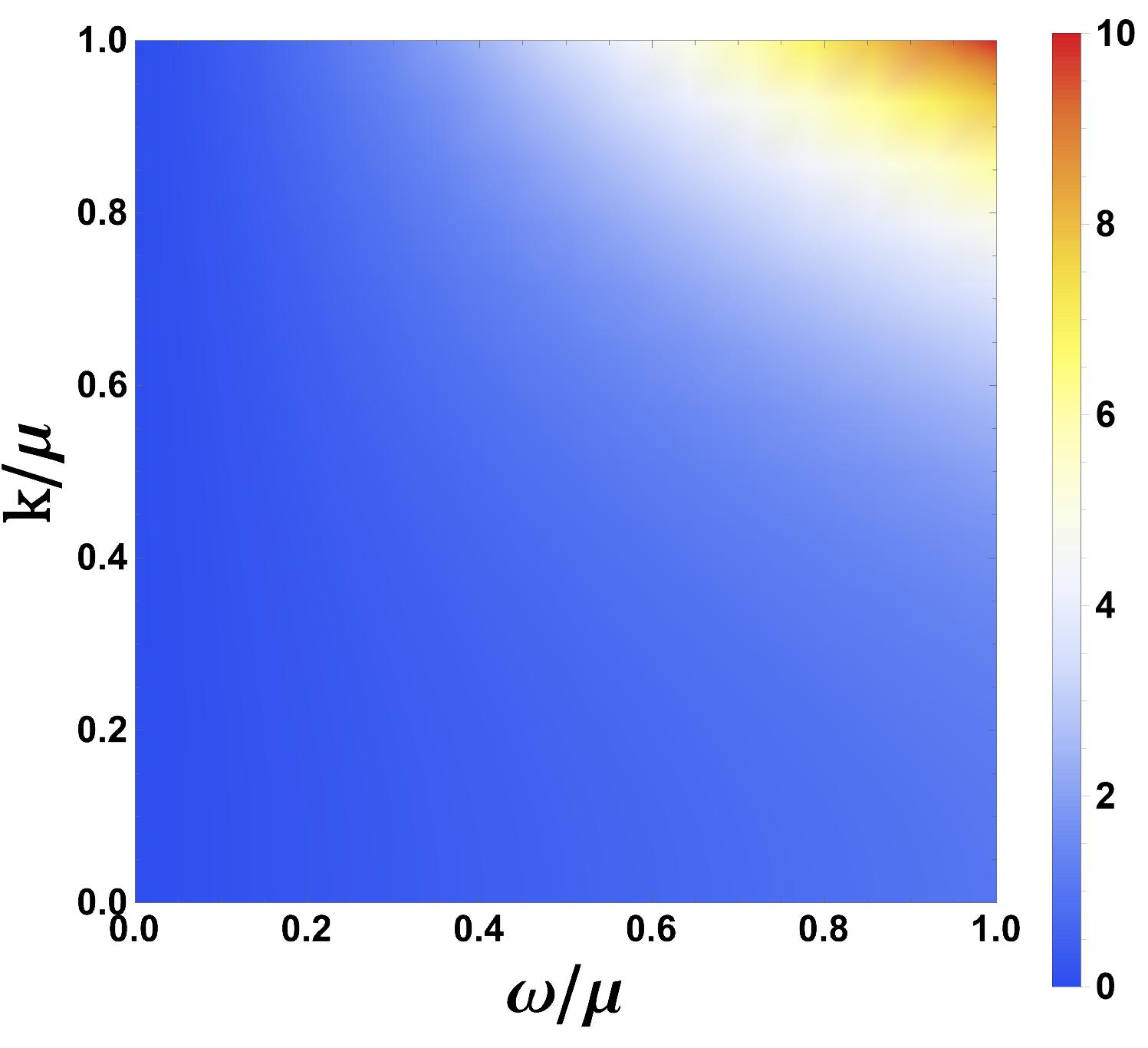}
		\caption{} \label{fig2:Plot_fullGIR_65}
	\end{subfigure}
	\hfill
	\begin{subfigure}{0.32\textwidth}
		\centering
		\includegraphics[width=\textwidth]{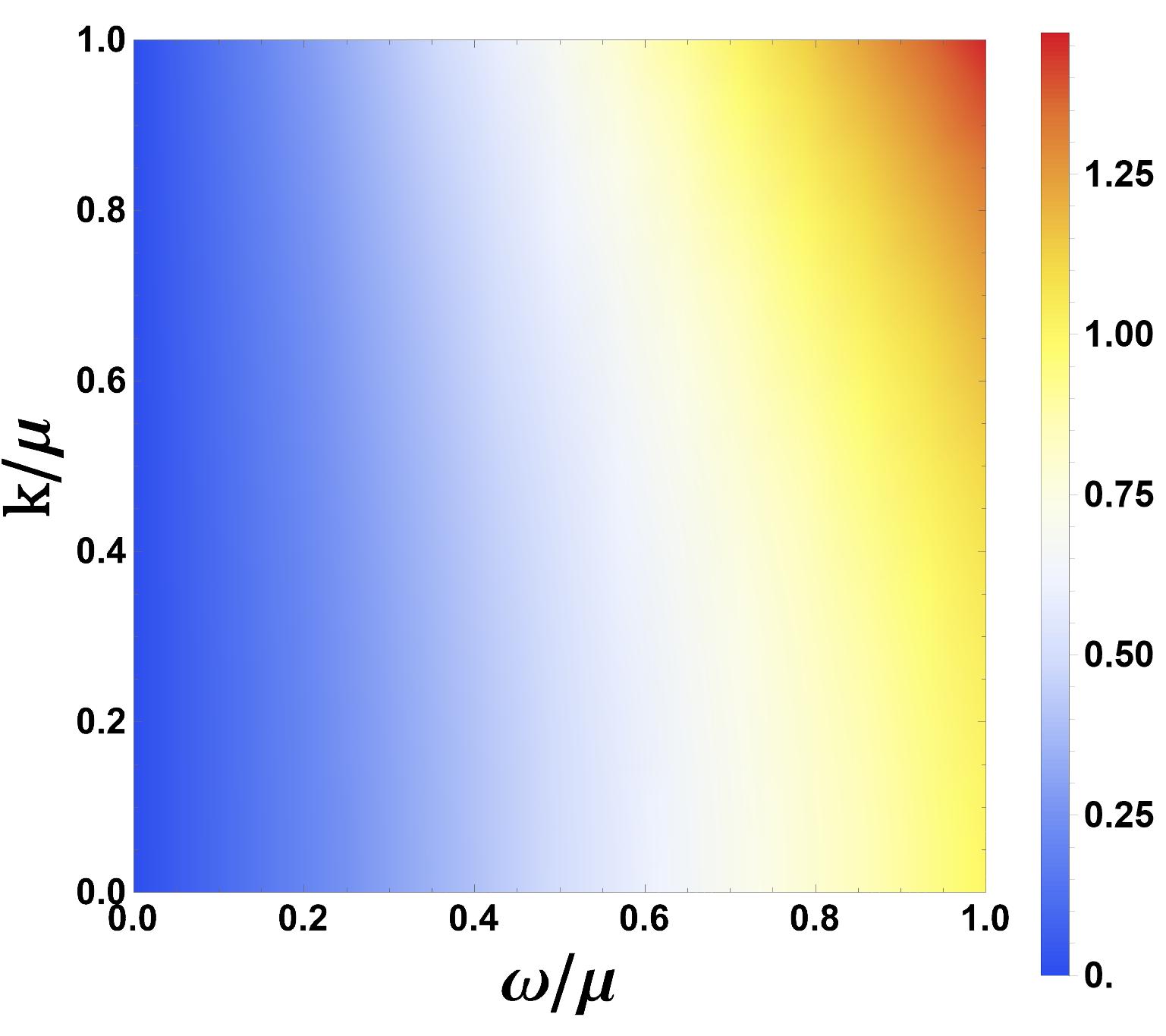}
		\caption{} \label{fig2:Plot_fullGIR_5}
	\end{subfigure}
	\caption{The imaginary part of the full-IR correlator \eqref{eq:trfullapprox} multiplied by $-T/(\mu\,\Sigma)$ as a function of $\omega/\mu$ and $k/\mu$. The isospin chemical potential is set to zero. The background space-time is characterized by $\mu_q/T=10^4$ (panel \ref{fig2:Plot_fullGIR_104}), $\mu_q/T=65$ (panel \ref{fig2:Plot_fullGIR_65}) and $\mu_q/T=5$ (panel \ref{fig2:Plot_fullGIR_5}).} \label{fig2:Plot_fullGIR}
\end{figure}

We start by first  showing the plots of the various approximations of the imaginary part of the polarization functions. In particular, in figure \ref{fig2:Plot_fullGIR} we plot $-\text{Im}\Pi^{\perp}_\text{full-IR}T/(\Sigma\mu)$ as a function of $\omega/\mu$ and $k/\mu$, for  $\mu_q/T=10^4$ (panel \ref{fig2:Plot_fullGIR_104}), $\mu_q/T=65$ (panel \ref{fig2:Plot_fullGIR_65}) and $\mu_q/T=5$ (panel \ref{fig2:Plot_fullGIR_5}). In all these plots $\mu_3= 0$.

The plots show how the magnitude of the imaginary part increases as $\mu_q/T$ grows. For $\mu_q/T=5$ (panel \ref{fig2:Plot_fullGIR_5}), the variation is smooth and the dependence on $\omega/\mu$ is relatively mild across the full range.

For $\mu_q/T=65$ (panel \ref{fig2:Plot_fullGIR_65}), a clear gradient develops toward larger values of both $\omega/\mu$ and $k/\mu$, with the enhancement becoming more localized in the upper-right region of the plot.

For $\mu_q/T=10^4$ (panel \ref{fig2:Plot_fullGIR_104}), this effect is much more pronounced: the largest values are sharply concentrated near $\omega/\mu\sim 1$ and $k/\mu\sim 1$, while the rest of the plot remains comparatively suppressed. This shows that increasing $\mu_q/T$ leads to a stronger growth at larger $\omega/\mu$ and $k/\mu$, together with a progressive localization of the response in that region. The normalization by $\mu$ is introduced only to keep the overall scale comparable between panels.

The growth of the correlator with increasing $\mu_q/T$ has a clear physical origin. At fixed $\omega/\mu$, increasing $\mu_q/T$ effectively increases $\omega/T$, so the system is probed at higher frequencies relative to the temperature. As a result, the regime where $\omega \ll T$ (where the response is smooth and approximately linear in $\omega$) becomes progressively smaller in $\omega/\mu$. For large $\mu_q/T$, most of the plotted region instead corresponds to $\omega \gtrsim T$, where the response is no longer controlled by thermal effects but by the intrinsic IR scaling of the system. This leads to a stronger dependence on $\omega$ and $k$, producing the observed enhancement in the plots. In this sense, the growth reflects the transition from a thermally dominated regime to a regime governed by the underlying IR dynamics.

\begin{figure}[h]
	\centering

	\begin{subfigure}{0.45\textwidth}
		\centering
		\includegraphics[width=\textwidth]{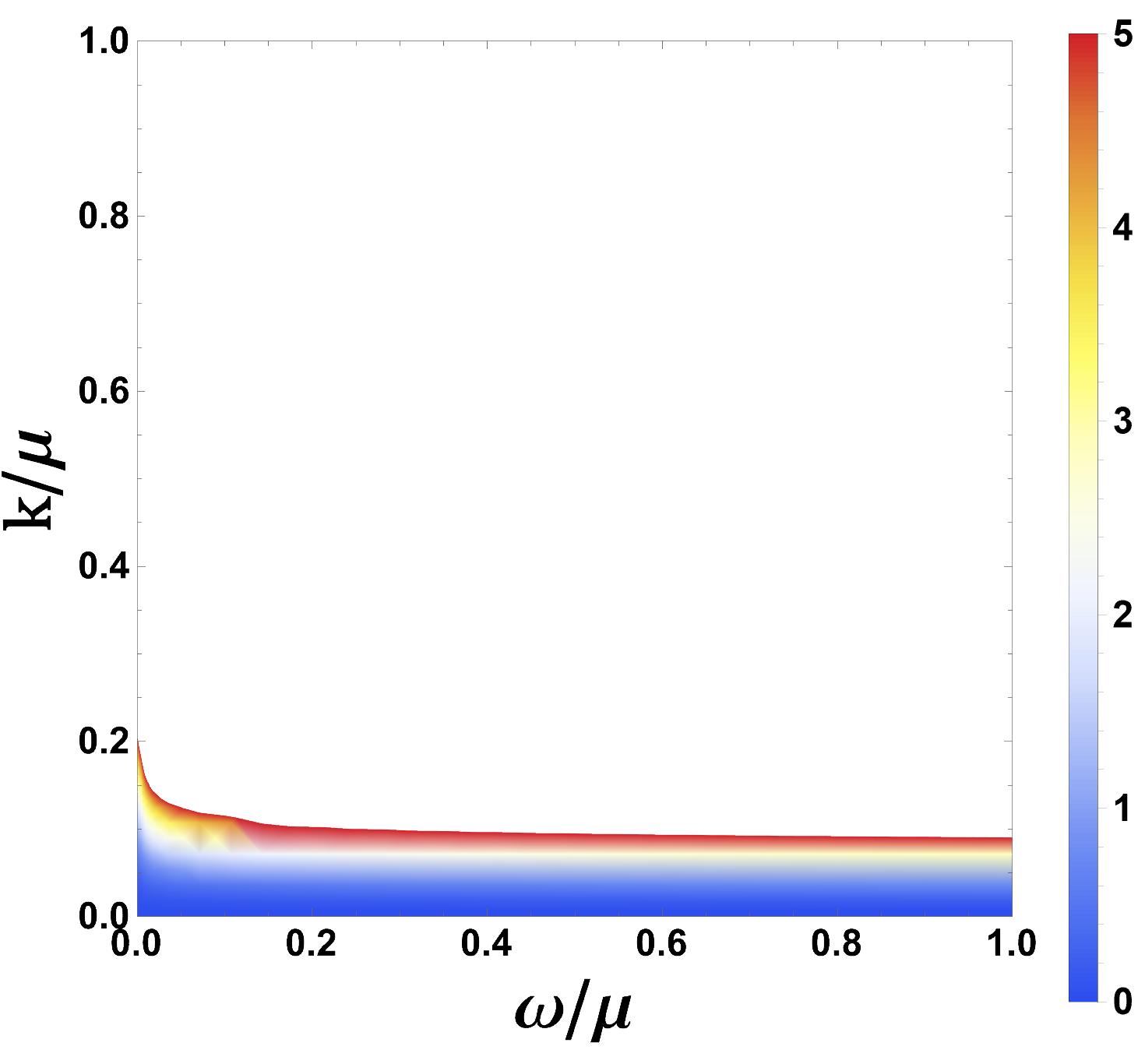}
		\caption{} \label{fig2:RelDiff_fullGIR_vs_hydro_104}
	\end{subfigure}
	\hfill
	\begin{subfigure}{0.45\textwidth}
		\centering
		\includegraphics[width=\textwidth]{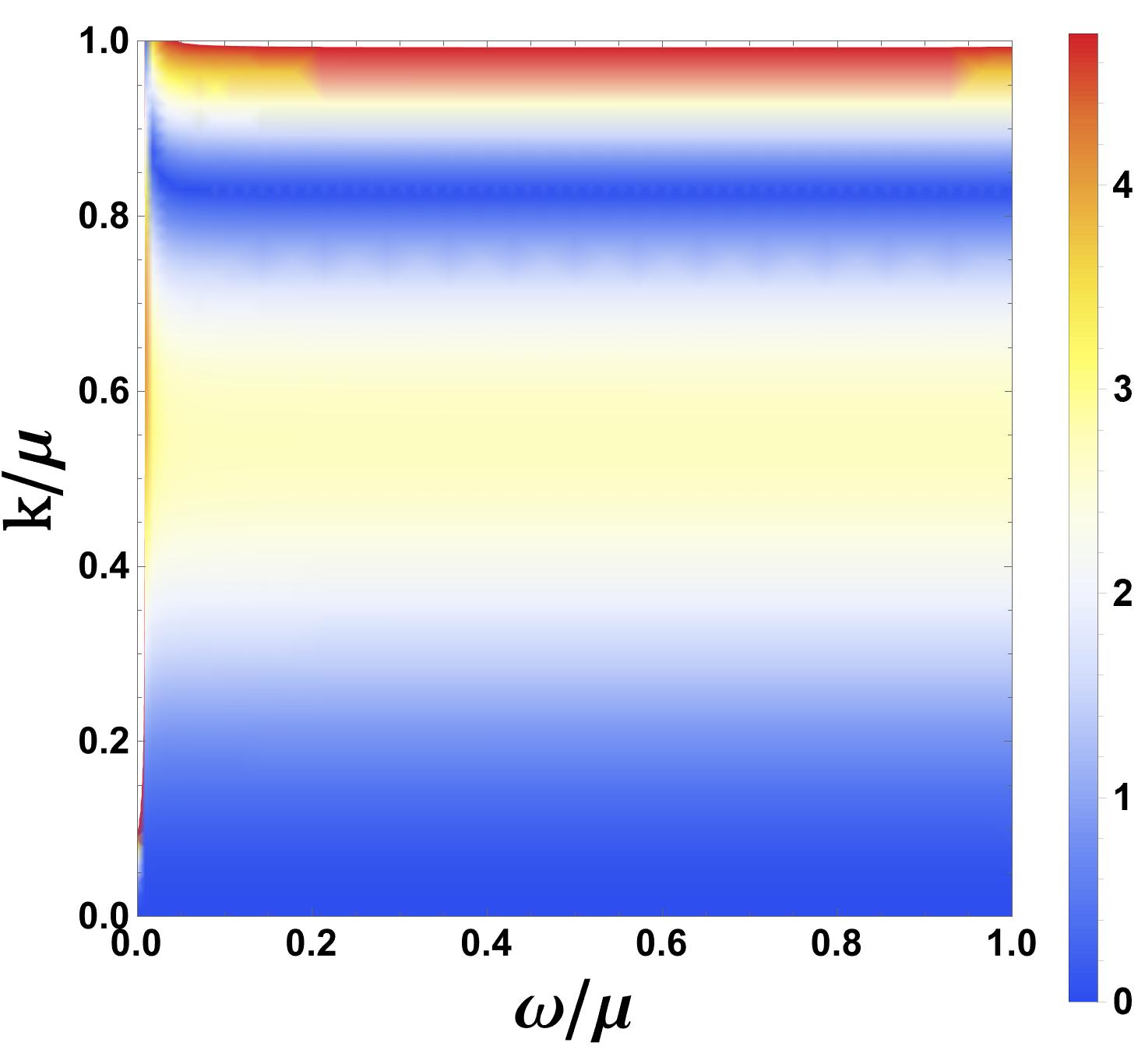}
		\caption{} \label{fig2:RelDiff_fullGIR_vs_exthydro_104}
	\end{subfigure}
	\hfill

	\begin{subfigure}{0.45\textwidth}
		\centering
		\includegraphics[width=\textwidth]{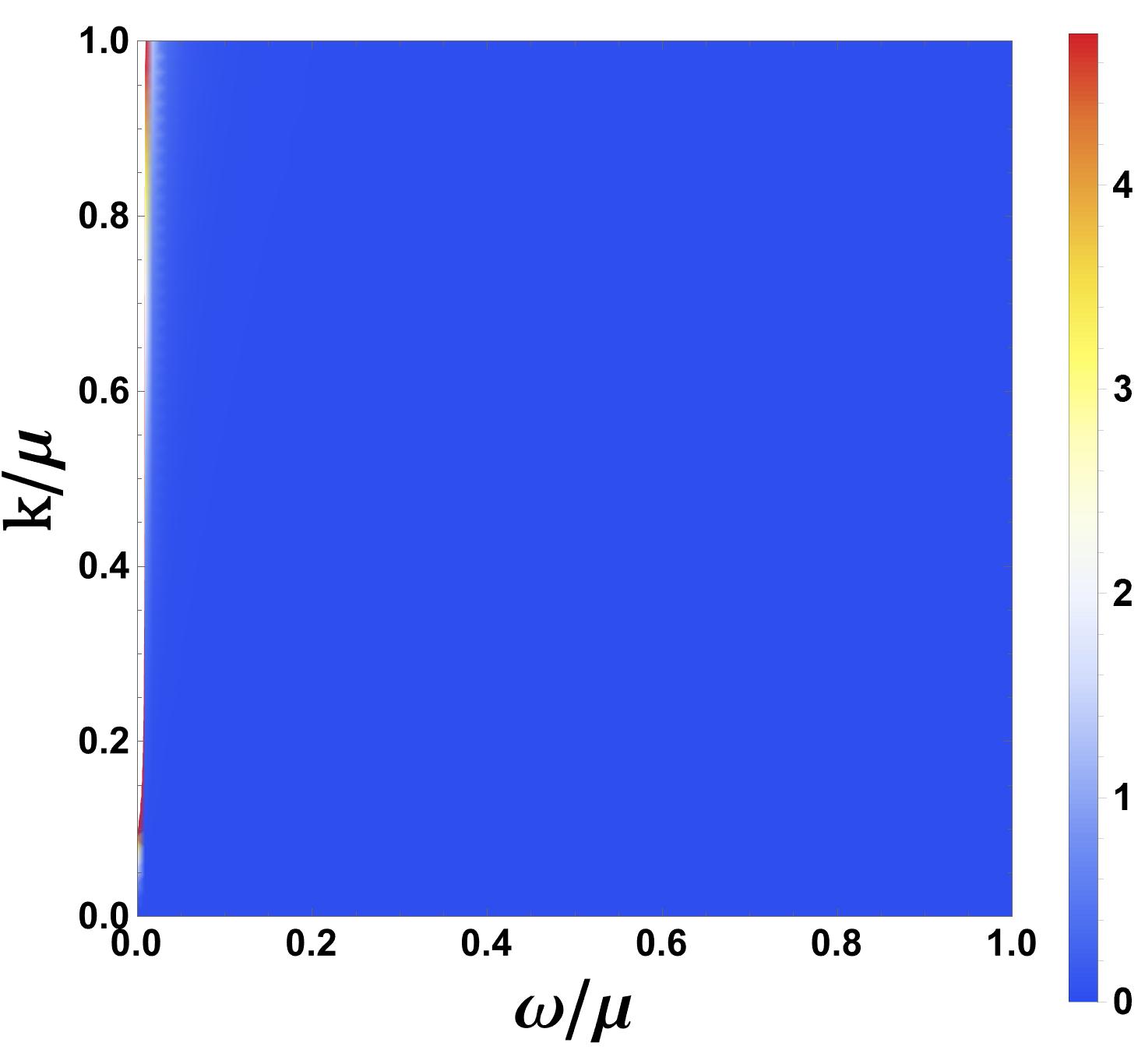}
		\caption{} \label{fig2:RelDiff_fullGIR_vs_improvedexthydro_104}
	\end{subfigure}
	\hfill
	\begin{subfigure}{0.45\textwidth}
		\centering
		\includegraphics[width=\textwidth]{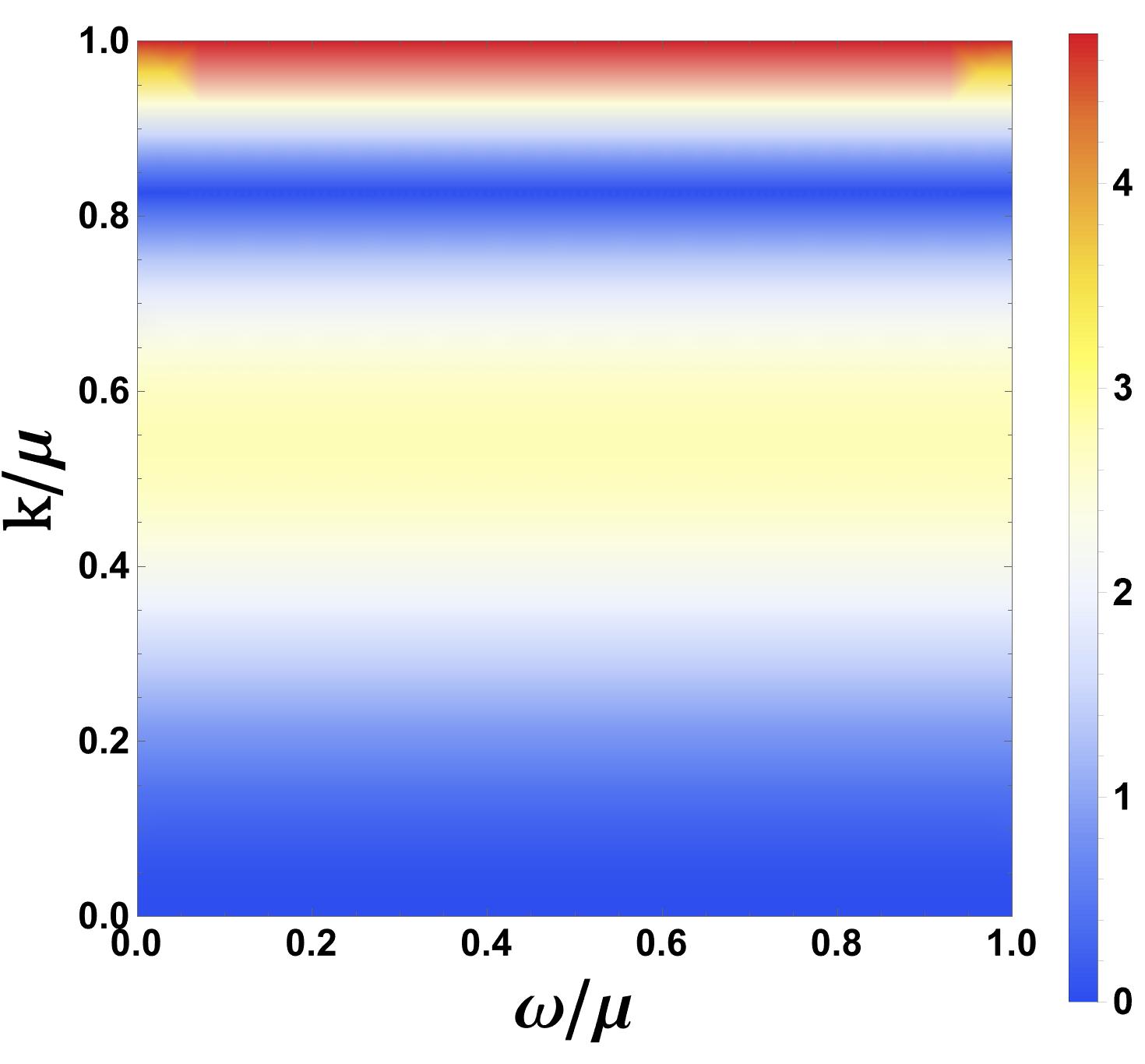}
		\caption{} \label{fig2:RelDiff_fullGIR_vs_normGIR_104}
	\end{subfigure}
	\caption{The relative difference \eqref{eq:percIR}. This is defined as 100 times the absolute value of the difference between the imaginary part of the full IR-AdS$_2$ correlator and the hydrodynamic approximation (panel \ref{fig2:RelDiff_fullGIR_vs_hydro_104}) the extended hydrodynamic approximation (panel \ref{fig2:RelDiff_fullGIR_vs_exthydro_104}) the improved extended hydrodynamic approximation (panel \ref{fig2:RelDiff_fullGIR_vs_improvedexthydro_104}) and the imaginary part of the normalized IR-AdS$_2$ correlator (panel \ref{fig2:RelDiff_fullGIR_vs_normGIR_104}). The background is characterized by $\mu_q/T=10^4$ and $\mu_3=0$.} \label{fig2:Rel_corr_104}
\end{figure}

\begin{figure}[h]
	\centering

	\begin{subfigure}{0.45\textwidth}
		\centering
		\includegraphics[width=\textwidth]{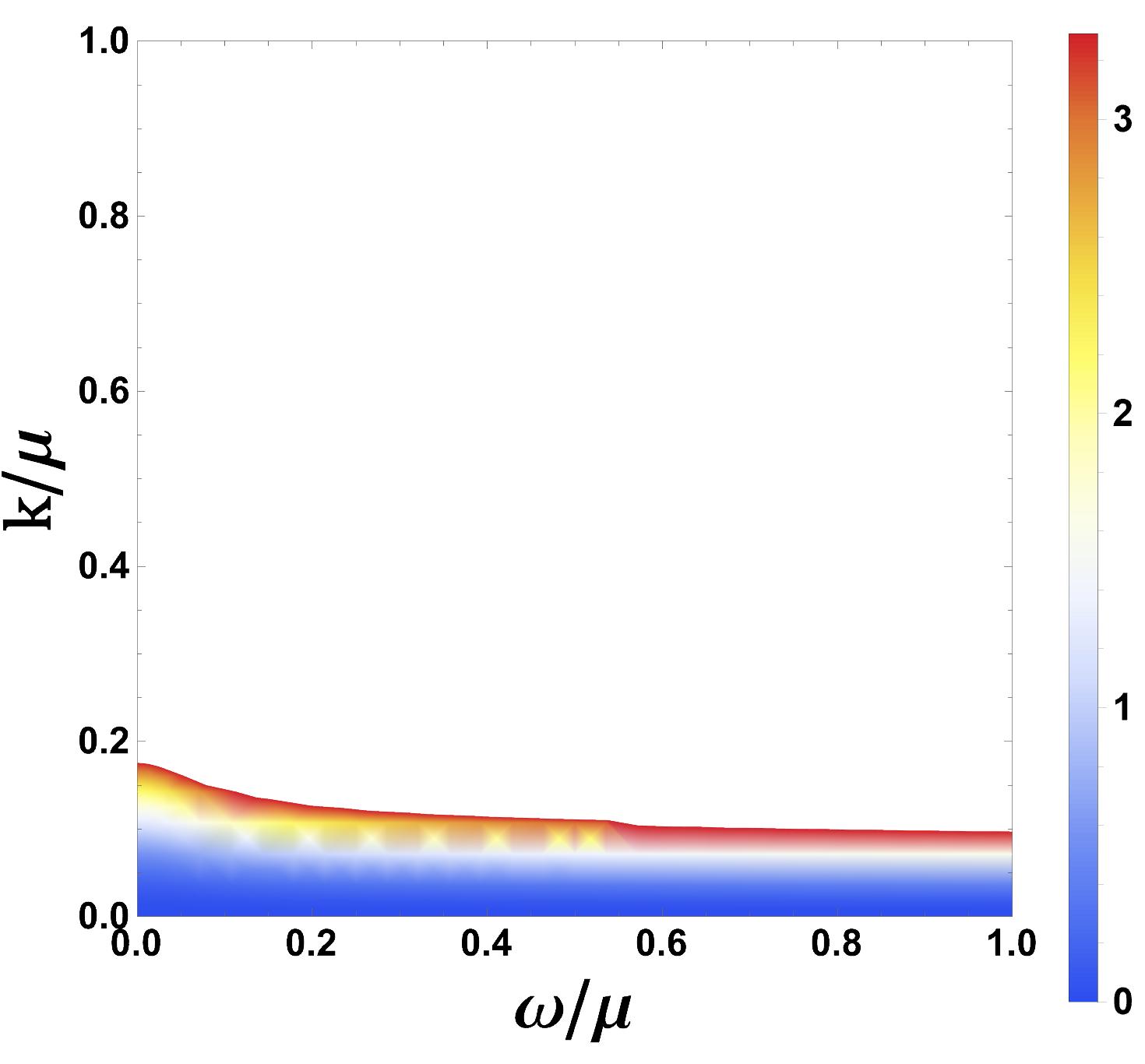}
		\caption{} \label{fig2:RelDiff_fullGIR_vs_hydro_65}
	\end{subfigure}
	\hfill
	\begin{subfigure}{0.45\textwidth}
		\centering
		\includegraphics[width=\textwidth]{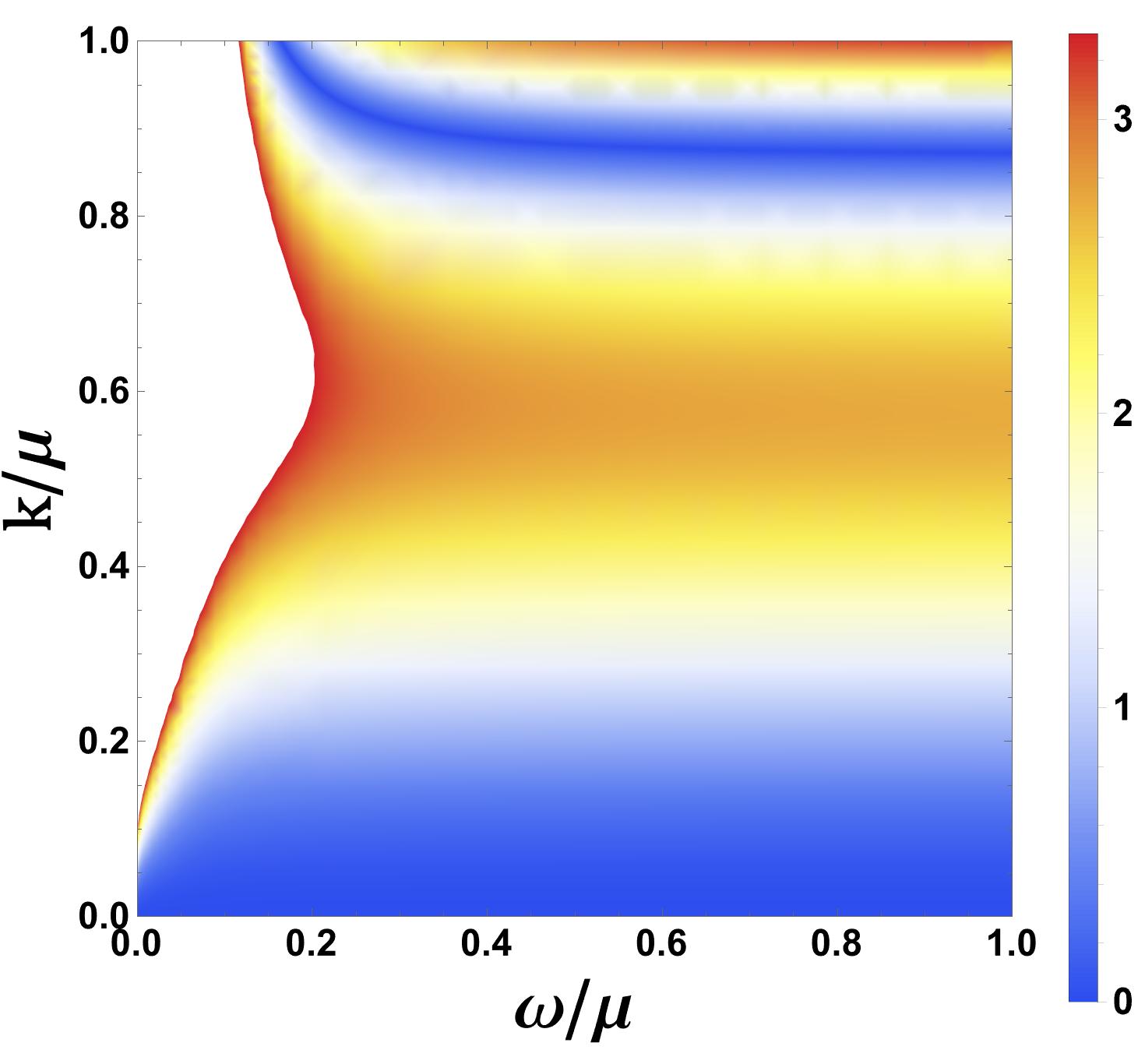}
		\caption{} \label{fig2:RelDiff_fullGIR_vs_exthydro_65}
	\end{subfigure}
	\hfill

	\begin{subfigure}{0.45\textwidth}
		\centering
		\includegraphics[width=\textwidth]{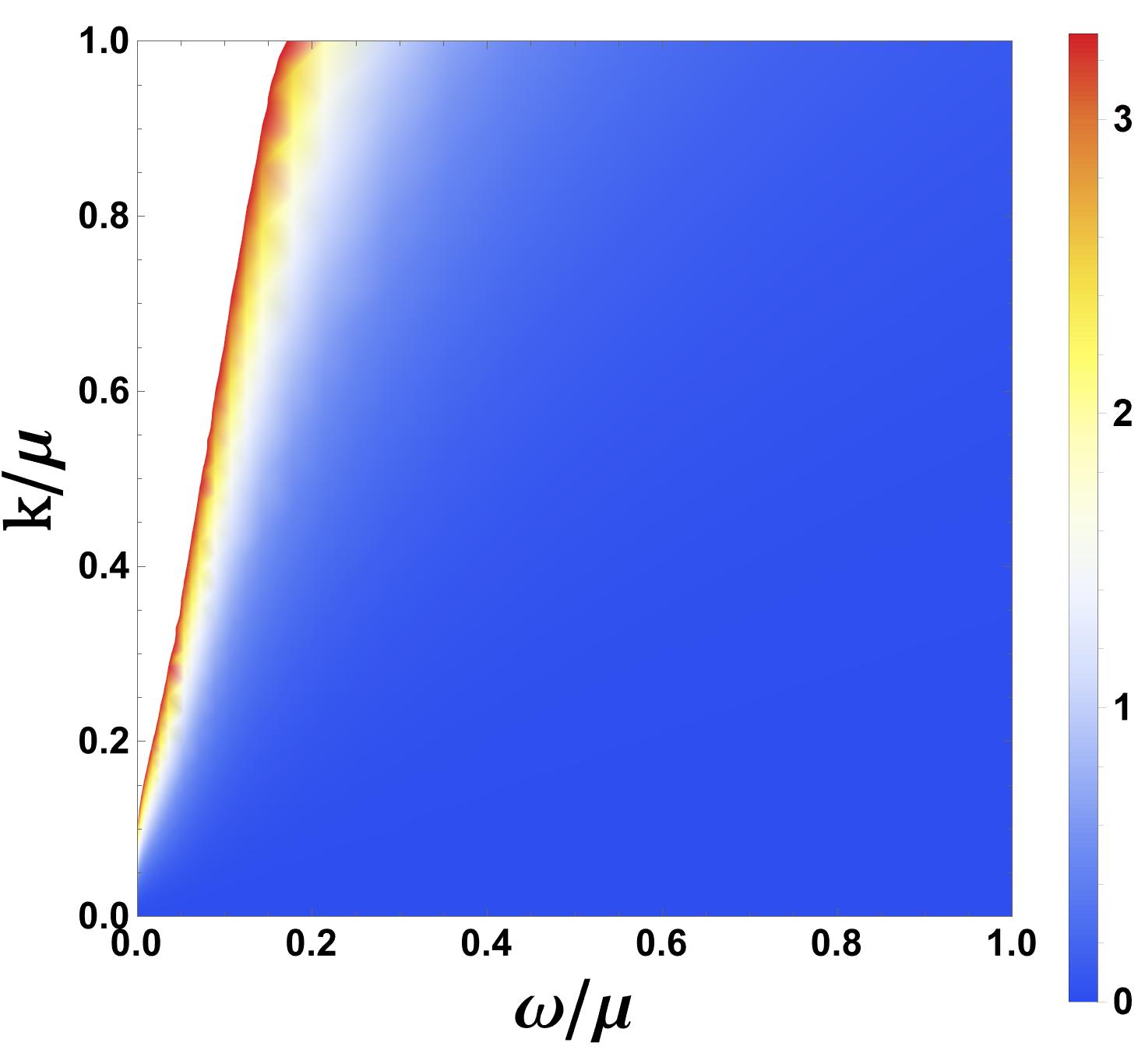}
		\caption{} \label{fig2:RelDiff_fullGIR_vs_improvedexthydro_65}
	\end{subfigure}
	\hfill
	\begin{subfigure}{0.45\textwidth}
		\centering
		\includegraphics[width=\textwidth]{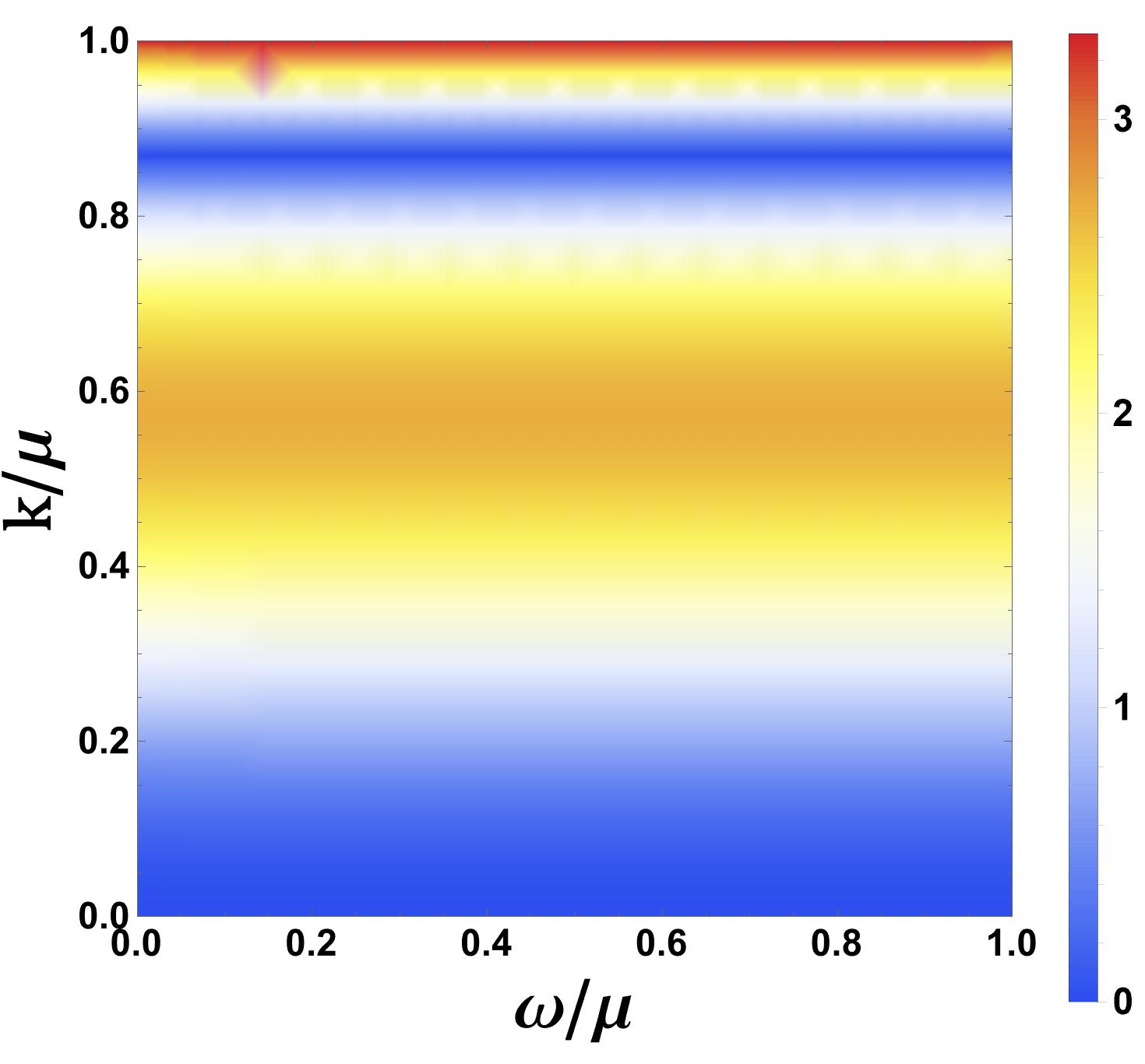}
		\caption{} \label{fig2:RelDiff_fullGIR_vs_normGIR_65}
	\end{subfigure}

	\caption{The relative difference \eqref{eq:percIR}. This is defined as 100 times the absolute value of the difference between the imaginary part of the full IR-AdS$_2$ correlator and the hydrodynamic approximation (panel \ref{fig2:RelDiff_fullGIR_vs_hydro_65}) the extended hydrodynamic approximation (panel \ref{fig2:RelDiff_fullGIR_vs_exthydro_65}) the improved extended hydrodynamic approximation (panel \ref{fig2:RelDiff_fullGIR_vs_improvedexthydro_65}) the imaginary part of the normalized IR-AdS$_2$ correlator (panel \ref{fig2:RelDiff_fullGIR_vs_normGIR_65}). The background is characterized by $\mu_q/T=65$ and $\mu_3=0$.} \label{fig2:Rel_corr_65}
\end{figure}

\begin{figure}[h]
	\centering
	\begin{subfigure}{0.45\textwidth}
		\centering
		\includegraphics[width=\textwidth]{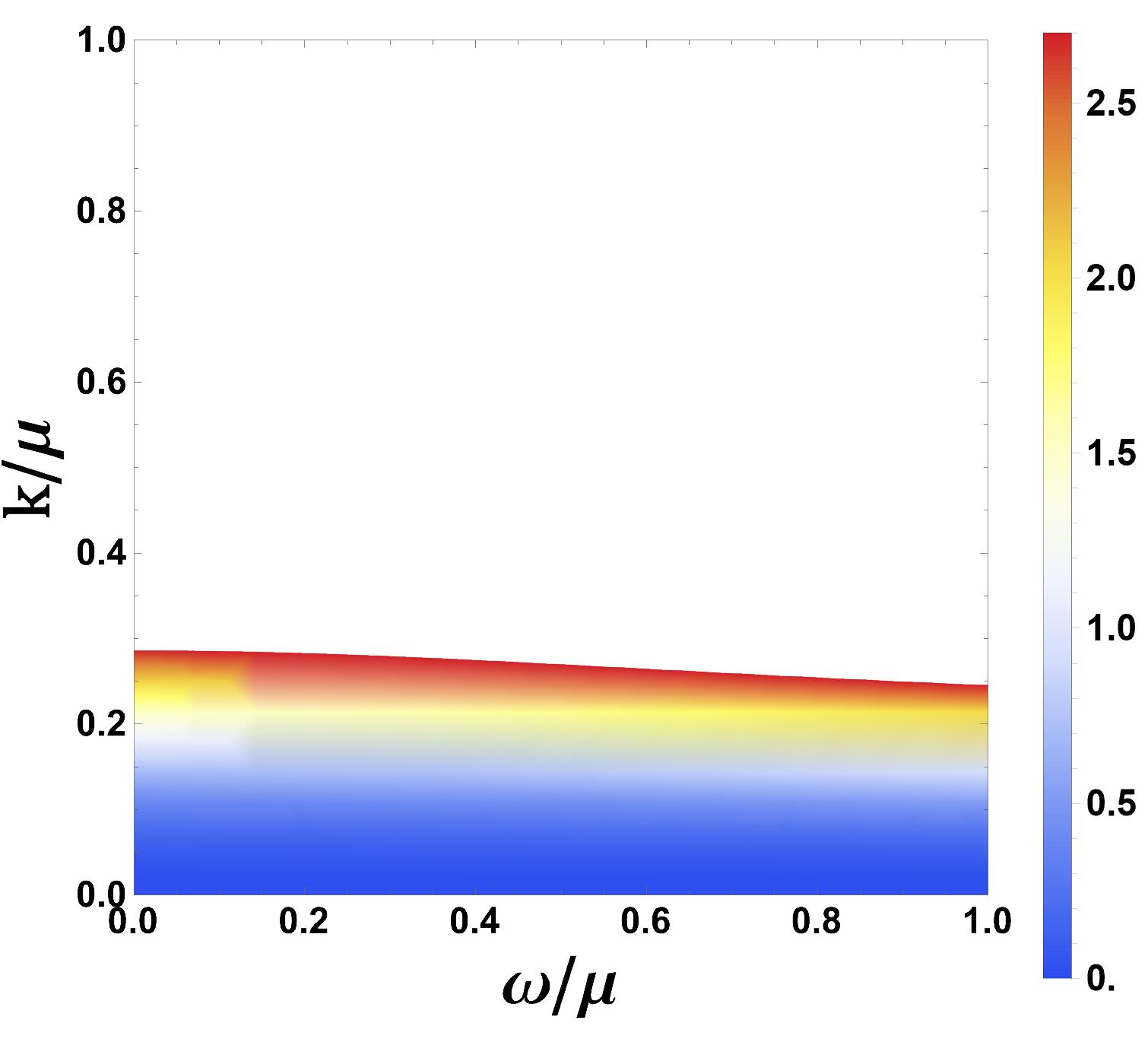}
		\caption{} \label{fig2:RelDiff_fullGIR_vs_hydro_5}
	\end{subfigure}
	\hfill
	\begin{subfigure}{0.45\textwidth}
		\centering
		\includegraphics[width=\textwidth]{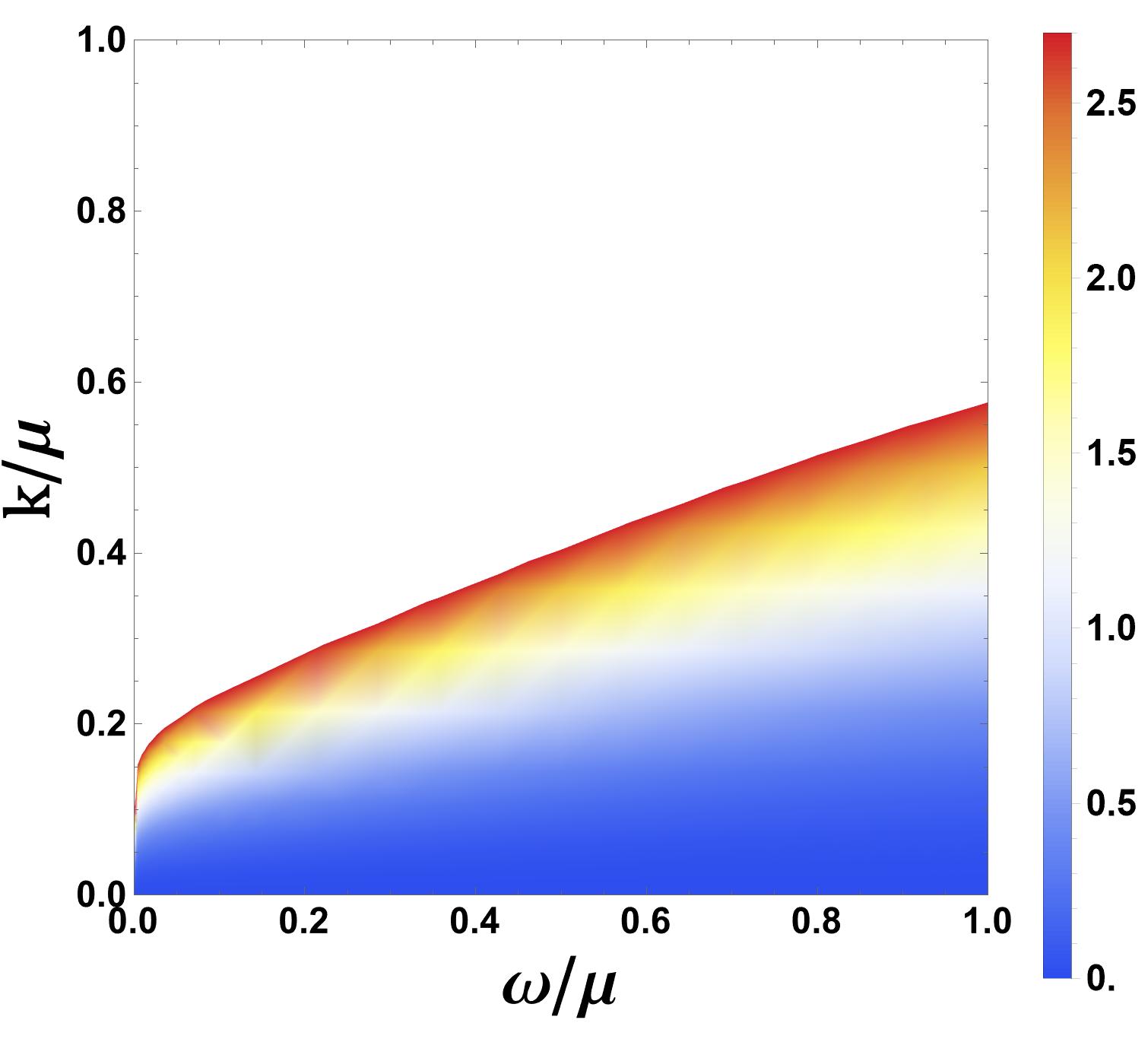}
		\caption{} \label{fig:RelDiff_fullGIR_vs_exthydro_5}
	\end{subfigure}
	\hfill

	\begin{subfigure}{0.45\textwidth}
		\centering
		\includegraphics[width=\textwidth]{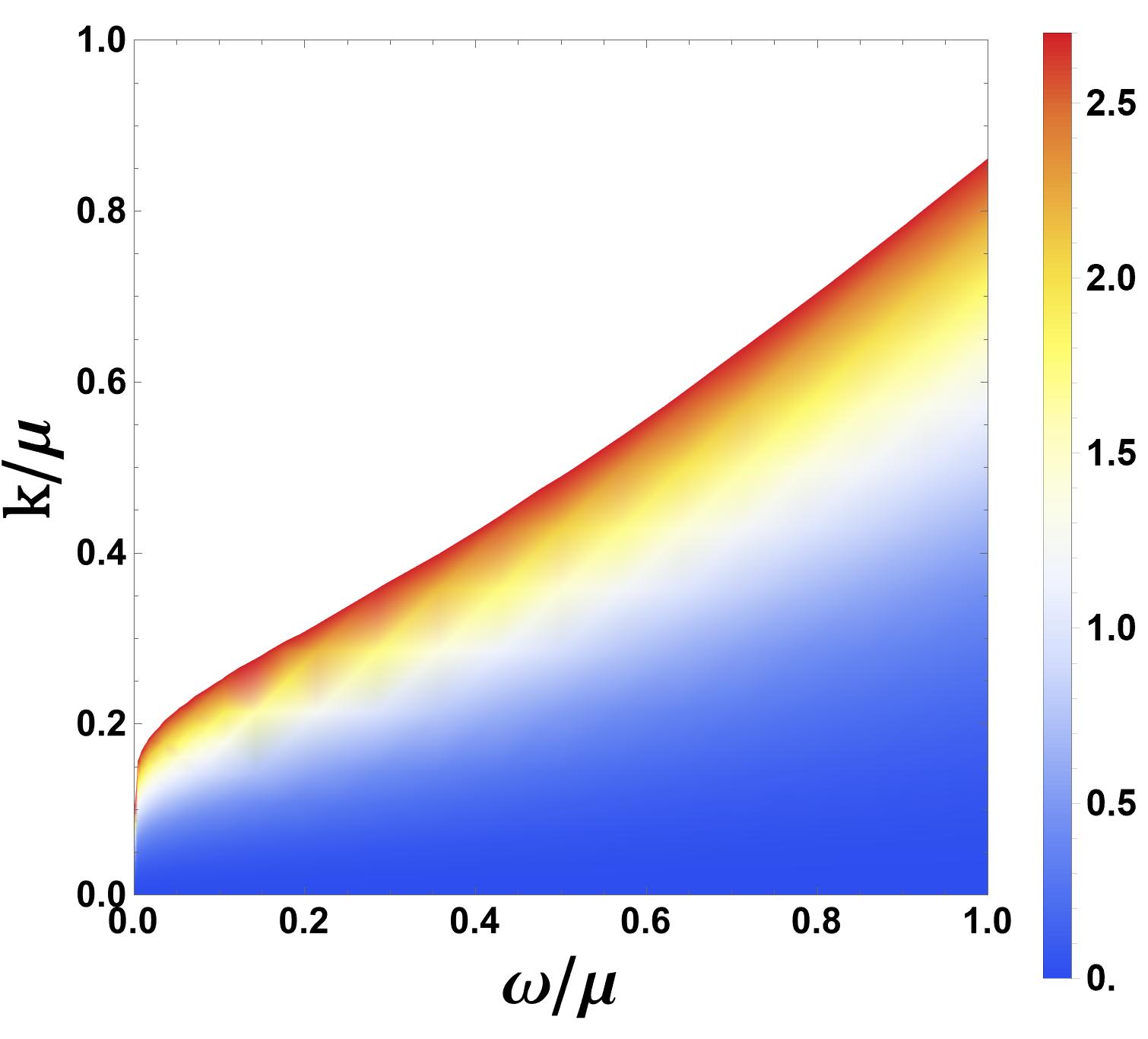}
		\caption{} \label{fig:RelDiff_fullGIR_vs_improvedexthydro_5}
	\end{subfigure}
	\hfill
	\begin{subfigure}{0.45\textwidth}
		\centering
		\includegraphics[width=\textwidth]{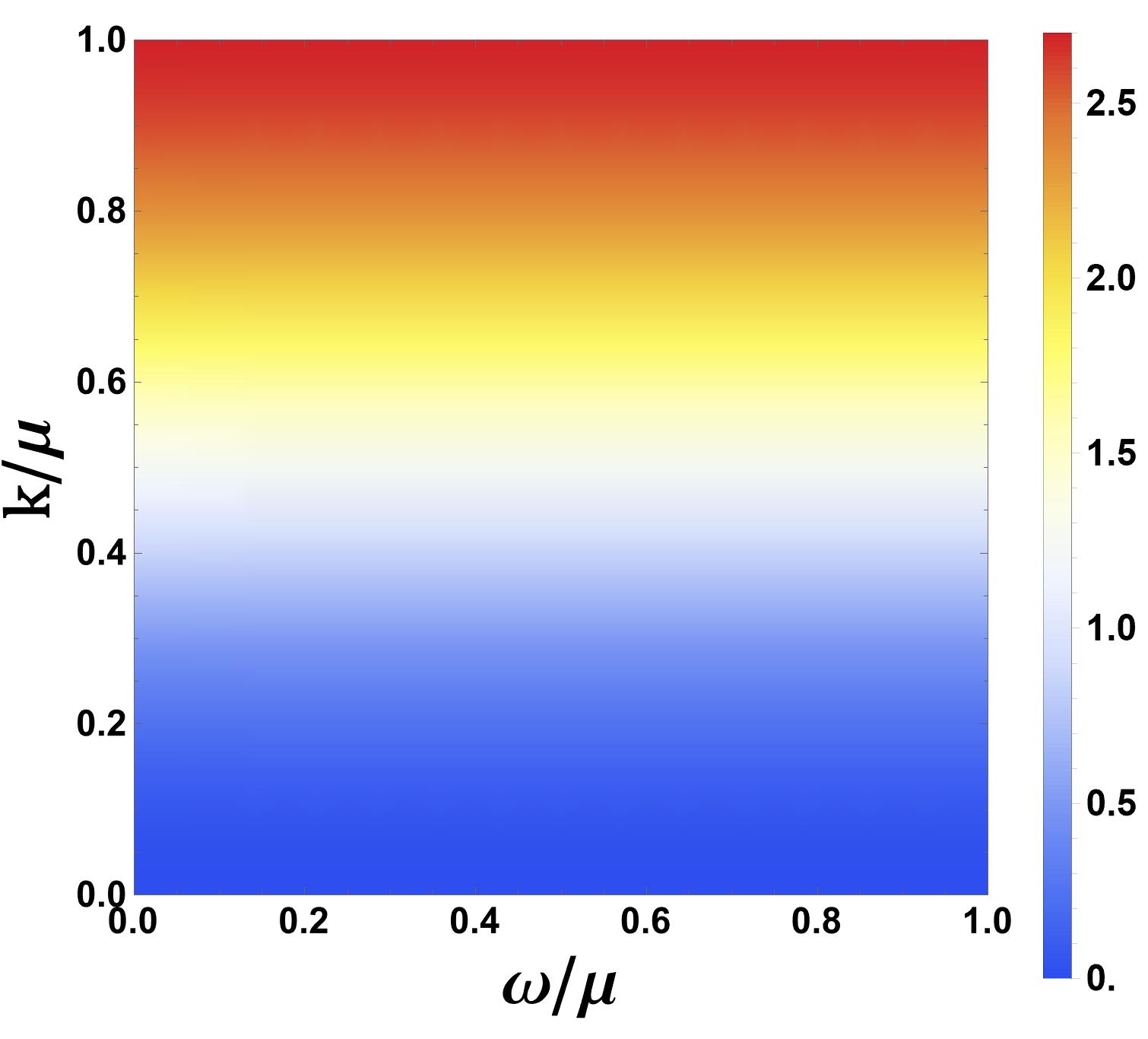}
		\caption{} \label{fig:RelDiff_fullGIR_vs_GIRnorm_5}
	\end{subfigure}

	\caption{The relative difference \eqref{eq:percIR}. This is defined as 100 times the absolute value of the difference between the imaginary part of the full IR-AdS$_2$ correlator and the hydrodynamic approximation (panel \ref{fig2:RelDiff_fullGIR_vs_hydro_5}) the extended hydrodynamic approximation (panel \ref{fig:RelDiff_fullGIR_vs_exthydro_5}) the improved extended hydrodynamic approximation (panel \ref{fig:RelDiff_fullGIR_vs_improvedexthydro_5}) the imaginary part of the normalized IR-AdS$_2$ correlator (panel \ref{fig:RelDiff_fullGIR_vs_GIRnorm_5}). The background is characterized by $\mu_q/T=5$ and $\mu_3=0$.} \label{fig2:Rel_corr_5}
\end{figure}

Then, we compare the exact IR-AdS$_2$ \eqref{eq:qq3} result with the other approximations outlined in this section, by plotting the percentage relative difference with respect to it. In particular, we plot
\be\label{eq:percIR}
\text{relative difference} = 100\bigg|1- \dfrac{\text{Im}\Pi^\perp_i}{\text{Im}\Pi^\perp_\text{full-IR}}\bigg|\,,
\ee
expressed in percentage, where $i$ runs over the other approximations $\{$hydro, ext-hydro, improved, norm$\}$. We plot this percentage relative difference in figures \ref{fig2:Rel_corr_104}, for $\mu_q/T = 10^4$, figure \ref{fig2:Rel_corr_65}, for $\mu_q/T=65$ and figure \ref{fig2:Rel_corr_5}, for $\mu_q/T=5$. In all the plots $\mu_3=0$, and the quantities are functions of $\omega/\mu$ and $k/\mu$ in the extended hydrodynamic regime, which extends over  $\omega/\mu\in\{0,1\}$ and $k/\mu\in\{0,1\}$.

For $\mu_q/T=10^4$, the relative-difference plots make clear that the hydrodynamic approximation only approximate well the full IR correlator in a very small corner of the larger parameter space. In panel \ref{fig2:RelDiff_fullGIR_vs_hydro_104}, which compares the full IR-AdS$_2$ result with the linear hydrodynamic approximation, the plot is almost entirely white because at such large $\mu_q/T$ one has $\omega/T=10^4\omega/\mu$, so even moderate values of $\omega/\mu$ already correspond to $\omega/T\gg1$; the truly hydrodynamic region is therefore compressed into a tiny neighborhood of $\omega/\mu\simeq0$, where the error becomes small.

Panel \ref{fig2:RelDiff_fullGIR_vs_exthydro_104} shows that the extended hydrodynamic approximation captures the overall $\omega$-dependence much better, but still exhibits a sizable mismatch that depends mainly on $k/\mu$, visible as nearly horizontal bands; this reflects the fact that the power law in $\omega/\mu$ is correct while its overall coefficient is not.

Panel \ref{fig2:RelDiff_fullGIR_vs_improvedexthydro_104} confirms that including the improved large-$\omega/T$ coefficient gives a much better approximation over most of the plotted region, with the remaining discrepancy localized near very small $\omega/\mu$, where thermal effects are still important.

By contrast, panel \ref{fig2:RelDiff_fullGIR_vs_normGIR_104} remains non-negligible over the whole plane because the normalized correlator is constructed to have unit coefficient in the asymptotic power-law regime, and therefore does not reproduce the exact overall normalization of the full IR-AdS$_2$ correlator at finite $T$. In this sense, the figure \ref{fig2:Rel_corr_104} shows that for very large $\mu_q/T$ the improved extended hydrodynamic approximation is the closest one to the full IR-AdS$_2$ result across the extended hydrodynamic region.

For smaller values of $\mu_q/T$, shown in figures \ref{fig2:Rel_corr_65} and \ref{fig2:Rel_corr_5}, the qualitative picture continues to hold. However, the top-right panels \ref{fig2:RelDiff_fullGIR_vs_hydro_65}, and \ref{fig2:RelDiff_fullGIR_vs_hydro_5} show that the hydrodynamic approximation reproduces better the full IR-AdS$_2$ correlator for larger values of $\omega/\mu$ and $k/\mu$. Moreover, the improved extended hydrodynamic approximation (see panels \ref{fig2:RelDiff_fullGIR_vs_improvedexthydro_65}, \ref{fig:RelDiff_fullGIR_vs_improvedexthydro_5}) still works well but not as good as at $\mu_q/T=10^4$ since now the region $\omega/T\gg1$ shows up at larger $\omega/\mu$.

\clearpage

\section{Computation of the residues of the IR-AdS$_2$ poles}\label{app:residues}
This appendix is devoted to the detailed computation of the residues associated with the AdS$_2$ poles of the correlator \eqref{eq:q3}. For the reader's convenience,
 we report here the expression of the IR-AdS$_2$ correlator
\begin{align}\label{eq:app:q3}
	\GG^\pm_\text{IR}(\omega, k) =&(4\pi T)^{2\D( k,\mu_3)-1}\frac{\Gamma(1\!-\!2\D(k,\mu_3))\Gamma(\D(k,\mu_3)\!-\!\frac{i\omega}{2\pi T}\pm \frac{i}{6}r_H\mu_3)}{\Gamma(2\D(k,\mu_3)\!-\!1)\Gamma(1\!-\!\D(k,\mu_3)\!-\!\frac{i\omega}{2\pi T}\pm \frac{i}{6}r_H\mu_3)}\times\\ \non
	&\times\dfrac{\Gamma(\D(k,\mu_3)\mp \frac{i}{6}r_H\mu_3)}{\Gamma(1\!-\!\D(k,\mu_3)\mp \frac{i}{6}r_H\mu_3)},
\end{align}
with $\D(k,\mu_3)$ the IR conformal dimension computed from the near-boundary ($\zeta\to\infty$) asymptotics of \eqref{eq:q2}
\begin{equation}
	\D(k,\mu_3) = \frac{1}{2} + \frac{1}{2}\sqrt{1+\frac{1}{3}r_H^2 k^2 - \frac{1}{9}r_H^2\mu_3^2} \, .
\end{equation}

As explained in section \ref{sec:IRAdS2CORR}, the AdS$_2$ poles then correspond to the poles of the second $\Gamma$-function in the numerator of \eqref{eq:app:q3}. They are given by
\begin{equation}
	\label{eq:app:q5} \omega^{\pm,(n)}_{\text{AdS}_2} = -i2\pi T(\D(k,\mu_3) + n) \pm \frac{1}{3} r_H \pi T \mu_3 \sp n \in \mathbb{N} \, ,
\end{equation}
where the superscript $(n)$ corresponds to the $n^\text{th}$ pole.

To characterize the strength of the poles in \eqref{eq:app:q5}, we compute their residues. For a simple pole at $\omega=\omega^{\pm,(n)}_{\text{AdS}_2}$, the residue is defined by
\be
\underset{\omega=\omega^{\pm,(n)}_{\text{AdS}_2}}{\mathrm{Res}}\,
\GG^\pm_\text{IR}(\omega,k)
=
\lim_{\omega\to\omega^{\pm,(n)}_{\text{AdS}_2}}
\left(\omega-\omega^{\pm,(n)}_{\text{AdS}_2}\right)
\GG^\pm_\text{IR}(\omega,k)\,.
\ee
as the finite coefficient that remains after multiplying the correlator by $\omega-\omega^{\pm,(n)}_{\text{AdS}_2}$ and taking the limit $\omega\to\omega^{\pm,(n)}_{\text{AdS}_2}$.

For generic values of $\D(k,\mu_3)$, the Gamma function in the denominator of \eqref{eq:app:q3} is finite at the frequencies \eqref{eq:app:q5}. At such a frequency, the Gamma function in the numerator satisfies
\be
\lim_{\omega\to\omega^{\pm,(n)}_{\text{AdS}_2}}
\left(\omega-\omega^{\pm,(n)}_{\text{AdS}_2}\right)
\Gamma\left(
\D(k,\mu_3)
-\frac{i\omega}{2\pi T}
\pm\frac{i}{6}r_H\mu_3
\right)
=
\frac{(-1)^n}{n!}\,i\,2\pi T\,.
\ee
Substituting this result into \eqref{eq:app:q3} gives
\begin{align}\non
	\underset{\omega=\omega^{\pm,(n)}_{\text{AdS}_2}}{\mathrm{Res}}\,
	\GG^\pm_\text{IR}(\omega,k)
	=
	&i\,2\pi T\,(4\pi T)^{2\D(k,\mu_3)-1}
	\frac{
		(-1)^n\Gamma\left(1-2\D(k,\mu_3)\right)
	}{
		n!\,\Gamma\left(2\D(k,\mu_3)-1\right)
		\Gamma\left(1-2\D(k,\mu_3)-n\right)
	}\times
	\\ \label{eq:q6}
	&\times\frac{
		\Gamma\left(\D(k,\mu_3)\mp\frac{i}{6}r_H\mu_3\right)
	}{
		\Gamma\left(1-\D(k,\mu_3)\mp\frac{i}{6}r_H\mu_3\right)
	}\,.
\end{align}
The generic residues do determine the complete pole-part of the IR-AdS$_2$ correlator \eqref{eq:app:q3}. More precisely, for generic $\D(k,\mu_3)$ the correlator may be represented as a sum over the poles \eqref{eq:app:q5}, supplemented by a function analytic in $\omega$. This analytic contribution is not fixed by the pole data; in holography, its scheme-dependent part corresponds to local contact terms.

A direct, unsubtracted sum over the poles, reproducing the full IR-AdS$_2$ correlator converges only for Re$\Delta(k,\mu_3) <1/2$. For Re$\Delta(k,\mu_3) \geq 1/2$, the same representation must be understood through analytic continuation, or equivalently as a subtracted Mittag--Leffler expansion. The required subtractions modify only the analytic part of the correlator and hence do not affect either the pole positions or the residues derived above.

In the following, we discuss special cases when $2\D(k,\mu_3)-1$ is a non-negative integer, for which the residues formula \eqref{eq:q6} require separate treatment. When $\D(k,\mu_3)=1/2$,\footnote{Note that this value is not physical since it could be reached only if
	\begin{equation}
		1+\frac{1}{3}r_H^2 k^2 - \frac{1}{9}r_H^2\mu_3^2=0\,,
	\end{equation}
	which can never be satisfied by a real $\mu_3$ for the value of $w_0$ used in the present work. It is however interesting to show what happens to the IR-AdS$_2$ correlator in this unphysical case.} the two Gamma functions that depend on $\omega$ in \eqref{eq:app:q3} have identical arguments and cancel. In addition,
\be
\lim_{\D(k,\mu_3)\to1/2}
\frac{\Gamma\left(1-2\D(k,\mu_3)\right)}
{\Gamma\left(2\D(k,\mu_3)-1\right)}
=-1\,.
\ee
The last ratio of Gamma functions in \eqref{eq:app:q3} tends to one. Therefore, in the normalization of \eqref{eq:app:q3},
\be
\lim_{\D(k,\mu_3)\to1/2}
\GG^\pm_\text{IR}(\omega,k)
=-1\,,
\ee
and there is no tower of frequency poles.

For the higher resonant values
\begin{equation}\label{eq:q7}
	2\D(k,\mu_3)-1=m\,,\qquad m\in\mathbb{N}\,,
\end{equation}
the expression \eqref{eq:q6} requires a separate treatment. These values do not signal additional bulk modes. Rather, they correspond to a resonance of the two independent near-boundary solutions of \eqref{eq:q2}. Indeed, for $\zeta\to\infty$ (boundary of the infrared AdS$_2$ region) these solutions behave as
\begin{equation}
	y(\zeta)=A\,\zeta^{\D(k,\mu_3)-1}\left(1+\cdots\right)
	+B\,\zeta^{-\D(k,\mu_3)}\left(1+\cdots\right)\,,
	\label{eq:q8}
\end{equation}
with $A,B$ integration constants, and the difference between the two exponents is $2\D(k,\mu_3)-1$. When this difference is a positive integer, the Frobenius expansion is resonant and the second solution generically develops a logarithmic term. In holographic renormalization, this logarithm is associated with divergent local terms in the on-shell action, which must be removed by local counterterms \cite{Skenderis:2002wp}.

To see explicitly how the renormalization arises, we regulate the resonant value by writing
\begin{equation}
	\D(k,\mu_3)=\frac{m+1}{2}+\epsilon\,,
	\qquad \epsilon\to0\,.
	\label{eq:q9}
\end{equation}
This analytic continuation is only a regulator; after subtracting the divergent local term, the limit $\epsilon\to0$ defines the renormalized correlator.

The frequency-independent prefactor in \eqref{eq:app:q3} behaves as
\begin{align}\non
	&(4\pi T)^{2\D(k,\mu_3)-1}
	\frac{\Gamma\left(1-2\D(k,\mu_3)\right)}
	{\Gamma\left(2\D(k,\mu_3)-1\right)}\,\dfrac{\Gamma(\D(k,\mu_3)\mp \frac{i}{6}r_H\mu_3)}{\Gamma(1\!-\!\D(k,\mu_3)\mp \frac{i}{6}r_H\mu_3)}=\\
	&=
	\frac{(-1)^{m+1}(4\pi T)^m}
	{2m!(m-1)!}\,\dfrac{\Gamma(\frac{m+1}{2}\mp \frac{i}{6}r_H\mu_3)}{\Gamma(\frac{1-m}{2}\mp \frac{i}{6}r_H\mu_3)}\,
	\frac{1}{\epsilon}
	+\mathcal{O}(\epsilon^0)\,.
	\label{eq:q10}
\end{align}
On the other hand, defining
\begin{equation}
	z^\pm(\omega)
	\equiv
	\frac{1-m}{2}
	-\frac{i\omega}{2\pi T}
	\pm\frac{i}{6}r_H\mu_3\,,
	\label{eq:q11}
\end{equation}
the frequency-dependent ratio in \eqref{eq:app:q3} becomes
\begin{align}\non
	&\frac{
		\Gamma\left(
		\D(k,\mu_3)-\frac{i\omega}{2\pi T}
		\pm\frac{i}{6}r_H\mu_3
		\right)}
	{
		\Gamma\left(
		1-\D(k,\mu_3)-\frac{i\omega}{2\pi T}
		\pm\frac{i}{6}r_H\mu_3
		\right)}=
	\frac{\Gamma\left(z^\pm(\omega)+m+\epsilon\right)}
	{\Gamma\left(z^\pm(\omega)-\epsilon\right)}\\
	& =
	\frac{\Gamma\left(z^\pm(\omega)+m\right)}
	{\Gamma\left(z^\pm(\omega)\right)}
	\left[
	1+\epsilon\left(
	\psi\left(z^\pm(\omega)+m\right)
	+\psi\left(z^\pm(\omega)\right)
	\right)
	+\mathcal{O}(\epsilon^2)
	\right]\,,
	\label{eq:q12}
\end{align}
where $\psi(z)=\Gamma'(z)/\Gamma(z)$ is the digamma function. At $\epsilon=0$,
\begin{equation}
	\frac{\Gamma\left(z^\pm(\omega)+m\right)}
	{\Gamma\left(z^\pm(\omega)\right)}
	=
	\prod_{j=0}^{m-1}
	\left(z^\pm(\omega)+j\right)\,,
	\label{eq:13}
\end{equation}
which is a polynomial in $\omega$. Therefore, the divergent term obtained by combining \eqref{eq:q10} and \eqref{eq:q12} is analytic in $\omega$ and corresponds to a local contact term. It can be subtracted by a local counterterm.

The $\mathcal{O}(\epsilon^0)$ terms that are analytic in $\omega$ remain scheme dependent. The non-analytic, scheme-independent part of the renormalized correlator is
\begin{align}\non
	\GG^\pm_{\mathrm{IR,ren}}(\omega,k)
	={}&
	\frac{(-1)^{m+1}(4\pi T)^m}
	{2m!(m-1)!}\,
	\frac{
		\Gamma\left(\frac{m+1}{2}
		\mp\frac{i}{6}r_H\mu_3\right)}
	{
		\Gamma\left(\frac{1-m}{2}
		\mp\frac{i}{6}r_H\mu_3\right)}
	\frac{
		\Gamma\left(z^\pm(\omega)+m\right)}
	{\Gamma\left(z^\pm(\omega)\right)}\times
	\\
	&\times
	\left[
	\psi\left(z^\pm(\omega)+m\right)
	+\psi\left(z^\pm(\omega)\right)
	\right]
	+\text{analytic polynomial in $\omega$}\,.
	\label{eq:q14}
\end{align}
The polynomial in the last line depends on the choice of finite counterterms and does not affect the pole locations or their residues.

The poles of \eqref{eq:q14} arise from the digamma functions. Both digamma functions have a pole whenever
\begin{equation}
	z^\pm(\omega)=-m-n\,,
	\qquad m,n\in\mathbb{N}\,,
	\label{eq:q15}
\end{equation}
which reproduces \eqref{eq:app:q5} with
\begin{equation}
	\D(k,\mu_3)=\frac{m+1}{2}\,.
\end{equation}
Near such a pole,
\begin{equation}
	\psi\left(z^\pm(\omega)+m\right)
	+\psi\left(z^\pm(\omega)\right)
	=
	-\frac{2}{z^\pm(\omega)+m+n}
	+\mathcal{O}(1)\,,
	\label{eq:q16}
\end{equation}
Using
\begin{equation}
	z^\pm(\omega)+m+n
	=
	-\frac{i}{2\pi T}
	\left(\omega-\omega^{\pm,(n)}\mathrm{IR}\right)\,,
	\label{eq:q17}
\end{equation}
we obtain the resonant residues
\begin{align}
	\underset{\omega=\omega^{\pm,(n)}_{\text{AdS}_2}}{\mathrm{Res}},
	\GG^\pm_{\mathrm{IR,ren}}(\omega,k)
	={}&
	i\,2\pi T\,(4\pi T)^m
	\frac{(m+n)!}{n!\,m!\,(m-1)!}
	\,
	\frac{
		\Gamma\left(\frac{m+1}{2}
		\mp\frac{i}{6}r_H\mu_3\right)}
	{
		\Gamma\left(\frac{1-m}{2}
		\mp\frac{i}{6}r_H\mu_3\right)}\,.
	\label{eq:q18}
\end{align}
Equivalently, \eqref{eq:q18} follows by taking the regulated limit of the generic residue formula \eqref{eq:q6}. In particular,
\begin{equation}
	\lim_{\D(k,\mu_3)\to\frac{m+1}{2}}
	\frac{
		(-1)^n\Gamma\left(1-2\D(k,\mu_3)\right)}
	{
		\Gamma\left(1-2\D(k,\mu_3)-n\right)}
	=
	\frac{(m+n)!}{m!}\,.
	\label{eq:q19}
\end{equation}
Thus, \eqref{eq:q6} remains valid at the resonant values only when it is understood as a limit in $\D(k,\mu_3)$, rather than as a formula in which the resonant value is substituted from the outset.

Finally, none of the remaining Gamma functions in \eqref{eq:app:q3} produces additional poles in the frequency plane. The factor $\Gamma\left(1-2\D(k,\mu_3)\right)$ is independent of $\omega$; its singularities occur only at the resonant values discussed above and are removed by renormalization. The Gamma function in the frequency-dependent denominator appears through its reciprocal, $1/\Gamma(z)$, which is entire and has zeros rather than poles. The final charge-dependent denominator has the same property. Consequently, these factors can cancel poles at special parameter values, but they cannot generate additional poles in $\omega$.

It is useful to relate the cancellations mentioned above to the
phenomenon of pole skipping. Pole skipping occurs at isolated points
$(\omega_\star,k_\star)$ in the complexified frequency--momentum
plane where a line of poles intersects a line of zeros of the retarded
correlator. Equivalently, the residue of the pole vanishes as the
pole-skipping momentum is approached, while the value of the
correlator at the intersection is not unique and depends on the
direction from which the point is approached
\cite{Blake:2019otz,Natsuume:2020snz,Yuan:2023tft}.

To identify such points in the present IR correlator \eqref{eq:app:q3}, we first recall that the $n^\text{th}$ AdS$_2$ pole of the correlator satisfies
\begin{equation}
		\D(k,\mu_3)-\frac{i\omega}{2\pi T}
	\pm\frac{i}{6}r_H\mu_3=-n.
\end{equation}
On the other hand, the frequency-independent factor
of the correlator \eqref{eq:app:q3} ($1/\Gamma(1\!-\!\D(k,\mu_3)\mp \frac{i}{6}r_H\mu_3)$) has a zero whenever
\begin{equation}
	1-\D(k_\star,\mu_3)\mp\frac{i}{6}r_H\mu_3=-p,
	\qquad p\in\mathbb{N}_0.
\end{equation}
Thus,
\begin{equation}
	\D(k_\star,\mu_3)=1+p\mp\frac{i}{6}r_H\mu_3,
	\qquad
	\omega_\star^{(n,p)}
	=-i\,2\pi T(n+p+1),
\end{equation}
and, using the explicit expression for the IR conformal dimension,
\begin{equation}
	k_{\star,\pm}^{\,2}
	=
	\frac{12p(p+1)}{r_H^2}
	\mp\,\frac{2i(2p+1)\mu_3}{r_H}.
\end{equation}
Near one of these points, the correlator behaves schematically as
\begin{equation}
	\GG^\pm_{\mathrm{IR}}(\omega,k)
	=
	\mathcal{C}_{n,p}^{\pm}\,
	\frac{\delta\D}
	{\delta\D-i\,\delta\omega/(2\pi T)}
	\left[
	1+\mathcal{O}(\delta\D,\delta\omega)
	\right],
	\label{eq:localPS}
\end{equation}
where $\mathcal{C}_{n,p}^{\pm}$ is the finite, generically non-vanishing product of all the remaining Gamma-function factors evaluated at
$(\omega_{\star}^{(n,p)},k_{\star,\pm})$. Moreover, we have defined
\begin{equation}
	\omega=\omega_{\star}^{(n,p)}+\delta\omega \sp 	\delta\D
	\equiv
	\D(k,\mu_3)-\D(k_{\star,\pm},\mu_3)\,,
\end{equation}
that are small displacement from one of the pole-skipping points.

Equation \eqref{eq:localPS} exhibits the intersection of a zero line,
\begin{equation}
	\delta\D=0,
\end{equation}
and a pole line,
\begin{equation}
	\delta\D-\frac{i\,\delta\omega}{2\pi T}=0.
\end{equation}
The value of the correlator at their intersection is consequently not uniquely defined. For example, approaching the point along a trajectory
\begin{equation}
	\delta\omega=\lambda\,\delta\D
\end{equation}
gives
\begin{equation}
	\lim_{\delta\D\to0}
	\GG^\pm_{\mathrm{IR}}
	=
	\frac{\mathcal{C}_{n,p}^{\pm}}
	{1-i\lambda/(2\pi T)},
\end{equation}
which depends on the slope $\lambda$ of the chosen trajectory.

Equivalently, at fixed momentum close to $k_{\star,\pm}$, the pole is located at
\begin{equation}
	\omega^{\pm,(n)}_{\text{AdS}_2}
	=
	\omega_{\star}^{(n,p)}
	-i\,2\pi T\,\delta\D
	+\mathcal{O}(\delta\D^2),
\end{equation}
and its residue behaves as
\begin{equation}
	\underset{\omega=\omega^{\pm,(n)}_{\text{AdS}_2}}
	{\mathrm{Res}}\,
	\GG^\pm_{\mathrm{IR}}(\omega,k)
	=
	i\,2\pi T\,\mathcal{C}_{n,p}^{\pm}\,
	\delta\D
	+\mathcal{O}(\delta\D^2).
\end{equation}
The residue therefore vanishes in the limit
$k^2\to k_{\star,\pm}^{\,2}$, as expected at a pole-skipping point.

We stress that the resonant condition
$2\D-1\in\mathbb{N}$ discussed above does not, by itself, imply pole
skipping. At those values the singular normalization of the
correlator must first be removed by holographic renormalization, and
the resulting digamma poles generally have the non-vanishing
residues given in \eqref{eq:q18}. Pole skipping occurs only when a
genuine zero of the renormalized correlator also intersects the pole.

\section{Schr\"odinger potentials}

\label{sec:SP}

We rewrite here the second order fluctuation equations \eqref{EoMVhtpm} and \eqref{Eqvf} in Schr\"odinger form, and extract the corresponding Schr\"odinger potentials. As mentioned in section \ref{sec:IRcorr-approx}, the properties of these potentials are relevant for the application of the product formula.

We start from the transverse equation
\begin{equation}
\label{sp1} \partial_r\left(\frac{f(r)}{r}\partial_r\mathcal{L}_i^{\perp,\pm}\right) + \frac{1}{rf(r)}\left(\Omega_\pm(r)^2-\vec{k}^2f(r)\right)\mathcal{L}_i^{\perp,\pm}=0\, .
\end{equation}
Wo then introduce the tortoise coordinate $z$, such that $\intd z/\intd r = f(r)^{-1}$, for which \eqref{sp1} can be written as
\begin{equation}
\label{sp2} \partial^2_z\mathcal{L}_i^{\perp,\pm} + p_\perp(z) \partial_z\mathcal{L}_i^{\perp,\pm} + q_\perp(z)\mathcal{L}_i^{\perp,\pm} = 0\, ,
\end{equation}
\begin{equation}
\label{sp3} p_\perp(z) \equiv - \frac{r'(z)}{r(z)} \sp q_\perp(z) \equiv \Omega_\pm(z)^2-\vec{k}^2f(z) \, .
\end{equation}
This can now be written in Schr\"odinger form with the standard change of variable:
\begin{equation}
\label{sp4} \mathcal{L}_i^{\perp,\pm}(z) = \Psi_\perp^\pm(z)\ex^{-\frac{1}{2}\int^z p_\perp(z')\intd z'} \, .
\end{equation}
Substituting \eqref{sp4} into \eqref{sp2} gives
\begin{equation}
\label{sp5} (\Psi_\perp^\pm)''(z) + \left(\omega^2 - V_\perp(z)\right)\Psi_\perp^\pm(z) = 0 \, ,
\end{equation}
with the transverse Schr\"odinger potential given by
\begin{equation}
\label{sp6} V_\perp(z) = \mp 2\omega \Phi_3(z) - \Phi_3(z)^2 + k^2 f(z) - \frac{f'(z)}{2r(z)} + \frac{3f(z)^2}{4r(z)^2} \, .
\end{equation}
Note that this potential is regular for $r>0$.

We now discuss the longitudinal fluctuation equation \eqref{Eqvf}, which reads in tortoise coordinates
\begin{equation}
\label{sp7} r(z)\partial_z\left(\frac{ \Omega_\pm(z)^2 \pa_z\varphi^\pm}{r(z) \big(\Omega_\pm(z)^2- f(z)k^2\big)}\right) + \Omega_\pm(z)^2\varphi^\pm = 0 \, .
\end{equation}
We first introduce a new variable
\begin{equation}
\label{sp8} \phi^\pm = \frac{ \Omega_\pm(z)^2 \pa_z\varphi^\pm}{r(z) \big(\Omega_\pm(z)^2- f(z)k^2\big)} \, ,
\end{equation}
which obeys a simpler equation
\begin{equation}
\label{sp9} \frac{\Omega_\pm(z)^2}{r(z)}\pa_z\left(\frac{r(z)}{\Omega_\pm(z)^2}\pa_z\phi^\pm\right) + (\Omega_\pm(z)^2 - k^2f(z))\phi^\pm = 0 \, .
\end{equation}
This can be rewritten in the same form as \eqref{sp2}:
\begin{equation}
\label{sp10} \partial^2_z\phi^\pm + p_\parallel(z) \partial_z\phi^\pm + q_\parallel(z)\phi^\pm = 0\, ,
\end{equation}
\begin{equation}
\label{sp11} p_\parallel(z) \equiv  \frac{r'(z)}{r(z)} - 2\frac{\Omega_\pm'(z)}{\Omega_\pm(z)} \sp q_\parallel(z) \equiv \Omega_\pm(z)^2-\vec{k}^2f(z) \, .
\end{equation}
Now considering the same kind of change of variable as \eqref{sp4},
\begin{equation}
\label{sp12} \phi^\pm(z) = \Psi_\parallel^\pm(z) \ex^{-\frac{1}{2}\int^z p_\parallel(z')\intd z'} \, ,
\end{equation}
we finally obtain the longitudinal Schr\"odinger equation
\begin{equation}
\label{sp13} (\Psi_\parallel^\pm)''(z) + \left(\omega^2 - V_\parallel(z)\right)\Psi_\parallel^\pm(z) = 0 \, ,
\end{equation}
with the longitudinal potential given by
\begin{align}
\nn V_\parallel(z) = &\mp 2\omega \Phi_3(z) - \Phi_3(z)^2 + k^2 f(z) -\frac{f(z)^2}{4r(z)^2} + \frac{f'(z)}{2r(z)}+\\
\label{sp14} &+ 2\left(\frac{\Omega_\pm'(z)}{\Omega_\pm(z)}\right)^2 - \frac{f(z)}{r(z)}\frac{\Omega_\pm'(z)}{\Omega_\pm(z)} - \frac{\Omega_\pm''(z)}{\Omega_\pm(z)} .
\end{align}
This potential has singularities in the bulk whenever $\Omega_\pm(r) = \omega \pm \Phi_3(r)$ vanishes.

\newpage

\section{Details on the numerical computation of the correlators}\label{app:numerical}

In this appendix we describe the numerical procedure used to compute the imaginary part of the  transverse and longitudinal polarization functions. The fluctuation equations themselves are given in the main text and will not be repeated here, see equations \eqref{EoMVhtpm}, \eqref{EpEoM}.

The numerical integration is performed using the dimensionless radial coordinate
\begin{equation}
	u = 1-\frac{r}{r_H}\,,
\end{equation}
so that the horizon is located at $u=0$, while the asymptotic AdS$_5$ boundary is at $u=1$. To avoid carrying explicit powers of the horizon radius in the numerical equations, the frequency, the momentum, and the isospin chemical potential are expressed in dimensionless form,
\begin{equation}
	\omega \to  r_H\,\omega\,,
	\qquad
	k \to r_H\,k\,,
	\qquad
	\mu_3 \to r_H\,\mu_{3}\,.
\end{equation}
We use the same symbols for the dimensionless variables throughout the this appendix. We report here  the definition of the horizon radius
\begin{equation}
	r_H =
	\frac{2}{\pi T}
	\left[
	1+
	\sqrt{
		1+\frac{w_0^2}{3N_c\pi^2\ell^4}
		\left(
		\frac{\mu_q^2+\mu_3^2}{T^2}
		\right)}
	\right]^{-1}\,
\end{equation}
where the bulk parameters used for the numerics are
\begin{equation}
	N_c=3\sp \ell=1\sp
	w_0=6\sqrt{\frac{5}{13}}\,,
\end{equation}
as discussed in Appendix~\ref{app:thermo}. Then, the values of the quark and isospin chemical potential are scanned over the following values, as extensively discussed in the main text:
\be
\dfrac{\mu_q}{T}\in\{10^4,65,5\}\sp\dfrac{\mu_3}{\mu_q}\in\{-0.5,-0.1,0\}\,.
\ee

Direct integration of the original fluctuation fields is inconvenient because their near-horizon behavior contains the rapidly varying ingoing phase. We therefore introduce the tortoise coordinate $z$, defined by
\begin{equation}
	\frac{d z}{d r}=\frac{1}{f(r)}\,,
\end{equation}
as in equation \eqref{EF4}, and factor the universal phase out of each fluctuation,
\begin{equation}
	F\to
	\exp\left(\frac{i\omega}{r_H}z(r)\right)F(u).
\end{equation}
with $F = \{\mathcal{L}_i^{\perp,+},E^{\parallel,+}\}$ being the generic plus-charged fluctuation. With the Fourier convention employed in the main text, this factor implements the ingoing boundary condition at the future horizon. The remaining function $F(u)$ is regular at $u=0$, and the fluctuation problem becomes an initial-value problem for a regular complex function. The resulting equations of motion are written in \eqref{EF8} and \eqref{EF9}.

The initial conditions are not imposed exactly at the horizon, where the coefficients of the differential equation are singular, but at a small radial cutoff
\begin{equation}
	u=\delta,\qquad \delta=10^{-6}.
\end{equation}
They are obtained from a Frobenius expansion,
\begin{equation}
	F(u)=\sum_{n=0}^{N_H}a_n u^n,
	\qquad a_0=1.
	\label{eq:horizon_series_numerics}
\end{equation}
The choice $a_0=1$ fixes the otherwise arbitrary overall normalization of the linear solution. It has no effect on the final correlator, which is computed as a ratio of the response and source coefficients of the near-boundary expansion of the solutions.

The substitution of equation~\eqref{eq:horizon_series_numerics} into the transverse or longitudinal equation gives a set of algebraic relations that determine $a_1,\ldots,a_{N_H}$ recursively. The truncated series and its derivative are then evaluated at $u=\delta$:
\begin{equation}
	F(\delta)=\sum_{n=0}^{N_H}a_n\delta^n,
	\qquad
	F'(\delta)=\sum_{n=1}^{N_H}n a_n\delta^{n-1}.
\end{equation}
Separate recursion relations are generated for the transverse and longitudinal channels. The numerical results presented in this work use $N_H=8$.

For every pair $(\omega,k)$, the corresponding complex ordinary differential equation is integrated from
\begin{equation}
	u=\delta
	\quad\text{to}\quad
	u=1-\epsilon,
	\qquad
	\epsilon=10^{-6},
\end{equation}
using \texttt{NDSolve}. The calculation is carried out with approximately $25$ digits of working precision. In the longitudinal channel, where the equation can become comparatively stiff in some regions of parameter space, the maximum number of integration steps is left unrestricted.

Since the fluctuation equations are linear, no ultraviolet boundary condition is imposed during this step. The integration produces the unique solution satisfying the ingoing condition and the normalization $F(0)=1$. Its boundary value subsequently determines the source coefficient, while the subleading ultraviolet behavior determines the response.

The scan over frequency and momentum is parallelized. The result associated with each point is stored as a triplet
\begin{equation}
	\{\omega,k,\mathcal{P}(\omega,k)\},
\end{equation}
where $\mathcal{P}$ denotes the imaginary part of the appropriately normalized response-to-source ratio which will be defined in a few lines.

In asymptotically AdS geometries, the near-boundary expansion of a gauge-field fluctuation contains a logarithmic term (see the near-boundary expansion of the fluctuations in equations \eqref{Ltnb} and \eqref{Elnb}). Consequently, a direct evaluation of the radial derivative at $u=1-\epsilon$ converges slowly and includes the logarithmic ultraviolet contribution. We subtract this contribution before extracting the UV normalizable coefficient.

For both channels, the numerical implementation constructs the combination
\begin{equation}
	\mathcal{R}(u)=
	i\omega F(u)-F'(u)
	+\left[(\omega + \mu_3)^2-k^2\right]
	(1-u)\log(1-u)\,F(u)\,,
	\label{eq:renormalized_numerical_momentum}
\end{equation}
where we recall that we are focusing on the plus-charged fluctuation. The last term removes the logarithmic piece dictated by the ultraviolet asymptotic expansion. Equation~\eqref{eq:renormalized_numerical_momentum} is evaluated at $N_{\mathrm{fit}}+1$ equally spaced points in a narrow interval close to the boundary,
\begin{equation}
	u_i\in[0.99999,1-\epsilon],
	\qquad N_{\mathrm{fit}}=100.
\end{equation}
The resulting data are fitted to a linear function,
\begin{equation}
	\mathcal{R}(u)=a\,u+b.
	\label{eq:linear_uv_fit}
\end{equation}
The coefficient $a$ obtained from this local fit is then the finite response coefficient. Extracting it from several points, rather than from a single numerical derivative at the cutoff, substantially reduces sensitivity to $\epsilon$ and to residual subleading terms.

The source coefficient is represented numerically by
\begin{equation}
	F_{\mathrm{UV}}=F(1-\epsilon).
\end{equation}
Up to the overall channel-dependent normalization and possible real contact terms specified in the holographic renormalization prescription, the retarded correlator is therefore obtained from
\begin{equation}
	G^R(\omega,k)\propto \frac{a}{F_{\mathrm{UV}}}.
	\label{eq:numerical_retarded_ratio}
\end{equation}
The numerical arrays used for the plots retain the quantity
\begin{equation}
	\mathcal{P}(\omega,k)
	=
	-\operatorname{Im}
	\left[
	\frac{a}{F(1-\epsilon)}
	\right].
	\label{eq:numerical_spectral_quantity}
\end{equation}
Any overall prefactor, as well as the conventional factor relating
$-\operatorname{Im}G^R$ to the spectral density, is restored separately. Real local counterterms do not affect equation~\eqref{eq:numerical_spectral_quantity} since the focus is on the imaginary part of the correlator.

Different grids are employed in the two channels (transverse and longitudinal) because their relevant structures occur on different scales. For the transverse channel, a uniform two-dimensional grid is used:
\begin{equation}
	\omega_i =
	\frac{i}{N_{\mathrm{it}}},
	\qquad
	k_j =
	\frac{j}{N_{\mathrm{it}}},
	\qquad
	i,j=0,\ldots,N_{\mathrm{it}},
\end{equation}
with $N_{\mathrm{it}}=100$. This gives a square grid of $(N_{\mathrm{it}}+1)^2$ points.

For the longitudinal channel, the low-frequency region requires finer resolution. The frequency grid is
\begin{equation}
	\omega_i =
	\frac{i}{N_{\mathrm{it}}}
	\label{eq:longitudinal_frequency_grid}
\end{equation}
whereas the momentum grid is chosen as
\begin{equation}
	k_{ij}=
	\left(\frac{j}{N_{\mathrm{it}}}\right)^2
	\sqrt{\omega_i}\,.
	\label{eq:longitudinal_momentum_grid}
\end{equation}
The scaling $k\propto\sqrt{\omega}$ is adapted to the diffusive regime, in which the relevant combination is $\omega/k^2$, while the quadratic dependence on $j/N_{\mathrm{it}}$ increases the density of points at small momentum. The zero-density row is added separately using
\begin{equation}
	\mathcal{P}_{\parallel}(0,k)=0.
\end{equation}

The dominant numerical systematics originate from four sources: truncation of the horizon series, the infrared and ultraviolet cutoffs, the working precision of the integration, and the finite fitting window used to extract the ultraviolet response. The convergence of the final data is assessed by repeating representative portions of the scan while varying
\begin{equation}
	N_H,\qquad
	\delta,\qquad
	\epsilon,\qquad
	N_{\mathrm{fit}},
\end{equation}
as well as the working precision and lower endpoint of the fitting interval. A reliable point is required to remain stable under these variations and under a moderate refinement of the $(\omega,k)$ grid.

\newpage

\section{Testing other IR-based approximations of the charged current correlators}\label{app:otherapprox65}

In this appendix, in addition to the (near-extremal) hydrodynamic and the extended hydrodynamic approximations \eqref{app:eq:trhydroapprox}-\eqref{app:eq:longexthydroapprox}, we present the comparison of the exact, numerical charged current polarization function, with the other IR-based approximations presented in the appendix \ref{app:IRcorr-approx}. All of them are given in equations \eqref{app:eq:trhydroapprox}, \eqref{app:eq:trexthydroapprox}, \eqref{eq:trimprexthydroapprox}, \eqref{eq:trnormapprox} and \eqref{eq:trfullapprox} for the transverse correlator; and in \eqref{app:eq:longhydroapprox}, \eqref{app:eq:longexthydroapprox}, \eqref{eq:longimprexthydroapprox}, \eqref{eq:longnormapprox} and \eqref{eq:longfullapprox} for the longitudinal correlator.

 We do this for a background space-time characterized by $\mu_q/T=65$ and $\mu_3=0$. As in section \ref{sec:exactresults}, we present the plots of the percentage relative difference \eqref{eq:percrelfdiff} for the various approximations listed above. We also provide, for all the approximations, the results for the coarse-grained $\chi_{ij}$ and fine-grained $s_{ij}$ relative difference defined in \eqref{eq:chiRL} and \eqref{eq:chiCELL} respectively. In appendices \ref{app:coarsegrained} and \ref{app:finegrained}, we present the results for those observables associated with all the analytic approximations presented in appendix \ref{app:IRcorr-approx} for other values of the quark and isospin chemical potentials.

\subsection{Transverse correlator}

\begin{figure}[htb]
	\centering
	\begin{subfigure}{0.4\textwidth}
		\centering
		\includegraphics[width=\textwidth]{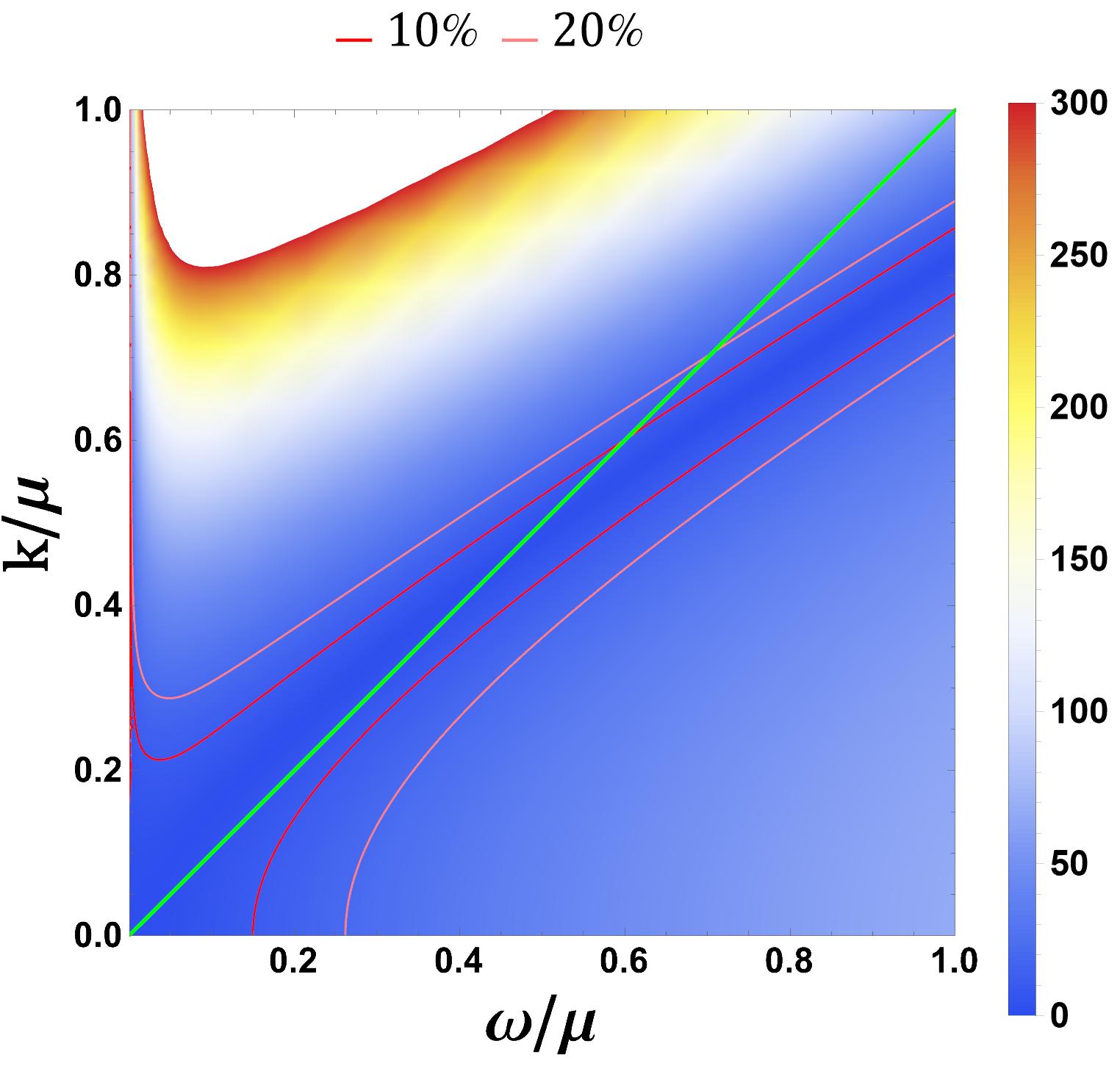}
	\end{subfigure}
	\hspace{1cm}
	\begin{subfigure}{0.4\textwidth}
		\centering
		\includegraphics[width=\textwidth]{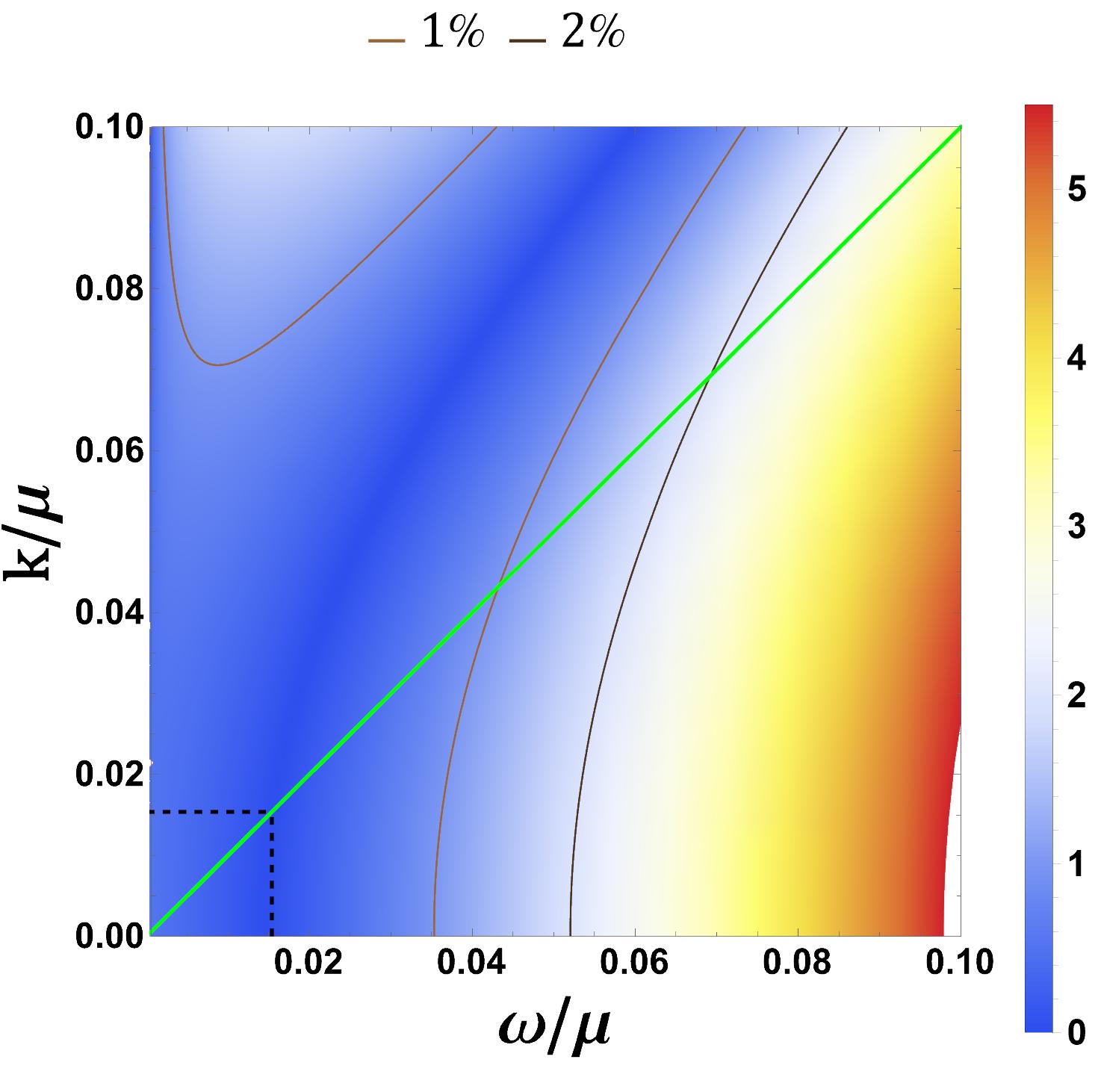}
	\end{subfigure}
	
	\caption{The percentage relative difference \eqref{eq:percrelfdiff} of the imaginary part of the transverse charged current polarization function with respect to the improved extended hydrodynamic approximation \eqref{eq:trimprexthydroapprox}, for $\mu_q/T= 65$ and $\mu_3=0$. The green line shows the locus $\omega = k$. The right plot shows a subregion of the left
	plot. The dashed square in this right plot represents the standard hydrodynamic region $\omega/\mu,k/\mu \in [0,T/\mu]$.} \label{fig:RelDiffNum_tr_mu65mu30_imprext}
\end{figure}

\begin{figure}[htb]
	\centering
	\begin{subfigure}{0.4\textwidth}
		\centering
		\includegraphics[width=\textwidth]{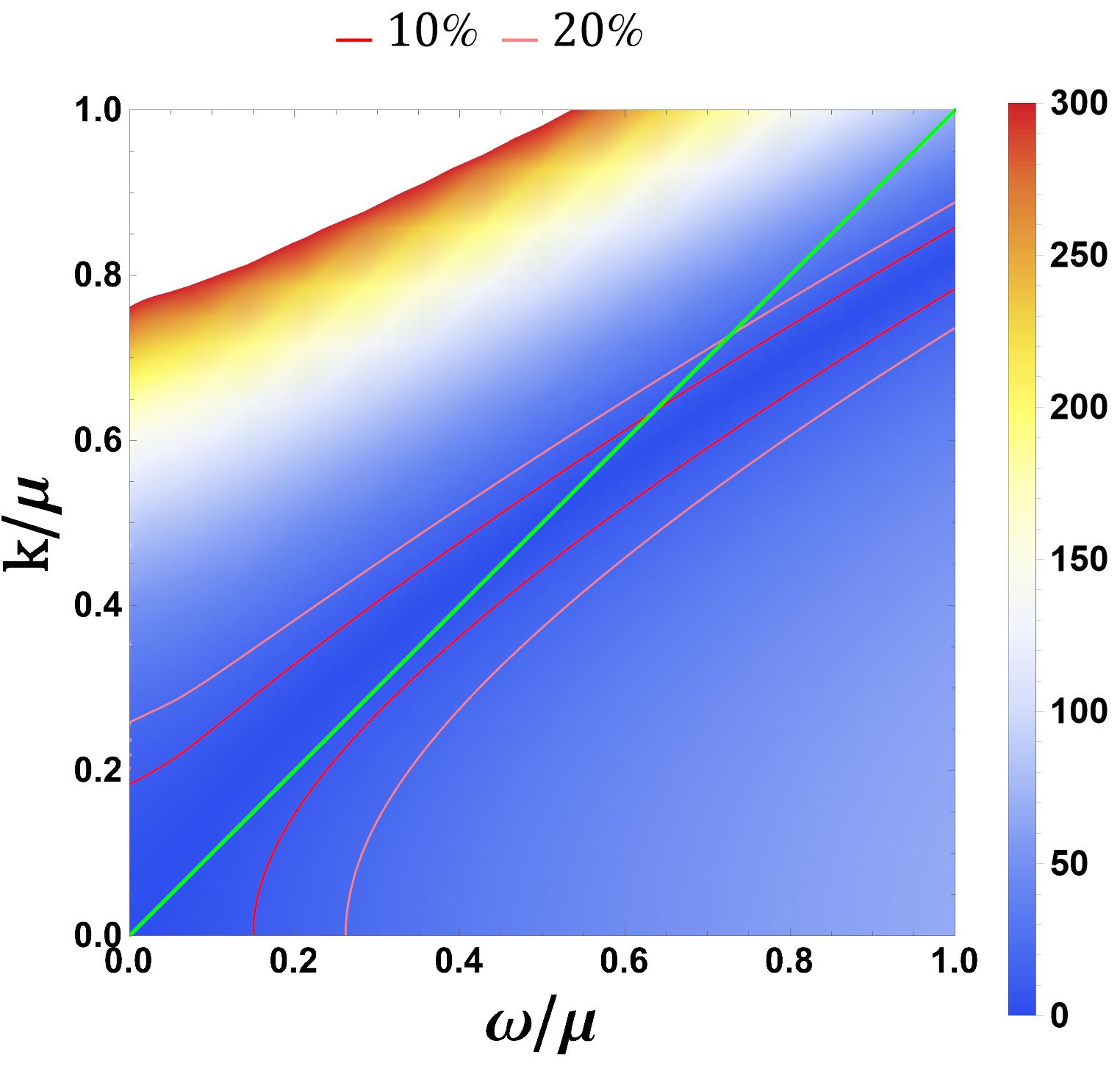}
	\end{subfigure}
	\hspace{1cm}
	\begin{subfigure}{0.4\textwidth}
		\centering
		\includegraphics[width=\textwidth]{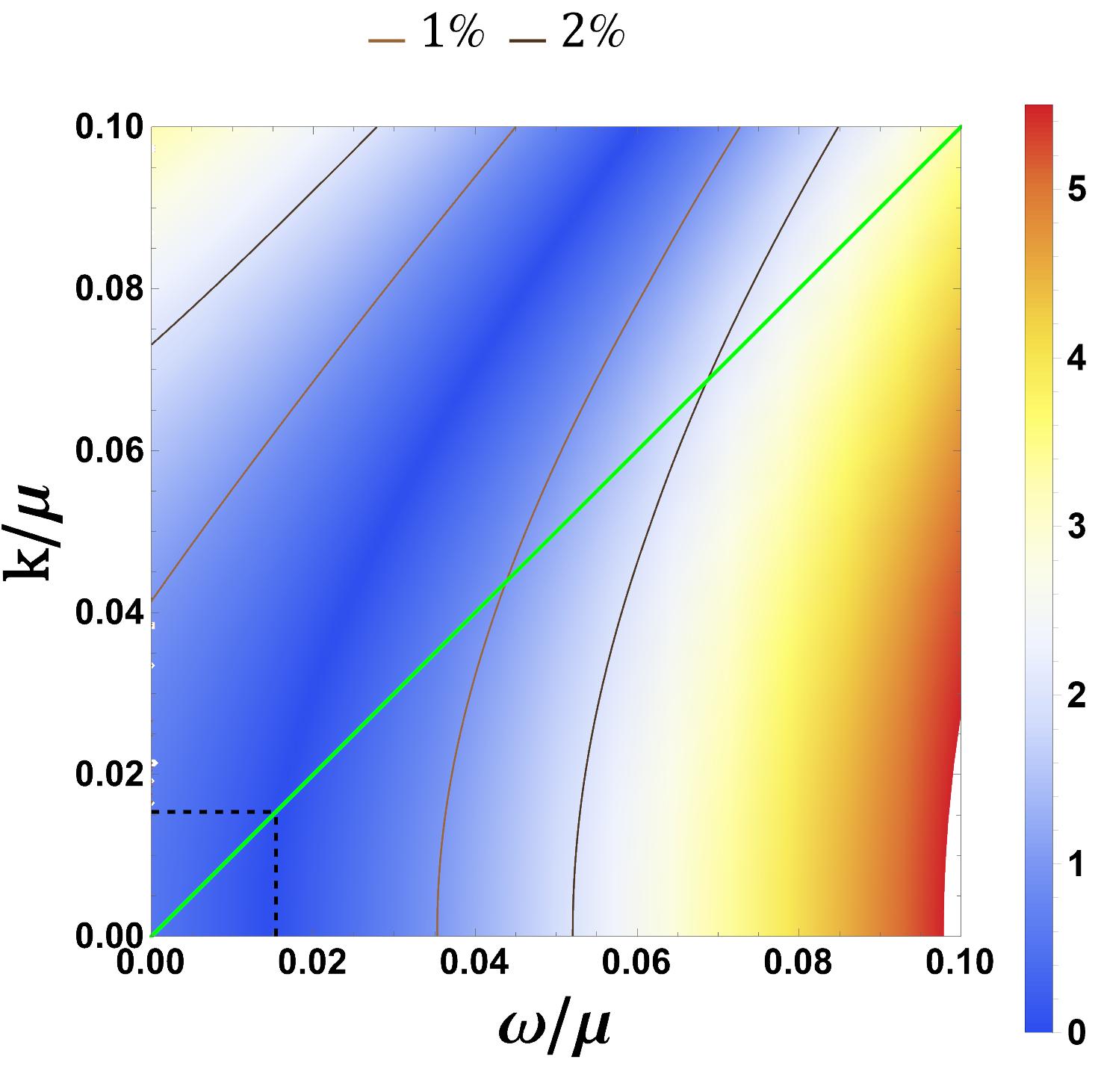}
	\end{subfigure}
	
	\caption{As figure~\ref{fig:RelDiffNum_tr_mu65mu30_imprext} but for the normalized correlator \eqref{eq:trnormapprox} instead of the improved extended hydrodynamic approximation.}
	\label{fig:RelDiffNum_tr_mu65mu30_Gnorm}
\end{figure}

\begin{figure}[htb]
	\centering
	\begin{subfigure}{0.4\textwidth}
		\centering
		\includegraphics[width=\textwidth]{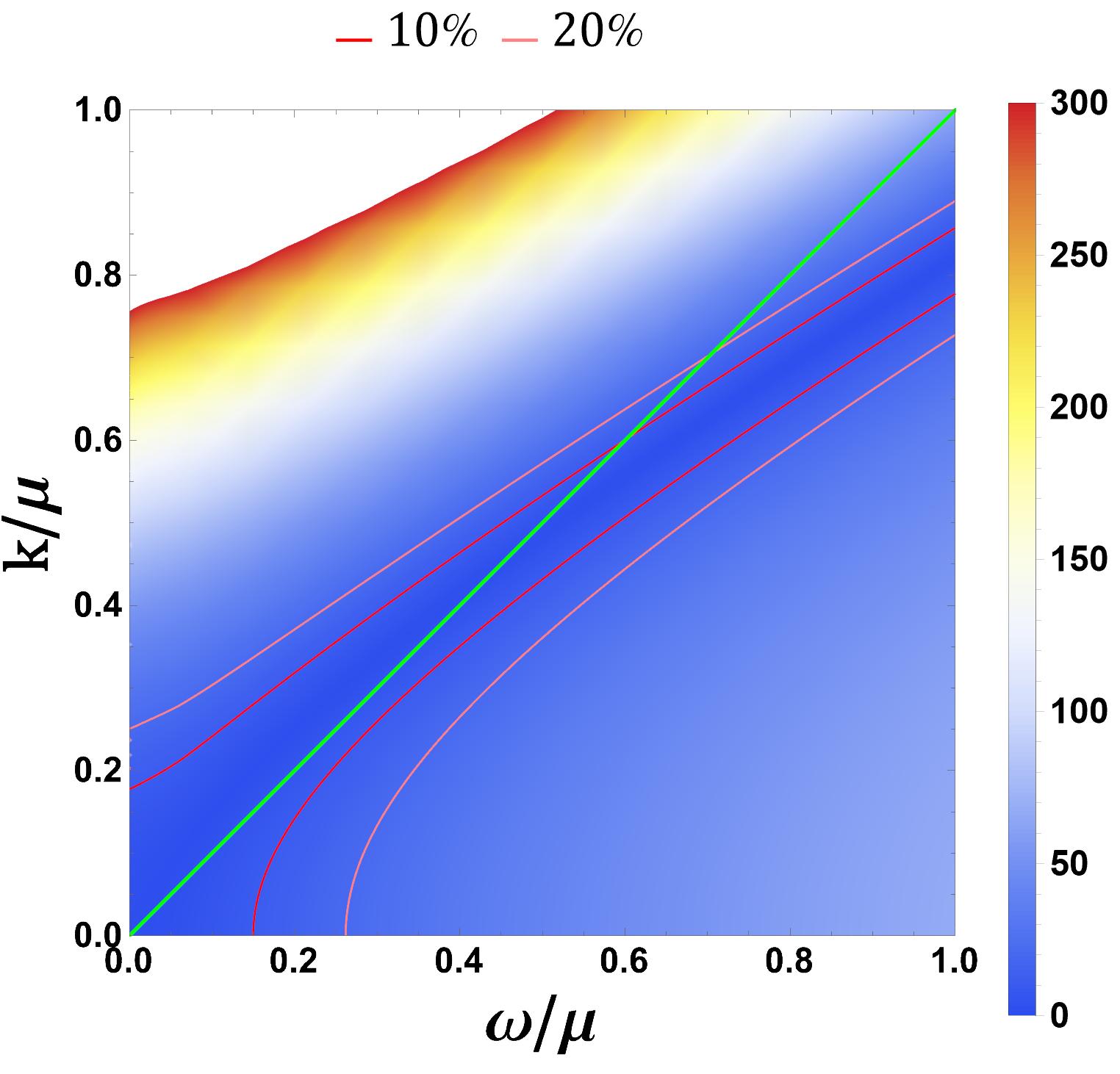}
	\end{subfigure}
	\hspace{1cm}
	\begin{subfigure}{0.4\textwidth}
		\centering
		\includegraphics[width=\textwidth]{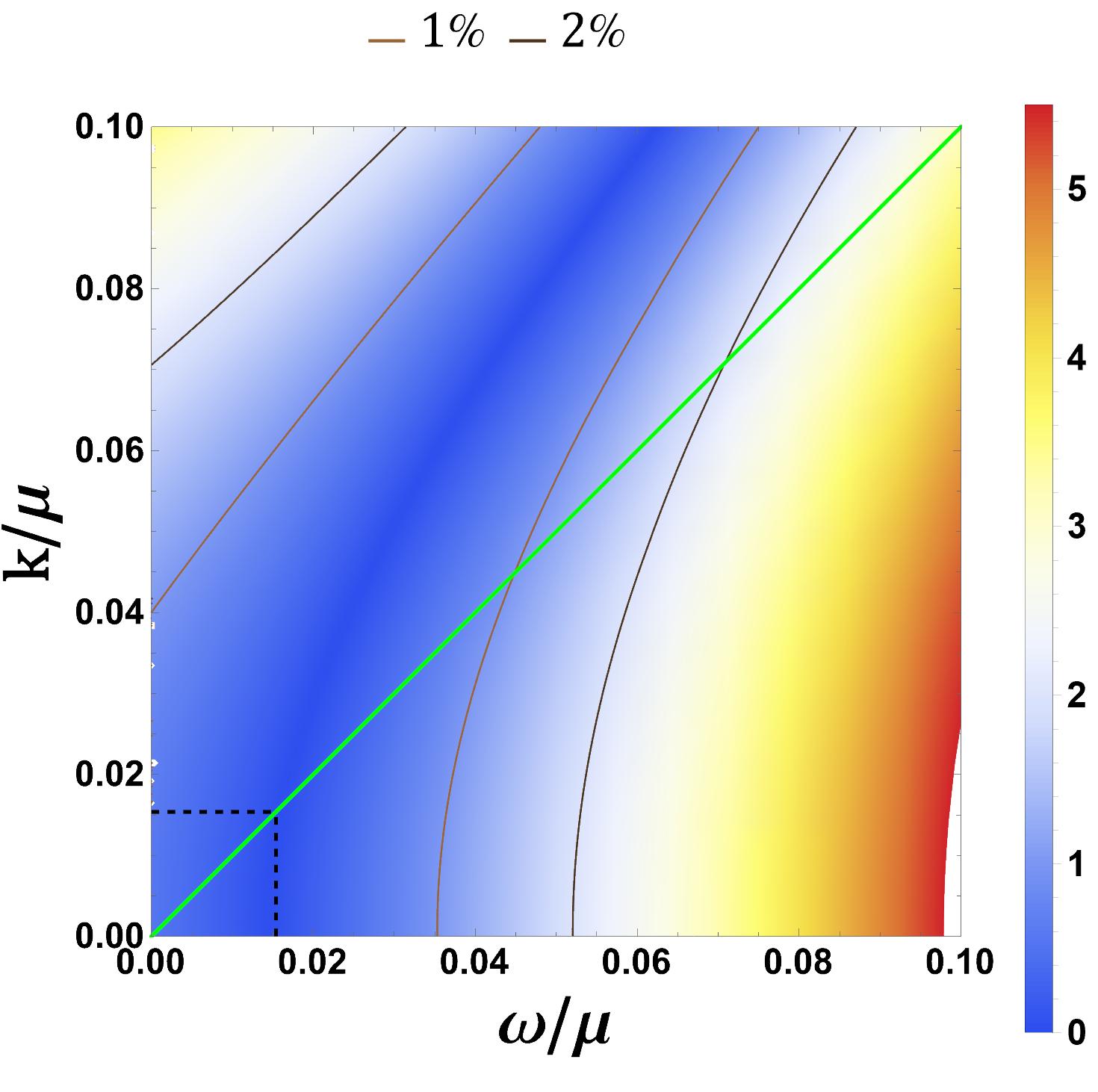}
	\end{subfigure}
	
	\caption{As in figure~\ref{fig:RelDiffNum_tr_mu65mu30_imprext} but for the full correlator \eqref{eq:trfullapprox} instead of the improved extended hydrodynamic approximation.}	\label{fig:RelDiffNum_tr_mu65mu30_Gfull}
\end{figure}

\begin{table}[htb]
	\centering
	{
		\small
		\begin{tabular}{c|cccc}
			\multicolumn{5}{c}{(a) hydrodynamic}\\
			$1/2$ & 0.38 & 0.30 & 0.26 & 0.25 \\
			$3/8$ & 0.18 & 0.15 & 0.15 & 0.18 \\
			$1/4$ & 0.07 & 0.08 & 0.12 & 0.17 \\
			$1/8$ & 0.03 & 0.07 & 0.13 & 0.18 \\\hline
			& $1/8$ & $1/4$ & $3/8$ & $1/2$ \\
		\end{tabular}\hspace{1cm}
		\begin{tabular}{c|cccc}
			\multicolumn{5}{c}{(b) extended-hydrodynamic}\\
			$1/2$ & 0.18 & 0.17 & 0.17 & 0.18 \\
			$3/8$ & 0.09 & 0.09 & 0.12 & 0.16 \\
			$1/4$ & 0.04 & 0.06 & 0.11 & 0.16 \\
			$1/8$ & 0.02 & 0.07 & 0.13 & 0.18 \\\hline
			& $1/8$ & $1/4$ & $3/8$ & $1/2$ \\
		\end{tabular}
		
		\begin{tabular}{c|cccc}
			\multicolumn{5}{c}{(c) improved}\\
			$1/2$ & 0.20 & 0.18 & 0.18 & 0.19 \\
			$3/8$ & 0.10 & 0.10 & 0.12 & 0.16 \\
			$1/4$ & 0.04 & 0.07 & 0.11 & 0.16 \\
			$1/8$ & 0.02 & 0.07 & 0.13 & 0.18 \\\hline
			& $1/8$ & $1/4$ & $3/8$ & $1/2$ \\
		\end{tabular}\hspace{1cm}
		\begin{tabular}{c|cccc}
			\multicolumn{5}{c}{(d) normalized}\\
			$1/2$ & 0.22 & 0.19 & 0.18 & 0.19 \\
			$3/8$ & 0.11 & 0.10 & 0.12 & 0.16 \\
			$1/4$ & 0.05 & 0.07 & 0.12 & 0.17 \\
			$1/8$ & 0.03 & 0.07 & 0.13 & 0.18 \\\hline
			& $1/8$ & $1/4$ & $3/8$ & $1/2$ \\
		\end{tabular}
		
		\begin{tabular}{c|cccc}
			\multicolumn{5}{c}{(e) full-IR}\\
			$1/2$ & 0.24 & 0.21 & 0.19 & 0.20 \\
			$3/8$ & 0.12 & 0.11 & 0.13 & 0.16 \\
			$1/4$ & 0.05 & 0.07 & 0.11 & 0.17 \\
			$1/8$ & 0.03 & 0.07 & 0.13 & 0.18 \\\hline
			& $1/8$ & $1/4$ & $3/8$ & $1/2$ \\
		\end{tabular}
	}
	\caption{Values $\chi_{ij}$ (formula \eqref{eq:chiRL}) in the transverse sector for $\mu_q/T=65$ and $\mu_3=0$.}
	\label{tab:app:chi650_tr}
\end{table}

\begin{table}[htb]
	\centering
	{
		\small
		\begin{tabular}{c|cccc}
			\multicolumn{5}{c}{(a) hydrodynamic}\\
			$1/2$ & 0.97 & 0.54 & 0.22 & 0.07 \\
			$3/8$ & 0.41 & 0.17 & 0.07 & 0.19 \\
			$1/4$ & 0.12 & 0.05 & 0.17 & 0.29 \\
			$1/8$ & 0.03 & 0.12 & 0.23 & 0.34 \\\hline
			& $1/8$ & $1/4$ & $3/8$ & $1/2$ \\
		\end{tabular}\hspace{1cm}
		\begin{tabular}{c|cccc}
			\multicolumn{5}{c}{(b) extended-hydrodynamic}\\
			$1/2$ & 0.45 & 0.35 & 0.16 & 0.07 \\
			$3/8$ & 0.20 & 0.10 & 0.07 & 0.20 \\
			$1/4$ & 0.05 & 0.06 & 0.18 & 0.30 \\
			$1/8$ & 0.02 & 0.12 & 0.24 & 0.34 \\\hline
			& $1/8$ & $1/4$ & $3/8$ & $1/2$ \\
		\end{tabular}
		
		\begin{tabular}{c|cccc}
			\multicolumn{5}{c}{(c) improved}\\
			$1/2$ & 0.49 & 0.39 & 0.18 & 0.07 \\
			$3/8$ & 0.22 & 0.11 & 0.07 & 0.19 \\
			$1/4$ & 0.06 & 0.06 & 0.18 & 0.29 \\
			$1/8$ & 0.02 & 0.12 & 0.23 & 0.34 \\\hline
			& $1/8$ & $1/4$ & $3/8$ & $1/2$ \\
		\end{tabular}\hspace{1cm}
		\begin{tabular}{c|cccc}
			\multicolumn{5}{c}{(d) normalized}\\
			$1/2$ & 0.56 & 0.36 & 0.16 & 0.07 \\
			$3/8$ & 0.24 & 0.10 & 0.07 & 0.20 \\
			$1/4$ & 0.07 & 0.06 & 0.18 & 0.30 \\
			$1/8$ & 0.03 & 0.12 & 0.24 & 0.34 \\\hline
			& $1/8$ & $1/4$ & $3/8$ & $1/2$ \\
		\end{tabular}
		
		\begin{tabular}{c|cccc}
			\multicolumn{5}{c}{(e) full-IR}\\
			$1/2$ & 0.60 & 0.39 & 0.18 & 0.07 \\
			$3/8$ & 0.26 & 0.11 & 0.07 & 0.19 \\
			$1/4$ & 0.07 & 0.06 & 0.17 & 0.29 \\
			$1/8$ & 0.03 & 0.12 & 0.23 & 0.34 \\\hline
			& $1/8$ & $1/4$ & $3/8$ & $1/2$ \\
		\end{tabular}
	}
	\caption{Values $s_{ij}$ (formula \eqref{eq:chiCELL}) in the transverse sector for $\mu_q/T=65$ and $\mu_3=0$.}
	\label{tab:app:s650_tr}
\end{table}

In this section, we present the results for the imaginary part of the transverse polarization function. The figures \ref{fig:RelDiffNum_tr_mu65mu30_imprext}, \ref{fig:RelDiffNum_tr_mu65mu30_Gnorm}, and \ref{fig:RelDiffNum_tr_mu65mu30_Gfull} show the percentage relative difference defined as in formula \eqref{eq:percrelfdiff} for the improved extended hydrodynamic approximation \eqref{eq:trimprexthydroapprox}, the approximation using the normalized IR-AdS$_2$ correlator \eqref{eq:trnormapprox}, and the approximation using the full IR-AdS$_2$ correlator \eqref{eq:trfullapprox} respectively.

We can confront these plots with the ones presented in section \ref{sec:exactresults} regarding the hydrodynamic (top row of figure \ref{fig:RelDiffNum_tr_mu65mu30}) and extended hydrodynamic (bottom row of figure \ref{fig:RelDiffNum_tr_mu65mu30}) approximation.

From these plots, we can infer that all the IR-based approximations presented in this section improve the hydrodynamic approximation. However, none of them is significantly better than the extended hydrodynamic approximation even though the improved extended hydrodynamic approximation is the most similar to the extended one. Overall, once the hydrodynamic approximation \eqref{app:eq:trhydroapprox} is improved to \eqref{app:eq:trexthydroapprox}, \eqref{eq:trimprexthydroapprox}, \eqref{eq:trnormapprox}, or \eqref{eq:trfullapprox} in order to incorporate more of the infrared analytic structure, the agreement with the exact numerical correlator becomes uniformly better over the full extended hydrodynamic regime.  Among these improved descriptions, the extended hydrodynamic approximation \eqref{app:eq:trexthydroapprox} appears to offer the best compromise between simplicity and
accuracy, since it performs as well as the more sophisticated constructions while remaining technically the easiest to handle. In particular, we can observe that the improved approximation (figure \ref{fig:RelDiffNum_tr_mu65mu30_imprext}) close to $\omega=0$ shows a relative difference of $10\%$ for $k/\mu$ at least 0.1. This behavior is analogous to the one showed by the extended hydrodynamic approximation.

We support these observations with the integrated relative differences
\eqref{eq:chiRL} and \eqref{eq:chiCELL}, shown respectively in
tables~\ref{tab:app:chi650_tr} and \ref{tab:app:s650_tr}, for the five approximations
\eqref{app:eq:trhydroapprox}, \eqref{app:eq:trexthydroapprox},
\eqref{eq:trimprexthydroapprox}, \eqref{eq:trnormapprox}, and
\eqref{eq:trfullapprox}.

The two tables show the same qualitative pattern. The
hydrodynamic approximation gives the largest errors, especially in the first column,
where the frequency is smallest and the logs in $\omega$ due to the AdS$_2$ poles are most important .
The improvement due to the IR-based approximations is therefore most pronounced
at small $\omega/\mu$ and moderate or large $k/\mu$. This is visible in the first
column of both tables, where all IR-based approximations are substantially better
than the purely hydrodynamic one. Among them, the extended-hydrodynamic
approximation is slightly favored in table~\ref{tab:app:chi650_tr}: it gives the
smallest or essentially smallest values in most entries, in particular in the first
column.

The improved approximation gives very similar results, but it does not
produce a systematic further improvement over the extended-hydrodynamic one. The
normalized and full correlator approximations are also better than hydrodynamics,
but their errors are usually slightly larger than those of the extended and improved
approximations.

At larger values of $\omega/\mu$, the differences between the approximations become
less pronounced. This is clear from the last column of both tables, where all
approximations give very close values. In this region the IR corrections do not
change the result as much, and the five approximations tend to become comparable.
The same happens in the bottom row, corresponding to the smallest value of
$k/\mu$, where the hydrodynamic and IR-based approximations nearly coincide.
This indicates that, for this background and in the transverse sector, the main
advantage of the IR-based approximations is concentrated at small frequency and
larger momentum, while at small momentum or large frequency the different
approximations become much harder to distinguish.

\subsection{Longitudinal correlator}

\begin{figure}[htb]
	\centering
	\begin{subfigure}{0.4\textwidth}
		\centering
		\includegraphics[width=\textwidth]{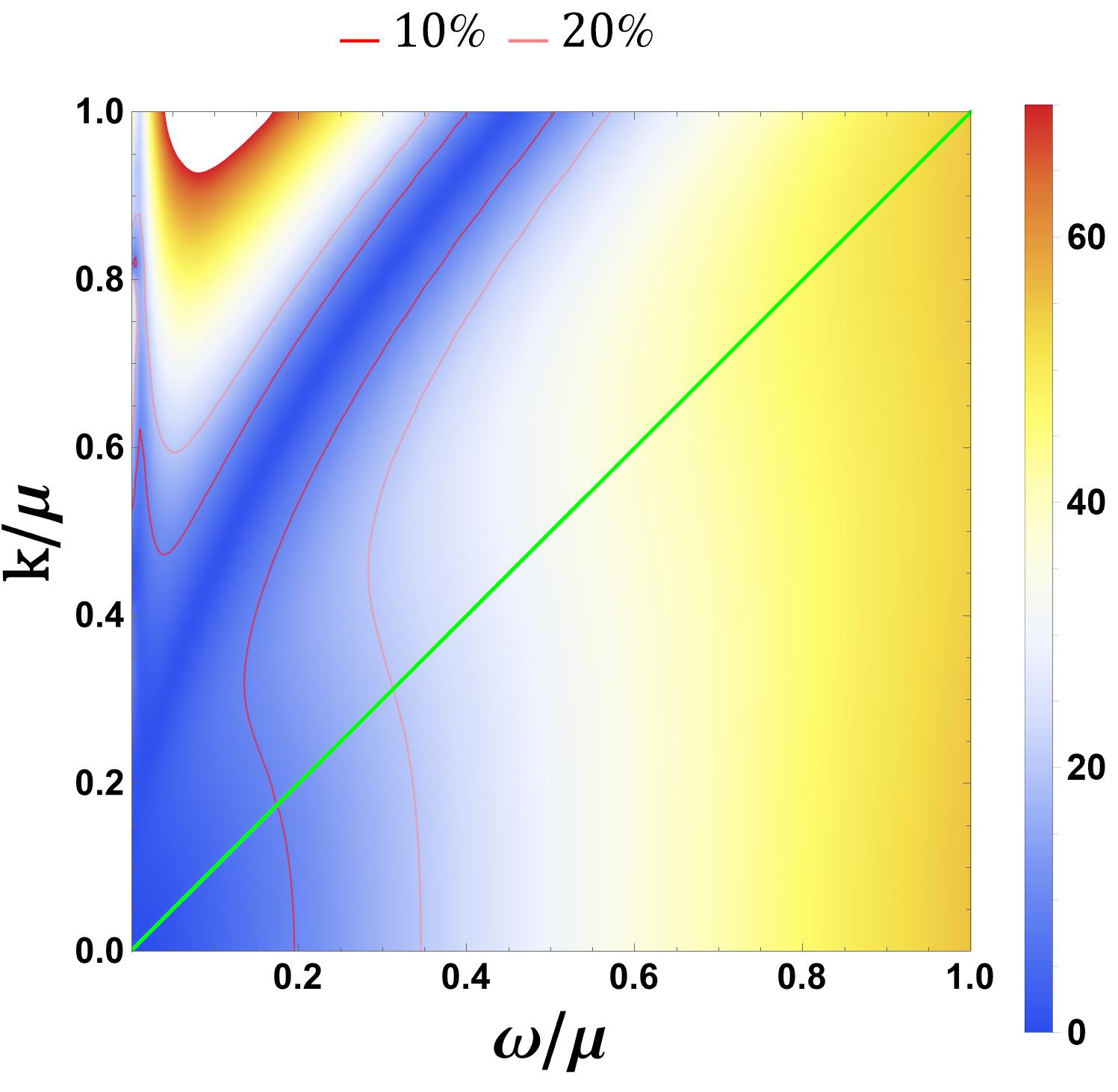}
	\end{subfigure}
	\hspace{1cm}
	\begin{subfigure}{0.4\textwidth}
		\centering
		\includegraphics[width=\textwidth]{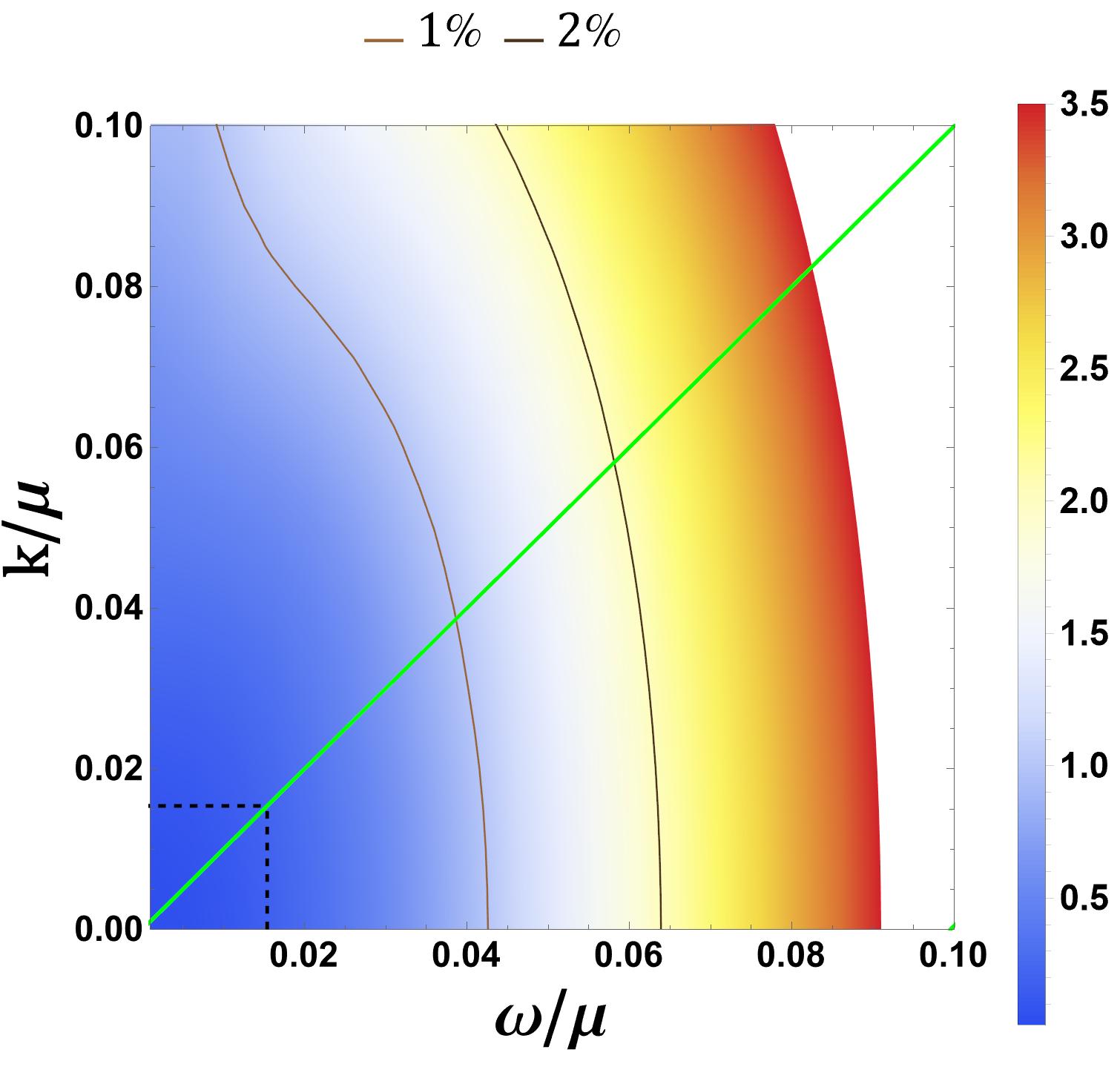}
	\end{subfigure}
	
	\caption{The percentage relative difference \eqref{eq:percrelfdiff} of the imaginary part of the transverse charged current polarization function with respect to the improved extended hydrodynamic approximation \eqref{eq:longimprexthydroapprox}, for $\mu_q/T= 65$ and $\mu_3=0$. The green line shows the locus $\omega = k$. The right plot shows a subregion of the left
		plot. The dashed square in this right plot represents the standard hydrodynamic region $\omega/\mu,k/\mu \in [0,T/\mu]$.} \label{fig:RelDiffNum_long_mu65mu30_imprext}
\end{figure}

\begin{figure}[htb]
	\centering
	\begin{subfigure}{0.4\textwidth}
		\centering
		\includegraphics[width=\textwidth]{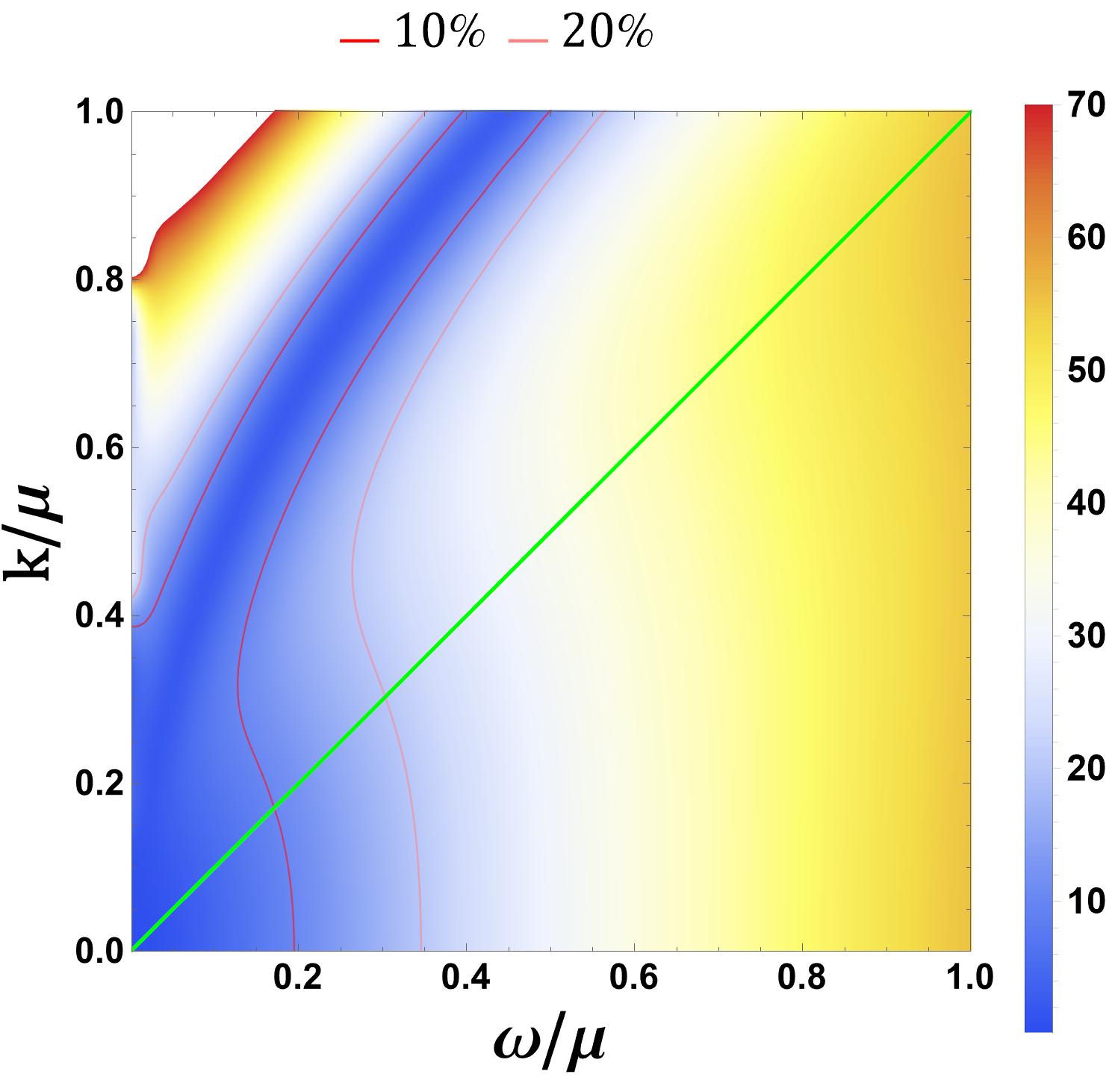}
	\end{subfigure}
	\hspace{1cm}
	\begin{subfigure}{0.4\textwidth}
		\centering
		\includegraphics[width=\textwidth]{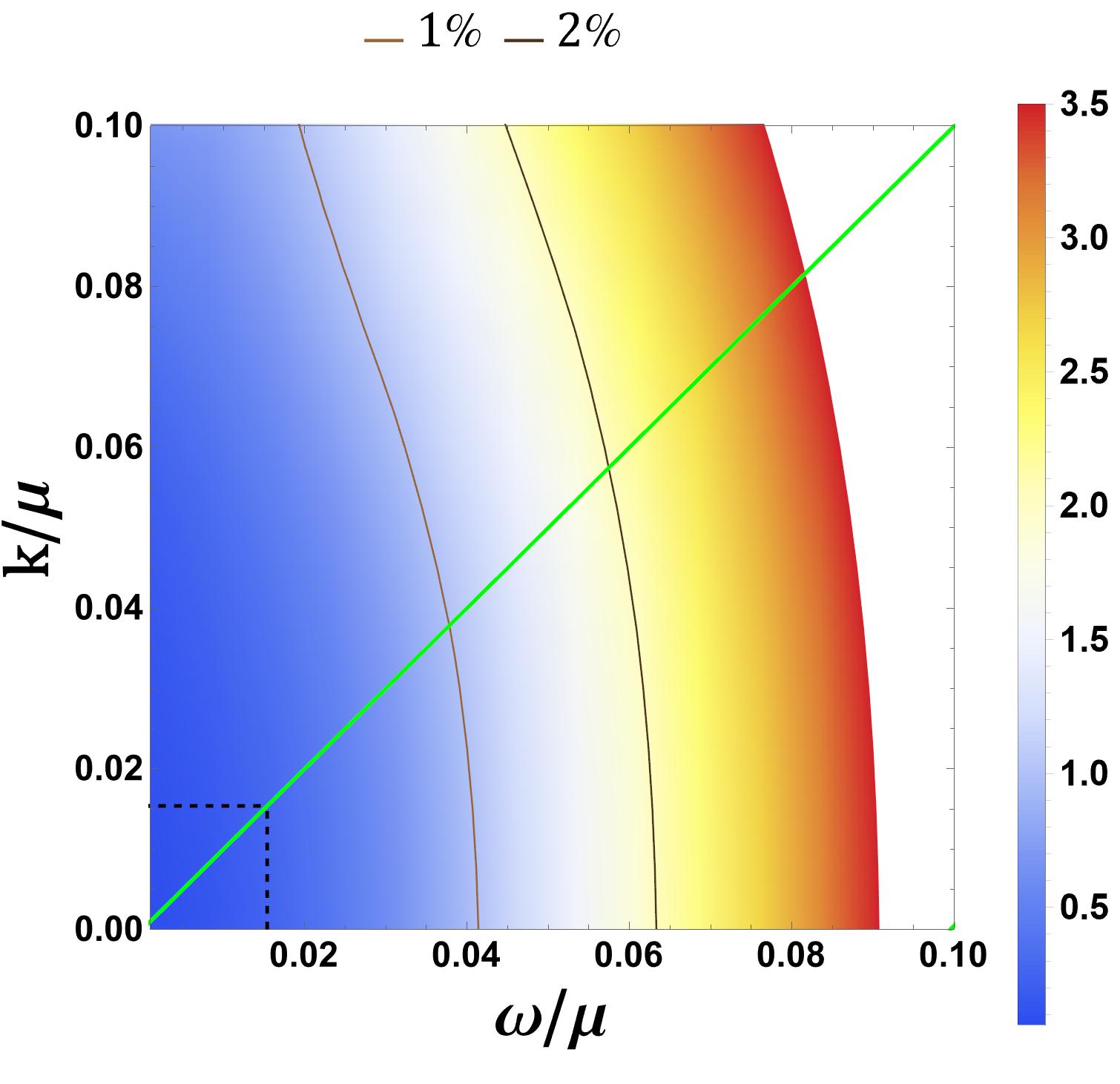}
	\end{subfigure}
	
	\caption{As figure~\ref{fig:RelDiffNum_long_mu65mu30_imprext} but for the normalized correlator \eqref{eq:trnormapprox} instead of the improved extended hydrodynamic approximation.}	\label{fig:RelDiffNum_long_mu65mu30_Gnorm}
\end{figure}

\begin{figure}[htb]
	\centering
	\begin{subfigure}{0.4\textwidth}
		\centering
		\includegraphics[width=\textwidth]{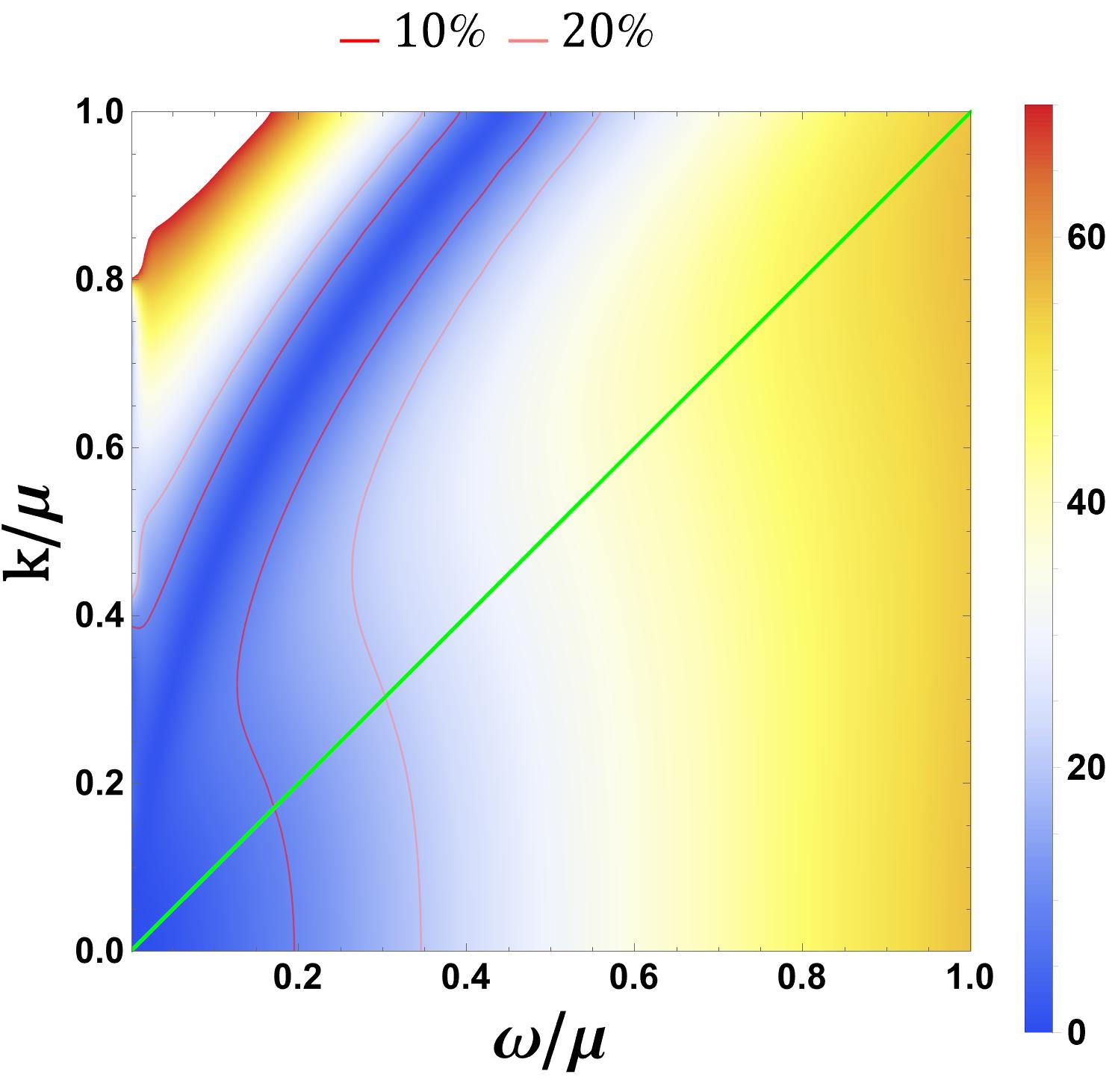}
	\end{subfigure}
	\hspace{1cm}
	\begin{subfigure}{0.4\textwidth}
		\centering
		\includegraphics[width=\textwidth]{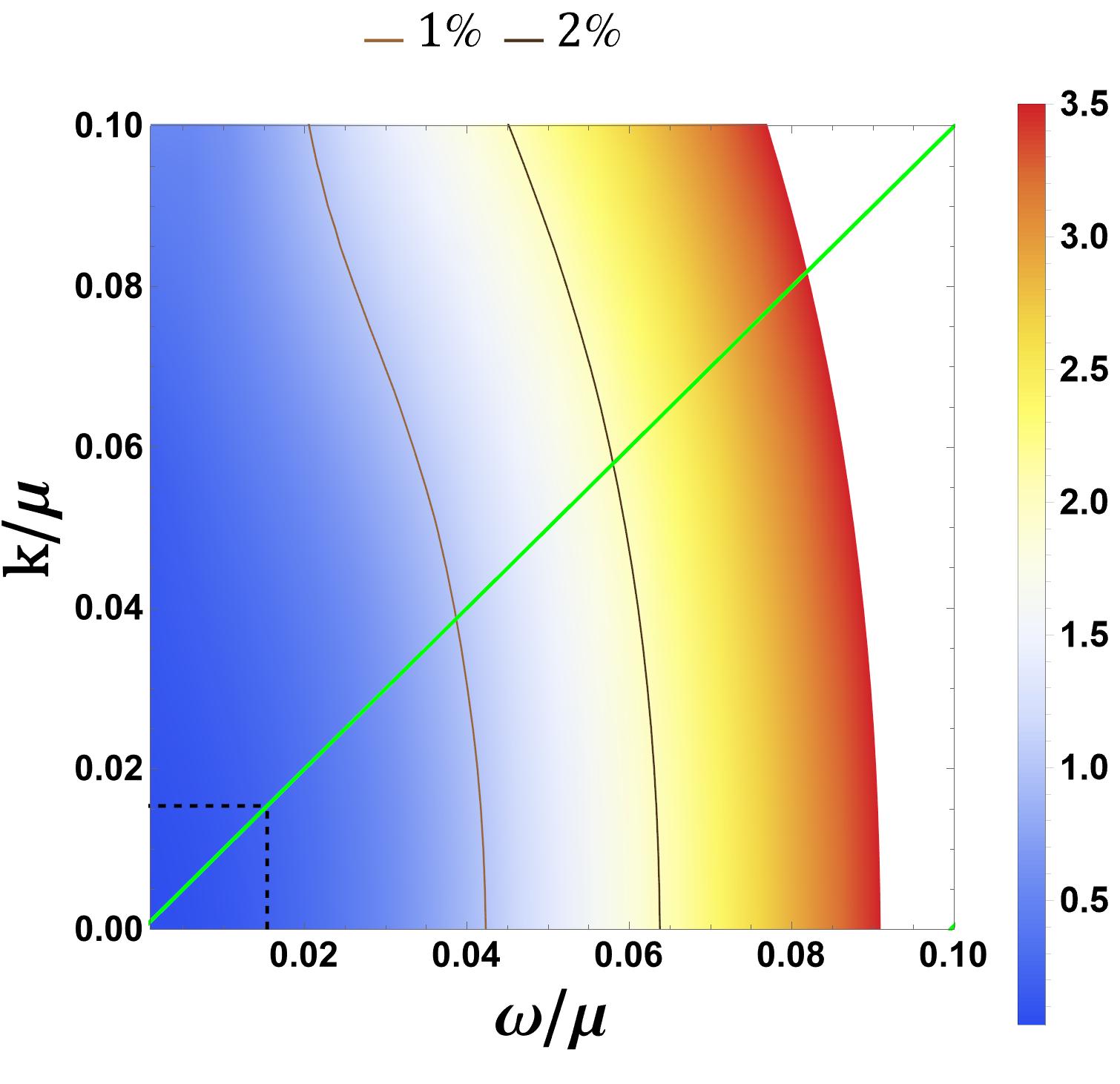}
	\end{subfigure}
	
	\caption{As figure~\ref{fig:RelDiffNum_long_mu65mu30_imprext} but for the full correlator \eqref{eq:longfullapprox} instead of the improved extended hydrodynamic approximation.}\label{fig:RelDiffNum_long_mu65mu30_Gfull}
\end{figure}

\begin{table}[h!]
	\centering
	{
		\small
		\begin{tabular}{c|cccc}
			\multicolumn{5}{c}{(a) hydrodynamic}\\
			$1/2$ & 0.09 & 0.09 & 0.12 & 0.16 \\
			$3/8$ & 0.04 & 0.07 & 0.11 & 0.15 \\
			$1/4$ & 0.03 & 0.06 & 0.10 & 0.14 \\
			$1/8$ & 0.02 & 0.06 & 0.10 & 0.14 \\\hline
			& $1/8$ & $1/4$ & $3/8$ & $1/2$ \\
		\end{tabular}\hspace{1cm}
		\begin{tabular}{c|cccc}
			\multicolumn{5}{c}{(b) extended-hydrodynamic}\\
			$1/2$ & 0.04 & 0.08 & 0.12 & 0.16 \\
			$3/8$ & 0.04 & 0.08 & 0.12 & 0.15 \\
			$1/4$ & 0.03 & 0.07 & 0.11 & 0.14 \\
			$1/8$ & 0.02 & 0.06 & 0.10 & 0.14 \\\hline
			& $1/8$ & $1/4$ & $3/8$ & $1/2$ \\
		\end{tabular}
		
		\begin{tabular}{c|cccc}
			\multicolumn{5}{c}{(c) improved}\\
			$1/2$ & 0.04 & 0.08 & 0.12 & 0.15 \\
			$3/8$ & 0.04 & 0.07 & 0.11 & 0.15 \\
			$1/4$ & 0.03 & 0.07 & 0.11 & 0.14 \\
			$1/8$ & 0.02 & 0.06 & 0.10 & 0.14 \\\hline
			& $1/8$ & $1/4$ & $3/8$ & $1/2$ \\
		\end{tabular}\hspace{1cm}
		\begin{tabular}{c|cccc}
			\multicolumn{5}{c}{(d) normalized}\\
			$1/2$ & 0.05 & 0.08 & 0.12 & 0.16 \\
			$3/8$ & 0.04 & 0.08 & 0.12 & 0.15 \\
			$1/4$ & 0.03 & 0.07 & 0.11 & 0.14 \\
			$1/8$ & 0.02 & 0.06 & 0.10 & 0.14 \\\hline
			& $1/8$ & $1/4$ & $3/8$ & $1/2$ \\
		\end{tabular}
		
		\begin{tabular}{c|cccc}
			\multicolumn{5}{c}{(e) full-IR}\\
			$1/2$ & 0.05 & 0.08 & 0.12 & 0.16 \\
			$3/8$ & 0.04 & 0.07 & 0.11 & 0.15 \\
			$1/4$ & 0.03 & 0.07 & 0.11 & 0.14 \\
			$1/8$ & 0.02 & 0.06 & 0.10 & 0.14 \\\hline
			& $1/8$ & $1/4$ & $3/8$ & $1/2$ \\
		\end{tabular}
	}
	\caption{Values $\chi_{ij}$ (formula \eqref{eq:chiRL}) in the longitudinal sector for $\mu_q/T=65$ and $\mu_3=0$.}
	\label{tab:app:chi650_long}
\end{table}

\begin{table}[h!]
	\centering
	{
		\small
		\begin{tabular}{c|cccc}
			\multicolumn{5}{c}{(a) hydrodynamic}\\
			$1/2$ & 0.23 & 0.07 & 0.18 & 0.27 \\
			$3/8$ & 0.08 & 0.09 & 0.18 & 0.27 \\
			$1/4$ & 0.03 & 0.10 & 0.18 & 0.26 \\
			$1/8$ & 0.02 & 0.10 & 0.18 & 0.26 \\\hline
			& $1/8$ & $1/4$ & $3/8$ & $1/2$ \\
		\end{tabular}\hspace{1cm}
		\begin{tabular}{c|cccc}
			\multicolumn{5}{c}{(b) extended-hydrodynamic}\\
			$1/2$ & 0.04 & 0.12 & 0.18 & 0.27 \\
			$3/8$ & 0.04 & 0.10 & 0.20 & 0.27 \\
			$1/4$ & 0.03& 0.10 & 0.18 & 0.26 \\
			$1/8$ & 0.03 & 0.10 & 0.18 & 0.26 \\\hline
			& $1/8$ & $1/4$ & $3/8$ & $1/2$ \\
		\end{tabular}
		
		\begin{tabular}{c|cccc}
			\multicolumn{5}{c}{(c) improved}\\
			$1/2$ & 0.05 & 0.12 & 0.22 & 0.28 \\
			$3/8$ & 0.04 & 0.12 & 0.20 & 0.27 \\
			$1/4$ & 0.04 & 0.11 & 0.19 & 0.26 \\
			$1/8$ & 0.03 & 0.10 & 0.18 & 0.26 \\\hline
			& $1/8$ & $1/4$ & $3/8$ & $1/2$ \\
		\end{tabular}\hspace{1cm}
		\begin{tabular}{c|cccc}
			\multicolumn{5}{c}{(d) normalized}\\
			$1/2$ & 0.08 & 0.13 & 0.23 & 0.30 \\
			$3/8$ & 0.05 & 0.14 & 0.21 & 0.28 \\
			$1/4$ & 0.04 & 0.11 & 0.19 & 0.26 \\
			$1/8$ & 0.03 & 0.10 & 0.18 & 0.26 \\\hline
			& $1/8$ & $1/4$ & $3/8$ & $1/2$ \\
		\end{tabular}
		
		\begin{tabular}{c|cccc}
			\multicolumn{5}{c}{(e) full-IR}\\
			$1/2$ & 0.09 & 0.12 & 0.21 & 0.28 \\
			$3/8$ & 0.05 & 0.13 & 0.20 & 0.27 \\
			$1/4$ & 0.04 & 0.11 & 0.18 & 0.26 \\
			$1/8$ & 0.03 & 0.10 & 0.18 & 0.26 \\\hline
			& $1/8$ & $1/4$ & $3/8$ & $1/2$ \\
		\end{tabular}
	}
	\caption{Values $s_{ij}$ (formula \eqref{eq:chiCELL}) in the longitudinal sector for $\mu_q/T=65$ and $\mu_3=0$.}
	\label{tab:app:s650_long}
\end{table}

In this section, we present the results for the longitudinal spectral function. The figures \ref{fig:RelDiffNum_long_mu65mu30_imprext}, \ref{fig:RelDiffNum_long_mu65mu30_Gnorm}, and \ref{fig:RelDiffNum_long_mu65mu30_Gfull} show the percentage relative difference \eqref{eq:percrelfdiff} for the improved extended hydrodynamic approximation \eqref{eq:longimprexthydroapprox}, the approximation using the normalized IR-AdS$_2$ correlator \eqref{eq:longnormapprox}, and the approximation using the full IR-AdS$_2$ correlator \eqref{eq:longfullapprox} respectively.

We can confront these plots with the ones presented in section \ref{sec:exactresults} regarding the hydrodynamic (top row of figure \ref{fig:RelDiffNum_long_mu65mu30}) and extended hydrodynamic (bottom row of figure \ref{fig:RelDiffNum_long_mu65mu30}) approximation. As for the transverse case, all the IR-based approximation presented in this section improve the hydrodynamic approximation. None of them does significantly better than the extended hydrodynamic approximation. Therefore, also for the longitudinal sector, the extended hydrodynamic approximation \eqref{app:eq:longexthydroapprox} results in the best compromise between simplicity an accuracy, among the IR-based hydrodynamic approximations.

For the longitudinal correlator, we provide the results for the integrated
relative differences \eqref{eq:chiRL} and \eqref{eq:chiCELL} for the five
approximations \eqref{app:eq:longhydroapprox}, \eqref{app:eq:longexthydroapprox},
\eqref{eq:longimprexthydroapprox}, \eqref{eq:longnormapprox}, and
\eqref{eq:longfullapprox}. These results are shown in
tables~\ref{tab:app:chi650_long} and \ref{tab:app:s650_long}. The two tables give
a consistent picture: the hydrodynamic approximation is less accurate mainly at
small frequencies and large momenta, while the differences between the
approximations become much smaller at low momentum and at larger values of
$\omega/\mu$.

In table~\ref{tab:app:chi650_long}, the improvement over hydrodynamics is most
visible in the first column. For example, for
the hydrodynamic approximation $\chi_{11}=0.09$, while the extended and
improved approximations reduce this value to $0.04$. Away from this
small-frequency region, however, the five approximations become very close. In
particular, in the bottom row, corresponding to the smallest value of $k/\mu$, all
approximations essentially coincide, and the same is true in the last column.

Table~\ref{tab:app:s650_long} shows the same main low-frequency feature, but the
pattern at larger $\omega/\mu$ is more delicate. In the first column, the extended
hydrodynamic approximation gives the clearest improvement over hydrodynamics. The improved, normalized, and full
approximations also improve over hydrodynamics in this small-frequency region, but
not as uniformly as the extended one. For the remaining columns of table~\ref{tab:app:s650_long}, instead, the
IR-based approximations do not systematically improve the result. In
the third column, and especially in the last column, all approximations give very
similar values. In this region the hydrodynamic
approximation is already as good as the IR-based ones, and in several entries it is
slightly better. This is why the advantage of the extended hydrodynamic
approximation in the $s_{ij}$ table should be understood as a low-frequency effect,
rather than as a uniform improvement over the whole grid.

Among the IR-based approximations, the extended and improved extended
hydrodynamic approximations are the most accurate overall. The improved
approximation is slightly favored in table~\ref{tab:app:chi650_long}, especially in
the first two rows, while in table~\ref{tab:app:s650_long} the extended
hydrodynamic approximation performs marginally better in the first column. The
normalized and full correlator approximations do not lead to a systematic
improvement over the extended or improved approximations. They remain close to
them in many entries, but they tend to be slightly worse in the region of larger
$k/\mu$ and intermediate or large $\omega/\mu$. Overall, for the longitudinal sector
at $\mu_q/T=65$ and $\mu_3=0$, the gain from the IR-based approximations is
concentrated at small frequency and large momentum, while at small momentum all
approximations become essentially indistinguishable.

\clearpage

\section{Numerical results for the charged current correlators at other values of $\mu_q/T$ and $\mu_3/\mu_q$}\label{app:otherexactresults}
In this appendix, we extend the discussion of section \ref{sec:exactresults} where we presented the numerical, exact results for the imaginary part of the charged polarization functions to backgrounds characterized by $\mu_q/T\in\{10^4,5\}$. As explained in the main text, these values are relevant for application to neutron star physics since $\mu_q/T=10^4$ is a typical value for old/cold neutron stars, while $\mu_q/T = 5$ can be reached in neutron star mergers \cite{Glendenning:1992vb,Jaikumar:2002vg,Buballa:2003qv,Steiner:2004fi,Jaikumar:2005hy,Alford:2007xm,Dexheimer:2008ax,Belvedere:2012qn,Alford:2018lhf,Roark:2018uls}.

It is also interesting to consider these two values, since the former describe a system very close to extremality, while the latter probes the system rather away from extremality, where thermal effects could be more relevant. These two values of $\mu_q/T$ will be considered  at zero isospin chemical potential and finite isospin chemical $\mu_3/\mu_q=-0.1$. In section \ref{app:mu305} of this appendix, we shall push the value of the isospin chemical potential to $\mu_3/\mu_q= -0.5$ to show the departure from hydrodynamics even at small frequencies and momenta due to the presence of a $\mu_3$ which is not much smaller than the hard scale $\mu$.

This appendix will be organized as follows: sections \ref{app:mu30}, \ref{app:mu301}, and \ref{app:mu305} are associated with the values of isospin chemical potential $\mu_3/\mu_q=0$, $\mu_3/\mu_q=-0.1$, and $\mu_3/\mu_q=-0.5$ respectively. In each one of these sections, we first present the plots for the transverse and longitudinal imaginary part of the charged current polarization function as a function of $\omega/\mu$ and $k/\mu$. Then, we present the plots of the relative difference (see formula \eqref{eq:percrelfdiff} in the main text) with the (near-extremal) hydrodynamic and extended hydrodynamic approximations \eqref{eq:trhydroapprox}-\eqref{eq:longexthydroapprox}. For each section, we present the results for backgrounds characterized by $\mu_q/T\in\{10^4,5\}$. In section \ref{app:mu305}, we also consider the case of $\mu_q/T=65$ at $\mu_3/\mu_q=-0.5$, which was not discussed in the main text.

\subsection{Numerical results at $\mu_3=0$}\label{app:mu30}
\begin{figure}[H]
	\centering
	\begin{subfigure}{0.48\textwidth}
		\centering
		\includegraphics[width=\textwidth]{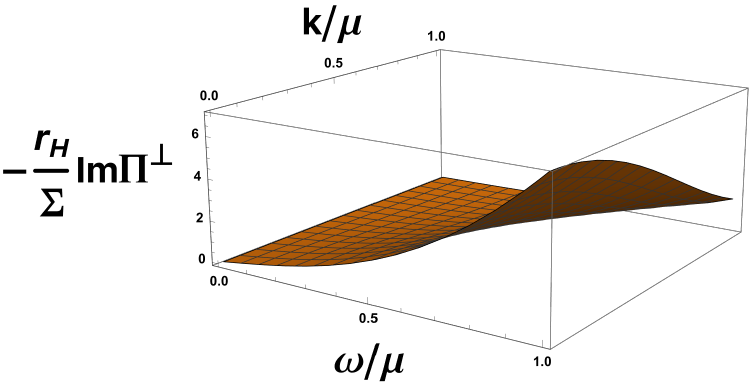}
	\end{subfigure}
	\hfill
	\begin{subfigure}{0.48\textwidth}
		\centering
		\includegraphics[width=\textwidth]{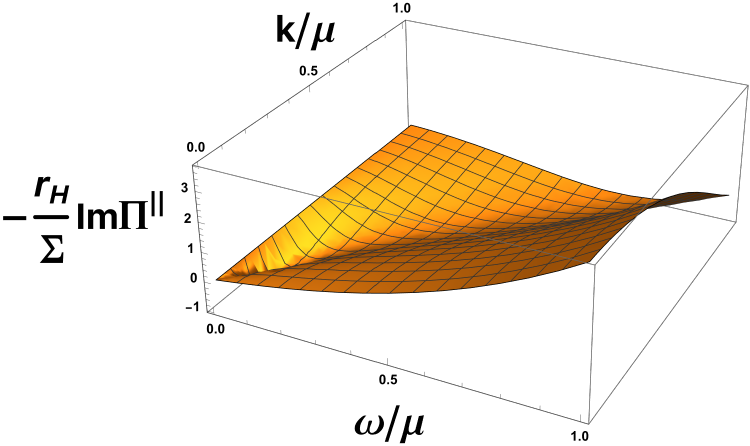}
	\end{subfigure}
	\caption{Imaginary part of the transverse (left panel) and longitudinal (right panel) charged current retarded polarization function (normalized by $-\Sigma/r_H$) for $\mu_q/T = 10^4$, and $\mu_3=0$. The $\pm$ notation has been suppressed since we are at zero isospin chemical potential.} \label{fig:pol1040}
\end{figure}

In this section, we present the results for the imaginary part of the charged current polarization function at zero isospin chemical potential for $\mu_q/T\in\{10^4,5\}$. First, we present the plots for the imaginary part of the polarization functions in figures \ref{fig:pol1040} ($\mu_q/T=10^4$) and \ref{fig:pol50} ($\mu_q/T=5$). Then, we show the plots of the relative difference \eqref{eq:percrelfdiff} at $\mu_3=0$ for  $\mu_q/T =10^4$ (figure \ref{fig:RelDiffNum_tr_mu104mu30}, and \ref{fig:RelDiffNum_long_mu104mu30} for the transverse and longitudinal sector respectively), and $\mu_q/T=5$ (figure \ref{fig:RelDiffNum_tr_mu5mu30}, and \ref{fig:RelDiffNum_long_mu5mu30} for the transverse and longitudinal sector respectively). The top and bottom rows show respectively the relative difference with respect to the hydrodynamic and the extended hydrodynamic approximation.
As for the case of $\mu_q/T=65$ in the main text in section \ref{sec:exactresults}, these plots show that, overall, the extended hydrodynamic approximation approximates better the exact, numerical results, compared to the  (near-extremal) hydrodynamic approximation.

\subsubsection{Background at $\mu_q/T=10^4$}

In figure \ref{fig:pol1040}, we present the plots for the imaginary part of the transverse (left panel) and longitudinal (right panel) polarization functions at $\mu_q/T=10^4$ and zero isospin chemical potential. The qualitative behavior of these plots is the same of the one displayed for $\mu_q/T=65$ and $\mu_3=0$ in section \ref{sec:exactresults}. Both the transverse and longitudinal polarization functions grow for large $\omega$, while in the longitudinal sector, we find the usual diffusive pole. Note that, the magnitude of the polarization functions is somewhat  larger with respect to the case of $\mu_q/T = 65$ (see figures \ref{fig:transversepol}, \ref{fig:longitudinalpol}).

In the following, we present the results for the comparison of the exact numerical results with the hydrodynamic and the extended hydrodynamic approximation.

\subsubsection*{Transverse correlator}
\begin{figure}[H]
	\begin{subfigure}{\textwidth}
		\centering
		\begin{subfigure}{0.4\textwidth}
			\centering
			\includegraphics[width=\textwidth]{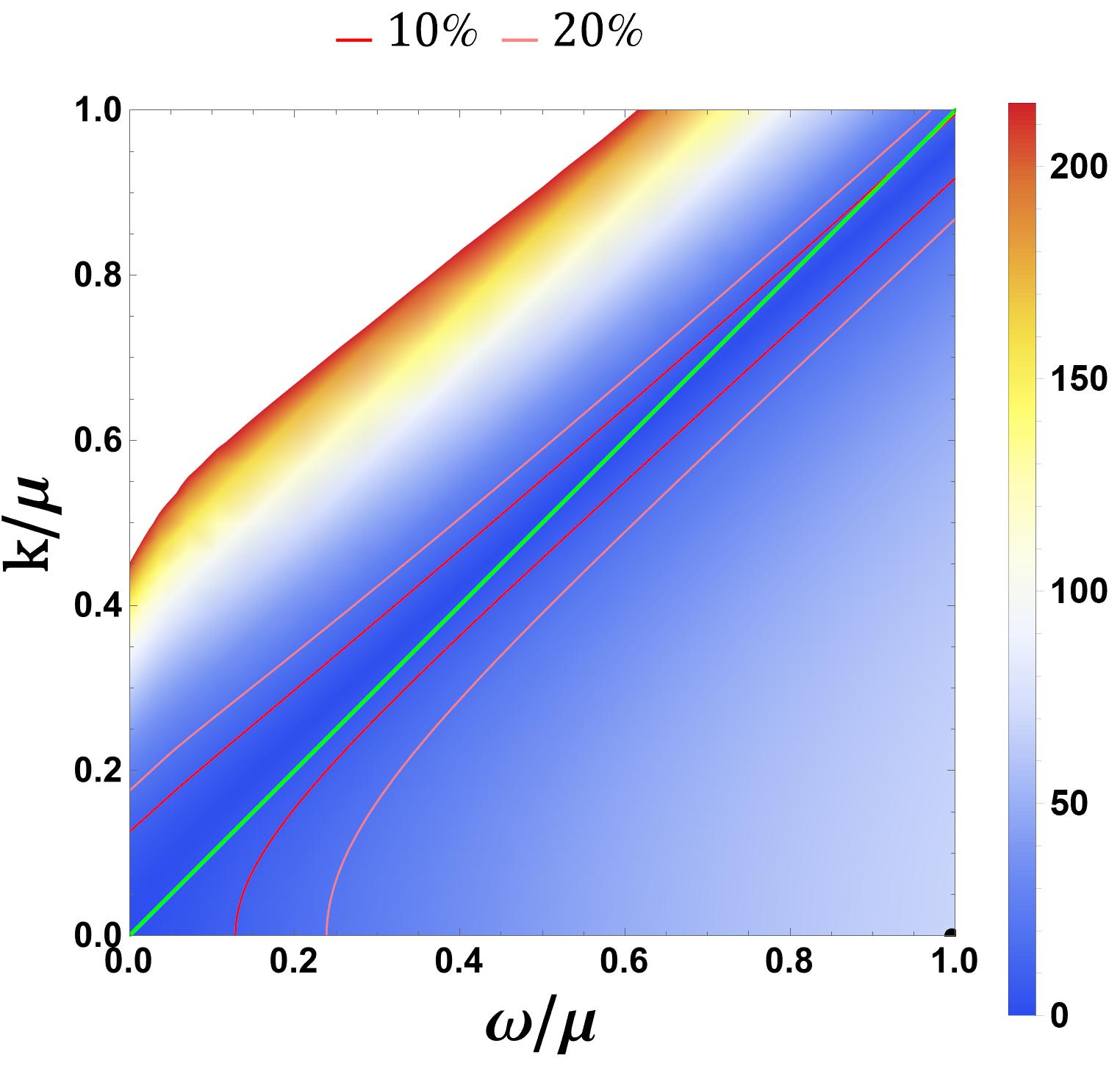}
		\end{subfigure}
		\hspace{1cm}
		\begin{subfigure}{0.4\textwidth}
			\centering
			\includegraphics[width=\textwidth]{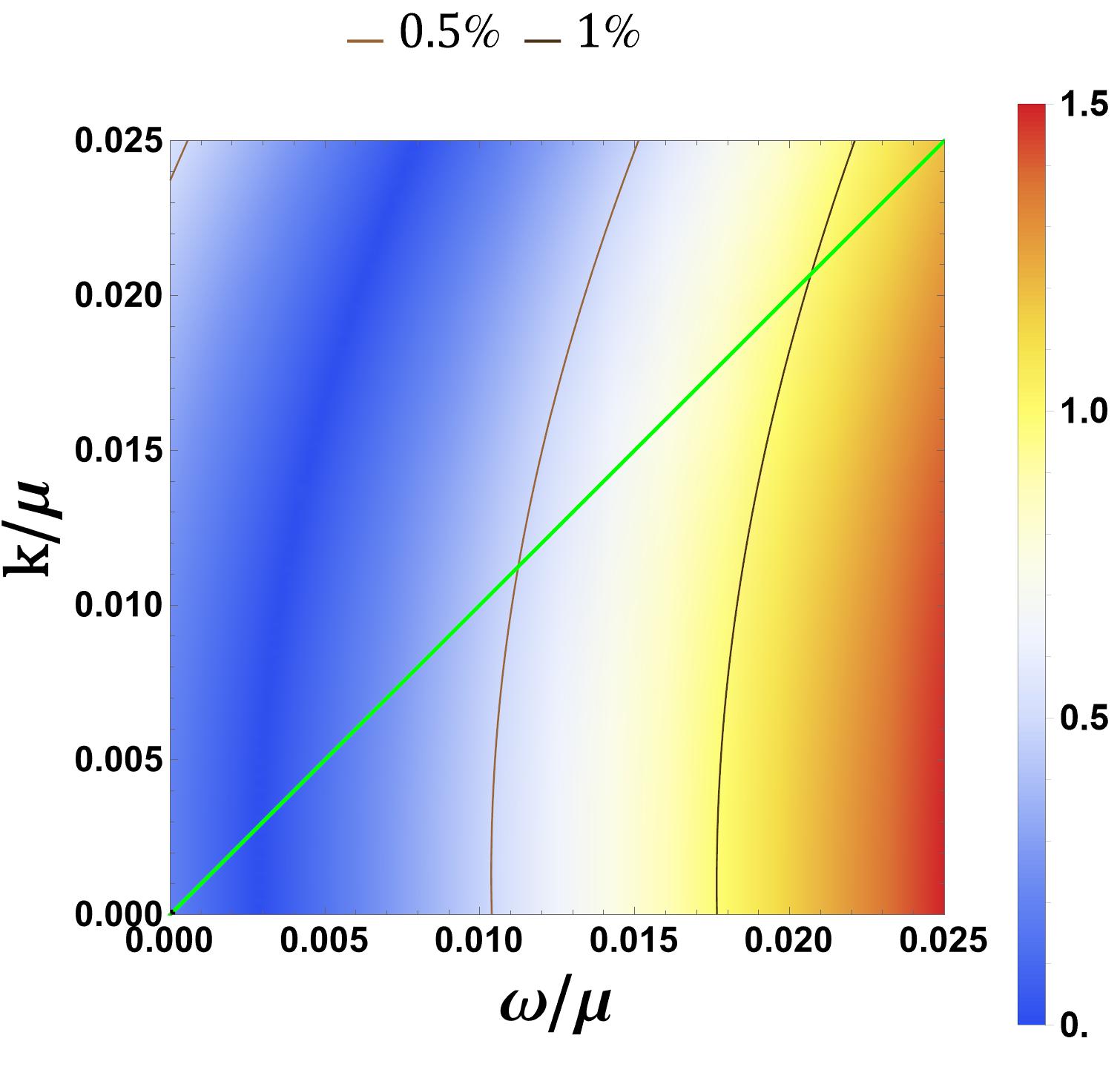}
		\end{subfigure}
		\label{fig:RelDiffNum_tr_mu104mu30_hydro}
	\end{subfigure}
	\begin{subfigure}{\textwidth}
		\centering
		\begin{subfigure}{0.4\textwidth}
			\centering
			\includegraphics[width=\textwidth]{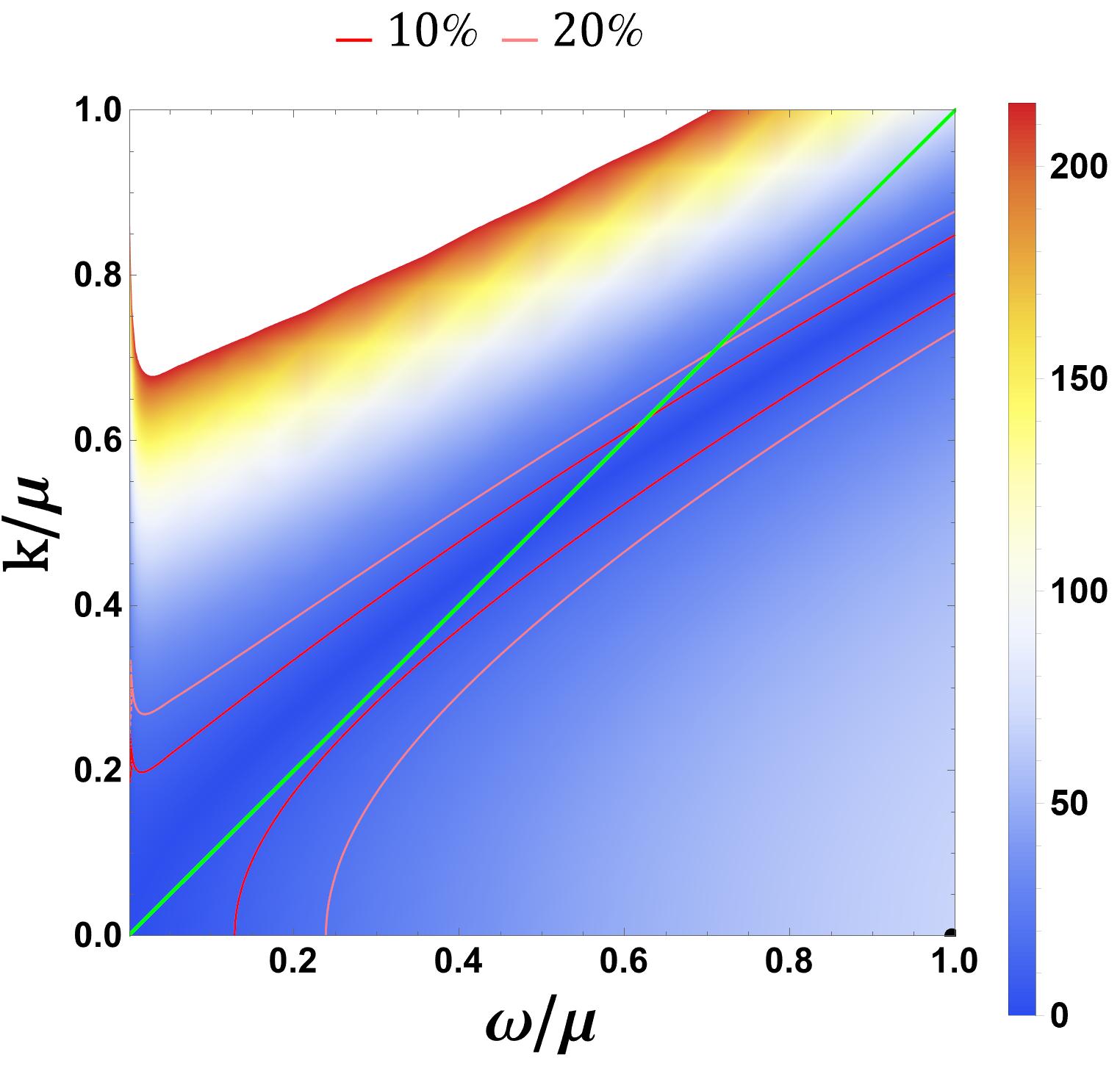}
		\end{subfigure}
		\hspace{1cm}
		\begin{subfigure}{0.4\textwidth}
			\centering
			\includegraphics[width=\textwidth]{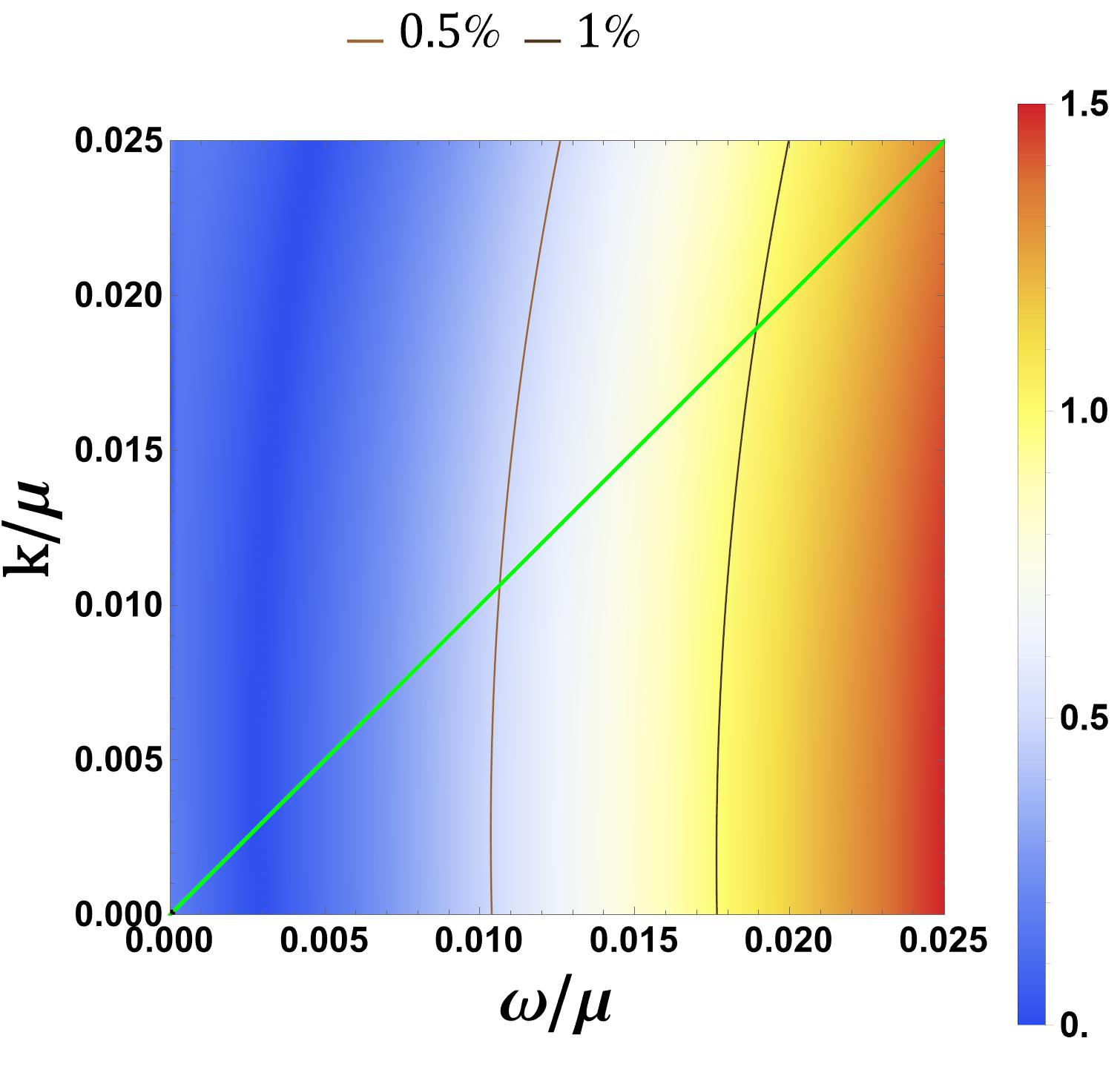}
		\end{subfigure}
		\label{fig:RelDiffNum_tr_mu104mu30_exthydro}
	\end{subfigure}
	\caption{The percentage relative difference \eqref{eq:percrelfdiff} of the imaginary part of the transverse charged current polarization function with respect to the hydrodynamic approximation \eqref{eq:trhydroapprox} (top row) and the extended hydrodynamic approximation \eqref{eq:trexthydroapprox} (bottom row), for $\mu_q/T= 10^4$ and zero $\mu_3$. The right plots shows a subregion of the left
		plots. The green line shows the locus $\omega = k$. The black dot in bottom-right corner of the plots in the left column is the real part of the first AdS$_5$ QNM.} \label{fig:RelDiffNum_tr_mu104mu30}
\end{figure}

In Figure~\ref{fig:RelDiffNum_tr_mu104mu30}, we present the relative difference plots for the hydrodynamic and extended hydrodynamic approximation for the transverse correlator at $\mu_q/T=10^4$ and $\mu_3=0$. The hydrodynamic approximation, shown on the top two plots, already works very well only in a small region around the origin, but this region is still much larger than the naive scale set by $T/\mu \sim 10^{-4}$. In the full-range plot (top-left plot), the hydrodynamic relative difference reaches order one along the line $\omega/\mu = k/\mu -0.3$. The 10\% and 20\% contours cover the slanted band which follows the locus $\omega = k$.

The zoomed plot (top-right plot) shows that the error stays below the percent level for $\omega/\mu\leq 0.01$ and $k/\mu\leq 0.025$, with the contours mainly controlled by $\omega/\mu$ while they remain constant in $k/\mu$.

The extended hydrodynamic approximation (bottom row of figure \ref{fig:RelDiffNum_tr_mu104mu30}) gives a clear improvement in the full-range plot, especially at low frequencies: the large-error region is pushed to larger values of $k/\mu\simeq 0.6$, and the 10\% and 20\% contours move upward in $k/\mu$. In the zoomed region (bottom-right plot), however, the change is much milder, and both approximations have very similar percent-level contours.

For the values of $\mu_q/T=10^4,\mu_3=0$ the first AdS$_5$ pole enters the extended hydrodynamic region, being (barely) visible at the bottom-right corner of the left column of figure \ref{fig:RelDiffNum_tr_mu104mu30}.

\subsubsection*{Longitudinal correlator}
\begin{figure}[H]
	\begin{subfigure}{\textwidth}
		\centering
		\begin{subfigure}{0.4\textwidth}
			\centering
			\includegraphics[width=\textwidth]{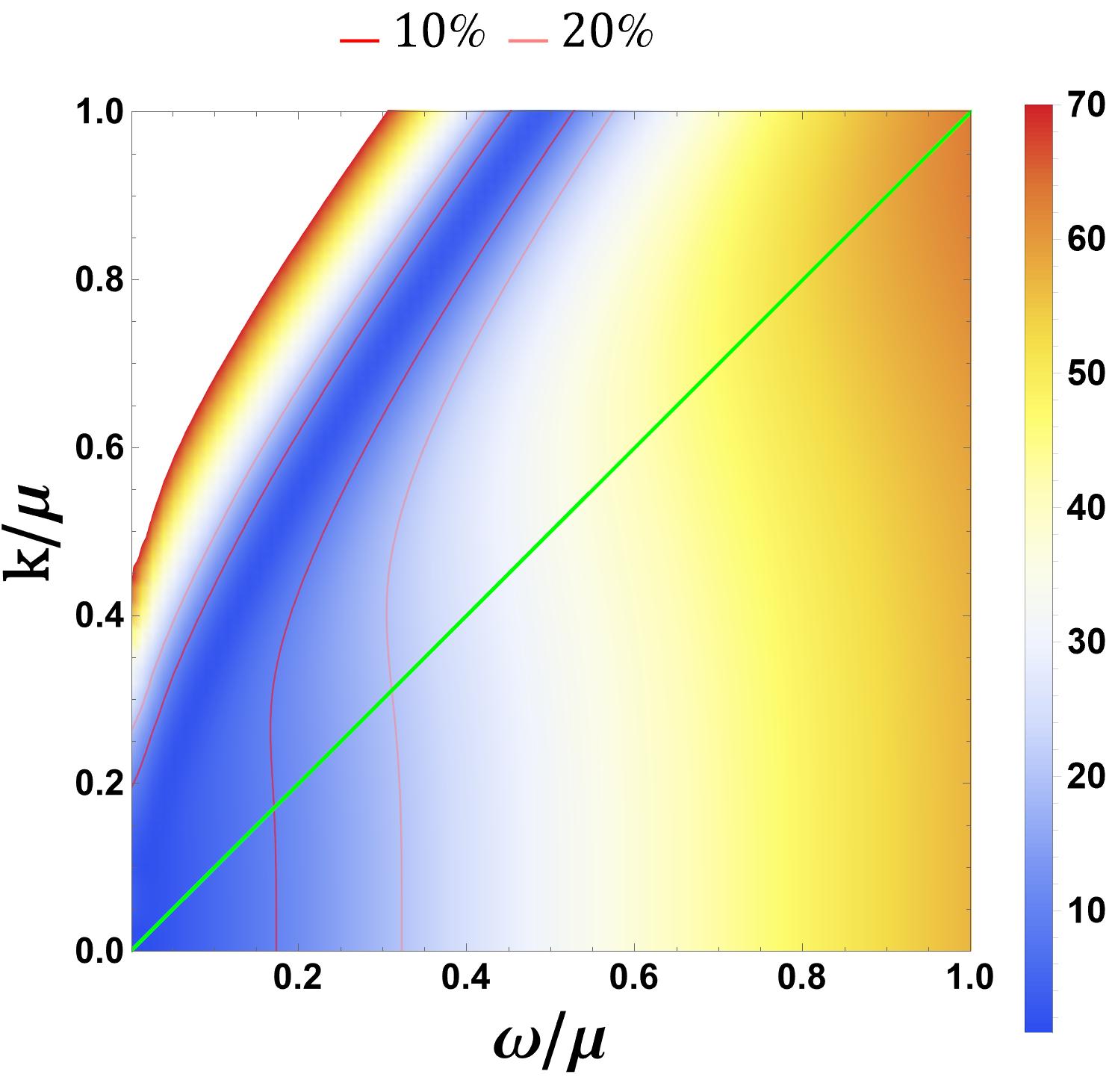}
		\end{subfigure}
		\hspace{1cm}
		\begin{subfigure}{0.4\textwidth}
			\centering
			\includegraphics[width=\textwidth]{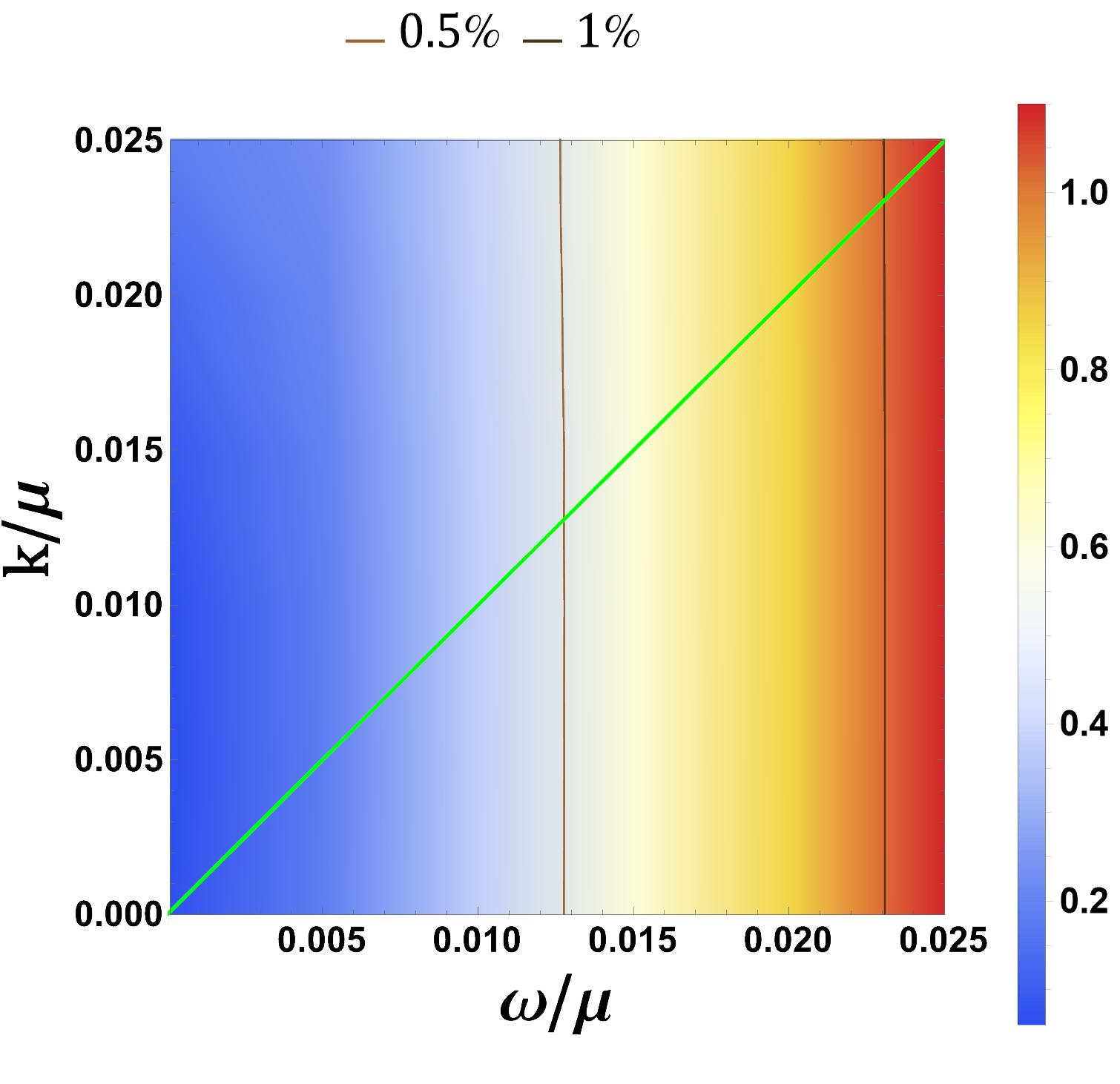}
		\end{subfigure}
	\end{subfigure}

	\begin{subfigure}{\textwidth}
		\centering
		\begin{subfigure}{0.4\textwidth}
			\centering
			\includegraphics[width=\textwidth]{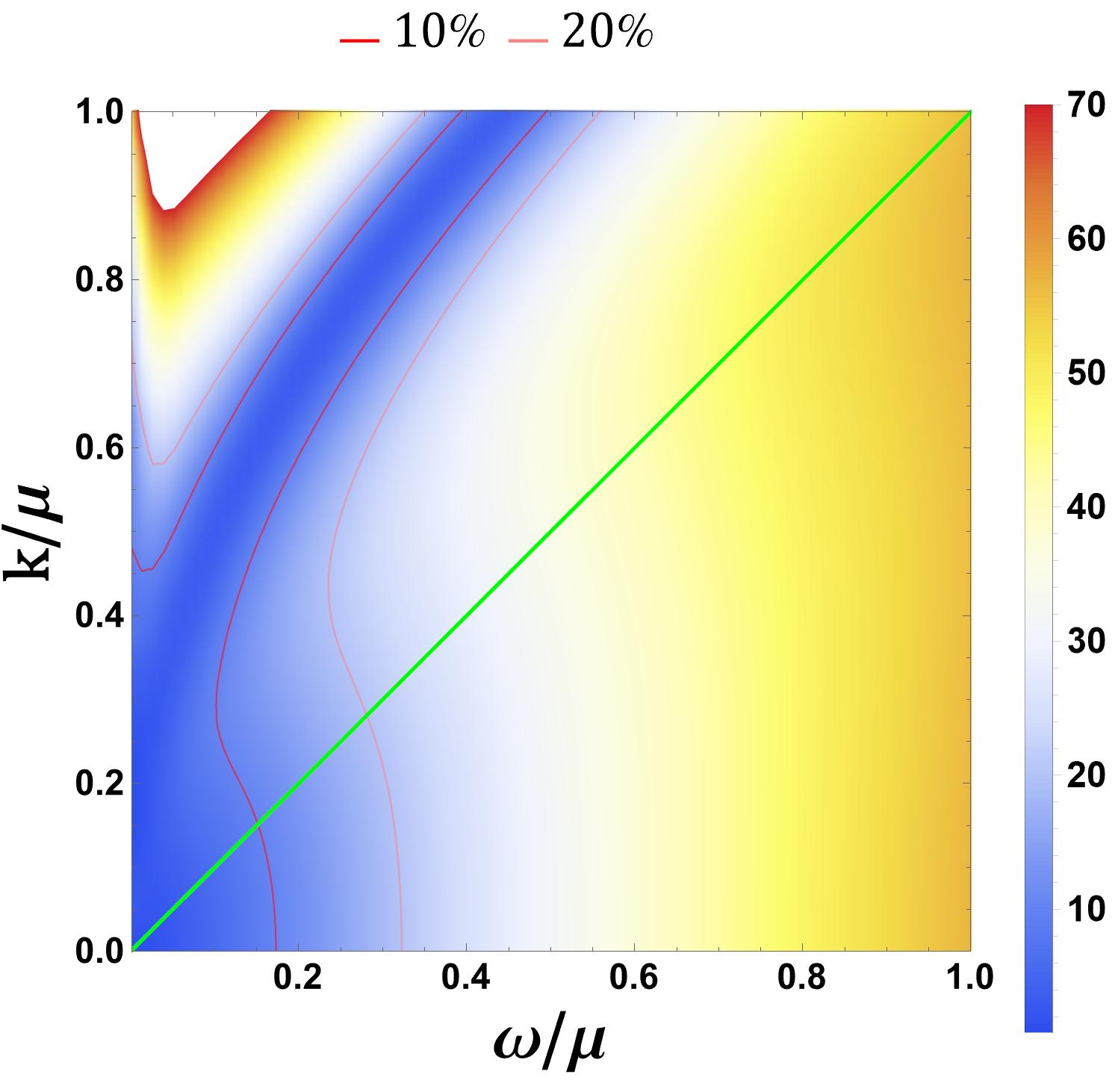}
		\end{subfigure}
		\hspace{1cm}
		\begin{subfigure}{0.4\textwidth}
			\centering
			\includegraphics[width=\textwidth]{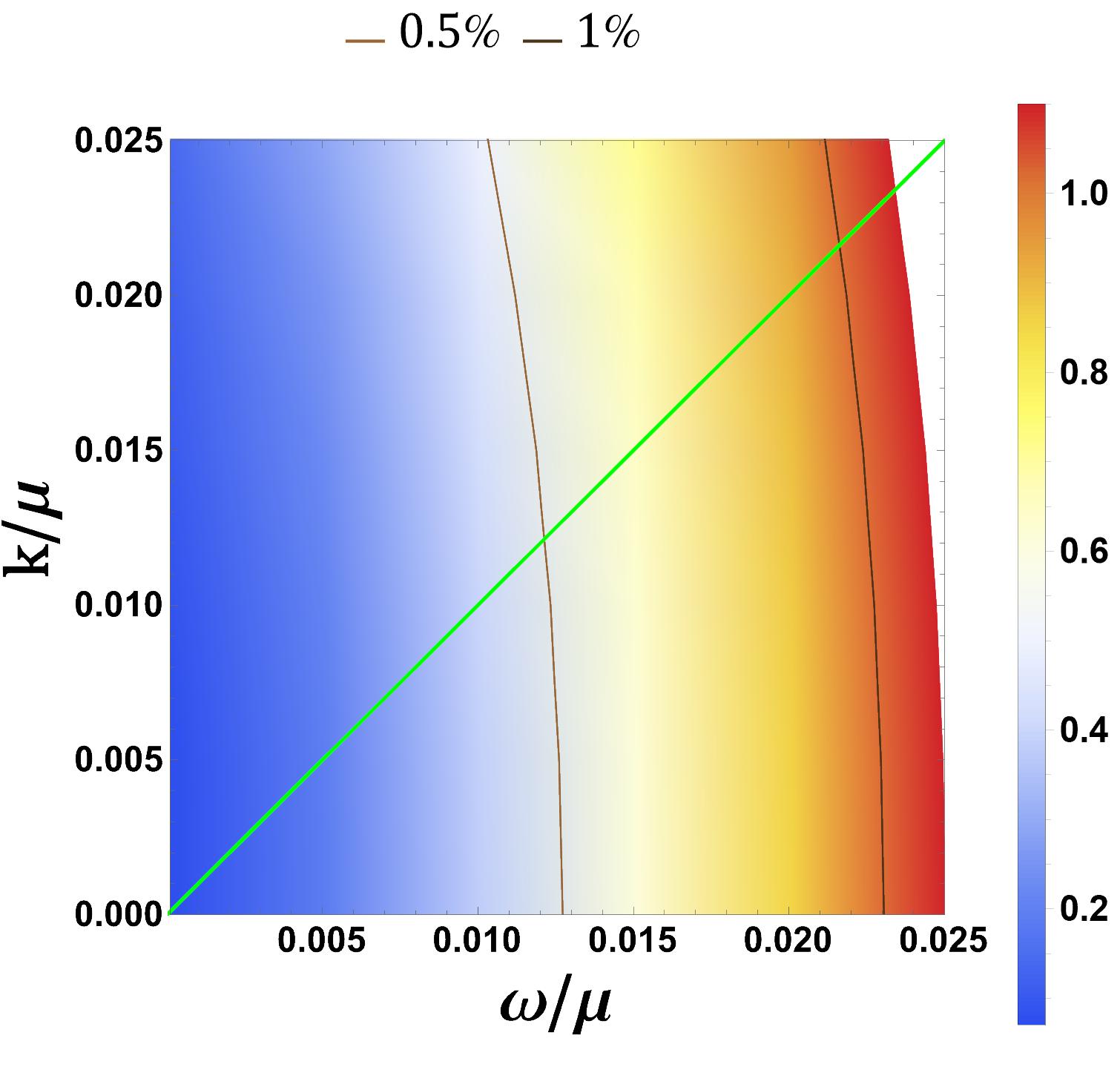}
		\end{subfigure}
	\end{subfigure}

	\caption{As figure \ref{fig:RelDiffNum_tr_mu104mu30} but for the longitudinal polarization function instead of the transverse polarization function.}	\label{fig:RelDiffNum_long_mu104mu30}
\end{figure}

Figure~\ref{fig:RelDiffNum_long_mu104mu30} shows a somewhat different pattern from the transverse case. In the full-range (top-left) plot, the hydrodynamic error grows mainly as $\omega/\mu$ increases, with a high-error band also appearing at large $k/\mu$ and small $\omega/\mu$. In particular, the relative difference for $\omega\simeq 0$ and $k/\mu\gtrsim 0.4$ is larger than $70\%$. The region below the 10\% contour is broadest at small $k/\mu\lesssim 0.3$, while at larger momentum the contour bends toward right and becomes narrower.

The extended hydrodynamic approximation improves the low-frequency region: the region where the error is around $70\%$ at $\omega\simeq 0$ is now pushed to the top-left corner of the plot $k/\mu\simeq 1$ (see the bottom-left plot). Also the 10\% and 20\% regions cover a much wider range, close to $\omega \simeq 0$. At larger frequencies $\omega/\mu\gtrsim 0.5$, both approximations become poor over most values of $k/\mu$.

In the zoomed plots (right column) the error is below the percent level only for $\omega/\mu\lesssim 0.01$--$0.015$, almost independently of $k/\mu$\footnote{This is still an extension of the hydrodynamic validity by a factor of a 100 in frequency as in the same units, $T=10^{-4}$.}, so the limitation is controlled mostly by frequency rather than momentum. There is not much difference between the two approximations apart for value of $k/\mu\gtrsim 0.01$ for which the extended hydrodynamic approximation seems to work slightly worse.

\subsubsection{Background at $\mu_q/T=5$}

\begin{figure}[htb]
	\centering
	\begin{subfigure}{0.48\textwidth}
		\centering
		\includegraphics[width=\textwidth]{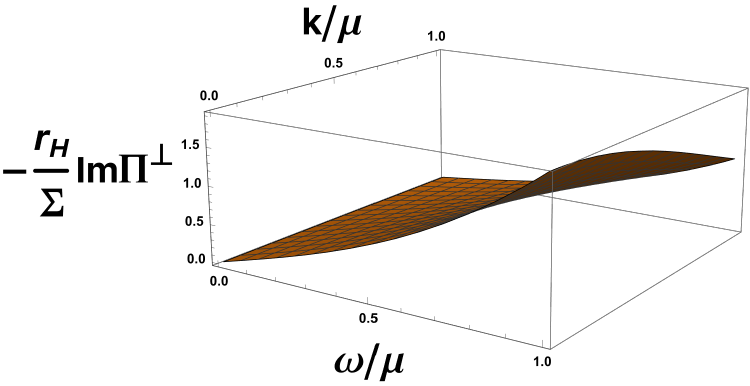}
	\end{subfigure}
	\hfill
	\begin{subfigure}{0.5\textwidth}
		\centering
		\includegraphics[width=\textwidth]{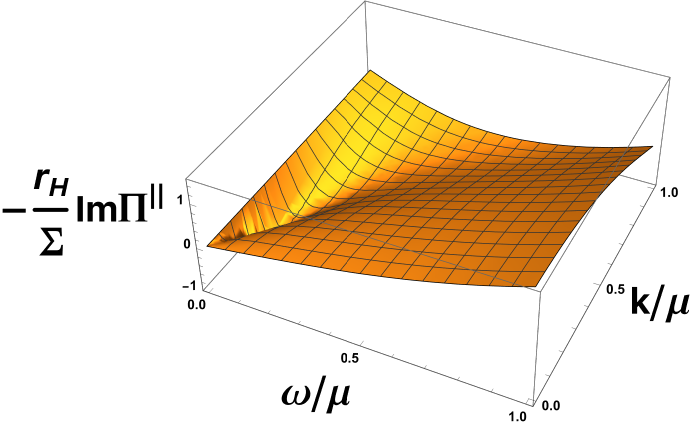}
	\end{subfigure}
	\caption{Imaginary part of the transverse (left panel) and longitudinal (right panel) charged current retarded polarization function (normalized by $-\Sigma/r_H$) for $\mu_q/T = 5$, and $\mu_3=0$. The $\pm$ notation has been suppressed since we are at zero isospin chemical potential.} \label{fig:pol50}
\end{figure}

In figure \ref{fig:pol50} we present the plots for the imaginary part of the transverse (left panel) and longitudinal (right panel) polarization functions at $\mu_q/T=5$ and zero isospin chemical potential. For this value of parameters, we witness the same qualitative behavior of the polarization functions as in the previous cases for $\mu_q/T\in\{10^4,65\}$.
The transverse correlator grows with  frequency and the longitudinal correlator develops the diffusive pole. If we compare these plots with the corresponding ones at $\mu_q/T \in\{10^4,65\}$ (figure \ref{fig:pol1040}, \ref{fig:transversepol650}, and \ref{fig:longitudinalpol650}), we can actually observe that quantitatively, as we lower $\mu_q/T$, the size of the polarization function decreases.

In the following, we present the results for the comparison of the exact numerical results with the hydrodynamic and the extended hydrodynamic approximation.

\subsubsection*{The transverse correlator}

\begin{figure}[H]
	\begin{subfigure}{\textwidth}
		\centering
		\begin{subfigure}{0.4\textwidth}
			\centering
			\includegraphics[width=\textwidth]{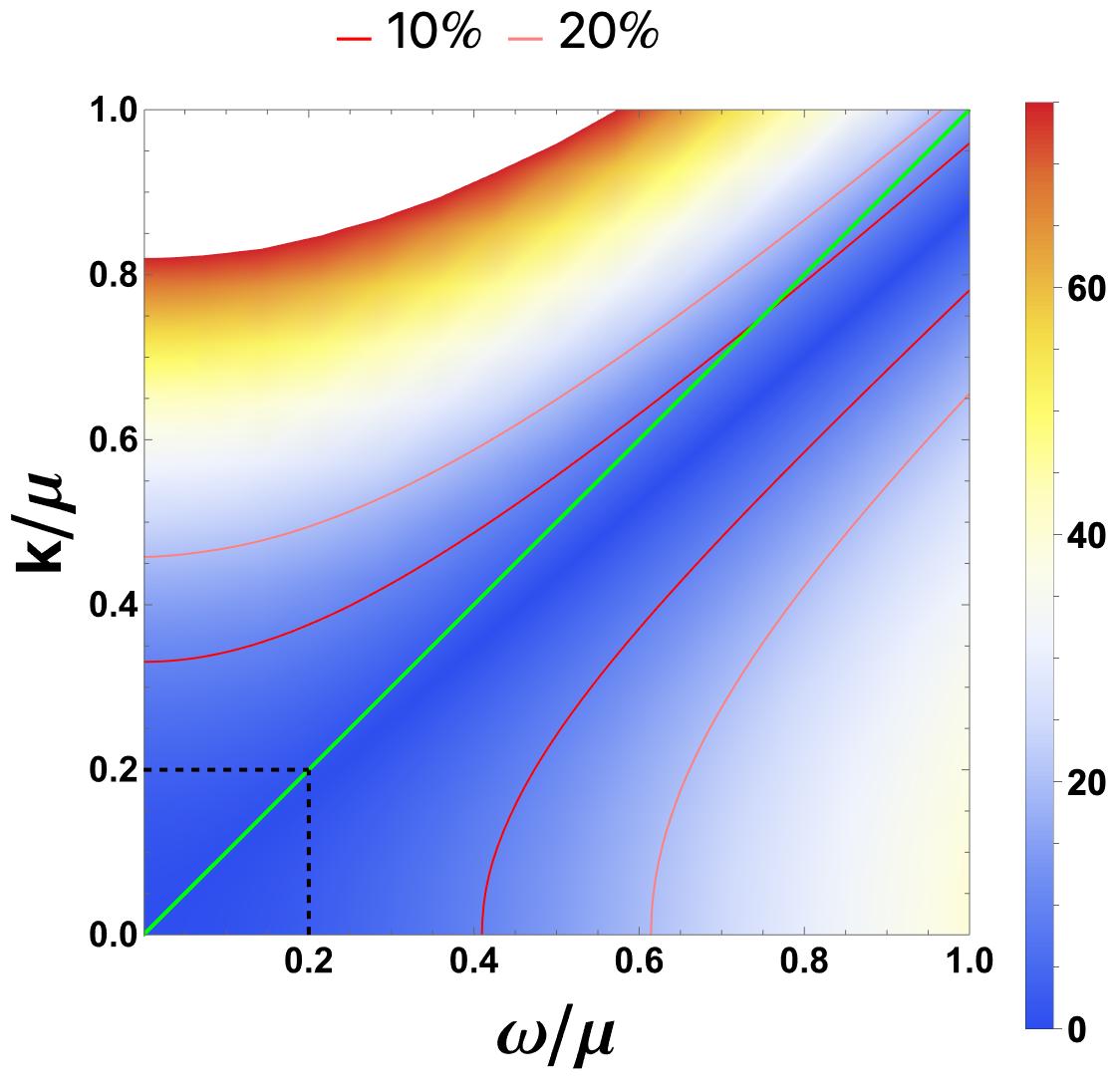}
		\end{subfigure}
		\hspace{1cm}
		\begin{subfigure}{0.4\textwidth}
			\centering
			\includegraphics[width=\textwidth]{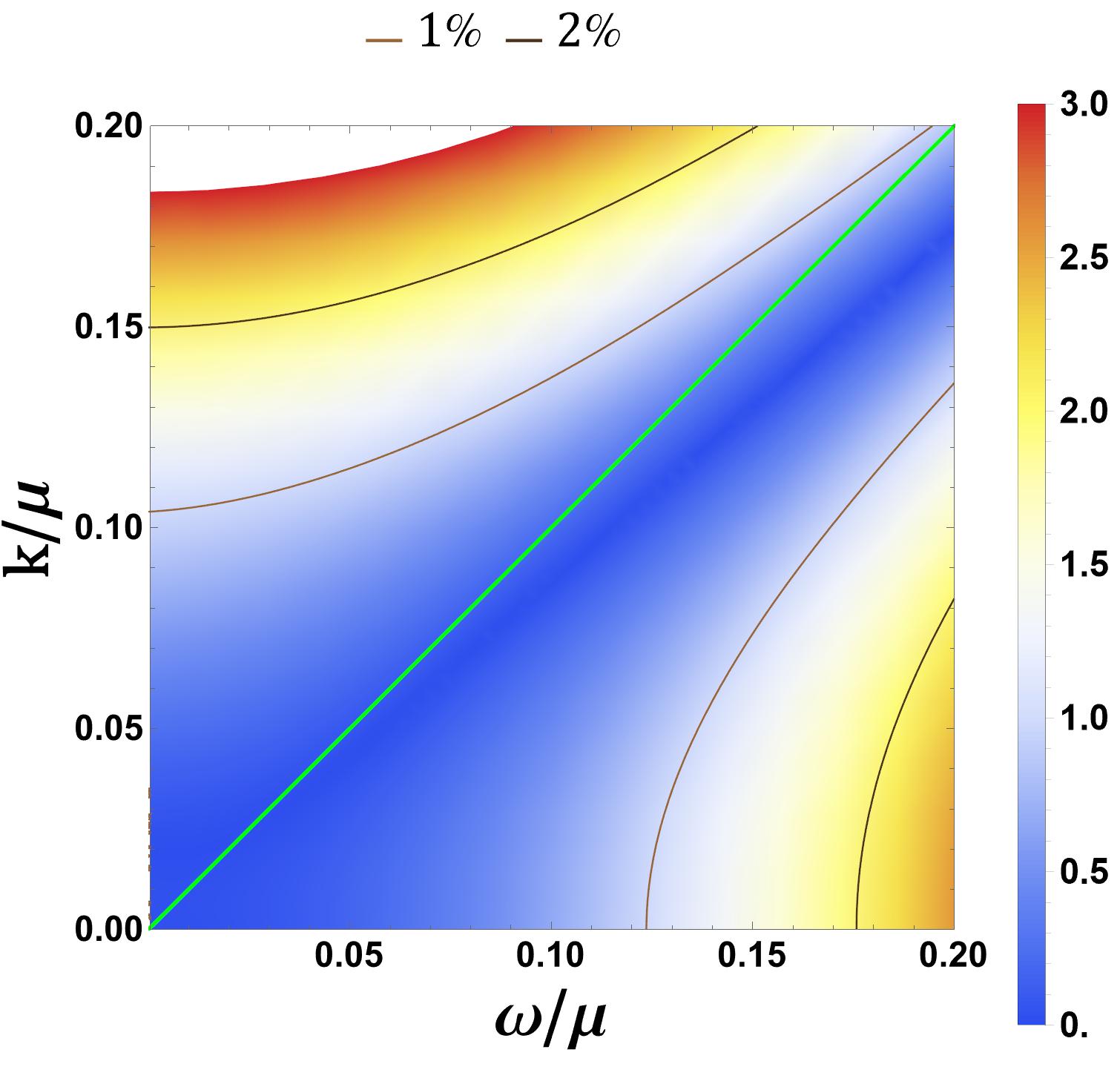}
		\end{subfigure}
	\end{subfigure}

	\begin{subfigure}{\textwidth}
		\centering
		\begin{subfigure}{0.4\textwidth}
			\centering
			\includegraphics[width=\textwidth]{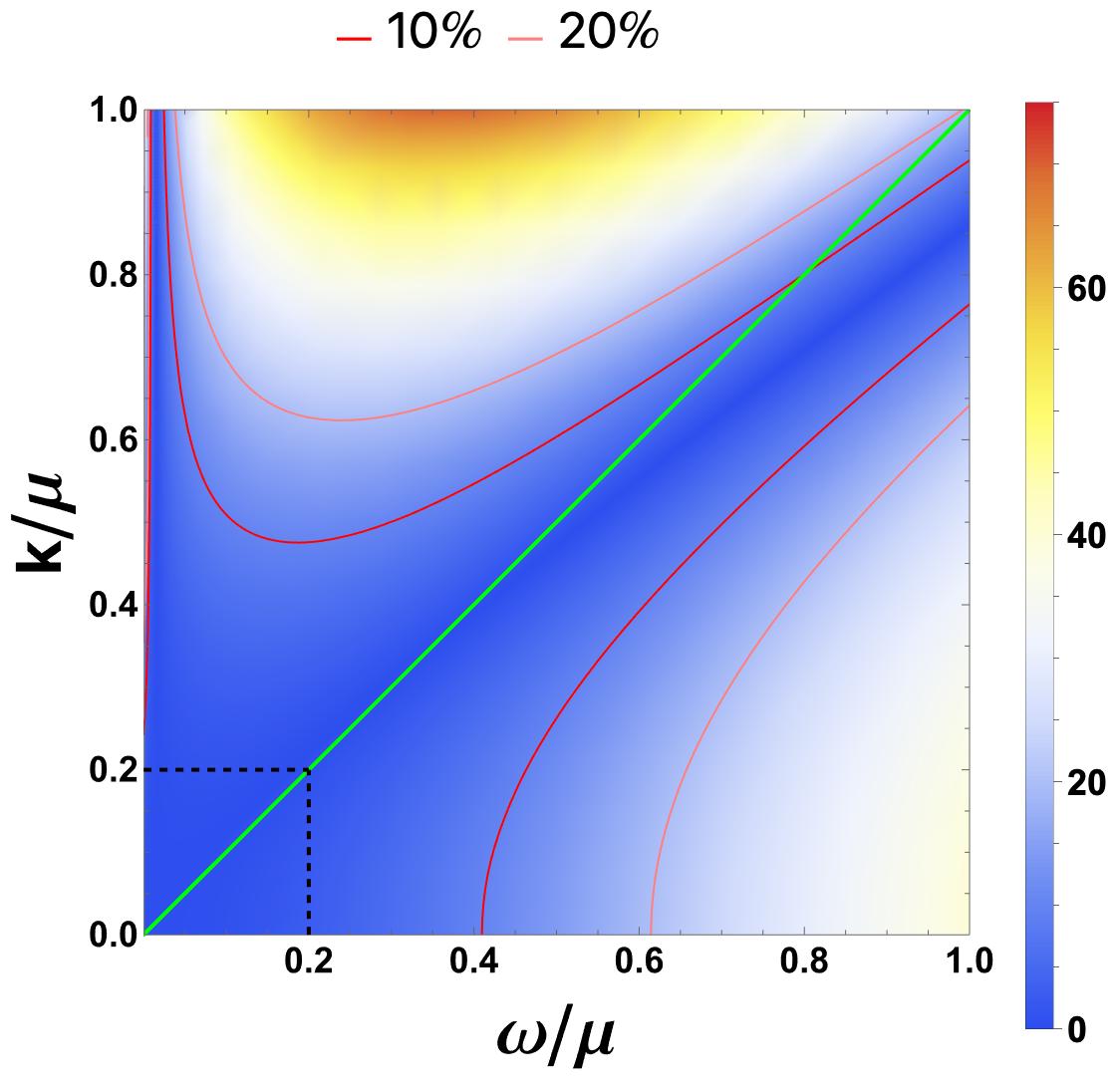}
		\end{subfigure}
		\hspace{1cm}
		\begin{subfigure}{0.4\textwidth}
			\centering
			\includegraphics[width=\textwidth]{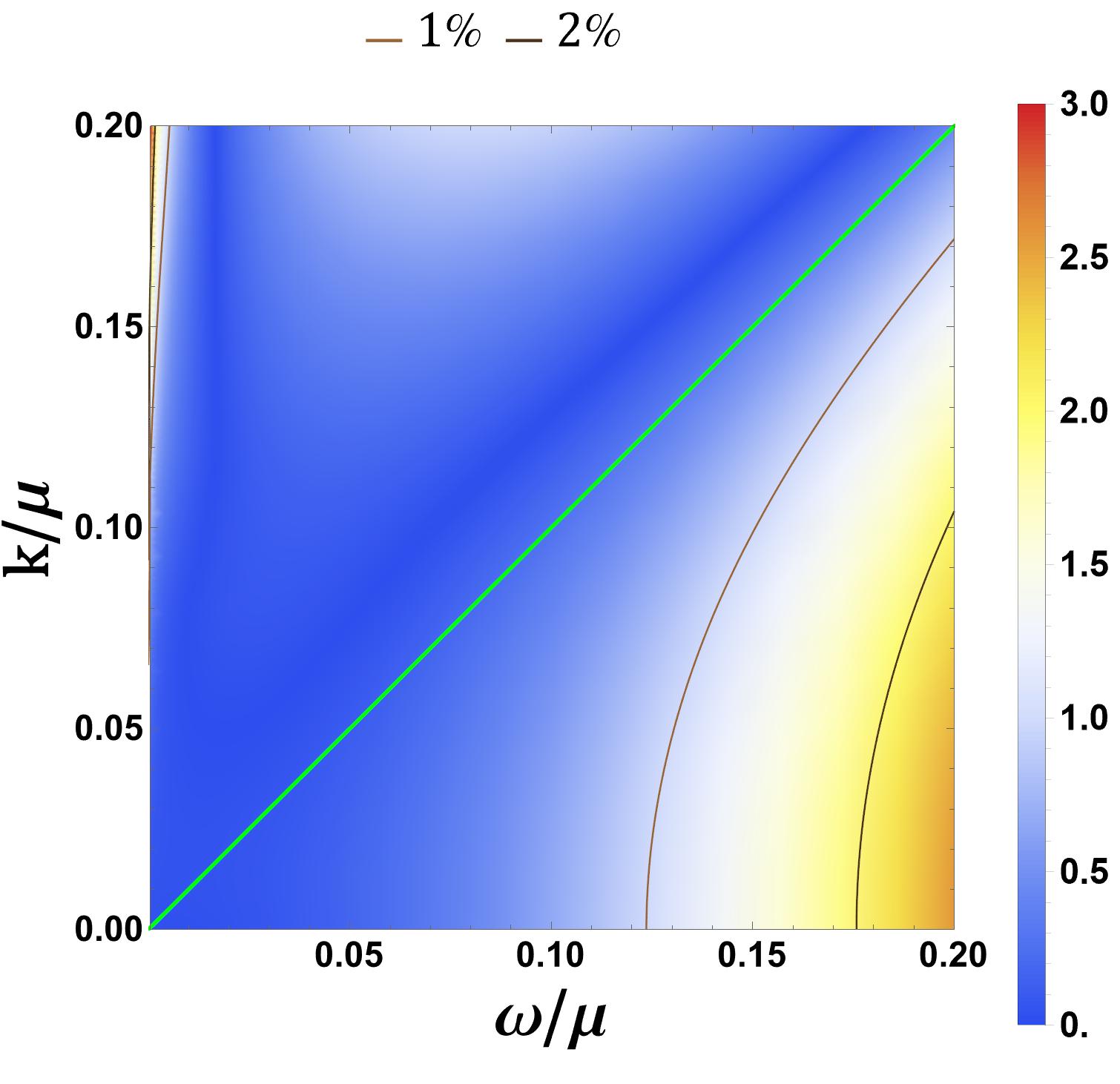}
		\end{subfigure}
	\end{subfigure}
	
	\caption{The percentage relative difference \eqref{eq:percrelfdiff} of the imaginary part of the transverse charged current polarization function with respect to the hydrodynamic approximation \eqref{eq:trhydroapprox} (top row) and the extended hydrodynamic approximation \eqref{eq:trexthydroapprox} (bottom row), for $\mu_q/T= 5$ and zero $\mu_3$.  The right plots shows a subregion of the left
		plots. The green line shows the locus $\omega = k$.} \label{fig:RelDiffNum_tr_mu5mu30}
\end{figure}

In figure~\ref{fig:RelDiffNum_tr_mu5mu30}, the relative difference plots for the hydrodynamic and extended hydrodynamic approximation is shown for $\mu_q/T = 5$ and $\mu_3=0$. If we compare this figure to figure~\ref{fig:RelDiffNum_tr_mu104mu30} we observe that the errors are smaller than in the near-extremal case, but the qualitative structure is the same. Looking at the top-left plot of  figure \ref{fig:RelDiffNum_tr_mu5mu30}, associated to the hydrodynamic approximation, the error is small around the diagonal $\omega \simeq k$ region being inside the 10\% and 20\%, while it becomes large both at large $k/\mu(\simeq 0.6)$ and small $\omega/\mu$ (relative difference around 30\%), and also toward the large-frequency and small momentum edge (relative difference around 40\%).

The extended hydrodynamic approximation (bottom-left plot) improves the upper part of the plot in the extended hydrodynamic region, pushing the 10\% and 20\% contours to larger values of $k/\mu$ for a sizable range of frequencies ($\omega\lesssim 0.3$). The zoomed plots show that the percent-level region is also enlarged by the extended approximation, especially in the rectangle $0<\omega/\mu\lesssim 0.1, 0<k/\mu <0.2$. Nevertheless close to $\omega\simeq 0$ and at larger $k/\mu$ the extended approximation still develops a visible strip with relative difference greater than 1\%.

\subsubsection*{Longitudinal correlator}

\begin{figure}[H]
	\begin{subfigure}{\textwidth}
		\centering
		\begin{subfigure}{0.4\textwidth}
			\centering
			\includegraphics[width=\textwidth]{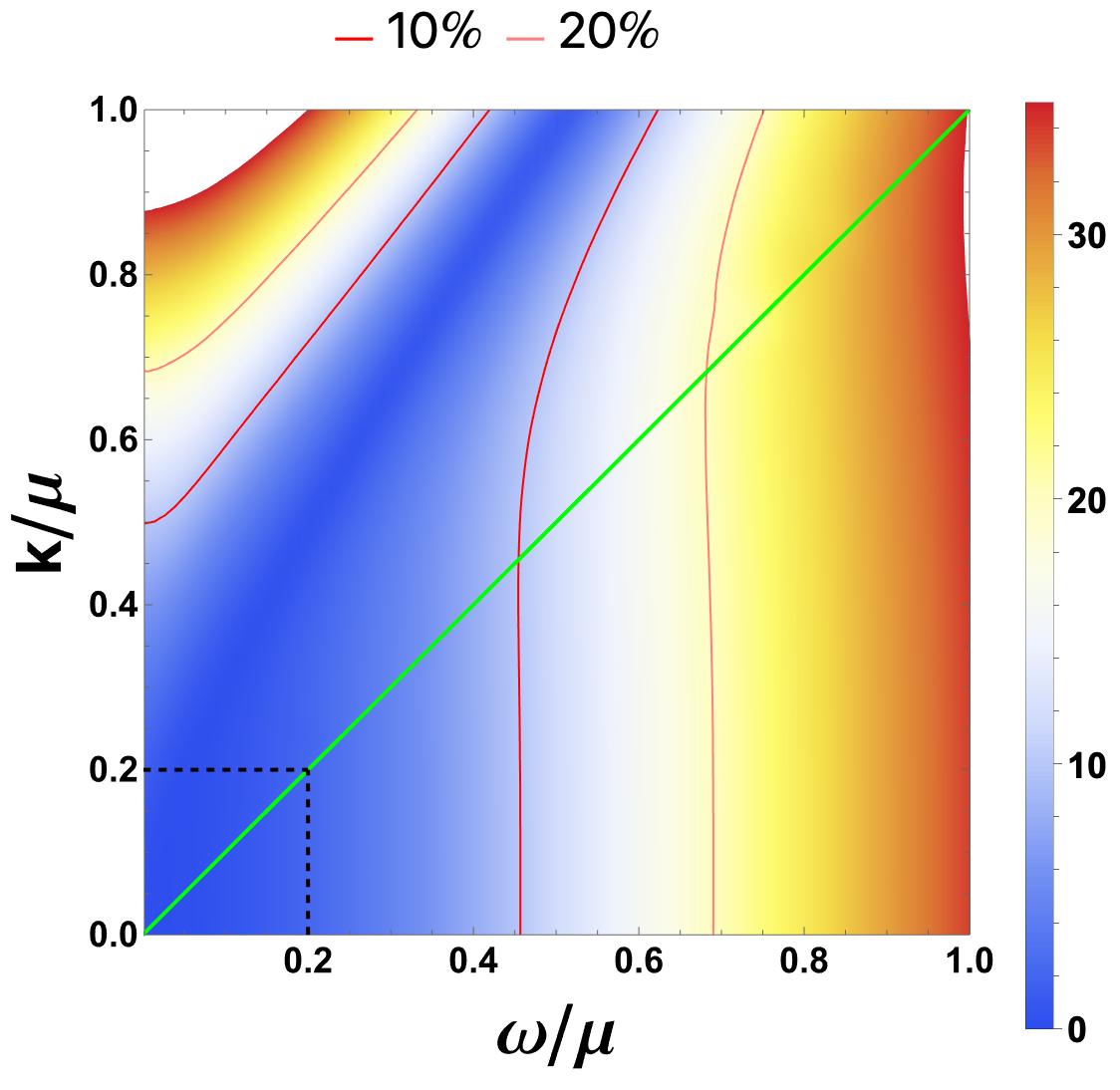}
		\end{subfigure}
		\hspace{1cm}
		\begin{subfigure}{0.4\textwidth}
			\centering
			\includegraphics[width=\textwidth]{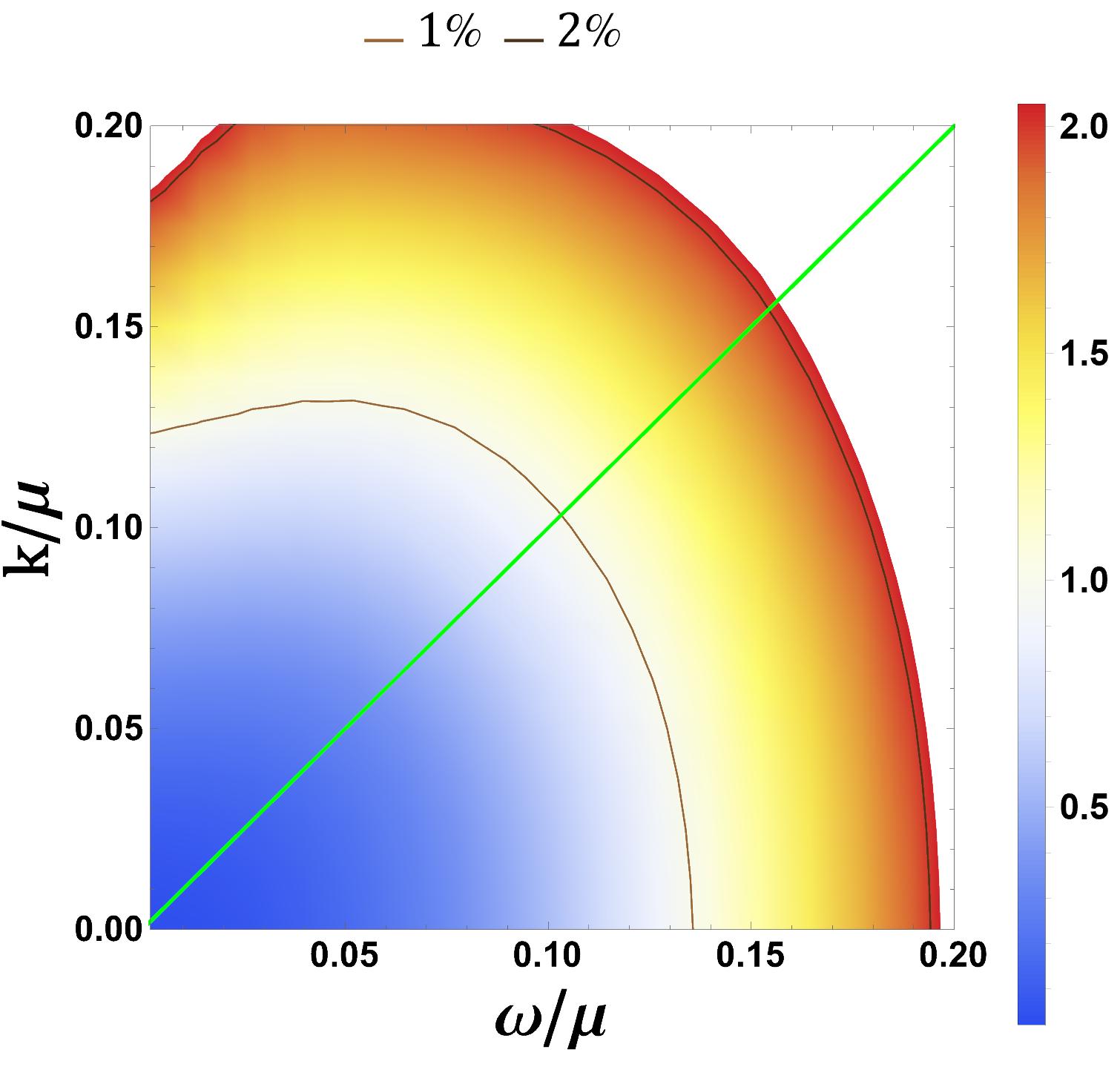}
		\end{subfigure}
	\end{subfigure}

	\begin{subfigure}{\textwidth}
		\centering
		\begin{subfigure}{0.4\textwidth}
			\centering
			\includegraphics[width=\textwidth]{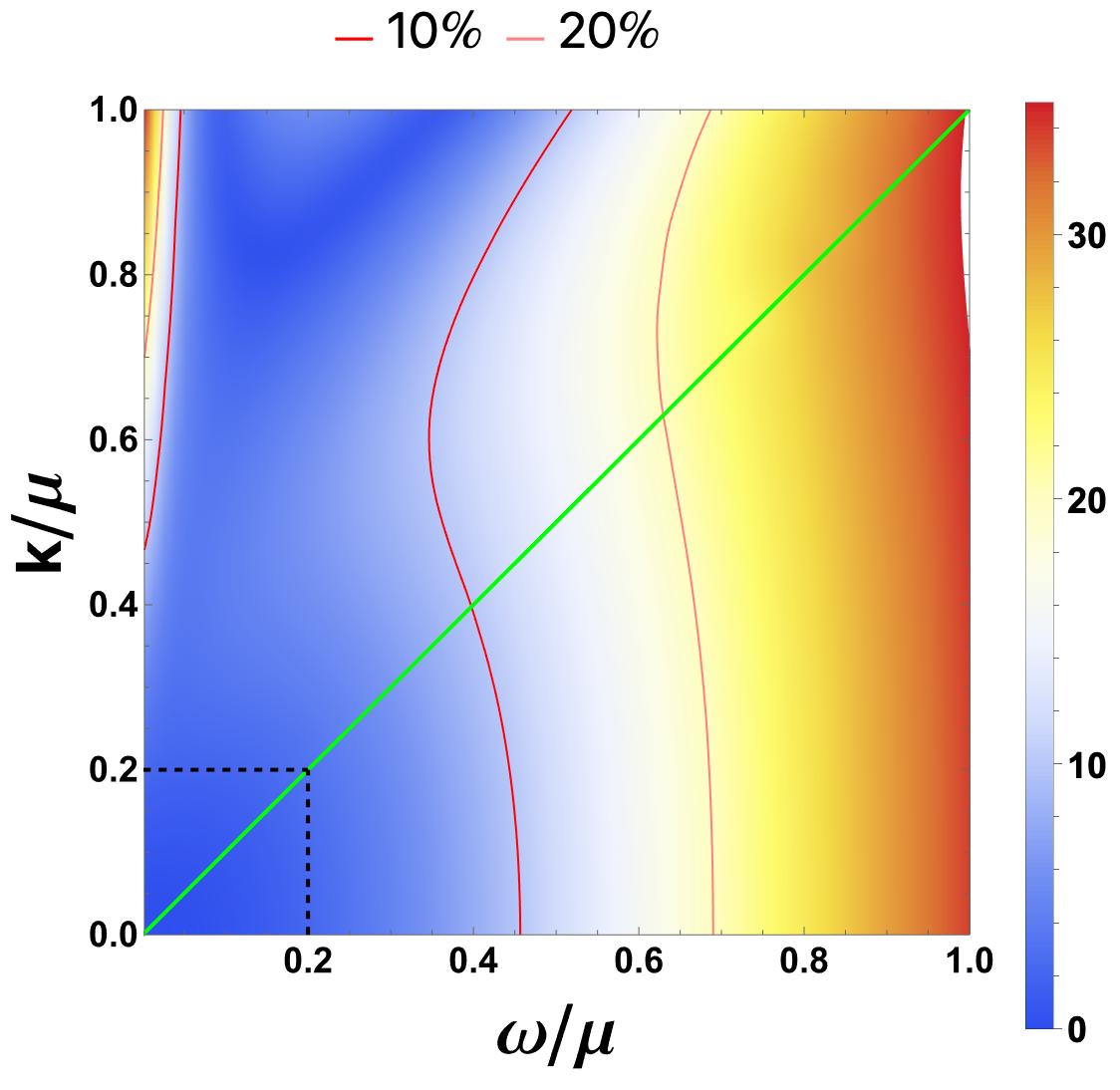}
		\end{subfigure}
		\hspace{1cm}
		\begin{subfigure}{0.4\textwidth}
			\centering
			\includegraphics[width=\textwidth]{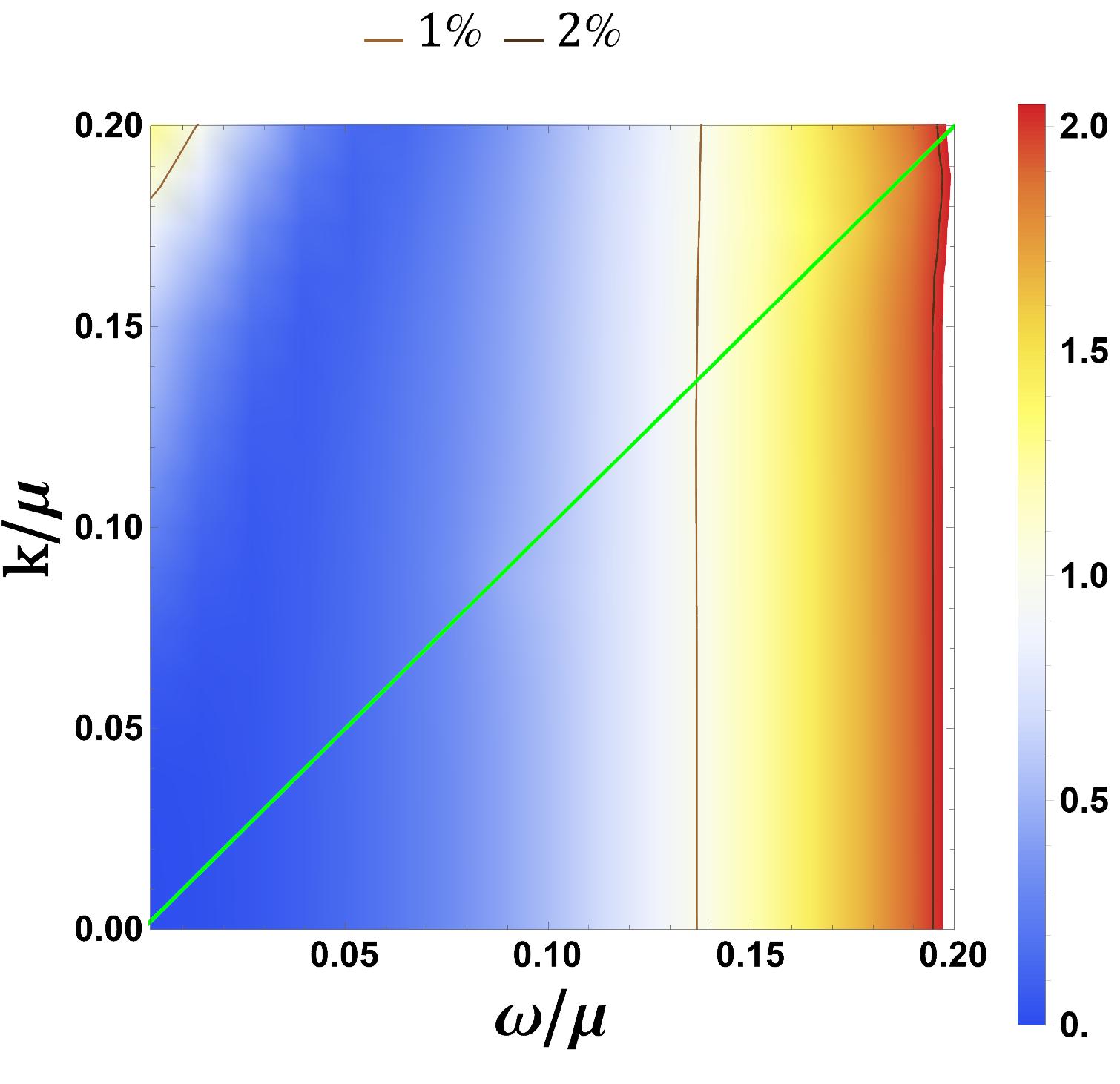}
		\end{subfigure}
	\end{subfigure}

	\caption{As figure \ref{fig:RelDiffNum_tr_mu5mu30} but for the longitudinal polarization function instead of the transverse polarization function.}\label{fig:RelDiffNum_long_mu5mu30}
\end{figure}

In figure~\ref{fig:RelDiffNum_long_mu5mu30}, we present the relative difference plots for the hydrodynamic and extended hydrodynamic approximation for the longitudinal correlator at $\mu_q/T=5$ and $\mu_3=0$. The longitudinal relative difference is organized mostly in vertical bands, meaning that the error is largely controlled by $\omega/\mu$. Looking at the top-left plot of the figure, the hydrodynamic approximation is best at small and intermediate frequencies, while the error grows toward the right side of the plot. The extended hydrodynamic approximation slightly widens the low-error region inside the 10\% and 20\% contours at large $k/\mu$ and small $\omega/\mu\lesssim 0.4$.

In the zoomed plots at right, the hydrodynamic approximation shows a rounded 1\%--2\% contour, whereas the extended approximation produces a more vertical boundary. The latter improves the error overall in the zoomed regime (right column plots) but especially at low frequency and large momentum.

\subsection{Numerical results at $\mu_3/\mu_q=-0.1$}\label{app:mu301}
In this section, we present the results for the imaginary part of the charged current polarization function at finite isospin chemical potential $\mu_3/\mu_q=-0.1$ for $\mu_q/T\in\{10^4,5\}$. First, we present the plots for the imaginary part of the polarization functions in figures \ref{fig:pol10401} ($\mu_q/T=10^4$) and \ref{fig:pol501} ($\mu_q/T=5$). Then, we show the plots of the relative difference \eqref{eq:percrelfdiff} at $\mu_3=-0.1$ for  $\mu_q/T =10^4$ (figure \ref{fig:RelDiffNum_tr_mu104mu301}, and \ref{fig:RelDiffNum_long_mu104mu301} for the transverse and longitudinal sector respectively), and $\mu_q/T=5$ (figure \ref{fig:RelDiffNum_tr_mu5mu301}, and \ref{fig:RelDiffNum_long_mu5mu301} for the transverse and longitudinal sector respectively). The top and bottom rows show respectively the relative difference with respect to the hydrodynamic and the extended hydrodynamic approximation. The right columns are zoomed plots of the left columns.

\subsubsection{Background at $\mu_q/T=10^4$}
\begin{figure}[H]
	\centering
	\begin{subfigure}{0.48\textwidth}
		\centering
		\includegraphics[width=\textwidth]{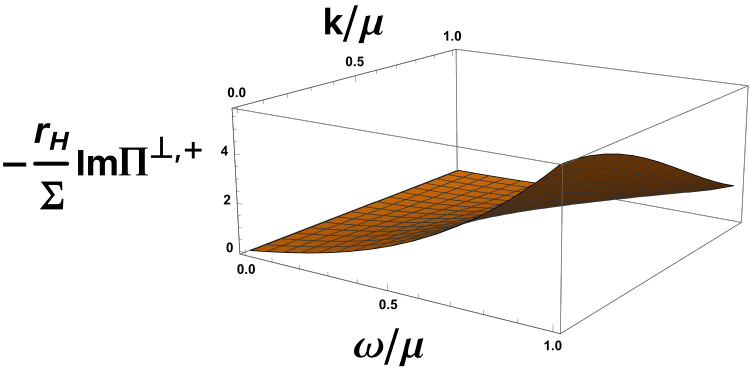}
	\end{subfigure}
	\hfill
	\begin{subfigure}{0.5\textwidth}
		\centering
		\includegraphics[width=\textwidth]{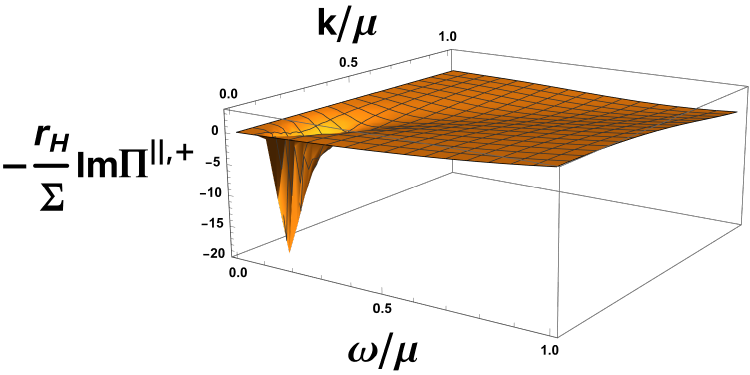}
	\end{subfigure}
	\caption{Imaginary part of the transverse (left panel) and longitudinal (right panel) charged current retarded polarization function (normalized by $-\Sigma/r_H$) for $\mu_q/T = 10^4$, and $\mu_3/\mu_q = -0.1$.} \label{fig:pol10401}
\end{figure}

In figure \ref{fig:pol10401} we present the plots for the imaginary part of the transverse (left panel) and longitudinal (right panel) polarization functions at $\mu_q/T=10^4$ and finite isospin chemical potential $\mu_3/\mu_q=-0.1$. The transverse polarization function is not significantly affected by the presence of the isospin chemical potential, while looking at the longitudinal polarization function is clear how the diffusive pole is shifted at real positive frequencies. We also observe that the presence of a negative $\mu_3$ generally decrease the magnitude of the polarization functions, as already observed in section \ref{sec:exactresults}.

In the following, we present the results for the comparison of the exact numerical results with the hydrodynamic and the extended hydrodynamic approximation.

\subsubsection*{The transverse correlator}

\begin{figure}[H]
	\begin{subfigure}{\textwidth}
		\centering
		\begin{subfigure}{0.4\textwidth}
			\centering
			\includegraphics[width=\textwidth]{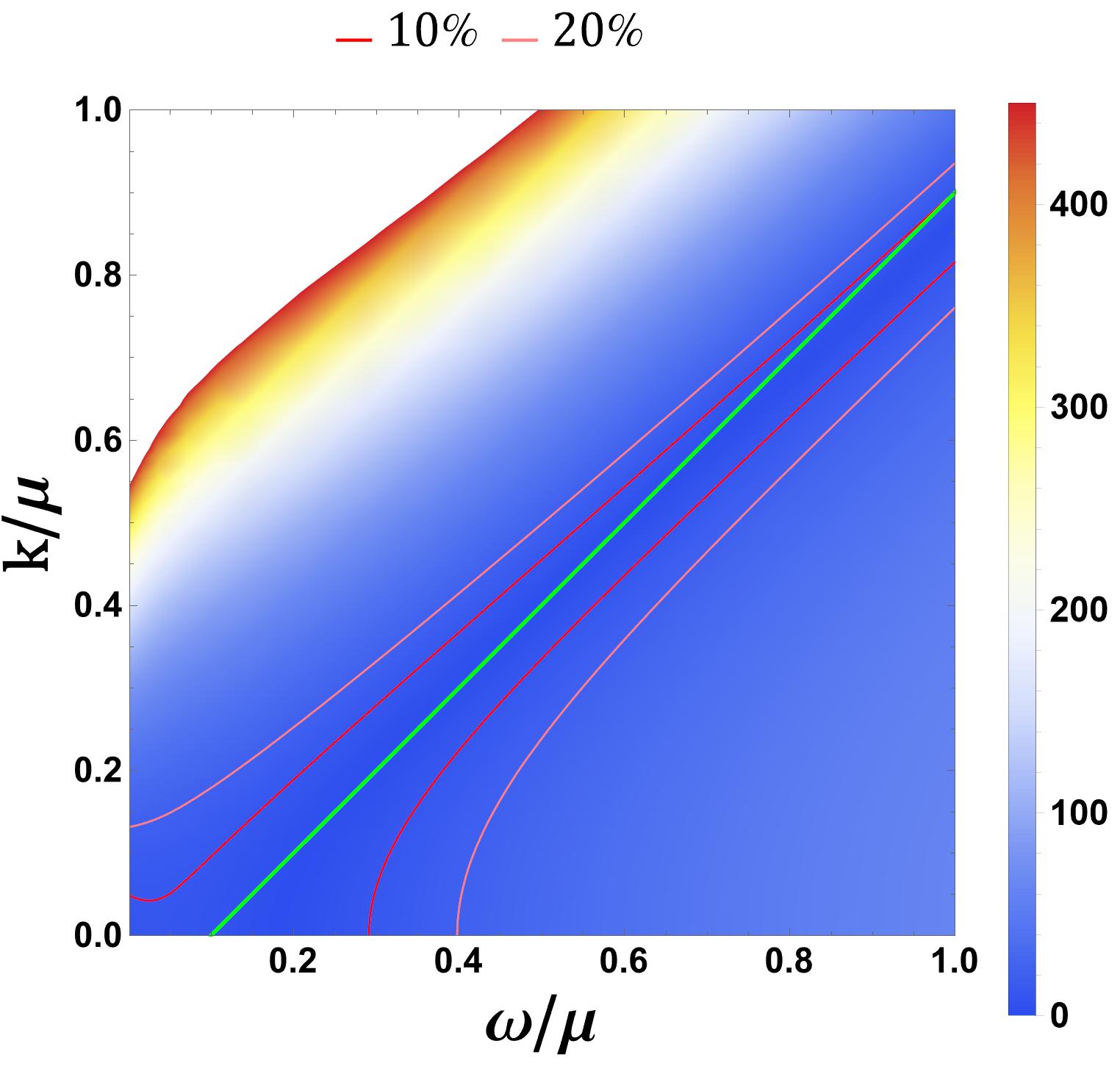}
		\end{subfigure}
		\hspace{1cm}
		\begin{subfigure}{0.4\textwidth}
			\centering
			\includegraphics[width=\textwidth]{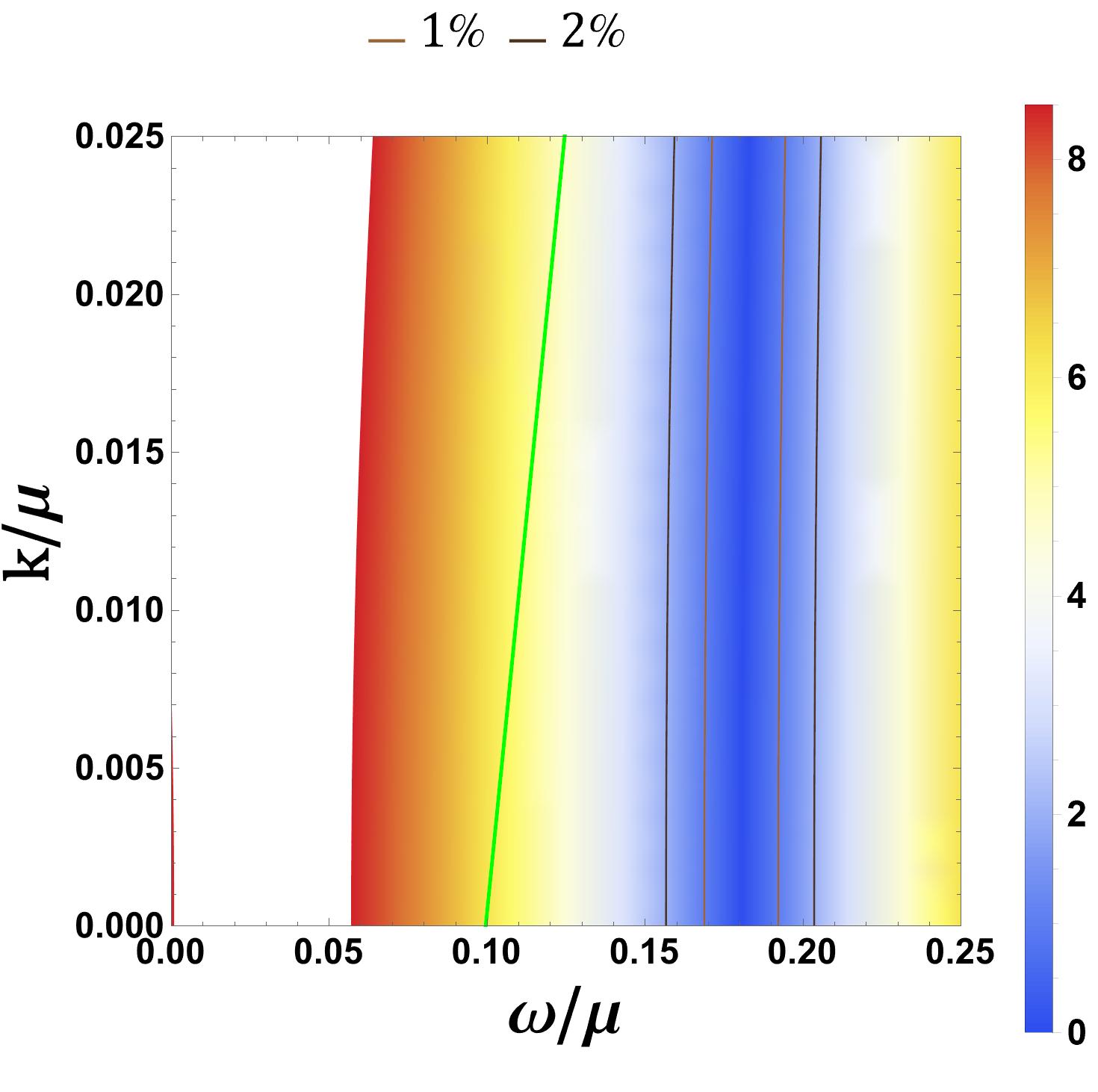}
		\end{subfigure}
	\end{subfigure}

	\begin{subfigure}{\textwidth}
		\centering
		\begin{subfigure}{0.4\textwidth}
			\centering
			\includegraphics[width=\textwidth]{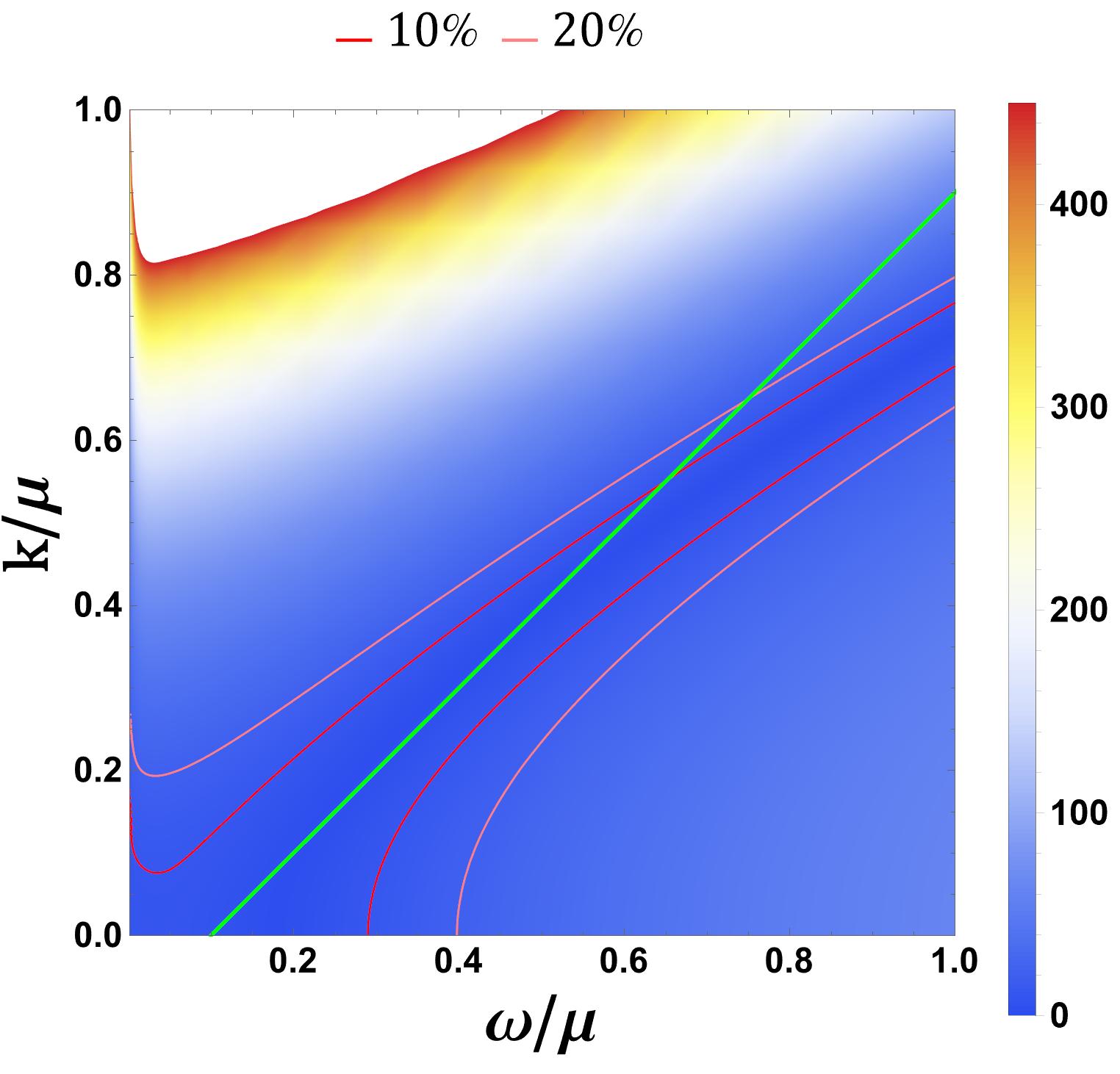}
		\end{subfigure}
		\hspace{1cm}
		\begin{subfigure}{0.4\textwidth}
			\centering
			\includegraphics[width=\textwidth]{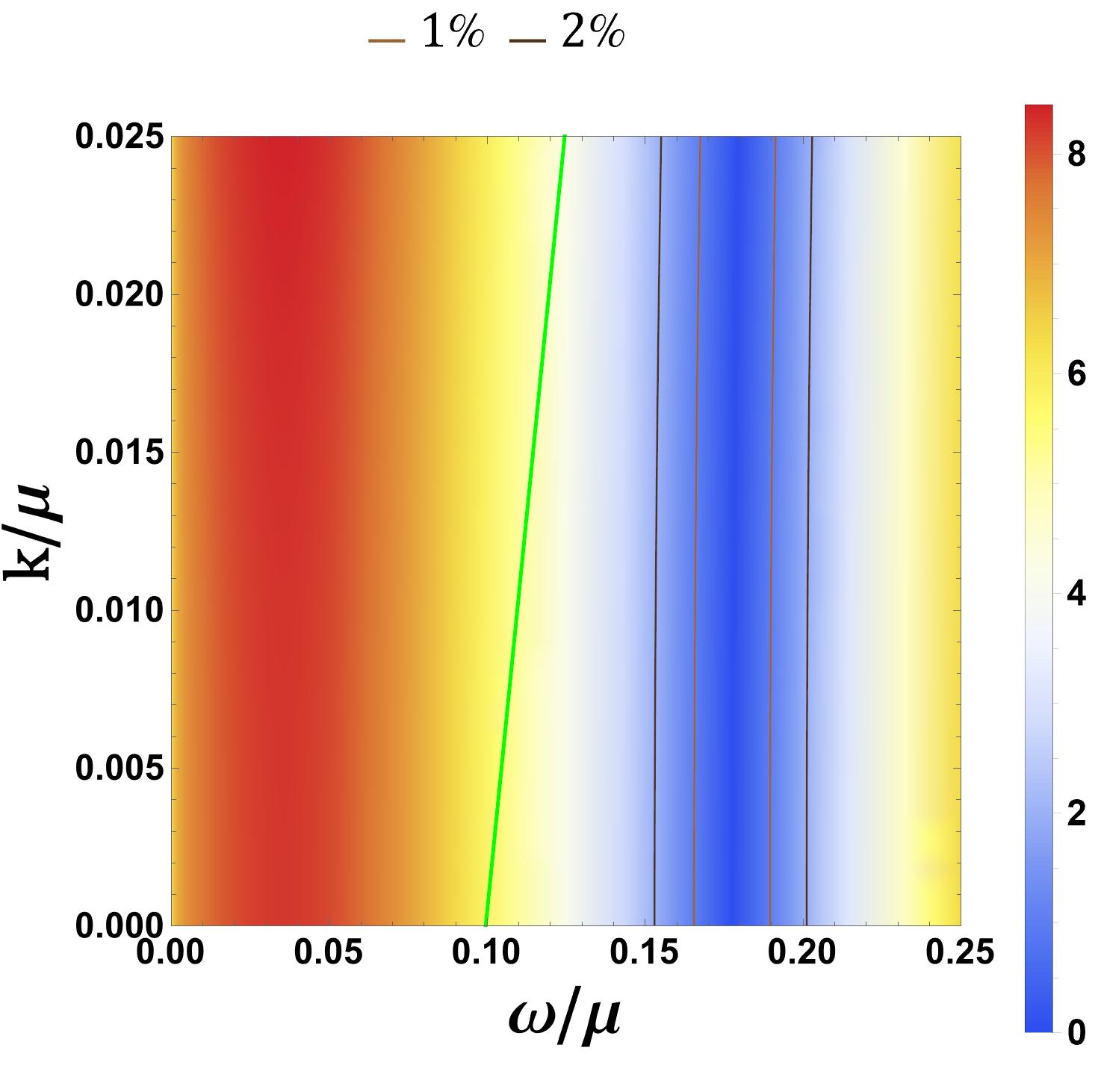}
		\end{subfigure}
	\end{subfigure}
	
	\caption{The percentage relative difference \eqref{eq:percrelfdiff} of the imaginary part of the transverse charged current polarization function with respect to the hydrodynamic approximation \eqref{eq:trhydroapprox} (top row) and the extended hydrodynamic approximation \eqref{eq:trexthydroapprox} (bottom row), for $\mu_q/T= 10^4$ and $\mu_3/\mu_q=-0.1$.  The right plots shows a subregion of the left
		plots. The green line shows the locus $\omega = k+\mu_3$. } \label{fig:RelDiffNum_tr_mu104mu301}
\end{figure}

Figure~\ref{fig:RelDiffNum_tr_mu104mu301} shows the first clear effect of turning on a non-zero isospin chemical potential, in this case $\mu_3/\mu_q=-0.1$. In the full-range plots (left column), the extended hydrodynamic approximation still improves the hydrodynamic one, especially by lowering the error in the low-frequency region and pushing the 10\% and 20\% contours upward in $k/\mu$. However, the zoomed plots (right column) reveal a new structure which was absent at $\mu_3=0$: the error is no longer minimized simply near the origin, but develops a strong dependence on $\omega/\mu$ with vertical bands. In particular, the relative difference is large in a band at small positive frequency (bigger than 8\% for the hydrodynamic approximation and only slightly less than 8\% for the extended hydrodynamic approximation), and becomes smaller only after moving away from that band. The extended approximation does not remove this feature, even if the error, close to $\omega=0$, decreases.

\subsubsection*{The longitudinal correlator}

\begin{figure}[H]
	\begin{subfigure}{\textwidth}
		\centering
		\begin{subfigure}{0.4\textwidth}
			\centering
			\includegraphics[width=\textwidth]{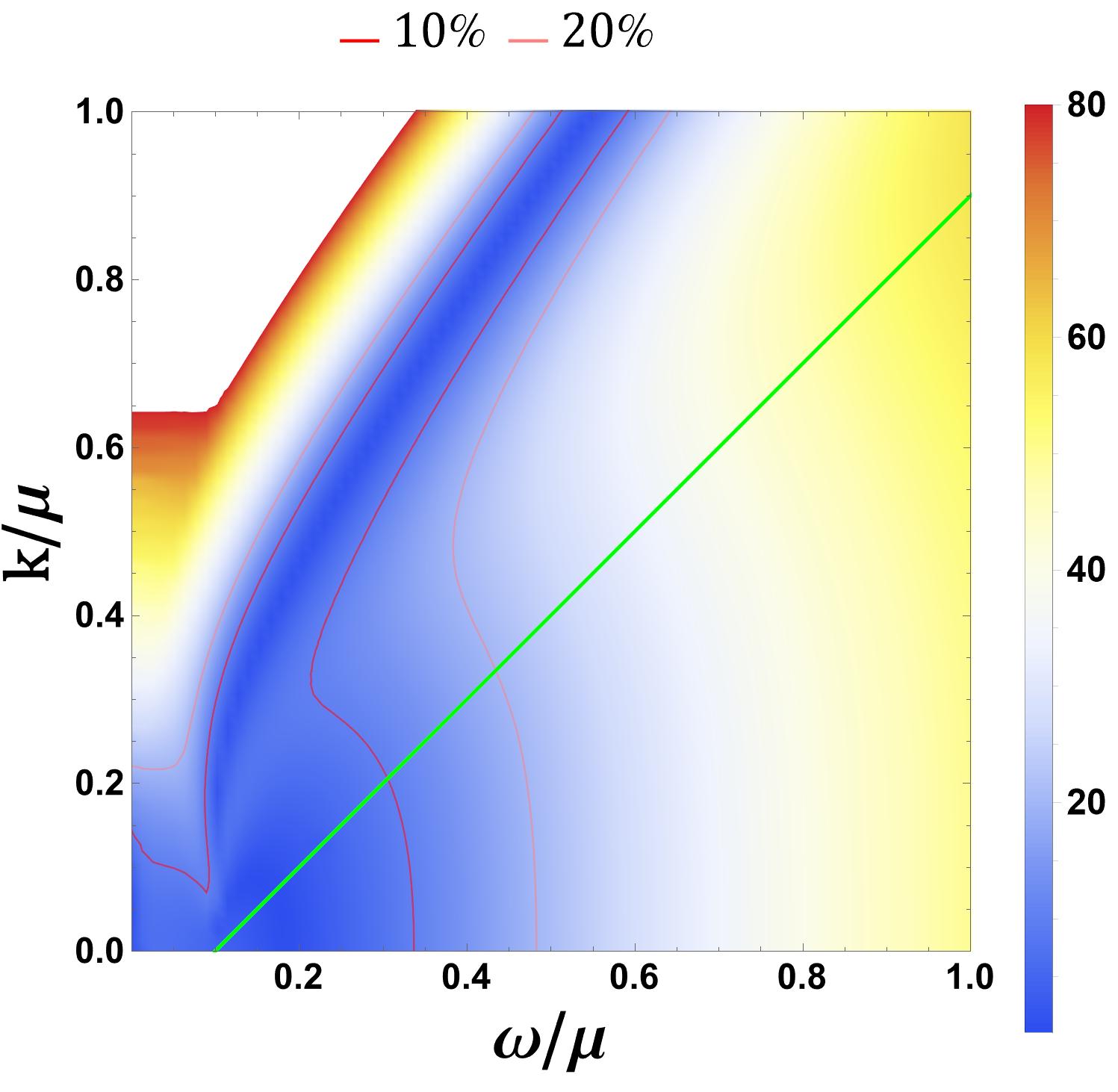}
		\end{subfigure}
		\hspace{1cm}
		\begin{subfigure}{0.4\textwidth}
			\centering
			\includegraphics[width=\textwidth]{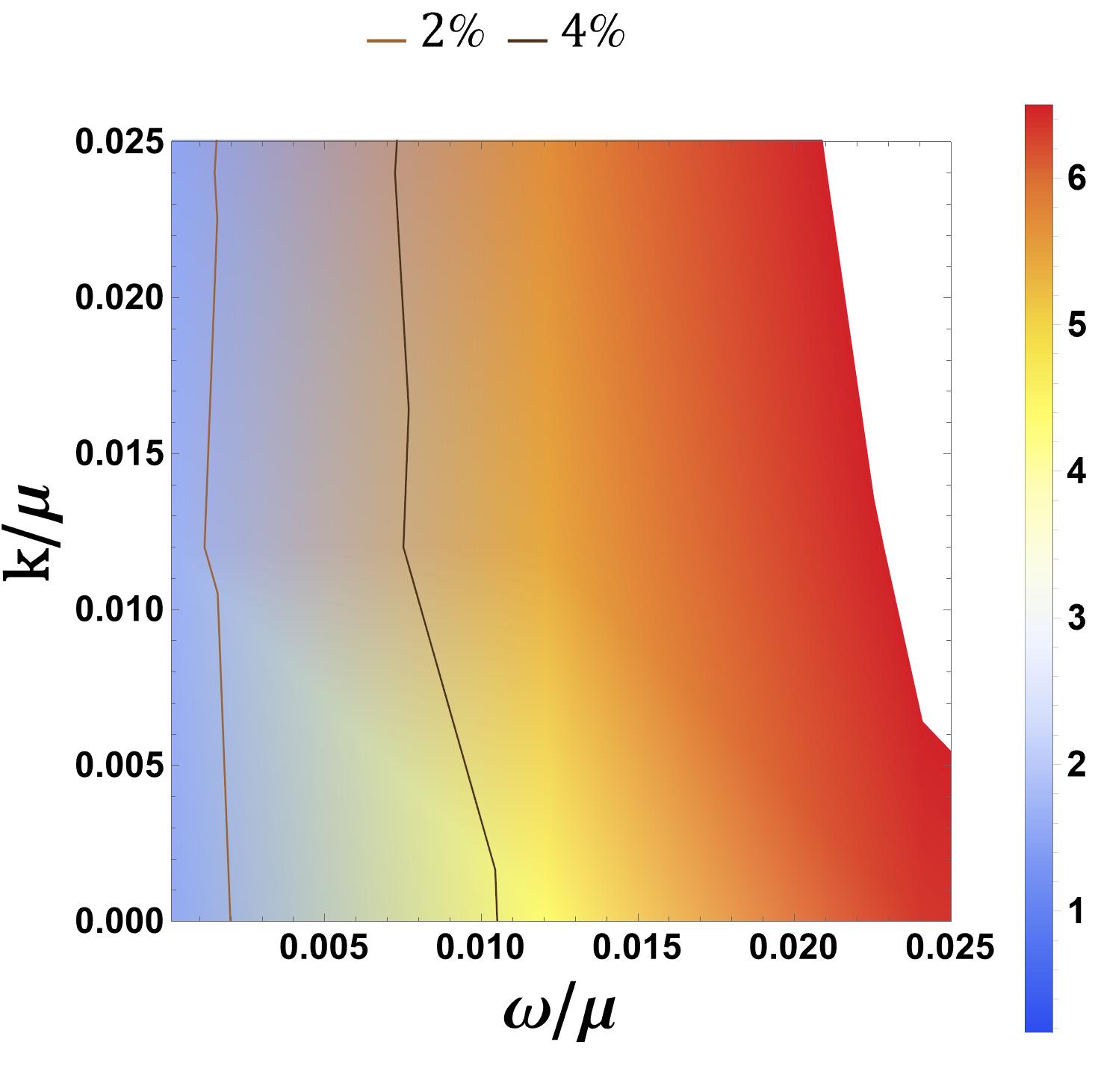}
		\end{subfigure}
	\end{subfigure}

	\begin{subfigure}{\textwidth}
		\centering
		\begin{subfigure}{0.4\textwidth}
			\centering
			\includegraphics[width=\textwidth]{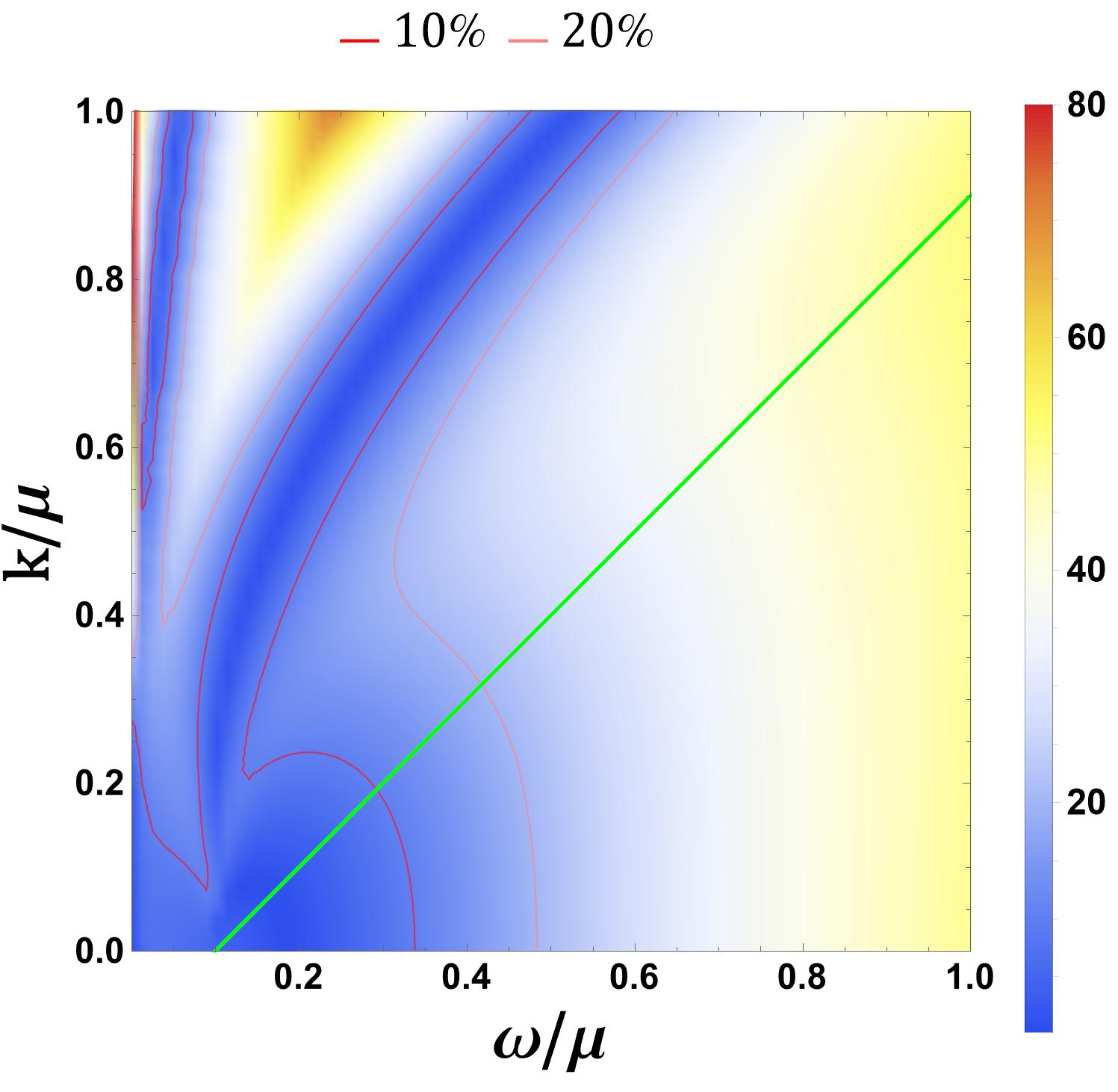}
		\end{subfigure}
		\hspace{1cm}
		\begin{subfigure}{0.4\textwidth}
			\centering
			\includegraphics[width=\textwidth]{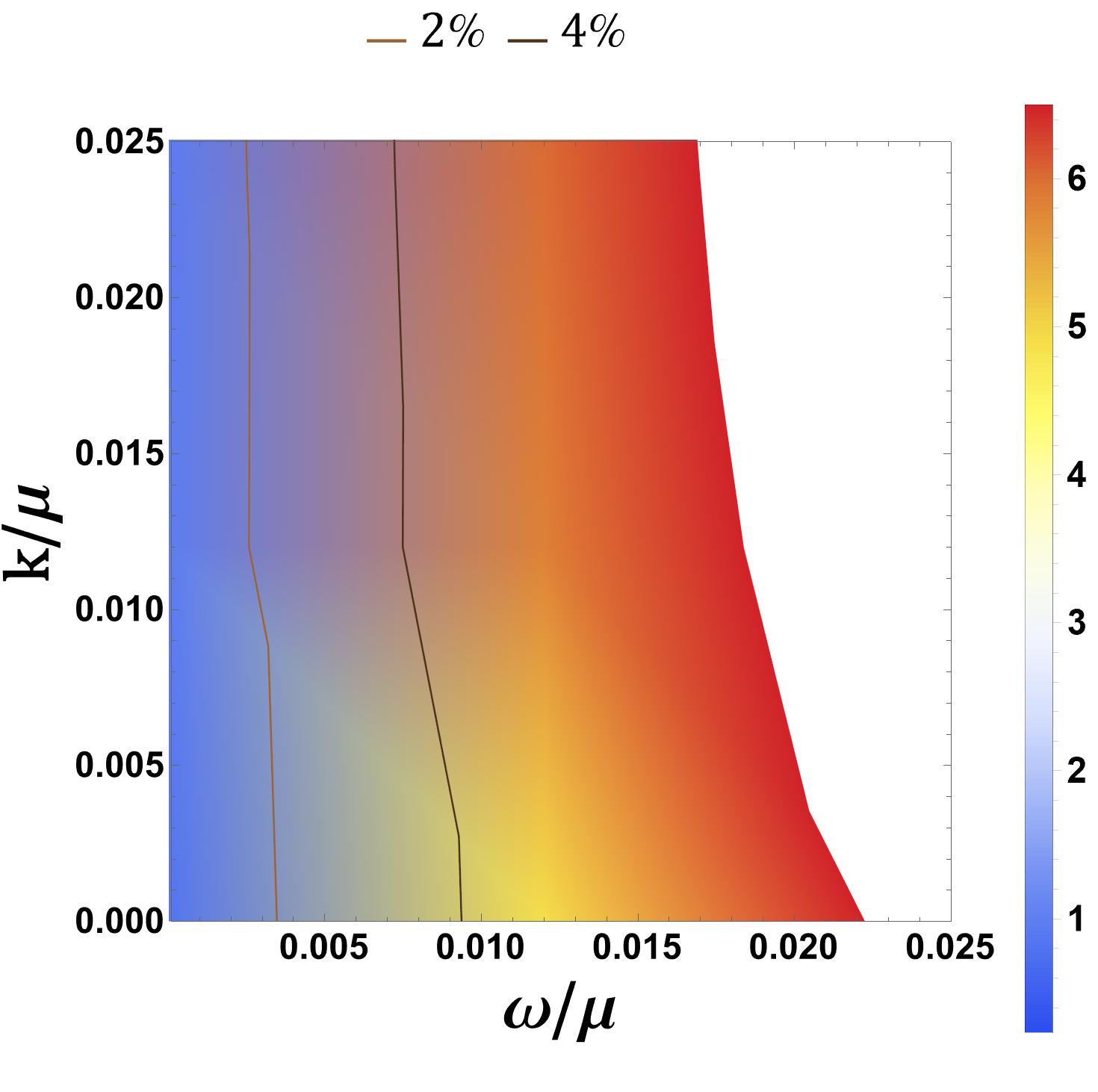}
		\end{subfigure}
	\end{subfigure}

	\caption{As figure \ref{fig:RelDiffNum_tr_mu104mu301} but for the longitudinal polarization function instead of the transverse polarization function.}\label{fig:RelDiffNum_long_mu104mu301}
\end{figure}

In figure~\ref{fig:RelDiffNum_long_mu104mu301}, the full-range plots (left column) still resemble the zero-isospin longitudinal case, with a low-error valley at small-to-intermediate $\omega/\mu, k/\mu$ and a large-error region at large frequencies and momenta. However, the $10\%$ and $20\%$ contours are now centered around $\omega\simeq \mu_3$. The extended hydrodynamic approximation produces a modest improvement in the region of small frequencies and large momenta, in particular for $k/\mu\gtrsim0.4$.

The effect of finite value of $\mu_3$ is also visible in the zoomed plots (right column): the contours shown are now at the 2\%--4\% level rather than at the sub-percent level, so the IR accuracy is already noticeably worse than at $\mu_3=0$ (see figure \ref{fig:RelDiffNum_long_mu104mu30}). In these plots, the extended hydrodynamic approximation slightly changes the shape and position of the low-error region. It lowers the error close to $\omega=0$, while it does worse than the hydrodynamic approximation for $k/\mu\gtrsim 0.015$. As in the transverse sector, the finite isospin chemical potential still reduces the size of the truly hydrodynamic region.

\subsubsection{Background at $\mu_q/T=5$}
\begin{figure}[htb]
	\centering
	\begin{subfigure}{0.48\textwidth}
		\centering
		\includegraphics[width=\textwidth]{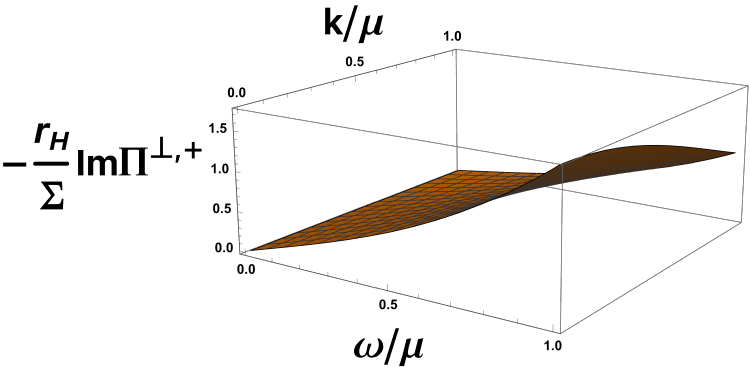}
	\end{subfigure}
	\hfill
	\begin{subfigure}{0.5\textwidth}
		\centering
		\includegraphics[width=\textwidth]{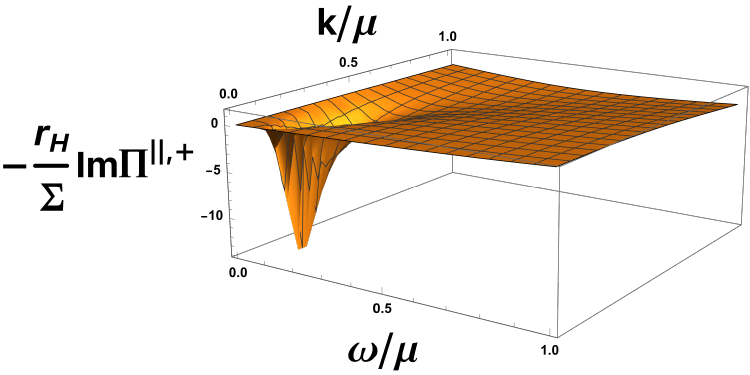}
	\end{subfigure}
	\caption{Imaginary part of the transverse (left panel) and longitudinal (right panel) charged current retarded polarization function (normalized by $-\Sigma/r_H$) for $\mu_q/T = 5$, and $\mu_3=-0.1$.} \label{fig:pol501}
\end{figure}

In figure \ref{fig:pol501} we present the plots for the imaginary part of the transverse (left panel) and longitudinal (right panel) polarization functions at $\mu_q/T=5$ and finite isospin chemical potential $\mu_3/\mu_q=-0.1$. The qualitative features are the same as the polarization functions for $\mu_q/T = 10^4$ (figure  \ref{fig:pol10401}) with the transverse growing with the frequency and the longitudinal developing the diffusive pole at real frequencies. Also for this value of $\mu_3$, as we lower $\mu_q/T$, the magnitude of the polarization function is also lowered.

In the following, we present the results for the comparison of the exact numerical results with the hydrodynamic and the extended hydrodynamic approximation.

\begin{figure}[htb]
	\begin{subfigure}{\textwidth}
		\centering
		\begin{subfigure}{0.4\textwidth}
			\centering
			\includegraphics[width=\textwidth]{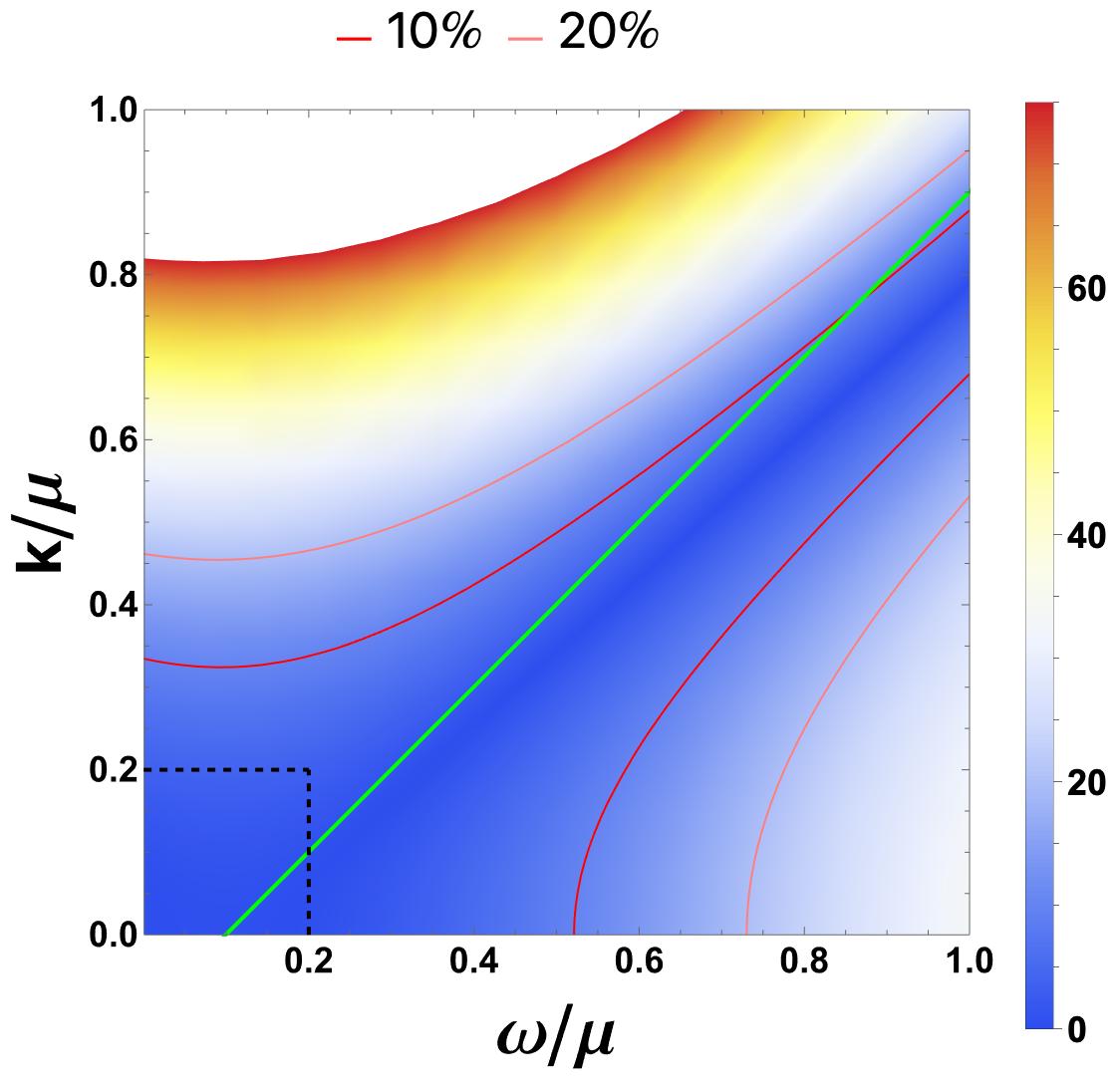}
		\end{subfigure}
		\hspace{1cm}
		\begin{subfigure}{0.4\textwidth}
			\centering
			\includegraphics[width=\textwidth]{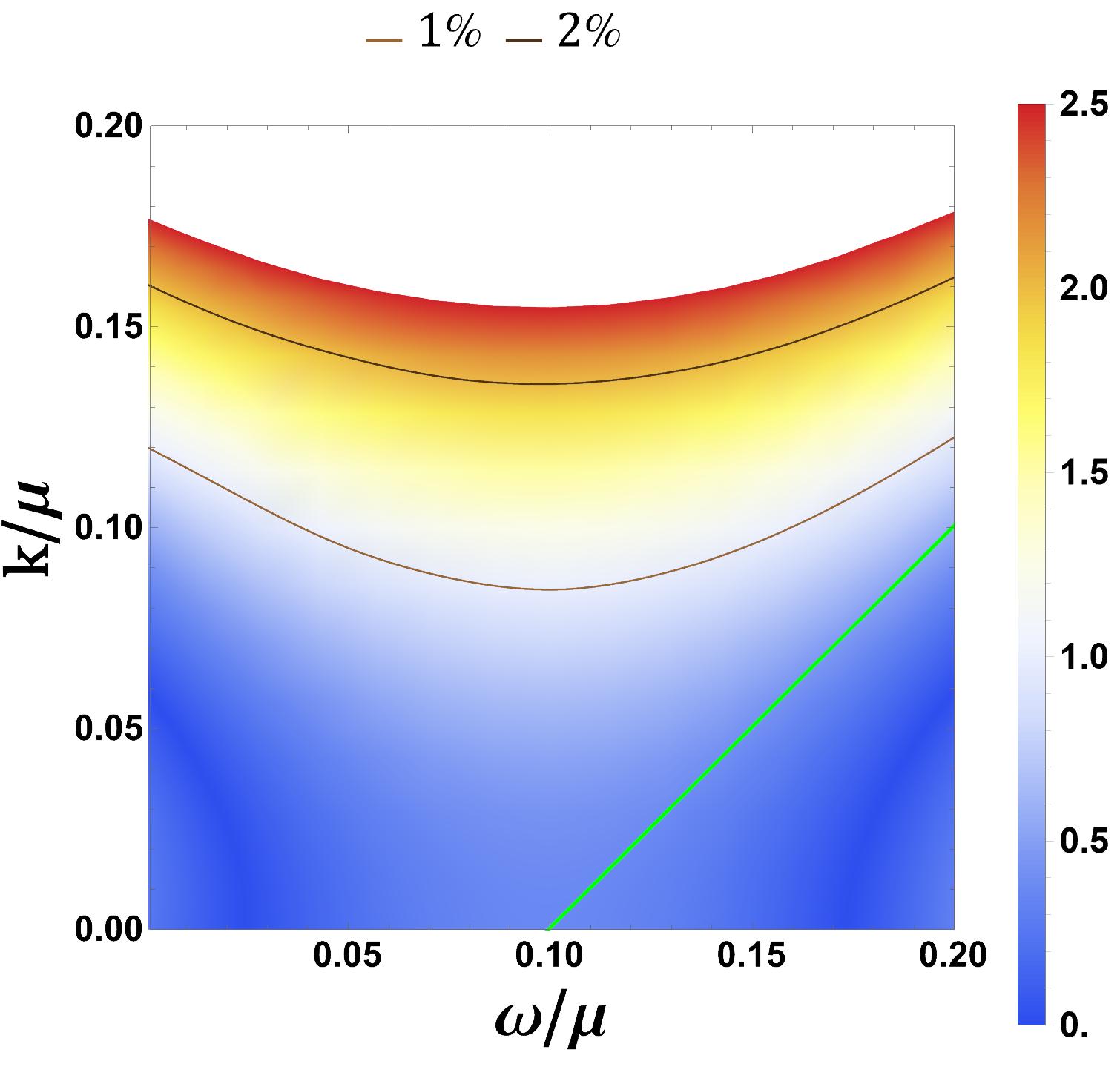}
		\end{subfigure}
	\end{subfigure}

	\begin{subfigure}{\textwidth}
		\centering
		\begin{subfigure}{0.4\textwidth}
			\centering
			\includegraphics[width=\textwidth]{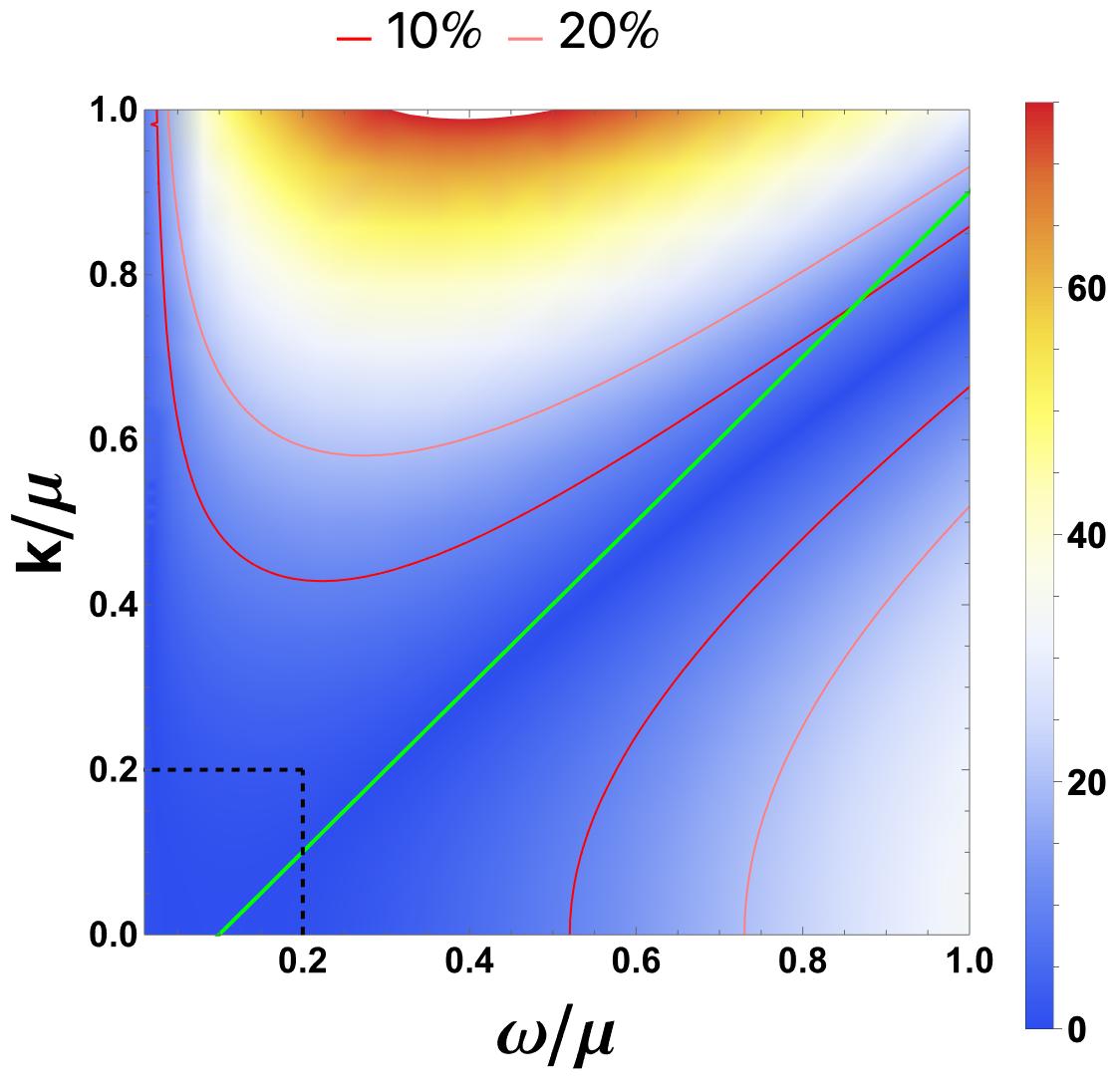}
		\end{subfigure}
		\hspace{1cm}
		\begin{subfigure}{0.4\textwidth}
			\centering
			\includegraphics[width=\textwidth]{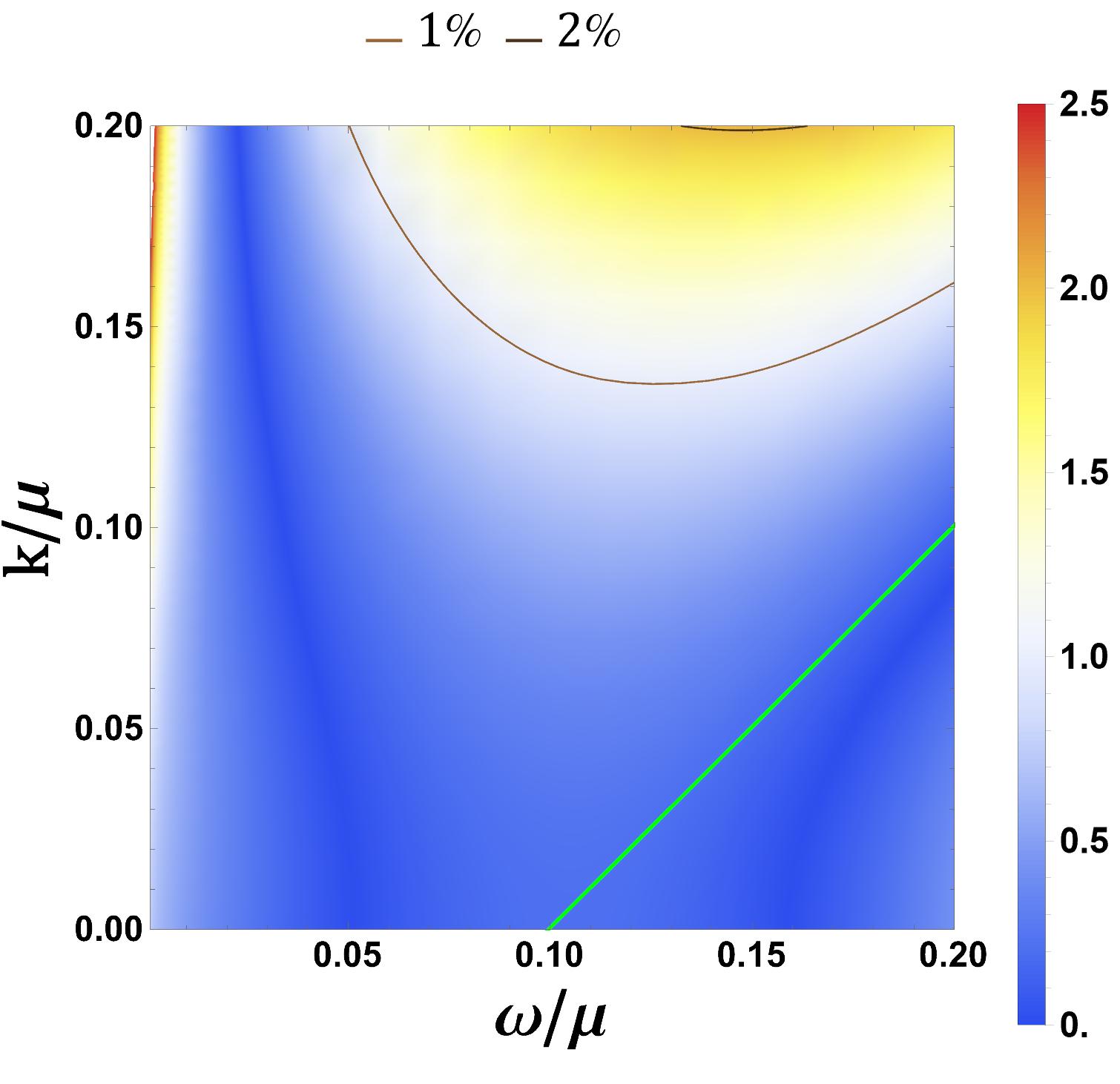}
		\end{subfigure}
	\end{subfigure}
	
	\caption{The percentage relative difference \eqref{eq:percrelfdiff} of the imaginary part of the transverse charged current polarization function with respect to the hydrodynamic approximation \eqref{eq:trhydroapprox} (top row) and the extended hydrodynamic approximation \eqref{eq:trexthydroapprox} (bottom row), for $\mu_q/T= 5$ and $\mu_3/\mu_q=-0.1$.  The right plots shows a subregion of the left
		plots. The green line shows the locus $\omega = k+\mu_3$.} \label{fig:RelDiffNum_tr_mu5mu301}
\end{figure}

\subsubsection*{The transverse correlator}

In figure~\ref{fig:RelDiffNum_tr_mu5mu301}, where the temperature is now larger and comparable to $\mu$, the relative difference is smaller than at $\mu_q/T=10^4$ as observed at $\mu_3=0$. The top-left plot of the figure shows that for the hydrodynamic approximation the $10\%$ contour starts around $k/\mu\simeq 0.3$--$0.35$ at $\omega/\mu\simeq 0$ and bends upward as $\omega/\mu$ increases. For example, around $\omega/\mu\simeq 0.5$, the $10\%$ contour is at $k/\mu\simeq 0.55$, while for $\omega/\mu\simeq 0.8$ it is close to $k/\mu\simeq 0.75$. The extended hydrodynamic approximation (bottom-left plot) moves this contour upward in most of the plot, especially for $0.1\lesssim \omega/\mu\lesssim 0.6$, showing a clear improvement over the hydrodynamic approximation in the intermediate-frequency region. At large $k/\mu$ (close to 1) the two approximations do not differ.

The zoomed plots (right column plots) show that the $1\%$--$2\%$ region is rather large. For the hydrodynamic approximation, the error remains below about $1\%$ for $k/\mu\lesssim 0.08$--$0.10$, depending on the value of $\omega/\mu$, while the $2\%$ contour lies around $k/\mu\simeq 0.14$--$0.16$. With the extended hydrodynamic approximation, the low-error region is enlarged for most values of $\omega/\mu$, but the improvement is less visible for small $k/\mu$. In particular, the extended approximation performs best for $k/\mu\gtrsim 0.07$.

\subsubsection*{The longitudinal correlator}

\begin{figure}[H]
	\begin{subfigure}{\textwidth}
		\centering
		\begin{subfigure}{0.4\textwidth}
			\centering
			\includegraphics[width=\textwidth]{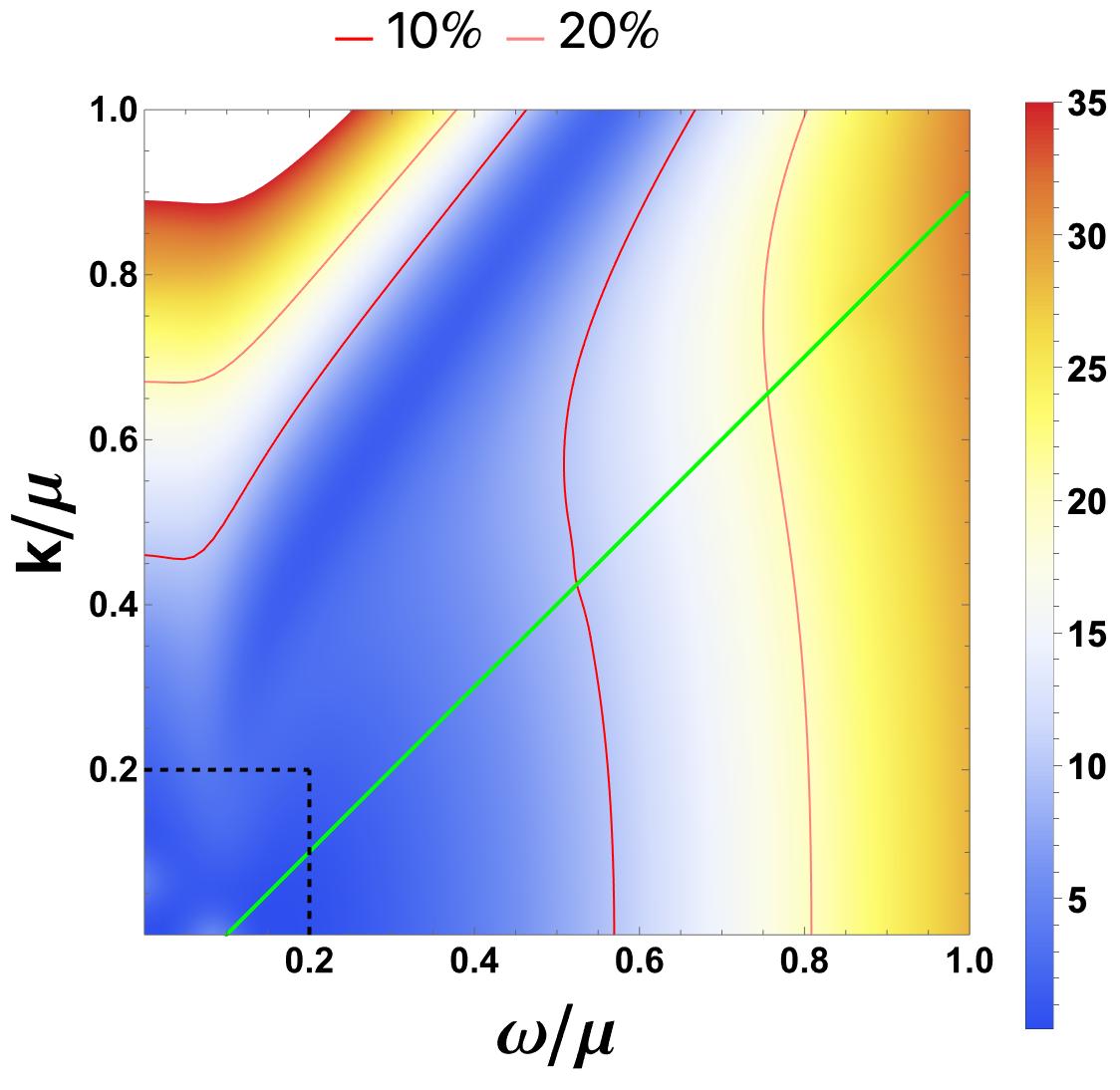}
		\end{subfigure}
		\hspace{1cm}
		\begin{subfigure}{0.4\textwidth}
			\centering
			\includegraphics[width=\textwidth]{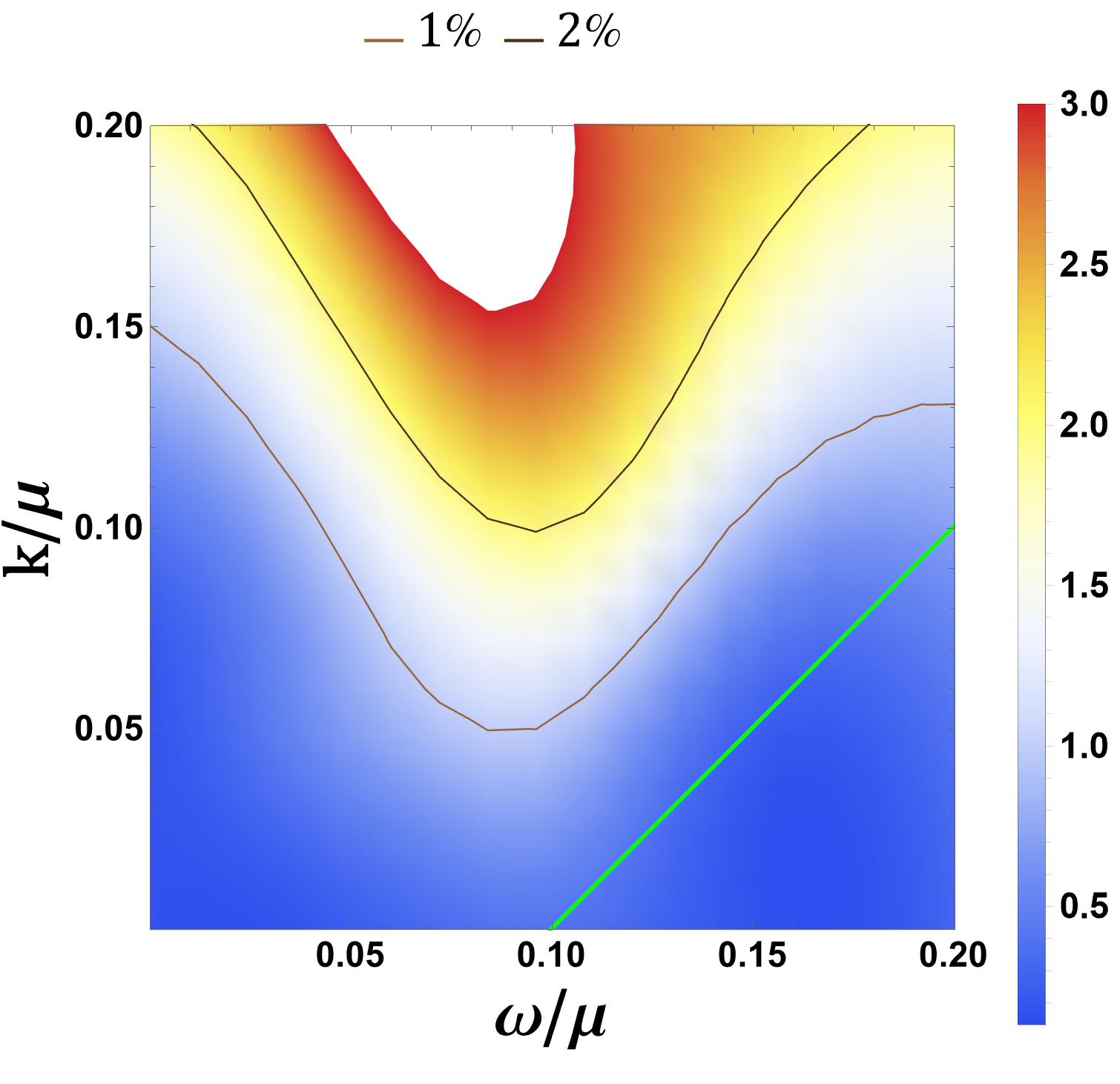}
		\end{subfigure}
	\end{subfigure}

	\begin{subfigure}{\textwidth}
		\centering
		\begin{subfigure}{0.4\textwidth}
			\centering
			\includegraphics[width=\textwidth]{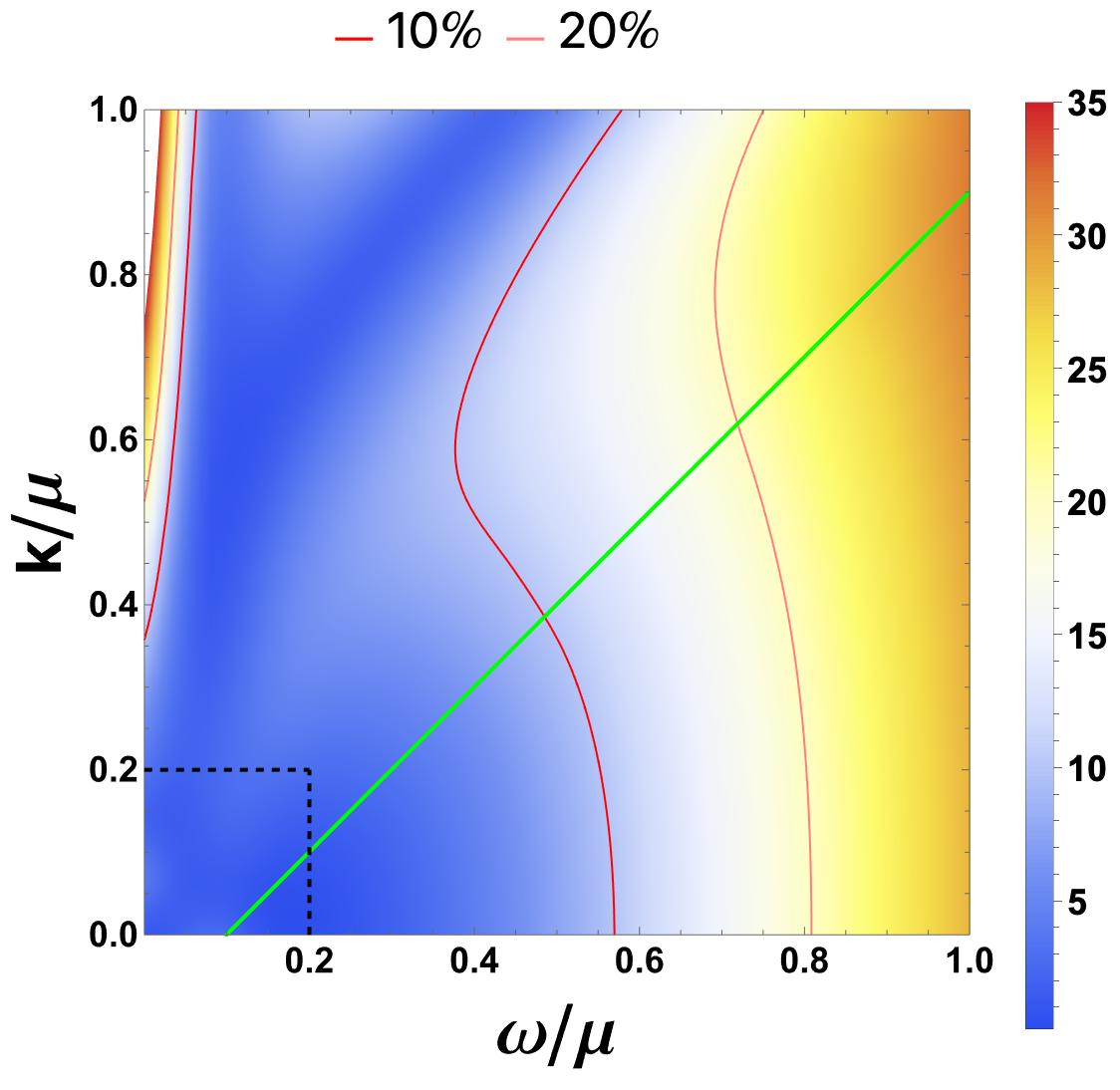}
		\end{subfigure}
		\hspace{1cm}
		\begin{subfigure}{0.4\textwidth}
			\centering
			\includegraphics[width=\textwidth]{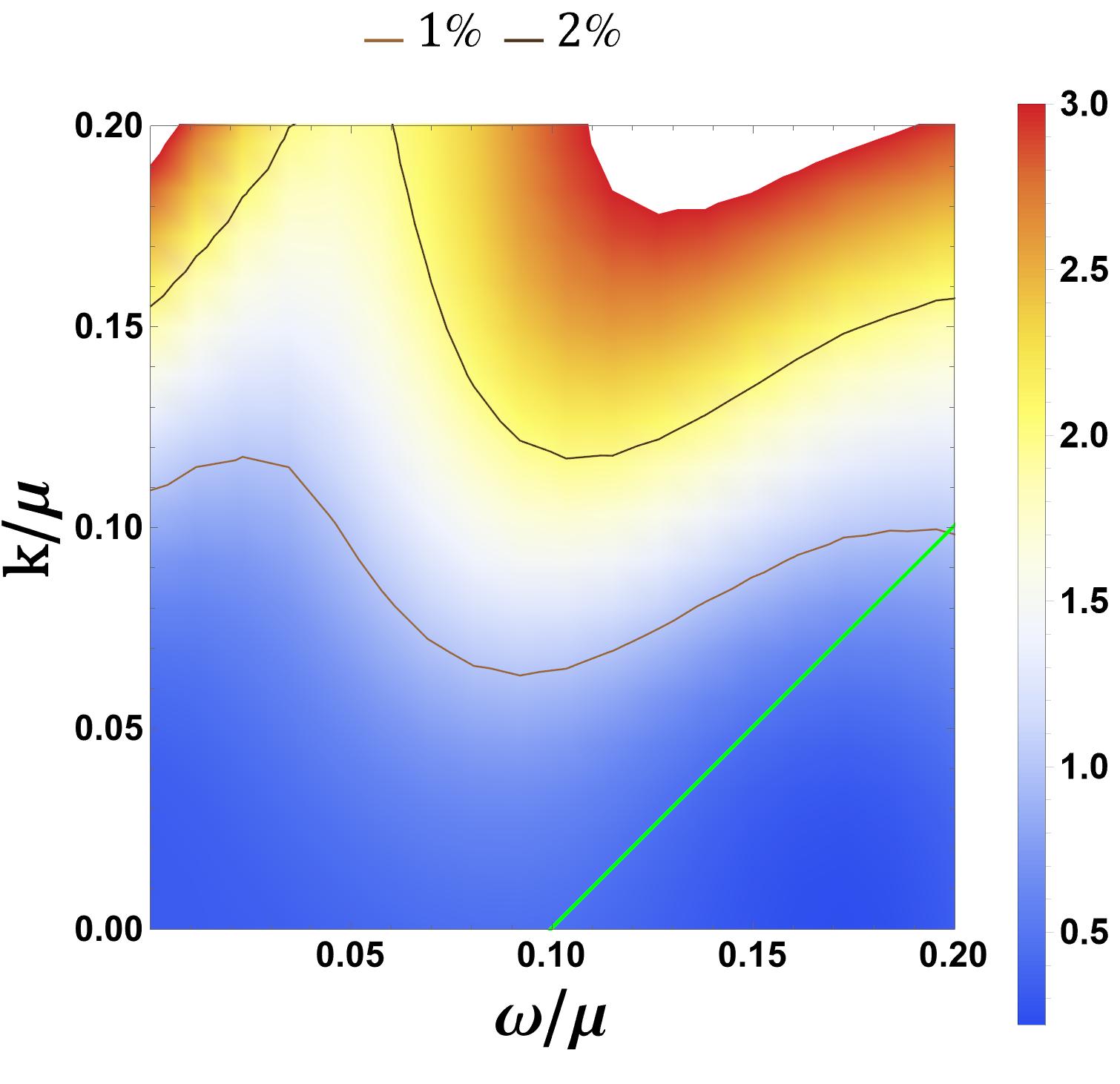}
		\end{subfigure}
	\end{subfigure}

	\caption{As figure \ref{fig:RelDiffNum_tr_mu5mu301} but for the longitudinal polarization function instead of the transverse polarization function.}\label{fig:RelDiffNum_long_mu5mu301}
\end{figure}

In figure~\ref{fig:RelDiffNum_long_mu5mu301}, the longitudinal correlator at $\mu_q/T=5$ has a broad low-error band in the extended hydrodynamic regime (left column plots). For the hydrodynamic approximation, the error is below $10\%$ in a sizeable region around intermediate values of $\omega/\mu$, while it increases at large $\omega/\mu$ and also in the upper-left corner of the plot. The extended hydrodynamic approximation bends the 10\% line towards smaller $\omega/\mu$ around $k/\mu\simeq 0.6$, improving the approximation at small and intermediate frequencies. It does not however remove the increase of the error at large $\omega/\mu$.

In the zoomed plots (right column plots), the hydrodynamic approximation displays a characteristic cap-shaped high-error region around intermediate $\omega/\mu(\simeq 0.1)$ and relatively large $k/\mu(\gtrsim 0.15)$. The extended approximation reshapes this region rather than removing it: it lowers the error in part of the lower-left region, but the upper part of the zoom remains at the percent level. Therefore, the extended approximation helps locally but does not give a uniform improvement in the IR.

\subsection{Numerical results at $\mu_3/\mu_q=-0.5$}\label{app:mu305}
In this section, we present the results for the imaginary part of the charged current polarization function at non-zero isospin chemical potential $\mu_3/\mu_q=-0.5$ for $\mu_q/T\in\{10^4,65,5\}$. First, we present the plots for the imaginary part of the polarization functions in figures \ref{fig:pol10405} ($\mu_q/T=10^4$), \ref{fig:pol6505} ($\mu_q/T=65$) and \ref{fig:pol505} ($\mu_q/T=5$). Then, we show the plots of the relative difference \eqref{eq:percrelfdiff} at $\mu_3=-0.5$ for  $\mu_q/T =10^4$ (figure \ref{fig:RelDiffNum_tr_mu104mu305}, and \ref{fig:RelDiffNum_long_mu104mu305} for the transverse and longitudinal sector respectively), $\mu_q/T=65$ (figure \ref{fig:RelDiffNum_tr_mu65mu305}, and \ref{fig:RelDiffNum_long_mu65mu305} for the transverse and longitudinal sector respectively), and $\mu_q/T=5$ (figure \ref{fig:RelDiffNum_tr_mu5mu305}, and \ref{fig:RelDiffNum_long_mu5mu305} for the transverse and longitudinal sector respectively). The top and bottom rows show respectively the relative difference with respect to the hydrodynamic and the extended hydrodynamic approximation.

This value of isospin is well beyond what is needed in neutron star physics. We analyze  it however, as we want to show how such large values depart from the hydro and extended hydro regimes, especially at small values of $\omega/\mu$ and $k/\mu$.

\subsubsection{Background at $\mu_q/T=10^4$}
\begin{figure}[H]
	\centering
	\begin{subfigure}{0.48\textwidth}
		\centering
		\includegraphics[width=\textwidth]{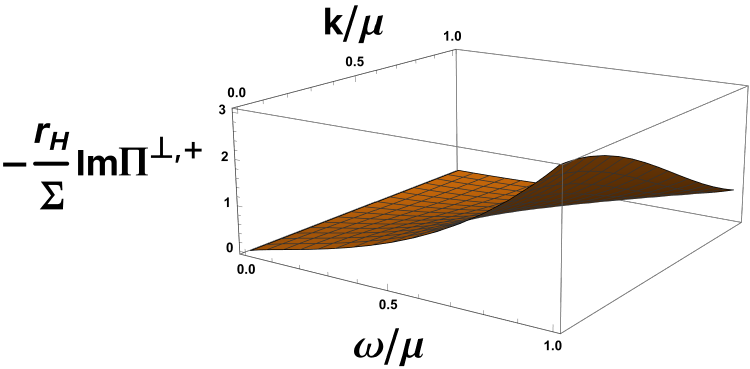}
	\end{subfigure}
	\hfill
	\begin{subfigure}{0.5\textwidth}
		\centering
		\includegraphics[width=\textwidth]{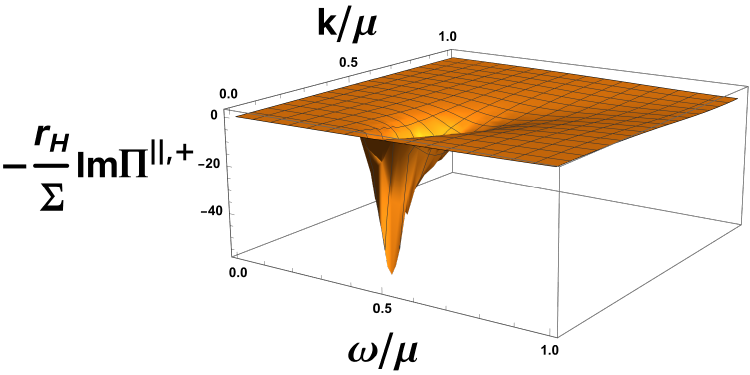}
	\end{subfigure}
	\caption{Imaginary part of the transverse (left panel) and longitudinal (right panel) charged current retarded polarization function (normalized by $-\Sigma/r_H$) for $\mu_q/T = 10^4$, and $\mu_3/\mu_q = -0.5$.} \label{fig:pol10405}
\end{figure}

In figure \ref{fig:pol10405} we present the plots for the imaginary part of the transverse (left panel) and longitudinal (right panel) polarization functions at $\mu_q/T=10^4$ and finite isospin chemical potential $\mu_3/\mu_q=-0.5$. The transverse polarization function is still qualitatively affected by the presence of the isospin chemical potential, and it increases as the frequency increases. Also in this case, the longitudinal polarization function develops a diffusive pole with a real part $\sim -\mu_3$. In the presence of an even more negative value of $\mu_3$, the magnitude of the polarization functions further  decreases.\footnote{We checked that for positive isospin chemical potential, the magnitude of the polarization functions increases instead of decreasing, with respect to the case of $\mu_3=0$. This is due to the relativistic dispersion relation of the plus-charged correlator which reads $(\omega+\mu_3)^2-\vec k^2$.}

In the following, we present the results for the comparison of the exact numerical results with the hydrodynamic and the extended hydrodynamic approximation.

\subsubsection*{The transverse correlator}
\begin{figure}[H]
	\begin{subfigure}{\textwidth}
		\centering
		\begin{subfigure}{0.4\textwidth}
			\centering
			\includegraphics[width=\textwidth]{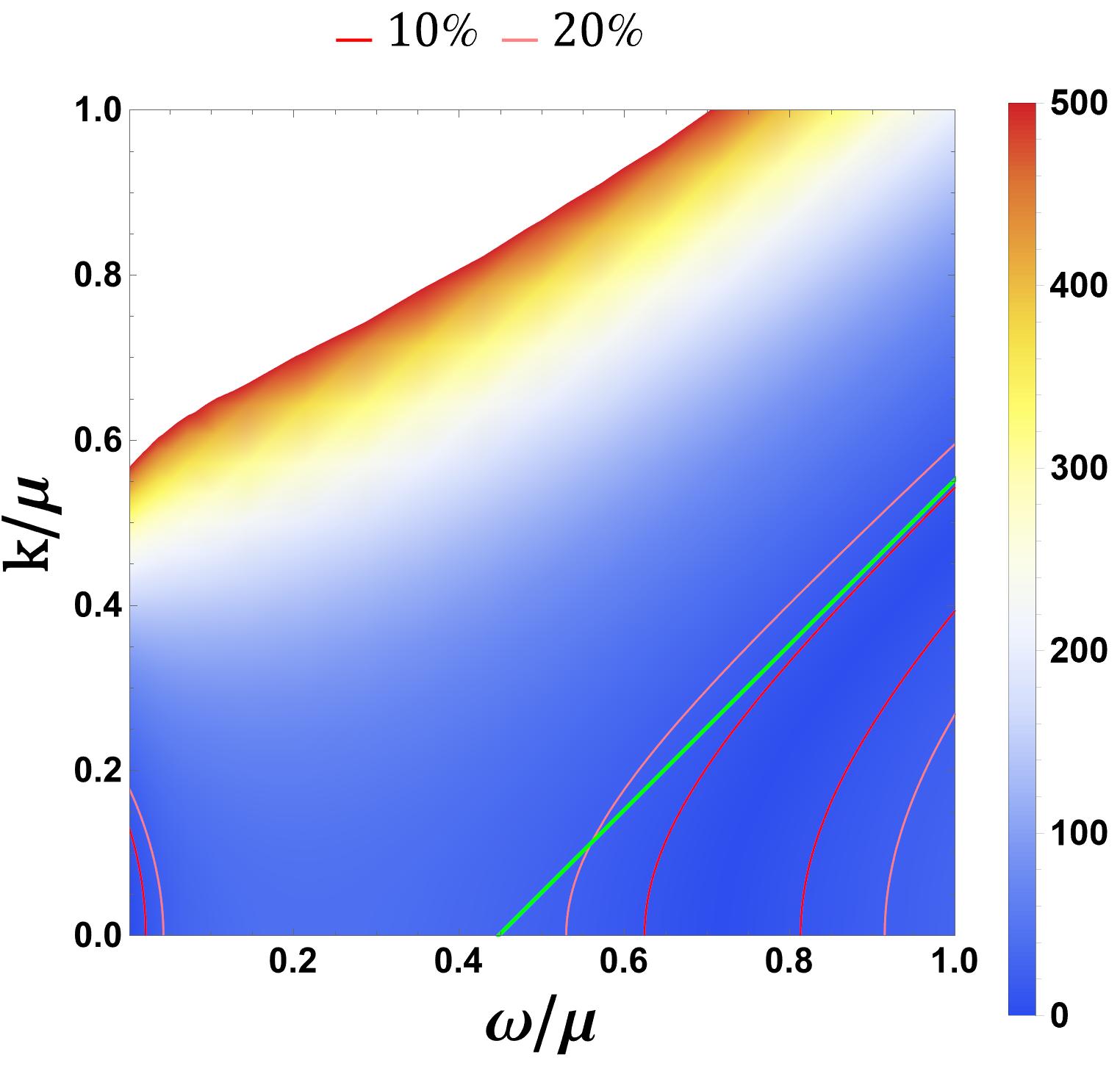}
		\end{subfigure}
		\hspace{1cm}
		\begin{subfigure}{0.4\textwidth}
			\centering
			\includegraphics[width=\textwidth]{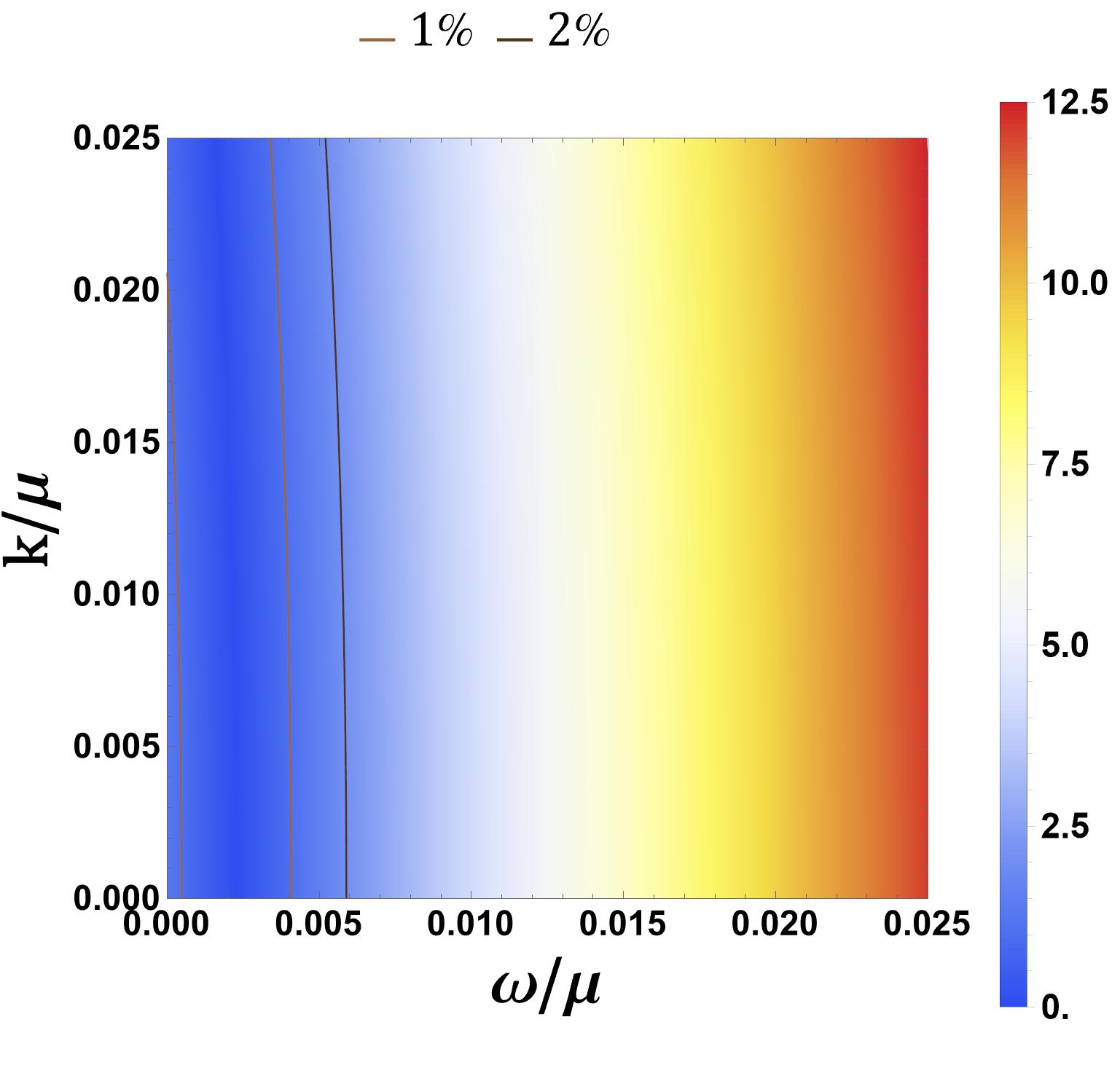}
		\end{subfigure}
	\end{subfigure}

	\begin{subfigure}{\textwidth}
		\centering
		\begin{subfigure}{0.4\textwidth}
			\centering
			\includegraphics[width=\textwidth]{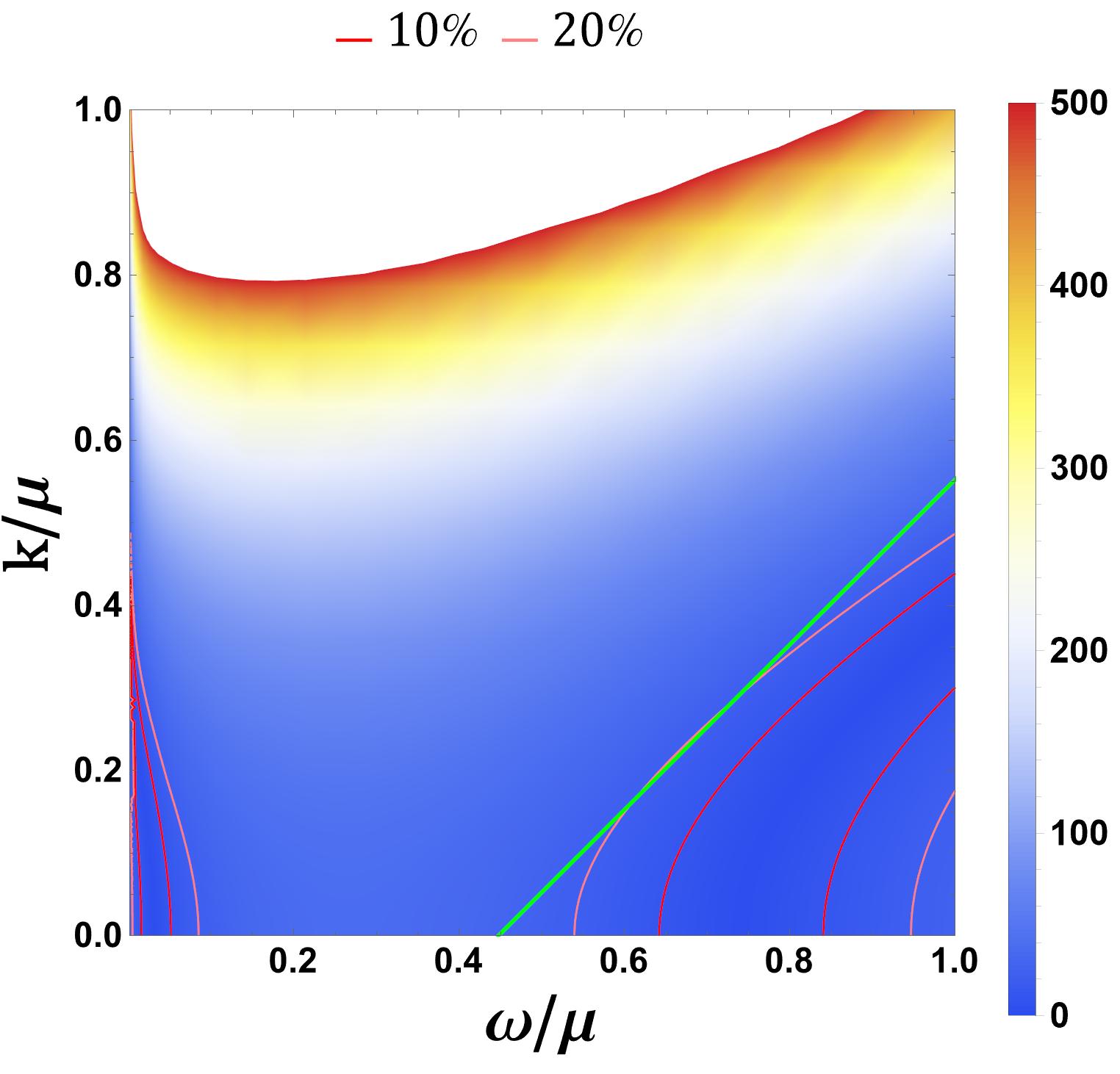}
		\end{subfigure}
		\hspace{1cm}
		\begin{subfigure}{0.4\textwidth}
			\centering
			\includegraphics[width=\textwidth]{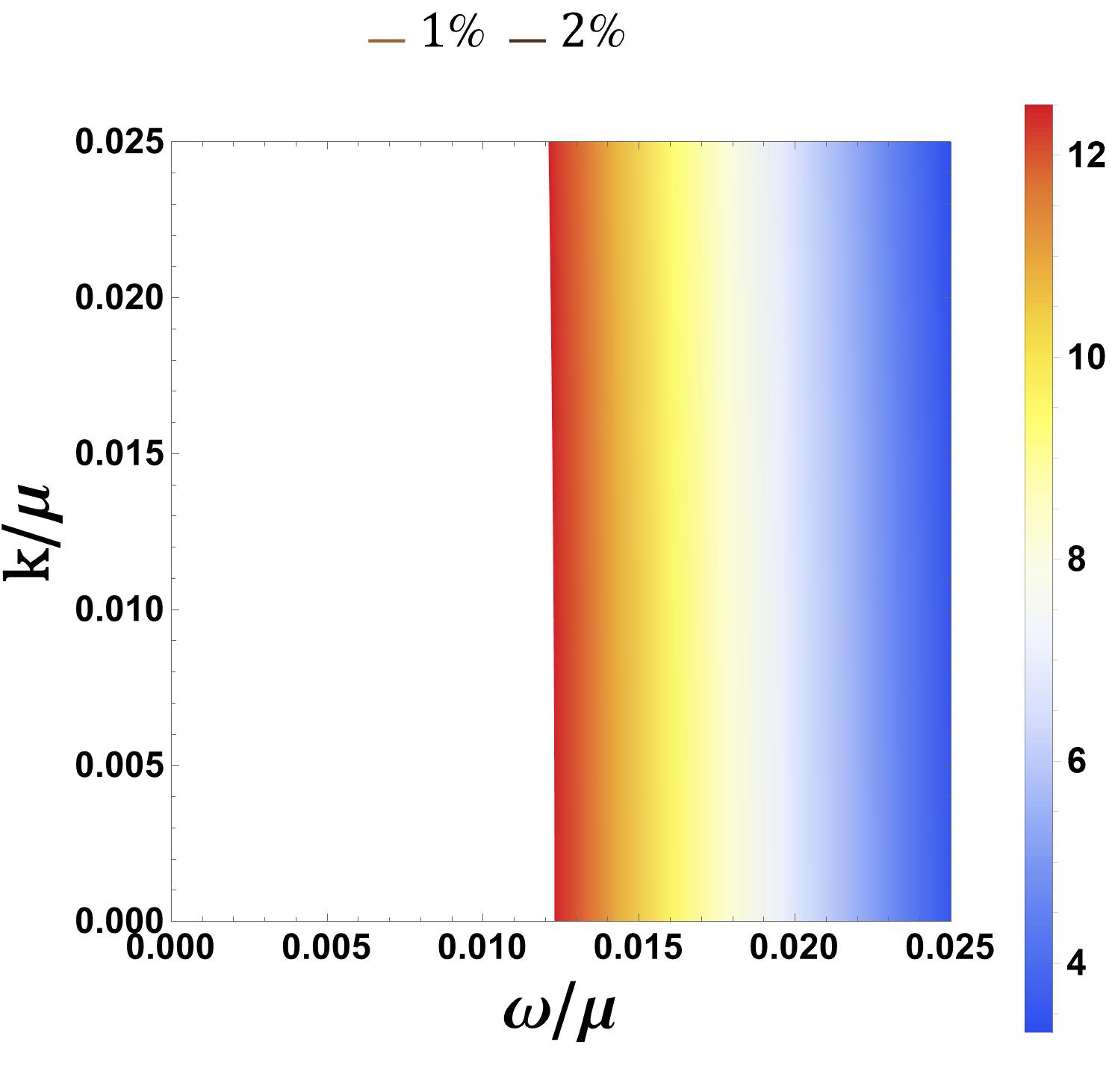}
		\end{subfigure}
	\end{subfigure}
	
	\caption{The percentage relative difference \eqref{eq:percrelfdiff} of the imaginary part of the transverse charged current polarization function with respect to the hydrodynamic approximation \eqref{eq:trhydroapprox} (top row) and the extended hydrodynamic approximation \eqref{eq:trexthydroapprox} (bottom row), for $\mu_q/T= 10^4$ and $\mu_3=-0.5$. The right plots shows a subregion of the left
		plots. The green line shows the locus $\omega = k+\mu_3$.} \label{fig:RelDiffNum_tr_mu104mu305}
\end{figure}

Figure~\ref{fig:RelDiffNum_tr_mu104mu305} shows a strong departure from the hydrodynamic regime once $\mu_3/\mu_q$ is increased to $-0.5$. In the full-range plots (left column), both approximations have very large errors over a substantial part of the plane, especially at large $k/\mu$, where the color scale reaches several hundred percent (at $\omega\simeq 0$ and $k/\mu$ around 0.6 the percentage error is bigger than 500\%). The $10\%$ and $20\%$ are now more limited and do not reach values of $k/\mu$ bigger than $0.6$ in the extended hydrodynamic region.

The extended approximation still improves the position of the 10\% and 20\% contours in the region of small frequencies of the left plots, but the improvement is much less decisive than for smaller $|\mu_3|$.

The zoomed plots (right column) are particularly important: even in the very small region $\omega/\mu,k/\mu\lesssim 0.025$, the error is already several percent and is organized in almost vertical bands. For the hydrodynamic approximation, the error increases from values of order a few percent at $\omega/\mu\lesssim 0.005$ to values of order $10\%$ or larger by $\omega/\mu\simeq 0.02$--$0.025$, even for $k/\mu\leq 0.025$.

For the extended hydrodynamic approximation, the zoomed plot does not show a region with $1\%$--$2\%$ accuracy; instead, the visible range starts at errors of order several percent, and a vertical structure appears around $\omega/\mu\simeq 0.012$--$0.015$. The extended approximation does not restore a small-error hydrodynamic corner; instead, it shifts the vertical bands and leaves a sizeable error throughout the zoom. Here we are witnessing a decisive departure from the hydrodynamic regime signaled by the high values of relative difference.

\subsubsection*{The longitudinal correlator}

\begin{figure}[H]
	\begin{subfigure}{\textwidth}
		\centering
		\begin{subfigure}{0.4\textwidth}
			\centering
			\includegraphics[width=\textwidth]{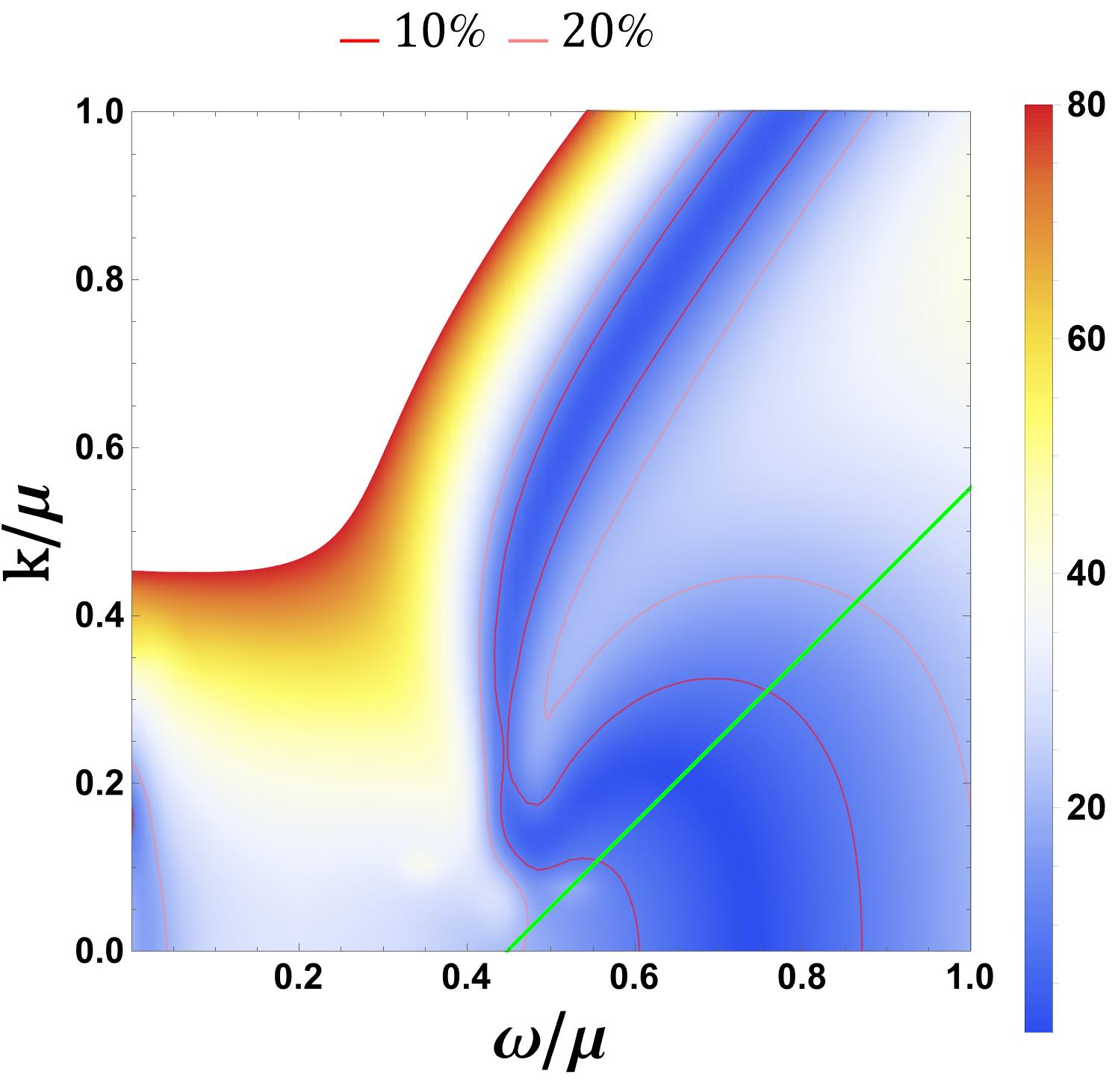}
		\end{subfigure}
		\hspace{1cm}
		\begin{subfigure}{0.4\textwidth}
			\centering
			\includegraphics[width=\textwidth]{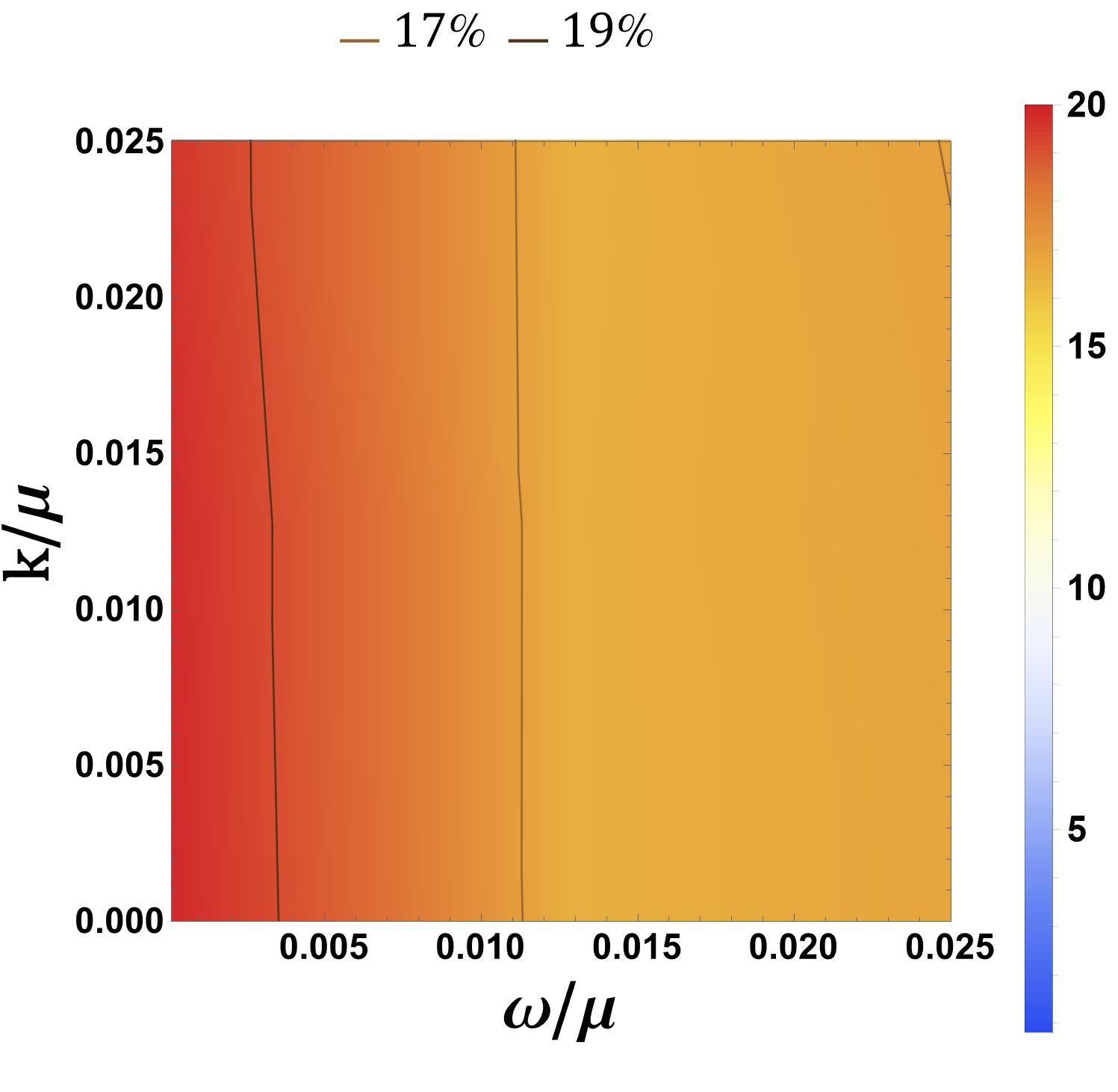}
		\end{subfigure}
	\end{subfigure}

	\begin{subfigure}{\textwidth}
		\centering
		\begin{subfigure}{0.4\textwidth}
			\centering
			\includegraphics[width=\textwidth]{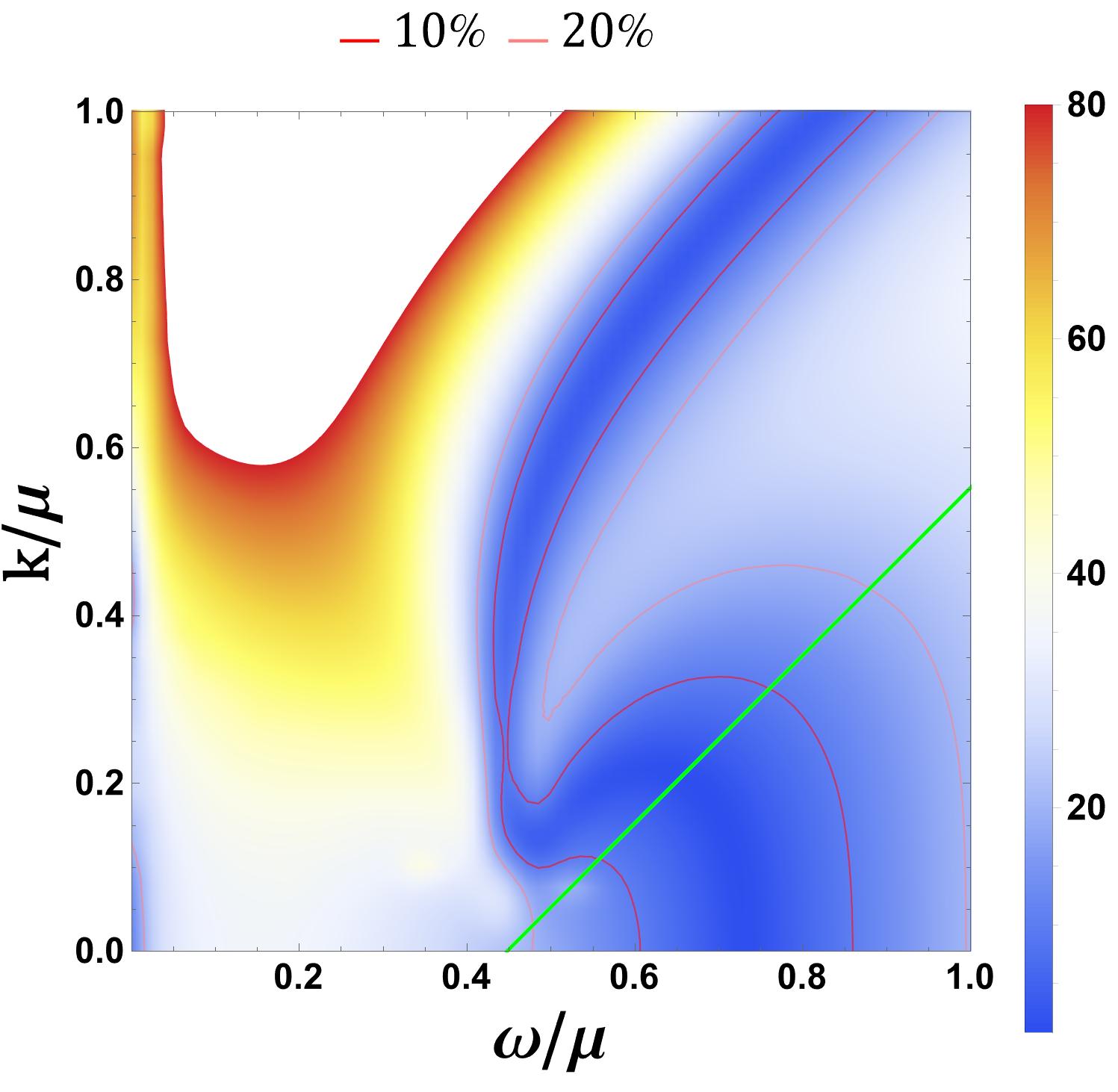}
		\end{subfigure}
		\hspace{1cm}
		\begin{subfigure}{0.4\textwidth}
			\centering
			\includegraphics[width=\textwidth]{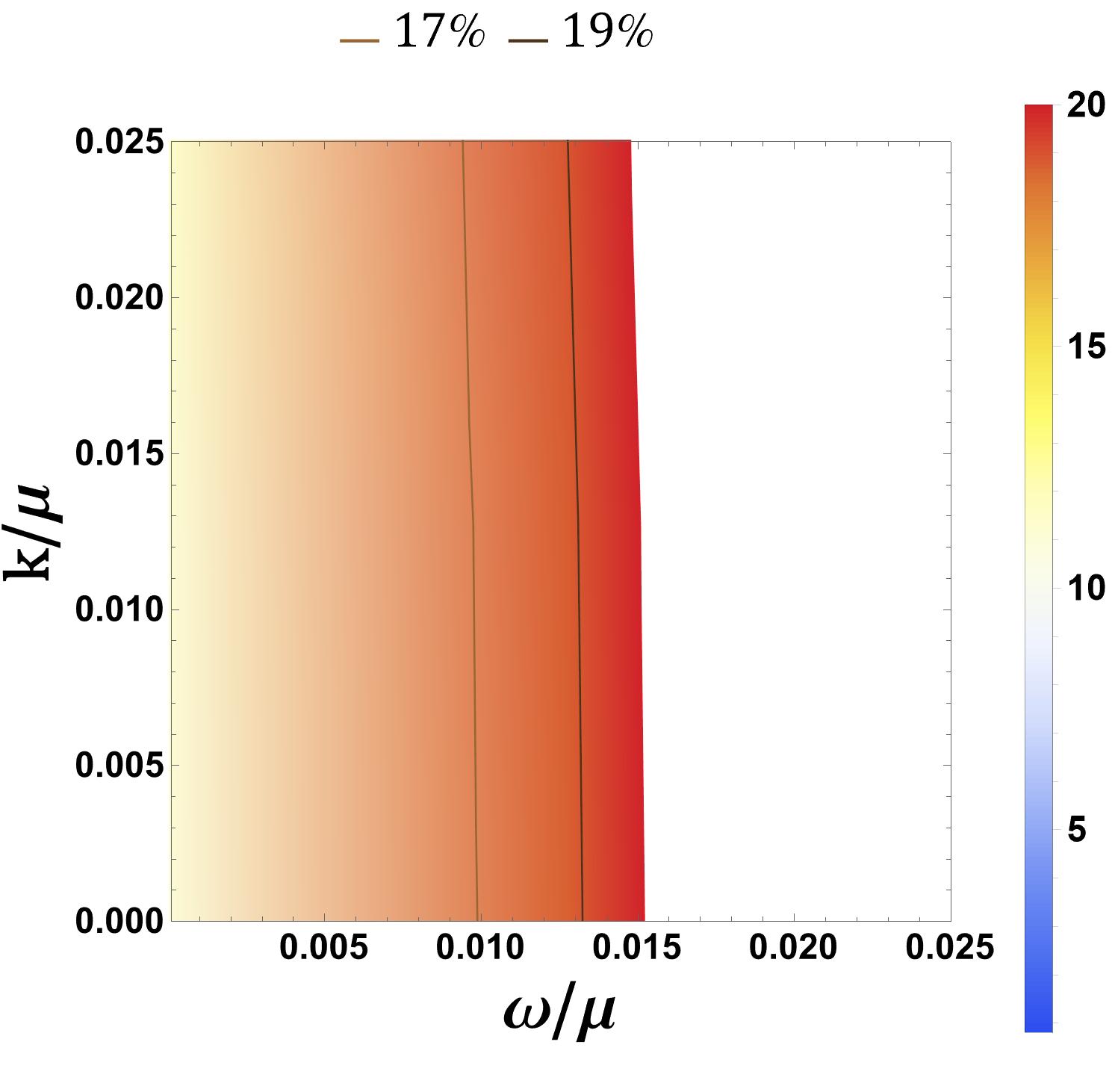}
		\end{subfigure}
	\end{subfigure}

	\caption{As figure \ref{fig:RelDiffNum_tr_mu104mu305} but for the longitudinal polarization function instead of the transverse polarization function.}\label{fig:RelDiffNum_long_mu104mu305}
\end{figure}

In figure~\ref{fig:RelDiffNum_long_mu104mu305}, the longitudinal sector also shows a strong breakdown of the simple hydrodynamic picture. In the top-left plot, associated with the hydrodynamic approximation, the error is not monotonic in either variable. There is a low-error valley around $\omega/\mu\simeq 0.6$--$0.8$ and $k/\mu\lesssim 0.3$, but the error grows strongly for large $k/\mu$ at smaller frequencies, especially around $\omega/\mu\simeq 0.2$--$0.5$. The extended hydrodynamic approximation changes the shape of this structure but does not eliminate it. In particular, it reduces the relative error for small frequencies and high momenta, but a large-error band remains at intermediate $\omega/\mu$ and large $k/\mu$.

The zoomed plots are even more striking: the contours are at the 17\%--19\% level, so the error is already large in the region that would normally be considered deep IR. The extended approximation does not improve this region in a useful way; it mainly shifts the high-error band in $\omega/\mu$. This indicates that for large $|\mu_3|$ the finite isospin scale dominates over the usual hydrodynamic expansion. The left column plots signal that a small relative difference is only recovered at frequencies of order the isospin chemical potential $\omega\simeq -\mu_3$. One can therefore argue that the would-be hydrodynamic region is shifted around $\omega\simeq -\mu_3$.

\subsubsection{Background at $\mu_q/T=65$}
\begin{figure}[htb]
	\centering
	\begin{subfigure}{0.48\textwidth}
		\centering
		\includegraphics[width=\textwidth]{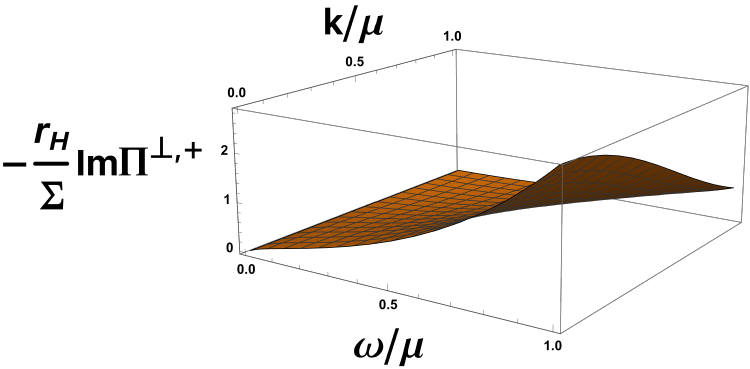}
	\end{subfigure}
	\hfill
	\begin{subfigure}{0.5\textwidth}
		\centering
		\includegraphics[width=\textwidth]{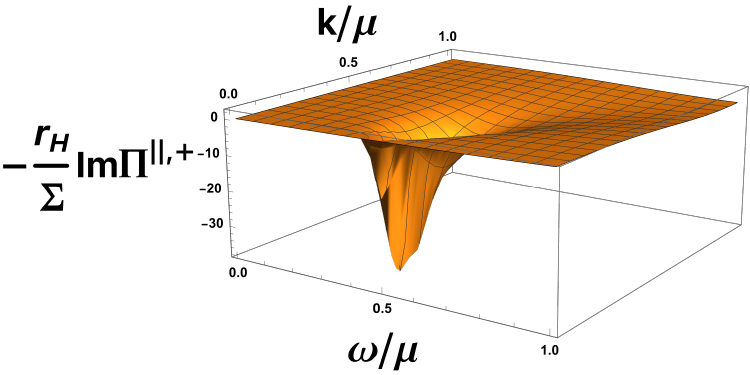}
	\end{subfigure}
	\caption{Imaginary part of the transverse (left panel) and longitudinal (right panel) charged current retarded polarization function (normalized by $-\Sigma/r_H$) for $\mu_q/T = 65$, and $\mu_3=-0.5$.} \label{fig:pol6505}
\end{figure}

In figure \ref{fig:pol6505} we present the plots for the imaginary part of the transverse (left panel) and longitudinal (right panel) polarization functions at $\mu_q/T=65$ and finite isospin chemical potential $\mu_3/\mu_q=-0.5$. These polarization functions can be compared with the corresponding one at $\mu_q/T=65$ in the main text, figure \ref{fig:transversepol} (transverse) and \ref{fig:longitudinalpol} (longitudinal). We can observe how the presence of a larger (in absolute value) $\mu_3$ decreases the magnitude of the polarization function and shifts the diffusive pole to a larger positive frequency.

In the following, we present the results for the comparison of the exact numerical results with the hydrodynamic and the extended hydrodynamic approximation.

In figure~\ref{fig:RelDiffNum_tr_mu65mu305}, the transverse sector at $\mu_q/T=65$ and $\mu_3/\mu_q=-0.5$ shows the same qualitative departure from hydrodynamics as at $\mu_q/T=10^4$, but the zoomed region is now large enough to display explicitly the square $0\leq \omega/\mu,k/\mu\leq T/\mu$. In the large hydrodynamic plot, the $10\%$ and $20\%$ contours lie near the lower-right part of the figure, while the error is large over most of the region with large $k/\mu$ and small or moderate $\omega/\mu$. The extended hydrodynamic approximation gives only a small improvement in the large plot; the position of the $10\%$ and $20\%$ contours changes mildly, but the overall pattern remains essentially the same.

\subsubsection*{The transverse correlator}

\begin{figure}[H]
	\begin{subfigure}{\textwidth}
		\centering
		\begin{subfigure}{0.4\textwidth}
			\centering
			\includegraphics[width=\textwidth]{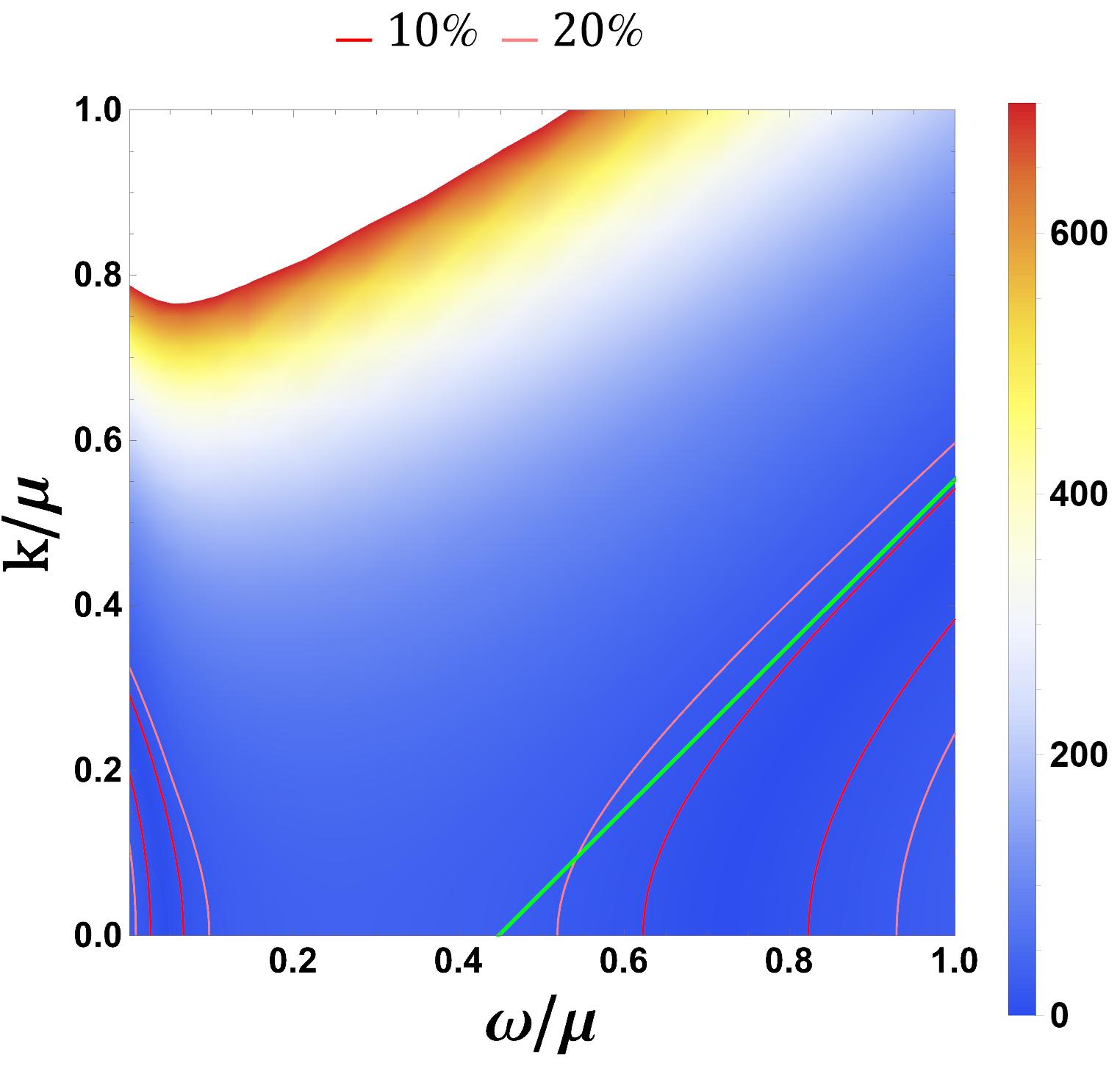}
		\end{subfigure}
		\hspace{1cm}
		\begin{subfigure}{0.4\textwidth}
			\centering
			\includegraphics[width=\textwidth]{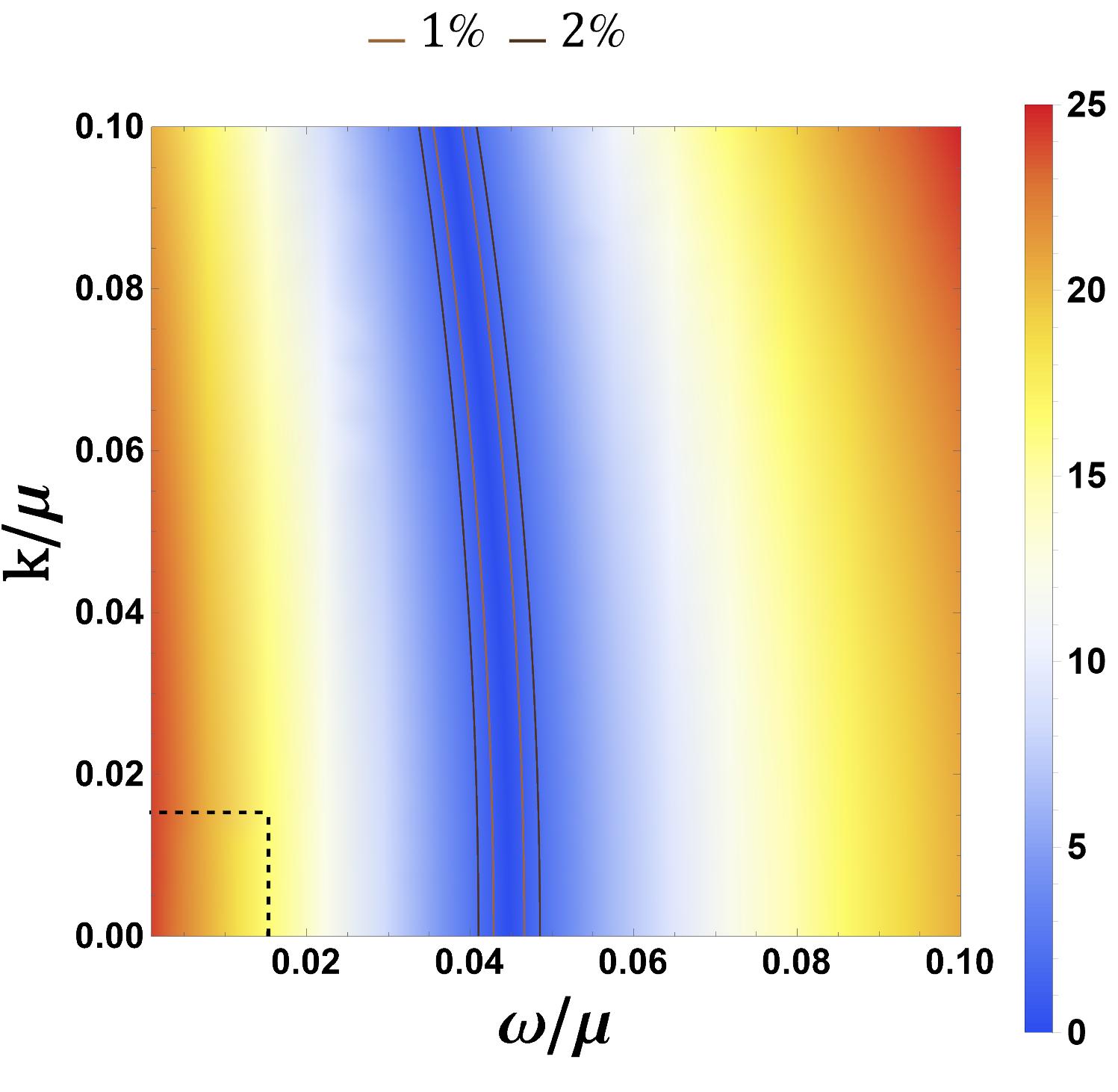}
		\end{subfigure}
	\end{subfigure}

	\begin{subfigure}{\textwidth}
		\centering
		\begin{subfigure}{0.4\textwidth}
			\centering
			\includegraphics[width=\textwidth]{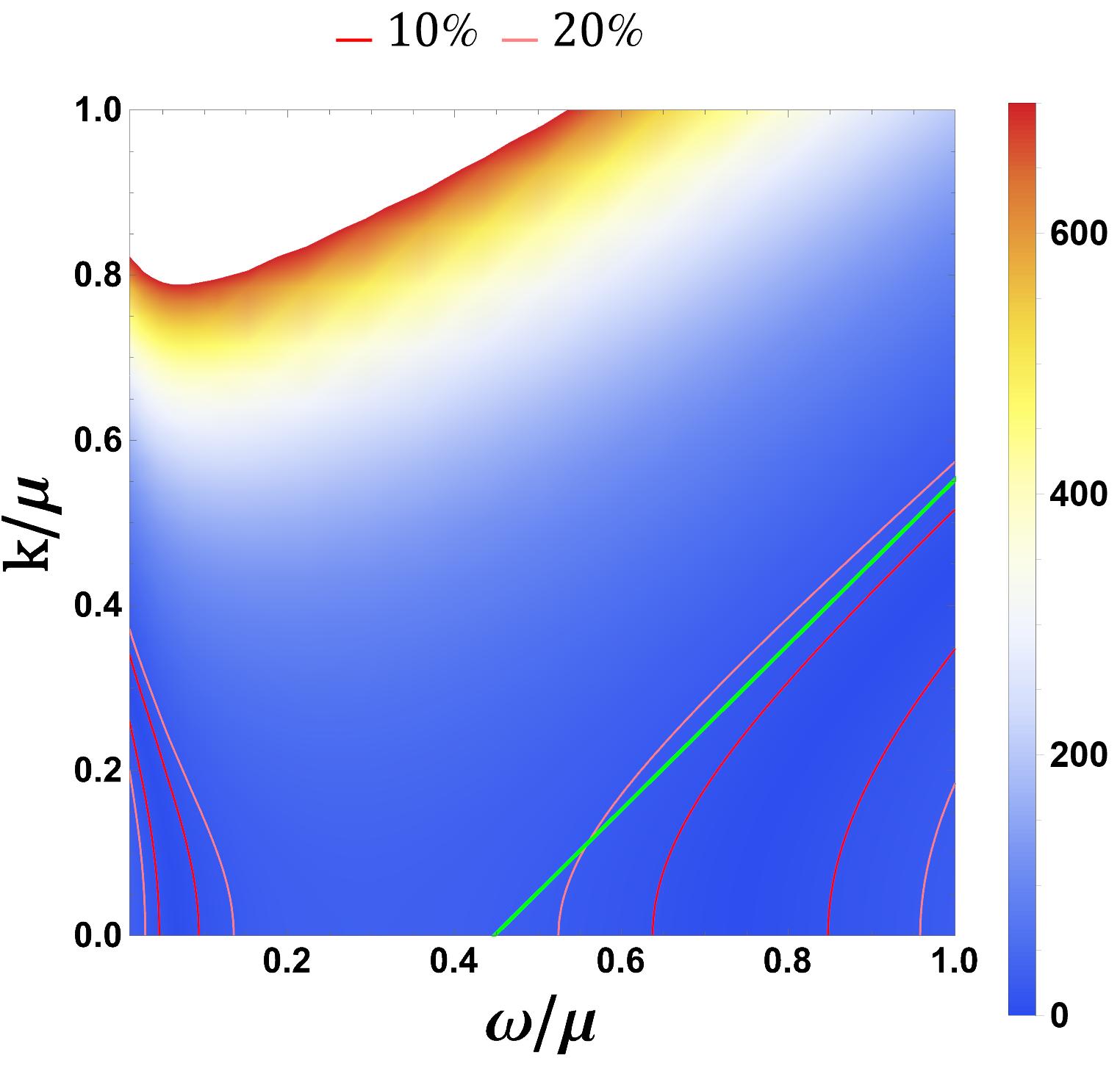}
		\end{subfigure}
		\hspace{1cm}
		\begin{subfigure}{0.4\textwidth}
			\centering
			\includegraphics[width=\textwidth]{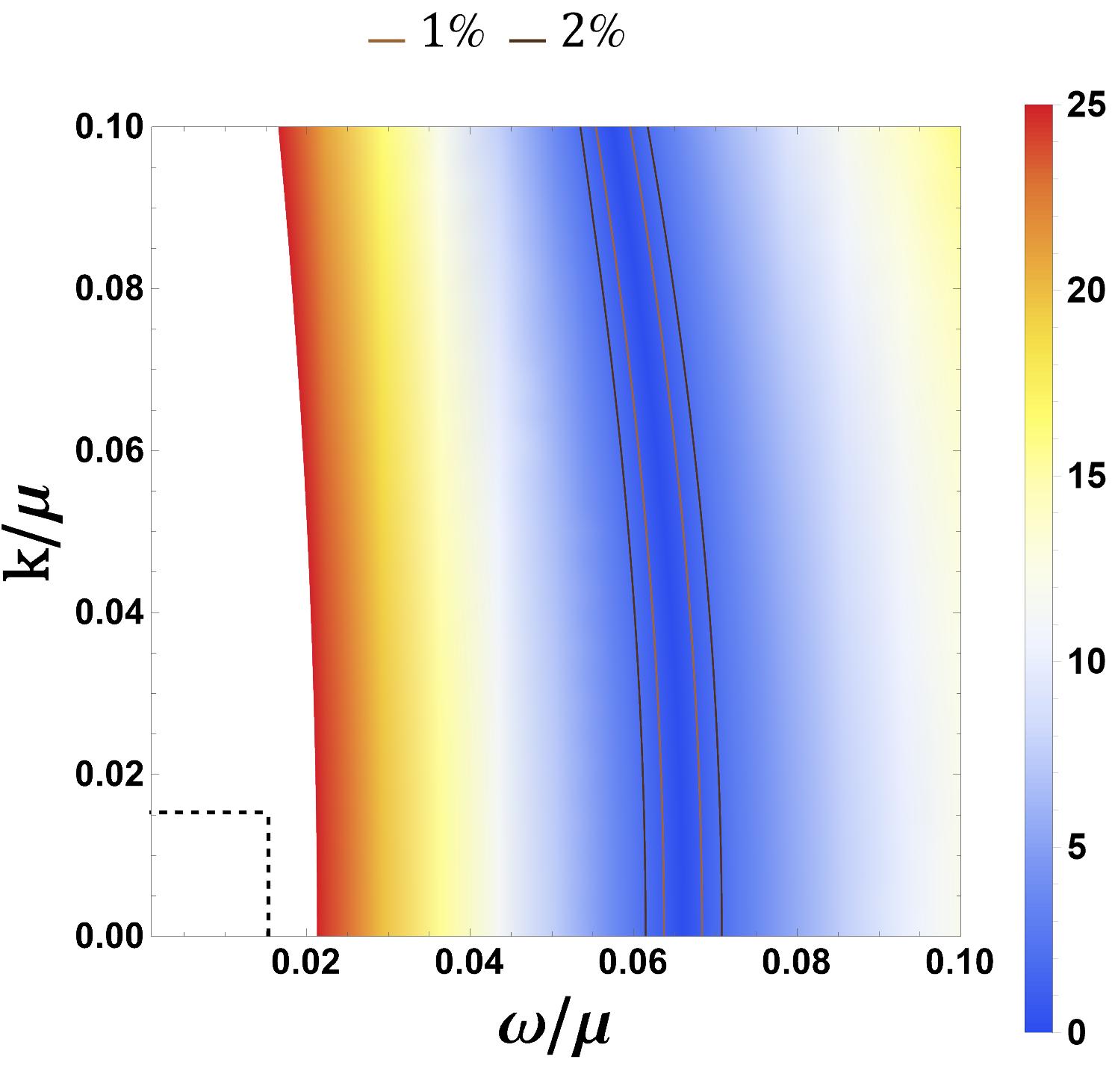}
		\end{subfigure}
	\end{subfigure}
	
	\caption{The percentage relative difference \eqref{eq:percrelfdiff} of the imaginary part of the transverse charged current polarization function with respect to the hydrodynamic approximation \eqref{eq:trhydroapprox} (top row) and the extended hydrodynamic approximation \eqref{eq:trexthydroapprox} (bottom row), for $\mu_q/T= 65$ and $\mu_3=-0.5$. The right plots shows a subregion of the left
		plots. The dashed square in the right column represents the hydrodynamic region $\omega/\mu,k/\mu\in [0,T/\mu]$. The green line shows the locus $\omega = k+\mu_3$. } \label{fig:RelDiffNum_tr_mu65mu305}
\end{figure}

In figure~\ref{fig:RelDiffNum_tr_mu65mu305}, the left column plots are very similar to the near-extremal case $\mu_q/T= 10^4$ with the same value of $\mu_3/\mu_q$. The error is extremely large at high $k/\mu$ (it reaches more than $700\%$ in the top-left corner), while the 10\% and 20\% contours enclose a small corner at low frequency and momentum and they appear again for $\omega/\mu\gtrsim 0.5$ at $k\simeq 0$ bending towards $\omega/k\simeq 1$ at $k/\mu\lesssim 0.6$.  The extended hydrodynamic approximation gives only a small improvement: the position of the $10\%$ and $20\%$ contours changes mildly, but the overall pattern remains essentially the same.

In the zoomed plots, the dashed square marks the nominal hydrodynamic region, but the error is not small inside or near it. Instead, the relative difference forms nearly vertical bands in $\omega/\mu$, with a narrow low-error band (1\% and 2\% lines) trough around $\omega/\mu\simeq 0.04$--$0.05$ for the hydrodynamic approximation in the top-right plot. The extended approximation shifts this trough to slightly larger $\omega/\mu\simeq 0.055$--$0.065$, but it does not create a broad hydrodynamic domain. This is a clear example where the scale set by $\mu_3$ spoils the naive expectation that for smaller values of $\omega$ and $k$ the approximations become  better.

\begin{figure}[H]
	\begin{subfigure}{\textwidth}
		\centering
		\begin{subfigure}{0.4\textwidth}
			\centering
			\includegraphics[width=\textwidth]{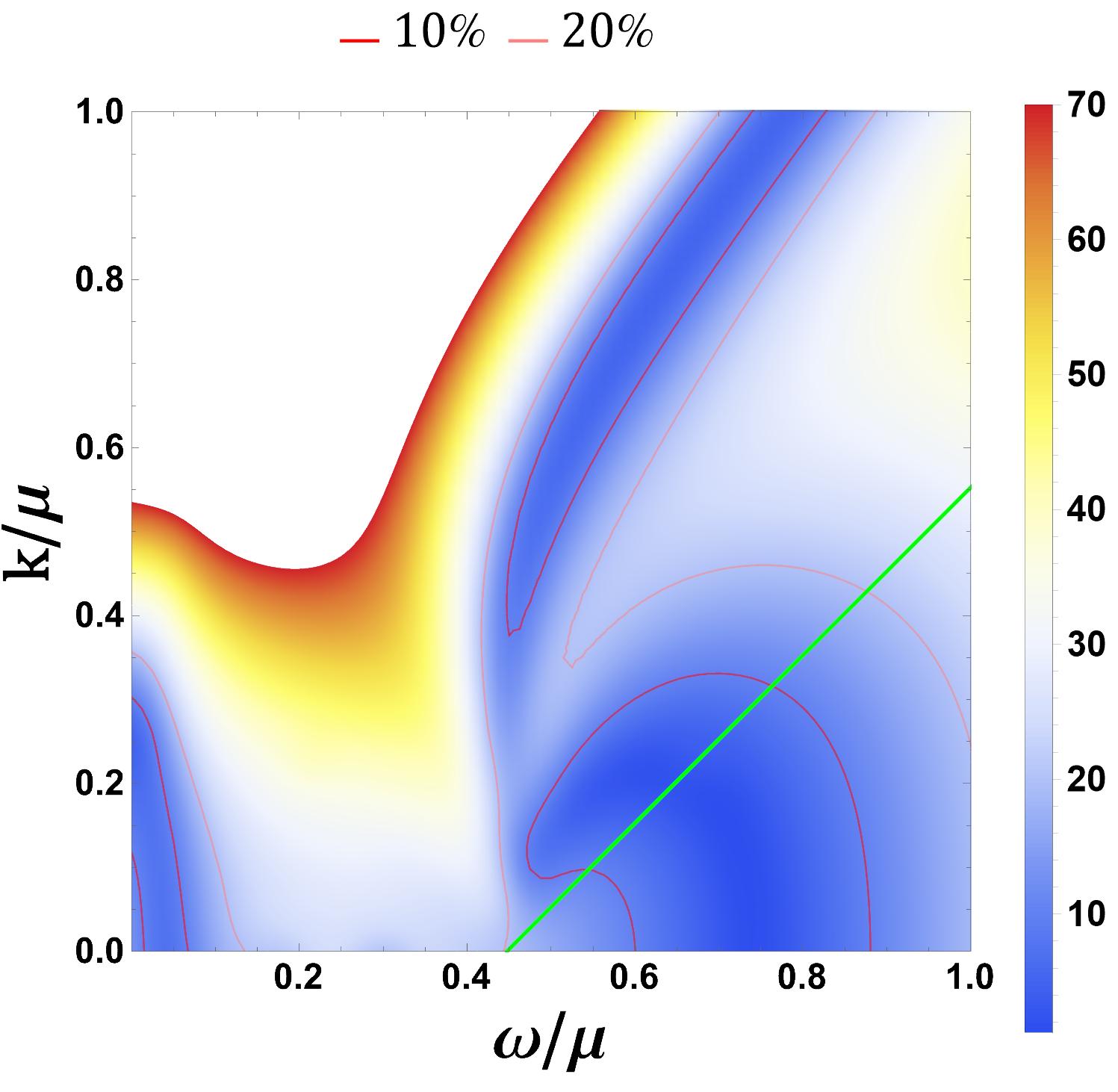}
		\end{subfigure}
		\hspace{1cm}
		\begin{subfigure}{0.4\textwidth}
			\centering
			\includegraphics[width=\textwidth]{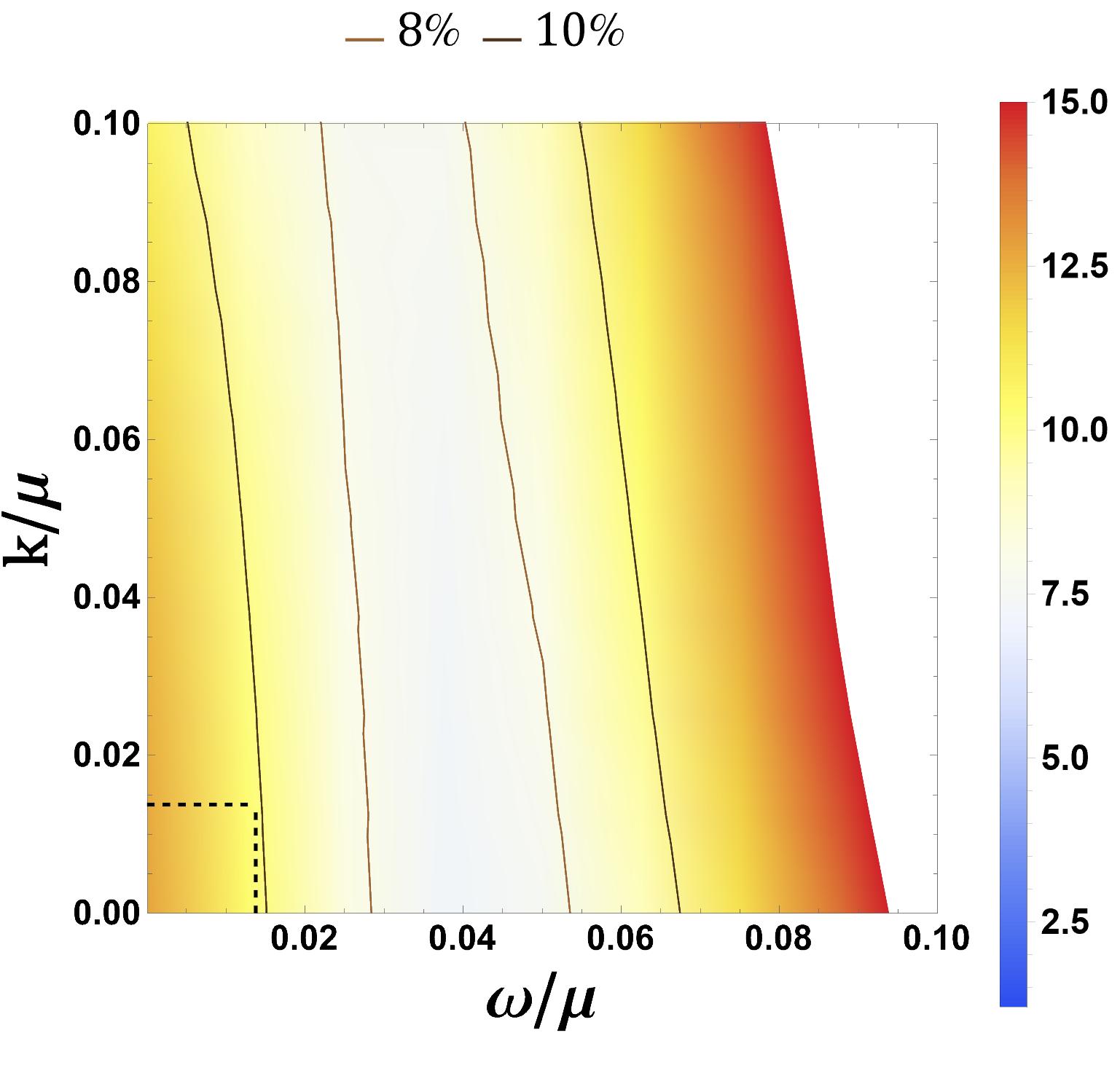}
		\end{subfigure}
	\end{subfigure}

	\begin{subfigure}{\textwidth}
		\centering
		\begin{subfigure}{0.4\textwidth}
			\centering
			\includegraphics[width=\textwidth]{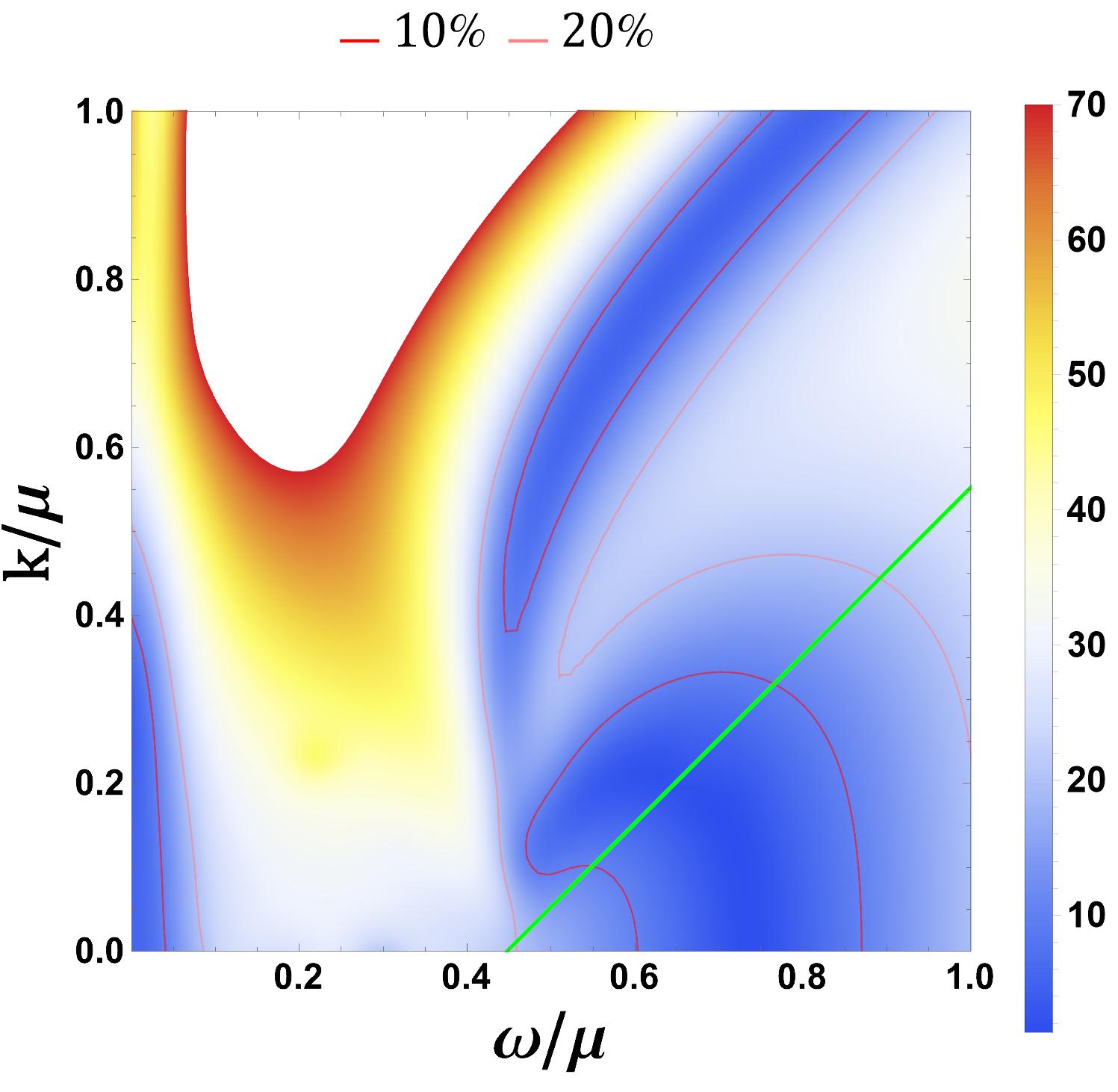}
		\end{subfigure}
		\hspace{1cm}
		\begin{subfigure}{0.4\textwidth}
			\centering
			\includegraphics[width=\textwidth]{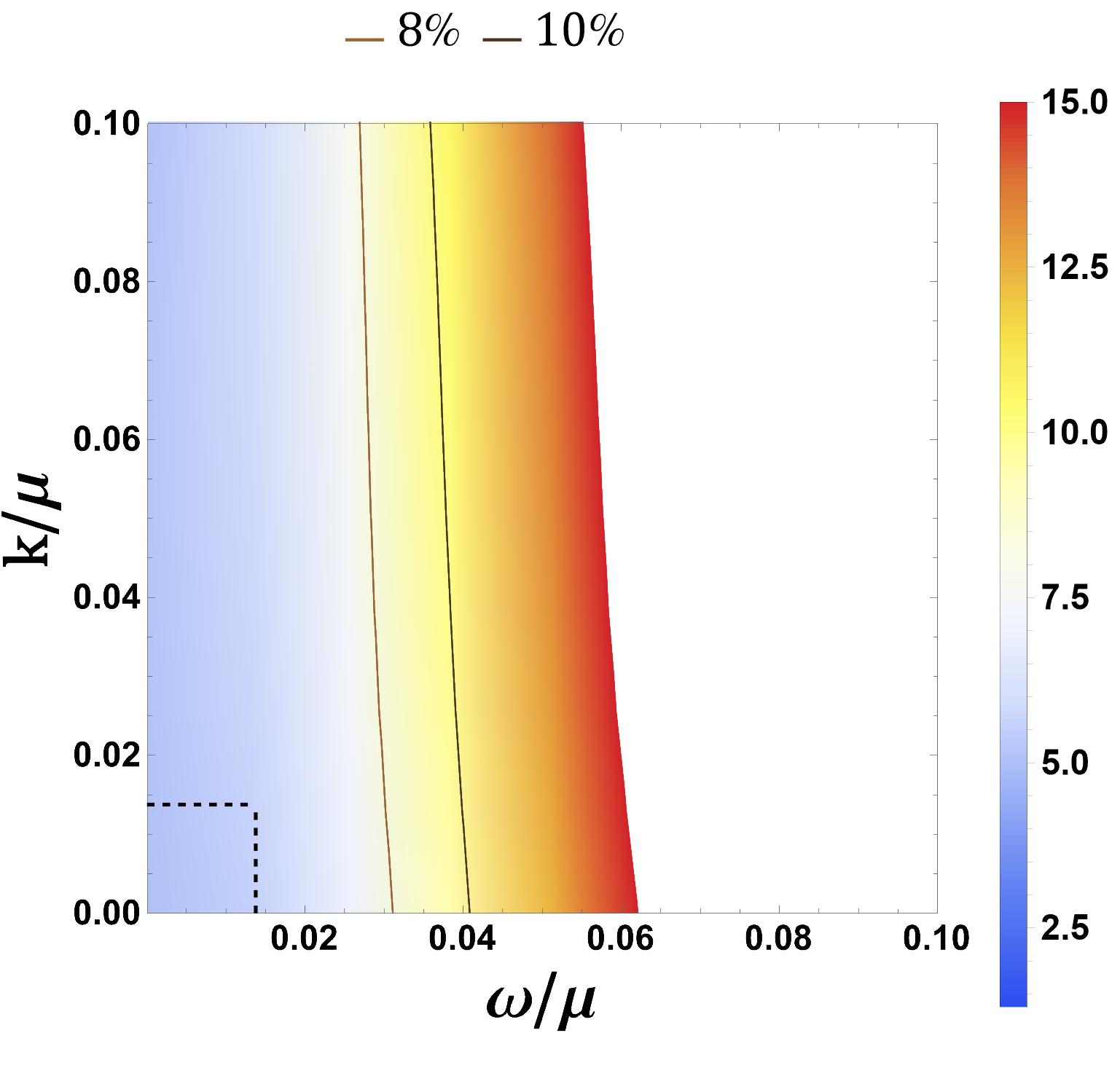}
		\end{subfigure}
	\end{subfigure}	
	\caption{As figure \ref{fig:RelDiffNum_tr_mu65mu305} but for the longitudinal polarization function instead of the transverse polarization function.}\label{fig:RelDiffNum_long_mu65mu305}
\end{figure}

\subsubsection*{The longitudinal correlator}

Figure~\ref{fig:RelDiffNum_long_mu65mu305} shows a complicated longitudinal error pattern, with a curved high-error region at small-to-intermediate $\omega/\mu(\simeq 0.2)$ and large $k/\mu(\gtrsim 0.5 $ for the hydrodynamic approximation and $\gtrsim 0.6$ for the extended hydrodynamic approximation), and a low-error valley at larger $\omega/\mu$ of order the isospin chemical potential. The extended hydrodynamic approximation only mildly changes the plot in the extended hydrodynamic regime (left column): the same ridge and valley structure remains visible.

The zoomed plots are again quite instructive. In the hydrodynamic approximation (top-right plot), the region inside the dashed square has an error of order $10\%$--$12\%$, and the $8\%$--$10\%$ contours are located at frequencies of order $\omega/\mu\simeq 0.02$--$0.08$, slightly depending on $k/\mu$. The extended hydrodynamic approximation improves the very small-frequency part of the zoom: for $\omega/\mu\lesssim 0.02$ the error is visibly smaller than in the hydrodynamic plot. However, the error then increases quickly as $\omega/\mu$ approaches $0.04$--$0.06$. Therefore, in this longitudinal case, the extended hydrodynamic approximation gives a real improvement near the origin, but the improvement is confined to a relatively narrow range of frequencies.

\subsubsection{Background at $\mu_q/T=5$}
\begin{figure}[H]
	\centering
	\begin{subfigure}{0.48\textwidth}
		\centering
		\includegraphics[width=\textwidth]{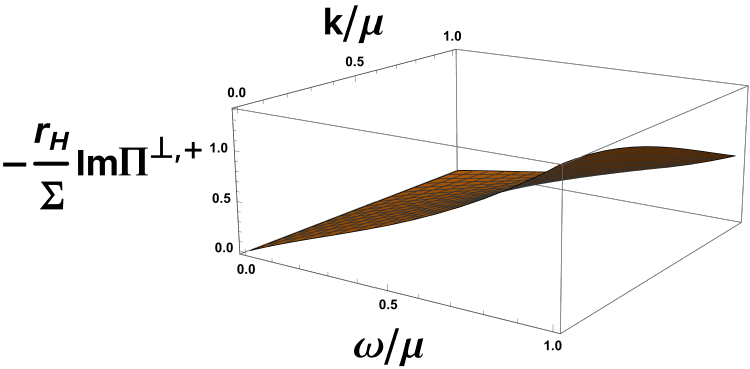}
	\end{subfigure}
	\hfill
	\begin{subfigure}{0.5\textwidth}
		\centering
		\includegraphics[width=\textwidth]{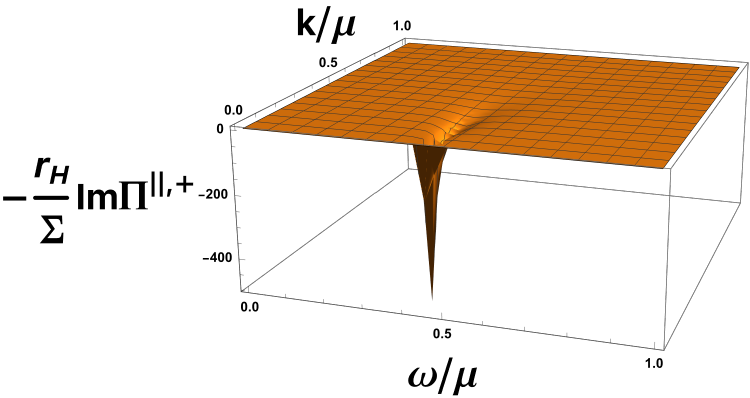}
	\end{subfigure}
	\caption{Imaginary part of the transverse (left panel) and longitudinal (right panel) charged current retarded polarization function (normalized by $-\Sigma/r_H$) for $\mu_q/T = 5$, and $\mu_3=-0.5$.} \label{fig:pol505}
\end{figure}

Finally, in figure \ref{fig:pol505} we present the plots for the imaginary part of the transverse (left panel) and longitudinal (right panel) polarization functions at $\mu_q/T=5$ and finite isospin chemical potential $\mu_3/\mu_q=-0.5$. The qualitative picture of a polarization function growing with $\omega$ and a diffusive pole with a real part $\sim -\mu_3$ in the longitudinal sector continues to hold for these values of the parameter. As already observed for other values of $\mu_3$, as we lower $\mu_q/T$, the magnitude of the polarization function decreases.

In the following, we present the results for the comparison of the exact numerical results with the hydrodynamic and the extended hydrodynamic approximation.

\subsubsection*{The transverse correlator}

\begin{figure}[H]
	\begin{subfigure}{\textwidth}
		\centering
		\begin{subfigure}{0.4\textwidth}
			\centering
			\includegraphics[width=\textwidth]{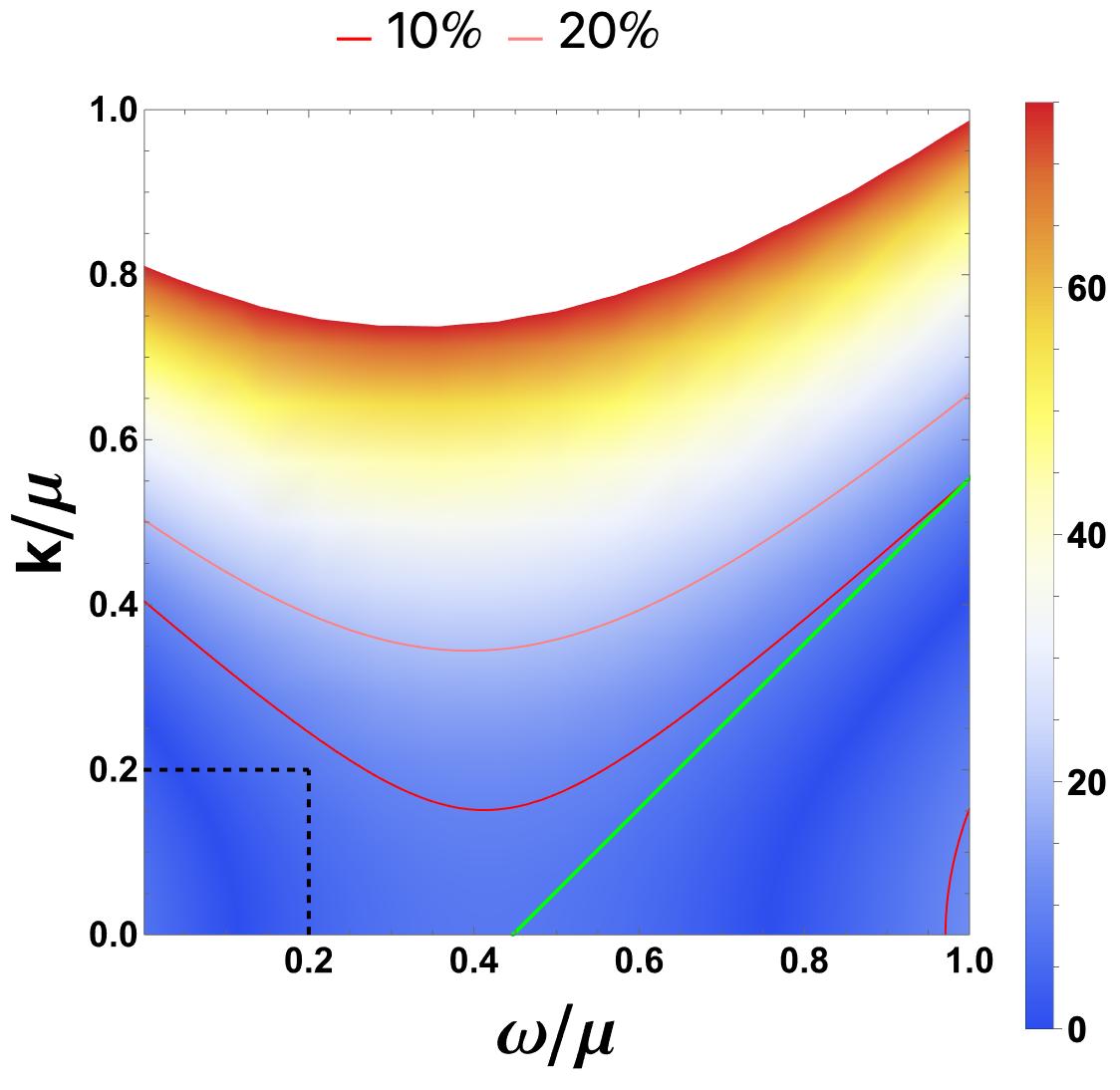}
		\end{subfigure}
		\hspace{1cm}
		\begin{subfigure}{0.4\textwidth}
			\centering
			\includegraphics[width=\textwidth]{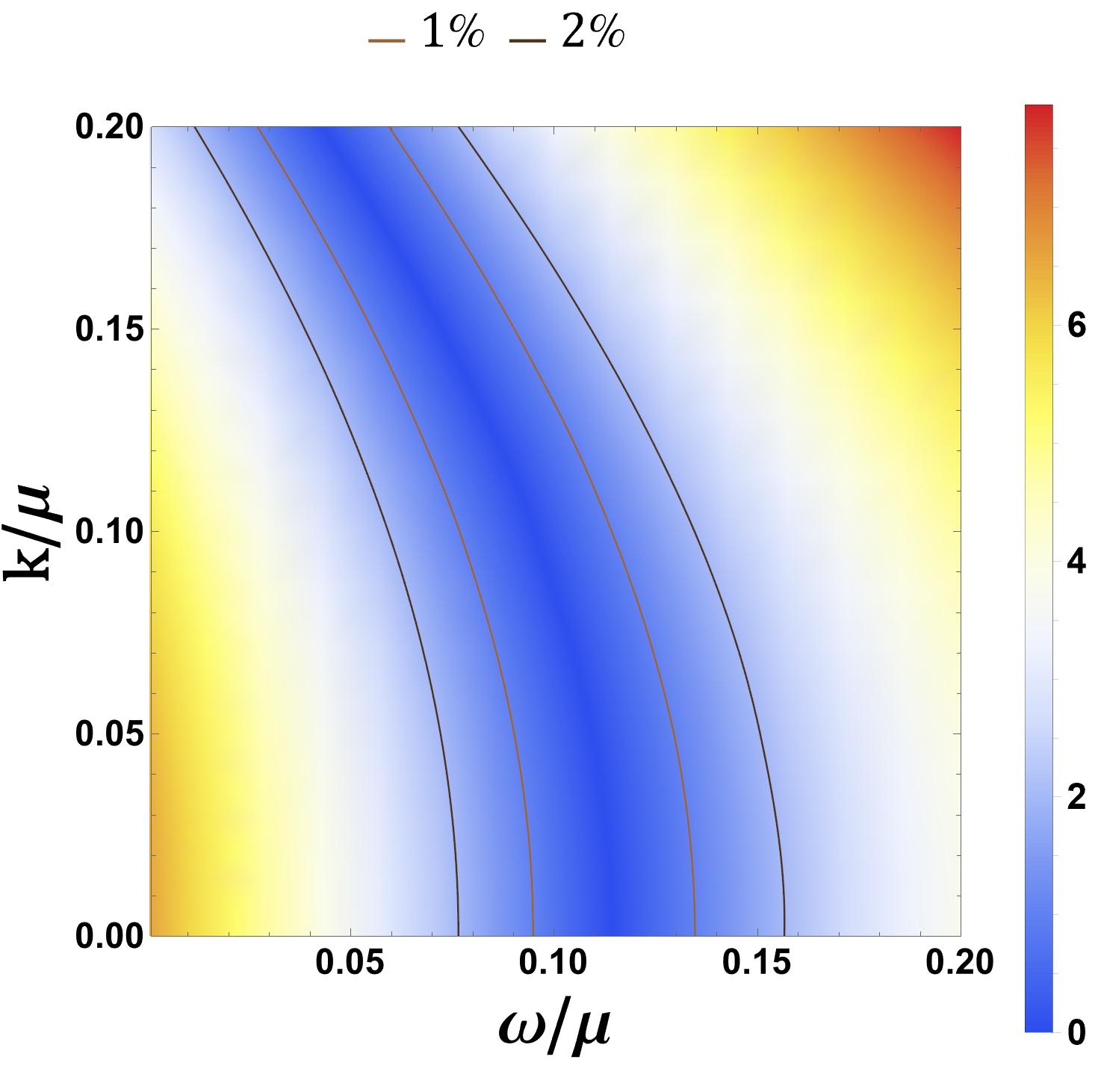}
		\end{subfigure}
	\end{subfigure}

	\begin{subfigure}{\textwidth}
		\centering
		\begin{subfigure}{0.4\textwidth}
			\centering
			\includegraphics[width=\textwidth]{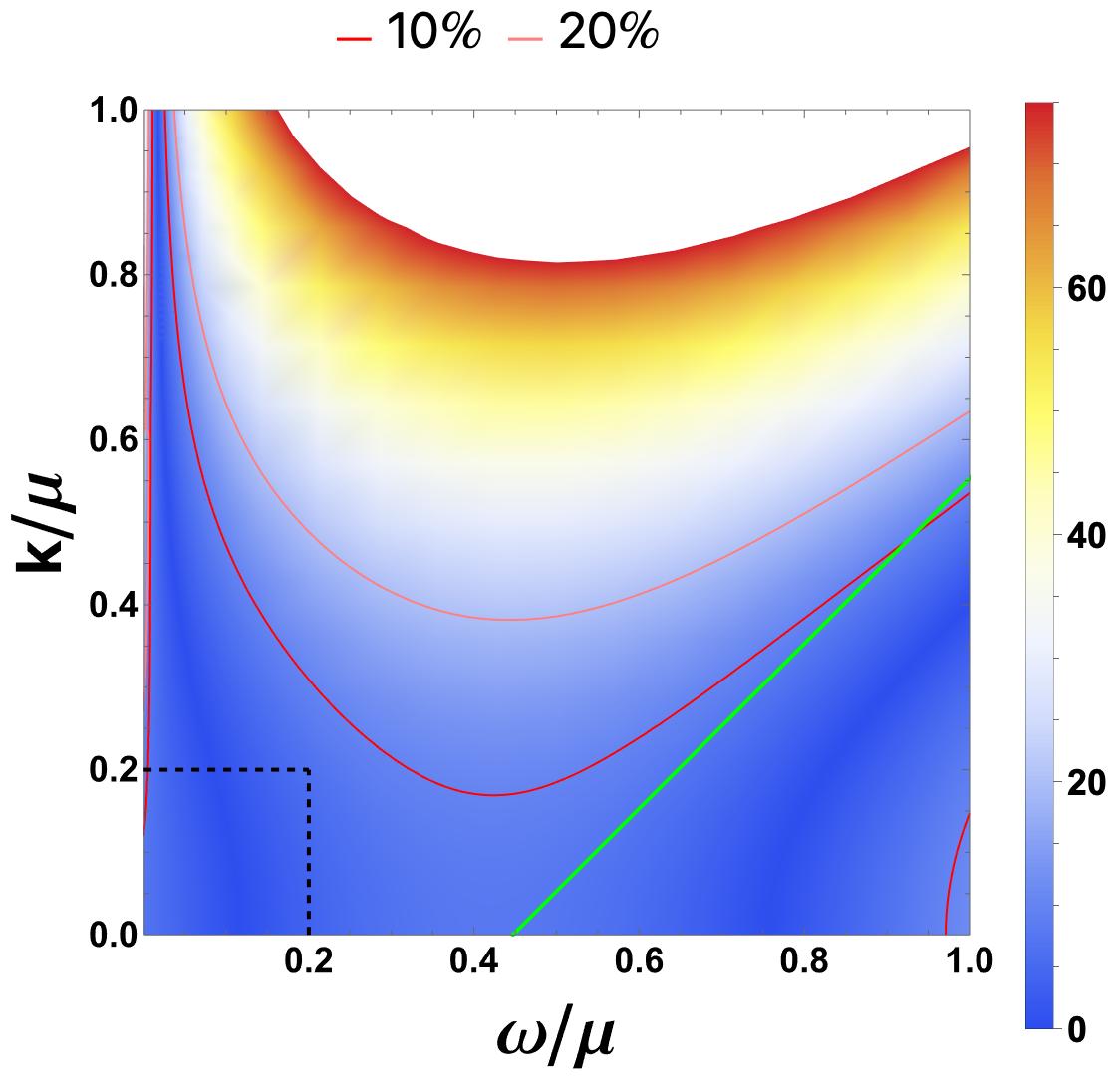}
		\end{subfigure}
		\hspace{1cm}
		\begin{subfigure}{0.4\textwidth}
			\centering
			\includegraphics[width=\textwidth]{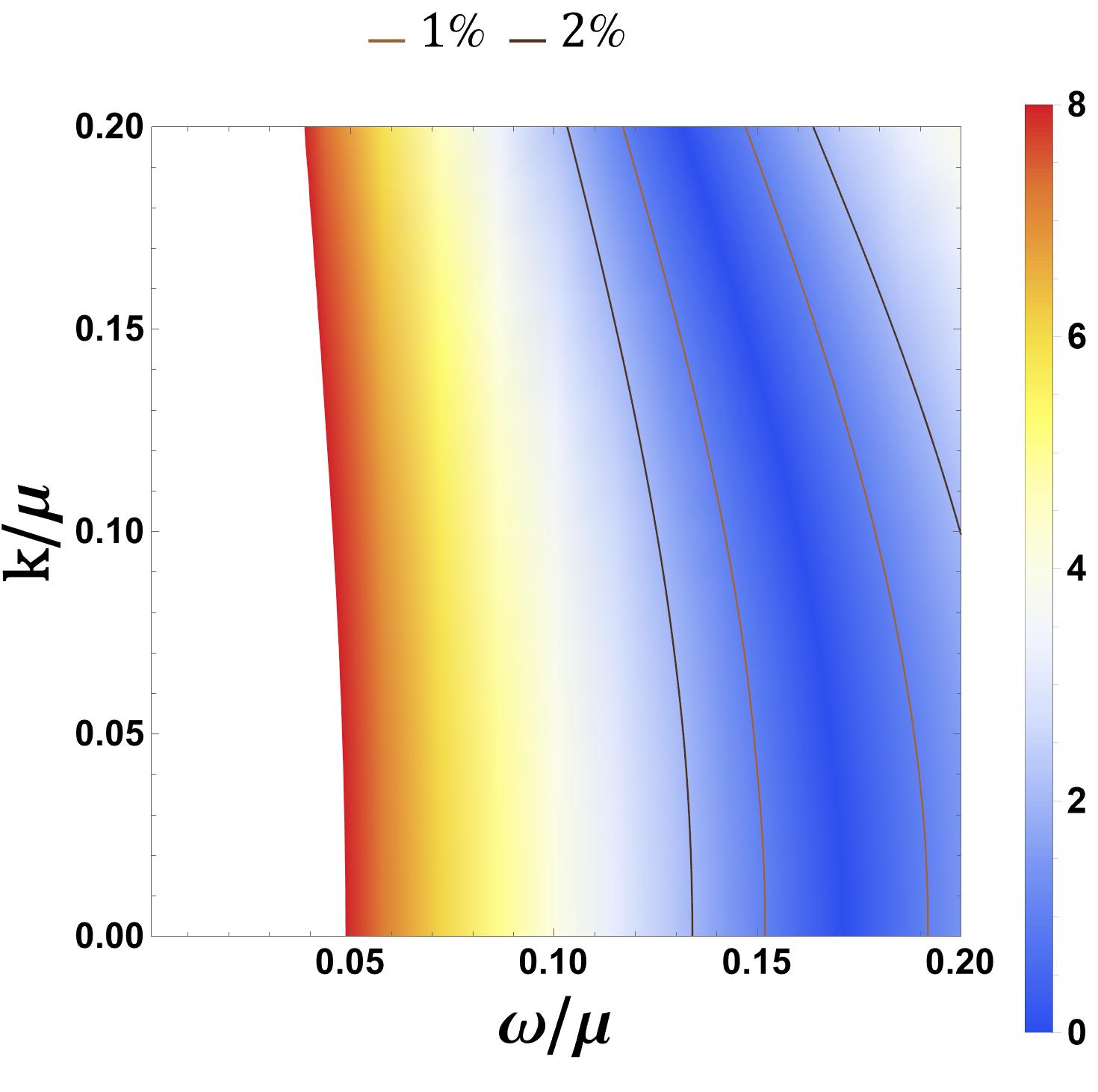}
		\end{subfigure}
	\end{subfigure}
	
	\caption{The percentage relative difference \eqref{eq:percrelfdiff} of the imaginary part of the transverse charged current polarization function with respect to the hydrodynamic approximation \eqref{eq:trhydroapprox} (top row) and the extended hydrodynamic approximation \eqref{eq:trexthydroapprox} (bottom row), for $\mu_q/T= 5$ and $\mu_3=-0.5$.  The right plots shows a subregion of the left
		plots. The green line shows the locus $\omega = k+\mu_3$. } \label{fig:RelDiffNum_tr_mu5mu305}
\end{figure}

In figure~\ref{fig:RelDiffNum_tr_mu5mu305}, the lower value of $\mu_q/T$ reduces the maximum errors compared to the near-extremal plots, but the qualitative effect of large $|\mu_3|$ remains. In the full-range (left) plots, the hydrodynamic approximation has a broad low-error basin at small $k/\mu$ and intermediate $\omega/\mu$, while the error grows strongly at large $k/\mu$.

The extended hydrodynamic approximation reshapes the plot rather than uniformly improving it: it lowers the error in some high-frequency regions, but creates a strong vertical high-error band at small $\omega/\mu$.

The zoomed plots make this more apparent. The hydrodynamic approximation has a curved low-error band passing through the middle of the zoom, whereas the extended approximation moves the low-error region toward larger $\omega/\mu$ and performs worse near the left side. Therefore, at large isospin chemical potential, the extended approximation is not systematically better in the transverse sector.

\subsubsection*{The longitudinal correlator}

\begin{figure}[H]
	\begin{subfigure}{\textwidth}
		\centering
		\begin{subfigure}{0.4\textwidth}
			\centering
			\includegraphics[width=\textwidth]{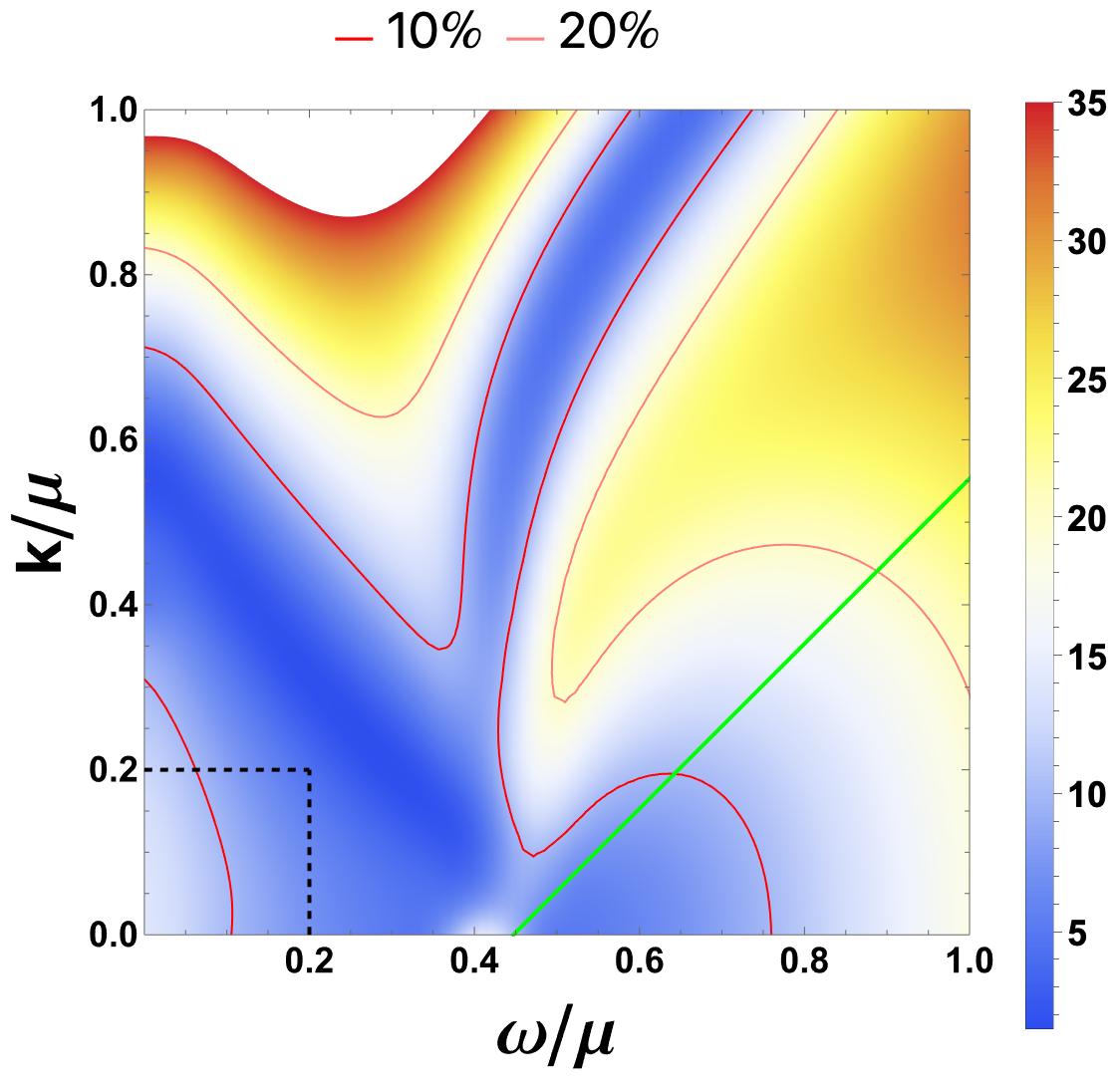}
		\end{subfigure}
		\hspace{1cm}
		\begin{subfigure}{0.4\textwidth}
			\centering
			\includegraphics[width=\textwidth]{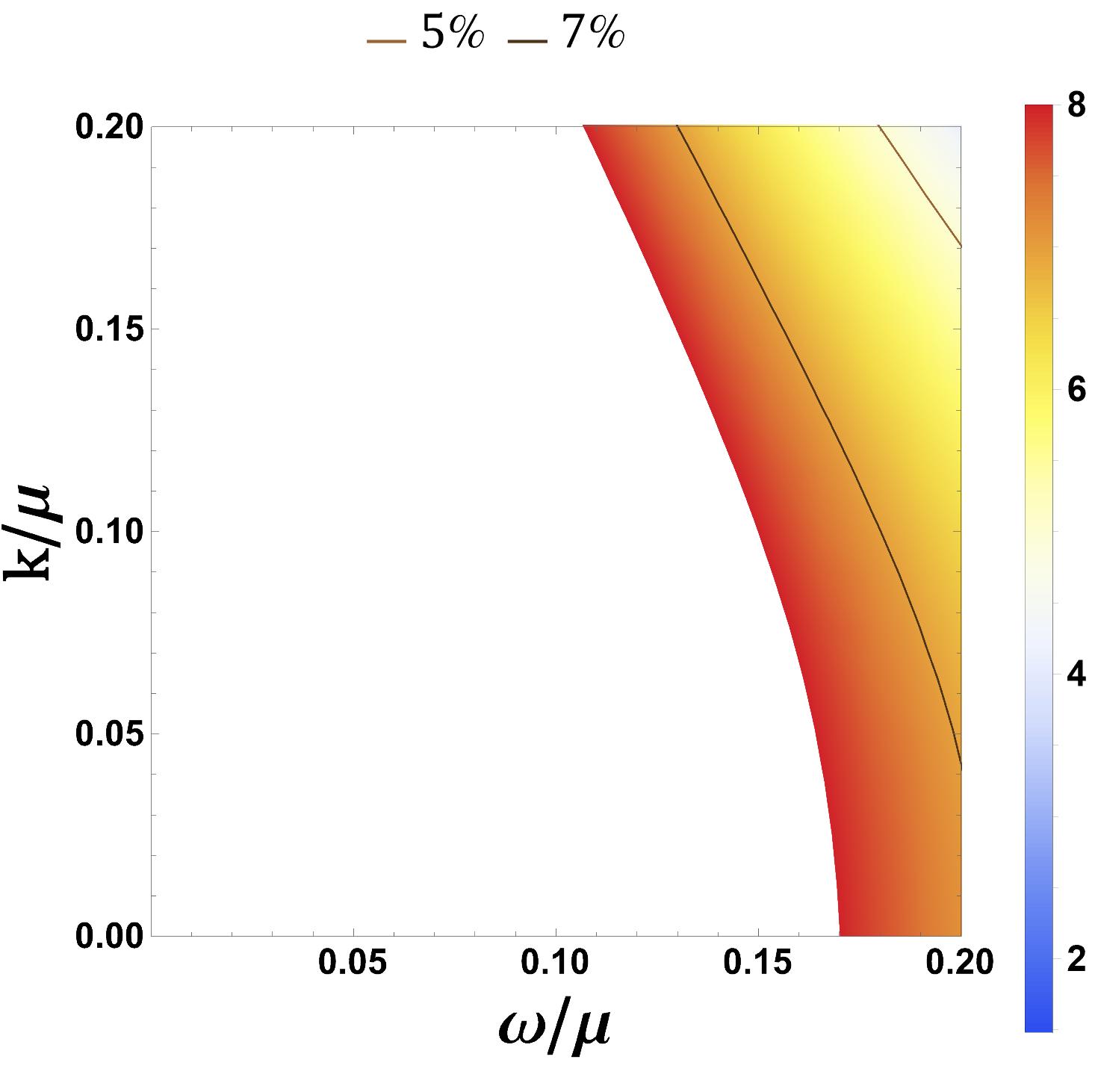}
		\end{subfigure}
	\end{subfigure}

	\begin{subfigure}{\textwidth}
		\centering
		\begin{subfigure}{0.4\textwidth}
			\centering
			\includegraphics[width=\textwidth]{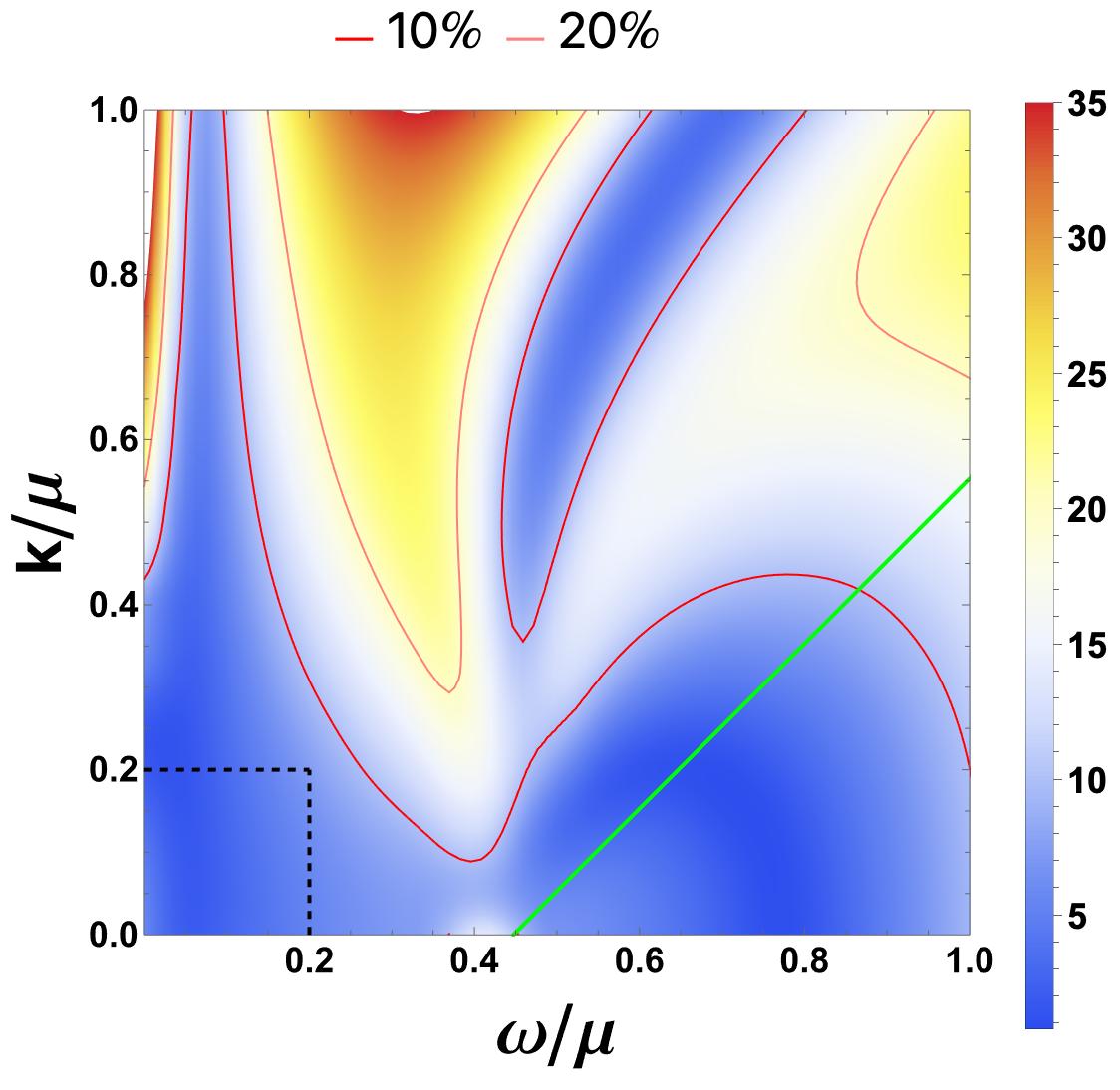}
		\end{subfigure}
		\hspace{1cm}
		\begin{subfigure}{0.4\textwidth}
			\centering
			\includegraphics[width=\textwidth]{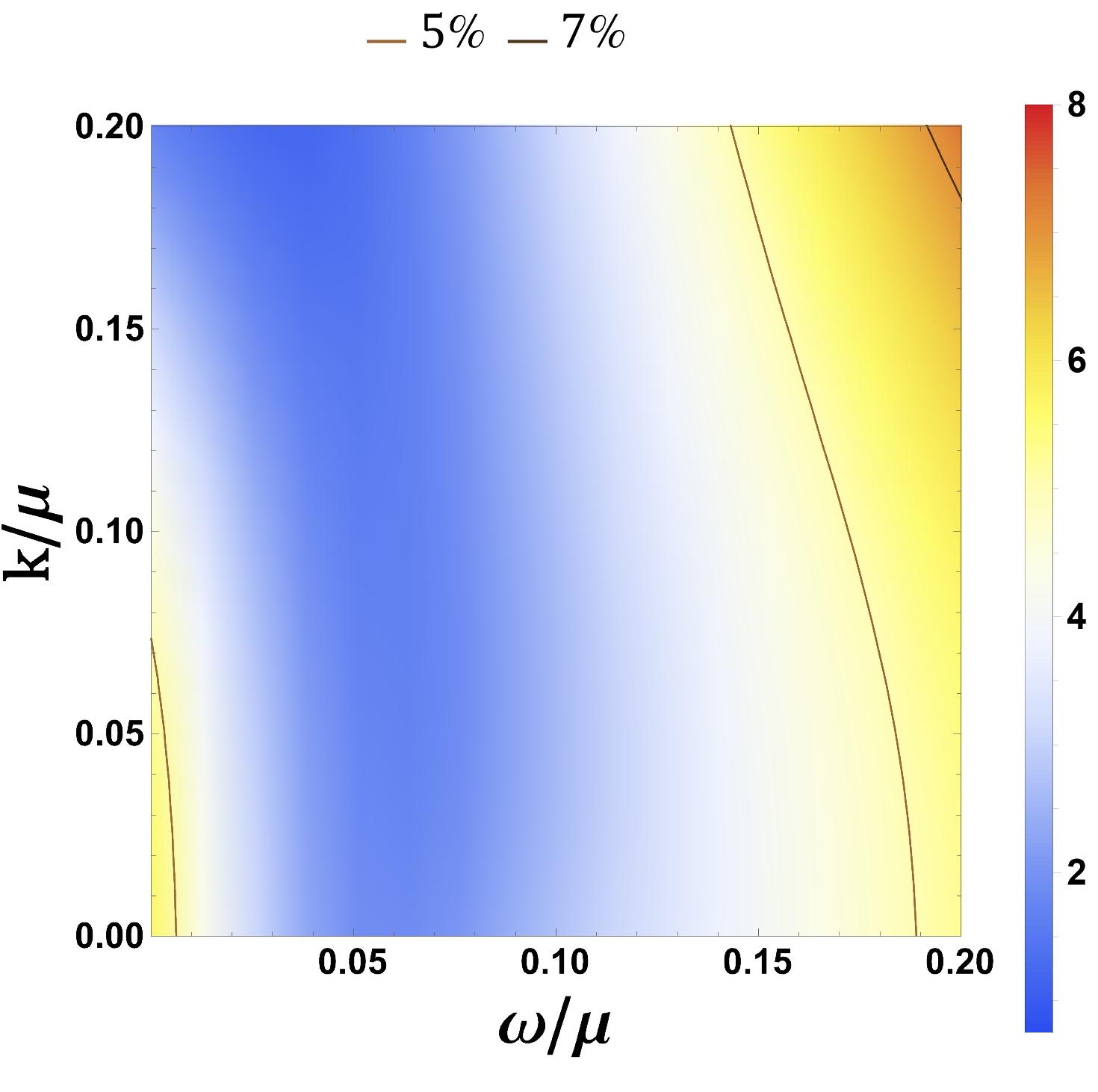}
		\end{subfigure}
	\end{subfigure}

	\caption{As figure \ref{fig:RelDiffNum_tr_mu5mu305} but for the longitudinal polarization function instead of the transverse polarization function.}\label{fig:RelDiffNum_long_mu5mu305}
\end{figure}

In figure~\ref{fig:RelDiffNum_long_mu5mu305}, the longitudinal sector has a broad region of moderate error and a more complicated dependence on $\omega/\mu$ than the transverse sector (see left column). The error increases for large $k/\mu$ and smaller frequencies, and it also grows again near the right side of the plot. The extended hydrodynamic approximation reduces the error on the small-frequency side, but it does not remove the increase at larger $k/\mu$.

In the zoomed hydrodynamic plot, the lowest displayed errors are not centered at $\omega=0$. Instead, for the hydrodynamic approximation (top-right plot) the errors of order $5\%$--$7\%$ appear for $\omega/\mu\gtrsim 0.12$, with a contour bending from $\omega/\mu\simeq 0.16$ at small $k/\mu$ to slightly larger $\omega/\mu\simeq 0.12$ at larger $k/\mu$. The extended hydrodynamic approximation substantially changes the zoomed region: for $\omega/\mu\simeq 0.05$--$0.12$ the error is reduced compared with the hydrodynamic plot, while for $\omega/\mu\gtrsim 0.17$ the error increases again to several percent. Hence, at $\mu_q/T=5$ and $\mu_3/\mu_q=-0.5$, the extended approximation improves the longitudinal correlator in a finite band of small-to-intermediate frequencies, but it does not produce a uniformly accurate hydrodynamic region around the origin.

\clearpage

\section{Coarse-grained integrated relative difference}\label{app:coarsegrained}

In this appendix, we present the results for the observable $\chi_{ij}$ defined in formula \eqref{eq:chiRL} for other values of the $\mu_q/T$ and $\mu_3/\mu_q$ in addition to the already studied $\mu_q/T = 65$ and $\mu_3/\mu_q\in\{0,-0.1\}$ in section \ref{sec:exactresults} and appendix \ref{app:otherapprox65}. In particular, we extend our discussion to $\mu_q/T\in\{10^4,5\}$ and to the extreme value $\mu_3/\mu_q=-0.5$. We recall that the observable $\chi_{ij}$ is defined inside rectangles on the $(\omega/\mu,k/\mu)$-plane that are defined by
\be
0\leq {\omega\over \mu}\leq n_i\sp 0\leq {k\over \mu}\leq n_j
\ee
where $n_{i}$ takes the four values ${i\over 8}$, $i=1,2,3,4$ and the same for $n_j$.
We reproduce here for convenience the definition of $\chi_{ij}$,
\be\label{eq:chiRLa}
\chi_{ij}^{(l)} =\dfrac{1}{n_i\cdot n_j}\,\int_0^{n_i}\intd(\omega/\mu) \int_0^{n_j}\intd (k/\mu)\,\bigg|1-
\dfrac{\text{Im}\Pi^{\perp,\parallel,\pm}_l}{\text{Im}\Pi^{\perp,\parallel,\pm}_\text{numerical}}\bigg|\,,
\ee
As explained in the main text, the value of $\chi_{ij}$ gives a measure of how much a given approximation of the correlator differs from its exact result, inside the rectangle over which the integration is performed.

We considered the five approximations introduced in appendix \ref{app:IRcorr-approx}. We remind the reader that they are
\begin{itemize}
	\item the (near-extremal) hydrodynamic approximation, defined in (\ref{app:eq:trhydroapprox}) and (\ref{app:eq:longhydroapprox}),
	\item the extended-hydrodynamic approximation, defined in (\ref{app:eq:trexthydroapprox}) and (\ref{app:eq:longexthydroapprox}),
	\item the improved extended hydrodynamic approximation, defined in (\ref{eq:trimprexthydroapprox}) and (\ref{eq:longimprexthydroapprox}),
	\item the approximation using the normalized IR-AdS$_2$ correlator, defined in (\ref{eq:trnormapprox}) and (\ref{eq:longnormapprox}),
	\item the approximation using the full IR-AdS$_2$ correlator, defined in (\ref{eq:trfullapprox}) and (\ref{eq:longfullapprox}).
\end{itemize}

We present below the tables associated with the relative errors. In all the tables below, the horizontal axis refers to $\omega/\mu$ and the vertical axis to $k/\mu$. The four values on each axis correspond to the values of $n_{i,j}$. The entries multiplied by 100 give the \% average error in the given rectangle.

In appendix \ref{app:finegrained}, we  use the same grid choice in order  to present the local versions of these errors.

\subsection{Background at $\mu_q/T=10^4$ and $\mu_3=0$}

\subsection*{The transverse correlator}
\begin{table}[htb]
	\centering
	{
		\small
		\begin{tabular}{c|cccc}
			\multicolumn{5}{c}{(a) hydrodynamic}\\
			$1/2$ & 0.46 & 0.35 & 0.29 & 0.28 \\
			$3/8$ & 0.21 & 0.17 & 0.17 & 0.20 \\
			$1/4$ & 0.08 & 0.09 & 0.14 & 0.19 \\
			$1/8$ & 0.04 & 0.09 & 0.14 & 0.20 \\
			\hline
			& $1/8$ & $1/4$ & $3/8$ & $1/2$ \\
		\end{tabular}\hspace{1cm}
		\begin{tabular}{c|cccc}
			\multicolumn{5}{c}{(b) extended-hydrodynamic}\\
			$1/2$ & 0.22 & 0.20 & 0.19 & 0.20 \\
			$3/8$ & 0.11 & 0.11 & 0.13 & 0.17 \\
			$1/4$ & 0.05 & 0.08 & 0.13 & 0.18 \\
			$1/8$ & 0.04 & 0.09 & 0.14 & 0.20 \\
			\hline
			& $1/8$ & $1/4$ & $3/8$ & $1/2$ \\
		\end{tabular}
		
		\begin{tabular}{c|cccc}
			\multicolumn{5}{c}{(c) improved}\\
			$1/2$ & 0.24 & 0.21 & 0.20 & 0.21 \\
			$3/8$ & 0.12 & 0.11 & 0.13 & 0.17 \\
			$1/4$ & 0.05 & 0.08 & 0.13 & 0.18 \\
			$1/8$ & 0.04 & 0.09 & 0.14 & 0.20 \\
			\hline
			& $1/8$ & $1/4$ & $3/8$ & $1/2$ \\
		\end{tabular}\hspace{1cm}
		\begin{tabular}{c|cccc}
			\multicolumn{5}{c}{(d) normalized}\\
			$1/2$ & 0.23 & 0.20 & 0.19 & 0.20 \\
			$3/8$ & 0.11 & 0.11 & 0.13 & 0.17 \\
			$1/4$ & 0.05 & 0.08 & 0.13 & 0.18 \\
			$1/8$ & 0.04 & 0.09 & 0.15 & 0.20 \\
			\hline
			& $1/8$ & $1/4$ & $3/8$ & $1/2$ \\
		\end{tabular}
		
		\begin{tabular}{c|cccc}
			\multicolumn{5}{c}{(e) full-IR}\\
			$1/2$ & 0.24 & 0.21 & 0.20 & 0.21 \\
			$3/8$ & 0.12 & 0.11 & 0.13 & 0.17 \\
			$1/4$ & 0.05 & 0.08 & 0.13 & 0.18 \\
			$1/8$ & 0.04 & 0.09 & 0.14 & 0.20 \\
			\hline
			& $1/8$ & $1/4$ & $3/8$ & $1/2$ \\
		\end{tabular}
	}
	\caption{Values $\chi_{ij}$ (formula \eqref{eq:chiRL}) in the transverse sector for $\mu_q/T=10^4$ and $\mu_3=0$.}
	\label{tab:app:chi1040_tr}
\end{table}

In table~\ref{tab:app:chi1040_tr}, we observe that the extended hydrodynamic approximation is better than the hydrodynamic one across the board. It becomes substantially better at smaller values of $\omega$ (i.e. the first column). On the other hand for the smallest values of $k$ where the AdS$_2$ dimension is near one, it does not a substantial improvement.

The improved extended hydrodynamic approximation is mostly better than the hydro approximation except for the lowest values of $k$. It is also better than the extended hydrodynamic approximation across the board except for the lowest values of $k$ (bottom row). On the other hand the normalized approximation is worse than all three previous ones.

We also observe that approximation using the full correlator is a mixed bag compared to the extended and the improved approximations. It is somewhat better at low $k$ and larger $\omega$ than the improved approximation. Similar comments apply to the comparison with the extended hydrodynamic approximation.

\begin{table}[htb]
	\centering
	{
		\small
		\begin{tabular}{c|cccc}
			\multicolumn{5}{c}{(a) hydrodynamic}\\
			$1/2$ & 0.11 & 0.11 & 0.14 & 0.17 \\
			$3/8$ & 0.06 & 0.08 & 0.12 & 0.16 \\
			$1/4$ & 0.04 & 0.07 & 0.11 & 0.15 \\
			$1/8$ & 0.03 & 0.07 & 0.11 & 0.15 \\
			\hline
			& $1/8$ & $1/4$ & $3/8$ & $1/2$ \\
		\end{tabular}\hspace{1cm}
		\begin{tabular}{c|cccc}
			\multicolumn{5}{c}{(b) extended-hydrodynamic}\\
			$1/2$ & 0.05 & 0.10 & 0.14 & 0.18 \\
			$3/8$ & 0.05 & 0.09 & 0.13 & 0.17 \\
			$1/4$ & 0.05 & 0.08 & 0.12 & 0.16 \\
			$1/8$ & 0.04 & 0.07 & 0.11 & 0.15 \\
			\hline
			& $1/8$ & $1/4$ & $3/8$ & $1/2$ \\
		\end{tabular}
		
		\begin{tabular}{c|cccc}
			\multicolumn{5}{c}{(c) improved}\\
			$1/2$ & 0.05 & 0.09 & 0.13 & 0.17 \\
			$3/8$ & 0.05 & 0.09 & 0.13 & 0.16 \\
			$1/4$ & 0.04 & 0.08 & 0.12 & 0.16 \\
			$1/8$ & 0.04 & 0.07 & 0.11 & 0.15 \\
			\hline
			& $1/8$ & $1/4$ & $3/8$ & $1/2$ \\
		\end{tabular}\hspace{1cm}
		\begin{tabular}{c|cccc}
			\multicolumn{5}{c}{(d) normalized}\\
			$1/2$ & 0.06 & 0.10 & 0.14 & 0.18 \\
			$3/8$ & 0.05 & 0.09 & 0.13 & 0.17 \\
			$1/4$ & 0.05 & 0.08 & 0.12 & 0.16 \\
			$1/8$ & 0.04 & 0.07 & 0.11 & 0.15 \\
			\hline
			& $1/8$ & $1/4$ & $3/8$ & $1/2$ \\
		\end{tabular}
		
		\begin{tabular}{c|cccc}
			\multicolumn{5}{c}{(e) full-IR}\\
			$1/2$ & 0.06 & 0.09 & 0.13 & 0.17 \\
			$3/8$ & 0.05 & 0.09 & 0.13 & 0.17 \\
			$1/4$ & 0.04 & 0.08 & 0.12 & 0.16 \\
			$1/8$ & 0.04 & 0.07 & 0.11 & 0.15 \\
			\hline
			& $1/8$ & $1/4$ & $3/8$ & $1/2$ \\
		\end{tabular}
	}
	\caption{Values $\chi_{ij}$ (formula \eqref{eq:chiRL}) in the longitudinal sector for $\mu_q/T=10^4$ and $\mu_3=0$.}
	\label{tab:app:chi1040_long}
\end{table}

\begin{table}[htb]
	\centering
	{
		\small
		\begin{tabular}{c|cccc}
			\multicolumn{5}{c}{(a) hydrodynamic}\\
			$1/2$ & 0.36 & 0.29 & 0.23 & 0.20 \\
			$3/8$ & 0.23 & 0.18 & 0.14 & 0.13 \\
			$1/4$ & 0.14 & 0.11 & 0.08 & 0.09 \\
			$1/8$ & 0.10 & 0.07 & 0.06 & 0.08 \\
			\hline
			& $1/8$ & $1/4$ & $3/8$ & $1/2$ \\
		\end{tabular}\hspace{1cm}
		\begin{tabular}{c|cccc}
			\multicolumn{5}{c}{(b) extended-hydrodynamic}\\
			$1/2$ & 0.23 & 0.20 & 0.16 & 0.15 \\
			$3/8$ & 0.16 & 0.14 & 0.11 & 0.10 \\
			$1/4$ & 0.12 & 0.09 & 0.07 & 0.08 \\
			$1/8$ & 0.10 & 0.07 & 0.06 & 0.07 \\
			\hline
			& $1/8$ & $1/4$ & $3/8$ & $1/2$ \\
		\end{tabular}
		
		\begin{tabular}{c|cccc}
			\multicolumn{5}{c}{(c) improved}\\
			$1/2$ & 0.14 & 0.12 & 0.11 & 0.12 \\
			$3/8$ & 0.07 & 0.07 & 0.07 & 0.09 \\
			$1/4$ & 0.03 & 0.03 & 0.05 & 0.08 \\
			$1/8$ & 0.01 & 0.03 & 0.06 & 0.09 \\
			\hline
			& $1/8$ & $1/4$ & $3/8$ & $1/2$ \\
		\end{tabular}\hspace{1cm}
		\begin{tabular}{c|cccc}
			\multicolumn{5}{c}{(d) normalized}\\
			$1/2$ & 0.24 & 0.20 & 0.17 & 0.15 \\
			$3/8$ & 0.17 & 0.14 & 0.11 & 0.10 \\
			$1/4$ & 0.13 & 0.10 & 0.08 & 0.08 \\
			$1/8$ & 0.10 & 0.07 & 0.06 & 0.08 \\
			\hline
			& $1/8$ & $1/4$ & $3/8$ & $1/2$ \\
		\end{tabular}
		
		\begin{tabular}{c|cccc}
			\multicolumn{5}{c}{(e) full-IR}\\
			$1/2$ & 0.15 & 0.13 & 0.12 & 0.12 \\
			$3/8$ & 0.08 & 0.07 & 0.07 & 0.09 \\
			$1/4$ & 0.04 & 0.04 & 0.05 & 0.09 \\
			$1/8$ & 0.02 & 0.03 & 0.06 & 0.09 \\
			\hline
			& $1/8$ & $1/4$ & $3/8$ & $1/2$ \\
		\end{tabular}
	}
	\caption{Values $\chi_{ij}$ (formula \eqref{eq:chiRL}) in the transverse sector for $\mu_q/T=10^4$ and $\mu_3/\mu_q=-0.1$.}
	\label{tab:app:chi10401_tr}
\end{table}

\subsection*{The longitudinal correlator}

In table~\ref{tab:app:chi1040_long}, the hydrodynamic approximation is already rather accurate, with entries between $0.03$ and $0.17$. The extended hydrodynamic approximation improves the first column in the upper rows, for example reducing the entry at $k/\mu=1/2$ from $0.11$ to $0.05$. However, this improvement is not uniform over the table: in the last column the extended approximation gives $0.18$ in the top row, while hydrodynamics gives $0.17$, and similar small worsenings appear in the second and third rows.

The improved and full approximations are slightly better than the extended and normalized ones in several entries. The differences are nevertheless very small, and they do not indicate a strong separation between the IR-based approximations. The bottom row confirms this point. At $k/\mu=1/8$, all approximations give essentially the same sequence of values.

Therefore, in this longitudinal sector, the table shows only a mild advantage of the IR-based approximations at small $\omega/\mu$ and large $k/\mu$. Away from that corner, the hydrodynamic approximation is already competitive with the refined approximations.

In order again to understand the extension of the standard hydrodynamic regime, for this value of $T/\mu_q$, we realize that in units of the chemical potential, $T=10^{-4}$. Therefore, in the bottom-left entry, the hydrodynamic approximation displays  an average error of 3\%, whit $k,\omega$ which are 1250 times the temperature in the same units. The extended hydrodynamic approximation in the same bottom-left entry is less accurate, at 4\%, but this is since the quadrant is much bigger than the traditional hydro regime.

\subsection{Background at $\mu_q/T=10^4$ and $\mu_3/\mu_q=-0.1$}

\subsection*{The transverse correlator}

In table~\ref{tab:app:chi10401_tr}, turning on a small isospin chemical potential changes the hierarchy between the approximations. The extended hydrodynamic approximation still improves over hydrodynamics across the table, especially in the first column at small $\omega$. Also, the gain is clear in the upper rows, while it becomes small at lower values of $k/\mu$.

The improved extended hydrodynamic approximation gives a much stronger reduction. Its first column shows values considerably smaller than both the hydrodynamic and the extended hydrodynamic values. The same improvement persists in the first three columns, where the improved approximation is usually the best one or tied with the best one. This shows that, for this background, the improved approximation reproduces better the exact correlator than the extended hydrodynamic approximation alone.

The normalized approximation follows the extended hydrodynamic one rather closely and does not provide an additional improvement. The full approximation is very close to the improved one, with the improved approximation is therefore slightly favored, but the full approximation is essentially equivalent for most entries. The only region where the advantage becomes less clear is the lower-right part of the table, where the hydrodynamic error is already small and all approximations are close.

\subsection*{The longitudinal correlator}
\begin{table}[h]
	\centering
	{
		\small
		\begin{tabular}{c|cccc}
			\multicolumn{5}{c}{(a) hydrodynamic}\\
			$1/2$ & 0.21 & 0.13 & 0.13 & 0.14 \\
			$3/8$ & 0.14 & 0.10 & 0.10 & 0.12 \\
			$1/4$ & 0.10 & 0.07 & 0.08 & 0.10 \\
			$1/8$ & 0.07 & 0.04 & 0.06 & 0.09 \\
			\hline
			& $1/8$ & $1/4$ & $3/8$ & $1/2$ \\
		\end{tabular}\hspace{1cm}
		\begin{tabular}{c|cccc}
			\multicolumn{5}{c}{(b) extended-hydrodynamic}\\
			$1/2$ & 0.11 & 0.09 & 0.11 & 0.13 \\
			$3/8$ & 0.09 & 0.08 & 0.09 & 0.12 \\
			$1/4$ & 0.08 & 0.07 & 0.08 & 0.10 \\
			$1/8$ & 0.07 & 0.04 & 0.06 & 0.09 \\
			\hline
			& $1/8$ & $1/4$ & $3/8$ & $1/2$ \\
		\end{tabular}
		
		\begin{tabular}{c|cccc}
			\multicolumn{5}{c}{(c) improved}\\
			$1/2$ & 0.06 & 0.10 & 0.13 & 0.17 \\
			$3/8$ & 0.05 & 0.09 & 0.13 & 0.16 \\
			$1/4$ & 0.04 & 0.08 & 0.11 & 0.14 \\
			$1/8$ & 0.03 & 0.06 & 0.10 & 0.13 \\
			\hline
			& $1/8$ & $1/4$ & $3/8$ & $1/2$ \\
		\end{tabular}\hspace{1cm}
		\begin{tabular}{c|cccc}
			\multicolumn{5}{c}{(d) normalized}\\
			$1/2$ & 0.11 & 0.09 & 0.11 & 0.13 \\
			$3/8$ & 0.10 & 0.08 & 0.10 & 0.12 \\
			$1/4$ & 0.09 & 0.07 & 0.08 & 0.10 \\
			$1/8$ & 0.07 & 0.05 & 0.06 & 0.09 \\
			\hline
			& $1/8$ & $1/4$ & $3/8$ & $1/2$ \\
		\end{tabular}
		
		\begin{tabular}{c|cccc}
			\multicolumn{5}{c}{(e) full-IR}\\
			$1/2$ & 0.06 & 0.10 & 0.13 & 0.17 \\
			$3/8$ & 0.05 & 0.10 & 0.13 & 0.16 \\
			$1/4$ & 0.04 & 0.08 & 0.11 & 0.14 \\
			$1/8$ & 0.03 & 0.06 & 0.09 & 0.13 \\
			\hline
			& $1/8$ & $1/4$ & $3/8$ & $1/2$ \\
		\end{tabular}
	}
	\caption{Values $\chi_{ij}$ (formula \eqref{eq:chiRL}) in the longitudinal sector for $\mu_q/T=10^4$ and $\mu_3/\mu_q=-0.1$.}
	\label{tab:app:chi10401_long}
\end{table}

In table~\ref{tab:app:chi10401_long}, the extended hydrodynamic approximation gives the most balanced improvement over hydrodynamics. The improvement is strongest in the first column as well as in the upper rows and essentially absent in the bottom row. In the remaining columns the extended hydrodynamic approximation stays close to hydrodynamics.

The improved and full approximations behave differently. They are very good at small $\omega/\mu$: their first-column entries are the smallest values in that column. However, their errors grow as $\omega/\mu$ increases. In the last column, the improved approximation gives larger values than the extended hydrodynamic approximation. Therefore, the improved and full approximations are better on the left side of the table, but worse on the right side.

The normalized approximation remains close to the extended hydrodynamic approximation, with slightly larger values in a few cells. Overall, table~\ref{tab:app:chi10401_long} does not select a single approximation in all regions: the improved and full approximations are preferred at the smallest frequencies, whereas the extended hydrodynamic approximation is the most reliable global choice once the whole grid is considered.

\subsection{Background at  $\mu_q/T=10^4$ and $\mu_3/\mu_q=-0.5$}

\subsection*{The transverse correlator}
\begin{table}[htb]
	\centering
	{
		\small
		\begin{tabular}{c|cccc}
			\multicolumn{5}{c}{(a) hydrodynamic}\\
			$1/2$ & 0.87 & 0.87 & 0.84 & 0.77 \\
			$3/8$ & 0.54 & 0.59 & 0.59 & 0.55 \\
			$1/4$ & 0.36 & 0.43 & 0.44 & 0.42 \\
			$1/8$ & 0.26 & 0.34 & 0.36 & 0.35 \\
			\hline
			& $1/8$ & $1/4$ & $3/8$ & $1/2$ \\
		\end{tabular}\hspace{1cm}
		\begin{tabular}{c|cccc}
			\multicolumn{5}{c}{(b) extended-hydrodynamic}\\
			$1/2$ & 0.76 & 0.80 & 0.78 & 0.73 \\
			$3/8$ & 0.58 & 0.62 & 0.61 & 0.57 \\
			$1/4$ & 0.47 & 0.51 & 0.50 & 0.46 \\
			$1/8$ & 0.41 & 0.45 & 0.44 & 0.41 \\
			\hline
			& $1/8$ & $1/4$ & $3/8$ & $1/2$ \\
		\end{tabular}
		
		\begin{tabular}{c|cccc}
			\multicolumn{5}{c}{(c) improved}\\
			$1/2$ & 0.19 & 0.18 & 0.18 & 0.18 \\
			$3/8$ & 0.14 & 0.12 & 0.12 & 0.14 \\
			$1/4$ & 0.16 & 0.14 & 0.15 & 0.17 \\
			$1/8$ & 0.20 & 0.18 & 0.19 & 0.21 \\
			\hline
			& $1/8$ & $1/4$ & $3/8$ & $1/2$ \\
		\end{tabular}\hspace{1cm}
		\begin{tabular}{c|cccc}
			\multicolumn{5}{c}{(d) normalized}\\
			$1/2$ & 0.82 & 0.83 & 0.81 & 0.75 \\
			$3/8$ & 0.64 & 0.65 & 0.63 & 0.58 \\
			$1/4$ & 0.52 & 0.54 & 0.52 & 0.48 \\
			$1/8$ & 0.46 & 0.48 & 0.46 & 0.42 \\
			\hline
			& $1/8$ & $1/4$ & $3/8$ & $1/2$ \\
		\end{tabular}
		
		\begin{tabular}{c|cccc}
			\multicolumn{5}{c}{(e) full-IR}\\
			$1/2$ & 0.19 & 0.19 & 0.18 & 0.18 \\
			$3/8$ & 0.12 & 0.12 & 0.12 & 0.13 \\
			$1/4$ & 0.14 & 0.13 & 0.14 & 0.16 \\
			$1/8$ & 0.18 & 0.17 & 0.18 & 0.20 \\
			\hline
			& $1/8$ & $1/4$ & $3/8$ & $1/2$ \\
		\end{tabular}
	}
	\caption{Values $\chi_{ij}$ (formula \eqref{eq:chiRL}) in the transverse sector for $\mu_q/T=10^4$ and $\mu_3/\mu_q=-0.5$.}
	\label{tab:app:chi10405_tr}
\end{table}

In table~\ref{tab:app:chi10405_tr}, the increase of $|\mu_3|/\mu_q$ to $0.5$ produces a very clear separation between the approximations. The hydrodynamic approximation has large errors throughout the grid, ranging from $26\%$ to $87\%$. The extended hydrodynamic approximation does not solve the problem. It slightly improves the top row, but it becomes worse in the lower rows.

The normalized approximation is also poor. It is comparable to, or worse than, the hydrodynamic and extended hydrodynamic approximations.
By contrast, the improved and full approximations reduce the error by a large factor.

The full approximation is slightly better overall. This is most visible in the lower rows. The difference is small, but systematic in many entries. Therefore, for this transverse table, the hydrodynamic, extended hydrodynamic, and normalized approximations perform worse at the quantitative level than the improved and full approximations.

\subsection*{The longitudinal correlator}

In table~\ref{tab:app:chi10405_long}, the same qualitative hierarchy displayed in the transverse case appears in the longitudinal sector. The hydrodynamic approximation has large errors, especially at small $\omega/\mu$ and large $k/\mu$. The extended hydrodynamic approximation improves only the first row mildly, but it becomes worse in the lower rows.

The normalized approximation is again not so useful in this regime, since it is often worse than both hydrodynamics and extended hydrodynamics. The improved and full approximations, instead, reduce the error substantially. Their first rows shows much smaller values than the corresponding hydrodynamic and normalized approximation values.

A characteristic feature of the improved and full IR-AdS$_2$ approximations is that their errors are smallest in the first two columns and increase as $\omega/\mu$ grows. The full approximation is marginally better than the improved one in almost every entry, but the two are very close. Therefore, at large isospin chemical potential, the longitudinal table also favors the full and improved IR-AdS$_2$ approximations over the other ones.

\begin{table}[htb]
	\centering
	{
		\small
		\begin{tabular}{c|cccc}
			\multicolumn{5}{c}{(a) hydrodynamic}\\
			$1/2$ & 0.67 & 0.63 & 0.51 & 0.41 \\
			$3/8$ & 0.47 & 0.48 & 0.40 & 0.33 \\
			$1/4$ & 0.33 & 0.37 & 0.32 & 0.26 \\
			$1/8$ & 0.26 & 0.29 & 0.27 & 0.22 \\
			\hline
			& $1/8$ & $1/4$ & $3/8$ & $1/2$ \\
		\end{tabular}\hspace{1cm}
		\begin{tabular}{c|cccc}
			\multicolumn{5}{c}{(b) extended-hydrodynamic}\\
			$1/2$ & 0.59 & 0.58 & 0.47 & 0.39 \\
			$3/8$ & 0.51 & 0.51 & 0.42 & 0.35 \\
			$1/4$ & 0.45 & 0.45 & 0.38 & 0.31 \\
			$1/8$ & 0.40 & 0.39 & 0.35 & 0.28 \\
			\hline
			& $1/8$ & $1/4$ & $3/8$ & $1/2$ \\
		\end{tabular}
		
		\begin{tabular}{c|cccc}
			\multicolumn{5}{c}{(c) improved}\\
			$1/2$ & 0.13 & 0.12 & 0.18 & 0.26 \\
			$3/8$ & 0.14 & 0.14 & 0.20 & 0.27 \\
			$1/4$ & 0.18 & 0.18 & 0.23 & 0.28 \\
			$1/8$ & 0.21 & 0.21 & 0.24 & 0.28 \\
			\hline
			& $1/8$ & $1/4$ & $3/8$ & $1/2$ \\
		\end{tabular}\hspace{1cm}
		\begin{tabular}{c|cccc}
			\multicolumn{5}{c}{(d) normalized}\\
			$1/2$ & 0.64 & 0.61 & 0.49 & 0.40 \\
			$3/8$ & 0.56 & 0.54 & 0.44 & 0.36 \\
			$1/4$ & 0.49 & 0.47 & 0.40 & 0.32 \\
			$1/8$ & 0.45 & 0.42 & 0.36 & 0.30 \\
			\hline
			& $1/8$ & $1/4$ & $3/8$ & $1/2$ \\
		\end{tabular}
		
		\begin{tabular}{c|cccc}
			\multicolumn{5}{c}{(e) full-IR}\\
			$1/2$ & 0.12 & 0.12 & 0.18 & 0.25 \\
			$3/8$ & 0.12 & 0.13 & 0.19 & 0.26 \\
			$1/4$ & 0.16 & 0.17 & 0.22 & 0.27 \\
			$1/8$ & 0.18 & 0.20 & 0.23 & 0.27 \\
			\hline
			& $1/8$ & $1/4$ & $3/8$ & $1/2$ \\
		\end{tabular}
	}
	\caption{Values $\chi_{ij}$ (formula \eqref{eq:chiRL}) in the longitudinal sector for $\mu_q/T=10^4$ and $\mu_3/\mu_q=-0.5$.}
	\label{tab:app:chi10405_long}
\end{table}

\subsection{Background at  $\mu_q/T=65$ and $\mu_3=-0.1$}

\subsection*{The transverse correlator}
\begin{table}[!htb]
	\centering
	{
		\small
		\begin{tabular}{c|cccc}
			\multicolumn{5}{c}{(a) hydrodynamic}\\
			$1/2$ & 0.46 & 0.38 & 0.32 & 0.28 \\
			$3/8$ & 0.25 & 0.20 & 0.17 & 0.16 \\
			$1/4$ & 0.12 & 0.09 & 0.09 & 0.11 \\
			$1/8$ & 0.05 & 0.04 & 0.06 & 0.10 \\
			\hline
			& $1/8$ & $1/4$ & $3/8$ & $1/2$ \\
		\end{tabular}\hspace{1cm}
		\begin{tabular}{c|cccc}
			\multicolumn{5}{c}{(b) extended-hydrodynamic}\\
			$1/2$ & 0.25 & 0.24 & 0.21 & 0.20 \\
			$3/8$ & 0.15 & 0.14 & 0.12 & 0.13 \\
			$1/4$ & 0.09 & 0.07 & 0.07 & 0.10 \\
			$1/8$ & 0.05 & 0.04 & 0.06 & 0.10 \\
			\hline
			& $1/8$ & $1/4$ & $3/8$ & $1/2$ \\
		\end{tabular}
		
		\begin{tabular}{c|cccc}
			\multicolumn{5}{c}{(c) improved}\\
			$1/2$ & 0.17 & 0.17 & 0.17 & 0.17 \\
			$3/8$ & 0.09 & 0.09 & 0.10 & 0.13 \\
			$1/4$ & 0.05 & 0.06 & 0.09 & 0.14 \\
			$1/8$ & 0.06 & 0.08 & 0.12 & 0.16 \\
			\hline
			& $1/8$ & $1/4$ & $3/8$ & $1/2$ \\
		\end{tabular}\hspace{1cm}
		\begin{tabular}{c|cccc}
			\multicolumn{5}{c}{(d) normalized}\\
			$1/2$ & 0.37 & 0.31 & 0.26 & 0.23 \\
			$3/8$ & 0.24 & 0.19 & 0.16 & 0.15 \\
			$1/4$ & 0.16 & 0.11 & 0.10 & 0.12 \\
			$1/8$ & 0.11 & 0.07 & 0.08 & 0.11 \\
			\hline
			& $1/8$ & $1/4$ & $3/8$ & $1/2$ \\
		\end{tabular}
		
		\begin{tabular}{c|cccc}
			\multicolumn{5}{c}{(e) full-IR}\\
			$1/2$ & 0.25 & 0.22 & 0.20 & 0.19 \\
			$3/8$ & 0.13 & 0.11 & 0.11 & 0.14 \\
			$1/4$ & 0.05 & 0.05 & 0.09 & 0.13 \\
			$1/8$ & 0.05 & 0.05 & 0.09 & 0.14 \\
			\hline
			& $1/8$ & $1/4$ & $3/8$ & $1/2$ \\
		\end{tabular}
	}
	\caption{Values $\chi_{ij}$ (formula \eqref{eq:chiRL}) in the transverse sector for $\mu_q/T=65$ and $\mu_3=-0.1$.}
	\label{tab:app:chi6501_tr}
\end{table}

In table~\ref{tab:app:chi6501_tr}, the extended hydrodynamic approximation is better than the hydrodynamic one across the board. The improvement is particularly visible at small $\omega/\mu$, namely in the first column. On the other hand, at the smallest value of $k/\mu$, the improvement is not substantial.

The improved extended hydrodynamic approximation is mostly better than hydrodynamics, except for the lowest values of $k/\mu$. It also improves over the extended hydrodynamic approximation in most of the upper part of the grid. Therefore, the improved approximation is clearly advantageous in the first two rows, but in the bottom row it becomes worse than the extended hydrodynamic approximation, especially at larger $\omega/\mu$. The normalized approximation does not give a useful refinement in this table. It is worse than the extended and improved approximations in almost all entries, and in the lower rows it can even be worse than hydrodynamics.

The full approximation is a mixed case when compared with the extended and improved approximations. It is close to the extended hydrodynamic result in the top row, and it is better than the improved approximation at lower $k/\mu$. However, it is not uniformly better than the improved approximation in the upper rows.

\subsection*{The longitudinal correlator}
\begin{table}[!htb]
	\centering
	{
		\small
		\begin{tabular}{c|cccc}
			\multicolumn{5}{c}{(a) hydrodynamic}\\
			$1/2$ & 0.17 & 0.11 & 0.11 & 0.13 \\
			$3/8$ & 0.11 & 0.08 & 0.08 & 0.11 \\
			$1/4$ & 0.07 & 0.05 & 0.06 & 0.09 \\
			$1/8$ & 0.04 & 0.03 & 0.05 & 0.07 \\
			\hline
			& $1/8$ & $1/4$ & $3/8$ & $1/2$ \\
		\end{tabular}\hspace{1cm}
		\begin{tabular}{c|cccc}
			\multicolumn{5}{c}{(b) extended-hydrodynamic}\\
			$1/2$ & 0.08 & 0.07 & 0.09 & 0.12 \\
			$3/8$ & 0.06 & 0.06 & 0.08 & 0.10 \\
			$1/4$ & 0.05 & 0.05 & 0.06 & 0.09 \\
			$1/8$ & 0.04 & 0.03 & 0.05 & 0.07 \\
			\hline
			& $1/8$ & $1/4$ & $3/8$ & $1/2$ \\
		\end{tabular}
		
		\begin{tabular}{c|cccc}
			\multicolumn{5}{c}{(c) improved}\\
			$1/2$ & 0.05 & 0.09 & 0.12 & 0.16 \\
			$3/8$ & 0.05 & 0.09 & 0.12 & 0.15 \\
			$1/4$ & 0.05 & 0.08 & 0.11 & 0.14 \\
			$1/8$ & 0.05 & 0.07 & 0.10 & 0.13 \\
			\hline
			& $1/8$ & $1/4$ & $3/8$ & $1/2$ \\
		\end{tabular}\hspace{1cm}
		\begin{tabular}{c|cccc}
			\multicolumn{5}{c}{(d) normalized}\\
			$1/2$ & 0.13 & 0.10 & 0.10 & 0.12 \\
			$3/8$ & 0.11 & 0.08 & 0.09 & 0.11 \\
			$1/4$ & 0.09 & 0.06 & 0.07 & 0.09 \\
			$1/8$ & 0.08 & 0.05 & 0.05 & 0.08 \\
			\hline
			& $1/8$ & $1/4$ & $3/8$ & $1/2$ \\
		\end{tabular}
		
		\begin{tabular}{c|cccc}
			\multicolumn{5}{c}{(e) full-IR}\\
			$1/2$ & 0.07 & 0.10 & 0.12 & 0.15 \\
			$3/8$ & 0.05 & 0.09 & 0.11 & 0.14 \\
			$1/4$ & 0.04 & 0.07 & 0.10 & 0.13 \\
			$1/8$ & 0.04 & 0.05 & 0.08 & 0.12 \\
			\hline
			& $1/8$ & $1/4$ & $3/8$ & $1/2$ \\
		\end{tabular}
	}
	\caption{Values $\chi_{ij}$ (formula \eqref{eq:chiRL}) in the longitudinal sector for $\mu_q/T=65$ and $\mu_3=-0.1$.}
	\label{tab:app:chi6501_long}
\end{table}

In table~\ref{tab:app:chi6501_long}, the extended hydrodynamic approximation improves over hydrodynamics most clearly in the first column and in the upper rows.

The improved approximation has a different pattern. It is good at the smallest frequency, but it becomes worse as $\omega/\mu$ increases. Therefore, the improved approximation is not globally better in this longitudinal sector, even though it performs well in the first column.

The normalized approximation stays rather close to hydrodynamics and does not give a substantial improvement. The full approximation is intermediate: it improves over the improved approximation in some entries at larger $\omega/\mu$, but it is still generally worse than the extended hydrodynamic approximation.
\subsection{Background at  $\mu_q/T=65$ and $\mu_3/\mu_q=-0.5$}

\begin{table}[htb]
	\centering
	{
		\small
		\begin{tabular}{c|cccc}
			\multicolumn{5}{c}{(a) hydrodynamic}\\
			$1/2$ & 0.80 & 0.82 & 0.79 & 0.74 \\
			$3/8$ & 0.51 & 0.56 & 0.56 & 0.54 \\
			$1/4$ & 0.34 & 0.41 & 0.43 & 0.41 \\
			$1/8$ & 0.25 & 0.33 & 0.36 & 0.35 \\
			\hline
			& $1/8$ & $1/4$ & $3/8$ & $1/2$ \\
		\end{tabular}\hspace{1cm}
		\begin{tabular}{c|cccc}
			\multicolumn{5}{c}{(b) extended-hydrodynamic}\\
			$1/2$ & 0.70 & 0.75 & 0.74 & 0.70 \\
			$3/8$ & 0.54 & 0.59 & 0.59 & 0.55 \\
			$1/4$ & 0.45 & 0.49 & 0.49 & 0.46 \\
			$1/8$ & 0.39 & 0.44 & 0.43 & 0.41 \\
			\hline
			& $1/8$ & $1/4$ & $3/8$ & $1/2$ \\
		\end{tabular}
		
		\begin{tabular}{c|cccc}
			\multicolumn{5}{c}{(c) improved}\\
			$1/2$ & 0.17 & 0.17 & 0.17 & 0.16 \\
			$3/8$ & 0.13 & 0.11 & 0.11 & 0.12 \\
			$1/4$ & 0.16 & 0.13 & 0.13 & 0.15 \\
			$1/8$ & 0.19 & 0.17 & 0.17 & 0.18 \\
			\hline
			& $1/8$ & $1/4$ & $3/8$ & $1/2$ \\
		\end{tabular}\hspace{1cm}
		\begin{tabular}{c|cccc}
			\multicolumn{5}{c}{(d) normalized}\\
			$1/2$ & 1.10 & 1.00 & 0.96 & 0.88 \\
			$3/8$ & 0.91 & 0.83 & 0.77 & 0.70 \\
			$1/4$ & 0.76 & 0.70 & 0.65 & 0.59 \\
			$1/8$ & 0.68 & 0.63 & 0.59 & 0.53 \\
			\hline
			& $1/8$ & $1/4$ & $3/8$ & $1/2$ \\
		\end{tabular}
		
		\begin{tabular}{c|cccc}
			\multicolumn{5}{c}{(e) full-IR}\\
			$1/2$ & 0.28 & 0.24 & 0.21 & 0.20 \\
			$3/8$ & 0.14 & 0.12 & 0.11 & 0.12 \\
			$1/4$ & 0.06 & 0.06 & 0.07 & 0.10 \\
			$1/8$ & 0.03 & 0.06 & 0.08 & 0.11 \\
			\hline
			& $1/8$ & $1/4$ & $3/8$ & $1/2$ \\
		\end{tabular}
	}
	\caption{Values $\chi_{ij}$ (formula \eqref{eq:chiRL}) in the transverse sector for $\mu_q/T=65$ and $\mu_3/\mu_q=-0.5$.}
	\label{tab:app:chi6505_tr}
\end{table}

\subsection*{The transverse correlator}

In table~\ref{tab:app:chi6505_tr}, the hydrodynamic and extended hydrodynamic approximations both have large errors, due to the large absolute value of $\mu_3$. The extended approximation improves with respect to hydrodynamics the top row slightly, but it becomes worse in the lower rows.

The normalized approximation is clearly the worst one in this table. This shows that the normalized IR-AdS$_2$ correlator does not capture the relevant behavior for this transverse sector at $\mu_3/\mu_q=-0.5$.

The improved approximation gives a major improvement. It is especially good in the upper rows, where it is nearly uniform. The full approximation is even better overall, although not in every single cell. In the top row it is worse than the improved approximation, but in the lower rows it becomes much better. Therefore, the improved approximation is preferable at the largest $k/\mu$, while the full approximation gives the best global description, especially at smaller $k/\mu$.

\subsection*{The longitudinal correlator}
\begin{table}[htb]
	\centering
	{
		\small
		\begin{tabular}{c|cccc}
			\multicolumn{5}{c}{(a) hydrodynamic}\\
			$1/2$ & 0.27 & 0.35 & 0.38 & 0.33 \\
			$3/8$ & 0.19 & 0.27 & 0.31 & 0.28 \\
			$1/4$ & 0.14 & 0.22 & 0.25 & 0.24 \\
			$1/8$ & 0.12 & 0.19 & 0.22 & 0.21 \\
			\hline
			& $1/8$ & $1/4$ & $3/8$ & $1/2$ \\
		\end{tabular}\hspace{1cm}
		\begin{tabular}{c|cccc}
			\multicolumn{5}{c}{(b) extended-hydrodynamic}\\
			$1/2$ & 0.23 & 0.32 & 0.35 & 0.31 \\
			$3/8$ & 0.19 & 0.28 & 0.31 & 0.29 \\
			$1/4$ & 0.17 & 0.25 & 0.28 & 0.26 \\
			$1/8$ & 0.16 & 0.23 & 0.25 & 0.24 \\
			\hline
			& $1/8$ & $1/4$ & $3/8$ & $1/2$ \\
		\end{tabular}
		
		\begin{tabular}{c|cccc}
			\multicolumn{5}{c}{(c) improved}\\
			$1/2$ & 0.17 & 0.12 & 0.11 & 0.14 \\
			$3/8$ & 0.19 & 0.14 & 0.12 & 0.15 \\
			$1/4$ & 0.21 & 0.16 & 0.14 & 0.16 \\
			$1/8$ & 0.22 & 0.17 & 0.16 & 0.17 \\
			\hline
			& $1/8$ & $1/4$ & $3/8$ & $1/2$ \\
		\end{tabular}\hspace{1cm}
		\begin{tabular}{c|cccc}
			\multicolumn{5}{c}{(d) normalized}\\
			$1/2$ & 0.45 & 0.47 & 0.47 & 0.41 \\
			$3/8$ & 0.39 & 0.42 & 0.42 & 0.37 \\
			$1/4$ & 0.36 & 0.38 & 0.38 & 0.34 \\
			$1/8$ & 0.34 & 0.35 & 0.35 & 0.32 \\
			\hline
			& $1/8$ & $1/4$ & $3/8$ & $1/2$ \\
		\end{tabular}
		
		\begin{tabular}{c|cccc}
			\multicolumn{5}{c}{(e) full-IR}\\
			$1/2$ & 0.07 & 0.07 & 0.07 & 0.11 \\
			$3/8$ & 0.06 & 0.06 & 0.06 & 0.10 \\
			$1/4$ & 0.08 & 0.07 & 0.07 & 0.10 \\
			$1/8$ & 0.10 & 0.09 & 0.09 & 0.11 \\
			\hline
			& $1/8$ & $1/4$ & $3/8$ & $1/2$ \\
		\end{tabular}
	}
	\caption{Values $\chi_{ij}$ (formula \eqref{eq:chiRL}) in the longitudinal sector for $\mu_q/T=65$ and $\mu_3/\mu_q=-0.5$.}
	\label{tab:app:chi6505_long}
\end{table}

In table~\ref{tab:app:chi6505_long}, the hydrodynamic and extended hydrodynamic approximations are comparable, but neither is particularly accurate. The extended approximation improves the first row slightly, but it becomes worse in the lower rows.

The normalized approximation is again the worst one, being larger than the hydrodynamic and extended hydrodynamic approximations throughout the table. The improved approximation gives a visible reduction, lowering the errors to the range $0.11$--$0.22$.

The full approximation is clearly the best across the whole grid, since it improves over the improved approximation in every row. The advantage is especially pronounced in the upper rows. This table therefore gives a clean hierarchy: full is best, improved is next, hydrodynamic and extended hydrodynamic are worse and comparable, while normalized is the least accurate.

\subsection{Background at  $\mu_q/T=5$ and $\mu_3=0$}

\subsection*{The transverse correlator}
\begin{table}[htb]
	\centering
	{
		\small
		\begin{tabular}{c|cccc}
			\multicolumn{5}{c}{(a) hydrodynamic}\\
			$1/2$ & 0.07 & 0.07 & 0.06 & 0.06 \\
			$3/8$ & 0.04 & 0.04 & 0.04 & 0.05 \\
			$1/4$ & 0.02 & 0.02 & 0.03 & 0.04 \\
			$1/8$ & 0.004 & 0.01 & 0.03 & 0.05 \\
			\hline
			& $1/8$ & $1/4$ & $3/8$ & $1/2$ \\
		\end{tabular}\hspace{1cm}
		\begin{tabular}{c|cccc}
			\multicolumn{5}{c}{(b) extended-hydrodynamic}\\
			$1/2$ & 0.02 & 0.03 & 0.03 & 0.04 \\
			$3/8$ & 0.01 & 0.01 & 0.02 & 0.04 \\
			$1/4$ & 0.005 & 0.01 & 0.02 & 0.04 \\
			$1/8$ & 0.003 & 0.01 & 0.03 & 0.05 \\
			\hline
			& $1/8$ & $1/4$ & $3/8$ & $1/2$ \\
		\end{tabular}
		
		\begin{tabular}{c|cccc}
			\multicolumn{5}{c}{(c) improved}\\
			$1/2$ & 0.02 & 0.03 & 0.04 & 0.04 \\
			$3/8$ & 0.01 & 0.02 & 0.02 & 0.04 \\
			$1/4$ & 0.01 & 0.01 & 0.02 & 0.04 \\
			$1/8$ & 0.003 & 0.01 & 0.03 & 0.05 \\
			\hline
			& $1/8$ & $1/4$ & $3/8$ & $1/2$ \\
		\end{tabular}\hspace{1cm}
		\begin{tabular}{c|cccc}
			\multicolumn{5}{c}{(d) normalized}\\
			$1/2$ & 0.07 & 0.06 & 0.06 & 0.06 \\
			$3/8$ & 0.04 & 0.03 & 0.03 & 0.05 \\
			$1/4$ & 0.02 & 0.02 & 0.03 & 0.04 \\
			$1/8$ & 0.004 & 0.01 & 0.03 & 0.05 \\
			\hline
			& $1/8$ & $1/4$ & $3/8$ & $1/2$ \\
		\end{tabular}
		
		\begin{tabular}{c|cccc}
			\multicolumn{5}{c}{(e) full-IR}\\
			$1/2$ & 0.07 & 0.07 & 0.06 & 0.07 \\
			$3/8$ & 0.04 & 0.04 & 0.04 & 0.05 \\
			$1/4$ & 0.02 & 0.02 & 0.03 & 0.04 \\
			$1/8$ & 0.004 & 0.01 & 0.03 & 0.05 \\
			\hline
			& $1/8$ & $1/4$ & $3/8$ & $1/2$ \\
		\end{tabular}
	}
	\caption{Values $\chi_{ij}$ (formula \eqref{eq:chiRL}) in the transverse sector for $\mu_q/T=5$ and $\mu_3=0$.}
	\label{tab:app:chi50_tr}
\end{table}

In table~\ref{tab:app:chi50_tr}, all errors are small in absolute value with the hydrodynamic approximation already at the level of a few percent. Nevertheless, the extended hydrodynamic approximation gives a clear improvement in the upper rows.

The improvement becomes less relevant in the bottom row. Therefore, also in this hotter background, the extended hydrodynamic correction mainly affects larger values of $k/\mu$.

The improved approximation is very close to the extended hydrodynamic one, but it is slightly worse in a few entries, for example in the third column of the top row. The normalized and full approximations instead remain much closer to hydrodynamics. In particular, the full approximation essentially reproduces the hydrodynamic values and even gives $0.07$ in the last entry of the top row. Therefore, although all approximations are reasonably accurate, the extended hydrodynamic approximation is the most efficient one in this transverse sector.

\subsection*{The longitudinal correlator}
\begin{table}[htb]
	\centering
	{
		\small
		\begin{tabular}{c|cccc}
			\multicolumn{5}{c}{(a) hydrodynamic}\\
			$1/2$ & 0.02 & 0.02 & 0.03 & 0.04 \\
			$3/8$ & 0.01 & 0.01 & 0.03 & 0.04 \\
			$1/4$ & 0.004 & 0.01 & 0.02 & 0.04 \\
			$1/8$ & 0.003 & 0.01 & 0.02 & 0.04 \\
			\hline
			& $1/8$ & $1/4$ & $3/8$ & $1/2$ \\
		\end{tabular}\hspace{1cm}
		\begin{tabular}{c|cccc}
			\multicolumn{5}{c}{(b) extended-hydrodynamic}\\
			$1/2$ & 0.03 & 0.03 & 0.04 & 0.06 \\
			$3/8$ & 0.02 & 0.02 & 0.04 & 0.05 \\
			$1/4$ & 0.01 & 0.02 & 0.03 & 0.05 \\
			$1/8$ & 0.01 & 0.01 & 0.03 & 0.04 \\
			\hline
			& $1/8$ & $1/4$ & $3/8$ & $1/2$ \\
		\end{tabular}
		
		\begin{tabular}{c|cccc}
			\multicolumn{5}{c}{(c) improved}\\
			$1/2$ & 0.02 & 0.03 & 0.04 & 0.05 \\
			$3/8$ & 0.02 & 0.02 & 0.03 & 0.05 \\
			$1/4$ & 0.01 & 0.02 & 0.03 & 0.05 \\
			$1/8$ & 0.01 & 0.01 & 0.02 & 0.04 \\
			\hline
			& $1/8$ & $1/4$ & $3/8$ & $1/2$ \\
		\end{tabular}\hspace{1cm}
		\begin{tabular}{c|cccc}
			\multicolumn{5}{c}{(d) normalized}\\
			$1/2$ & 0.01 & 0.02 & 0.03 & 0.05 \\
			$3/8$ & 0.01 & 0.01 & 0.03 & 0.04 \\
			$1/4$ & 0.004 & 0.01 & 0.02 & 0.04 \\
			$1/8$ & 0.003 & 0.01 & 0.02 & 0.04 \\
			\hline
			& $1/8$ & $1/4$ & $3/8$ & $1/2$ \\
		\end{tabular}
		
		\begin{tabular}{c|cccc}
			\multicolumn{5}{c}{(e) full-IR}\\
			$1/2$ & 0.02 & 0.02 & 0.03 & 0.04 \\
			$3/8$ & 0.01 & 0.01 & 0.03 & 0.04 \\
			$1/4$ & 0.004 & 0.01 & 0.02 & 0.04 \\
			$1/8$ & 0.003 & 0.01 & 0.02 & 0.04 \\
			\hline
			& $1/8$ & $1/4$ & $3/8$ & $1/2$ \\
		\end{tabular}
	}
	\caption{Values $\chi_{ij}$ (formula \eqref{eq:chiRL}) in the longitudinal sector for $\mu_q/T=5$ and $\mu_3=0$.}
	\label{tab:app:chi50_long}
\end{table}

\begin{table}[h!]
	\centering
	{
		\small
		\begin{tabular}{c|cccc}
			\multicolumn{5}{c}{(a) hydrodynamic}\\
			$1/2$ & 0.08 & 0.08 & 0.07 & 0.07 \\
			$3/8$ & 0.04 & 0.04 & 0.04 & 0.04 \\
			$1/4$ & 0.02 & 0.02 & 0.02 & 0.03 \\
			$1/8$ & 0.01 & 0.01 & 0.01 & 0.02 \\
			\hline
			& $1/8$ & $1/4$ & $3/8$ & $1/2$ \\
		\end{tabular}\hspace{1cm}
		\begin{tabular}{c|cccc}
			\multicolumn{5}{c}{(b) extended-hydrodynamic}\\
			$1/2$ & 0.03 & 0.04 & 0.04 & 0.04 \\
			$3/8$ & 0.02 & 0.02 & 0.02 & 0.03 \\
			$1/4$ & 0.01 & 0.01 & 0.01 & 0.02 \\
			$1/8$ & 0.01 & 0.00 & 0.01 & 0.02 \\
			\hline
			& $1/8$ & $1/4$ & $3/8$ & $1/2$ \\
		\end{tabular}
		
		\begin{tabular}{c|cccc}
			\multicolumn{5}{c}{(c) improved}\\
			$1/2$ & 0.04 & 0.04 & 0.04 & 0.05 \\
			$3/8$ & 0.04 & 0.04 & 0.04 & 0.06 \\
			$1/4$ & 0.05 & 0.05 & 0.06 & 0.07 \\
			$1/8$ & 0.05 & 0.06 & 0.06 & 0.08 \\
			\hline
			& $1/8$ & $1/4$ & $3/8$ & $1/2$ \\
		\end{tabular}\hspace{1cm}
		\begin{tabular}{c|cccc}
			\multicolumn{5}{c}{(d) normalized}\\
			$1/2$ & 0.14 & 0.13 & 0.12 & 0.10 \\
			$3/8$ & 0.10 & 0.10 & 0.08 & 0.07 \\
			$1/4$ & 0.08 & 0.07 & 0.06 & 0.05 \\
			$1/8$ & 0.06 & 0.06 & 0.05 & 0.04 \\
			\hline
			& $1/8$ & $1/4$ & $3/8$ & $1/2$ \\
		\end{tabular}
		
		\begin{tabular}{c|cccc}
			\multicolumn{5}{c}{(e) full-IR}\\
			$1/2$ & 0.08 & 0.07 & 0.07 & 0.07 \\
			$3/8$ & 0.04 & 0.04 & 0.04 & 0.04 \\
			$1/4$ & 0.02 & 0.02 & 0.02 & 0.03 \\
			$1/8$ & 0.00 & 0.01 & 0.02 & 0.03 \\
			\hline
			& $1/8$ & $1/4$ & $3/8$ & $1/2$ \\
		\end{tabular}
	}
	\caption{Values $\chi_{ij}$ (formula \eqref{eq:chiRL}) in the transverse sector for $\mu_q/T=5$ and $\mu_3/\mu_q=-0.1$.}
	\label{tab:app:chi501_tr}
\end{table}

In table~\ref{tab:app:chi50_long}, the hydrodynamic approximation is already extremely accurate. The errors range from $0.003$ to $0.04$, and the values increase mainly as $\omega/\mu$ increases.

The extended hydrodynamic approximation is actually slightly worse than hydrodynamics in most entries. The improved approximation behaves similarly, although it is marginally better than the extended one in a few cells. The deterioration is small in absolute terms, but it shows that the extended and improved hydrodynamic corrections do not improve the longitudinal correlator in this regime.

The normalized and full approximations stay closest to hydrodynamics. The full approximation is essentially identical to hydrodynamics across the table, while the normalized approximation differs only mildly. The main conclusion is therefore that, for $\mu_q/T=5$ and $\mu_3=0$ in the longitudinal sector, the hydrodynamic approximation is already sufficient and the refinements do not bring a systematic gain.

\subsection{Background at  $\mu_q/T=5$ and $\mu_3/\mu_q=-0.1$}

\subsection*{The transverse correlator}

In table~\ref{tab:app:chi501_tr}, the extended hydrodynamic approximation gives the best overall performance. It improves over hydrodynamics in almost every entry, especially in the upper rows. At the smallest value of $k/\mu$, the improvement is smaller, but the extended approximation is still at least as good as hydrodynamics. The improved approximation does not improve the result in this case.

The normalized approximation is the least accurate one, being worse than hydrodynamics throughout most of the grid. The full approximation is much closer to hydrodynamics, and it is better than the improved and normalized approximations, but it does not outperform the extended hydrodynamic approximation. Therefore, for this transverse table, the extended hydrodynamic approximation is clearly preferred.

\subsection*{The longitudinal correlator}
\begin{table}[htb]
	\centering
	{
		\small
		\begin{tabular}{c|cccc}
			\multicolumn{5}{c}{(a) hydrodynamic}\\
			$1/2$ & 0.04 & 0.03 & 0.03 & 0.04 \\
			$3/8$ & 0.03 & 0.02 & 0.02 & 0.03 \\
			$1/4$ & 0.02 & 0.01 & 0.02 & 0.03 \\
			$1/8$ & 0.01 & 0.01 & 0.01 & 0.02 \\
			\hline
			& $1/8$ & $1/4$ & $3/8$ & $1/2$ \\
		\end{tabular}\hspace{1cm}
		\begin{tabular}{c|cccc}
			\multicolumn{5}{c}{(b) extended-hydrodynamic}\\
			$1/2$ & 0.01 & 0.02 & 0.03 & 0.04 \\
			$3/8$ & 0.01 & 0.02 & 0.03 & 0.04 \\
			$1/4$ & 0.01 & 0.01 & 0.02 & 0.03 \\
			$1/8$ & 0.01 & 0.01 & 0.01 & 0.02 \\
			\hline
			& $1/8$ & $1/4$ & $3/8$ & $1/2$ \\
		\end{tabular}
		
		\begin{tabular}{c|cccc}
			\multicolumn{5}{c}{(c) improved}\\
			$1/2$ & 0.05 & 0.06 & 0.07 & 0.08 \\
			$3/8$ & 0.05 & 0.06 & 0.07 & 0.08 \\
			$1/4$ & 0.05 & 0.06 & 0.06 & 0.07 \\
			$1/8$ & 0.05 & 0.05 & 0.06 & 0.07 \\
			\hline
			& $1/8$ & $1/4$ & $3/8$ & $1/2$ \\
		\end{tabular}\hspace{1cm}
		\begin{tabular}{c|cccc}
			\multicolumn{5}{c}{(d) normalized}\\
			$1/2$ & 0.09 & 0.06 & 0.04 & 0.04 \\
			$3/8$ & 0.07 & 0.05 & 0.04 & 0.03 \\
			$1/4$ & 0.06 & 0.05 & 0.03 & 0.03 \\
			$1/8$ & 0.05 & 0.05 & 0.04 & 0.03 \\
			\hline
			& $1/8$ & $1/4$ & $3/8$ & $1/2$ \\
		\end{tabular}
		
		\begin{tabular}{c|cccc}
			\multicolumn{5}{c}{(e) full-IR}\\
			$1/2$ & 0.04 & 0.03 & 0.04 & 0.05 \\
			$3/8$ & 0.03 & 0.03 & 0.03 & 0.04 \\
			$1/4$ & 0.02 & 0.02 & 0.02 & 0.04 \\
			$1/8$ & 0.01 & 0.01 & 0.02 & 0.03 \\
			\hline
			& $1/8$ & $1/4$ & $3/8$ & $1/2$ \\
		\end{tabular}
	}
	\caption{Values $\chi_{ij}$ (formula \eqref{eq:chiRL}) in the longitudinal sector for $\mu_q/T=5$ and $\mu_3/\mu_q=-0.1$.}
	\label{tab:app:chi501_long}
\end{table}

In table~\ref{tab:app:chi501_long}, the hydrodynamic approximation is already quite accurate with error up to 4\% at most. The extended hydrodynamic approximation gives a small but visible improvement at small $\omega/\mu$ and large $k/\mu$. In the lower rows the improvement is smaller, and in the last column the extended approximation is essentially equal to, or slightly worse than, hydrodynamics.

The improved approximation performs poorly in this table. It is worse than both hydrodynamics and extended hydrodynamics everywhere. The normalized approximation is also worse than hydrodynamics in most entries, especially in the first column. This is the opposite of the behavior of the extended hydrodynamic approximation, which is most useful precisely in that column.

The full approximation is intermediate. It remains close to hydrodynamics, and is much better than the improved and normalized approximations. However, it does not beat the extended hydrodynamic approximation in the upper-left part of the table. Hence the hierarchy is mild because the absolute errors are small, but the extended hydrodynamic approximation is still the most accurate global choice.

\subsection{Background at  $\mu_q/T=5$ and $\mu_3/\mu_q=-0.5$}

\subsection*{The transverse correlator}
\begin{table}[htb]
	\centering
	{
		\small
		\begin{tabular}{c|cccc}
			\multicolumn{5}{c}{(a) hydrodynamic}\\
			$1/2$ & 0.07 & 0.10 & 0.12 & 0.13 \\
			$3/8$ & 0.04 & 0.06 & 0.08 & 0.09 \\
			$1/4$ & 0.02 & 0.04 & 0.05 & 0.06 \\
			$1/8$ & 0.03 & 0.03 & 0.04 & 0.05 \\
			\hline
			& $1/8$ & $1/4$ & $3/8$ & $1/2$ \\
		\end{tabular}\hspace{1cm}
		\begin{tabular}{c|cccc}
			\multicolumn{5}{c}{(b) extended-hydrodynamic}\\
			$1/2$ & 0.04 & 0.08 & 0.10 & 0.12 \\
			$3/8$ & 0.03 & 0.06 & 0.08 & 0.09 \\
			$1/4$ & 0.02 & 0.05 & 0.07 & 0.07 \\
			$1/8$ & 0.02 & 0.04 & 0.06 & 0.06 \\
			\hline
			& $1/8$ & $1/4$ & $3/8$ & $1/2$ \\
		\end{tabular}
		
		\begin{tabular}{c|cccc}
			\multicolumn{5}{c}{(c) improved}\\
			$1/2$ & 0.23 & 0.20 & 0.18 & 0.17 \\
			$3/8$ & 0.24 & 0.22 & 0.20 & 0.19 \\
			$1/4$ & 0.24 & 0.23 & 0.21 & 0.21 \\
			$1/8$ & 0.25 & 0.23 & 0.22 & 0.21 \\
			\hline
			& $1/8$ & $1/4$ & $3/8$ & $1/2$ \\
		\end{tabular}\hspace{1cm}
		\begin{tabular}{c|cccc}
			\multicolumn{5}{c}{(d) normalized}\\
			$1/2$ & 0.37 & 0.38 & 0.39 & 0.39 \\
			$3/8$ & 0.32 & 0.34 & 0.34 & 0.34 \\
			$1/4$ & 0.29 & 0.30 & 0.31 & 0.31 \\
			$1/8$ & 0.27 & 0.28 & 0.29 & 0.29 \\
			\hline
			& $1/8$ & $1/4$ & $3/8$ & $1/2$ \\
		\end{tabular}
		
		\begin{tabular}{c|cccc}
			\multicolumn{5}{c}{(e) full-IR}\\
			$1/2$ & 0.06 & 0.06 & 0.06 & 0.06 \\
			$3/8$ & 0.04 & 0.04 & 0.04 & 0.04 \\
			$1/4$ & 0.05 & 0.04 & 0.03 & 0.04 \\
			$1/8$ & 0.07 & 0.05 & 0.05 & 0.05 \\
			\hline
			& $1/8$ & $1/4$ & $3/8$ & $1/2$ \\
		\end{tabular}
	}
	\caption{Values of $\chi_{ij}$ (defined in \eqref{eq:chiRL}) in the transverse sector for $\mu_q/T=5$ and $\mu_3/\mu_q=-0.5$.}
	\label{tab:app:chi505_tr}
\end{table}

In table~\ref{tab:app:chi505_tr}, the hydrodynamic and extended hydrodynamic approximations are close to each other. The extended approximation improves some large-$k/\mu$ entries, for example the top-left entry decreases from $0.07$ to $0.04$, but it also becomes slightly worse in parts of the lower rows, such as the third and fourth columns at $k/\mu=1/8$. Therefore the extended approximation does not give a uniform improvement in this regime, but it is definitely better in the top-left corner of the table.

The improved and normalized approximations clearly fail to approximate better the correlator for this background. These values are much bigger than the hydrodynamic and extended hydrodynamic errors.

The full approximation is the best overall. It is especially good at larger $\omega/\mu$, where it improves over both hydrodynamics and extended hydrodynamics. The only exceptions occur at very small $\omega/\mu$ and smaller $k/\mu$, where hydrodynamics can be slightly better.

\subsection*{The longitudinal correlator}

In Table~\ref{tab:app:chi505_long}, the hydrodynamic and extended hydrodynamic approximations are again comparable. The extended approximation improves the upper-left part of the table. However, the improvement is small, and in the lower rows the extended approximation is sometimes slightly worse than hydrodynamics.

The improved approximation and the normalized approximation are not reliable in this regime. These errors are several times larger than those of hydrodynamics, extended hydrodynamics, and the full approximation.

The full approximation improves over hydrodynamics in most of the right half of the table. At the smallest values of $\omega/\mu$, the full approximation is not always better than hydrodynamics, but its errors remain small.

\begin{table}[htb]
	\centering
	{
		\small
		\begin{tabular}{c|cccc}
			\multicolumn{5}{c}{(a) hydrodynamic}\\
			$1/2$ & 0.04 & 0.06 & 0.09 & 0.10 \\
			$3/8$ & 0.02 & 0.04 & 0.07 & 0.08 \\
			$1/4$ & 0.02 & 0.03 & 0.05 & 0.06 \\
			$1/8$ & 0.02 & 0.03 & 0.04 & 0.04 \\
			\hline
			& $1/8$ & $1/4$ & $3/8$ & $1/2$ \\
		\end{tabular}\hspace{1cm}
		\begin{tabular}{c|cccc}
			\multicolumn{5}{c}{(b) extended-hydrodynamic}\\
			$1/2$ & 0.02 & 0.05 & 0.08 & 0.09 \\
			$3/8$ & 0.02 & 0.04 & 0.07 & 0.08 \\
			$1/4$ & 0.02 & 0.04 & 0.06 & 0.07 \\
			$1/8$ & 0.02 & 0.03 & 0.05 & 0.05 \\
			\hline
			& $1/8$ & $1/4$ & $3/8$ & $1/2$ \\
		\end{tabular}
		
		\begin{tabular}{c|cccc}
			\multicolumn{5}{c}{(c) improved}\\
			$1/2$ & 0.21 & 0.18 & 0.16 & 0.16 \\
			$3/8$ & 0.21 & 0.19 & 0.17 & 0.17 \\
			$1/4$ & 0.21 & 0.20 & 0.18 & 0.18 \\
			$1/8$ & 0.21 & 0.20 & 0.19 & 0.19 \\
			\hline
			& $1/8$ & $1/4$ & $3/8$ & $1/2$ \\
		\end{tabular}\hspace{1cm}
		\begin{tabular}{c|cccc}
			\multicolumn{5}{c}{(d) normalized}\\
			$1/2$ & 0.27 & 0.30 & 0.32 & 0.30 \\
			$3/8$ & 0.25 & 0.27 & 0.29 & 0.27 \\
			$1/4$ & 0.24 & 0.25 & 0.27 & 0.25 \\
			$1/8$ & 0.23 & 0.24 & 0.25 & 0.24 \\
			\hline
			& $1/8$ & $1/4$ & $3/8$ & $1/2$ \\
		\end{tabular}
		
		\begin{tabular}{c|cccc}
			\multicolumn{5}{c}{(e) full-IR}\\
			$1/2$ & 0.03 & 0.04 & 0.05 & 0.06 \\
			$3/8$ & 0.03 & 0.03 & 0.03 & 0.05 \\
			$1/4$ & 0.04 & 0.03 & 0.03 & 0.04 \\
			$1/8$ & 0.05 & 0.04 & 0.03 & 0.04 \\
			\hline
			& $1/8$ & $1/4$ & $3/8$ & $1/2$ \\
		\end{tabular}
	}
	\caption{Values of $\chi_{ij}$ (defined in \eqref{eq:chiRL}) in the longitudinal sector for  $\mu_q/T=5$ and $\mu_3/\mu_q=-0.5$.}
	\label{tab:app:chi505_long}
\end{table}

\clearpage

\section{Fine-grained integrated relative difference}\label{app:finegrained}

In this appendix, we present the results for the observable $s_{ij}$ defined in formula \eqref{eq:chiCELL} for other values of the $\mu_q/T$ and $\mu_3/\mu_q$ in addition to the already studied $\mu_q/T = 65$ and $\mu_3/\mu_q=0$. In particular, we extend our discussion to $\mu_q/T\in\{10^4,5\}$ and $\mu_3/\mu_q=-0.5$. The observable $s_{ij}$ is defined inside a single cell in the $(\omega/\mu,k/\mu)$-plane defined by
\be
n_i\leq {\omega\over \mu}\leq n_i+\dfrac{1}{8}\sp n_j\leq {k\over \mu}\leq n_j+\dfrac{1}{8}\,.
\ee
where $n_{i}$ takes the four values ${i\over 8}$, $i=1,2,3,4$ and the same for $n_j$.
WE reproduce $s_{ij}$ here for convenience
\be\label{eq:chiCELLa}
s_{ij}^{(l)} =64\,\int_{n_i}^{n_i+1/8}\intd(\omega/\mu) \int_{n_j}^{n_j+1/8}\intd (k/\mu)\,\bigg|1- \dfrac{\text{Im}\Pi^{\perp,\parallel,\pm}_l}{\text{Im}\Pi^{\perp,\parallel,\pm}_\text{numerical}}\bigg|\,,
\ee
It is a local version of the error between an approximation and the exact correlator.

We considered the five approximations introduced in appendix \ref{app:IRcorr-approx} which are listed below
\begin{itemize}
	\item the (near-extremal) hydrodynamic approximation, defined in (\ref{app:eq:trhydroapprox}) and (\ref{app:eq:longhydroapprox}),
	\item the extended-hydrodynamic approximation, defined in (\ref{app:eq:trexthydroapprox}) and (\ref{app:eq:longexthydroapprox}),
	\item the improved extended hydrodynamic approximation, defined in (\ref{eq:trimprexthydroapprox}) and (\ref{eq:longimprexthydroapprox}),
	\item the approximation using the normalized IR-AdS$_2$ correlator, defined in (\ref{eq:trnormapprox}) and (\ref{eq:longnormapprox}),
	\item the approximation using the full IR-AdS$_2$ correlator, defined in (\ref{eq:trfullapprox}) and (\ref{eq:longfullapprox}).
\end{itemize}

We present below the tables associated with the relative errors. We remind the reader that the horizontal axis refers to $\omega/\mu$ and the vertical axis to $k/\mu$. The four values on each axis correspond to the values of $n_{i,j}$. The entries multiplied by 100 give the \% average error in the given parallelogram.

\subsection{Background at $\mu_q/T=10^4$ and $\mu_3=0$}

\subsection*{The transverse correlator}
\begin{table}[htb]
	\centering
	{
		\small
		\begin{tabular}{c|cccc}
			\multicolumn{5}{c}{(a) hydrodynamic}\\
			$1/2$ & 1.20 & 0.56 & 0.22 & 0.08 \\
			$3/8$ & 0.48 & 0.17 & 0.07 & 0.21 \\
			$1/4$ & 0.13 & 0.06 & 0.19 & 0.32 \\
			$1/8$ & 0.04 & 0.14 & 0.26 & 0.36 \\
			\hline
			& $1/8$ & $1/4$ & $3/8$ & $1/2$ \\
		\end{tabular}\hspace{1cm}
		\begin{tabular}{c|cccc}
			\multicolumn{5}{c}{(b) extended-hydrodynamic}\\
			$1/2$ & 0.56 & 0.36 & 0.16 & 0.07 \\
			$3/8$ & 0.23 & 0.09 & 0.08 & 0.21 \\
			$1/4$ & 0.06 & 0.07 & 0.20 & 0.32 \\
			$1/8$ & 0.04 & 0.14 & 0.26 & 0.36 \\
			\hline
			& $1/8$ & $1/4$ & $3/8$ & $1/2$ \\
		\end{tabular}
		
		\begin{tabular}{c|cccc}
			\multicolumn{5}{c}{(c) improved}\\
			$1/2$ & 0.60 & 0.40 & 0.18 & 0.07 \\
			$3/8$ & 0.25 & 0.11 & 0.07 & 0.20 \\
			$1/4$ & 0.06 & 0.07 & 0.19 & 0.31 \\
			$1/8$ & 0.04 & 0.14 & 0.26 & 0.36 \\
			\hline
			& $1/8$ & $1/4$ & $3/8$ & $1/2$ \\
		\end{tabular}\hspace{1cm}
		\begin{tabular}{c|cccc}
			\multicolumn{5}{c}{(d) normalized}\\
			$1/2$ & 0.57 & 0.36 & 0.16 & 0.07 \\
			$3/8$ & 0.24 & 0.09 & 0.08 & 0.21 \\
			$1/4$ & 0.06 & 0.07 & 0.20 & 0.32 \\
			$1/8$ & 0.04 & 0.14 & 0.26 & 0.36 \\
			\hline
			& $1/8$ & $1/4$ & $3/8$ & $1/2$ \\
		\end{tabular}
		
		\begin{tabular}{c|cccc}
			\multicolumn{5}{c}{(e) full-IR}\\
			$1/2$ & 0.61 & 0.40 & 0.18 & 0.07 \\
			$3/8$ & 0.26 & 0.11 & 0.07 & 0.20 \\
			$1/4$ & 0.06 & 0.07 & 0.19 & 0.31 \\
			$1/8$ & 0.04 & 0.14 & 0.26 & 0.36 \\
			\hline
			& $1/8$ & $1/4$ & $3/8$ & $1/2$ \\
		\end{tabular}
	}
	\caption{Values of $s_{ij}$ (defined in \eqref{eq:chiCELL}) in the transverse sector for $\mu_q/T=10^4$ and $\mu_3=0$.}
	\label{tab:app:s1040_tr}
\end{table}

In table~\ref{tab:app:s1040_tr}, we observe that the extended hydrodynamic approximation is substantially better than the hydrodynamic one in the upper-left part of the grid, namely at large $k/\mu$ and small $\omega/\mu$. The improvement is still visible in the second column of the upper row, from $0.56$ to $0.36$, but it becomes much less relevant as one moves either to smaller $k/\mu$ or to larger $\omega/\mu$.

On the other hand, for the smallest value of $k/\mu$ all the approximations essentially coincide. This shows that the IR-based corrections to the correlator do not affect the low-$k$ cells in a significant way. Among the IR-based approximations, the extended hydrodynamic and normalized approximations are almost identical, while the improved and full approximations are slightly worse in the top-left corner of the table. Therefore, in this table the main gain is the passage from hydrodynamics to the extended hydrodynamic approximation; the further IR refinements do not give a clear additional improvement.

\begin{table}[htb]
	\centering
	{
		\small
		\begin{tabular}{c|cccc}
			\multicolumn{5}{c}{(a) hydrodynamic}\\
			$1/2$ & 0.27 & 0.08 & 0.20 & 0.29 \\
			$3/8$ & 0.10 & 0.11 & 0.20 & 0.28 \\
			$1/4$ & 0.04 & 0.11 & 0.19 & 0.28 \\
			$1/8$ & 0.03 & 0.11 & 0.19 & 0.27 \\
			\hline
			& $1/8$ & $1/4$ & $3/8$ & $1/2$ \\
		\end{tabular}\hspace{1cm}
		\begin{tabular}{c|cccc}
			\multicolumn{5}{c}{(b) extended-hydrodynamic}\\
			$1/2$ & 0.07 & 0.16 & 0.25 & 0.31 \\
			$3/8$ & 0.06 & 0.16 & 0.22 & 0.29 \\
			$1/4$ & 0.06 & 0.13 & 0.20 & 0.28 \\
			$1/8$ & 0.04 & 0.11 & 0.19 & 0.28 \\
			\hline
			& $1/8$ & $1/4$ & $3/8$ & $1/2$ \\
		\end{tabular}
		
		\begin{tabular}{c|cccc}
			\multicolumn{5}{c}{(c) improved}\\
			$1/2$ & 0.07 & 0.14 & 0.23 & 0.30 \\
			$3/8$ & 0.06 & 0.15 & 0.22 & 0.29 \\
			$1/4$ & 0.05 & 0.12 & 0.20 & 0.28 \\
			$1/8$ & 0.04 & 0.11 & 0.19 & 0.28 \\
			\hline
			& $1/8$ & $1/4$ & $3/8$ & $1/2$ \\
		\end{tabular}\hspace{1cm}
		\begin{tabular}{c|cccc}
			\multicolumn{5}{c}{(d) normalized}\\
			$1/2$ & 0.07 & 0.16 & 0.25 & 0.31 \\
			$3/8$ & 0.06 & 0.16 & 0.22 & 0.29 \\
			$1/4$ & 0.06 & 0.13 & 0.20 & 0.28 \\
			$1/8$ & 0.04 & 0.11 & 0.19 & 0.28 \\
			\hline
			& $1/8$ & $1/4$ & $3/8$ & $1/2$ \\
		\end{tabular}
		
		\begin{tabular}{c|cccc}
			\multicolumn{5}{c}{(e) full-IR}\\
			$1/2$ & 0.07 & 0.14 & 0.23 & 0.30 \\
			$3/8$ & 0.06 & 0.15 & 0.22 & 0.29 \\
			$1/4$ & 0.05 & 0.12 & 0.20 & 0.28 \\
			$1/8$ & 0.04 & 0.11 & 0.19 & 0.28 \\
			\hline
			& $1/8$ & $1/4$ & $3/8$ & $1/2$ \\
		\end{tabular}
	}
	\caption{Values of $s_{ij}$ (defined in \eqref{eq:chiCELL}) in the longitudinal sector for $\mu_q/T=10^4$ and $\mu_3=0$.}
	\label{tab:app:s1040_long}
\end{table}

\FloatBarrier

\subsection*{The longitudinal correlator}

In table~\ref{tab:app:s1040_long}, the longitudinal sector shows slightly different features if compared with the transverse one. The main improvement of the extended hydrodynamic approximation over the hydrodynamic one is confined to the first column and especially to the upper row. A smaller improvement is also present at $k/\mu=3/8$. However, in the $k/\mu\in\{1/8,1/4\}$ rows the hydrodynamic approximation is better. Also for larger values of $\omega/\mu$, however, the extended hydrodynamic approximation becomes worse than the hydrodynamic one in several entries.

The normalized approximation follows the extended hydrodynamic one almost exactly. The improved and full approximations reduce these values slightly, but they still do not clearly outperform the hydrodynamic approximation away from the first column. Therefore, the best conclusion is not that one approximation dominates the table, but rather that the IR-based corrections help at small $\omega/\mu$ and large $k/\mu$, while the hydrodynamic approximation remains competitive, and often better, once $\omega/\mu$ increases.

\subsection{Background at $\mu_q/T=10^4$ and $\mu_3=-0.1$}

\subsection*{The transverse correlator}
\begin{table}[htb]
	\centering
	{
		\small
		\begin{tabular}{c|cccc}
			\multicolumn{5}{c}{(a) hydrodynamic}\\
			$1/2$ & 0.75 & 0.49 & 0.31 & 0.15 \\
			$3/8$ & 0.39 & 0.25 & 0.12 & 0.04 \\
			$1/4$ & 0.19 & 0.11 & 0.03 & 0.08 \\
			$1/8$ & 0.10 & 0.04 & 0.04 & 0.12 \\
			\hline
			& $1/8$ & $1/4$ & $3/8$ & $1/2$ \\
		\end{tabular}\hspace{1cm}
		\begin{tabular}{c|cccc}
			\multicolumn{5}{c}{(b) extended-hydrodynamic}\\
			$1/2$ & 0.42 & 0.35 & 0.24 & 0.11 \\
			$3/8$ & 0.25 & 0.19 & 0.09 & 0.03 \\
			$1/4$ & 0.15 & 0.09 & 0.03 & 0.09 \\
			$1/8$ & 0.10 & 0.04 & 0.04 & 0.12 \\
			\hline
			& $1/8$ & $1/4$ & $3/8$ & $1/2$ \\
		\end{tabular}
		
		\begin{tabular}{c|cccc}
			\multicolumn{5}{c}{(c) improved}\\
			$1/2$ & 0.33 & 0.26 & 0.15 & 0.05 \\
			$3/8$ & 0.16 & 0.10 & 0.04 & 0.08 \\
			$1/4$ & 0.06 & 0.02 & 0.07 & 0.16 \\
			$1/8$ & 0.01 & 0.04 & 0.11 & 0.19 \\
			\hline
			& $1/8$ & $1/4$ & $3/8$ & $1/2$ \\
		\end{tabular}\hspace{1cm}
		\begin{tabular}{c|cccc}
			\multicolumn{5}{c}{(d) normalized}\\
			$1/2$ & 0.44 & 0.35 & 0.24 & 0.11 \\
			$3/8$ & 0.26 & 0.19 & 0.09 & 0.03 \\
			$1/4$ & 0.15 & 0.09 & 0.03 & 0.09 \\
			$1/8$ & 0.10 & 0.04 & 0.04 & 0.12 \\
			\hline
			& $1/8$ & $1/4$ & $3/8$ & $1/2$ \\
		\end{tabular}
		
		\begin{tabular}{c|cccc}
			\multicolumn{5}{c}{(e) full-IR}\\
			$1/2$ & 0.34 & 0.26 & 0.15 & 0.05 \\
			$3/8$ & 0.17 & 0.10 & 0.04 & 0.08 \\
			$1/4$ & 0.06 & 0.02 & 0.07 & 0.16 \\
			$1/8$ & 0.02 & 0.04 & 0.11 & 0.19 \\
			\hline
			& $1/8$ & $1/4$ & $3/8$ & $1/2$ \\
		\end{tabular}
	}
	\caption{Values of $s_{ij}$ (defined in \eqref{eq:chiCELL}) in the transverse sector for $\mu_q/T=10^4$ and $\mu_3=-0.1$.}
	\label{tab:app:s10401_tr}
\end{table}

In table~\ref{tab:app:s10401_tr}, a non-zero isospin chemical potential $\mu_3/\mu_q=-0.1$ is turned on. The extended hydrodynamic approximation improves over hydrodynamics in most of the grid. The reduction is largest in the upper-left part. The improvement is more moderate at smaller $k/\mu$, and in the bottom row the extended approximation is essentially identical to the hydrodynamic one.

The improved approximation gives a further and much more pronounced reduction. The full approximation is almost identical to the improved one, differing only by $0.01$ in a few entries. The normalized approximation, instead, stays very close to the extended hydrodynamic approximation and therefore does not capture this additional improvement. The only important caveat is the bottom row at large $\omega/\mu$ for which the improved and full approximations are worse than hydrodynamics and extended hydrodynamics ones. Therefore, the improved and full approximations are clearly preferred in the upper and central part of the table, but not at the smallest value of $k/\mu$.

\subsection*{Longitudinal correlator}
\begin{table}[htb]
	\centering
	{
		\small
		\begin{tabular}{c|cccc}
			\multicolumn{5}{c}{(a) hydrodynamic}\\
			$1/2$ & 0.40 & 0.08 & 0.15 & 0.23 \\
			$3/8$ & 0.23 & 0.08 & 0.13 & 0.20 \\
			$1/4$ & 0.13 & 0.06 & 0.10 & 0.18 \\
			$1/8$ & 0.07 & 0.02 & 0.09 & 0.17 \\
			\hline
			& $1/8$ & $1/4$ & $3/8$ & $1/2$ \\
		\end{tabular}\hspace{1cm}
		\begin{tabular}{c|cccc}
			\multicolumn{5}{c}{(b) extended-hydrodynamic}\\
			$1/2$ & 0.15 & 0.09 & 0.19 & 0.25 \\
			$3/8$ & 0.11 & 0.12 & 0.16 & 0.21 \\
			$1/4$ & 0.10 & 0.07 & 0.11 & 0.18 \\
			$1/8$ & 0.07 & 0.02 & 0.09 & 0.17 \\
			\hline
			& $1/8$ & $1/4$ & $3/8$ & $1/2$ \\
		\end{tabular}
		
		\begin{tabular}{c|cccc}
			\multicolumn{5}{c}{(c) improved}\\
			$1/2$ & 0.08 & 0.14 & 0.25 & 0.30 \\
			$3/8$ & 0.06 & 0.18 & 0.22 & 0.27 \\
			$1/4$ & 0.05 & 0.15 & 0.18 & 0.25 \\
			$1/8$ & 0.03 & 0.09 & 0.16 & 0.24 \\
			\hline
			& $1/8$ & $1/4$ & $3/8$ & $1/2$ \\
		\end{tabular}\hspace{1cm}
		\begin{tabular}{c|cccc}
			\multicolumn{5}{c}{(d) normalized}\\
			$1/2$ & 0.15 & 0.09 & 0.19 & 0.25 \\
			$3/8$ & 0.12 & 0.12 & 0.16 & 0.21 \\
			$1/4$ & 0.11 & 0.07 & 0.11 & 0.18 \\
			$1/8$ & 0.07 & 0.02 & 0.09 & 0.17 \\
			\hline
			& $1/8$ & $1/4$ & $3/8$ & $1/2$ \\
		\end{tabular}
		
		\begin{tabular}{c|cccc}
			\multicolumn{5}{c}{(e) full-IR}\\
			$1/2$ & 0.08 & 0.14 & 0.25 & 0.30 \\
			$3/8$ & 0.07 & 0.18 & 0.22 & 0.27 \\
			$1/4$ & 0.06 & 0.15 & 0.18 & 0.25 \\
			$1/8$ & 0.03 & 0.09 & 0.16 & 0.24 \\
			\hline
			& $1/8$ & $1/4$ & $3/8$ & $1/2$ \\
		\end{tabular}
	}
	\caption{Values of $s_{ij}$ (defined in \eqref{eq:chiCELL}) in the longitudinal sector for $\mu_q/T=10^4$ and $\mu_3=-0.1$.}
	\label{tab:app:s10401_long}
\end{table}

In table~\ref{tab:app:s10401_long}, the longitudinal sector shows a more mixed pattern. The extended hydrodynamic approximation improves the hydrodynamic result mainly in the first column. At the smallest value of $k/\mu$, however, the first-column value remains the same, $0.07$, and the bottom row is essentially unchanged.

The improved and full approximations are even better in the first column, up to small differences. However, this improvement comes at the cost of larger errors at higher frequencies. The same trend appears in the lower rows: the improved and full approximations are good at small $\omega/\mu$, but they overshoot the error as $\omega/\mu$ increases. The normalized approximation is instead very close to the extended hydrodynamic one. Therefore, in this table, the preferred approximation depends strongly on the cell: improved and full are best in the first column, while hydrodynamics or extended hydrodynamics are better in several cells at larger $\omega/\mu$.

\subsection{Background at $\mu_q/T=10^4$ and $\mu_3/\mu_q=-0.5$}

\subsection*{The transverse correlator}
\begin{table}[htb]
	\centering
	{
		\small
		\begin{tabular}{c|cccc}
			\multicolumn{5}{c}{(a) hydrodynamic}\\
			$1/2$ & 1.90 & 1.60 & 1.30 & 1.00 \\
			$3/8$ & 0.91 & 0.93 & 0.80 & 0.63 \\
			$1/4$ & 0.45 & 0.57 & 0.52 & 0.40 \\
			$1/8$ & 0.26 & 0.42 & 0.40 & 0.30 \\
			\hline
			& $1/8$ & $1/4$ & $3/8$ & $1/2$ \\
		\end{tabular}\hspace{1cm}
		\begin{tabular}{c|cccc}
			\multicolumn{5}{c}{(b) extended-hydrodynamic}\\
			$1/2$ & 1.30 & 1.40 & 1.20 & 1.00 \\
			$3/8$ & 0.80 & 0.88 & 0.79 & 0.63 \\
			$1/4$ & 0.53 & 0.61 & 0.54 & 0.41 \\
			$1/8$ & 0.41 & 0.49 & 0.42 & 0.31 \\
			\hline
			& $1/8$ & $1/4$ & $3/8$ & $1/2$ \\
		\end{tabular}
		
		\begin{tabular}{c|cccc}
			\multicolumn{5}{c}{(c) improved}\\
			$1/2$ & 0.34 & 0.39 & 0.31 & 0.18 \\
			$3/8$ & 0.08 & 0.09 & 0.06 & 0.07 \\
			$1/4$ & 0.13 & 0.08 & 0.12 & 0.20 \\
			$1/8$ & 0.20 & 0.16 & 0.20 & 0.26 \\
			\hline
			& $1/8$ & $1/4$ & $3/8$ & $1/2$ \\
		\end{tabular}\hspace{1cm}
		\begin{tabular}{c|cccc}
			\multicolumn{5}{c}{(d) normalized}\\
			$1/2$ & 1.40 & 1.40 & 1.20 & 1.00 \\
			$3/8$ & 0.87 & 0.89 & 0.79 & 0.63 \\
			$1/4$ & 0.58 & 0.62 & 0.54 & 0.41 \\
			$1/8$ & 0.46 & 0.49 & 0.43 & 0.31 \\
			\hline
			& $1/8$ & $1/4$ & $3/8$ & $1/2$ \\
		\end{tabular}
		
		\begin{tabular}{c|cccc}
			\multicolumn{5}{c}{(e) full-IR}\\
			$1/2$ & 0.39 & 0.40 & 0.31 & 0.19 \\
			$3/8$ & 0.09 & 0.09 & 0.06 & 0.07 \\
			$1/4$ & 0.10 & 0.08 & 0.12 & 0.20 \\
			$1/8$ & 0.18 & 0.16 & 0.19 & 0.26 \\
			\hline
			& $1/8$ & $1/4$ & $3/8$ & $1/2$ \\
		\end{tabular}
	}
	\caption{Values of $s_{ij}$ (defined in \eqref{eq:chiCELL}) in the transverse sector for $\mu_q/T=10^4$ and $\mu_3/\mu_q=-0.5$.}
	\label{tab:app:s10405_tr}
\end{table}

In table~\ref{tab:app:s10405_tr}, the increase to $\mu_3/\mu_q=-0.5$ produces a strong departure from the hydrodynamic regime in the transverse sector. The hydrodynamic errors are large across the whole grid with order 100\% in the upper row. The extended hydrodynamic approximation only partially improves the first row, but it does not give a systematic improvement elsewhere. In the bottom row it is actually worse than hydrodynamics.

The normalized approximation behaves similarly to the extended hydrodynamic. The qualitative change comes from the improved and full approximations. The improved approximation reduces the upper and second rows significantly. The full approximation is very close to the improved one: it is slightly worse in the upper row, but slightly better in some lower-row entries.

\subsection*{The longitudinal correlator}
\begin{table}[htb]
	\centering
	{
		\small
		\begin{tabular}{c|cccc}
			\multicolumn{5}{c}{(a) hydrodynamic}\\
			$1/2$ & 1.30 & 0.93 & 0.33 & 0.15 \\
			$3/8$ & 0.73 & 0.67 & 0.24 & 0.20 \\
			$1/4$ & 0.41 & 0.47 & 0.25 & 0.07 \\
			$1/8$ & 0.26 & 0.32 & 0.23 & 0.09 \\
			\hline
			& $1/8$ & $1/4$ & $3/8$ & $1/2$ \\
		\end{tabular}\hspace{1cm}
		\begin{tabular}{c|cccc}
			\multicolumn{5}{c}{(b) extended-hydrodynamic}\\
			$1/2$ & 0.81 & 0.76 & 0.28 & 0.16 \\
			$3/8$ & 0.63 & 0.64 & 0.23 & 0.20 \\
			$1/4$ & 0.49 & 0.51 & 0.26 & 0.07 \\
			$1/8$ & 0.40 & 0.38 & 0.25 & 0.09 \\
			\hline
			& $1/8$ & $1/4$ & $3/8$ & $1/2$ \\
		\end{tabular}
		
		\begin{tabular}{c|cccc}
			\multicolumn{5}{c}{(c) improved}\\
			$1/2$ & 0.09 & 0.06 & 0.25 & 0.50 \\
			$3/8$ & 0.06 & 0.06 & 0.33 & 0.54 \\
			$1/4$ & 0.15 & 0.14 & 0.34 & 0.47 \\
			$1/8$ & 0.21 & 0.22 & 0.30 & 0.39 \\
			\hline
			& $1/8$ & $1/4$ & $3/8$ & $1/2$ \\
		\end{tabular}\hspace{1cm}
		\begin{tabular}{c|cccc}
			\multicolumn{5}{c}{(d) normalized}\\
			$1/2$ & 0.88 & 0.77 & 0.29 & 0.15 \\
			$3/8$ & 0.70 & 0.64 & 0.23 & 0.20 \\
			$1/4$ & 0.54 & 0.51 & 0.26 & 0.07 \\
			$1/8$ & 0.45 & 0.39 & 0.25 & 0.09 \\
			\hline
			& $1/8$ & $1/4$ & $3/8$ & $1/2$ \\
		\end{tabular}
		
		\begin{tabular}{c|cccc}
			\multicolumn{5}{c}{(e) full-IR}\\
			$1/2$ & 0.11 & 0.06 & 0.25 & 0.50 \\
			$3/8$ & 0.04 & 0.05 & 0.33 & 0.54 \\
			$1/4$ & 0.13 & 0.14 & 0.34 & 0.47 \\
			$1/8$ & 0.18 & 0.21 & 0.30 & 0.38 \\
			\hline
			& $1/8$ & $1/4$ & $3/8$ & $1/2$ \\
		\end{tabular}
	}
	\caption{Values of $s_{ij}$ (defined in \eqref{eq:chiCELL}) in the longitudinal sector for $\mu_q/T=10^4$ and $\mu_3/\mu_q=-0.5$.}
	\label{tab:app:s10405_long}
\end{table}

In table~\ref{tab:app:s10405_long}, the longitudinal sector at $\mu_3/\mu_q=-0.5$ is again highly non-uniform. The hydrodynamic approximation has very large errors in the first two columns, but it becomes much better in the last column. We argue that this is due to the shift of the hydrodynamic degrees of freedom around $\omega\sim-\mu_3$. The extended hydrodynamic approximation improves the first column in the upper rows, but it becomes worse than hydrodynamics in the lower rows of the same column and does not significantly improve the last two columns.

The improved and full approximations drastically reduce the errors in the first two columns. In the second row the full approximation is even better than improved one. However, these approximations become worse in the last column. Therefore, the improved and full approximations are essential to reduce the error at small $\omega/\mu$, but they are not uniformly better over the whole table (see the large $\omega/\mu$ column).

\subsection{Background at $\mu_q/T=65$ and $\mu_3/\mu_q=-0.1$}

\subsection*{The transverse correlator}
\begin{table}[htb]
	\centering
	{
		\small
		\begin{tabular}{c|cccc}
			\multicolumn{5}{c}{(a) hydrodynamic}\\
			$1/2$ & 1.10 & 0.76 & 0.44 & 0.19 \\
			$3/8$ & 0.50 & 0.34 & 0.14 & 0.06 \\
			$1/4$ & 0.19 & 0.11 & 0.05 & 0.15 \\
			$1/8$ & 0.05 & 0.03 & 0.10 & 0.21 \\
			\hline
			& $1/8$ & $1/4$ & $3/8$ & $1/2$ \\
		\end{tabular}\hspace{1cm}
		\begin{tabular}{c|cccc}
			\multicolumn{5}{c}{(b) extended-hydrodynamic}\\
			$1/2$ & 0.55 & 0.55 & 0.36 & 0.17 \\
			$3/8$ & 0.28 & 0.25 & 0.11 & 0.06 \\
			$1/4$ & 0.12 & 0.08 & 0.05 & 0.16 \\
			$1/8$ & 0.05 & 0.03 & 0.10 & 0.21 \\
			\hline
			& $1/8$ & $1/4$ & $3/8$ & $1/2$ \\
		\end{tabular}
		
		\begin{tabular}{c|cccc}
			\multicolumn{5}{c}{(c) improved}\\
			$1/2$ & 0.43 & 0.42 & 0.25 & 0.09 \\
			$3/8$ & 0.17 & 0.14 & 0.05 & 0.13 \\
			$1/4$ & 0.04 & 0.04 & 0.13 & 0.24 \\
			$1/8$ & 0.06 & 0.10 & 0.19 & 0.29 \\
			\hline
			& $1/8$ & $1/4$ & $3/8$ & $1/2$ \\
		\end{tabular}\hspace{1cm}
		\begin{tabular}{c|cccc}
			\multicolumn{5}{c}{(d) normalized}\\
			$1/2$ & 0.76 & 0.59 & 0.38 & 0.18 \\
			$3/8$ & 0.41 & 0.28 & 0.12 & 0.06 \\
			$1/4$ & 0.20 & 0.11 & 0.04 & 0.15 \\
			$1/8$ & 0.11 & 0.03 & 0.09 & 0.20 \\
			\hline
			& $1/8$ & $1/4$ & $3/8$ & $1/2$ \\
		\end{tabular}
		
		\begin{tabular}{c|cccc}
			\multicolumn{5}{c}{(e) full-IR}\\
			$1/2$ & 0.62 & 0.46 & 0.27 & 0.10 \\
			$3/8$ & 0.28 & 0.16 & 0.06 & 0.12 \\
			$1/4$ & 0.08 & 0.04 & 0.12 & 0.23 \\
			$1/8$ & 0.05 & 0.08 & 0.18 & 0.29 \\
			\hline
			& $1/8$ & $1/4$ & $3/8$ & $1/2$ \\
		\end{tabular}
	}
	\caption{Values of $s_{ij}$ (defined in \eqref{eq:chiCELL}) in the transverse sector for $\mu_q/T=65$ and $\mu_3/\mu_q=-0.1$.}
	\label{tab:app:s6501_tr}
\end{table}

In table~\ref{tab:app:s6501_tr}, the transverse sector at $\mu_q/T=65$ and $\mu_3/\mu_q=-0.1$ shows a hierarchy similar to the one at $\mu_q/T=10^4$, but with larger errors in the hydrodynamic approximation. The extended hydrodynamic approximation gives a clear improvement in the upper rows. The improvement is much smaller in the lower rows, and in the bottom row the extended approximation coincides with hydrodynamics.

The improved approximation reduces the error further in most of the upper and central cells, where it is better than both hydrodynamics and extended hydrodynamics in every column. However, the improved approximation becomes worse at the smallest value of $k/\mu=1/8$. The full approximation is a compromise: it is not as good as the improved one in the upper row, but it performs better than the improved approximation in several lower-row entries. The normalized approximation gives only a partial improvement and remains closer to hydrodynamics than to the improved result. Therefore, the improved approximation is favored at larger $k/\mu$, while the full approximation is more competitive once $k/\mu$ is lowered.

\begin{table}[htb]
	\centering
	{
		\small
		\begin{tabular}{c|cccc}
			\multicolumn{5}{c}{(a) hydrodynamic}\\
			$1/2$ & 0.35 & 0.07 & 0.14 & 0.21 \\
			$3/8$ & 0.18 & 0.07 & 0.12 & 0.19 \\
			$1/4$ & 0.09 & 0.05 & 0.09 & 0.17 \\
			$1/8$ & 0.04 & 0.02 & 0.08 & 0.16 \\
			\hline
			& $1/8$ & $1/4$ & $3/8$ & $1/2$ \\
		\end{tabular}\hspace{1cm}
		\begin{tabular}{c|cccc}
			\multicolumn{5}{c}{(b) extended-hydrodynamic}\\
			$1/2$ & 0.12 & 0.08 & 0.18 & 0.23 \\
			$3/8$ & 0.08 & 0.11 & 0.15 & 0.20 \\
			$1/4$ & 0.07 & 0.07 & 0.10 & 0.17 \\
			$1/8$ & 0.04 & 0.02 & 0.08 & 0.16 \\
			\hline
			& $1/8$ & $1/4$ & $3/8$ & $1/2$ \\
		\end{tabular}
		
		\begin{tabular}{c|cccc}
			\multicolumn{5}{c}{(c) improved}\\
			$1/2$ & 0.06 & 0.12 & 0.23 & 0.28 \\
			$3/8$ & 0.04 & 0.17 & 0.21 & 0.25 \\
			$1/4$ & 0.05 & 0.14 & 0.17 & 0.23 \\
			$1/8$ & 0.05 & 0.09 & 0.15 & 0.23 \\
			\hline
			& $1/8$ & $1/4$ & $3/8$ & $1/2$ \\
		\end{tabular}\hspace{1cm}
		\begin{tabular}{c|cccc}
			\multicolumn{5}{c}{(d) normalized}\\
			$1/2$ & 0.21 & 0.07 & 0.17 & 0.23 \\
			$3/8$ & 0.14 & 0.09 & 0.14 & 0.19 \\
			$1/4$ & 0.11 & 0.06 & 0.09 & 0.17 \\
			$1/8$ & 0.08 & 0.02 & 0.07 & 0.15 \\
			\hline
			& $1/8$ & $1/4$ & $3/8$ & $1/2$ \\
		\end{tabular}
		
		\begin{tabular}{c|cccc}
			\multicolumn{5}{c}{(e) full-IR}\\
			$1/2$ & 0.13 & 0.11 & 0.23 & 0.28 \\
			$3/8$ & 0.07 & 0.16 & 0.20 & 0.25 \\
			$1/4$ & 0.06 & 0.13 & 0.16 & 0.23 \\
			$1/8$ & 0.04 & 0.08 & 0.14 & 0.22 \\
			\hline
			& $1/8$ & $1/4$ & $3/8$ & $1/2$ \\
		\end{tabular}
	}
	\caption{Values of $s_{ij}$ (defined in \eqref{eq:chiCELL}) in the longitudinal sector for $\mu_q/T=65$ and $\mu_3/\mu_q=-0.1$.}
	\label{tab:app:s6501_long}
\end{table}

\subsection*{The longitudinal correlator}

In table~\ref{tab:app:s6501_long}, we present the results for the longitudinal sector. The extended hydrodynamic approximation gives an improvement concentrated almost entirely in the first column. In the bottom row, however, the extended and hydrodynamic approximations coincide in the first, second, and third columns. At larger $\omega/\mu$, the extended hydrodynamic approximation is no longer clearly better.

The improved approximation lowers the first-column entries even more, but it is significantly worse in the last two columns. The full approximation follows the same pattern. The normalized approximation lies between hydrodynamics and extended hydrodynamics: it improves some first-column entries but does not give a clear global gain. Therefore, the reliable conclusion is that the extended approximation improves the low-frequency cells, whereas hydrodynamics remains competitive, and often better, for larger $\omega/\mu$.

\subsection{Background at $\mu_q/T=65$ and $\mu_3/\mu_q=-0.5$}
\subsection*{The transverse correlator}
\begin{table}[htb]
	\centering
	{
		\small
		\begin{tabular}{c|cccc}
			\multicolumn{5}{c}{(a) hydrodynamic}\\
			$1/2$ & 1.70 & 1.50 & 1.20 & 0.99 \\
			$3/8$ & 0.84 & 0.88 & 0.78 & 0.62 \\
			$1/4$ & 0.43 & 0.56 & 0.52 & 0.41 \\
			$1/8$ & 0.25 & 0.42 & 0.40 & 0.32 \\
			\hline
			& $1/8$ & $1/4$ & $3/8$ & $1/2$ \\
		\end{tabular}\hspace{1cm}
		\begin{tabular}{c|cccc}
			\multicolumn{5}{c}{(b) extended-hydrodynamic}\\
			$1/2$ & 1.20 & 1.30 & 1.20 & 0.97 \\
			$3/8$ & 0.74 & 0.84 & 0.76 & 0.62 \\
			$1/4$ & 0.50 & 0.59 & 0.53 & 0.42 \\
			$1/8$ & 0.39 & 0.48 & 0.43 & 0.33 \\
			\hline
			& $1/8$ & $1/4$ & $3/8$ & $1/2$ \\
		\end{tabular}
		
		\begin{tabular}{c|cccc}
			\multicolumn{5}{c}{(c) improved}\\
			$1/2$ & 0.31 & 0.37 & 0.30 & 0.19 \\
			$3/8$ & 0.07 & 0.09 & 0.06 & 0.06 \\
			$1/4$ & 0.12 & 0.07 & 0.10 & 0.17 \\
			$1/8$ & 0.19 & 0.14 & 0.17 & 0.23 \\
			\hline
			& $1/8$ & $1/4$ & $3/8$ & $1/2$ \\
		\end{tabular}\hspace{1cm}
		\begin{tabular}{c|cccc}
			\multicolumn{5}{c}{(d) normalized}\\
			$1/2$ & 1.90 & 1.50 & 1.30 & 1.00 \\
			$3/8$ & 1.20 & 0.98 & 0.84 & 0.67 \\
			$1/4$ & 0.84 & 0.71 & 0.60 & 0.46 \\
			$1/8$ & 0.68 & 0.58 & 0.49 & 0.36 \\
			\hline
			& $1/8$ & $1/4$ & $3/8$ & $1/2$ \\
		\end{tabular}
		
		\begin{tabular}{c|cccc}
			\multicolumn{5}{c}{(e) full-IR}\\
			$1/2$ & 0.72 & 0.48 & 0.36 & 0.23 \\
			$3/8$ & 0.31 & 0.17 & 0.09 & 0.05 \\
			$1/4$ & 0.08 & 0.03 & 0.07 & 0.15 \\
			$1/8$ & 0.03 & 0.08 & 0.14 & 0.21 \\
			\hline
			& $1/8$ & $1/4$ & $3/8$ & $1/2$ \\
		\end{tabular}
	}
	\caption{Values of $s_{ij}$ (defined in \eqref{eq:chiCELL}) in the transverse sector for $\mu_q/T=65$ and $\mu_3/\mu_q=-0.5$.}
	\label{tab:app:s6505_tr}
\end{table}

In table~\ref{tab:app:s6505_tr}, the transverse sector at $\mu_3/\mu_q=-0.5$ again shows a strong separation between the approximations. Hydrodynamics has large errors throughout the table. The extended hydrodynamic approximation improves the upper-left cell, but it is not a systematic improvement: in the lower rows it increases the error. The normalized approximation is clearly the worst one even if it improves the first and second entries of the first column with respect to the hydrodynamic table.

The improved and full approximations are much better. The improved approximation is best in the upper part of the table while the full approximation becomes better in the lower rows.  Therefore, the improved approximation is preferred at larger $k/\mu$, while the full approximation is better at smaller $k/\mu$.

\subsection*{The longitudinal correlator}
\begin{table}[htb]
	\centering
	{
		\small
		\begin{tabular}{c|cccc}
			\multicolumn{5}{c}{(a) hydrodynamic}\\
			$1/2$ & 0.51 & 0.66 & 0.57 & 0.17 \\
			$3/8$ & 0.28 & 0.47 & 0.49 & 0.20 \\
			$1/4$ & 0.16 & 0.34 & 0.37 & 0.22 \\
			$1/8$ & 0.12 & 0.26 & 0.27 & 0.20 \\
			\hline
			& $1/8$ & $1/4$ & $3/8$ & $1/2$ \\
		\end{tabular}\hspace{1cm}
		\begin{tabular}{c|cccc}
			\multicolumn{5}{c}{(b) extended-hydrodynamic}\\
			$1/2$ & 0.33 & 0.56 & 0.52 & 0.16 \\
			$3/8$ & 0.24 & 0.45 & 0.47 & 0.20 \\
			$1/4$ & 0.18 & 0.36 & 0.38 & 0.23 \\
			$1/8$ & 0.16 & 0.30 & 0.29 & 0.21 \\
			\hline
			& $1/8$ & $1/4$ & $3/8$ & $1/2$ \\
		\end{tabular}
		
		\begin{tabular}{c|cccc}
			\multicolumn{5}{c}{(c) improved}\\
			$1/2$ & 0.10 & 0.08 & 0.05 & 0.25 \\
			$3/8$ & 0.16 & 0.03 & 0.02 & 0.28 \\
			$1/4$ & 0.20 & 0.09 & 0.07 & 0.25 \\
			$1/8$ & 0.22 & 0.12 & 0.13 & 0.19 \\
			\hline
			& $1/8$ & $1/4$ & $3/8$ & $1/2$ \\
		\end{tabular}\hspace{1cm}
		\begin{tabular}{c|cccc}
			\multicolumn{5}{c}{(d) normalized}\\
			$1/2$ & 0.61 & 0.66 & 0.57 & 0.17 \\
			$3/8$ & 0.46 & 0.53 & 0.52 & 0.21 \\
			$1/4$ & 0.38 & 0.44 & 0.42 & 0.24 \\
			$1/8$ & 0.34 & 0.37 & 0.33 & 0.24 \\
			\hline
			& $1/8$ & $1/4$ & $3/8$ & $1/2$ \\
		\end{tabular}
		
		\begin{tabular}{c|cccc}
			\multicolumn{5}{c}{(e) full-IR}\\
			$1/2$ & 0.11 & 0.14 & 0.08 & 0.23 \\
			$3/8$ & 0.03 & 0.04 & 0.04 & 0.26 \\
			$1/4$ & 0.07 & 0.04 & 0.04 & 0.23 \\
			$1/8$ & 0.10 & 0.08 & 0.10 & 0.17 \\
			\hline
			& $1/8$ & $1/4$ & $3/8$ & $1/2$ \\
		\end{tabular}
	}
	\caption{Values of $s_{ij}$ (defined in \eqref{eq:chiCELL}) in the longitudinal sector for $\mu_q/T=65$ and $\mu_3/\mu_q=-0.5$.}
	\label{tab:app:s6505_long}
\end{table}

In table~\ref{tab:app:s6505_long}, the longitudinal sector also favors the improved and full approximations over hydrodynamics, but the full approximation is the most stable considering large frequencies. The hydrodynamic and extended hydrodynamic approximations are comparable: the extended approximation improves some upper-left entries, but it becomes worse in the lower rows. The normalized approximation is worse than both and no visible advantage.

The improved approximation reduces the middle columns very efficiently. However, it is less good in the first column at smaller $k/\mu$. The full approximation lowers many of these entries further and gives the smallest average error. The only region where the improved approximation is sometimes better is the upper row, especially the second and third columns.

\subsection{Background at $\mu_q/T=5$ and $\mu_3=0$}
\subsection*{The transverse correlator}
\begin{table}[htb]
	\centering
	{
		\small
		\begin{tabular}{c|cccc}
			\multicolumn{5}{c}{(a) hydrodynamic}\\
			$1/2$ & 0.18 & 0.15 & 0.10 & 0.04 \\
			$3/8$ & 0.09 & 0.06 & 0.03 & 0.04 \\
			$1/4$ & 0.03 & 0.01 & 0.03 & 0.09 \\
			$1/8$ & 0.004 & 0.02 & 0.06 & 0.11 \\
			\hline
			& $1/8$ & $1/4$ & $3/8$ & $1/2$ \\
		\end{tabular}\hspace{1cm}
		\begin{tabular}{c|cccc}
			\multicolumn{5}{c}{(b) extended-hydrodynamic}\\
			$1/2$ & 0.05 & 0.08 & 0.06 & 0.02 \\
			$3/8$ & 0.02 & 0.03 & 0.01 & 0.05 \\
			$1/4$ & 0.01 & 0.01 & 0.04 & 0.09 \\
			$1/8$ & 0.003 & 0.02 & 0.06 & 0.11 \\
			\hline
			& $1/8$ & $1/4$ & $3/8$ & $1/2$ \\
		\end{tabular}
		
		\begin{tabular}{c|cccc}
			\multicolumn{5}{c}{(c) improved}\\
			$1/2$ & 0.06 & 0.09 & 0.07 & 0.03 \\
			$3/8$ & 0.03 & 0.03 & 0.02 & 0.05 \\
			$1/4$ & 0.01 & 0.01 & 0.04 & 0.09 \\
			$1/8$ & 0.003 & 0.02 & 0.06 & 0.11 \\
			\hline
			& $1/8$ & $1/4$ & $3/8$ & $1/2$ \\
		\end{tabular}\hspace{1cm}
		\begin{tabular}{c|cccc}
			\multicolumn{5}{c}{(d) normalized}\\
			$1/2$ & 0.17 & 0.14 & 0.10 & 0.04 \\
			$3/8$ & 0.08 & 0.06 & 0.02 & 0.04 \\
			$1/4$ & 0.03 & 0.01 & 0.03 & 0.09 \\
			$1/8$ & 0.004 & 0.02 & 0.06 & 0.11 \\
			\hline
			& $1/8$ & $1/4$ & $3/8$ & $1/2$ \\
		\end{tabular}
		
		\begin{tabular}{c|cccc}
			\multicolumn{5}{c}{(e) full-IR}\\
			$1/2$ & 0.18 & 0.16 & 0.11 & 0.05 \\
			$3/8$ & 0.09 & 0.07 & 0.03 & 0.04 \\
			$1/4$ & 0.03 & 0.01 & 0.03 & 0.08 \\
			$1/8$ & 0.004 & 0.02 & 0.06 & 0.11 \\
			\hline
			& $1/8$ & $1/4$ & $3/8$ & $1/2$ \\
		\end{tabular}
	}
	\caption{Values of $s_{ij}$ (defined in \eqref{eq:chiCELL}) in the transverse sector for $\mu_q/T=5$ and $\mu_3=0$.}
	\label{tab:app:s50_tr}
\end{table}

In table~\ref{tab:app:s50_tr}, all errors are much smaller than in the nearly extremal cases. Nevertheless, the extended hydrodynamic approximation still gives a clear improvement over the hydrodynamic approximation in the transverse sector. The improvement is most visible at larger $k/\mu$, while it becomes much weaker at the smallest value of $k/\mu$.

The improved approximation is very close to the extended hydrodynamic one, but it is usually slightly worse in the upper half of the table. The normalized and full approximation remain close to hydrodynamics. In the bottom row all approximations coincide. Therefore, at $\mu_q/T=5$ and $\mu_3=0$, the extended hydrodynamic approximation is the only refinement that gives a clear gain, and this gain is concentrated at larger $k/\mu$.

\subsection*{The longitudinal correlator}
\begin{table}[htb]
	\centering
	{
		\small
		\begin{tabular}{c|cccc}
			\multicolumn{5}{c}{(a) hydrodynamic}\\
			$1/2$ & 0.05 & 0.01 & 0.05 & 0.09 \\
			$3/8$ & 0.02 & 0.02 & 0.05 & 0.09 \\
			$1/4$ & 0.00 & 0.02 & 0.05 & 0.09 \\
			$1/8$ & 0.003 & 0.02 & 0.05 & 0.09 \\
			\hline
			& $1/8$ & $1/4$ & $3/8$ & $1/2$ \\
		\end{tabular}\hspace{1cm}
		\begin{tabular}{c|cccc}
			\multicolumn{5}{c}{(b) extended-hydrodynamic}\\
			$1/2$ & 0.05 & 0.06 & 0.08 & 0.12 \\
			$3/8$ & 0.03 & 0.04 & 0.07 & 0.11 \\
			$1/4$ & 0.02 & 0.03 & 0.06 & 0.10 \\
			$1/8$ & 0.01 & 0.02 & 0.05 & 0.09 \\
			\hline
			& $1/8$ & $1/4$ & $3/8$ & $1/2$ \\
		\end{tabular}
		
		\begin{tabular}{c|cccc}
			\multicolumn{5}{c}{(c) improved}\\
			$1/2$ & 0.04 & 0.05 & 0.07 & 0.11 \\
			$3/8$ & 0.03 & 0.04 & 0.06 & 0.10 \\
			$1/4$ & 0.02 & 0.03 & 0.05 & 0.10 \\
			$1/8$ & 0.01 & 0.02 & 0.05 & 0.09 \\
			\hline
			& $1/8$ & $1/4$ & $3/8$ & $1/2$ \\
		\end{tabular}\hspace{1cm}
		\begin{tabular}{c|cccc}
			\multicolumn{5}{c}{(d) normalized}\\
			$1/2$ & 0.04 & 0.02 & 0.05 & 0.10 \\
			$3/8$ & 0.01 & 0.02 & 0.05 & 0.10 \\
			$1/4$ & 0.01 & 0.02 & 0.05 & 0.09 \\
			$1/8$ & 0.003 & 0.02 & 0.05 & 0.09 \\
			\hline
			& $1/8$ & $1/4$ & $3/8$ & $1/2$ \\
		\end{tabular}
		
		\begin{tabular}{c|cccc}
			\multicolumn{5}{c}{(e) full-IR}\\
			$1/2$ & 0.05 & 0.01 & 0.05 & 0.09 \\
			$3/8$ & 0.01 & 0.02 & 0.05 & 0.09 \\
			$1/4$ & 0.00 & 0.02 & 0.05 & 0.09 \\
			$1/8$ & 0.00 & 0.02 & 0.05 & 0.09 \\
			\hline
			& $1/8$ & $1/4$ & $3/8$ & $1/2$ \\
		\end{tabular}
	}
	\caption{Values of $s_{ij}$ (defined in \eqref{eq:chiCELL}) in the longitudinal sector for $\mu_q/T=5$ and $\mu_3=0$.}
	\label{tab:app:s50_long}
\end{table}

In table~\ref{tab:app:s50_long}, the longitudinal sector is already very well described by hydrodynamics with errors are all below $10\%$. The extended hydrodynamic approximation does not improve the table; instead, it increases the error in many cells.

The improved approximation reduces the extended hydrodynamic errors slightly, but it still remains above hydrodynamics in most of the right part of the table. The normalized approximation stays much closer to hydrodynamics, and the full approximation is almost identical to hydrodynamics entry by entry.

\subsection{Background at $\mu_q/T=5$ and $\mu_3/\mu_q=-0.1$}
\subsection*{The transverse correlator}
\begin{table}[htb]
	\centering
	{
		\small
		\begin{tabular}{c|cccc}
			\multicolumn{5}{c}{(a) hydrodynamic}\\
			$1/2$ & 0.18 & 0.18 & 0.15 & 0.09 \\
			$3/8$ & 0.09 & 0.09 & 0.06 & 0.02 \\
			$1/4$ & 0.03 & 0.03 & 0.01 & 0.04 \\
			$1/8$ & 0.01 & 0.01 & 0.02 & 0.06 \\
			\hline
			& $1/8$ & $1/4$ & $3/8$ & $1/2$ \\
		\end{tabular}\hspace{1cm}
		\begin{tabular}{c|cccc}
			\multicolumn{5}{c}{(b) extended-hydrodynamic}\\
			$1/2$ & 0.06 & 0.10 & 0.10 & 0.06 \\
			$3/8$ & 0.03 & 0.05 & 0.04 & 0.02 \\
			$1/4$ & 0.01 & 0.02 & 0.01 & 0.04 \\
			$1/8$ & 0.01 & 0.00 & 0.02 & 0.06 \\
			\hline
			& $1/8$ & $1/4$ & $3/8$ & $1/2$ \\
		\end{tabular}
		
		\begin{tabular}{c|cccc}
			\multicolumn{5}{c}{(c) improved}\\
			$1/2$ & 0.03 & 0.05 & 0.04 & 0.02 \\
			$3/8$ & 0.03 & 0.01 & 0.02 & 0.05 \\
			$1/4$ & 0.04 & 0.04 & 0.06 & 0.10 \\
			$1/8$ & 0.05 & 0.06 & 0.08 & 0.12 \\
			\hline
			& $1/8$ & $1/4$ & $3/8$ & $1/2$ \\
		\end{tabular}\hspace{1cm}
		\begin{tabular}{c|cccc}
			\multicolumn{5}{c}{(d) normalized}\\
			$1/2$ & 0.24 & 0.23 & 0.19 & 0.14 \\
			$3/8$ & 0.15 & 0.14 & 0.10 & 0.05 \\
			$1/4$ & 0.09 & 0.08 & 0.05 & 0.02 \\
			$1/8$ & 0.06 & 0.05 & 0.02 & 0.02 \\
			\hline
			& $1/8$ & $1/4$ & $3/8$ & $1/2$ \\
		\end{tabular}
		
		\begin{tabular}{c|cccc}
			\multicolumn{5}{c}{(e) full-IR}\\
			$1/2$ & 0.18 & 0.17 & 0.13 & 0.08 \\
			$3/8$ & 0.09 & 0.08 & 0.04 & 0.02 \\
			$1/4$ & 0.03 & 0.02 & 0.02 & 0.06 \\
			$1/8$ & 0.00 & 0.01 & 0.04 & 0.08 \\
			\hline
			& $1/8$ & $1/4$ & $3/8$ & $1/2$ \\
		\end{tabular}
	}
	\caption{Values of $s_{ij}$ (defined in \eqref{eq:chiCELL}) in the transverse sector for $\mu_q/T=5$ and $\mu_3/\mu_q=-0.1$.}
	\label{tab:app:s501_tr}
\end{table}

In table~\ref{tab:app:s501_tr}, the transverse sector at $\mu_q/T=5$ and $\mu_3/\mu_q=-0.1$ we can see how the extended hydrodynamic approximation reduces the hydrodynamic error in the upper rows. At the smallest value of $k/\mu$, the improvement is much smaller.

The improved approximation is better than the extended approximation in the upper row. However, it becomes worse in the lower half of the table. The normalized approximation does not give better results here, since it is worse than hydrodynamics in almost every entry. The full approximation is instead close to hydrodynamics, with only small improvements in a few cells. Therefore, the extended approximation gives the best global description, while the improved approximation is advantageous only at large $k/\mu$.

\subsection*{The longitudinal correlator}
\begin{table}[htb]
	\centering
	{
		\small
		\begin{tabular}{c|cccc}
			\multicolumn{5}{c}{(a) hydrodynamic}\\
			$1/2$ & 0.09 & 0.02 & 0.04 & 0.07 \\
			$3/8$ & 0.05 & 0.03 & 0.04 & 0.06 \\
			$1/4$ & 0.03 & 0.02 & 0.03 & 0.06 \\
			$1/8$ & 0.01 & 0.01 & 0.02 & 0.05 \\
			\hline
			& $1/8$ & $1/4$ & $3/8$ & $1/2$ \\
		\end{tabular}\hspace{1cm}
		\begin{tabular}{c|cccc}
			\multicolumn{5}{c}{(b) extended-hydrodynamic}\\
			$1/2$ & 0.02 & 0.05 & 0.08 & 0.10 \\
			$3/8$ & 0.02 & 0.05 & 0.06 & 0.08 \\
			$1/4$ & 0.02 & 0.03 & 0.03 & 0.06 \\
			$1/8$ & 0.01 & 0.01 & 0.02 & 0.05 \\
			\hline
			& $1/8$ & $1/4$ & $3/8$ & $1/2$ \\
		\end{tabular}
		
		\begin{tabular}{c|cccc}
			\multicolumn{5}{c}{(c) improved}\\
			$1/2$ & 0.05 & 0.09 & 0.12 & 0.14 \\
			$3/8$ & 0.05 & 0.09 & 0.10 & 0.12 \\
			$1/4$ & 0.05 & 0.08 & 0.08 & 0.11 \\
			$1/8$ & 0.05 & 0.05 & 0.07 & 0.10 \\
			\hline
			& $1/8$ & $1/4$ & $3/8$ & $1/2$ \\
		\end{tabular}\hspace{1cm}
		\begin{tabular}{c|cccc}
			\multicolumn{5}{c}{(d) normalized}\\
			$1/2$ & 0.13 & 0.03 & 0.01 & 0.05 \\
			$3/8$ & 0.09 & 0.01 & 0.01 & 0.04 \\
			$1/4$ & 0.07 & 0.02 & 0.01 & 0.03 \\
			$1/8$ & 0.05 & 0.04 & 0.02 & 0.02 \\
			\hline
			& $1/8$ & $1/4$ & $3/8$ & $1/2$ \\
		\end{tabular}
		
		\begin{tabular}{c|cccc}
			\multicolumn{5}{c}{(e) full-IR}\\
			$1/2$ & 0.08 & 0.02 & 0.06 & 0.09 \\
			$3/8$ & 0.04 & 0.03 & 0.05 & 0.08 \\
			$1/4$ & 0.02 & 0.03 & 0.04 & 0.07 \\
			$1/8$ & 0.01 & 0.01 & 0.04 & 0.07 \\
			\hline
			& $1/8$ & $1/4$ & $3/8$ & $1/2$ \\
		\end{tabular}
	}
	\caption{Values of $s_{ij}$ (defined in \eqref{eq:chiCELL}) in the longitudinal sector for $\mu_q/T=5$ and $\mu_3/\mu_q=-0.1$.}
	\label{tab:app:s501_long}
\end{table}

In table~\ref{tab:app:s501_long}, no single approximation dominates the whole longitudinal grid. Hydrodynamics is already very accurate, with errors between 1\% and 9\%. The extended hydrodynamic approximation improves the first column at large $k/\mu$, reducing the upper-left entry from $9\%$ to $2\%$ and the second-row first entry from $5\%$ to $2\%$. However, it worsens the larger-$\omega/\mu$ entries in the upper rows.

The improved approximation is clearly the least useful among the refined approximations, since it raises the value of many entries. The normalized approximation has a complementary behavior: it is worse in the first column, but it is very good in the middle and right part of the table. The full approximation remains close to hydrodynamics but is slightly worse in the last two columns. Therefore, for this table, hydrodynamics is somewhat a balanced approximation; the extended approximation is useful only at small $\omega/\mu$ and large $k/\mu$, while the normalized approximation is useful mainly at intermediate and larger $\omega/\mu$.

\subsection{Background at $\mu_q/T=5$ and $\mu_3/\mu_q=-0.5$}

\subsection*{The transverse correlator}
\begin{table}[htb]
	\centering
	{
		\small
		\begin{tabular}{c|cccc}
			\multicolumn{5}{c}{(a) hydrodynamic}\\
			$1/2$ & 0.18 & 0.24 & 0.28 & 0.28 \\
			$3/8$ & 0.07 & 0.14 & 0.17 & 0.18 \\
			$1/4$ & 0.02 & 0.07 & 0.10 & 0.11 \\
			$1/8$ & 0.03 & 0.04 & 0.07 & 0.08 \\
			\hline
			& $1/8$ & $1/4$ & $3/8$ & $1/2$ \\
		\end{tabular}\hspace{1cm}
		\begin{tabular}{c|cccc}
			\multicolumn{5}{c}{(b) extended-hydrodynamic}\\
			$1/2$ & 0.08 & 0.19 & 0.24 & 0.25 \\
			$3/8$ & 0.05 & 0.12 & 0.16 & 0.17 \\
			$1/4$ & 0.03 & 0.08 & 0.11 & 0.12 \\
			$1/8$ & 0.02 & 0.06 & 0.09 & 0.09 \\
			\hline
			& $1/8$ & $1/4$ & $3/8$ & $1/2$ \\
		\end{tabular}
		
		\begin{tabular}{c|cccc}
			\multicolumn{5}{c}{(c) improved}\\
			$1/2$ & 0.20 & 0.11 & 0.07 & 0.06 \\
			$3/8$ & 0.22 & 0.17 & 0.14 & 0.13 \\
			$1/4$ & 0.24 & 0.20 & 0.18 & 0.17 \\
			$1/8$ & 0.25 & 0.22 & 0.20 & 0.20 \\
			\hline
			& $1/8$ & $1/4$ & $3/8$ & $1/2$ \\
		\end{tabular}\hspace{1cm}
		\begin{tabular}{c|cccc}
			\multicolumn{5}{c}{(d) normalized}\\
			$1/2$ & 0.51 & 0.55 & 0.55 & 0.52 \\
			$3/8$ & 0.38 & 0.42 & 0.42 & 0.39 \\
			$1/4$ & 0.30 & 0.34 & 0.34 & 0.32 \\
			$1/8$ & 0.27 & 0.30 & 0.31 & 0.28 \\
			\hline
			& $1/8$ & $1/4$ & $3/8$ & $1/2$ \\
		\end{tabular}
		
		\begin{tabular}{c|cccc}
			\multicolumn{5}{c}{(e) full-IR}\\
			$1/2$ & 0.13 & 0.16 & 0.16 & 0.13 \\
			$3/8$ & 0.03 & 0.05 & 0.06 & 0.04 \\
			$1/4$ & 0.04 & 0.01 & 0.01 & 0.03 \\
			$1/8$ & 0.07 & 0.04 & 0.04 & 0.05 \\
			\hline
			& $1/8$ & $1/4$ & $3/8$ & $1/2$ \\
		\end{tabular}
	}
	\caption{Values of $s_{ij}$ (defined in \eqref{eq:chiCELL}) in the transverse sector for $\mu_q/T=5$ and $\mu_3/\mu_q=-0.5$.}
	\label{tab:app:s505_tr}
\end{table}

In table~\ref{tab:app:s505_tr}, the transverse sector at $\mu_3/\mu_q=-0.5$ shows a clear advantage for the full approximation, although not in every single cell. The hydrodynamic approximation has moderate errors, and the extended hydrodynamic approximation gives a mild improvement in the upper-left cells. The normalized approximation is much worse than all the others.

The improved approximation has a very different structure: it improves the large-$\omega/\mu$ part of the upper row, but it becomes worse at small $\omega/\mu$ and smaller $k/\mu$. The full approximation gives the best overall result, and in the two middle rows it is especially accurate. The extended approximation remains slightly better in a few first-column entries, but the full approximation is clearly more reliable across the table as a whole.

\begin{table}[htb]
	\centering
	{
		\small
		\begin{tabular}{c|cccc}
			\multicolumn{5}{c}{(a) hydrodynamic}\\
			$1/2$ & 0.08 & 0.17 & 0.25 & 0.12 \\
			$3/8$ & 0.03 & 0.10 & 0.18 & 0.14 \\
			$1/4$ & 0.02 & 0.06 & 0.11 & 0.13 \\
			$1/8$ & 0.02 & 0.03 & 0.06 & 0.07 \\
			\hline
			& $1/8$ & $1/4$ & $3/8$ & $1/2$ \\
		\end{tabular}\hspace{1cm}
		\begin{tabular}{c|cccc}
			\multicolumn{5}{c}{(b) extended-hydrodynamic}\\
			$1/2$ & 0.04 & 0.13 & 0.22 & 0.11 \\
			$3/8$ & 0.02 & 0.09 & 0.18 & 0.14 \\
			$1/4$ & 0.02 & 0.07 & 0.12 & 0.13 \\
			$1/8$ & 0.02 & 0.05 & 0.08 & 0.07 \\
			\hline
			& $1/8$ & $1/4$ & $3/8$ & $1/2$ \\
		\end{tabular}
		
		\begin{tabular}{c|cccc}
			\multicolumn{5}{c}{(c) improved}\\
			$1/2$ & 0.21 & 0.12 & 0.05 & 0.17 \\
			$3/8$ & 0.21 & 0.15 & 0.09 & 0.18 \\
			$1/4$ & 0.21 & 0.17 & 0.13 & 0.18 \\
			$1/8$ & 0.21 & 0.19 & 0.17 & 0.19 \\
			\hline
			& $1/8$ & $1/4$ & $3/8$ & $1/2$ \\
		\end{tabular}\hspace{1cm}
		\begin{tabular}{c|cccc}
			\multicolumn{5}{c}{(d) normalized}\\
			$1/2$ & 0.34 & 0.41 & 0.47 & 0.25 \\
			$3/8$ & 0.28 & 0.34 & 0.40 & 0.23 \\
			$1/4$ & 0.25 & 0.28 & 0.32 & 0.22 \\
			$1/8$ & 0.23 & 0.26 & 0.26 & 0.21 \\
			\hline
			& $1/8$ & $1/4$ & $3/8$ & $1/2$ \\
		\end{tabular}
		
		\begin{tabular}{c|cccc}
			\multicolumn{5}{c}{(e) full-IR}\\
			$1/2$ & 0.04 & 0.10 & 0.15 & 0.10 \\
			$3/8$ & 0.01 & 0.04 & 0.09 & 0.13 \\
			$1/4$ & 0.03 & 0.01 & 0.02 & 0.12 \\
			$1/8$ & 0.05 & 0.03 & 0.02 & 0.07 \\
			\hline
			& $1/8$ & $1/4$ & $3/8$ & $1/2$ \\
		\end{tabular}
	}
	\caption{Values of $s_{ij}$ (defined in \eqref{eq:chiCELL}) in the longitudinal sector for $\mu_q/T=5$ and $\mu_3/\mu_q=-0.5$.}
	\label{tab:app:s505_long}
\end{table}

\subsection*{The longitudinal correlator}

In table~\ref{tab:app:s505_long}, the longitudinal sector again shows that the full approximation is the most effective one for $\mu_3/\mu_q=-0.5$. The hydrodynamic approximation has errors up to $25\%$, especially in the third column of the upper rows. The extended hydrodynamic approximation slightly improves the first column and some upper-row entries, but the improvement remains modest. The normalized approximation is clearly poor when compared with the others.

The improved approximation captures some of the large-$\omega/\mu$ behavior, lowering the upper-row third-column entry. However, it performs badly in the first column, and it is not uniformly better than hydrodynamics. The full approximation gives the best compromise. It reduces the upper row errors, and in the central lower part of the table it reaches very small errors. It is not always the best entry by entry, but it is the only approximation that keeps the errors small over most of the grid.

\newpage

\end{document}